# Using Open Data and Open-Source Software to Develop Spatial Indicators of Urban Design and Transport Features for Achieving Healthy and Sustainable Cities

Geoff Boeing, Carl Higgs, Shiqin Liu, Billie Giles-Corti, James F Sallis, Ester Cerin, Melanie Lowe, Deepti Adlakha, Erica Hinckson, Anne Vernez Moudon, Deborah Salvo, Marc A Adams, Ligia Vizeu Barrozo, Tamara Bozovic, Xavier Delclòs-Alió, Jan Dygrýn, Sara Ferguson, Klaus Gebel, Thanh Phuong Ho, Poh-Chin Lai, Joan Carles Martori, Kornsupha Nitvimol, Ana Queralt, Jennifer D Roberts, Garba H Sambo, Jasper Schipperijn, David Vale, Nico Van de Weghe, Guillem Vich, and Jonathan Arundel

**Abstract.** Benchmarking and monitoring urban design and transport features is critical to achieving local and international health and sustainability goals. However, most urban indicator frameworks use coarse spatial scales that only allow between-city comparisons or require expensive, technical, local spatial analyses for within-city comparisons. This study developed a reusable open-source urban indicator computational framework using open data to enable consistent local and global comparative analyses. We demonstrate this framework by calculating spatial indicators—for 25 diverse cities in 19 countries—of urban design and transport features that support health and sustainability. We link these indicators to cities' policy contexts and identify populations living above and below critical thresholds for physical activity through walking. Efforts to broaden participation in crowdsourcing data and to calculate globally consistent indicators are essential for planning evidence-informed urban interventions, monitoring policy impacts, and learning lessons from peer cities to achieve health, equity, and sustainability goals.

## 1. Introduction

The policies that determine cities' urban form, land use patterns, and transport opportunities in turn determine health and sustainability. Creating healthier and more sustainable cities is a global priority integral to achieving the United Nations' (UN) Sustainable Development Goals (SDGs) and the World Health Organization's (WHO) health equity goals.[1] Various indicator frameworks have been proposed to monitor progress towards such ends. However, most existing frameworks rely on citywide measures and focus on comparisons *between* cities. While such comparisons are useful for determining priorities and interventions at the international and national levels, *within*-city (i.e., neighbourhood-level) comparisons are key to unlocking the full potential of city planning to unmask and attenuate local urban health inequities.[2,3]

> **Box 1: Limitations of current guidelines for developing consistent spatial indicators**
>
> There are various guidelines for developing health and sustainability indicators, but each has limitations. The International Organization for Standardization (ISO) is an independent non-governmental organisation established in 1946 to develop consensus-based, market-relevant international standards to help solve global challenges. Its ISO37120 standard proposes indicators for city services and quality of life that can track and monitor city performance over time.[5] ISO advocates for standardised indicators to provide a uniform approach to what is measured and how that measurement is undertaken. However, its framework stops short of recommending targets, and it does not deal with the difficulties involved in sourcing necessary (and consistent) data. Further, the indicators tend to be citywide rather than neighbourhood-scale, limiting capacity for within-city comparisons.
>
> For example, the ISO indicator for public transport access (*19.6 – Percentage of population living within 0.5 km of public transit running at least every 20 minutes during peak periods*) suggests a single citywide measure and does not stipulate potential data sources or a target for the percentage of the population served at this level. The indicator makes no reference to the spatial boundary used to determine the population catchment, nor does it stipulate whether the distance threshold for evaluating access should be measured using Euclidean distance or transport network traversal. Depending on the network's connectivity, Euclidean distance often overestimates access. Thus, two analysts independently measuring the same city at the same time may generate substantially different results.
>
> The UN Habitat's New Urban Agenda Monitoring Framework is more comprehensive and addresses barriers to consistently measure and compare cities.[6] Taking a similar public transport indicator as an example (*14. Proportion of the population that has convenient access to public transport disaggregated by age group, sex, and persons with disabilities*), UN Habitat provides seven pages of guidance on how the indicator should be constructed and possible data sources. However, it suggests only a citywide measure and remains silent on targets.

**1.1. Within-city versus between-city spatial indicators**

Maps can help local and regional planners reveal the spatial distribution of health-promoting infrastructure and amenities within cities (e.g., walkable streets, public transport, daily living needs, green spaces) and identify inequities in access. Mapped neighbourhood-level spatial indicators facilitate comparisons within cities, highlight resource distribution and areas needing interventions, encourage accountability, and empower communities to advocate for improvements.[4] Growing access to big data and high-powered computing enables neighbourhood-level spatial indicators to be developed and disseminated more readily.

Urban policy targets are often set at a citywide level (e.g., percentage of city population with access to amenities).[3] Neighbourhood-level spatial indicators help planners identify differences in access to urban design and transport features that support healthy and sustainable lifestyles, to better target local interventions. However, planners also need a means of aggregating consistently measured neighbourhood-level spatial indicators to the city scale, in order to compare between cities (benchmarking) and over time (monitoring). These are crucial first steps towards achieving urban health and sustainability goals. Nevertheless, many prominent indicator guidelines do not address measurement standards, indicator targets, or data acquisition (see Box 1).

**1.2. Creating globally applicable city planning spatial indicators**

Creating high-quality, fine-grained spatial indicators to measure progress towards healthy and sustainable cities globally presents technical challenges that plague both between- and within-city comparisons.[7-9] While some cities collect and maintain high-quality,[10] fine-grained data on land use, transport infrastructure, and socioeconomic characteristics, many do not. Even when such data exist, they may not be publicly available to researchers and practitioners. Researchers undertaking comparative analyses—particularly international analyses—must account for region- and dataset-specific inconsistencies in assumptions, standards, scales, and timeliness. Data quality varies widely, as do digitisation standards and encodings, collection dates, local meanings of transport infrastructure or land use classifications, and spatial scales (e.g., defining the "city" as an incorporated municipality, urbanised area, or metropolitan area).[11]

Lack of access to software, training, and resources also constrains indicator creation. Closed-source, proprietary geographic information systems (GIS) are often expensive, do not lend themselves to open science or reproducibility, and fit poorly with modern data science practices.[12] Advanced spatial analysis requires extensive training, but such expertise can be uncommon in government agencies and difficult or expensive to procure from the private sector. Resource constraints pose a particular challenge in low- and middle-income countries (LMICs), where data availability and quality and local technical capacity may be lacking. These limitations thwart efforts to develop actionable indicators to track the creation of healthy and sustainable urban environments in our planet's most rapidly developing cities and constrain governments' capacity to develop evidence-based policies and monitor their impacts.[13]

**1.3. A 21st-century approach to calculating spatial indicators**

Given the importance of urban spatial indicators to benchmark and monitor cities and inform interventions, a better model for creating indicators would leverage emerging open-source software and open-data commons to build high-quality, accessible, free tools for calculating and visualising such indicators. Open-data sources with global scope offer opportunities to measure and analyse urban health and sustainability indicators in diverse geographical contexts.[14-18] In particular, OpenStreetMap (OSM) is a crowdsourced mapping project that provides open access to regularly updated spatial data worldwide, coded according to consistent and community-led guidelines.[19]

This article addresses the need to better measure, map, and compare urban design and transport features important for creating healthy and sustainable cities. We present an open-source software framework that uses open data to calculate spatial indicators within and between cities around the world, including in under-studied and under-resourced countries. Then we demonstrate the feasibility and utility of our approach by creating a cross-sectional snapshot of priority indicators recommended in the first Lancet Series on Urban Design, Transport and Health, showing between-city comparisons, and mapping within-city spatial inequities.[20] We link these to the local policy contexts identified by Lowe et al.[21] and identify populations living above and below the critical thresholds identified by Cerin et al.[22] We discuss the practical value of this tool and empirical findings for policy-making. The article concludes with a call for action: to build healthy and sustainable cities, we must measure city-building better and we must build healthy and sustainable cities for all—not just for some—by reducing within-city inequities.

## 2. Measuring spatial indicators of urban design and transport features for healthy and sustainable cities

### 2.1. International collaboration network

The Global Healthy and Sustainable City-Indicators Collaboration comprises a network of built environment and health researchers, formed to develop a framework for assessing the progress of urban design and transport features that support healthy and sustainable cities. The network comprises a core team of multidisciplinary researchers working with local experts, including academics and city planners, across 25 cities in 19 lower-middle- to high-income countries across six world regions (see the appendix for details). Lowe et al.[21] describe the characteristics and sampling methods for these cities. Investigators were contacted through international networks and at conferences to volunteer to lead participation of cities.

### 2.2. Open-source framework

We developed an open-source software framework to calculate spatial indicators using open data at both fine-grained and aggregated levels, supporting within- and between-city comparisons, as described in Liu et al.[23] Detailed descriptions of these methods, developed in conjunction with collaborators, appear in the appendix, including urban study region boundary definitions, source data to support comparable analyses of what we define as "local neighbourhood" features across the cities, and reproducible workflows for indicator estimation. We use the term "neighbourhood" here in a technical sense, referring to a walkable catchment within some distance threshold of a residential reference point, rather than the colloquial sense of social or political boundaries. We generally defined urban study regions using city administrative boundaries and the Global Human Settlements 2015 urban centres layer[24] and we derived pedestrian-accessible street networks and built environment features from OSM, with validation conducted with local collaborators. To support between-city analyses of urban neighbourhoods, we generated a 250-metre (m) grid associated with 2015 population estimates[25] to summarise the indicators' distribution at high-resolution for mapping with regard to population.

For each city, we calculated spatial indicators of urban design and transport features that support healthy and sustainable cities.[20,26] The following methods and definitions are detailed in the appendix. Indicators were calculated for sample points generated at regular 30m intervals along the pedestrian street network for populated areas in each city's urban region. These sample points represent an assumed spatial distribution of dwellings in each city to facilitate the measurement of local neighbourhood characteristics. The 1000m (approximately 13-minute walking time[27]) extent of the pedestrian network reachable from each sample point was intersected with the 250m urban neighbourhood grid to represent a local walkable catchment area: a computationally tractable approximation of the "sausage buffer" method for walkable catchments[28,29] as described in Liu et al.[23] We estimated population and street intersection densities in each sample point's local walkable catchment. The nearest distance to several features—healthy food markets, convenience stores, public transport stops, and public open space entry points—was estimated for each sample point and evaluated against an accessibility threshold[5] of 500m using a binary access score (equal to 1 if the estimated access distance was within 500m, and 0 otherwise). Access to public transport stops was evaluated against three criteria: 1) any; and where transport schedule data were retrievable, 2) average weekday daytime service every 30 minutes or less, and 3) serviced every 20 minutes or less. Two kinds of public open spaces were measured: 1) any; and 2) larger than 1.5 hectares. These public transport and open space typologies are associated with active transport behaviours, following those measured by Arundel et al.[4] We summarised each sample point's local walkable environment using two composite indicators: 1) a daily living score[4,30,31] for local access to a mix of amenities (summing equal-weighted binary access scores to a healthy food market, a convenience store, and a public transport stop within 500m); and 2) a local walkability index[4,31] that sums equal-weighted standardised scores of population density, street intersection density, and daily living score. These measures of well-serviced and walkable neighbourhoods are standard in the built environment and health literature[32,33] with well-established associations with physical activity and walking for transport.[4,30,31]

Residential point measures were aggregated and averaged to 250m hexagonal cells serving as empirically derived "neighbourhoods." These urban neighbourhoods were the spatial units used to characterise the within- and between-city distribution of indicators 1) in absolute terms relative to the city (within-city estimates), and 2) relative to all cities via z-scores (between-city estimates). We conducted a spatial analysis of population and intersection density indicators for two physical activity scenarios identified by Cerin et al.[22] Scenario A meets the threshold for 80% probability of walking for transport and Scenario B meets the threshold for reaching WHO's target of a ≥15% relative reduction in insufficient physical activity through walking.

## 3. How cities performed against the indicators

### 3.1. Population access to amenities

Table 1 presents estimates of the population percentage within a 500m walk to amenities, alongside citywide estimates of 2015 transport-sector particulate matter (PM2·5) emissions, where available.[24] Broad variation exists between cities, but those in middle-income countries tended to have lower estimates of population access to amenities than cities in high-income countries. In contrast, transport-sector PM2·5 emissions were higher in cities of middle-income countries (1621·8 tonnes/year on average) than high-income countries (333·1 tonnes/year).

The population percentage within a 500m walk to a healthy food market varied from 6% (Phoenix) to 70% (Bern). European cities had the highest estimates (53% on average), while all three US cities were in the lowest quartile, each below 20%. The three Australian cities, along with Maiduguri, Bangkok, and Chennai, also exhibited low access (less than a quarter of the population with such access). On average, access to convenience stores within 500m (40%) was slightly higher than for healthy food markets (36%). This was particularly the case for cities with low access to healthy food. For example, 21% of the Phoenix population had a convenience store within 500m, more than three times greater than for healthy food. An exception to this pattern was Chennai, with an estimate of 16% for convenience stores, compared with 20% for healthy food. It is likely that not all existing healthy food market locations were available through OSM, particularly in cities like Bangkok, Chennai, Hanoi, and Maiduguri where informal stalls may be important sources of healthy food.

Access to any public transport stop (e.g., bus, ferry, train, tram) within 500m was achieved for more than 60% of the population in most cities, with three cities in middle-income countries as exceptions: Maiduguri (10%), Mexico City (36%) and Chennai (39%). However, in these cities the local transport context must be considered, as informal collective transport options play an important role but lack spatial data to track them. Nevertheless, the disparity between estimated access to formal public transport infrastructure in middle- and high-income countries was notable and may be a factor in the observed trend of approximately five-fold higher PM2·5 emissions in middle-income versus high-income countries' cities, notwithstanding considerable between-city variation for both groups (Table 1).

Accounting for public transport service frequency for cities where such data were available substantially reduced estimates of accessibility. The average population percentage with access to any stops with service every 30 minutes was 70% (standard deviation [s.d.] 27%). For service every 20 minutes the average city estimate fell to 45% (s.d. 23%). For example, 87% of Melbourne's population had access to any public transport, but only 67% to stops with weekday service every 30 minutes and only 49% with service every 20 minutes—below the average for high-income countries' cities (55% [s.d. 15%]). In other cities there were modest reductions in population access when adjusting for service frequency, indicating broad consistency in proximity and service frequency. For example, São Paulo, Bern, and Lisbon had greater than 90% population access to public transport with average weekday service frequency of 20 minutes or less.

For most cities, the average population percentage with access to public open space within 500m walking distance was relatively high at 76% (s.d. 22%). Some policy settings mandate universal access and any estimates below 100% suggest equity gaps that need addressing. Some cities had low estimates (e.g., Maiduguri and Bangkok), although this may reflect limitations of measuring public open space in different cities using OSM data. Once access was restricted to public open spaces larger than 1.5 hectares, population access dropped by approximately 10 percentage points on average, to 66% (s.d. 26%). However, in the European cities of Bern, Vic, and Odense, 70% or more of the population had access within 500m to such larger public open space. Substantial inequities in access to public open space within 500m were apparent between cities in middle- and high-income countries: only 42% of the population in the former's cities had access, compared with 75% in the latter's cities.

**Table 1.** Population percentage estimates for proximal access to amenities.

| Country | City | Estimated population percent with access within 500m walking distance to a... | | | | | | | Total emission of PM 2·5 from the transport sector in 2015 (tonnes/year) |
|---|---|---|---|---|---|---|---|---|---|
| | | Healthy food market | Convenience store | Public transport stop (OSM or GTFS) | Public transport stop with regular service (30 mins) | Public transport stop with regular service (20 mins) | Public open space | Public open space larger than 1·5 hectares | |
| **Africa** | | | | | | | | | |
| NGA | Maiduguri | 17·4 | 27·4 | 9·6 | – | – | 1·9 | 0·5 | 7·5 |
| **America, North** | | | | | | | | | |
| MEX | Mexico City | 26·4 | 22·7 | 35·8 | 24·7 | 19·7 | 49·6 | 19·7 | 532·0 |
| USA | Baltimore | 14·3 | 29·1 | 63·1 | 51·3 | 42·8 | 62·5 | 39·2 | 324·8 |
| | Phoenix | 5·6 | 21·0 | 66·0 | 61·6 | 24·1 | 36·5 | 24·6 | 268·3 |
| | Seattle | 15·5 | 26·1 | 60·3 | 36·3 | 26·6 | 59·2 | 35·0 | 316·3 |
| **America, South** | | | | | | | | | |
| BRA | São Paulo | 35·2 | 36·7 | 96·1 | 95·7 | 94·2 | 71·7 | 15·5 | 2306·5 |
| **Asia** | | | | | | | | | |
| HKG | Hong Kong | 48·9 | 52·2 | 89·5 | 86·9 | 83·6 | 86·9 | 54·1 | 1903·7 |
| IND | Chennai | 19·7 | 15·6 | 39·1 | 3·2 | 3·2 | 41·1 | 11·3 | 657·9 |
| THA | Bangkok | 15·3 | 33·4 | 63·0 | 62·1 | 43·2 | 14·1 | 6·5 | 4163·9 |
| VNM | Hanoi | 38·0 | 46·4 | 65·5 | 21·9 | 11·2 | 26·7 | 14·1 | 2062·8 |
| **Australasia** | | | | | | | | | |
| AUS | Adelaide | 18·8 | 19·9 | 89·2 | 81·9 | 53·7 | 87·3 | 58·0 | 147·4 |
| | Melbourne | 20·7 | 29·6 | 86·7 | 67·2 | 49·4 | 88·2 | 63·3 | 1364·0 |
| | Sydney | 22·3 | 28·7 | 94·7 | 78·4 | 57·7 | 90·1 | 60·3 | 564·8 |
| NZL | Auckland | 31·2 | 47·9 | 91·0 | 81·4 | 55·7 | 90·6 | 64·9 | 340·3 |
| **Europe** | | | | | | | | | |
| AUT | Graz | 62·6 | 56·1 | 92·2 | – | – | 84·9 | 39·5 | 14·4 |
| BEL | Ghent | 49·5 | 44·1 | 86·5 | – | – | 92·7 | 62·7 | 47·4 |
| CZE | Olomouc | 37·2 | 43·7 | 88·8 | – | – | 90·4 | 46·0 | 2·6 |
| DNK | Odense | 43·7 | 36·1 | 84·4 | 66·1 | 59·0 | 92·9 | 73·4 | 3·9 |
| DEU | Cologne | 51·1 | 56·9 | 79·0 | 71·7 | 60·2 | 89·6 | 65·8 | 158·9 |
| PRT | Lisbon | 64·2 | 60·7 | 97·0 | 95·7 | 92·8 | 90·1 | 51·3 | 208·1 |
| ESP | Barcelona | 63·8 | 61·6 | 91·4 | 78·3 | 75·8 | 88·2 | 62·8 | 186·1 |
| | Valencia | 59·7 | 48·1 | 81·6 | 78·3 | 77·2 | 78·4 | 43·8 | 105·0 |
| | Vic | 50·7 | 40·1 | 57·7 | – | – | 81·4 | 74·8 | – |
| CHE | Bern | 69·3 | 73·5 | 94·8 | 93·6 | 91·8 | 98·9 | 80·0 | 10·3 |
| GBR | Belfast | 29·0 | 47·8 | 92·9 | 82·9 | 72·6 | 65·0 | 46·8 | 29·4 |
| **Mean (standard deviation) summaries of city results by country income group** | | | | | | | | | |
| **Middle income** | | 25·3 (9·5) | 30·4 (10·9) | 34·3 (36·7) | 34·2 (25·3) | 11·3 (6·9) | 51·5 (29·9) | 41·5 (37·0) | 1621·8 (1539·4) |
| **High income** | | 39·9 (19·9) | 43·3 (15·0) | 61·5 (21·0) | 81·8 (15·4) | 55·1 (14·7) | 83·5 (12·4) | 74·1 (15·8) | 333·1 (496·9) |
| **Total** | | 36·4 (18·9) | 40·2 (15·0) | 54·7 (27·5) | 70·4 (27·2) | 44·6 (23·2) | 75·8 (22·3) | 66·0 (26·1) | 655·3 (1014·3) |

**Table 2.** Percentage of population meeting or exceeding scenario thresholds for spatial indicators that support physical activity goals.

| Country | City | Local population per km² | | Local street intersections per km² | |
|---|---|---|---|---|---|
| | | *Scenario A, meeting or exceeding target threshold 95% CrI [4790, 6750]* | *Scenario B, meeting or exceeding target threshold 95% CrI [5677, 7823]* | *Scenario A, meeting or exceeding target threshold 95% CrI [90, 110]* | *Scenario B, meeting or exceeding target threshold 95% CrI [106, 156]* |
| **Africa** | | | | | |
| NGA | Maiduguri | 98·0 | 95·9 | 45·6 | 28·5 |
| **America, North** | | | | | |
| MEX | Mexico City | 98·9 | 98·1 | 89·6 | 78·6 |
| USA | Baltimore | 39·6 | 28·0 | 64·8 | 51·7 |
| | Phoenix | 30·1 | 15·7 | 74·4 | 51·0 |
| | Seattle | 10·9 | 6·4 | 61·3 | 43·2 |
| **America, South** | | | | | |
| BRA | São Paulo | 99·6 | 99·4 | 88·0 | 70·4 |
| **Asia** | | | | | |
| HKG | Hong Kong | 98·3 | 97·7 | 95·7 | 91·5 |
| IND | Chennai | 99·8 | 99·7 | 90·4 | 79·3 |
| THA | Bangkok | 98·2 | 97·0 | 61·5 | 39·7 |
| VNM | Hanoi | 95·7 | 93·0 | 67·6 | 56·3 |
| **Australasia** | | | | | |
| AUS | Adelaide | 3·7 | 0·0 | 38·4 | 12·6 |
| | Melbourne | 33·4 | 17·8 | 37·8 | 20·8 |
| | Sydney | 67·5 | 51·0 | 24·9 | 13·4 |
| NZL | Auckland | 47·9 | 22·3 | 26·6 | 14·5 |
| **Europe** | | | | | |
| AUT | Graz | 64·0 | 44·1 | 92·5 | 81·3 |
| BEL | Ghent | 0·0 | 0·0 | 67·5 | 54·8 |
| CZE | Olomouc | 0·0 | 0·0 | 69·0 | 54·2 |
| DNK | Odense | 6·0 | 0·0 | 94·8 | 85·3 |
| DEU | Cologne | 47·5 | 21·6 | 83·9 | 71·6 |
| PRT | Lisbon | 98·1 | 96·8 | 99·7 | 98·6 |
| ESP | Barcelona | 95·5 | 92·3 | 82·6 | 74·9 |
| | Valencia | 97·8 | 95·8 | 78·6 | 72·3 |
| | Vic | 47·1 | 24·3 | 65·3 | 56·4 |
| CHE | Bern | 82·1 | 58·3 | 99·3 | 98·2 |
| GBR | Belfast | 59·6 | 40·1 | 91·1 | 74·0 |
| **Mean (standard deviation) summaries of city results by country income group** | | | | | |
| | *Middle income* | 98·4 (1·5) | 97·2 (2·5) | 73·8 (18·5) | 58·8 (21·1) |
| | *High income* | 48·9 (35·0) | 37·5 (35·3) | 71·0 (24·1) | 59·0 (28·1) |
| | *Total* | 60·8 (37·2) | 51·8 (40·2) | 71·6 (22·6) | 58·9 (26·2) |

*Scenario A*: meeting threshold for 80% probability of walking for transport. *Scenario B*: meeting threshold for reaching the WHO's target of a ≥15% relative reduction in insufficient physical activity through walking

### 3.2. Percentage of population meeting thresholds for urban design and transport features to support active lifestyles

Cerin et al.[22] identified thresholds to support active lifestyles and achieve WHO physical activity targets. On average, less than half of the population (49%) in high-income countries' cities lived in neighbourhoods reaching the population density thresholds for 80% probability of walking for transport or meeting WHO's target for reducing insufficient physical activity through walking (38%), compared with 98% of middle-income countries' cities. Cities with the highest estimated population percentages living in neighbourhoods with population densities that support higher transport walking (see Table 2) were in Africa (Maiduguri), Asia (Bangkok, Chennai, Hanoi, Hong Kong) and Latin America (Mexico City, São Paulo). These all exceeded 90% of the population for both scenarios A and B, as did the Iberian cities of Lisbon, Barcelona, and Valencia. The European cities of Belfast and Graz and the Australian city of Sydney exceeded 50% for at least one of the two scenarios. However, the European cities of Ghent, Odense, and Olomouc and the Australian city of Adelaide had notably low population density estimates that met neither population threshold. In North American (Baltimore, Phoenix, Seattle) and other Australasian cities (Auckland, Melbourne) along with Cologne and Vic in Europe, less than 50% of the population met density thresholds. For Scenario B, on average 97% (s.d. 3%) of the population of cities in middle-income countries lived in neighbourhoods supporting densities meeting or exceeding the lower threshold for reaching WHO's target of a ≥15% relative reduction in insufficient physical activity through combined transport and recreational walking, compared with 38% (s.d. 35%) for high-income countries' cities, across which population density was more variable.

Estimates of the percentage of population living in neighbourhoods with intersection density meeting thresholds to support higher physical activity displayed a distinct pattern for many cities when compared with the marginal population density characteristics summarised above. There were not clear differences in intersection density estimates by country income classification, as observed with population density, with broad variation between cities regardless of country income grouping. Cities with high population estimates (>70% for both scenarios) of meeting or exceeding thresholds for population density and intersection density included Mexico City, São Paulo, Hong Kong, Chennai, Lisbon, Barcelona, and Valencia. Cities with moderate estimates for neighbourhood population density (>40%) but high for intersection density (>70%) included European cities of Bern, Belfast, and Graz. Cities with lower population density estimates, but moderate to high population exposure (>50%) to intersection density, included North American cities of Baltimore and Phoenix, but also European cities of Ghent, Olomouc, Odense, Cologne, and Vic. Seattle had lower population percentage exposure estimates for both population density and intersection density than the other two US cities in this study. Maiduguri, Bangkok, and Hanoi each had high neighbourhood population density exposures but moderate to low intersection density exposure. In Australasia, Sydney had moderate neighbourhood population density exposure but lower exposure for intersection density, while Auckland and Melbourne had lower population densities and lower intersection density population exposure estimates, and Adelaide had both low population density and intersection density (<40%).

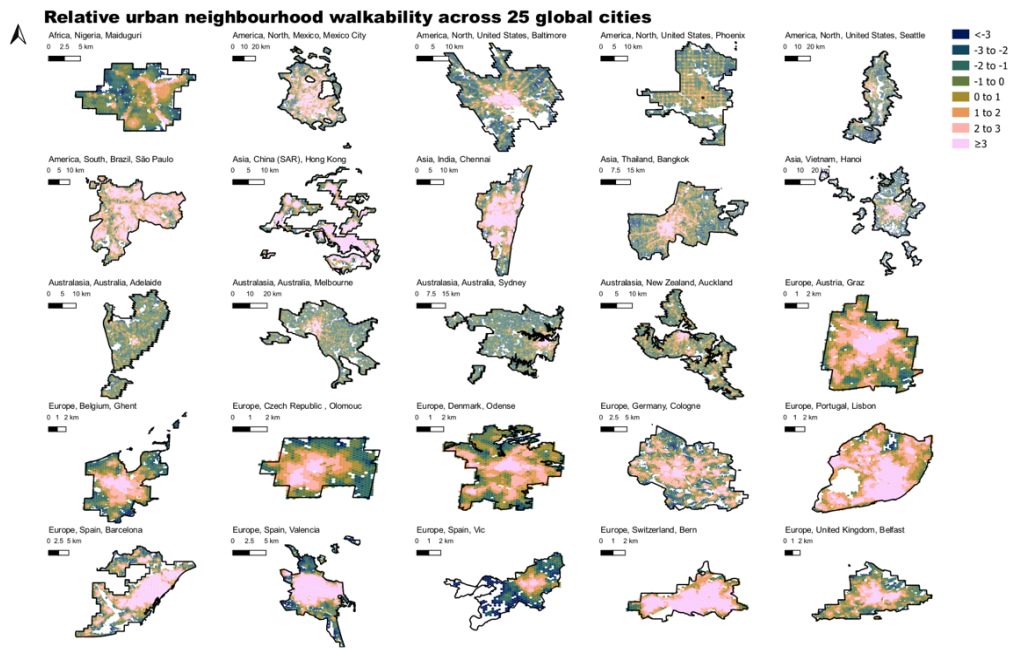

**Figure 1.** Relative walkability in 25 cities. Scale indicates standard deviations with zero as the global average.

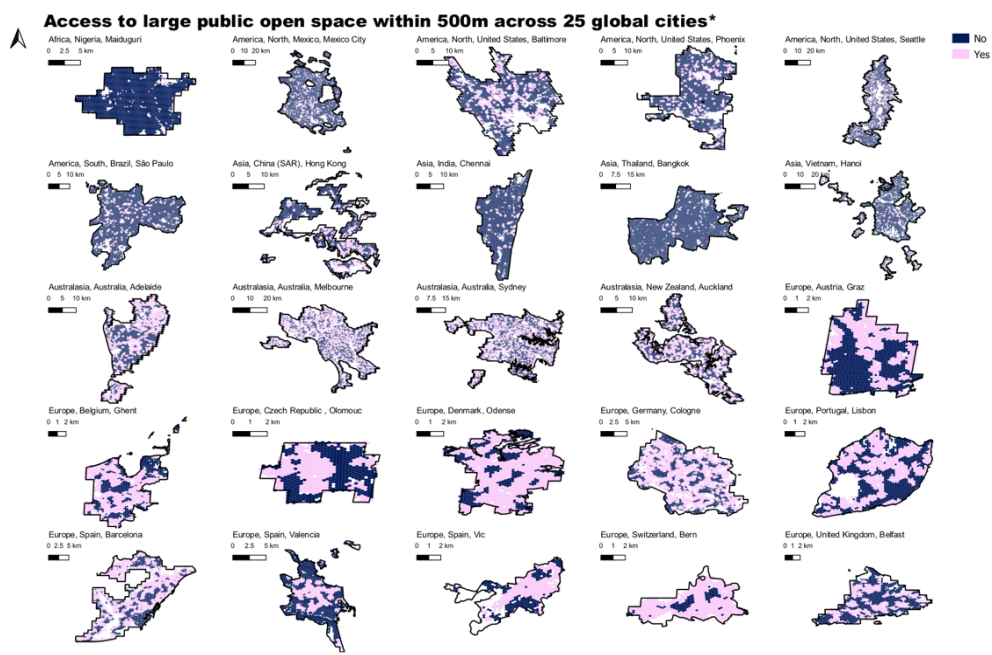

**Figure 2.** Access to large public open space within 500 m across 25 global cities. Access for urban neighbourhoods was considered achieved when at least half of the sampled walkable area was estimated to be located within 500 m of areas identified as public open space 1·5 hectares or larger.

### 3.3. Spatial distribution of walkable neighbourhoods and access to public open space

Neighbourhood-level results show spatial patterning and inequities within cities. Figure 1 presents the spatial distribution of urban walkability for the 25 cities. Access to large public open space within 500m (see Figure 2) offers a different conceptualisation of spaces that provide opportunities for physical activity—in addition to health benefits from nature, social connectedness, and heat-island mitigation—and demonstrates different spatial patterns than walkability. Achieving policy goals of access to both walkable neighbourhoods with local destinations and public open space requires evidence-driven city planning to reach a balance and prevent unintended negative consequences for the health and wellbeing of residents.[34-37]

Additional maps and visualisations of the population percentages meeting thresholds to support active and sustainable lifestyles for each city are included in the appendix. To summarise, while we identified walkable neighbourhoods across all cities, Australasian and US cities in particular exhibited sprawl outside their cores. Most of the cities in middle-income countries and the US were poorly served by public open space compared with the European and Australasian cities. The spatial distribution data from the project have been made publicly available and provide an opportunity for researchers to conduct their own analyses, which may be supplemented by using city-specific covariate data (e.g., sociodemographic characteristics, air pollutant distributions) which may not be publicly available globally.

## 4. An indicator framework for better city planning

This study developed an open-source urban indicator computational framework using open-data sources supporting international comparative analyses. It demonstrated applications of this framework by calculating spatial indicators of built environment features for 25 diverse cities. These data and this analytical workflow enabled comparisons of within- and between-city performance on the indicators. In general, the data were available and useful in all but the lowest-income cities, so broadening participation in crowdsourced data is essential to worldwide efforts to monitor urban indicators.

Many people do not have access to the urban design and transport features needed for healthy and sustainable cities. Our results show that older compact cities had better walkability, irrespective of economic development status. The worst-performing cities for walkability were in high-income countries including the US, Australia, and New Zealand. These cities developed primarily in the 20th century under a car-centric planning model.[38] Lowe et al. similarly found that Australian and US cities were the most likely to have contemporary urban design and transport policies that favour car use over active transport.[21] However, investment in public transport has made it a viable alternative to the car in some Australasian cities,[4] while US cities have fallen behind by global standards. In less-resourced cities, the economic necessity of providing mobility may take precedence over intersectoral planning that protects and promotes healthy lifestyles. While walking for transport is important, access to public open space is also critical for health and wellbeing,[39] particularly with a changing climate. Four of the five worst-performing cities for access to public open space were in middle-income countries, joined by the car-dominated city of Phoenix in the US. Inadequate policy frameworks and gaps between policy and implementation likely contributed to unequal access to health-supporting urban design and transport features.[21] Access inequities revealed by these spatial analyses point to areas requiring policy interventions to reduce health inequities between and within cities. Creating high-quality, fine-grained spatial indicators that incorporate evidence-based targets facilitates comparison not just between cities, but within cities. This evidence in turn provides a foundation for planning future interventions, monitoring policy impact, and harnessing lessons from comparable countries.

For example, São Paulo and Bangkok have similar populations and large proportions of residents in informal settlements. Yet more than 95% of São Paulo's residents were estimated to have access to frequent public transport, compared with 62% of Bangkok's. More generally, and across all indicators reported, São Paulo outperformed Bangkok. These divergences are best understood by looking not just at the between-city results, but the within-city results. Figure 1 shows, for São Paulo, that relative urban walkability was high across the majority of the urban area, whereas Bangkok achieved high walkability in the central city only. In São Paulo, the areas where walkability was lower followed the paths of the two major rivers that pass through the city, adjoined by highways and industrially zoned land. In Bangkok, outside the central city, walkability was achieved only in the areas near the highways that radiate outwards like spokes from the central hub. Between these spokes were the areas with the least walkable access to local amenities. Consistent with our findings, Lowe et al. found that Bangkok had the greatest policy gaps and limitations, while São Paulo's policy framework performed better than those of many cities in high-income countries.[21] The critical links between transport, land use, and health equity need to be recognised in future iterations and implementations of Bangkok's city plans, developed under the holistic approach advocated for by Peraphan and Sittha.[40] This applies similarly in other countries—including low-density, car-centric cities in high-income countries such as the US and Australia.

**4.1. Call for action**

This study demonstrated the feasibility of producing comparative spatial indicators to benchmark cities on urban design and transport features important for public health and sustainability. The workflow for creating the indicators has inherent value but is most useful if the urban policy and research community uses the open-source framework to continue monitoring cities' progress towards health and sustainability goals with periodic indicator scorecards. Regional, national, and global agencies can play important roles in incentivising such work, particularly when data collection is required to fill open-data gaps. Our open data and open-source methods allow anyone to freely replicate this study. The open-source philosophy posits that communities of research and practice should collectively build and share tools rather than develop individual ad hoc scripts that produce incomparable indicators. A potential benefit of using the methods presented here would be that consistent measures could be created and compared at different points in time with few barriers to participation.

Open data and open-source tools together create an opportunity that, for the first time, enables built environment, health, and policy researchers to quantify and monitor the progress of their city and compare local results across cities globally—if the data and tools are of sufficient quality. Data availability and quality vary both between and within cities, and this is a particular concern when using open data.[41,42] We therefore developed methods

> **Box 2: Global call for action**
>
> We urge the UN and WHO to:
> - incentivise and promote open data and open-source tools in the pursuit of meeting health, sustainability, and equity goals
> - support expansion of our indicators into a global observatory of within- and between-city indicators
> - provide practical guidance on identifying barriers to walking and wheelchair access
>
> We urge local, regional, and national governments to:
> - use the open-source framework we presented to create consistent measures that can inform policy and monitor cities' progress towards health and sustainability goals, using indicator scorecards to provide regular feedback and prompt targeted interventions
> - involve local populations in crowdsourcing data about their own communities
> - provide resources to fill gaps of missing data
> - map barriers to walking access to refine indicators
> - work together with academics and share best practices
> - use open commons and emerging standard open platforms like OSM to collect and contribute data
>
> We urge scientific societies to:
> - host city indicators for continual monitoring and comparison
> - provide critical resources, expertise, and tools to ensure longevity of the hosting platform
> - train students and involve local populations as citizen scientists to collect crowdsourced data about their own communities to identify and overcome data limitations through extensive consultation and validation with local collaborators throughout the process.[23] Data and tools will improve if researchers and practitioners contribute to common methods. We recommend that all cities participate in the open commons, using emerging standard open platforms like OSM to collect and contribute data, and adopt existing standard open-data platforms with easy access and consistent digitisation standards for local data collection. If the goal is the public good, then open-source should be the default for government data and analytics. But governments need not be solely responsible. Open data can be created through three mechanisms: government investment, commercial investment, and crowdsourcing. We encourage collaborations between academia and industry, alongside multilateral efforts to set standards and foster participation in the development and application of indicators. It is imperative that we actively encourage community engagement through crowdsourcing, and that these indicators are used for planning and advocacy to achieve health and sustainability goals while reducing inequities.[2]

In this article, we described a tool created by an international team and presented results for 25 international cities. But we also described a process that starts and finishes with local knowledge: gather data locally; calculate, analyse, and compare indicators globally; and then interpret the results locally using local context and knowledge to derive insights, plan interventions, and advocate for reform. As an international team with local collaborators, we identified study areas, developed tools, collected data, ran analyses, and validated results. This collaborative approach lowered the barriers—technical constraints, resource limitations, and costs—to conducting this kind of analysis. Better city planning around the world requires better monitoring by local governments, with an emphasis on local participation, local data, and local use. We have developed this framework, demonstrated its utility, and provided open-source tools to stimulate adoption and creation of common indicators that can be benchmarked and monitored to support healthy and sustainable cities.

To create an international system for monitoring spatial indicators of health, sustainability, and equity, cities should promote the crowdsourcing of data using the current indicators and thresholds as a foundation for global comparisons. They should provide technical assistance in data collection, analysis, and application. A growing number of data observatories focus on urban SDG indicators, but they tend to ignore spatial and population distributions that enable evidence-based planning for targeted local interventions. As our results demonstrate, beyond just a city-level focus, these indicators require a within-city focus to unpack heterogeneity. Maps allow such variations and inequities to be easily seen and understood. Organisations like the UN and WHO are well positioned to support the expansion of this work into a global observatory of within- and between-city indicators to promote better city planning that can be used by local decision-makers to benchmark and monitor progress.

## Supplementary material

The appendix provides additional detail on the analytical methods and presents a series of 379 maps prepared for the cities included in this study.

**Corresponding author:** Geoff Boeing, Department of Urban Planning and Spatial Analysis, Sol Price School of Public Policy, University of Southern California, Los Angeles, USA, boeing@usc.edu

**Affiliations:** Department of Urban Planning and Spatial Analysis, Sol Price School of Public Policy, University of Southern California, Los Angeles, CA, USA (G Boeing PhD); Healthy Liveable Cities Lab, RMIT University, Melbourne, VIC, Australia (C Higgs MPH, B Giles-Corti PhD, J Arundel PhD); School of Public Policy and Urban Affairs, Northeastern University, Boston, MA, USA (S Liu MS); School of Population Health, The University of Western Australia, Perth, WA, Australia (B Giles-Corti); Mary MacKillop Institute for Health Research, Australian Catholic University, Melbourne, VIC, Australia (J F Sallis PhD, E Cerin PhD); Herbert Wertheim School of Public Health and Human Longevity Science, University of California, San Diego, CA, USA (J F Sallis); School of Public Health (E Cerin) and Department of Geography (P-C Lai PhD), The University of Hong Kong, Hong Kong Special Administrative Region, China; Melbourne Centre for Cities (M Lowe PhD) and Transport, Health and Urban Design Research Lab, Melbourne School of Design (TP Ho MSc), University of Melbourne, Melbourne, VIC, Australia; Department of Landscape Architecture and Environmental Planning, Natural Learning Initiative, College of Design, North Carolina State University, Raleigh, NC, USA (D Adlakha PhD); Human Potential Centre, School of Sport and Recreation, Auckland University of Technology, Auckland, New Zealand (E Hinckson PhD, T Bozovic PhD); Department of Urban Design and Planning, Urban Form Lab, University of Washington, Seattle, WA, USA (A Vernez Moudon DSc); Prevention Research Center, Brown School, Washington University in St Louis, St Louis, MO, USA (D Salvo PhD); College of Health Solutions, Julie Ann Wrigley Global Futures Laboratory, Arizona State University, Phoenix, AZ, USA (MA Adams PhD); Department of Geography, School of Philosophy, Literature, and Human Sciences (L Vizeu Barrozo PhD) and Institute of Advanced Studies (L Vizeu Barrozo), University of São Paulo, São Paulo, Brazil; Barcelona Institute for Global Health, Barcelona, Spain (G Vich PhD); Department of Geography, Rovira i Virgili University, Vilaseca, Spain (G Vich, X Delclòs-Alió PhD); Faculty of Physical Culture, Palacký University Olomouc, Olomouc, Czech Republic (J Dygrýn PhD); School of Natural and Built Environment, Queen's University Belfast, Belfast, Northern Ireland, UK (S Ferguson PhD); Australian Centre for Public and Population Health Research, School of Public Health, University of Technology Sydney, Ultimo, NSW, Australia; Prevention Research Collaboration, School of Public Health, University of Sydney, Camperdown, NSW, Australia (K Gebel PhD), School of Public Health, Faculty of Health, University of Technology Sydney, Sydney, NSW, Australia; Department of Economics and Business, University of Vic—Central University of Catalonia, Vic, Spain (J Carles Martori); Office of the Permanent Secretary for the Bangkok Metropolitan Administration, Bangkok, Thailand (K Nitvimol MA); AFIPS Research Group, Department of Nursing, University of Valencia, Valencia, Spain (A Queralt PhD); Department of Kinesiology, School of Public Health, University of Maryland, College Park, MD, USA (J D Roberts DrPH); Department of Geography, University of Maiduguri, Maiduguri, Nigeria (G H Sambo PhD); Department of Sport Science and Clinical Biomechanics, University of Southern Denmark, Odense, Denmark (J Schipperijn PhD); Research Centre for Architecture, Urbanism and Design, Lisbon School of Architecture, University of Lisbon, Lisbon, Portugal (D Vale PhD); Department of Geography, Ghent University, Ghent, Belgium (N Van de Weghe PhD)

**Contributors:** GB, CH, SL, and JA were part of the study executive team, contributed to the development of the software and analytics, contributed sections of the original draft, and contributed to the conceptualisation, reviewing, and editing of the paper. BG-C, AVM, ML, EC, EH, DA, DS, and JFS were part of the study executive team and contributed to the conceptualisation, writing, review, and editing critically for important intellectual content. MAA, XD-A, LVB, TB, JD, SF, KG, TPH, P-CL, JCM, KN, AQ, JDR, GHS, JS, DV, NVdW, and GV contributed to the writing, review, and editing critically for important intellectual content.

**Declaration of interests:** GB reports grants from The Public Good Projects during the conduct of the study. BG-C reports Senior Principal Research Fellowship (GNT1107672) and grant support (numbers 1061404 and 9100003) from the Australian National Health and Medical Research Council (NHMRC) during the conduct of the study. CH reports grant support (numbers 1061404 and 9100003) from NHMRC. SL reports an experiential fellowship from the College of Social Science and Humanities, Northeastern University during the conduct of this study. JFS reports personal fees from SPARK programmes of Gopher Sport, and serving on the Board of Directors for Rails to Trails Conservancy, outside the submitted work. JFS also has a copyright on SPARK physical activity programmes with royalties paid by Gopher Sport. EC and JFS report support from the Australian Catholic University during the conduct of this study. DS reports support from Washington University in St Louis, Center for Diabetes Translation Research (number P30DK092950 from the US National Institute of Diabetes and Digestive and Kidney Diseases [NIDDK] and the US National Institutes of Health [NIH]) and from the Centers for Disease Control and Prevention (cooperative agreement number U48DP006395) during the conduct of this study. MAA reports grants from the US National Cancer Institute at the NIH (R01CA198915) during the conduct of the study. LVB reports grants from Conselho Nacional de Desenvolvimento Científico e Tecnológico (grant number 304636/2020-7) during the conduct of the study. All other authors report no competing interests. The content of this article is solely the responsibility of the authors and does not represent the official views of any of the NIDDK, NIH, CDC, or of any of the funding agencies supporting this work.

**Acknowledgments:** The authors wish to thank Sien Benoit, MSc, Department of Geography, Ghent University, Ghent, Belgium; Marcel Beyeler, MSc, GIS Competence Centre, Land Surveying Office, Bern, Switzerland; Nicholas Cerdera, Sol Price School of Public Policy, University of Southern California, Los Angeles, USA; Hannah Hook, College of Health Solutions, Arizona State University, Phoenix, USA; Ruth Hunter, PhD, School of Medicine, Dentistry and Biomedical Sciences, Queen's University Belfast, Belfast,

Northern Ireland; Oliver Konrad, MA, Town Planning Office, Graz, Austria; Lea Maňáková, MSc, Department of Transport and Urban Development, Municipality of Olomouc, Czech Republic; Suzanne Mavoa, PhD, Department of Forest and Conservation Sciences, Faculty of Forestry, University of British Columbia, Vancouver, Canada; Carme Miralles, PhD, Department of Geography, Autonomous University of Barcelona, Bellaterra, Spain; David Moctezuma, Sol Price School of Public Policy, University of Southern California, Los Angeles, USA; Javier Molina-García, PhD, Department of Teaching of Musical, Visual and Corporal Expression, University of Valencia, Valencia, Spain; Anne Luise Müller, Dipl Ing, Town Planning Office, City of Cologne, Germany; Thu Nguyen, Sol Price School of Public Policy, University of Southern California, Los Angeles, USA; Babatunji Omotara, PhD, Department of Community Medicine, University of Maiduguri, Maiduguri, Nigeria; Adetoyeje Oyeyemi, DHSc, Department of Physiotherapy, University of Maiduguri, Maiduguri, Nigeria; Adewale L Oyeyemi, PhD, Department of Physiotherapy, University of Maiduguri, Maiduguri, Nigeria; Adamu Ahmad Rufai, PhD, Department of Physiotherapy, University of Maiduguri, Maiduguri, Nigeria; Spatial Vision, Melbourne, Australia; Sylvia Titze, PhD, Institute of Sport Science, University of Graz, Graz, Austria; Minh Hieu Trinh, ICRSL Subproject Management Unit, The Ministry of Planning and Investment, Vietnam; Claudia Viana, MSc, Centre for Geographical Studies, Institute of Geography and Spatial Planning, University of Lisbon, Lisbon, Portugal; and Yuquan Zhou, Sol Price School of Public Policy, University of Southern California, Los Angeles, USA.

**Using open data and open-source software to develop spatial indicators of urban design and transport features for achieving healthy and sustainable cities: a 25-city study**

**Technical and Map Appendices**

Global Healthy & Sustainable City-Indicators Collaboration, 2022

# Appendix A: Technical documentation

**Contents**

# A1. Preliminaries

## A1.1. Spatial data collection

Collaborators from 19 countries (Table A) in the International Physical Activity and Environment Network (IPEN, https://www.ipenproject.org/) and other collaborators expressed interest in the Global Healthy and Sustainable City-Indicators Collaboration. They nominated local representatives with spatial expertise to complete a Google Forms survey (Table B) detailing available resources for representing their region in the study.

Seventeen cities had available street network data, with datasets publicly available in 15 cities. However, only six cities had publicly available data on urban design and transport features of interest, with commercially licensed data being more commonplace. Some datasets were provided by local collaborators during the survey, but they were inconsistent in terms of availability, formats, comparability, and coverage. Thus, for consistency we decided to use open data with global coverage in the first instance.

This 'open data first' approach meant that we were able to expand our analysis to include the full set of indicators for all 25 cities. For sources of open data for each studied city, see Appendix A2·1. The limitations of open data were identified through consultation with collaborators throughout the indicator estimation workflow, and we developed methods to mitigate these limitations where required by incorporating custom data and methods to improve the validity of our analysis. Details on the data validation process were published in the software framework method paper.[1]

Due to the lack of detailed information on street networks and walkability barriers, we recognise two important assumptions when measuring the considered walkable distances: 1) streets do not present barriers to walking or wheelchair access; and 2) people's walking ranges are higher.

**Table A. Study region characteristics**

| | City | Urban study region area (km²) | Population estimate (2015) | Population per km² | GNI per capita classification (2021) | Sample points analysed | Sample points analysed per km² |
|---|---|---|---|---|---|---|---|
| **Africa** | | | | | | | |
| Nigeria (NGA) | Maiduguri | 125 | 1,092,447 | 8,722 | Lower middle | 62,358 | 498 |
| **America, North** | | | | | | | |
| Mexico (MEX) | Mexico City | 2,312 | 20,216,501 | 8,744 | Upper middle | 1,516,235 | 656 |
| United States (USA) | Baltimore | 741 | 1,381,445 | 1,865 | High | 429,005 | 579 |
| | Phoenix | 772 | 1,320,016 | 1,710 | | 466,658 | 605 |
| | Seattle | 1,885 | 2,199,327 | 1,167 | | 1,038,975 | 551 |
| **America, South** | | | | | | | |
| Brazil (BRA) | São Paulo | 1,018 | 11,718,034 | 11,512 | Upper middle | 676,239 | 664 |
| **Asia** | | | | | | | |
| Hong Kong SAR (HKG) | Hong Kong | 373 | 7,325,576 | 19,665 | High | 269,800 | 724 |
| India (IND) | Chennai | 425 | 6,602,769 | 15,549 | Lower middle | 276,913 | 652 |
| Thailand (THA) | Bangkok | 1,190 | 9,337,076 | 7,844 | Upper middle | 695,113 | 584 |
| Vietnam (VNM) | Hanoi | 1,220 | 5,938,818 | 4,866 | Lower middle | 375,743 | 308 |
| **Australasia** | | | | | | | |
| Australia (AUS) | Adelaide | 541 | 985,647 | 1,822 | High | 307,819 | 569 |
| | Melbourne | 1,657 | 3,741,467 | 2,258 | | 929,061 | 561 |
| | Sydney | 1,334 | 4,082,229 | 3,061 | | 676,664 | 507 |
| New Zealand (NZL) | Auckland | 468 | 1,234,554 | 2,638 | High | 232,260 | 496 |
| **Europe** | | | | | | | |
| Austria (AUT) | Graz | 69 | 283,101 | 4,121 | High | 72,894 | 1,061 |
| Belgium (BEL) | Ghent | 75 | 174,411 | 2,339 | High | 50,852 | 682 |
| Czech Republic (CZE) | Olomouc | 27 | 88,044 | 3,275 | High | 26,775 | 996 |
| Denmark (DNM) | Odense | 56 | 157,018 | 2,791 | High | 49,196 | 874 |
| Germany (DEU) | Cologne | 348 | 1,118,442 | 3,218 | High | 225,412 | 648 |
| Portugal (PRT) | Lisbon | 85 | 583,347 | 6,867 | High | 79,941 | 941 |
| Spain (ESP) | Barcelona | 359 | 3,259,527 | 9,068 | High | 262,708 | 731 |
| | Valencia | 86 | 682,752 | 7,937 | | 76,410 | 888 |
| | Vic | 31 | 43,813 | 1,433 | | 11,499 | 376 |
| Switzerland (CHE) | Bern | 32 | 158,179 | 4,898 | High | 46,801 | 1,449 |
| Northern Ireland (GBR) | Belfast | 98 | 400,731 | 4,084 | High | 67,244 | 685 |
| **Summary** | | | | | | | |
| | **min** | 27 | 43,813 | 1,167 | | 11,499 | 308 |
| | **25th percentile** | 85 | 400,731 | 2,339 | | 67,244 | 561 |
| | **median** | 373 | 1,234,554 | 4,084 | | 262,708 | 652 |
| | **75th percentile** | 1,018 | 4,082,229 | 7,937 | | 466,658 | 731 |
| | **max** | 2,312 | 20,216,501 | 19,665 | | 1,516,235 | 1,449 |
| | **interquartile range** | 933 | 3,681,497 | 5,599 | | 399,414 | 170 |
| | **mean** | 613 | 3,365,011 | 5,658 | | 356,903 | 691 |
| | **sd** | 654 | 4,736,826 | 4,643 | | 379,168 | 240 |

**Table B. Spatial data survey items**

| # | Item |
|---|---|
| 1 | Timestamp |
| 2 | Email address |
| 3 | Name |
| 4 | What is the city for which you are uploading data? |
| 5 | Country |
| 6 | City boundary data (study region) |
| 7 | Is an alternative routable street network data set to OpenStreetMap preferred for indicator calculation in your city? If yes: file upload; Data source; URL or citation; Reference date; Spatial reference EPSG code; Licence type (e.g. "CC BY 4.0", "none specified", etc.); Please detail any restrictions, limitations or other important points to note on the use of this data in this project |
| 8 | *Do you have access to General Transit Feed Specification (GTFS) data for your city? If yes: file upload; Data source(s); URL(s) or citation(s); Reference date(s); License type (e.g. "CC BY 4.0", "none specified", etc.); Please detail any restrictions, limitations or other important points to note on the use of this data in this project* |
| 9 | Are alternative or supplementary Points of Interest (POIs) / destinations / activity centre data preferred for indicator calculation in your city, instead of or in addition to those sourced from OpenStreetMap? |
| 10 | *Population per city area table. If available: file upload; Data source; URL or citation; Reference date; License type (e.g. "CC BY 4.0", "none specified", etc.); Please detail any restrictions, limitations or other important points to note on the use of this data* |
| 11 | Area boundaries. If available: file upload; Attribute column used for linkage with the population per area CSV; Data source; URL or citation; Reference date; Spatial reference EPSG code; License type (e.g. "CC BY 4.0", "none specified", etc); Please detail any restrictions, limitations or other important points to note on the use of this data in this project |
| 12 | *Please upload a CSV or Excel file with the area ID (or name) and count of dwellings. If available: file upload; Data source; URL or citation; Reference date; License type (e.g. "CC BY 4.0", "none specified", etc); Please detail any restrictions, limitations or other important points to note on the use of this data* |
| 13 | Are the area boundaries associated with dwelling counts the same as those previously uploaded for population? If yes: file upload; Area boundaries; Attribute column used for linkage with the dwellings per area CSV; Data source; URL or citation; Reference date; Spatial reference EPSG code; License type (e.g. "CC BY 4.0", "none specified", etc.); Please detail any restrictions, limitations or other important points to note on the use of this data in this project |
| 14 | *Please upload a CSV or Excel file with the area ID (or name) of usual residence and usual place of work and joint counts. If available: file upload; Data source; URL or citation; Reference date; License type (e.g. "CC BY 4.0", "none specified", etc.); Please detail any restrictions, limitations or other important points to note on the use of this data* |
| 15 | Are the area boundaries associated with journey to work counts the same as those previously uploaded for population or dwellings? If no: file upload; Attribute column used for linkage with the journey to work CSV; Data source; URL or citation; Reference date; Spatial reference EPSG code; License type (e.g. "CC BY 4.0", "none specified", etc.); Please detail any restrictions, limitations or other important points to note on the use of this data in this project; Can you provide Digital Elevation Model (DEM) data for your study region superior to that which will be made available under NASADEM as described at https://earthdata.nasa.gov/community/community-data-system-programs/measures-projects/nasadem ? |
| 16 | *Can you provide Digital Elevation Model (DEM) data for your study region superior to that which will be made available under NASADEM as described at https://earthdata.nasa.gov/community/community-data-system-programs/measures-projects/nasadem? If yes: file upload; Data source; URL or citation; Reference date; Spatial reference EPSG code; License type (e.g. "CC BY 4.0", "none specified", etc); Please detail any restrictions, limitations or other important points to note on the use of this data in this project; Do you have an open source residential address data set available for your city which is appropriate for use in this project?; Residential address points* |
| 17 | Do you have an open source residential address data set available for your city which is appropriate for use in this project? If yes: file upload; Data source; URL or citation; Reference date; Spatial reference EPSG code; License type (e.g. "CC BY 4.0", "none specified", etc.); Please detail any restrictions, limitations or other important points to note on the use of this data in this project; Do you have a source of open space / green space data to be preferred over use of OpenStreetMap data?; Open Space data |
| 18 | *Do you have a source of open space / green space data to be preferred over use of OpenStreetMap data? If yes: file upload; Definition of open space; Data source; URL or citation; Reference date; Spatial reference EPSG code; License type (e.g. "CC BY 4.0", "none specified", etc.); Please detail any restrictions, limitations or other important points to note on the use of this data in this project* |

**A1.2. Defining study regions**

The analysis area for each city included in our study was conceptually framed as the urban portion of a city's broad administrative boundaries. Administrative boundaries were either supplied by collaborators via the preliminary spatial data survey or acquired by the researchers independently. The urban study region was defined as the intersection of this boundary with that of corresponding urban centres identified by the Global Human Settlements project (UCDB R2019A v1·2).[2] The use of this global data source for urban centres helps to ensure that the analysis focuses on exposure for urban populations across all cities, but not for lower-density rural settings on the urban fringe, which may otherwise fall within an administrative boundary.

**A1.3. Processing administrative boundaries**

A curated dataset of study region administrative boundaries was manually prepared. Administrative boundaries provided by collaborators for their city were used in the first instance, with alternative open-data sources for boundaries identified for the remaining cities. Boundaries for some cities were revised through consultation with collaborators following review of the preliminary data validation reports generated in the pre-processing phase. The final city-specific boundary methods are detailed below.

**A1.3.1 General methods**

Where city boundaries were acquired as a single valid feature, these were output in the relevant projected CRS for that region to a geopackage file. City boundaries supplied as a series of administrative polygons (e.g. Belfast) were dissolved using QGIS 3·8·1-Zanzibar.[3] Any gaps that incorrectly remained were removed using the "delete holes" tool, and saved to the geopackage file. For cities where boundaries were either not supplied (e.g., Auckland) or not supplied in a usable format (Hanoi, as lines and points—not areas) alternative boundaries were sourced. For cities with geometry type of PolygonZ (Ghent, Odense), these were exported as a regular polygon geometry, discarding Z-dimension data, which may cause difficulties with subsequent processing of multipolygons.

**A1.3.2 City-specific methods**

*GHS urban boundaries: Baltimore, Seattle (United States); Maiduguri (Nigeria)*

It was decided that for these cities the best solution for study region boundaries was to use the urban region as defined in the Global Human Settlements dataset (UCDB R2019A v1·2). It was decided not to do this for all regions, as a number of collaborators indicated that the predicted urban centre as defined in the GHS dataset did not reliably include urban-fringe growth areas, which were considered important (e.g. Australian cities, Mexico City). As such, where boundaries were approved by collaborators, these existing ones were used. Elsewhere, on approval from collaborators, the GHS urban centre was used.

*Auckland, New Zealand*

New Zealand administrative boundaries (statsnzmeshblock-higher-geographies-2019-generalised-SHP.zip) were retrieved from the Stats NZ geographic data portal (https://datafinder.stats.govt.nz/data/category/annual-boundaries/2019/). The Auckland city boundary was processed by selecting the Mesh Blocks with UR_2019_V_1 value of 'Auckland' and with IUR2019__1 classed as major urban. The boundaries were dissolved and saved to the boundaries geopackage.

*Barcelona, Spain*

The initial identified boundary was revised using a collaborator-supplied shape file following preliminary data validation. Boundaries were dissolved and exported following the general method.

*Bern, Switzerland*

A boundary zip file was retrieved from the Geoportal of the Canton of Bern (https://www.geo.apps.be.ch/index.php?option=com_easysdi_shop&task=download.direct&id=195). Using the shapefile "GENGRZ5_GEN2G5", the feature of the name "Bern" was exported as Bern to the boundaries geopackage.

*Chennai, India*

An administrative boundary for Chennai from OpenStreetMap was retrieved in geojson format (https://nominatim.openstreetmap.org/search.php?q=Chennai+India&polygon_geojson=1&format=geojson). The QGIS delete holes processing tool was used to include a gap identified as representing the St Thomas Mount cantonment (http://www.cbstm.org.in:9080/Cantonment/web/Home.html) in the Chennai urban study region. The derived Chennai boundary was then exported to the boundaries geopackage.

*Cologne, Germany*

The boundary was retrieved from the Köln (Cologne) open-data portal (https://offenedaten-koeln.de/dataset/stadtteile). The 'Fix geometries'' tool was run, then the Dissolve algorithm, the ''Delete holes'' algorithm, and finally was exported as Cologne to the boundaries geopackage.

*Ghent, Belgium*

A boundary for the city of Ghent omitting the North Sea port industrial area was supplied by collaborators, following the advice received in the preliminary data survey that this portion of the city should be excluded from consideration as 'urban'. Boundaries were dissolved and exported following the general method.

*Graz, Austria*

Boundaries were retrieved from the Austrian government open-data portal (https://www.data.gv.at/katalog/dataset/land-stmk_bezirksgrenzen/resource/086b453d-c474-4458-ba31-2b5408f5b999). The field BEZNAM with value 'Graz Stadt' was exported as Graz to the boundaries geopackage.

*Hanoi, Vietnam*

Administrative boundaries for Vietnam (vnm_admbnda_2018_shp.zip) were retrieved from Humanitarian Data Exchange (https://data.humdata.org/dataset/viet-nam-administrative-boundaries-polygon-polylinev), and the level 1 boundary layer (vnm_admbnda_adm1_2018_v2) feature corresponding to ADM1_EN value of 'Ha Noi' was saved to the boundaries geopackage.

*Hong Kong SAR, China*

A boundary was retrieved from ESRI China (http://opendata.esrichina.hk/datasets/hong-kong-18-districts/data). The 'fix geometries' algorithm was used to resolve validity issues, boundaries were dissolved, and 'delete holes' was run prior to saving to the curated geopackage.

*Mexico City, Mexico*

Administrative boundary data for Mexico City were retrieved from https://data.humdata.org/dataset/mexican-administrative-level-0-country-1-estado-and-2-municipio-boundary-polygons. Following advice from collaborators, the Mexico City boundary was constructed to reflect Greater Mexico City, in the sense of Zona Metropolitana del Valle de México (https://en.wikipedia.org/wiki/Greater_Mexico_City#Metropolitan_Area_of_the_Valley_of_Mexico). This involved the combination of: 16 municipalities (alcaldías) of Ciudad de México (Azcapotzalco, Álvaro Obregón, Benito Juárez, Coyoacán, Cuajimalpa de Morelos, Cuauhtémoc, Gustavo A. Madero, Iztacalco, Iztapalapa, Magdalena Contreras, Miguel Hidalgo, Milpa Alta, Tláhuac, Tlalpan, Venustiano Carranza, Xochimilco); the 40 municipalities of the State of Mexico (Acolman, Atenco, Atizapán de Zaragoza, Chalco, Chiautla, Chicoloapan, Chiconcuac, Chimalhuacán, Coacalco de Berriozábal, Cocotitlán, Coyotepec, Cuautitlán, Cuautitlán Izcalli, Ecatepec de Morelos, Huehuetoca, Huixquilucan, Ixtapaluca, Jaltenco, La Paz, Melchor Ocampo, Naucalpan de Juárez, Nextlalplan, Nezahualcoyotl, Nicolás Romero, Papalotla, San Martín de las Pirámides, Tecámac, Temamatla, Teoloyucán, Teotihuacán, Tepetlaoxtoc, Tepotzotlán, Texcoco, Tezoyuca, Tlamanalco, Tlalnepantla de Baz, Tultepec, Tultitlán, Valle de Chalco Solidaridad, Zumpango); one conurbation municipality of the State of Hidalgo (Tizayuca); and additional municipalities (Amecameca, Apaxco, Atlautla, Axapusco, Ayapango, Ecatzingo, Hueypoxtla, Isidro Fabela, Jilotzingo, Juchitepec, Nopaltepec, Otumba, Ozumba, Temascalapa, Tenango del Aire, Tepetlixpa, Tequixquiac, Villa del Carbón), and Tonanitla.

Noting that some of the ADM2_ES names in the source data were not unique within the country of Mexico, the ADM1_ES state names were limited to ''Distrito Federal'', ''Hidalgo'' and ''Mexico''. The required zones were identified using the query, which involved manually amending names to match their recorded representation in the source data (i.e. some missing characters, truncation, and switched characters). Matched provinces were verified to be in their correct location using a subsequent audit.

ADM2_ES IN ('Azcapotzalco', 'Álvaro Obregón', 'Benito Julrez', 'Coyoacán', 'Cuajimalpa de Morelos', 'Cuauhtémoc', 'Gustavo A. Madero', 'Iztacalco', 'Iztapalapa', 'La Magdalena Contreras', 'Miguel Hidalgo', 'Milpa Alta', 'Tláhuac', 'Tlalpan', 'Venustiano Carranza', 'Xochimilco', 'Acolman', 'Atenco', 'Atizapán de Zaragoza', 'Chalco', 'Chiautla', 'Chicoloapan', 'Chiconcuac', 'Chimalhuacán', 'Coacalco de Berrioz', 'Cocotitlán', 'Coyotepec', 'Cuautitlán', 'Cuautitlán Izcalli', 'Ecatepec de Morelos', 'Huehuetoca', 'Huixquilucan', 'Ixtapaluca', 'Jaltenco', 'La Paz', 'Melchor Ocampo', 'Naucalpan de JuIrez', 'Nextlalpan', 'Nezahualcoyotl', 'Nicolts Romero', 'Papalotla', 'San Martín de las Pirámides', 'Tec', 'Temamatla', 'Teoloyucan', 'Teotihuacan', 'Tepetlaoxtoc', 'Tepotzotlán', 'Texcoco', 'Tezoyuca', 'Tlalmanalco', 'Tlalnepantla de Baz', 'Tultepec', 'Tultitlán', 'Valle de Chalco Solidaridad', 'Zumpango', 'Tizayuca', 'Amecameca', 'Apaxco', 'Atlautla', 'Axapusco', 'Ayapango', 'Ecatzingo', 'Hueypoxtla', 'Isidro Fabela', 'Jilotzingo', 'Juchitepec', 'Nopaltepec', 'Otumba', 'Ozumba', 'Temascalapa', 'Tenango del Aire', 'Tepetlixpa', 'Tequixquiac', 'Villa del Carb', 'Tonanitla') AND ADM1_ES IN ('DISTRITO FEDERAL', 'HIDALGO', 'MEXICO')

The final query returned 76 municipalities correctly representing the de facto extent of Greater Mexico City. The retrieved polygons' boundaries were dissolved, reprojected to EPSG 32614 (UTM14N), and saved to the boundaries.gpkg file as the layer 'Mexico City'.

### *Phoenix, United States*

A boundary shapefile was retrieved from a US open-data portal (https://catalog.data.gov/dataset/tiger-line-shapefile-2015-state-arizona-current-place-state-based), and the feature with the name of 'Phoenix City' was exported as Phoenix to boundaries geopackage.

### *São Paulo, Brazil*

Administrative boundaries for Brazil were retrieved from Humanitarian Data Exchange (https://data.humdata.org/dataset/brazil-administrative-level-0-boundaries), and the feature with 'name_2' value of 'Sao Paolo' was saved to the boundaries geopackage.

## A2. Reproducible analytic workflow for estimating indicators

Following the preliminaries stage, we developed a generalised workflow for calculating within- and between-city indicator estimates. The following supplementary materials elaborate on how our open-source framework was applied to the analysis of the 25 cities presented in this study. This material is intended to provide guidance for researchers and practitioners seeking to reproduce our results or extend them for different study regions and time points.

Figure A below illustrates the general workflow of the framework, which comprises a four-step process from data acquisition to indicator aggregation. Researchers and practitioners seeking to reproduce the indicator results for their cities are advised to look at the open-source framework repository at https://github.com/global-healthy-liveable-cities/global-indicators.git, and then consult the following details for calculating indicators.

**Figure A. Indicator calculation workflow**

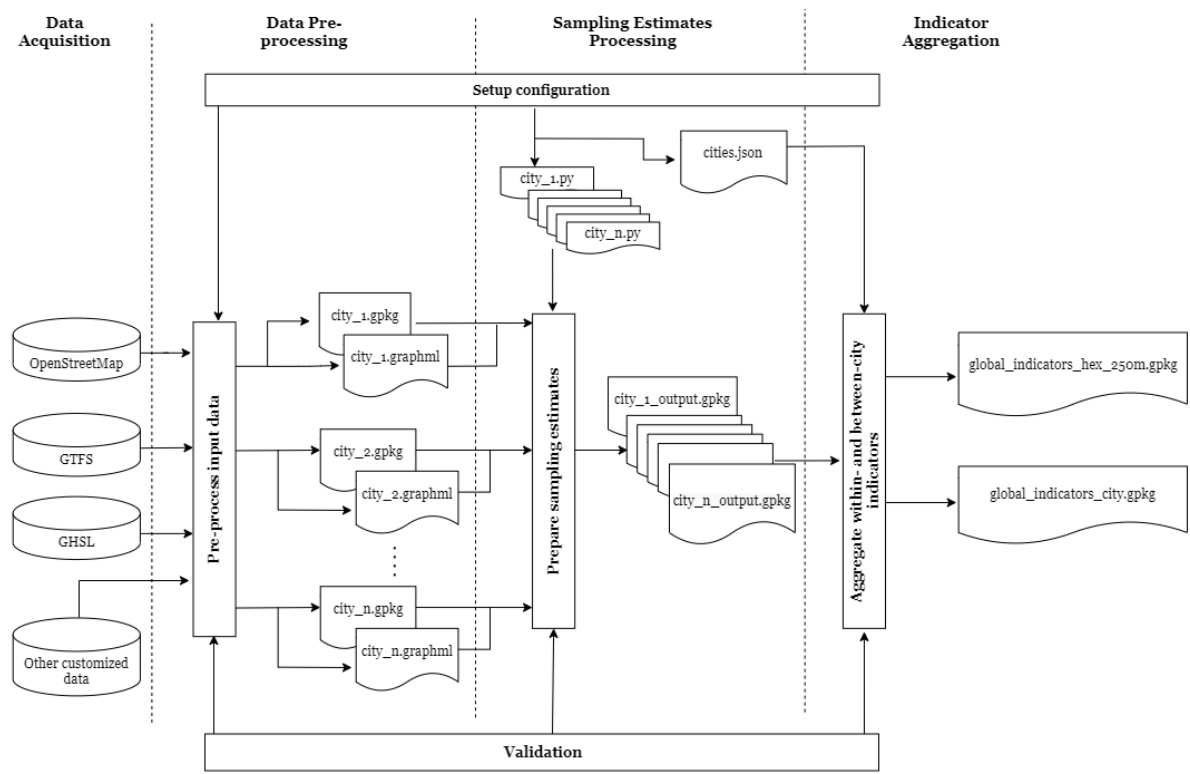

### A2.1. Data acquisition

Other than the study region boundaries described in the preliminaries, most data were sourced from open datasets with global scope. These global open-data sources are summarised in Table C. The other exceptions were a set of custom destinations developed by collaborators for Maiduguri, and public transport schedule datasets in General Transit Feed Specification (GTFS) format developed by specific agencies, many of which were hosted on OpenMobilityData. Data sources of GTFS feeds available for the studied cities are summarised in Table D.

**Table C. Input data sources for the 25 study regions**

| Input data | Source | URL | Comments |
|---|---|---|---|
| Study region boundaries | Global Human Settlements Layer Urban Centers Database, UCDB R2019A | https://jeodpp.jrc.ec.europa.eu/ftp/jrc-opendata/GHSL/GHS_STAT_UCDB2015MT_GLOBE_R2019A/ | Data were directly downloaded from the GHSL database. |
| Population | Global Human Settlements Layer, population distribution raster data, GHS-POP R2019A | https://cidportal.jrc.ec.europa.eu/ftp/jrc-opendata/GHSL/GHS_POP_MT_GLOBE_R2019A/ | Data were directly downloaded from the GHSL database. |
| Transport-sector PM2.5 emissions (tonnes per annum) | Global Human Settlements Layer, GHS UCDB 2015 | https://cidportal.jrc.ec.europa.eu/ftp/jrc-opendata/GHSL/GHS_POP_MT_GLOBE_R2019A/ | Data were directly downloaded from the GHSL database. Note that "transport-sector" encompasses road transport as well as aviation and shipping. |
| Pedestrian street network | OpenStreetMap planet archive file dated 3 August 2020 | https://planet.osm.org/pbf/planet-200803.osm.pbf | Data were retrieved from OpenStreetMap via the Overpass API for buffered boundaries of study regions using the Python package OSMnx. |
| Healthy food outlets, supermarkets, and convenience stores | OpenStreetMap | https://taginfo.openstreetmap.org/ | These spatial features were obtained following the OpenStreetMap tagging guidelines and collaborator feedback, along with the OpenStreetMap Taginfo for the appropriate key-value pair tag, detailed in Table G. |
| Public transportation | OpenStreetMap | https://transitfeeds.com/; https://taginfo.openstreetmap.org/ | Data were obtained from OpenStreetMap following the tagging guidelines and collaborator feedback detailed in Table G, and from various data sources for GTFS format. A more comprehensive list of GTFS data sources is shown in Table D. |
| Public open space | OpenStreetMap | https://taginfo.openstreetmap.org/ | Data were obtained from OpenStreetMap following the tagging guidelines and collaborator feedback detailed in Appendix A 2.3.3. |

**Table D. General Transit Feed Specification (GTFS) data sources**

| Country | City | Feed # | Agency / Provider | Year | Analysis period (yyyymmdd) Start | Analysis period (yyyymmdd) End |
|---|---|---|---|---|---|---|
| Mexico | Mexico City | 1 | FederalDistrictGovernment* | 2019 | 20190405 | 20190605 |
| United States | Baltimore | 1 | MarylandMTA* | 2019 | 20190405 | 20190605 |
| United States | Phoenix | 1 | Valleymetro* | 2019 | 20190405 | 20190605 |
| United States | Seattle | 1 | KingCountyMetro* | 2019 | 20190405 | 20190605 |
| Brazil | São Paulo | 1 | SPTrans* | 2019 | 20191008 | 20191205 |
| Hong Kong SAR | Hong Kong | 1 | data.gov.hk | 2019 | 20190405 | 20190605 |
| India | Chennai | 1 | MTC | 2010 | 20100405 | 20100605 |
| India | Chennai | 2 | J Kishore Kumar | 2016 | 20161008 | 20161205 |
| Thailand | Bangkok | 1 | Namtang Open Data | 2021 | 20210405 | 20210605 |
| Vietnam | Hanoi | 1,2,3 | World Bank | 2018 | 20180405 | 20180605 |
| Australia | Adelaide | 1 | AdelaideMetro* | 2019 | 20191008 | 20191205 |
| Australia | Melbourne | 1 | PublicTransportVictoria* | 2019 | 20191008 | 20191205 |
| Australia | Sydney | 1 | Transport for NSW | 2019 | 20191008 | 20191205 |
| New Zealand | Auckland | 1 | AucklandTransport* | 2019 | 20191008 | 20191205 |
| Denmark | Odense | 1 | Rejseplanen* | 2019 | 20190405 | 20190605 |
| Germany | Cologne | 1 | VRS* | 2018 | 20180405 | 20180605 |
| Portugal | Lisbon | 1 | Carris* | 2019 | 20190405 | 20190605 |
| Portugal | Lisbon | 2 | Metro de lisboa* | 2019 | 20190405 | 20190605 |
| Portugal | Lisbon | 3 | Fertagus* | 2019 | 20190405 | 20190605 |
| Portugal | Lisbon | 4 | MTS* | 2019 | 20190405 | 20190605 |
| Portugal | Lisbon | 5 | Soflusa* | 2019 | 20190405 | 20190605 |
| Portugal | Lisbon | 6 | Transtejo* | 2019 | 20190405 | 20190605 |
| Portugal | Lisbon | 7 | CP* | 2019 | 20190405 | 20190605 |
| Spain | Barcelona | 1 | AMB* | 2019 | 20190405 | 20190605 |
| Spain | Barcelona | 2 | TMB* | 2019 | 20190405 | 20190605 |
| Spain | Barcelona | 3 | TMB* | 2019 | 20190405 | 20190605 |
| Spain | Valencia | 1 | MetroValencia* | 2019 | 20190405 | 20190605 |
| Spain | Valencia | 2 | EMT* | 2019 | 20190405 | 20190605 |
| Switzerland | Bern | 1 | opentransportdata.swiss | 2019 | 20190405 | 20190605 |
| UK | Belfast | 1 | Translink | 2017 | 20170405 | 20170605 |

* Data sourced from GTFS feed aggregator OpenMobilityData (https://transitfeeds.com)

**Table E. Project parameters**

| Parameters | | Description | Value |
|---|---|---|---|
| *Project settings* | folderPath | The folder where data resources for the project are located | /home/jovyan/work/process/data |
| | units | Units used by the coordinate reference system | m |
| | units_full | Full name for the units | metres |
| | study_buffer | Study region buffer, to account for edge effects | 1600 |
| | buffered_study_region_name | Buffered study region's name for map display purposes | 1600 m study region buffer |
| | hex_diag | Hexagon diagonal length and buffer distance (metres) | 250 |
| | hex_buffer | Hexagon buffer distance, to account for edge effects | 250 |
| | multiprocessing | Number of processors to use in multiprocessing scripts | 6 |
| | no_forward_edge_issues | Used to flag and mitigate potential geometry discrepancies | 0 |
| | year | The year being targeted by this analysis | 2020 |
| *SQL* | admin_db | SQL settings to connect to Postgresql+Postgis Docker container | postgres |
| | db_host | as above | 127.0.0.1 |
| | db_port | as above | 5433 |
| *OpenStreetMap* | osm_data | Downloaded OpenStreetMap (OSM) data | planet-latest_2020-08-13.osm.pbf |
| | osm_date | Date at which OSM download was current | 20200813 |
| | osmnx_retain_all | If false, only retain main connected network when retrieving OSM roads | FALSE |
| *Sample points* | points_id | sampling points unique id | location_id |
| | points | name of point data locations used for sampling | sampling_points |
| | point_sampling_interval | interval in metres along which network is to be sampled | 30 |
| *Roads* | intersection_tolerance | tolerance in metres for cleaning intersections | 12 |
| | network_edges | as above | edges |
| | network_junctions | as above | nodes |
| *Network analysis* | distance | sausage buffer network size -- in units specified above | 1600 |
| | tolerance | search tolerance (in units specified above) | 500 |
| | line_buffer | buffer distance for network lines as sausage buffer | 50 |
| | limit | distance is a limit beyond which not to search for destinations | 1600 |
| | aos_threshold | Distance within which all Areas of Open Space are sought | 1600 |
| *Population* | population_data | population raster dataset (defined in datasets worksheet) | population |
| | population_target | target year for population data | 2015 |
| | pop_min_threshold | hex grid cell estimated population inclusion threshold | 5 |

Table F. Study region parameters used for pre-processing

| City | EPSG code | Region boundary data | Additional region-specific parameter notes |
|---|---|---|---|
| Maiduguri | 32633 | GHS:UC_NM_MN='Maiduguri' | 1 |
| Mexico City | 32614 | ./data/boundaries.gpkg:Mexico City | |
| Baltimore | 32618 | GHS:UC_NM_MN='Baltimore' | |
| Phoenix | 32612 | ./data/boundaries.gpkg:Phoenix | |
| Seattle | 32610 | GHS:UC_NM_MN='Seattle' | |
| São Paulo | 32723 | ./data/boundaries.gpkg:Sao Paulo | |
| Hong Kong | 32650 | ./data/boundaries.gpkg:Hong Kong | 2,3,4 |
| Chennai | 32644 | ./data/boundaries.gpkg:Chennai | |
| Bangkok | 32647 | ./data/boundaries.gpkg:Bangkok | |
| Hanoi | 32648 | ./data/boundaries.gpkg:Hanoi | |
| Adelaide | 7845 | ./data/boundaries.gpkg:Adelaide | |
| Melbourne | 7845 | ./data/boundaries.gpkg:Melbourne | |
| Sydney | 7845 | ./data/boundaries.gpkg:Sydney | |
| Auckland | 2193 | ./data/boundaries.gpkg:Auckland | |
| Graz | 32633 | ./data/boundaries.gpkg:Graz | |
| Ghent | 32631 | ./data/boundaries.gpkg:Ghent | |
| Olomouc | 32633 | ./data/boundaries.gpkg:Olomouc | |
| Odense | 32632 | ./data/boundaries.gpkg:Odense | |
| Cologne | 32631 | ./data/boundaries.gpkg:Cologne | |
| Lisbon | 3763 | ./data/boundaries.gpkg:Lisbon | |
| Barcelona | 25831 | ./data/boundaries.gpkg:Barcelona | |
| Valencia | 25830 | ./data/boundaries.gpkg:Valencia | |
| Vic | 25831 | ./data/boundaries.gpkg:Vic | 5 |
| Bern | 32633 | ./data/boundaries.gpkg:Bern | |
| Belfast | 29902 | ./data/boundaries.gpkg:Belfast | |

*Notes on parameters accounting for study region exceptions to generic processing workflow*

1. *Custom destinations:* Custom destination data to use, in addition to those identified using OSM. Use a comma-separated list specifying file name (located in study region folder, once generated), category plain name field, category full name field, Y coordinate, X coordinate, EPSG number. Specifically for Maiduguri: *Maiduguri_shops_convenience_complete_2020-07-13_categorised_final.csv,dest_name,dest_name_full,Latitude,Longitude,4326*

2. *No buffered study region:* Instead of using buffered study region, use regular study region for excerpting network from OSM. This may allow for looping over true islands to extract individual networks (e.g. Hong Kong), which may not be possible with the buffered region (which results in only retaining larger network segment). For Hong Kong, this is set to 'TRUE', for other cities it has been left blank.

3. *Network polygon iteration:* Iterate over polygons for network retrieval, and then combine. This is useful for cities spanning multiple islands, in conjunction with the '*No buffered study region*' parameter. For Hong Kong, this is set to 'TRUE', for other cities it has been left blank.

4. *Network connection threshold:* Minimum threshold distance in units as per project parameters for disconnected networks (e.g. road networks for cities spanning multiple islands) to be included in the combined network for a city. For Hong Kong, this is set to 200, a value determined through manual inspection of the geographic context; for other cities, it has been left blank.

5. *Not GHS urban intersection*: For cities where intersection with the GHS urban area is not valid. For Vic which does not have a corresponding urban area in the GHS dataset), this variable was set to 'TRUE', indicating that the provided boundary is to be used.

### A2.2. Setup configuration

Project and study region configuration parameters were defined using configuration files and were used to facilitate our software framework's semi-automated workflow to calculate spatial indicators of the 25 study regions. Tables E and F summarise project parameters and study region–specific parameters used during pre-processing. Other parameters used for sampling estimate processing and indicator calculation were compiled into a series of configuration files within the project's Github repository, which can be constructed with a single command (https://github.com/global-healthy-liveable-cities/global-indicators/blob/7852424e4852c99b5b0f7b65982d289db28e2ed0/process/setup_config.py). Researchers can draw upon them as templates and adapt for a new project as required.

### A2.3. Pre-processing input data

Data for each study region was pre-processed with a series of Python scripts, drawing upon the configuration parameters in Tables E and F for analyses, with output to city-specific PostgreSQL (PostGIS) databases, before final output in geopackage format as used in the main analyses. Generalised data pre-processing for urban boundaries, pedestrian street network, population, spatial features, and sample points has been described in our methods paper.[1] Specific details on deriving points of interest, regularly serviced public transport locations, and public open space access points are provided below.

### A2.3.1 Points of interest

Destination categories were coded using a conceptual mapping of key-value pair terms (Table G) from OpenStreetMap point, line, and polygon features, which were identified by reviewing coding guidelines (general, and community/country specific) and following a preliminary dataset validation survey of local collaborators, as shown in Table B). This approach helped us to ensure that the definition of built environment features and points of interest was consistent across study regions and matched the understandings of local experts. Where OpenStreetMap representation was considered inadequate based on the validation findings, we asked local collaborators to nominate usage of custom data for the study regions. In our study, this was only required for Maiduguri, Nigeria, because we found that, among the 25 study regions, Maiduguri is the only notable outlier regarding the representation of daily living amenities on OpenStreetMap (Table H). The custom destination data for Maiduguri were collected directly from the field by trained field assistants using handheld GPS, and the locations recorded in Microsoft Excel.

**Table G. OpenStreetMap tags (key-value pairs) used to identify urban features**

| Variable | Description | OpenStreetMap tag | |
|---|---|---|---|
| | | *Key* | *Value* |
| healthy_food_market | Healthy Food / Market | | |
| | | shop | supermarket |
| | | supermarket | supermarket |
| | | amenity | supermarket |
| | | building | supermarket |
| | | shop | grocery |
| | | shop | bakery |
| | | shop | pastry |
| | | name | Tortillería |
| | | shop | butcher |
| | | shop | seafood |
| | | shop | fishmonger |
| | | shop | greengrocer |
| | | shop | fruit |
| | | shop | fruits |
| | | shop | vegetables |
| | | shop | deli |
| | | shop | cheese |
| | | amenity | marketplace |
| | | amenity | market |
| | | amenity | market_place |
| | | amenity | public_market |
| | | shop | marketplace |
| | | shop | market |
| convenience | Convenience | | |
| | | shop | convenience |
| | | amenity | fuel |
| | | shop | kiosk |
| | | shop | newsagent |
| | | shop | newsagency |
| | | amenity | newsagency |
| pt_any | Public transport stop (any) | | |
| | | public_transport | platform |
| | | public_transport | stop_position |
| | | highway | bus_stop |
| | | highway | platform |
| | | railway | platform |
| | | public_transport | station |
| | | amenity | ferry_terminal |
| | | railway | tram_stop |
| | | railway | stop |

**Table H. Preliminary OpenStreetMap destination data audit results**

| City | Population estimate[1] | Urban area (km²) | Population per km² | OpenStreetMap destination count | | |
|---|---|---|---|---|---|---|
| | | | | Healthy food | Convenience | Public transport |
| Maiduguri[2] | 1,077,912 | 125 | 8,606 | 23 | 2 | 2 |
| Mexico City | 20,217,799 | 2,312 | 8,745 | 1,491 | 1,684 | 1,765 |
| Baltimore | 621,588 | 229 | 2,709 | 78 | 3,577 | 292 |
| Phoenix | 1,335,215 | 772 | 1,730 | 101 | 4,274 | 711 |
| Seattle | 922,474 | 551 | 1,675 | 371 | 5,984 | 695 |
| São Paulo | 11,770,758 | 1,018 | 11,564 | 1,562 | 5,104 | 1,819 |
| Hong Kong | 7,287,172 | 373 | 19,562 | 748 | 9,366 | 935 |
| Chennai | 6,502,693 | 425 | 15,314 | 209 | 853 | 151 |
| Bangkok | 9,301,270 | 1,190 | 7,814 | 654 | 1,247 | 1,667 |
| Hanoi | 5,936,947 | 1,220 | 4,865 | 537 | 915 | 1,127 |
| Adelaide | 995,195 | 541 | 1,840 | 352 | 5,985 | 297 |
| Melbourne | 3,753,083 | 1,657 | 2,265 | 1,530 | 8,968 | 1,707 |
| Sydney | 4,091,396 | 1,334 | 3,068 | 922 | 7,693 | 1,037 |
| Auckland | 1,247,659 | 468 | 2,666 | 659 | 5,028 | 776 |
| Graz | 280,642 | 69 | 4,085 | 285 | 2,122 | 187 |
| Ghent | 242,180 | 107 | 2,265 | 221 | 1,316 | 162 |
| Olomouc | 86,237 | 27 | 3208 | 60 | 340 | 64 |
| Odense | 151,224 | 56 | 2,688 | 97 | 244 | 65 |
| Cologne | 1,126,218 | 348 | 3,240 | 874 | 3,020 | 889 |
| Lisbon | 585,346 | 85 | 6,891 | 368 | 3,211 | 264 |
| Barcelona | 3,253,794 | 359 | 9,052 | 1,949 | 7,527 | 887 |
| Valencia | 729,856 | 86 | 8,485 | 395 | 1,537 | 152 |
| Vic[3] | 43,920 | 31 | 1,436 | 60 | 91 | 14 |
| Bern | 144,427 | 32 | 4,472 | 149 | 611 | 139 |
| Belfast | 385,650 | 98 | 3,931 | 95 | 298 | 138 |
| **Percentile summary** | | | | | | |
| 0 | 43,920 | 27 | 1,436 | 23 | 2 | 2 |
| 25 | 385,650 | 86 | 2,666 | 101 | 853 | 151 |
| (median) 50 | 1,077,912 | 359 | 3,931 | 368 | 2,122 | 297 |
| 75 | 4,091,396 | 772 | 8,485 | 748 | 5,104 | 935 |
| 100 | 20,217,799 | 2,312 | 19,562 | 1,949 | 9,366 | 1,819 |

[1] Population estimate was derived from the Global Human Settlement 2015 modelled population layer. The value corresponded to the population estimate within the identified urban portion of the city, identified using the Global Human Settlement dataset (with exception of Maiduguri and Vic). This was used as a common reference point for population data for all cities, as it was available at both a recent time point and high resolution (approximately 250sqm grid). The population data was used spatially in the project to indicate relative population density. Values in absolute terms may differ from other estimates, including those officially available, or more recent for the greater city areas which extend beyond urban centres.

[2] It was found that Maiduguri had relatively lower OSM destination counts than other cities of comparable urban size and population in the study. This observation led to the development of methods to include custom destinations, and an independent data-collection effort for Maiduguri.

[3] Vic appeared to be well represented with OSM destinations for its size and population. However, unlike all other cities it did not intersect a 2015 'urban area' identified by the Global Human Settlements dataset. In this regard, it should be noted that Vic is exceptional compared with other cities, whose urban study regions were defined using, or in conjunction with, the GHS urban layer; Vic's study region used in the above is its own administrative boundary (supplied by collaborators).

### A2.3.2 Public transport stops with regular daytime weekday service

The usual daytime weekday frequency of public transport service is a marker of its utility, beyond proximity: if a bus stop is close, but rarely served, its usefulness is more limited than a regularly served stop. There are many factors that could be considered as markers of utility for public transport which were beyond the scope of our study—for example, the cost or time taken to reach a particular destination. For this study, we sought to estimate the percentage of population with access to a public transport stop served frequently enough that an average wait during weekday daytime hours (7am to 7pm) would be approximately 10 minutes; that is, a headway of 20 minutes, with average wait time conceptualised as half of the headway between departures.[4]

Public transport schedule data were obtained from transport agencies in General Transit Feed Specification (GTFS) format. GTFS (https://gtfs.org/, https://developers.google.com/transit/gtfs/reference) provides a format that has been broadly adopted by public transport agencies around the world for publishing and disseminating regularly updated transport schedules since 2005 (https://beyondtransparency.org/chapters/part-2/pioneering-open-data-standards-the-

gtfs-story/). It is an established standard for representing public transportation schedules used for urban transport research in diverse contexts.[5-7] The GTFS dataset is a zipped folder comprising a series of text files that contain comma-delimited data for routes, stop times, dates of service, and agency metadata, which can be merged to provide granular detail of a transport agency's services at each of its stop locations for a particular period.

We identified cities with available GTFS data targeting 2019 for public transport stop location headway analysis. We developed a headway analysis method broadly following Arundel et al.,[8] except that we used headway instead of median inter-arrival time. Methodological details of the analysis are presented as open-source code available at the repository. For fair comparisons, we sought to include cities in the analysis with complete GTFS coverage for the buffered study regions, and with feeds schedules for a time in Spring 2019 when most schools were not on holidays (5 April to 5 June for Northern Hemisphere cities; 8 October to 5 December for those in the Southern Hemisphere). Not all schedules neatly met this criterion. Specifically, we approximated coverage for Chennai using a feed dated to 2010 to capture rail service representation in particular; this compromise was preferable to exclusion, but we recognised it as an important caveat on presenting the results of this city.

Public transport stop locations were analysed to identify those with normal weekday operation time from 7am to 7pm during the time period of interest for that city. The analysis time period was specified for each study region in Table D. Average headways were calculated for each stop location based on the number of daily departures during the analysis time period. Subsequently, three sets of public transport locations were exported to support the accessibility analysis of sample points: 1) any public transport stops (combining GTFS and OSM data); 2) stops with a headway of 20 minutes; and 3) stops with a headway of 30 minutes (for use as a sensitivity analysis for the effect of threshold choice). The modes of public transport in the GTFS specification include tram, metro, rail, bus, ferry, cable tram, aerial lift, funicular, trolleybus, and monorail. Due to some inconsistencies in the GTFS specification by different transport agencies, some cities required custom parameterisation to match the specific implementation used by their transport agencies. For example, Bern and Sydney each required custom specification of route numbers in the configuration of feeds for these cities, to accurately represent their transport modes.
A summary of the transport stop headway analysis is shown in Table I.

**Table I. GTFS stop count within 500m of urban study region boundaries (all modes), where data was identified and retrievable**

| City | GTFS stops within 500m of urban study region boundary | | | | |
|---|---|---|---|---|---|
| | N | Headway | | | |
| | | n ≤ 30 mins | n ≤ 20 mins | % ≤ 30 mins | % ≤ 20 mins |
| Maiduguri | – | – | – | – | – |
| Mexico City | 5,575 | 3,758 | 2,707 | 67.4 | 48.6 |
| Baltimore | 4,145 | 2,863 | 2,171 | 69.1 | 52.4 |
| Phoenix | 4,285 | 3,932 | 1,053 | 91.8 | 24.6 |
| Seattle | 7,225 | 4,531 | 2,837 | 62.7 | 39.3 |
| São Paulo | 20,063 | 19,550 | 17,712 | 97.4 | 88.3 |
| Hong Kong | 4,308 | 4,102 | 3,141 | 95.2 | 72.9 |
| Chennai | 58 | 35 | 35 | 60.3 | 60.3 |
| Bangkok | 6,218 | 6,111 | 3,759 | 98.3 | 60.5 |
| Hanoi | 6,871 | 451 | 152 | 6.6 | 2.2 |
| Adelaide | 6,064 | 4,865 | 2,611 | 80.2 | 43.1 |
| Melbourne | 17,085 | 11,289 | 6,858 | 66.1 | 40.1 |
| Sydney | 21,679 | 12,297 | 6,982 | 56.7 | 32.2 |
| Auckland | 10,044 | 8,252 | 4,416 | 82.2 | 44.0 |
| Graz | – | – | – | – | – |
| Ghent | – | – | – | – | – |
| Olomouc | – | – | – | – | – |
| Odense | 545 | 334 | 231 | 61.3 | 42.4 |
| Cologne | 973 | 881 | 649 | 90.5 | 66.7 |
| Lisbon | 2,119 | 1,878 | 1,481 | 88.6 | 69.9 |
| Barcelona | 3,138 | 2,828 | 2,485 | 90.1 | 79.2 |
| Valencia | 1,175 | 1,155 | 1,127 | 98.3 | 95.9 |
| Vic | – | – | – | – | – |
| Bern | 410 | 372 | 321 | 90.7 | 78.3 |
| Belfast | 1,550 | 1,159 | 875 | 74.8 | 56.5 |

**A2.3.3 Public open space**

Parks, nature reserves, plazas, and squares could all be considered areas of public open space: open areas where people may gather for leisure. The identification of public open space using OpenStreetMap is a distinct challenge to other kinds of destinations, which are usually localised as discrete 'points'. Public open spaces are areas, not points, and so more complex topological relationships must be considered in addition to mapping of classification tags. In terms of representation on OpenStreetMap, open spaces may be located next to each other; in effect, a jointly larger area than contiguous polygonal boundaries taken individually would suggest. They may also be nested within larger areas of open space, which may or may not be publicly accessible.

A series of logical queries were used to first identify areas of open space; meeting any one of these was grounds for consideration as a potential public open space. These are summarised in Table J.

A layer of areas that were categorically not to be considered as public open space was created, and any portions of potential areas of open space that overlap areas marked for exclusion were also excluded. For example, if there was an area coded to suggest it could be a natural area that might potentially be an open space (e.g. 'boundary=nature_reserve'), but was located entirely within an area with a military or industrial land use, or was otherwise tagged to indicate that access was not public (e.g. for employees or staff only, private, or otherwise inaccessible), this was not considered an area of public open space and was excluded. In addition, areas of public open space that were not larger than a minimum size threshold (10 $m^2$) were excluded to ensure incidental slivers of public land like small nature strips did not contribute spuriously to estimates of access to public open space.

Once areas of public open space had been identified, proxy locations for possible entry points were created at regular intervals (every 20 metres) on the sections of the boundaries of those areas of public open space which are within 30 metres of the road network.[9] These pseudo-entry points can be used to evaluate distance to the nearest public open space, regardless of the fact that our data cannot be used to infer where the true entry locations are (and in many cases, there may be no formal entry points). Sets of pseudo-entry points were respectively exported for areas of public open space of any size, and those with a public area of greater than 1.5 hectares. The final sets of access points were used in origin–destination network analyses evaluating network distance to near public open space.

**Table J. Processing parameters for areas of open space**

| Description | Values |
|---|---|
| *Required tags* | |
| These tag keys are assumed to be present on OSM features in order to evaluate values (null or otherwise); if they don't exist, they are created with null values when setting up OSM data. | beach,river,water,waterway,wetland,access,leisure,natural,sport,landuse,playground,boundary,recreation_ground,golf,military,agricultural,forestry,tourism,shop,supermarket,amenity,building,community_centre,place_of_worship,tourism,cuisine,gambling,place,highway,swimming_pool,garden:type |
| *Specific inclusion criteria* | |
| These are specific inclusion criteria which are to be joined as an OR query, using a specific table alias. | leisure IS NOT NULL,beach IS NOT NULL,place = 'square',highway = 'pedestrian' |
| *Land use considerations* | |
| These tags are to be joined in a comma-separated list, once they have been enclosed in single quotation marks. | common,conservation,forest,garden,leisure,park,recreation_ground,sport,trees,village_green,winter_sports,wood,dog_park,nature_reserve,off_leash,sports_centre,riverbank,beach |
| *Boundary considerations* | |
| These tags are to be joined in a comma-separated list, once they have been enclosed in single quotation marks. | national_park,nature_reserve,forest,state_forest,state_park,regional_park,park,county_park |
| *Exclusion on keys* | |
| Tags are joined using OR logic, and are used to define exclusion criteria where values are not null. | military,agricultural,forestry |
| *Exclusion on values* | |
| Where the keys in this json snippet are found to have values in their associated lists, these are grounds for exclusion. This snippet is used to format exclusion criteria using OR logic. | |
| *Water features* | |
| These tags are to be joined in a comma-separated list, once they have been enclosed in single quotation marks. These value tags indicate areas of blue space, which will be excluded when determining both access to and size of parks. | atoll,awash_rock,bay,coastal,coastline,coastline_old,glacier,high-water,hot_spring,island,islet,lake,marsh,oasis,old_coastline_import,peninsula,pond,river,river_terrace,riverbank,riverbed,shoal,spring,strait,stream,swamp,swimming_pool,underwater_rock,unprotected_spring,unprotected_well,water,water_park,waterfall,waterhole,waterway,wetland |
| *Water sports* | |
| These tags are to be joined in a comma-separated list, once they have been enclosed in single quotation marks. These tags are indicative of water features, which will be excluded from consideration. | swimming,surfing,canoe,scuba_diving,rowing,sailing,fishing,water_ski,water_sports,diving,windsurfing,canoeing,kayak |

**Table J. (cont.) Processing parameters for areas of open space**

| Description | Criteria |
|---|---|
| *Linear features* | |
| These tags are to be joined in a comma-separated list, once they have been enclosed in single quotation marks. These are linear features which may or may not be public open space, but need to be treated with care so they do not link together to form large areas of open space. | river,riverbank,riverbed,strait,waterway,stream,ditch,river,drain,canal,rapids,drystream,brook,derelict_canal,fairway |
| *Linear feature criteria* | |
| This is an SQL expression used to define a linear feature based on morphological or attribute criteria. | (area_ha > 0.5 AND roundness < 0.25) OR (waterway IS NOT NULL OR river IS NOT NULL) |
| *Tags to exclude* | |
| These tags are to be joined in a comma-separated list, once they have been enclosed in single quotation marks. | addr:city,addr:full,addr:place,addr:postcode,addr:province,addr:street,website,wikipedia,description,addr:housenumber,addr:interpolation,designation,email,phone,ref:capad2014_osm,nswlpi:cadid,wikidata,url |
| *Tags to retain* | |
| Keys to be joined in a comma-separated list (already in double quotation marks). | "os_id","area_ha","beach","river","public_access","within_public","amenity","access","boundary","golf","landuse","leisure","natural","playground","recreation_ground","sport","tourism","water","wetland","waterway","wood","water_feature","min_bounding_circle_area","min_bounding_circle_diameter","roundness","linear_feature","acceptable_linear_feature","highway","place" |
| *Exclusion criteria for public spaces* | |
| Where the keys in this json snippet are found to have values in their associated lists, these are used to indicate areas that are not to be flagged as public themselves, although they may be located inside larger areas of public open space. | {"amenity":["aged_care","animal_boarding","allotments","animal_boarding","bank","bar","biergarten","boatyard","carpark","childcare","casino","church","club","club_house","college","conference_centre","embassy","fast_food","garden_centre","grave_yard","hospital","gym","kindergarten","monastery","motel","nursing_home","parking","parking_space","prison","retirement","retirement_home","retirement_village","school","scout_hut","university"],"leisure":["garden","golf_course","horse_riding","pitch","racetrack","summer_camp","sports_club","stadium","sports_centre"],"building":["yes"],"area":["school"],"natural":["fell","bay","bog","cliff","geyser","reef","scrub","sinkhole","strait","volcano","wetland","wood","water"],"recreation_ground":["showground","school_playing_field","horse_racing","show_grounds","school_playing_fields"],"sport":["archery","badminton","bocce","boules","bowls","croquet","dog_racing","equestrian","futsal","gokarts","golf","greyhound_racing","horse_racing","karting","lacross","lacrosse","lawn_bowls","motocross","motor","motorcycle","polo","shooting","snooker","trugo"],"access":["customers","private","no"],"tourism":["alpine_hut","apartment","aquarium","bed_and_breakfast","caravan_site","chalet","gallery","guest_house","hostelsd","hotel","information","motel","museum","theme_park","zoo"],"garden:type":["residential","residental","private","commercial","pub","school","roof_garden"]} |

## A2.4. Processing sampling estimates, and aggregating indicators

Sample points estimates, urban neighbourhood grid, and city summaries of spatial indicators were processed following our open-source software framework. Specific methods applied to the case study cities are described in the current manuscript, and more details for the software framework are presented in our method paper.

A completed list of descriptive variables and spatial indicators is shown in Table K.

**Table K. Descriptive variables and spatial indicators of urban design and transport features that promote health and sustainability**

| Indicator | Units | Comments | Data notes |
|---|---|---|---|
| **Descriptive variables** | | | |
| *Urban study region area* | km² | Urban study region boundaries | 1 2 |
| *Urban study region population estimate* | population counts | Overall urban population estimate for the study region | 1 2 3 |
| *GNI per capita classification (2021)* | Lower / Lower-middle / Upper-middle / High | The World Bank 2021 fiscal year GNI per capita classification (Atlas method) | 7 |
| *Population per square kilometre* | | Urban population density estimate | 1 2 3 |
| *Sample point count* | points | Sample points generated every 30m along pedestrian network for 250m diagonal urban hexagonal grid regions with population estimates greater than 5 | 1 2 3 4 |
| **Percentage of population with access to a …** | | Pedestrian network distance indicators of access within 500m, evaluated for hexagonal grid cells and weighted for population percentage estimate | |
| *Healthy food market or supermarket* | % | Including supermarket, fresh food grocers, and food markets that sell fresh food ingredients that support options for healthier choices | 1 2 3 4 5 |
| *Convenience store* | % | Including newsagencies and convenience stores where basic staples and prepared food items may be purchased | 1 2 3 4 5 |
| *Public transport stop* | | | |
|     *any (OSM or GTFS)* | % | The best estimate of access using either OpenStreetMap or GTFS sourced data (noting that both are imperfect) | 1 2 3 4 5 6 |
|     *with regular service (≤30 mins)* | % | Accounting for 30 minute or less average weekday daytime service frequency | 1 2 3 4 5 |
|     *with regular service (≤20 mins)* | % | Accounting for 20 minute or less average weekday daytime service frequency | 1 2 3 4 5 |
| *Public open space* | | | |
|     *any* | % | Includes squares, parks, and other publicly accessible areas of natural land for leisure and recreation purposes. | 1 2 3 4 5 |
|     *1.5 hectares or larger* | % | As above, restricted to areas with large publicly accessible areas | 1 2 3 4 5 |
| **Local walkability indicators** | | Unweighted (spatial) and population-weighted city summary measures, both within city (in absolute units), and relative to all cities (as Z-scores) | |
| *Population per square kilometre* | density | Average of estimated population density within 1,000m local walkable catchments | 1 2 3 4 |
| *Intersections per square kilometre* | density | Average of estimated intersection density within 1,000m local walkable catchments | 1 2 3 4 |
| *Daily living score* | Sum of binary scores | Sum of binary access indicator scores to supermarket, convenience store, and public transport (OSM or GTFS) servicing, as a proxy of land use mix | 1 2 3 4 5 6 |
| *Walkability index* | sum of z-scores | Sum of Z-scores (within city, and between city) of population density, intersection density, and daily living score | 1 2 3 4 5 6 |

1 GHS UCDB 2015 (See Table C)
2 Custom boundaries (see Appendix A1.3)
3 GHS POP 2015 (See Table C)
4 OpenStreetMap derived pedestrian network, using OSMnx (See Table C)
5 OpenStreetMap derived points of interest (see Appendix A2.3.1)
6 GTFS (mixed sources, targeting 2019; see Appendix A2.3.2)
7 World Bank list of economies (June 2020), http://databank.worldbank.org/data/download/site-content/CLASS.xls , last accessed 21 April 2021

## A3. Mapping

In addition to some automated mapping using Python with Geopandas and Matplotlib, we also prepared bespoke maps for our publication using Giulio Fattori's QGIS multi-map plug in .[10] Our maps made use of Fabio Crameri's Scientific Colour Maps Batlow colour scale .[11] Please see our extensive supplementary map document in Appendix B for detailed maps for each city.

# Appendix B: Map reports

*(City map report sections are identified in the PDF bookmarks)*

**Contents**

# Relative urban neighbourhood walkability across 25 global cities   *(click a city to view map report)*

| <-3 |
| -3 to -2 |
| -2 to -1 |
| -1 to 0 |
| 0 to 1 |
| 1 to 2 |
| 2 to 3 |
| ≥3 |

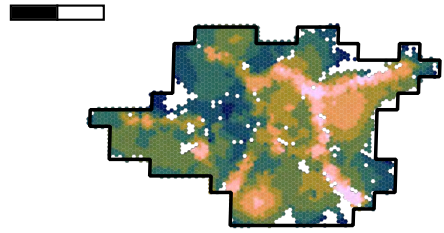
Africa, Nigeria, Maiduguri

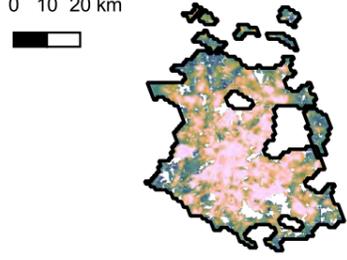
America, North, Mexico, Mexico City

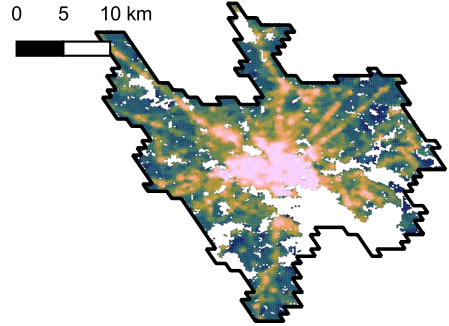
America, North, United States, Baltimore

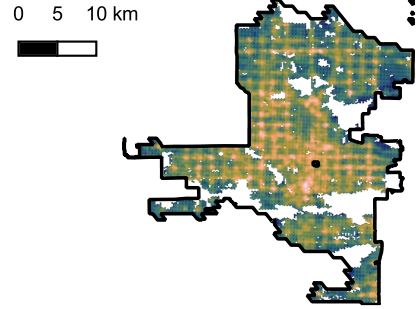
America, North, United States, Phoenix

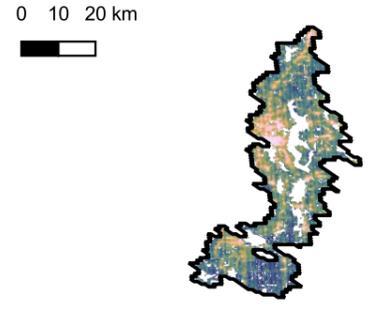
America, North, United States, Seattle

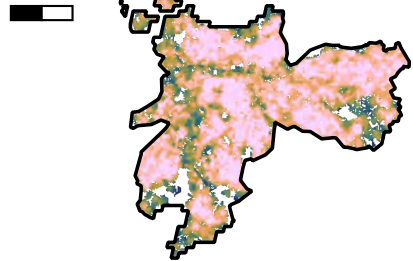
America, South, Brazil, São Paulo

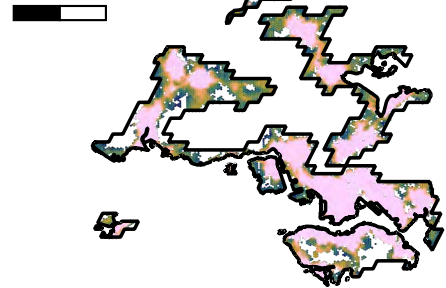
Asia, China (SAR), Hong Kong

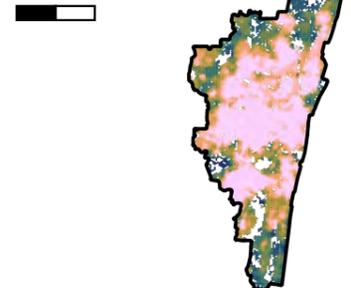
Asia, India, Chennai

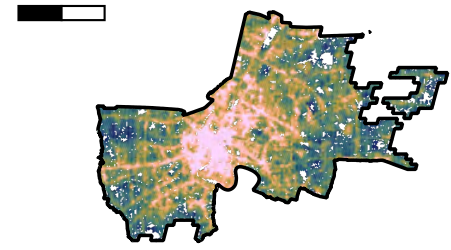
Asia, Thailand, Bangkok

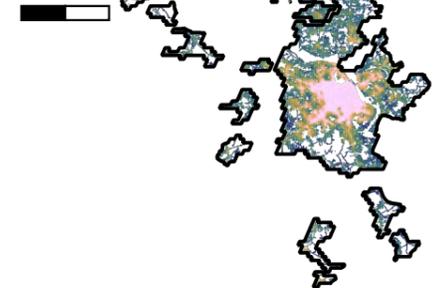
Asia, Vietnam, Hanoi

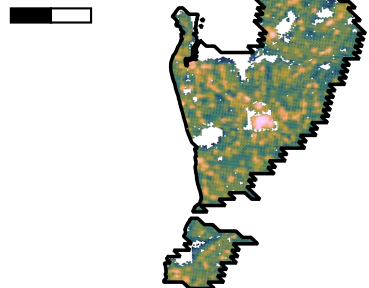
Australasia, Australia, Adelaide

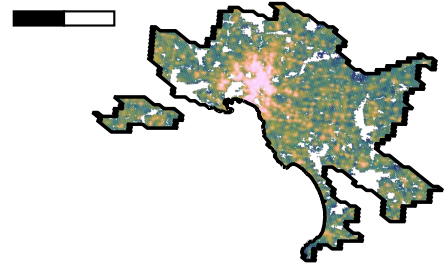
Australasia, Australia, Melbourne

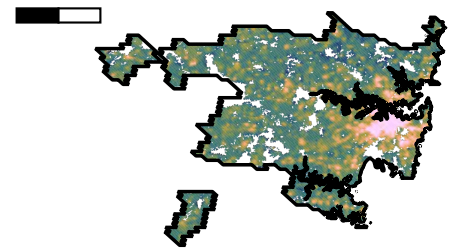
Australasia, Australia, Sydney

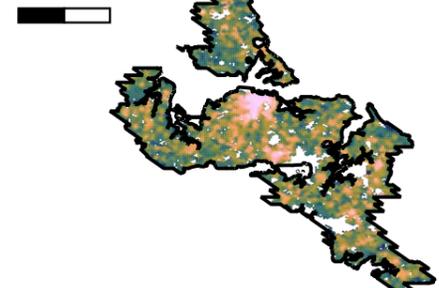
Australasia, New Zealand, Auckland

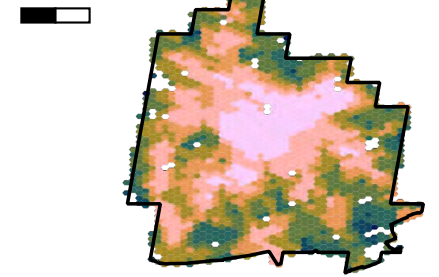
Europe, Austria, Graz

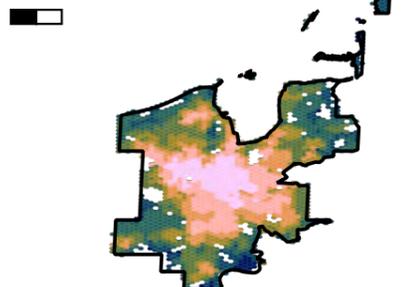
Europe, Belgium, Ghent

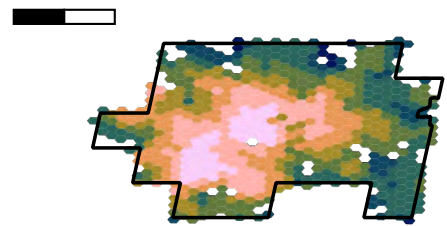
Europe, Czech Republic, Olomouc

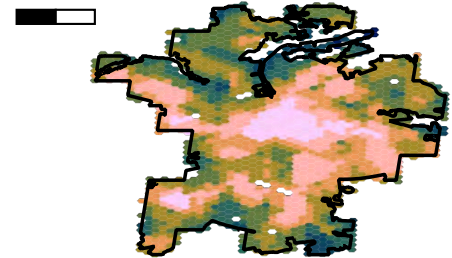
Europe, Denmark, Odense

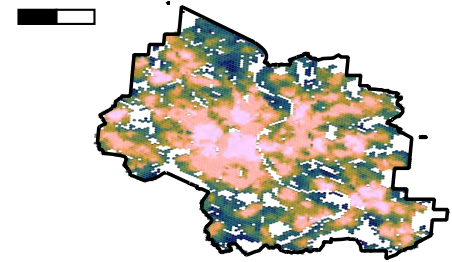
Europe, Germany, Cologne

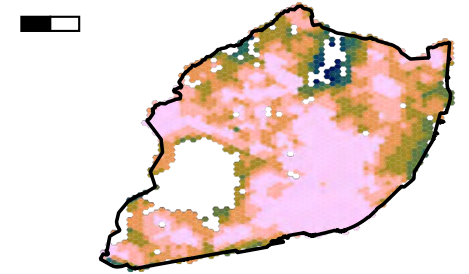
Europe, Portugal, Lisbon

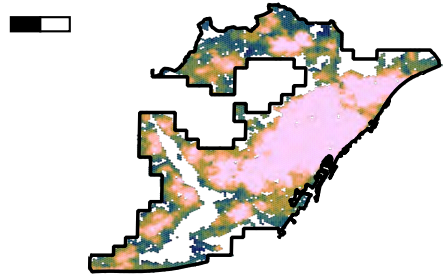
Europe, Spain, Barcelona

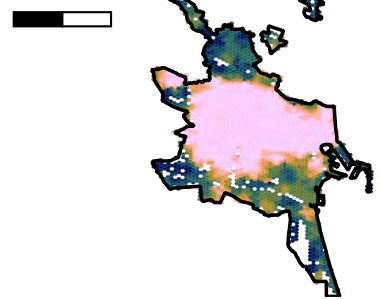
Europe, Spain, Valencia

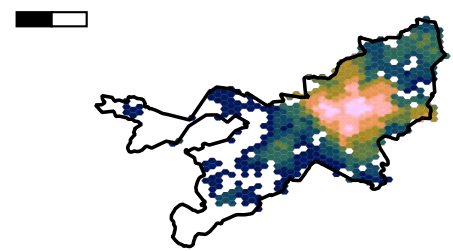
Europe, Spain, Vic

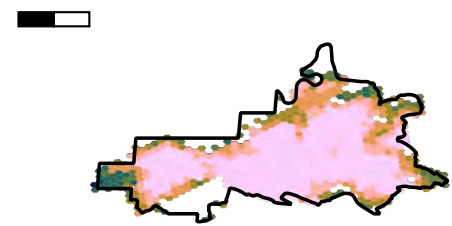
Europe, Switzerland, Bern

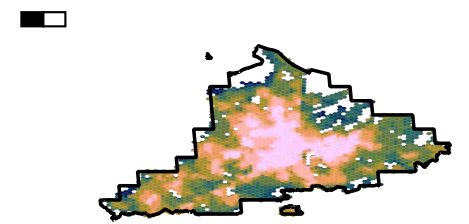
Europe, United Kingdom, Belfast

# Access to large public open space within 500m across 25 global cities*

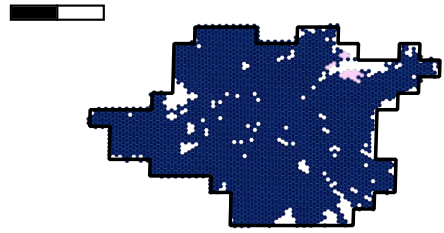
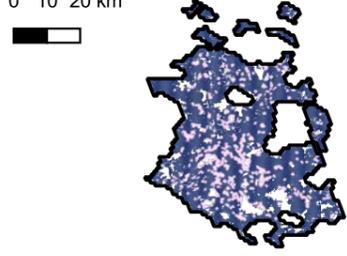
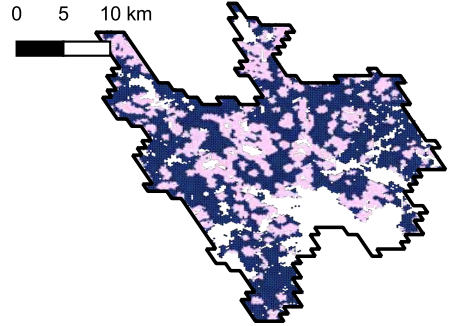
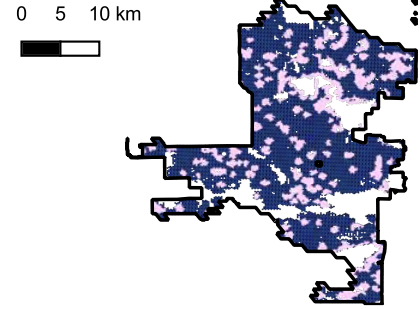
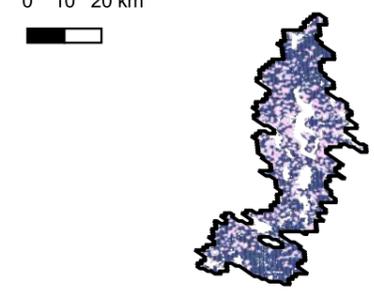
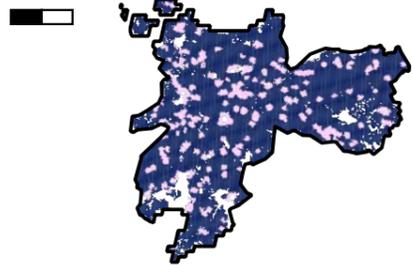
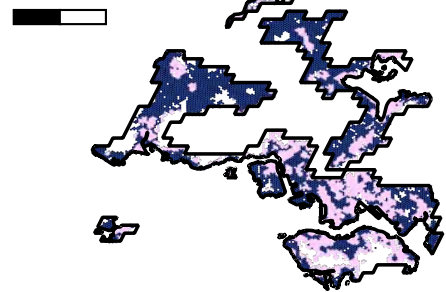
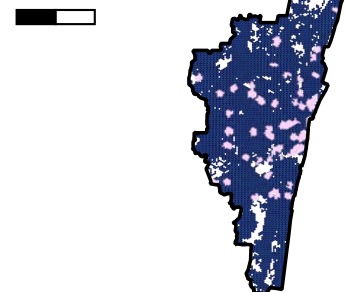
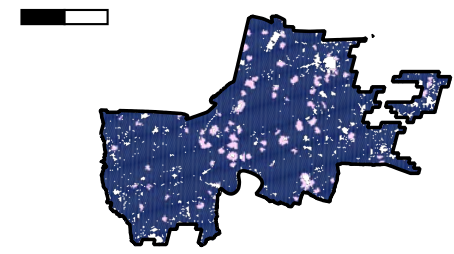
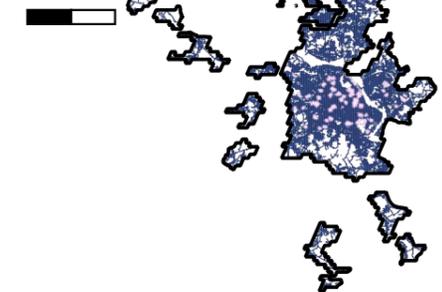
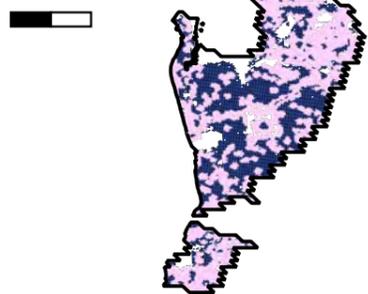
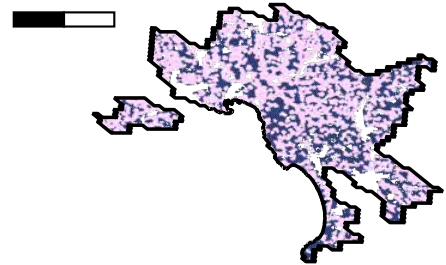
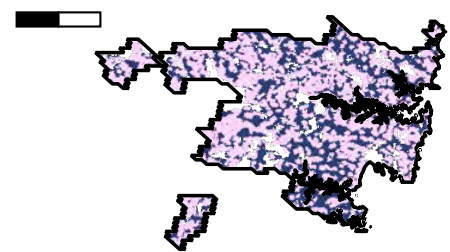
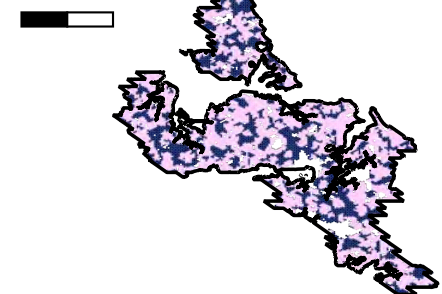
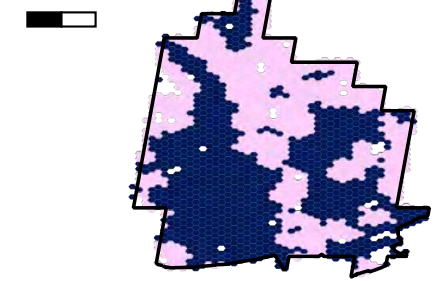
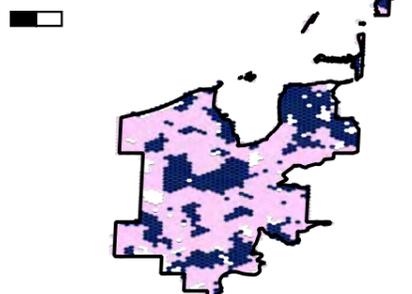
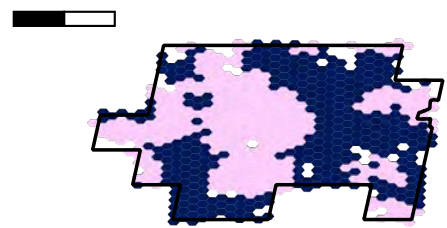
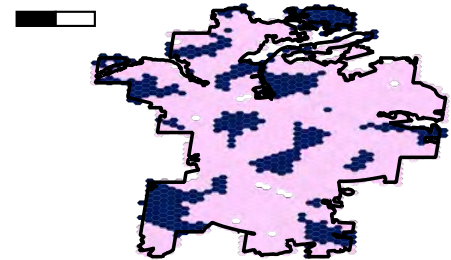
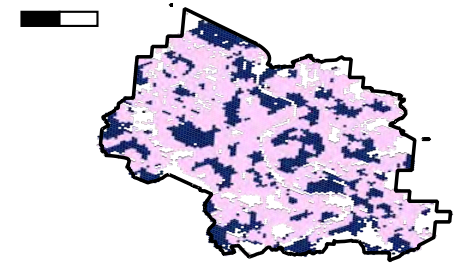
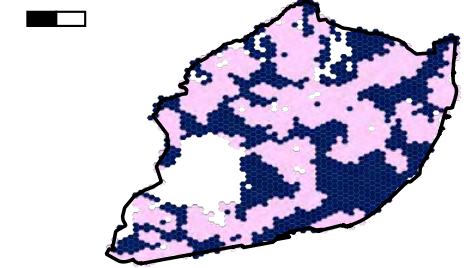
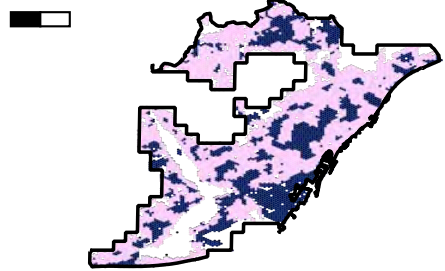
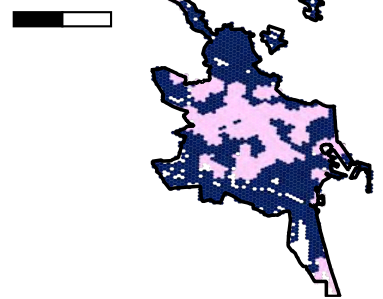
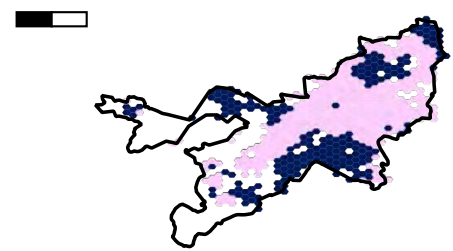
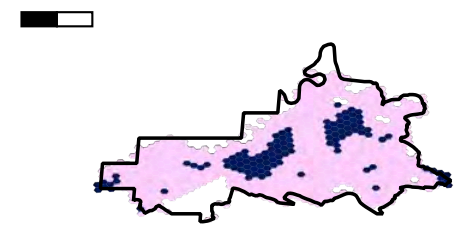
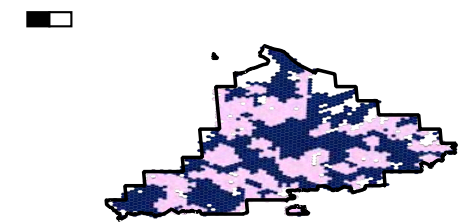

\* In this map access for urban neighbourhoods was considered achieved when at least half of the sampled walkable area was estimated to within be located within 500m of areas identified as public open space of size 1.5 hectares or larger.

# Access to a fresh food outlet (fruit and vegetable grocer, market, or supermarket) within 500m across 25 global cities*

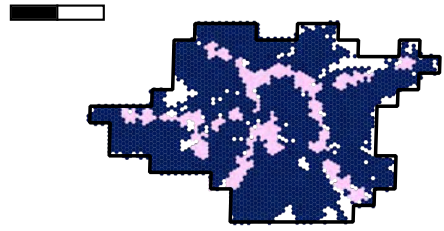
Africa, Nigeria, Maiduguri

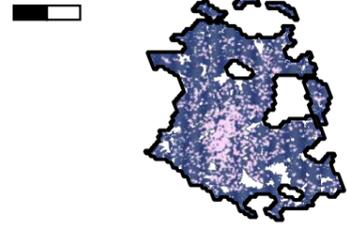
America, North, Mexico, Mexico City

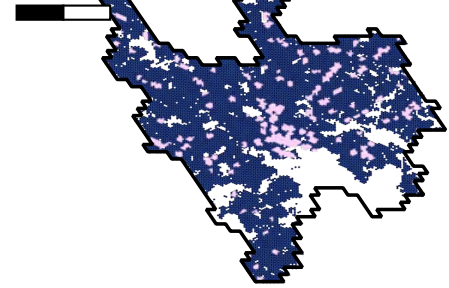
America, North, United States, Baltimore

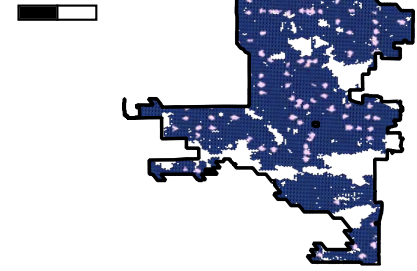
America, North, United States, Phoenix

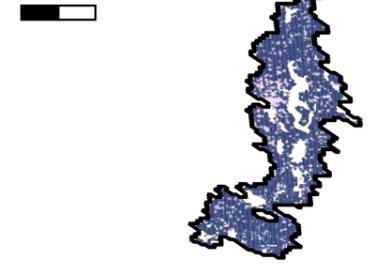
America, North, United States, Seattle

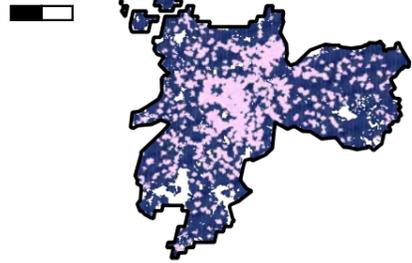
America, South, Brazil, São Paulo

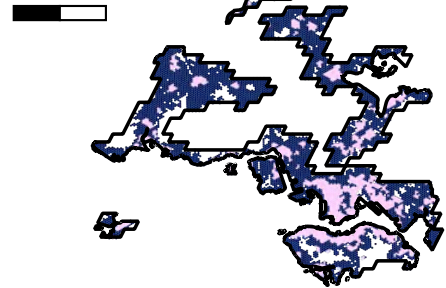
Asia, China (SAR), Hong Kong

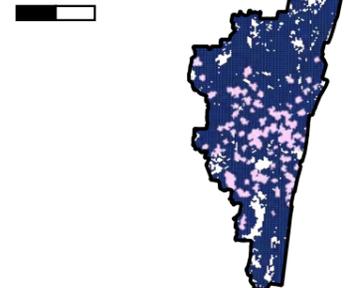
Asia, India, Chennai

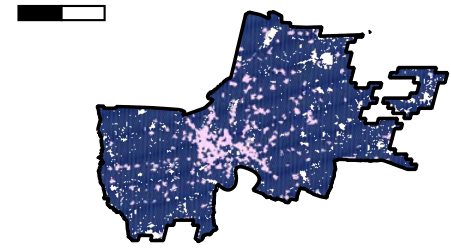
Asia, Thailand, Bangkok

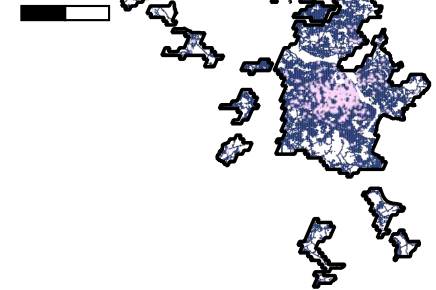
Asia, Vietnam, Hanoi

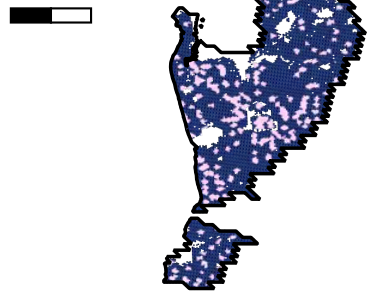
Australasia, Australia, Adelaide

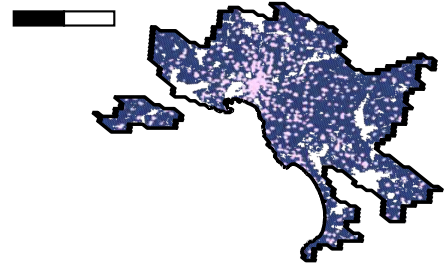
Australasia, Australia, Melbourne

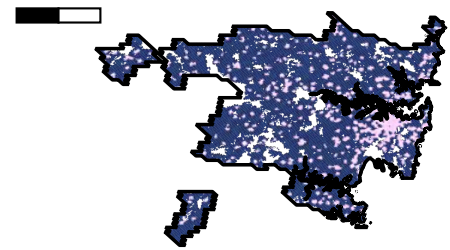
Australasia, Australia, Sydney

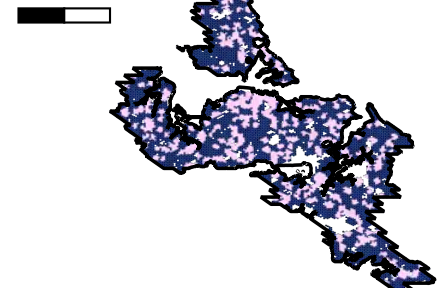
Australasia, New Zealand, Auckland

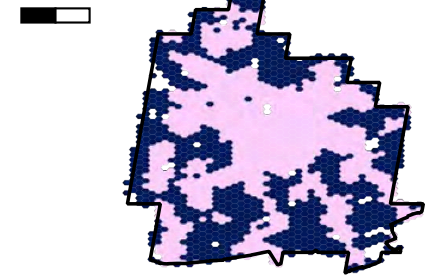
Europe, Austria, Graz

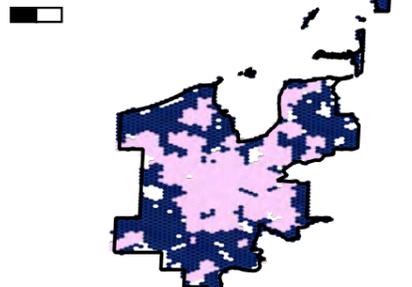
Europe, Belgium, Ghent

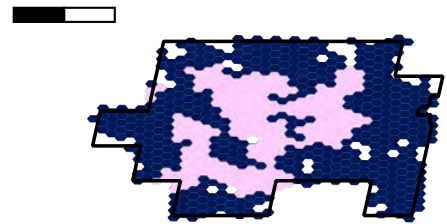
Europe, Czech Republic, Olomouc

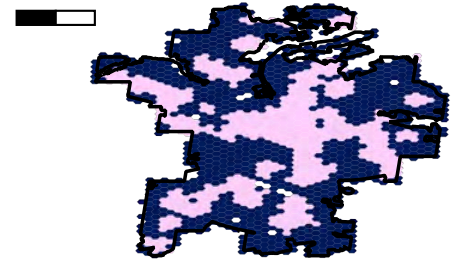
Europe, Denmark, Odense

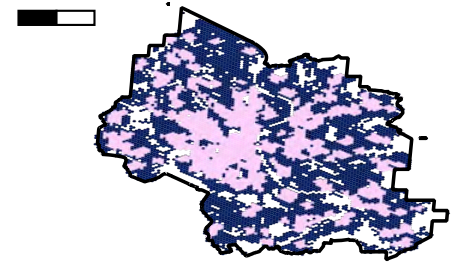
Europe, Germany, Cologne

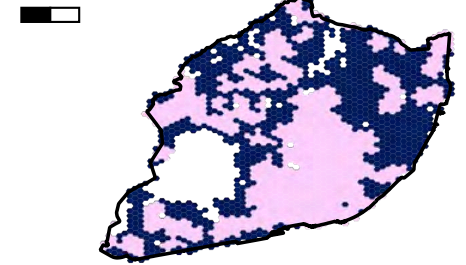
Europe, Portugal, Lisbon

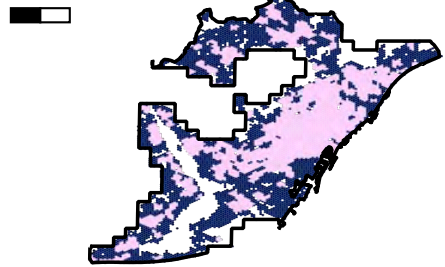
Europe, Spain, Barcelona

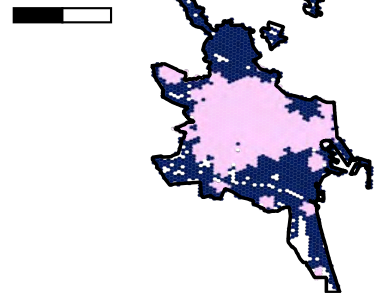
Europe, Spain, Valencia

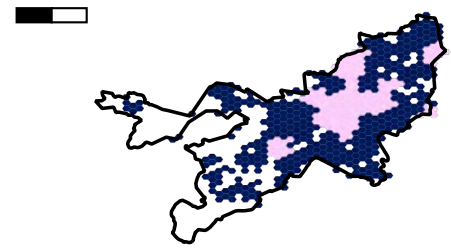
Europe, Spain, Vic

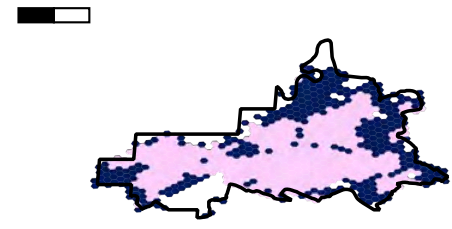
Europe, Switzerland, Bern

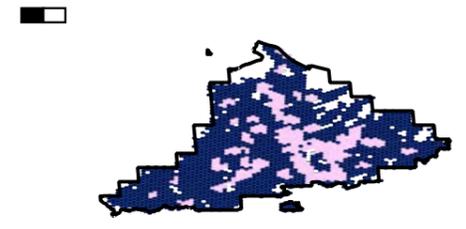
Europe, United Kingdom, Belfast

Legend: No / Yes

* In this map access for urban neighbourhoods was considered achieved when at least half of the sampled walkable area was estimated to within be located within 500m of identified features of interest.

# Access to any public transport within 500m across 25 global cities*

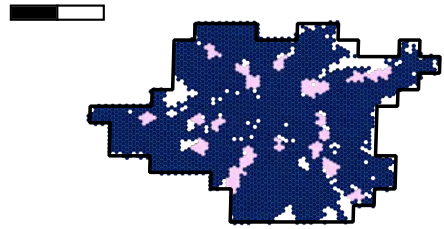
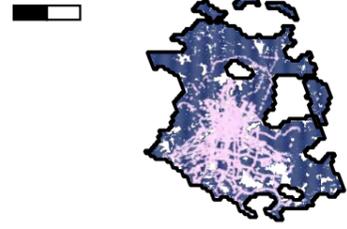
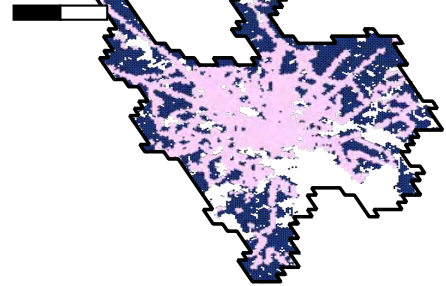
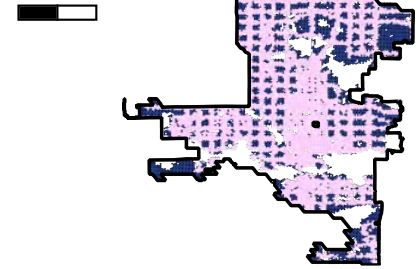
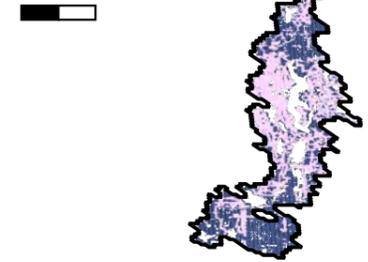
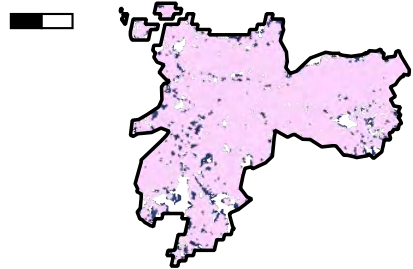
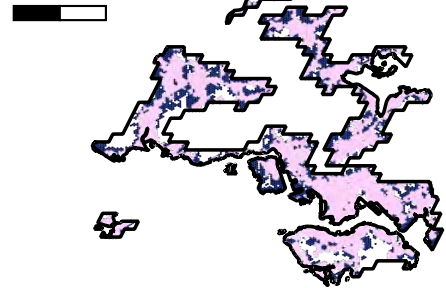
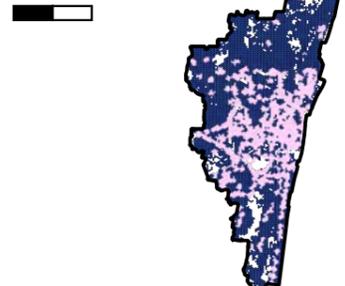
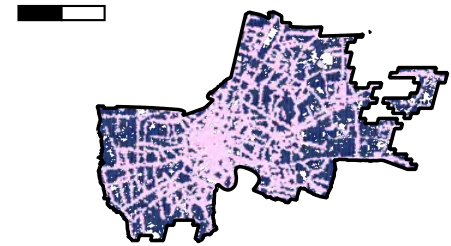
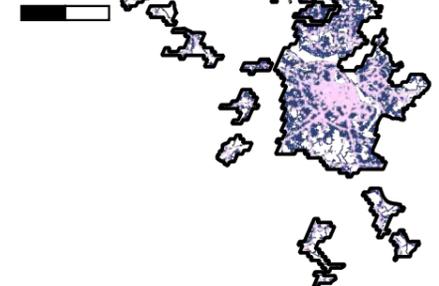
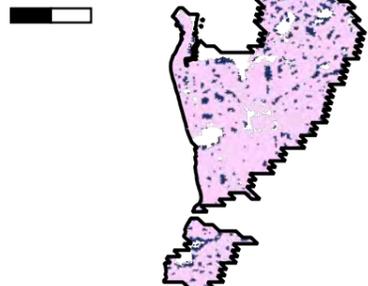
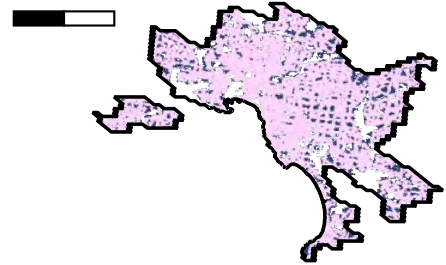
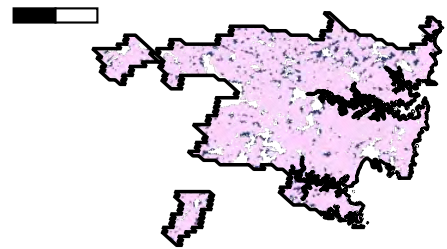
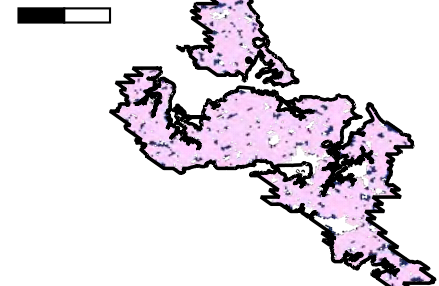
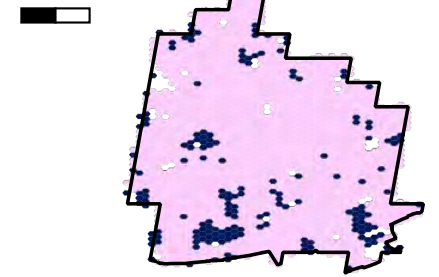
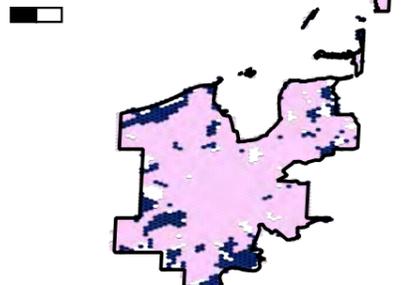
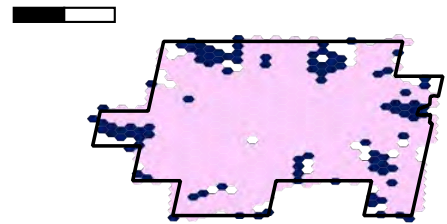
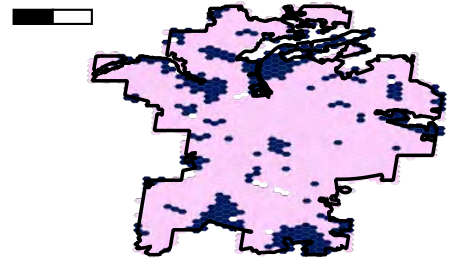
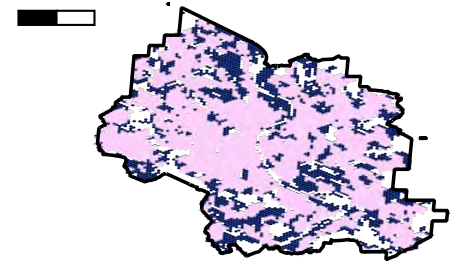
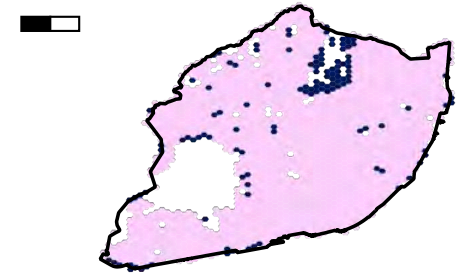
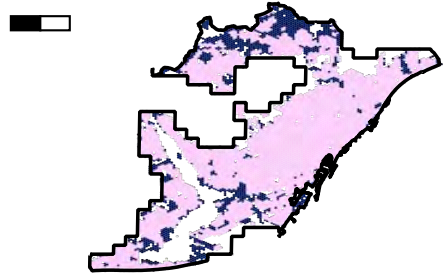
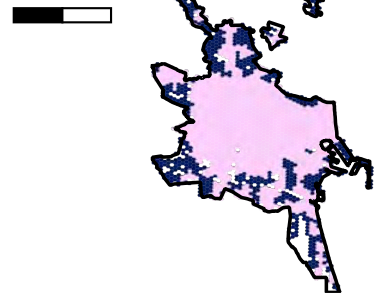
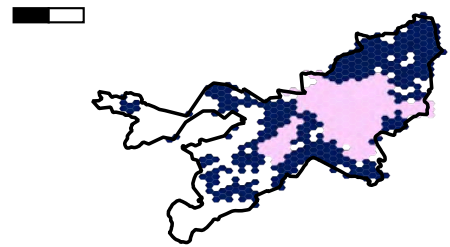
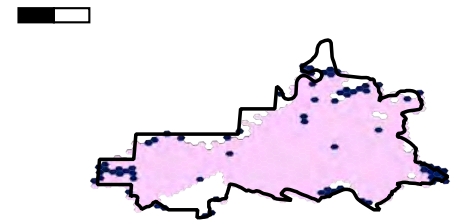
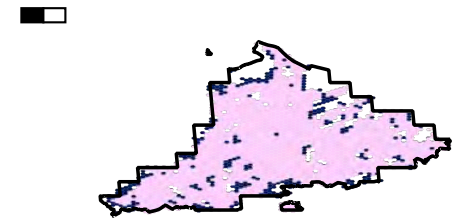

Legend: No / Yes

* In this map access for urban neighbourhoods was considered achieved when at least half of the sampled walkable area was estimated to within be located within 500m of identified features of interest.

# Access to public transport with day time weekday service frequency of 20 minutes or better within 500m across 25 global cities, where GTFS data were available*

Legend: No (dark blue), Yes (pink)

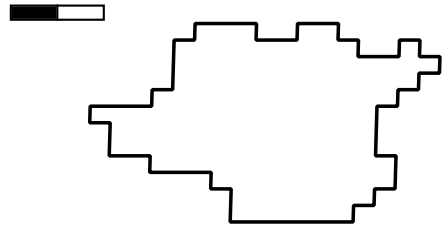
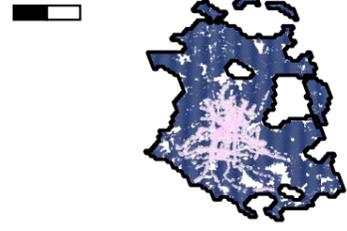
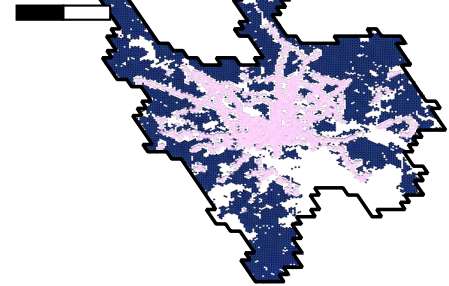
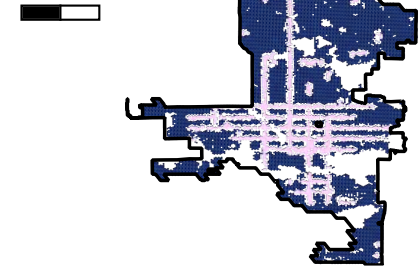
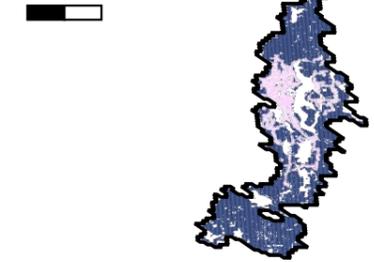
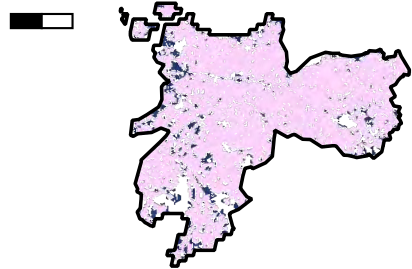
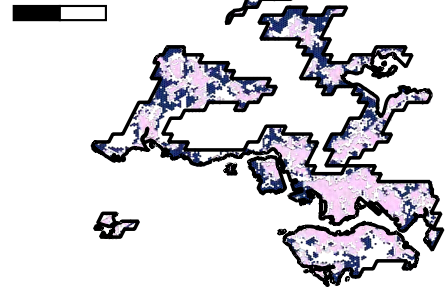
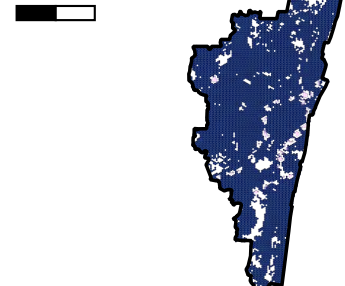
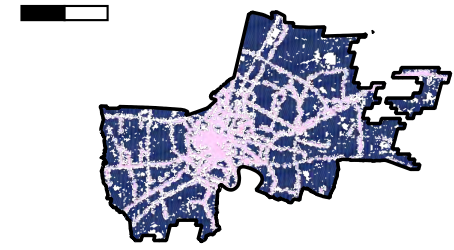
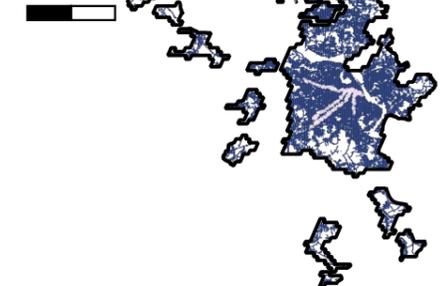
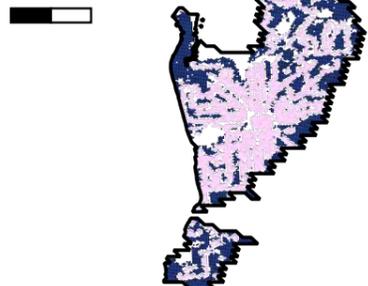
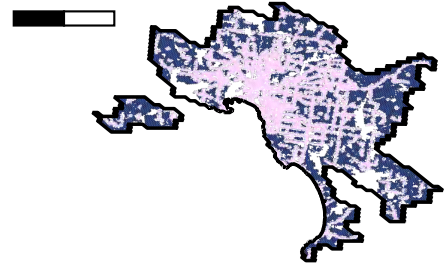
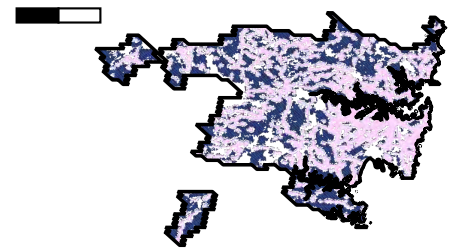
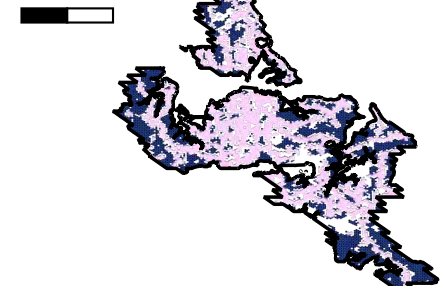
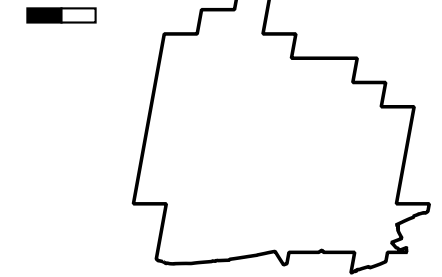
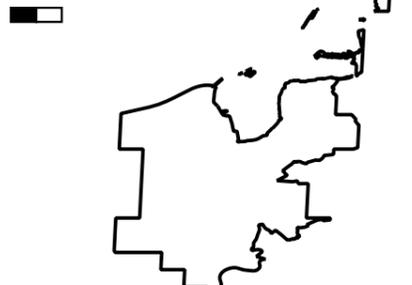
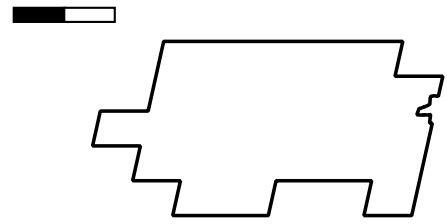
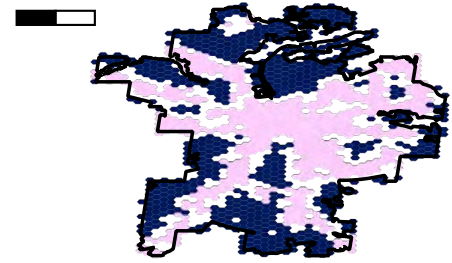
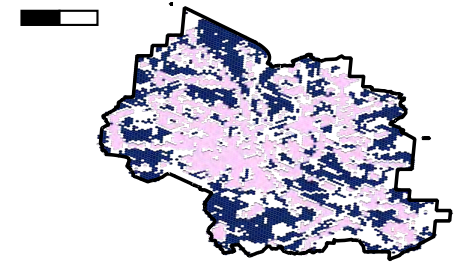
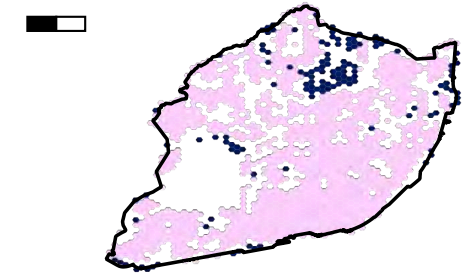
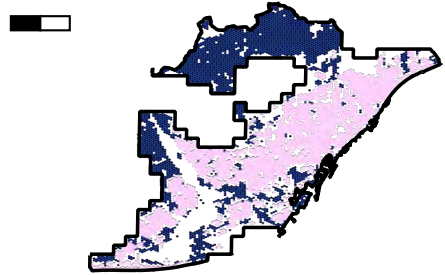
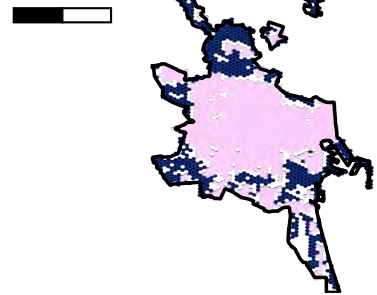
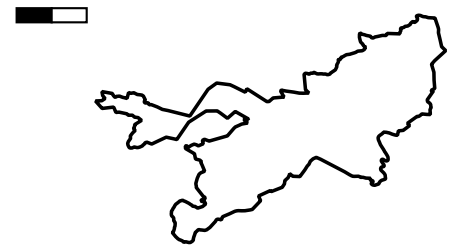
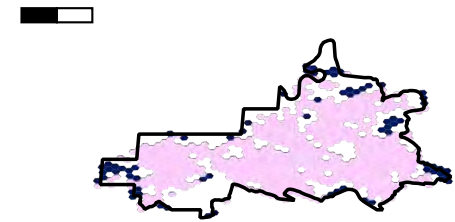
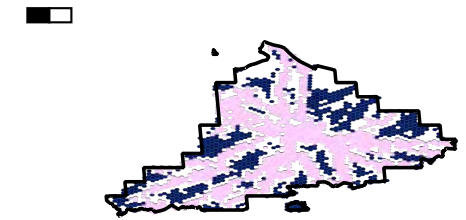

City panels (row by row):
- Africa, Nigeria, Maiduguri
- America, North, Mexico, Mexico City
- America, North, United States, Baltimore
- America, North, United States, Phoenix
- America, North, United States, Seattle
- America, South, Brazil, São Paulo
- Asia, China (SAR), Hong Kong
- Asia, India, Chennai
- Asia, Thailand, Bangkok
- Asia, Vietnam, Hanoi
- Australasia, Australia, Adelaide
- Australasia, Australia, Melbourne
- Australasia, Australia, Sydney
- Australasia, New Zealand, Auckland
- Europe, Austria, Graz
- Europe, Belgium, Ghent
- Europe, Czech Republic, Olomouc
- Europe, Denmark, Odense
- Europe, Germany, Cologne
- Europe, Portugal, Lisbon
- Europe, Spain, Barcelona
- Europe, Spain, Valencia
- Europe, Spain, Vic
- Europe, Switzerland, Bern
- Europe, United Kingdom, Belfast

* In this map access for urban neighbourhoods was considered achieved when at least half of the sampled walkable area was estimated to within be located within 500m of identified features of interest.

# Africa, Nigeria, Maiduguri

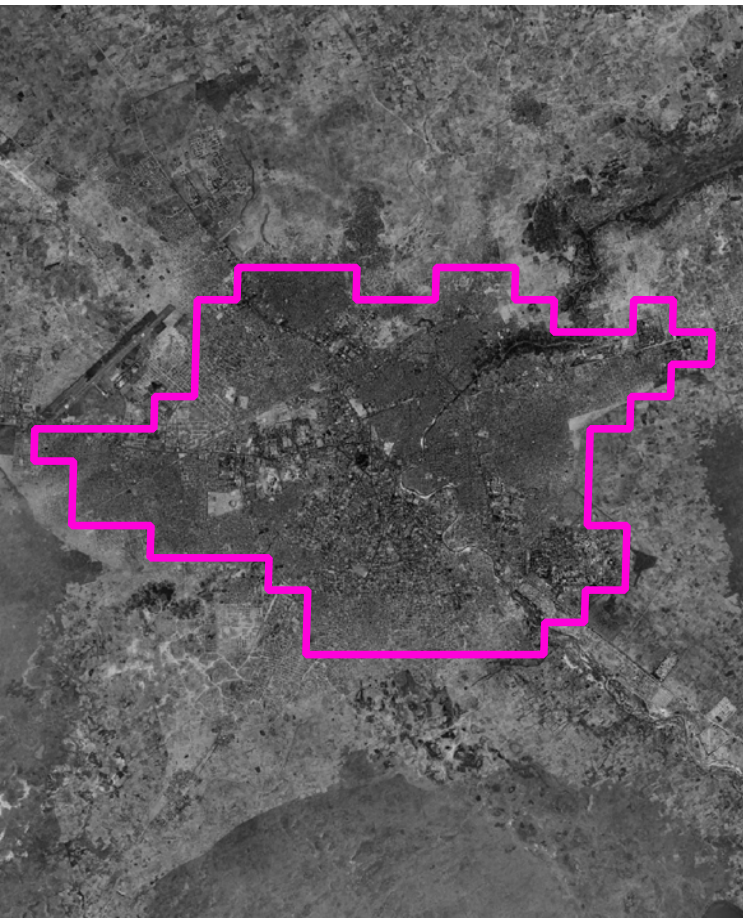
Satellite imagery of urban study region (Bing)

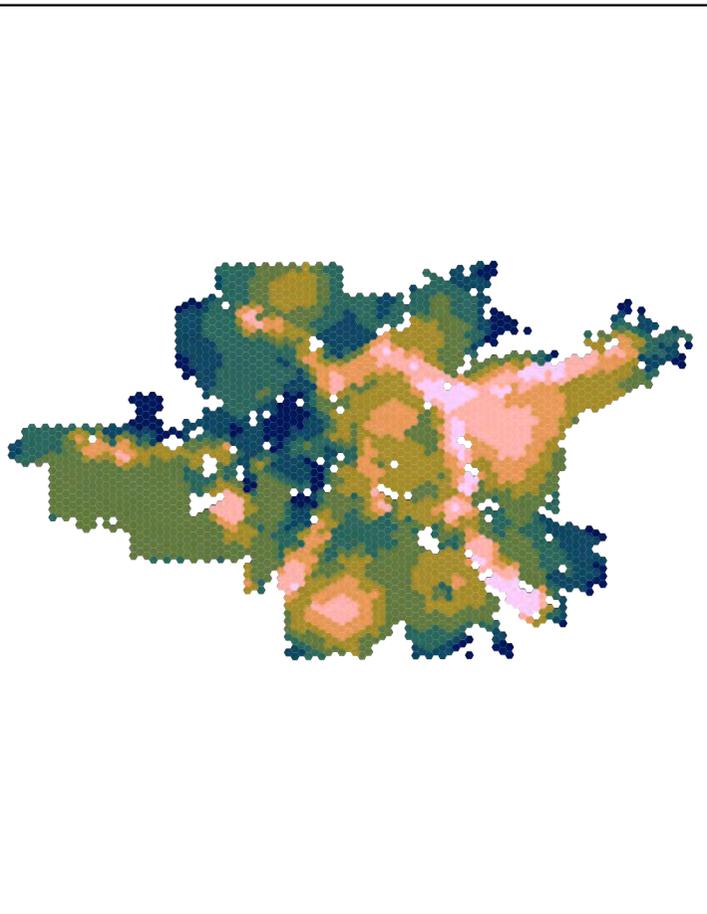
Walkability, relative to city

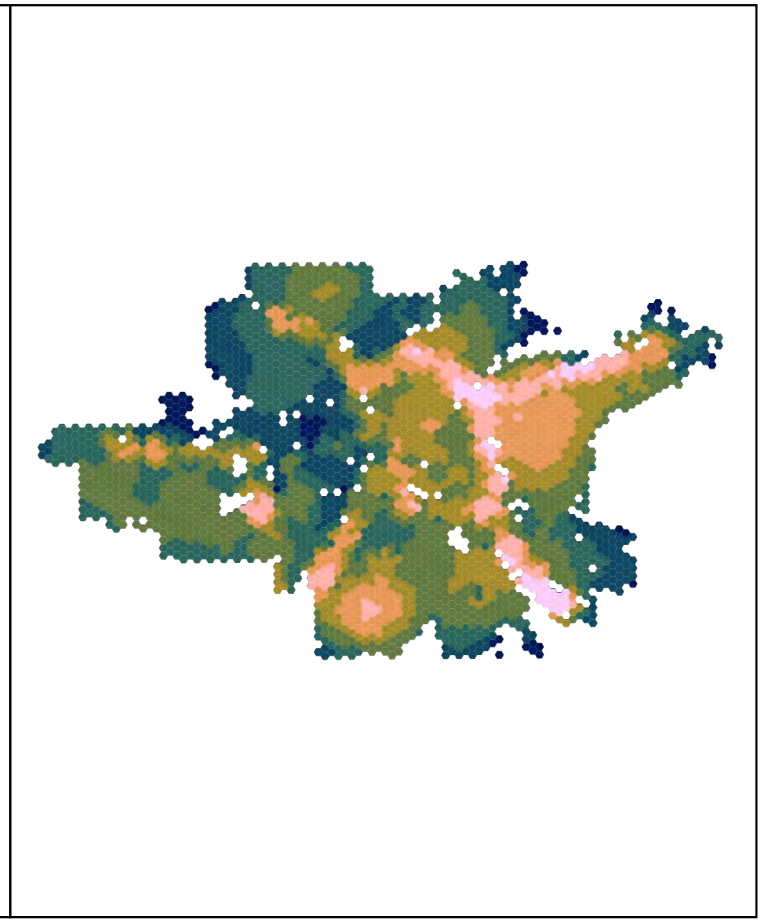
Walkability, relative to 25 global cities

Urban boundary

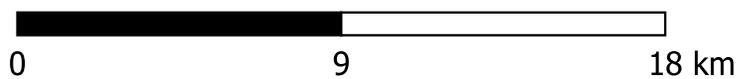
0 — 9 — 18 km

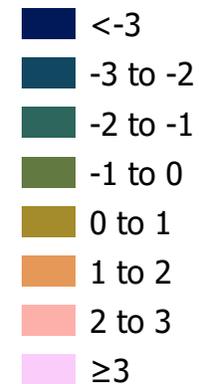
Walkability score
- <-3
- -3 to -2
- -2 to -1
- -1 to 0
- 0 to 1
- 1 to 2
- 2 to 3
- ≥3

Walkability relative to all cities by component variables (2D histograms), and overall (histogram)

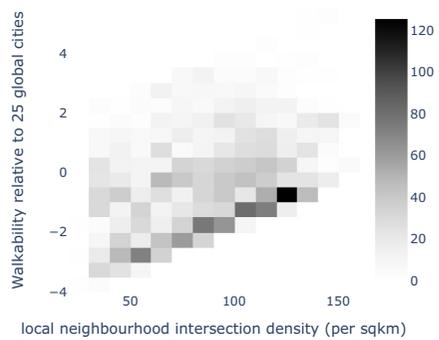
local neighbourhood intersection density (per sqkm)

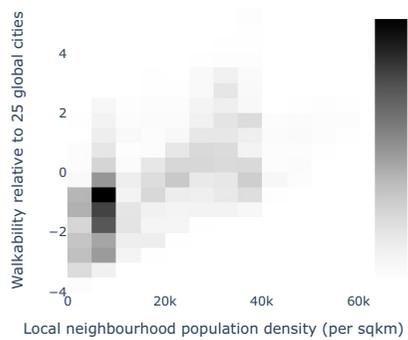
Local neighbourhood population density (per sqkm)

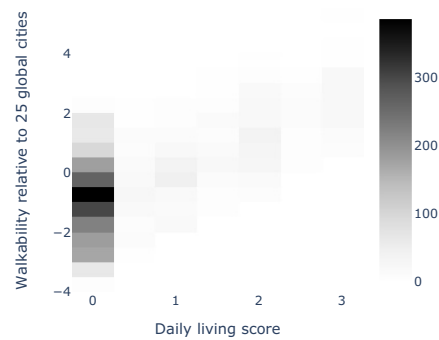
Daily living score

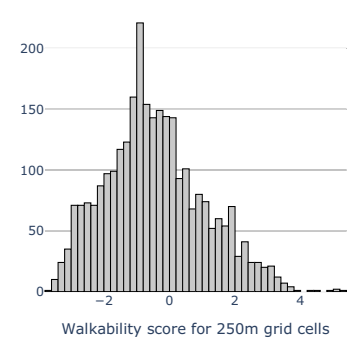
Walkability score for 250m grid cells

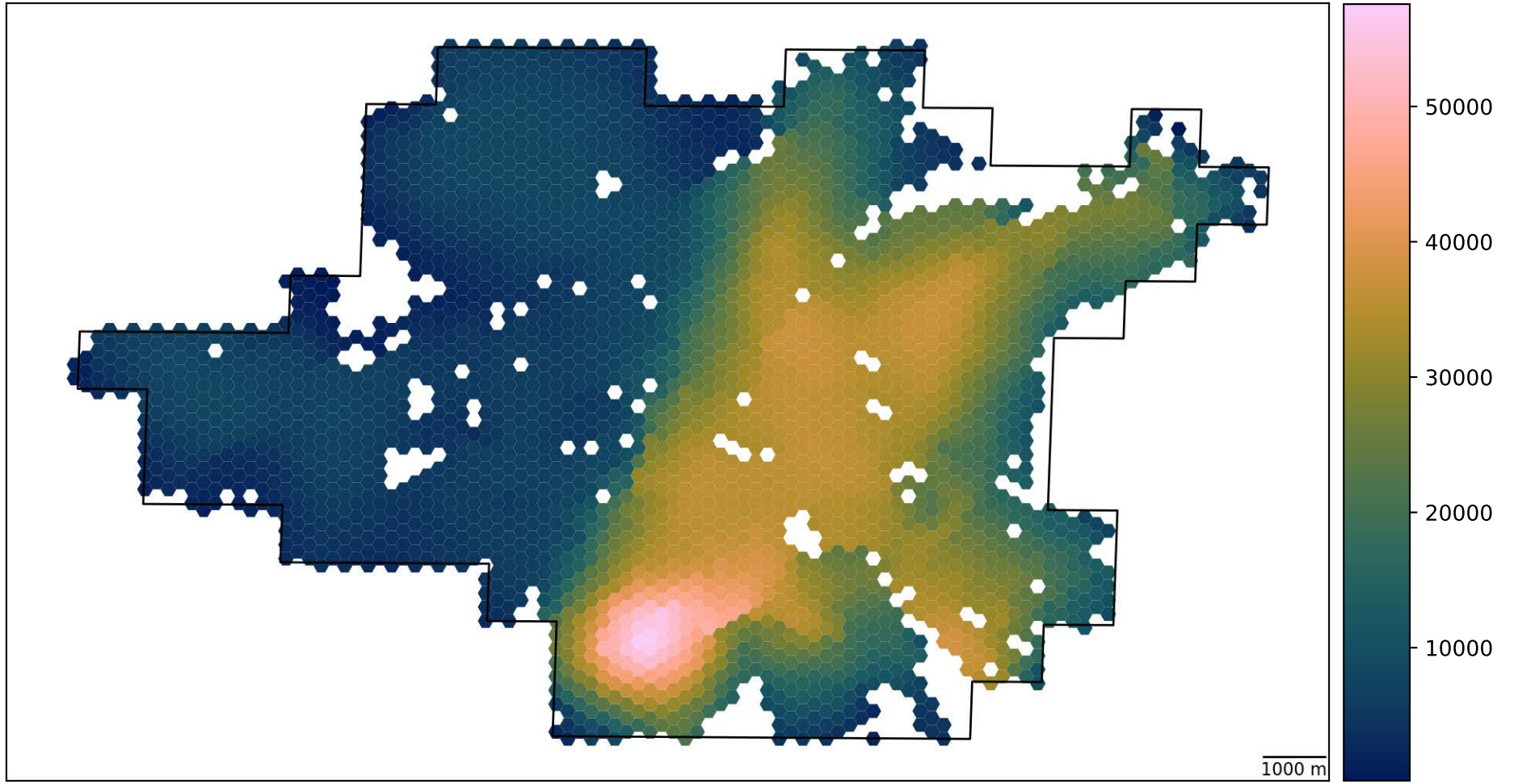

Mean 1000 m neighbourhood population per km²

A: Estimated Mean 1000 m neighbourhood population per km² requirement for ≥80% probability of engaging in walking for transport

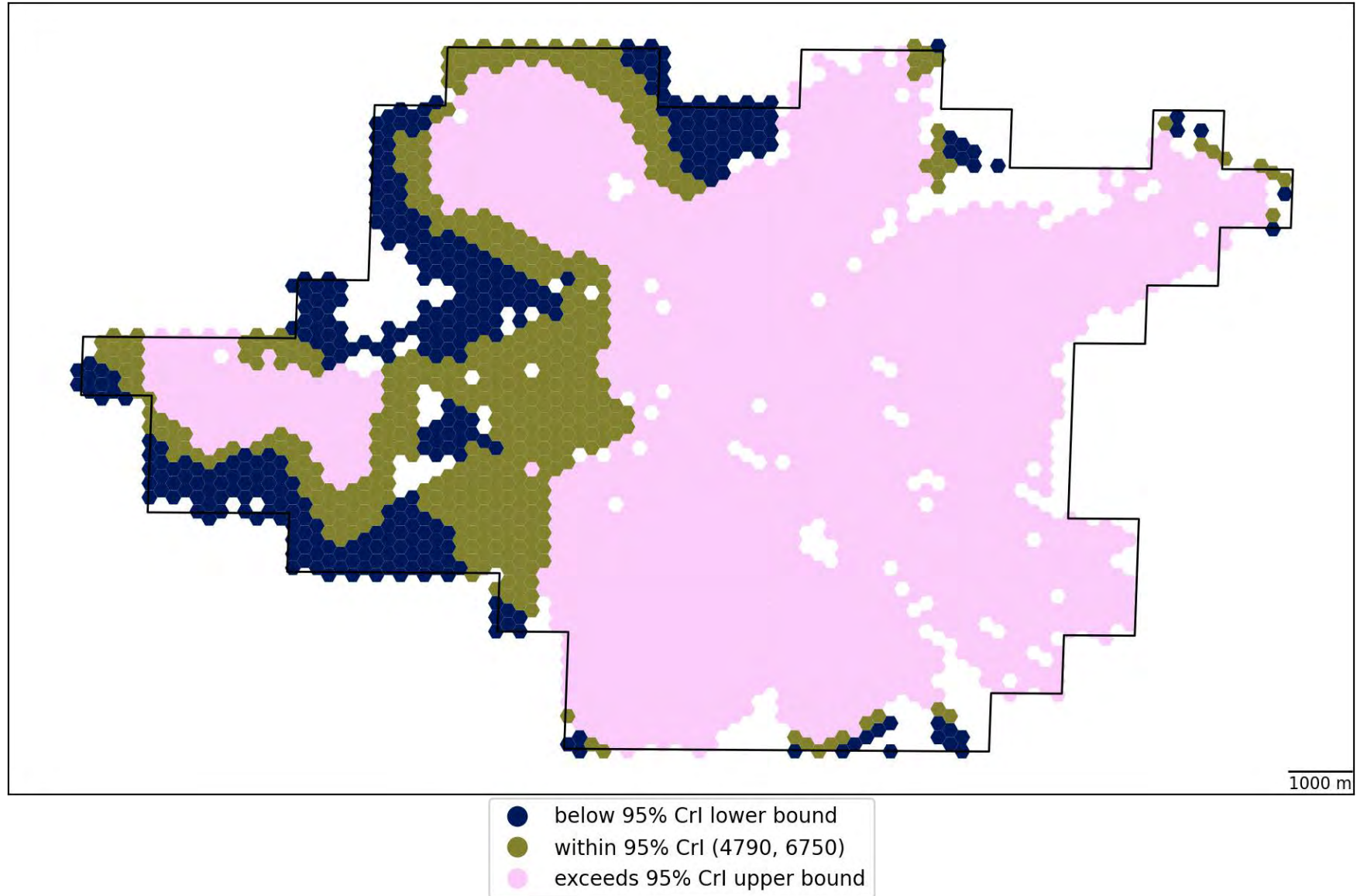

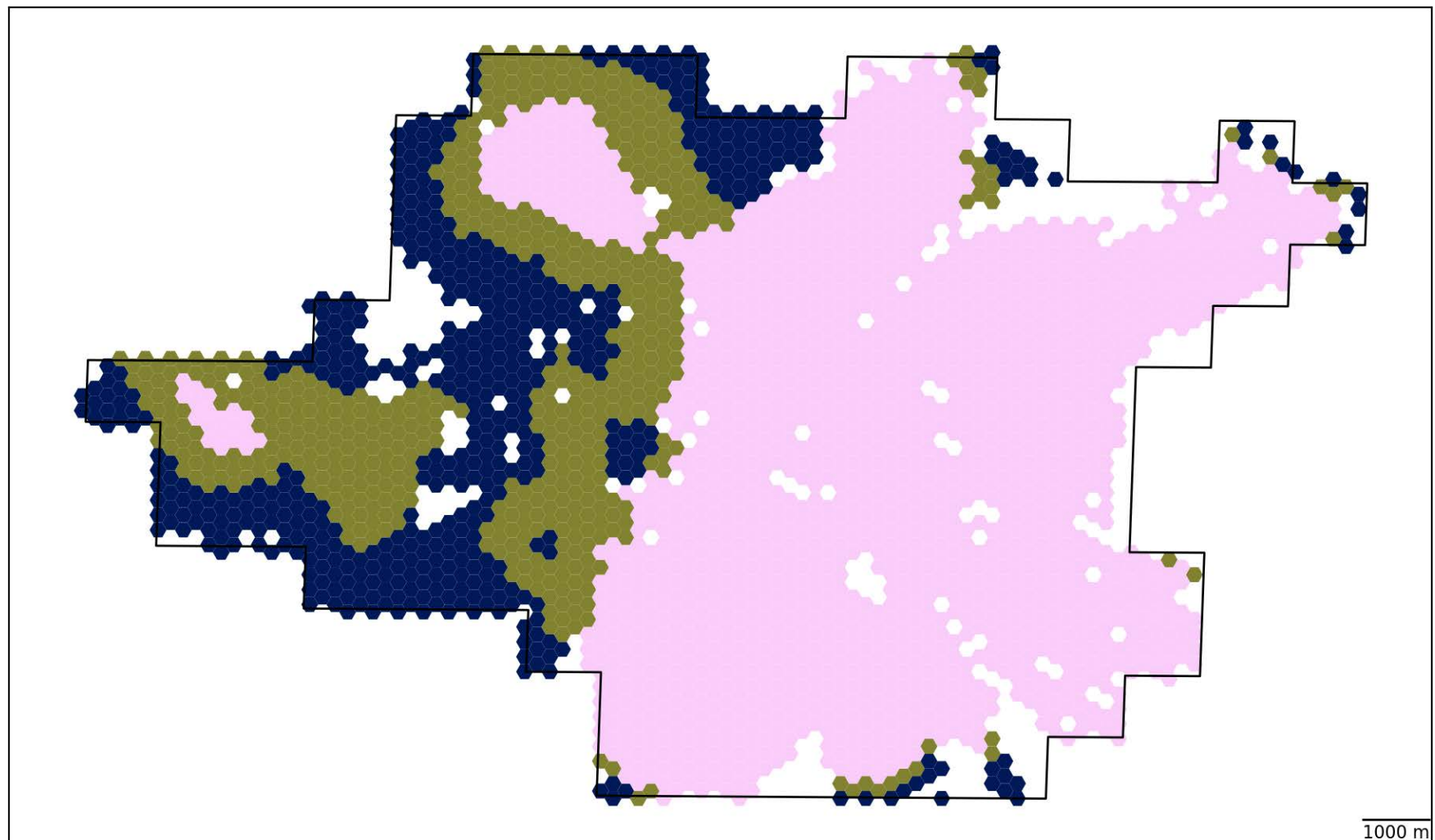

B: Estimated Mean 1000 m neighbourhood population per km² requirement for reaching the WHO's target of a ≥15% relative reduction in insufficient physical activity through walking

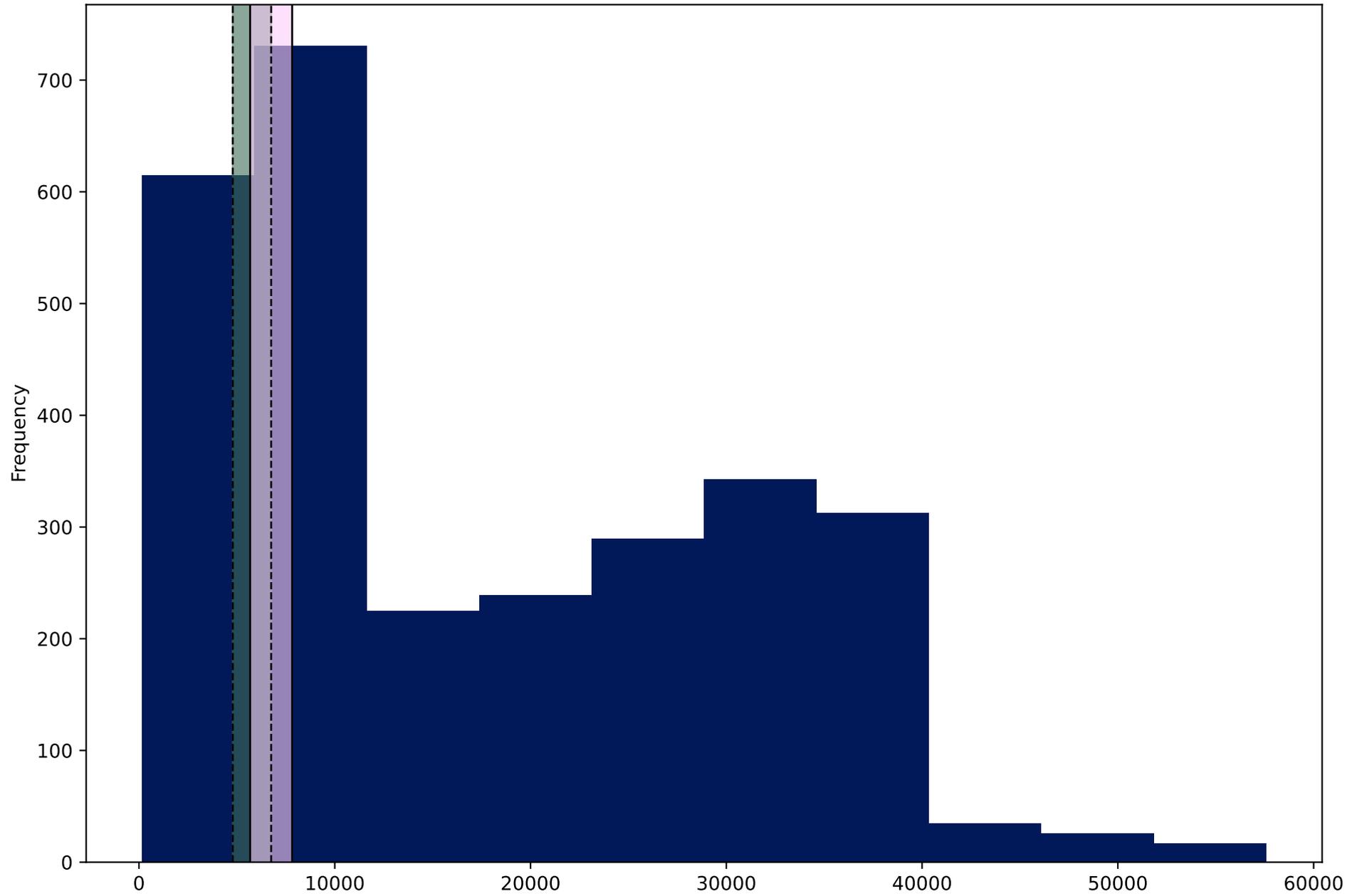

Mean 1000 m neighbourhood street intersections per km²

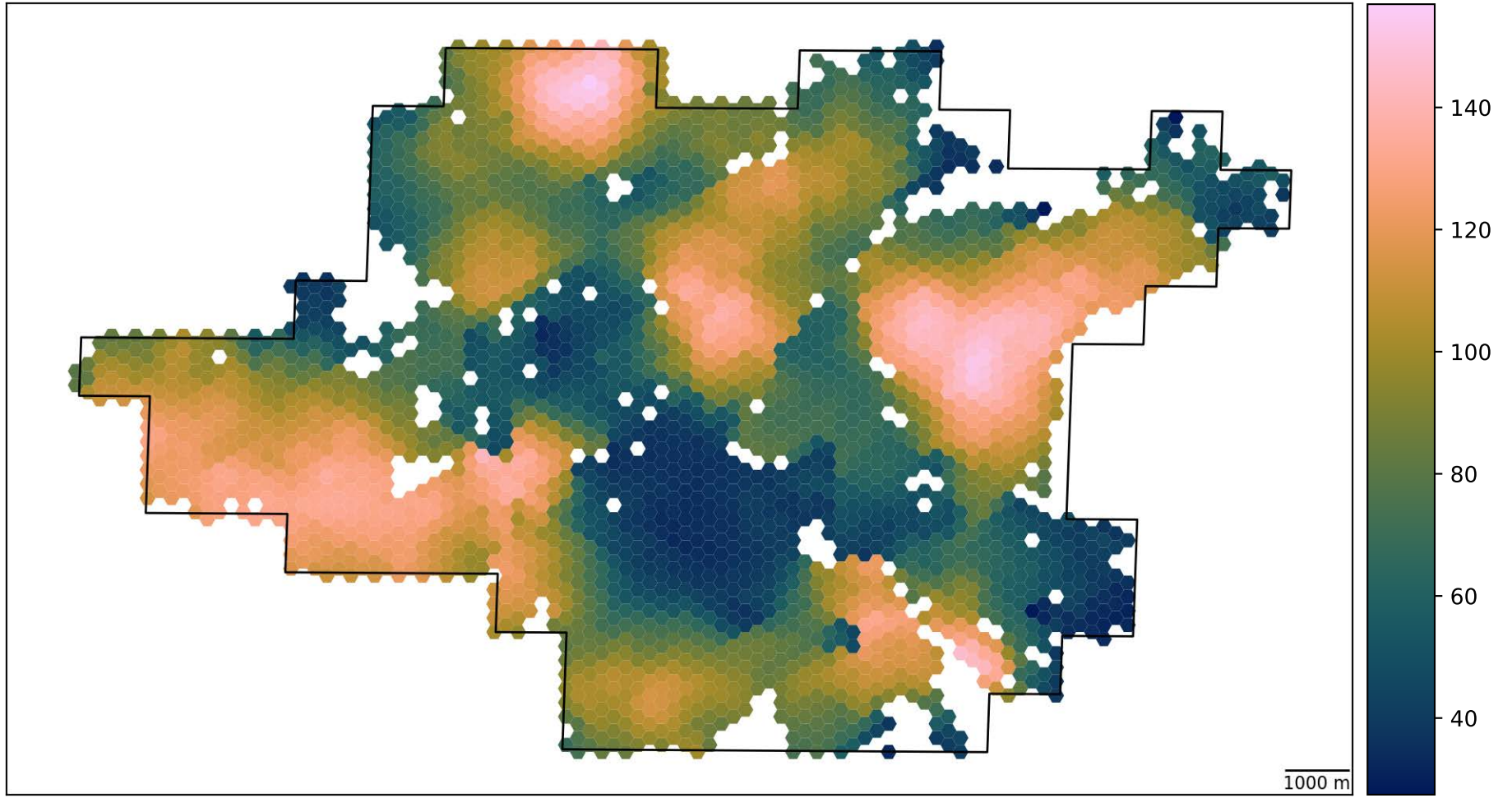

A: Estimated Mean 1000 m neighbourhood street intersections per km² requirement for ≥80% probability of engaging in walking for transport

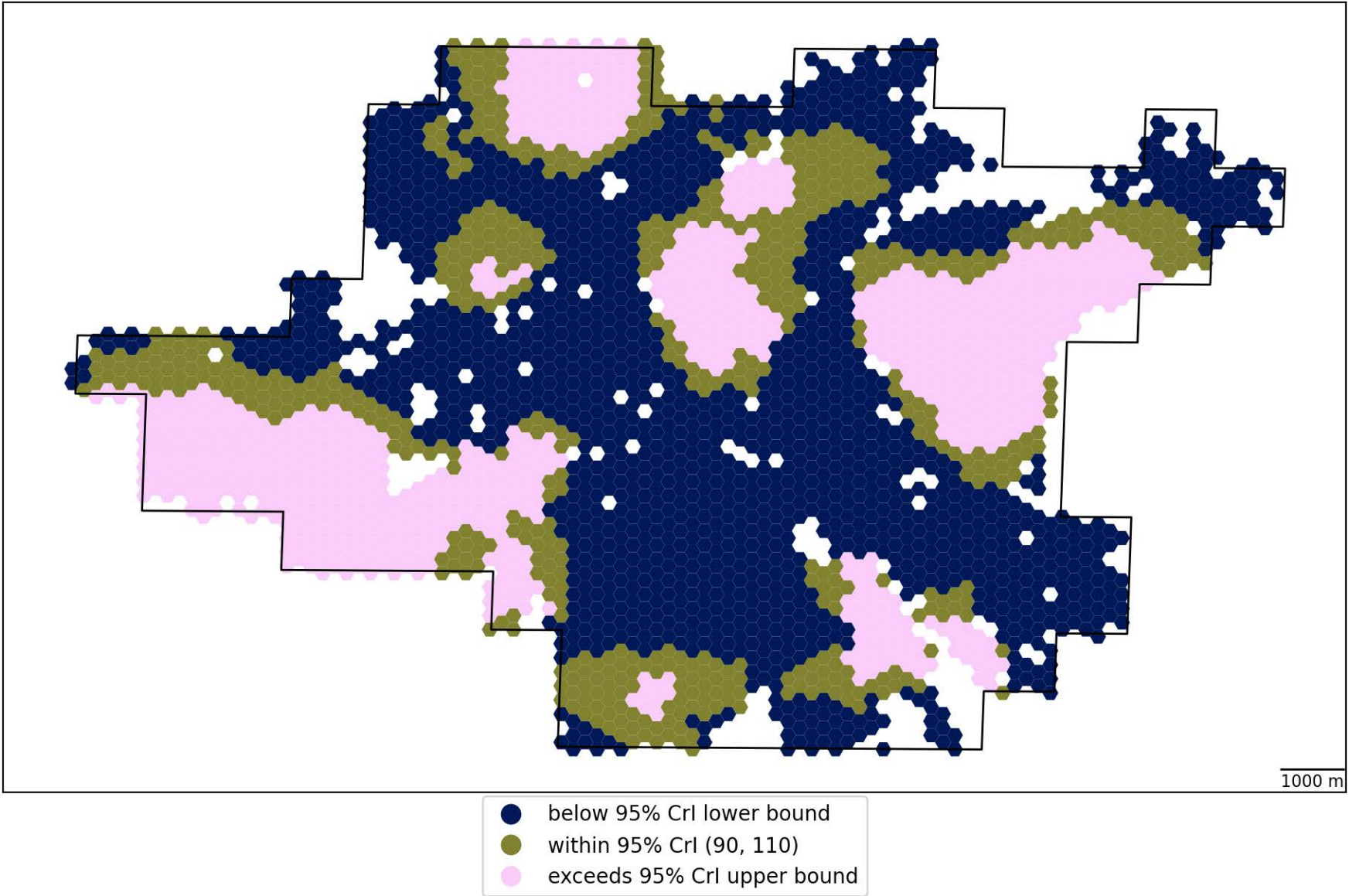

B: Estimated Mean 1000 m neighbourhood street intersections per km² requirement for reaching the WHO's target of a ≥15% relative reduction in insufficient physical activity through walking

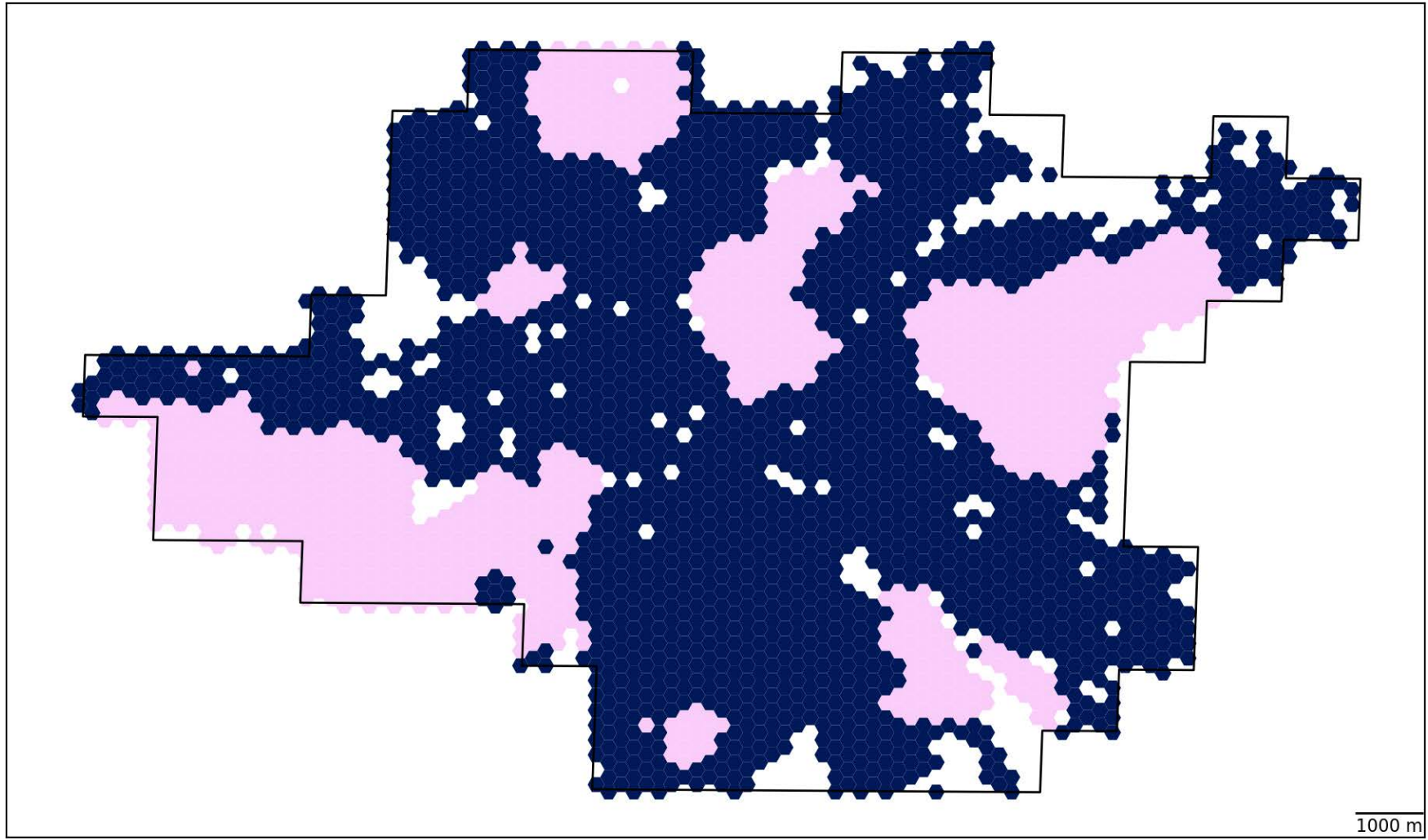

- below 95% CrI lower bound
- within 95% CrI (106, 156)

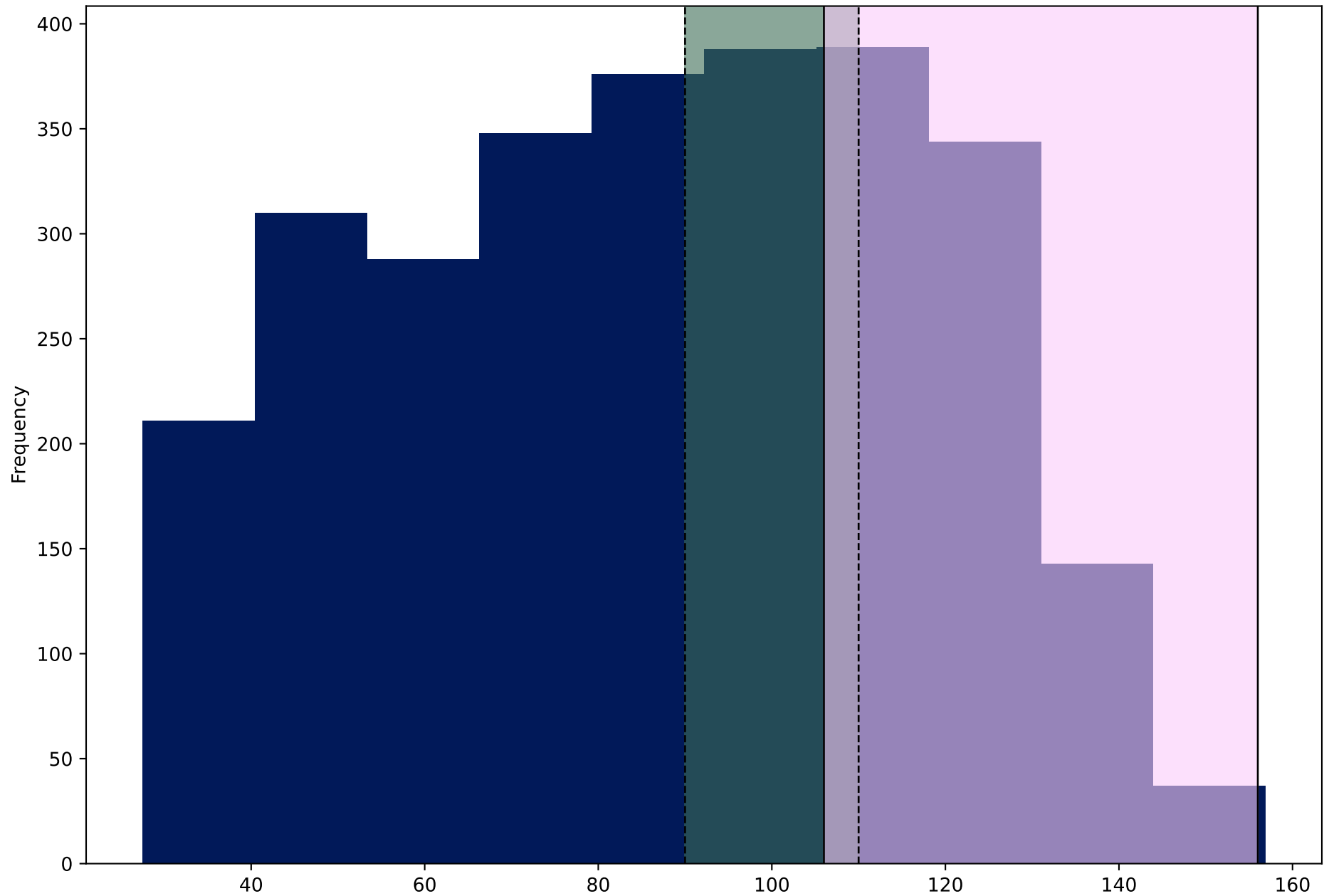

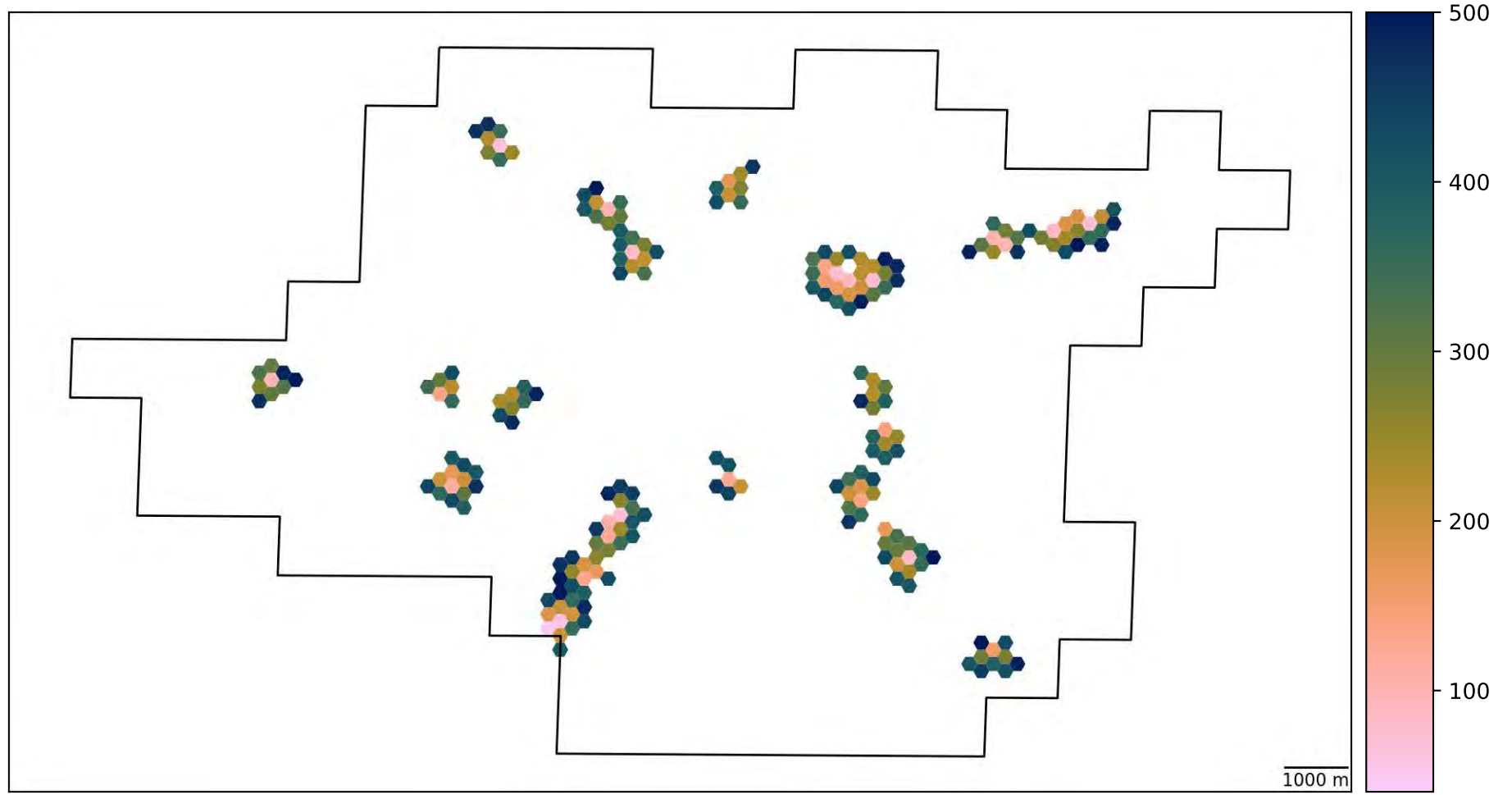

Distance to nearest public transport stops (m; up to 500m)

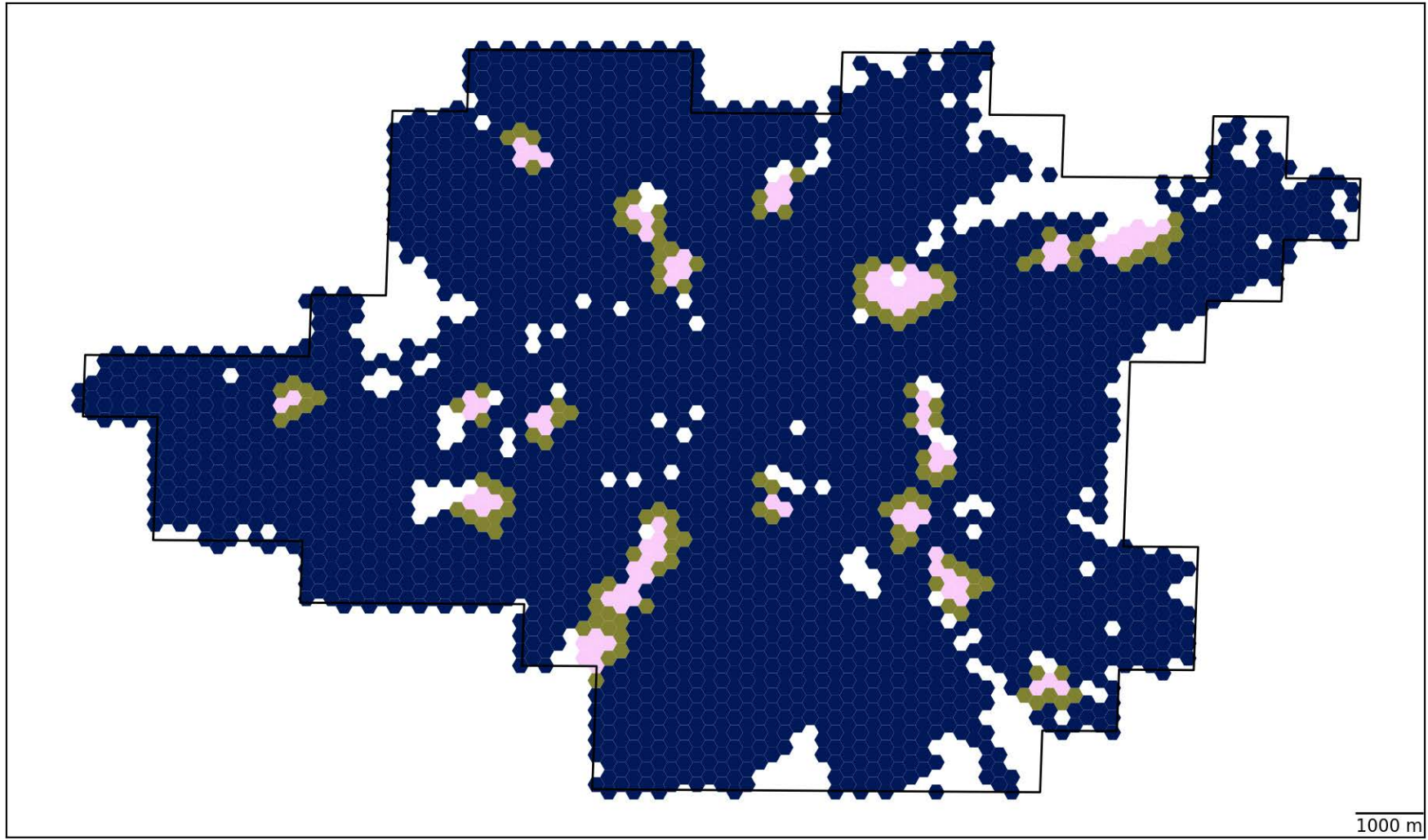

distances: Estimated Distance to nearest public transport stops (m; up to 500m) requirement for distances to destinations, measured up to a maximum distance target threshold of 500 metres

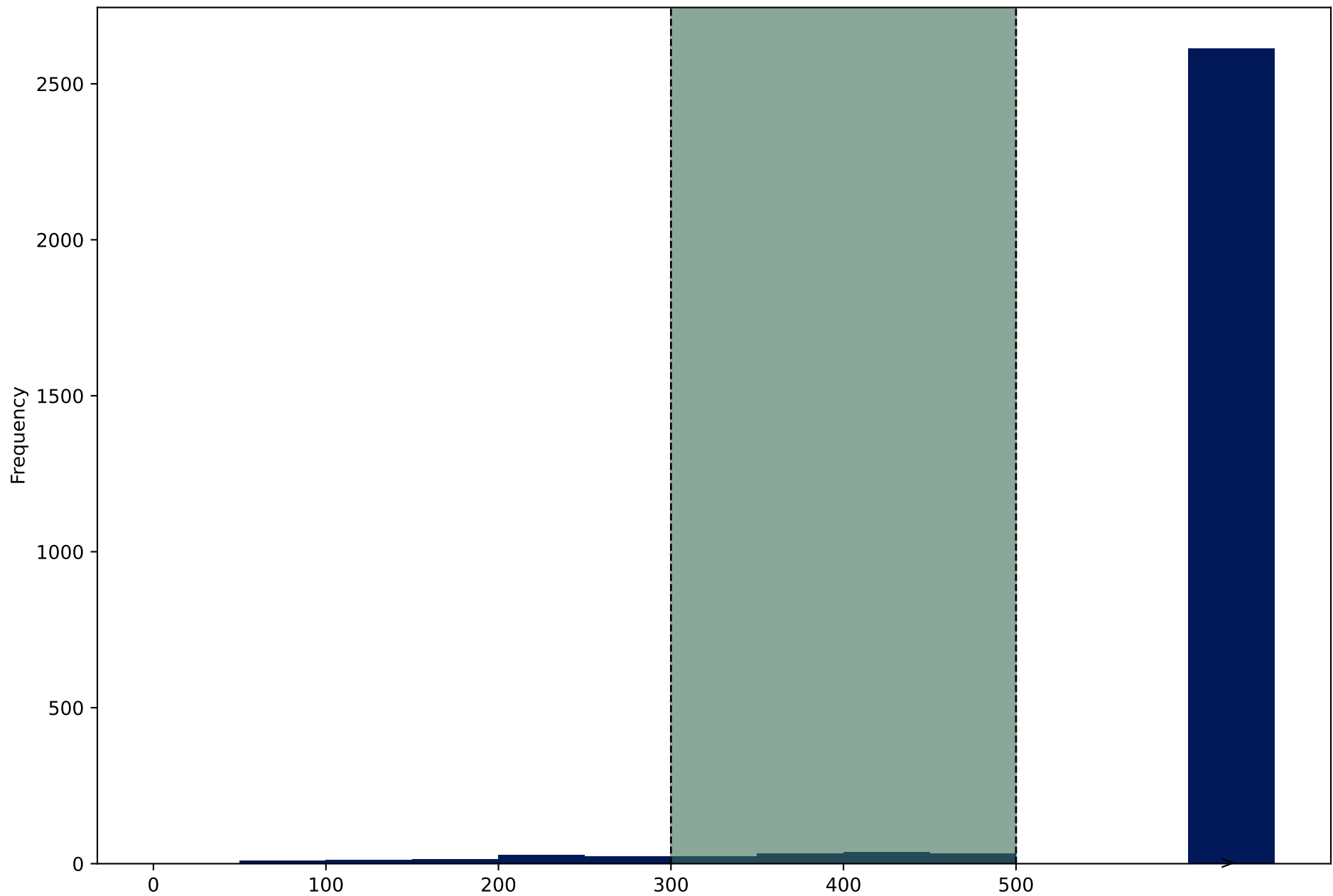

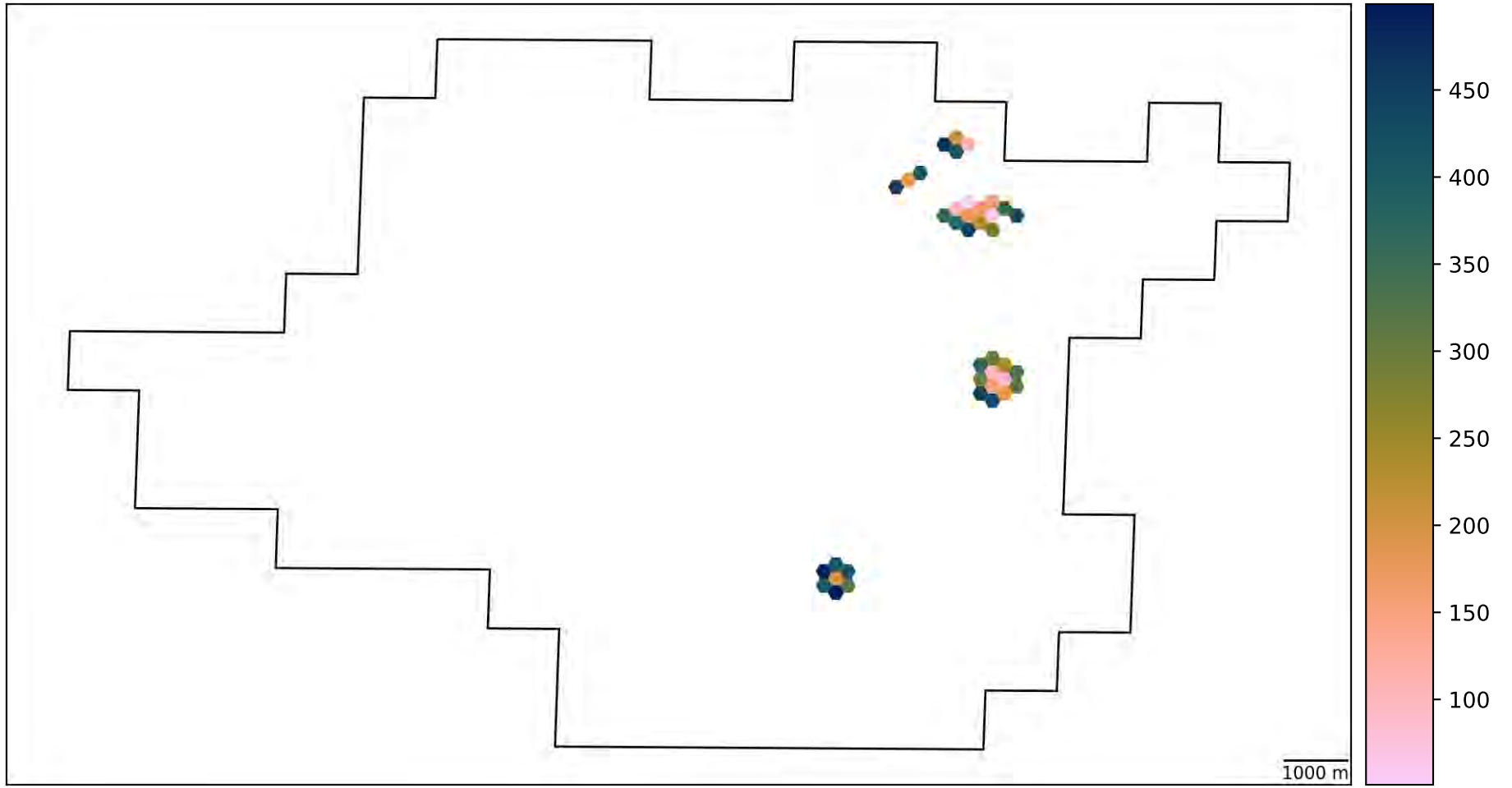

Distance to nearest park (m; up to 500m)

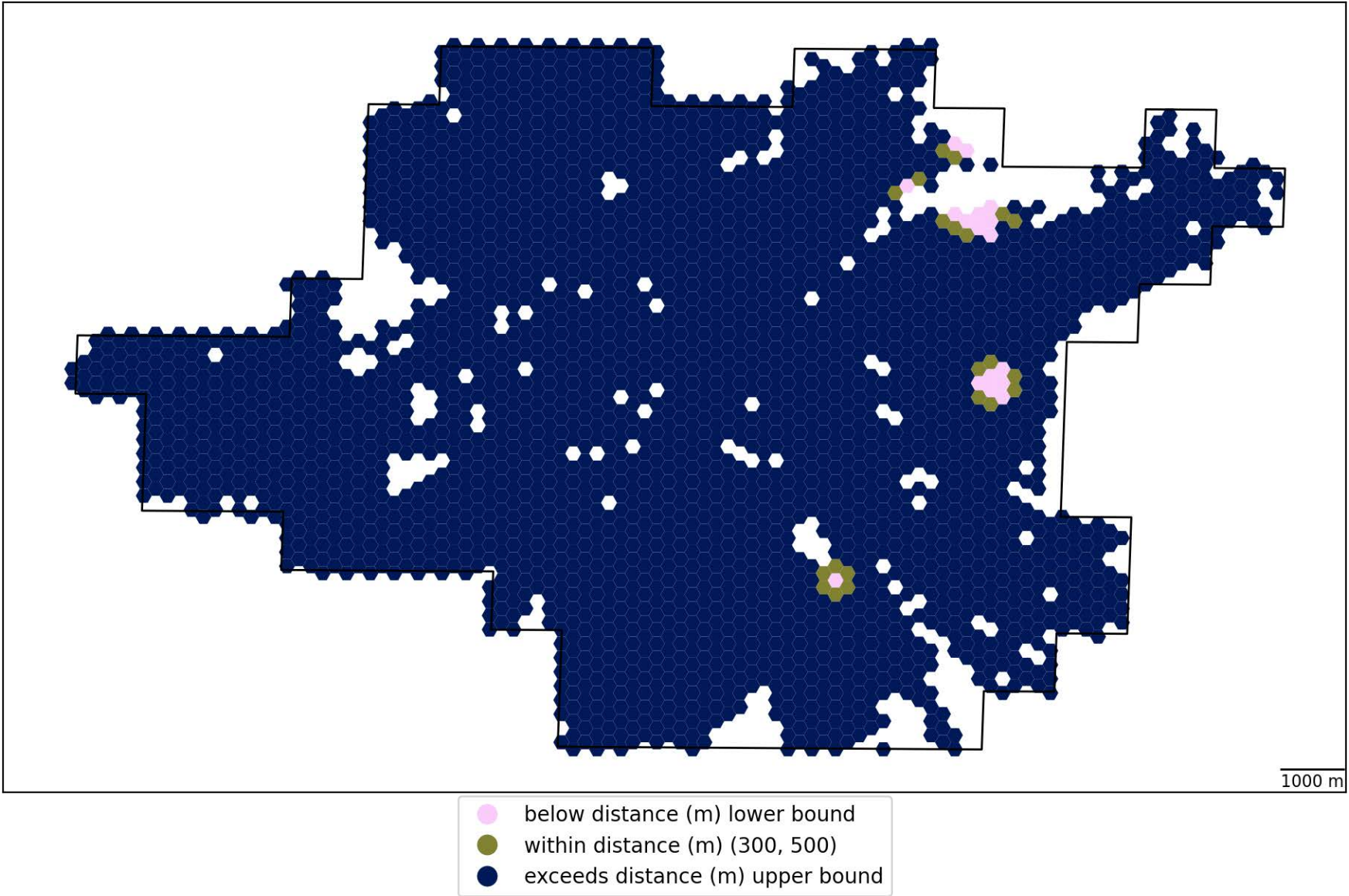

distances: Estimated Distance to nearest park (m; up to 500m) requirement for distances to destinations, measured up to a maximum distance target threshold of 500 metres

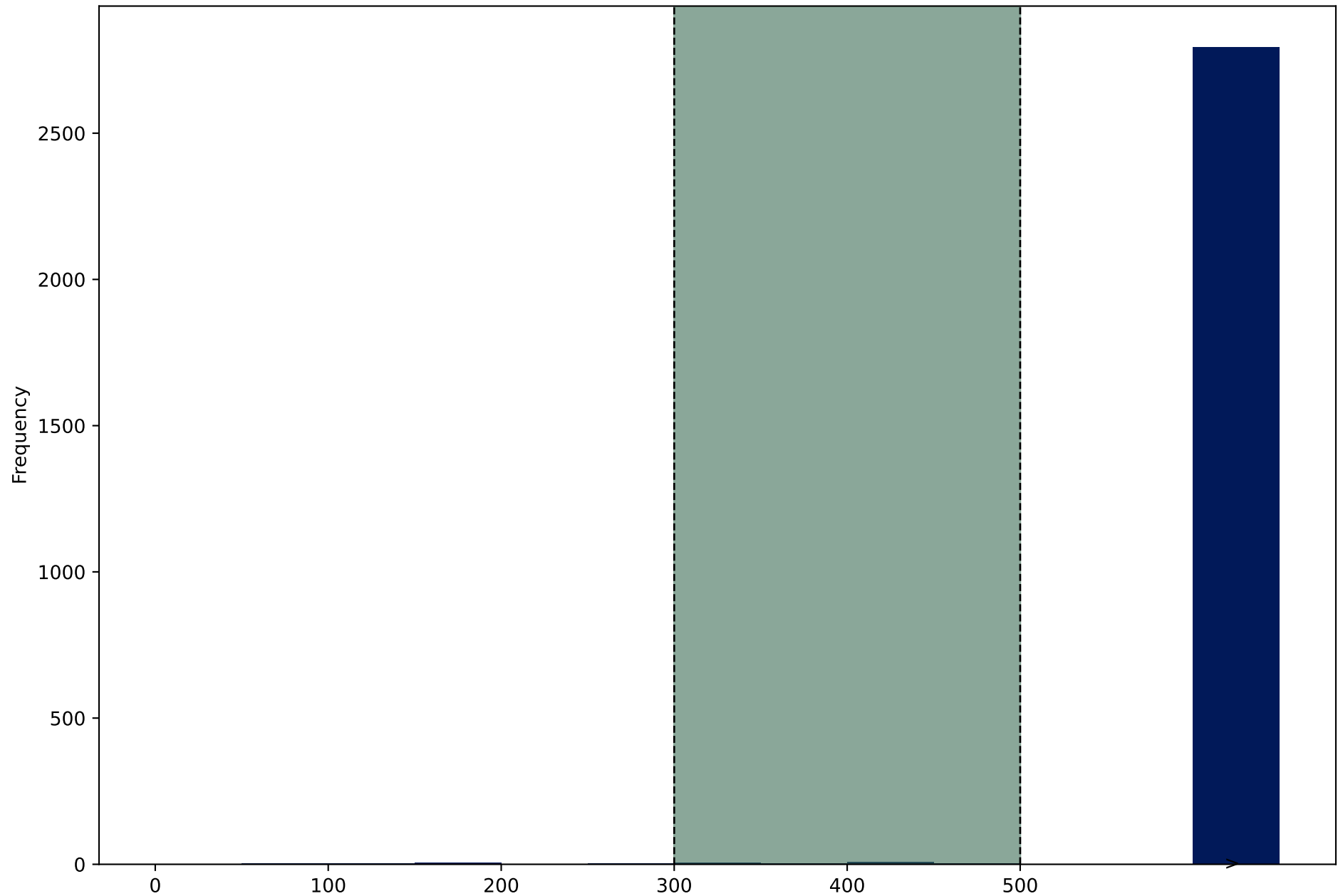

# America, North, Mexico, Mexico City

Satellite imagery of urban study region (Bing) | Walkability, relative to city | Walkability, relative to 25 global cities

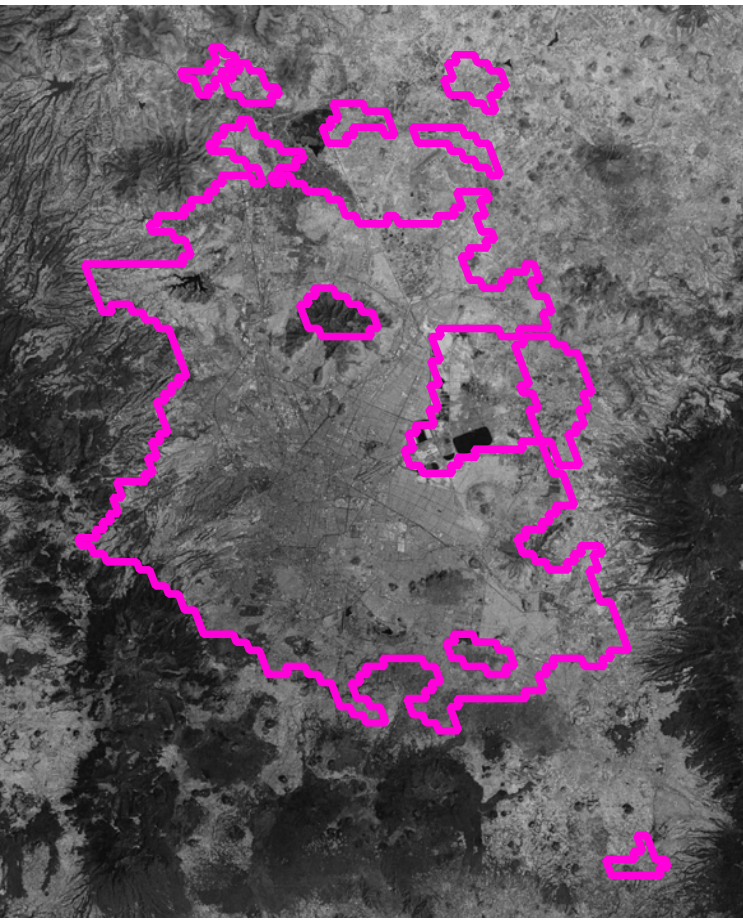
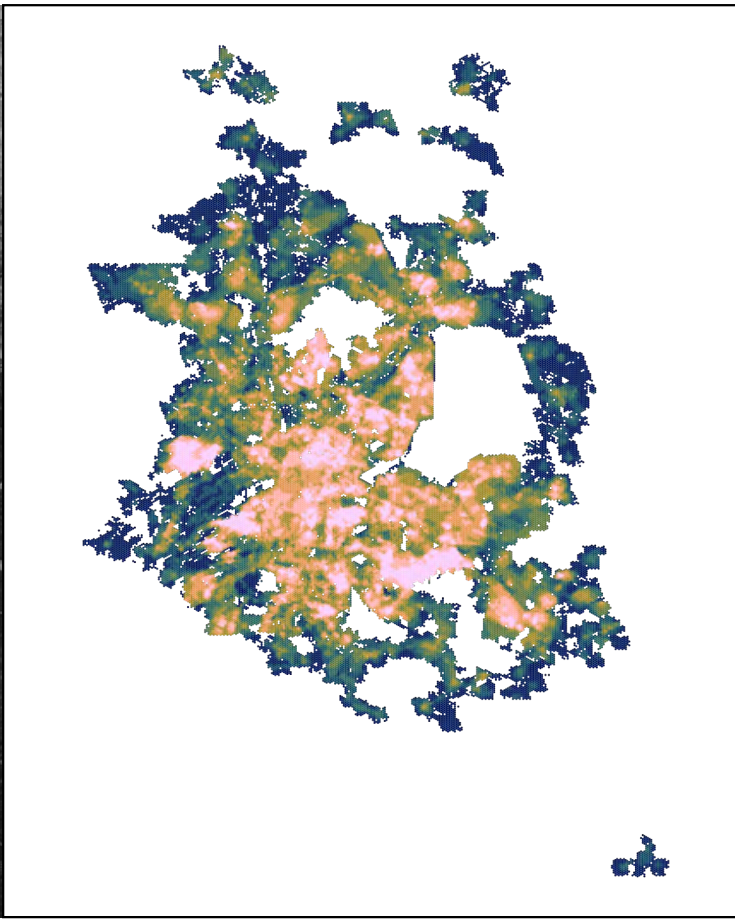
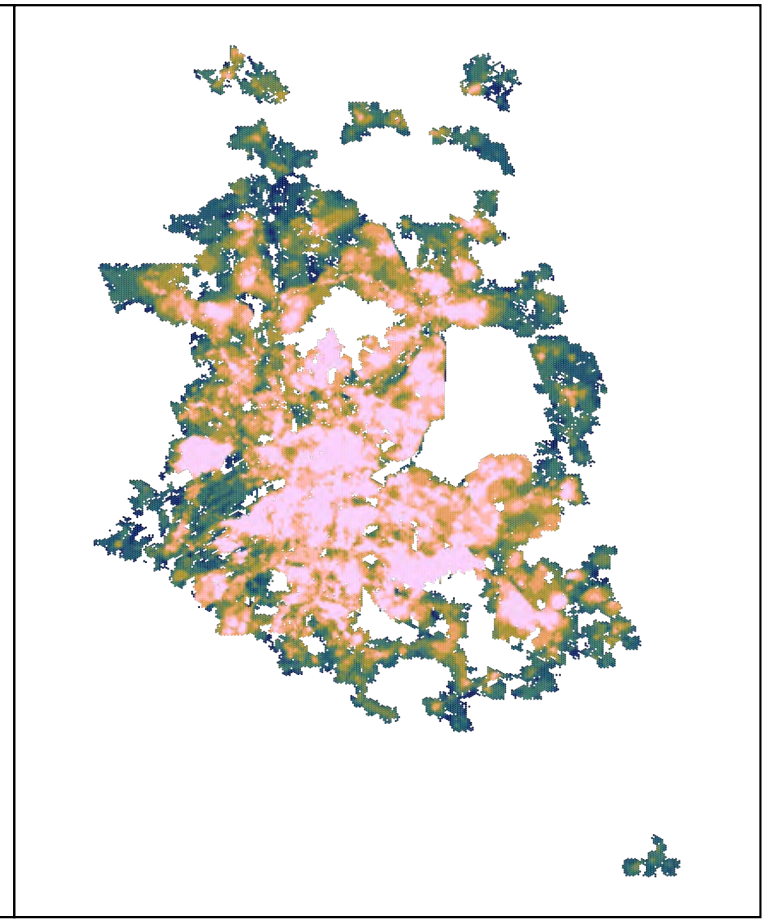

Urban boundary

0   30   60 km

Walkability score
- <-3
- -3 to -2
- -2 to -1
- -1 to 0
- 0 to 1
- 1 to 2
- 2 to 3
- ≥3

Walkability relative to all cities by component variables (2D histograms), and overall (histogram)

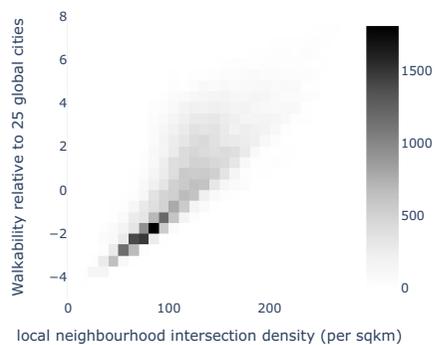
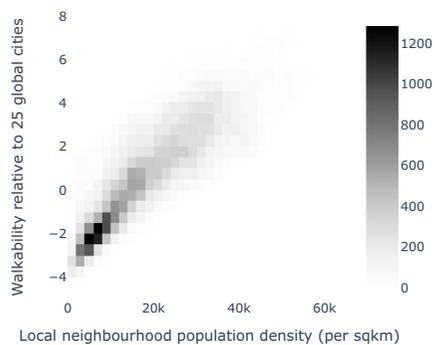
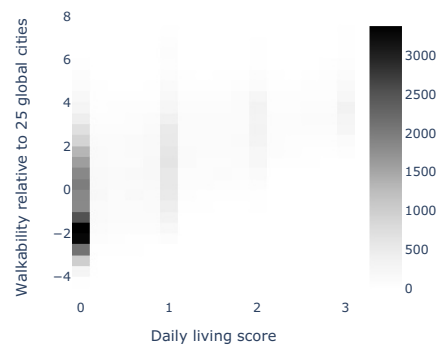
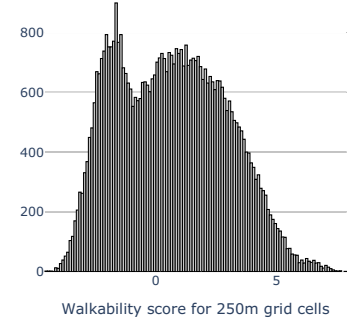

Mean 1000 m neighbourhood population per km²

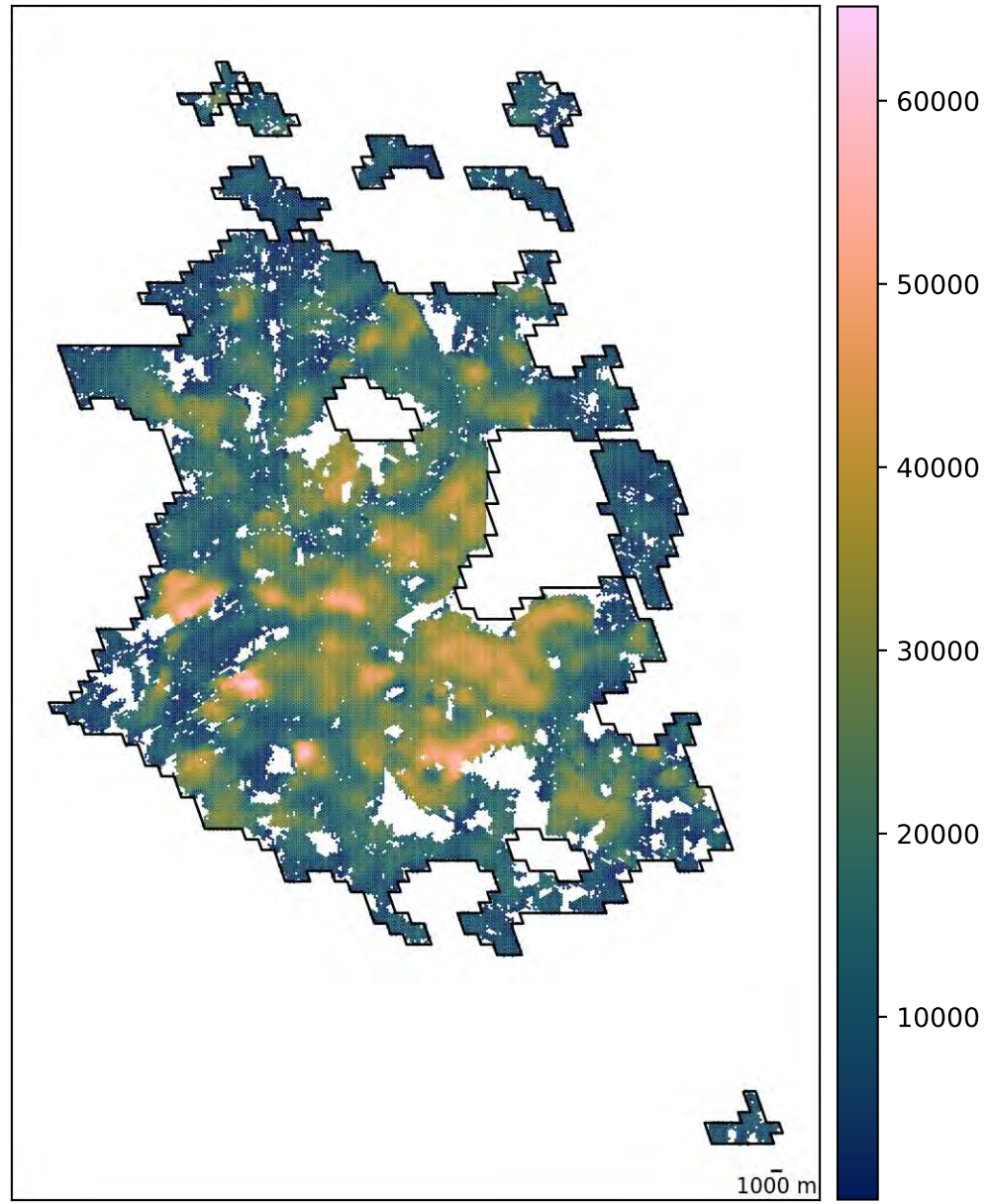

A: Estimated Mean 1000 m neighbourhood population per km² requirement for ≥80% probability of engaging in walking for transport

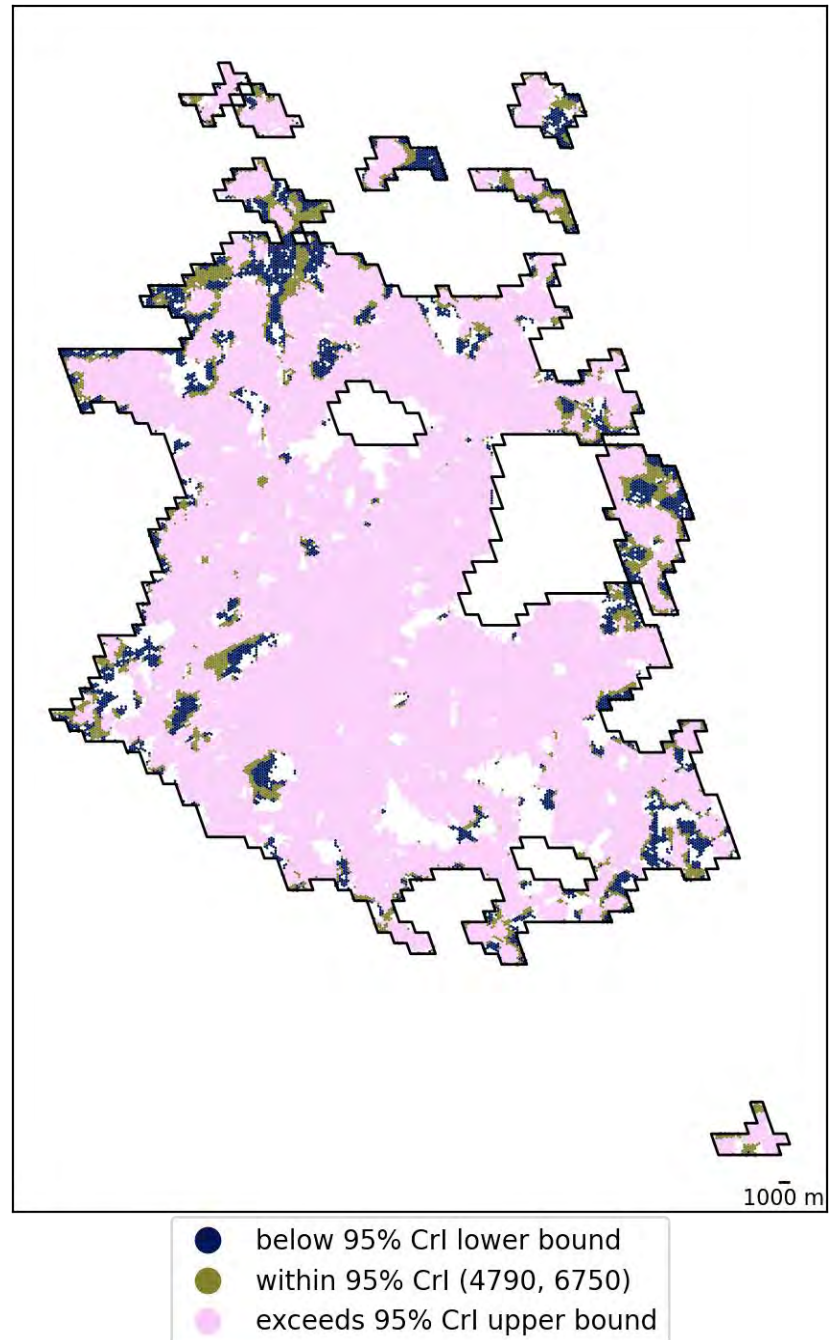

B: Estimated Mean 1000 m neighbourhood population per km² requirement for reaching the WHO's target of a ≥15% relative reduction in insufficient physical activity through walking

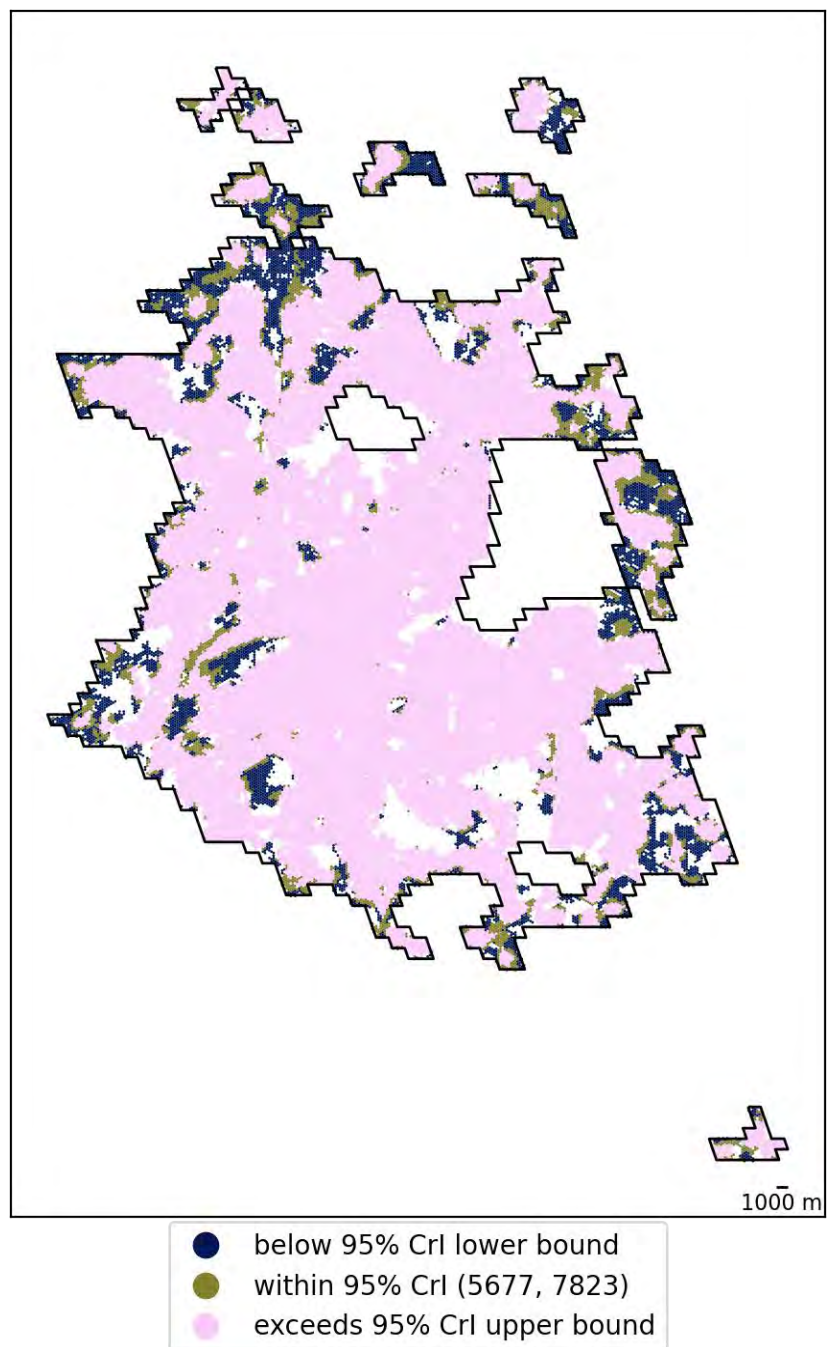

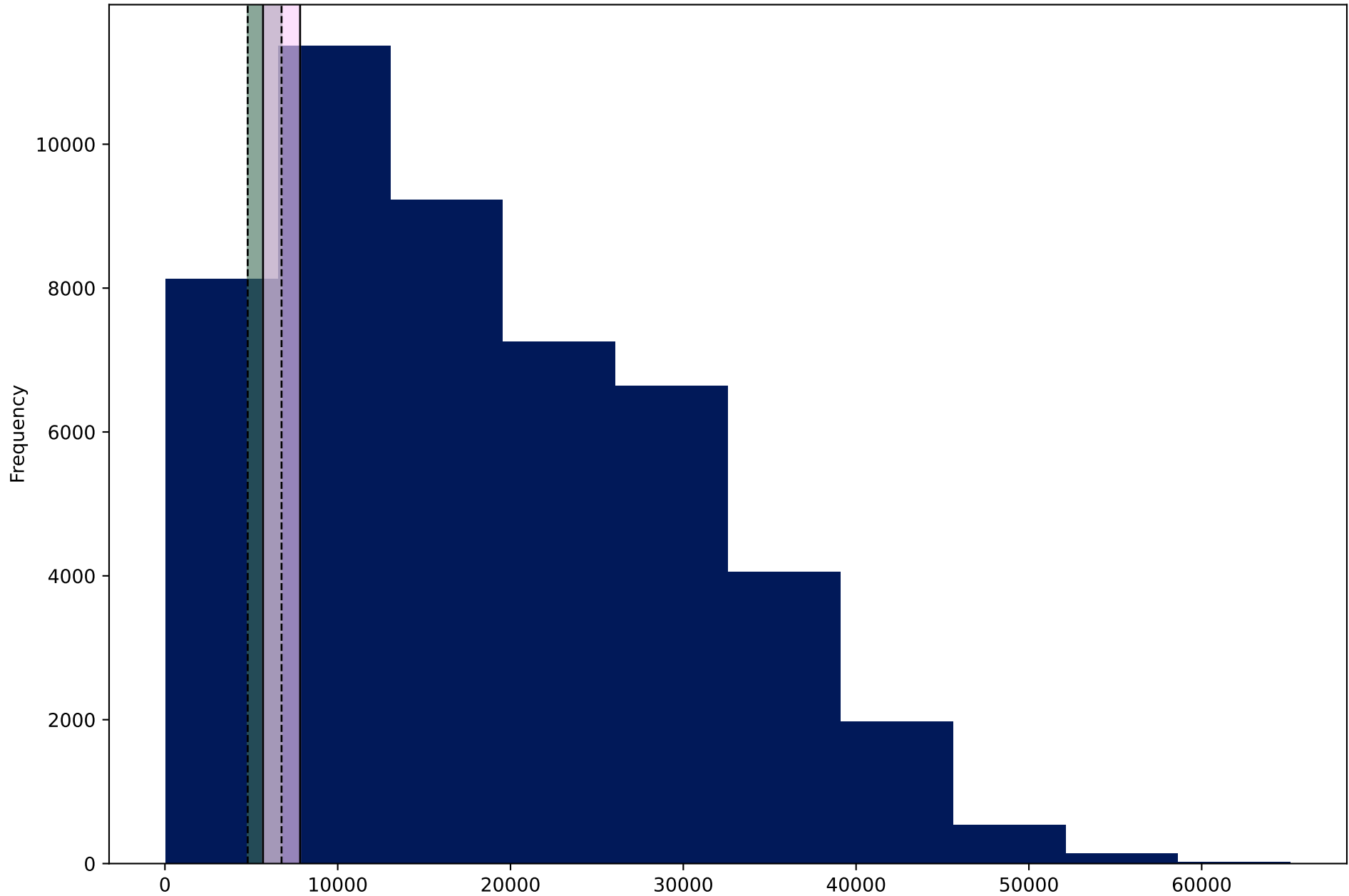

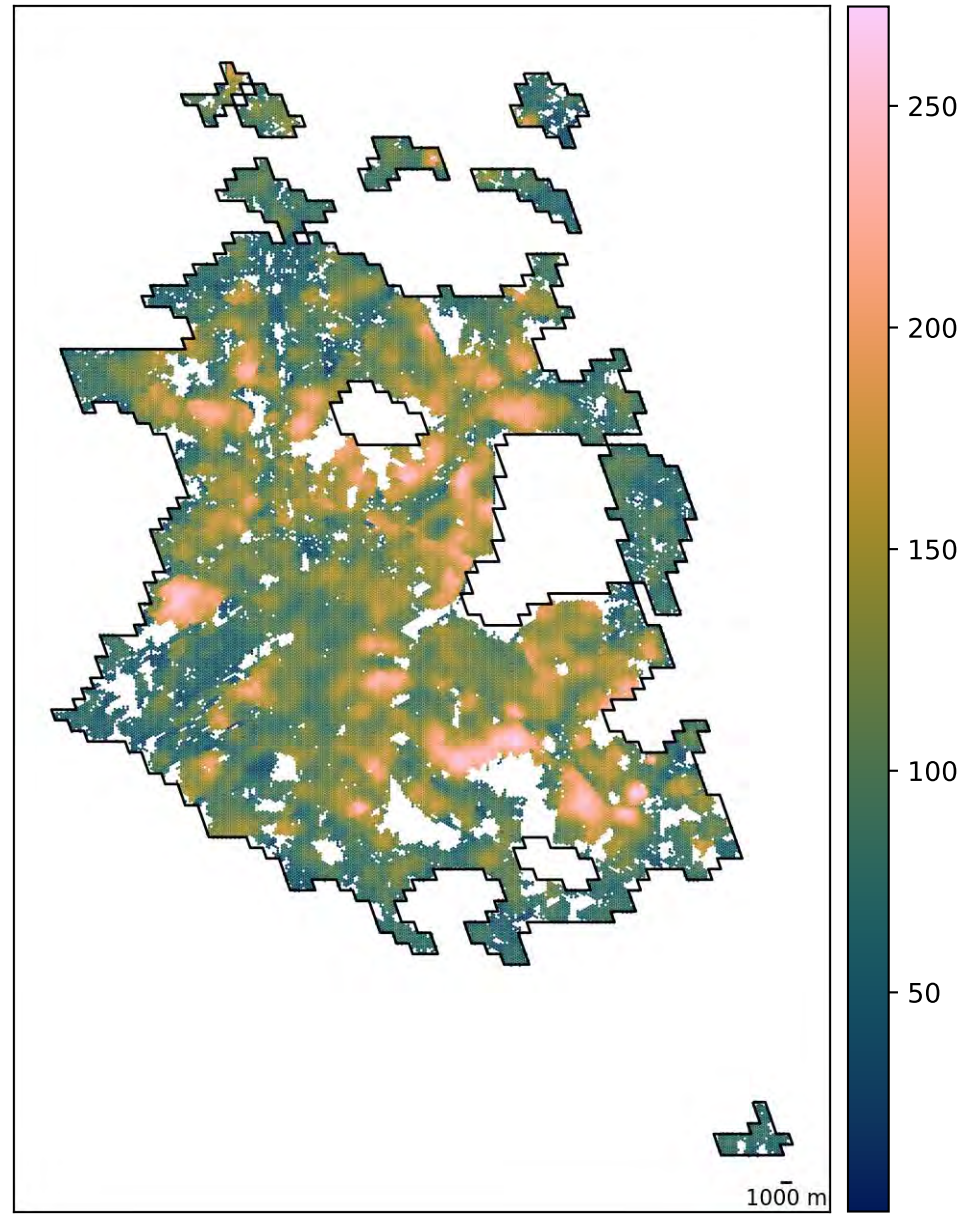

# A: Estimated Mean 1000 m neighbourhood street intersections per km² requirement for ≥80% probability of engaging in walking for transport

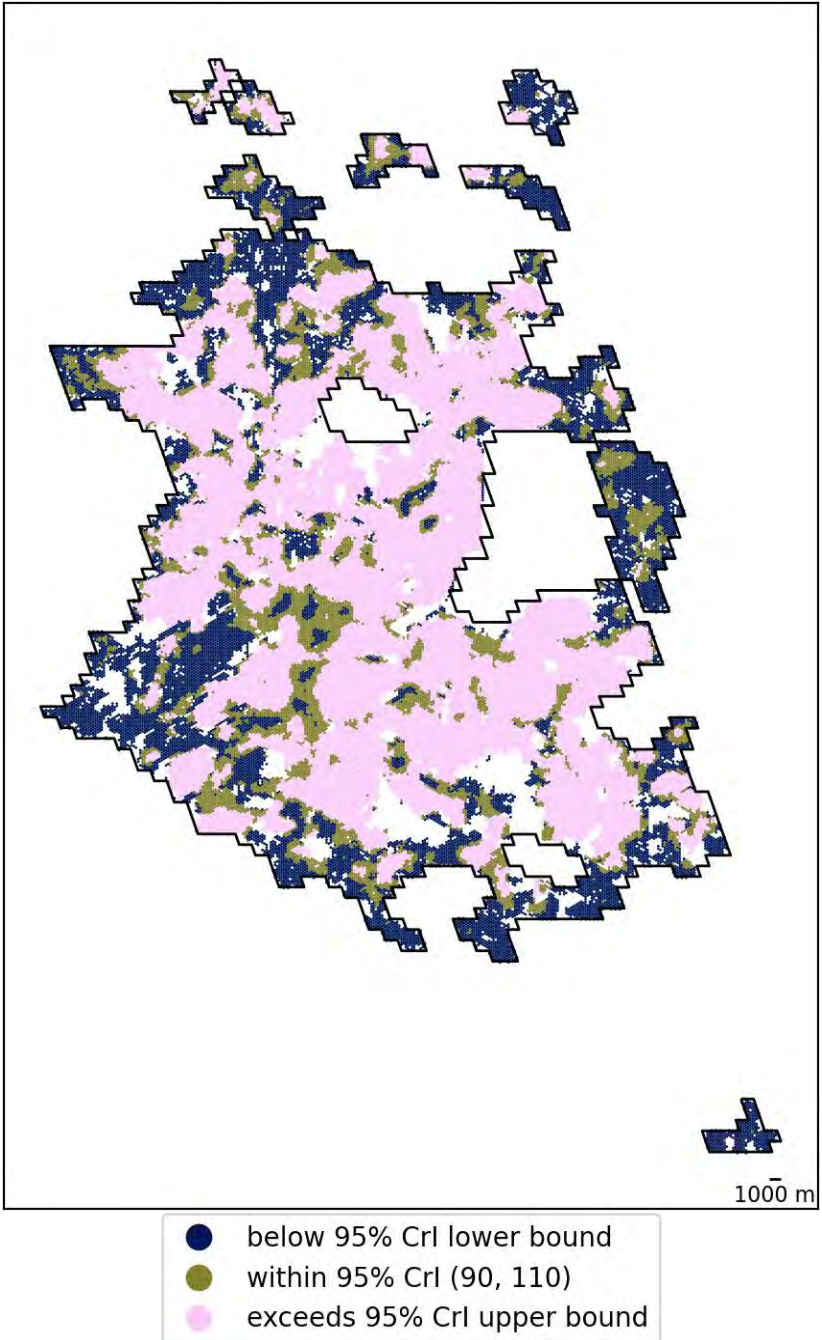

B: Estimated Mean 1000 m neighbourhood street intersections per km² requirement for reaching the WHO's target of a ≥15% relative reduction in insufficient physical activity through walking

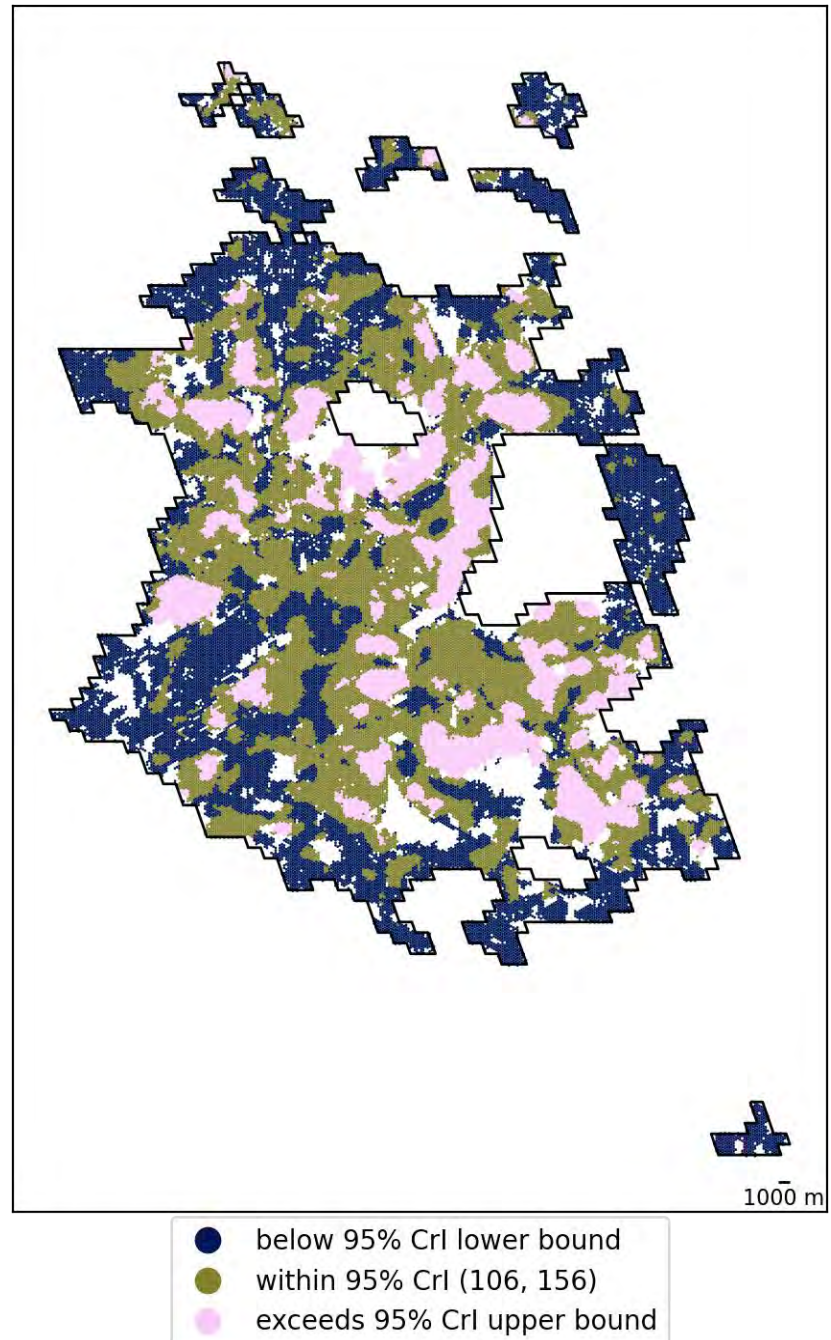

- below 95% CrI lower bound
- within 95% CrI (106, 156)
- exceeds 95% CrI upper bound

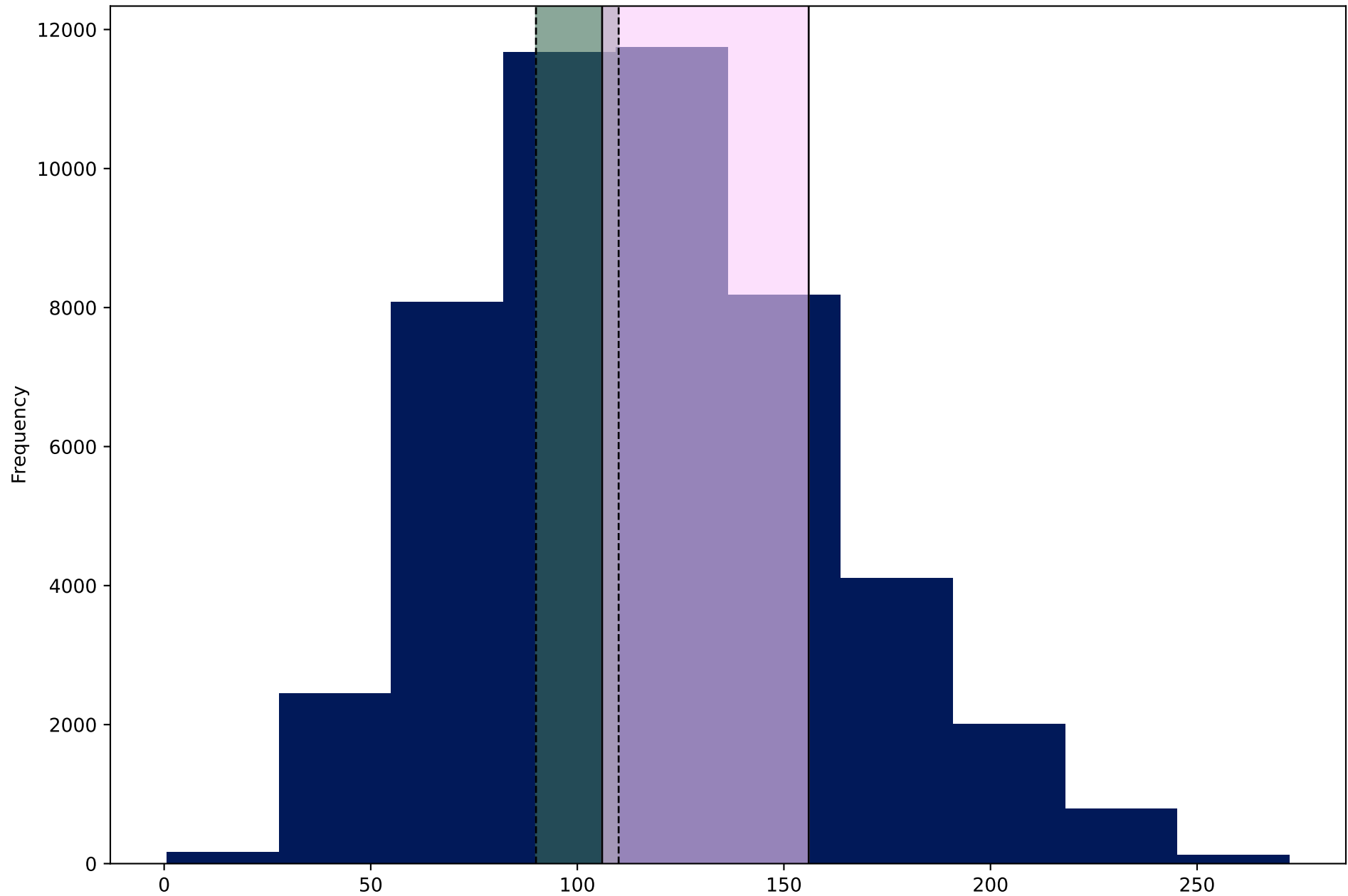

Distance to nearest public transport stops (m; up to 500m)

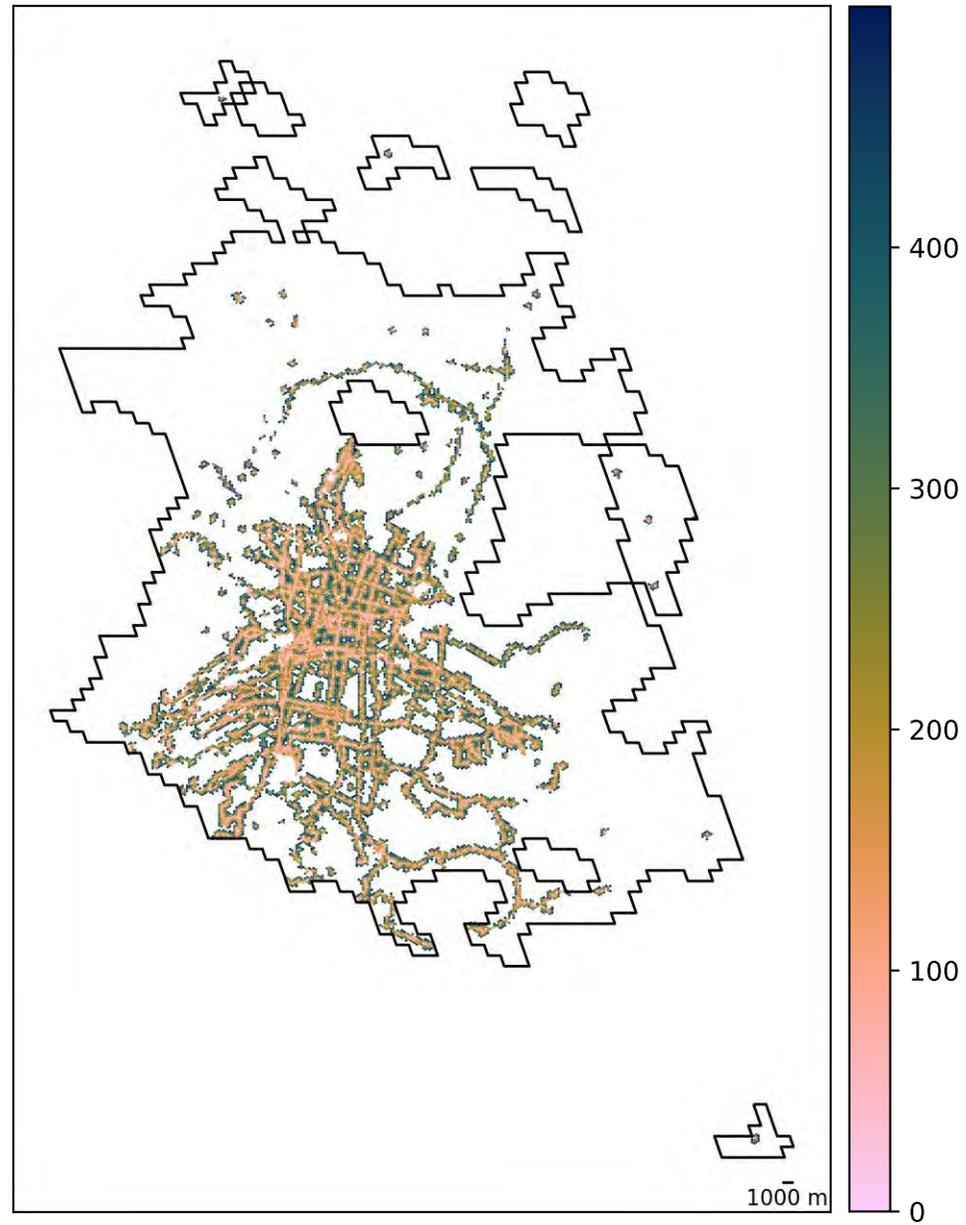

distances: Estimated Distance to nearest public transport stops (m; up to 500m) requirement for distances to destinations, measured up to a maximum distance target threshold of 500 metres

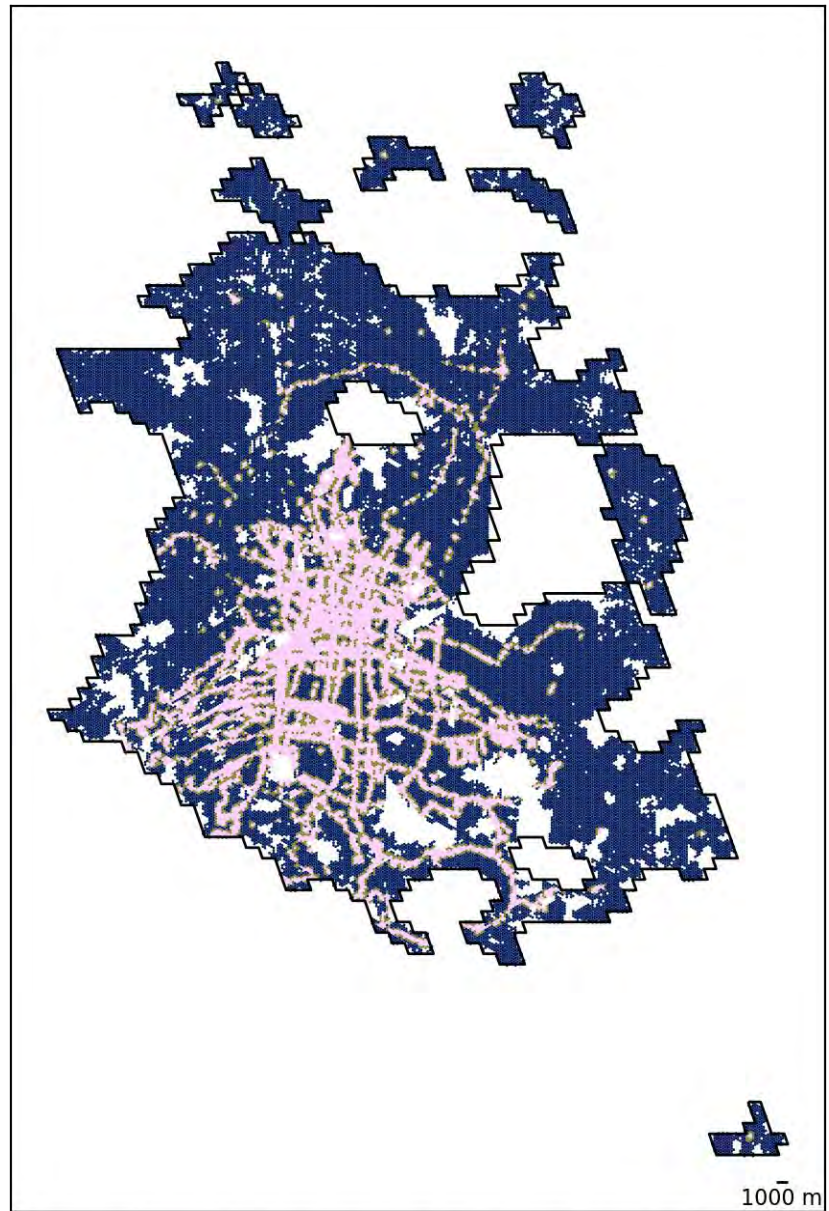

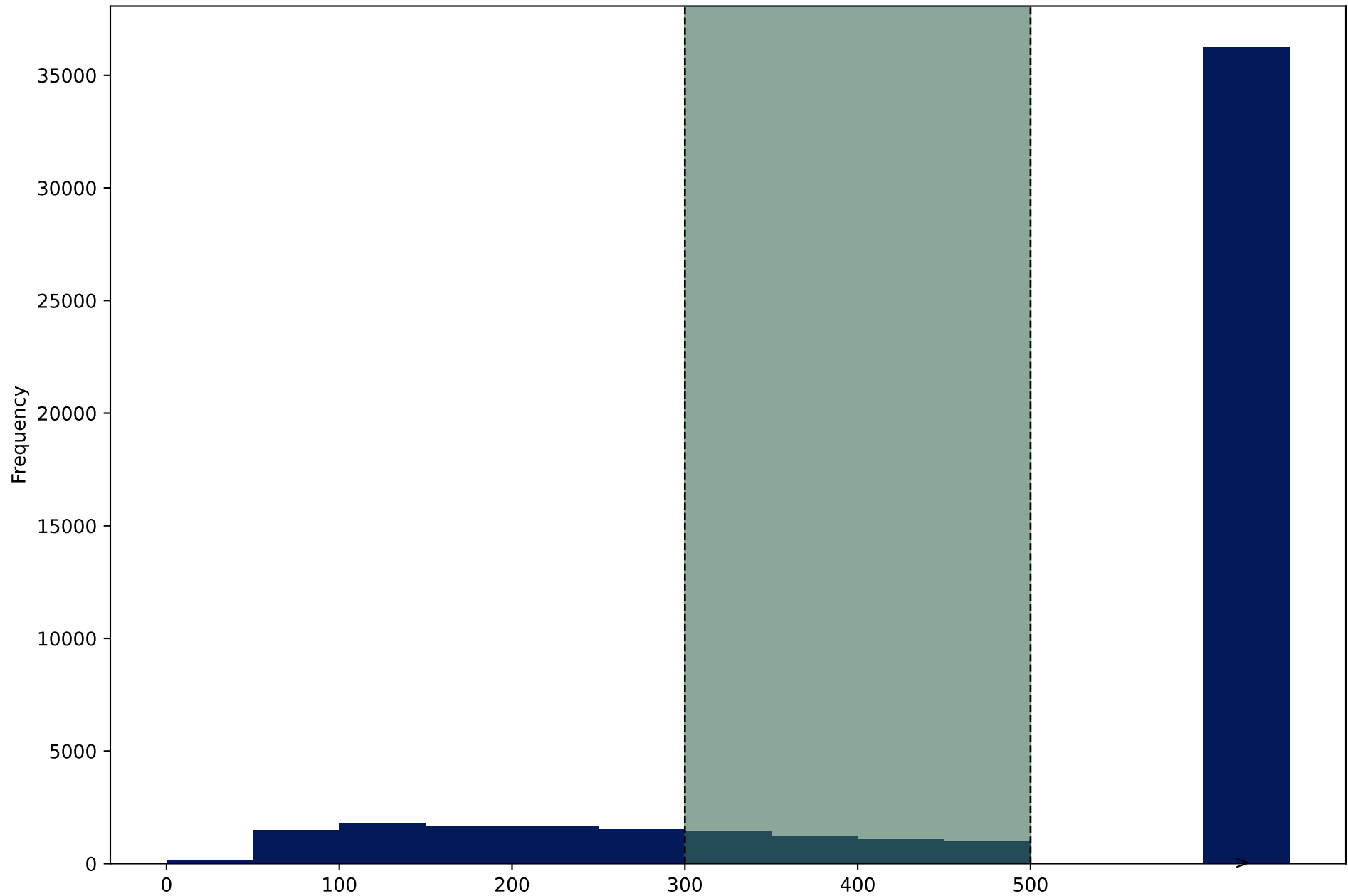

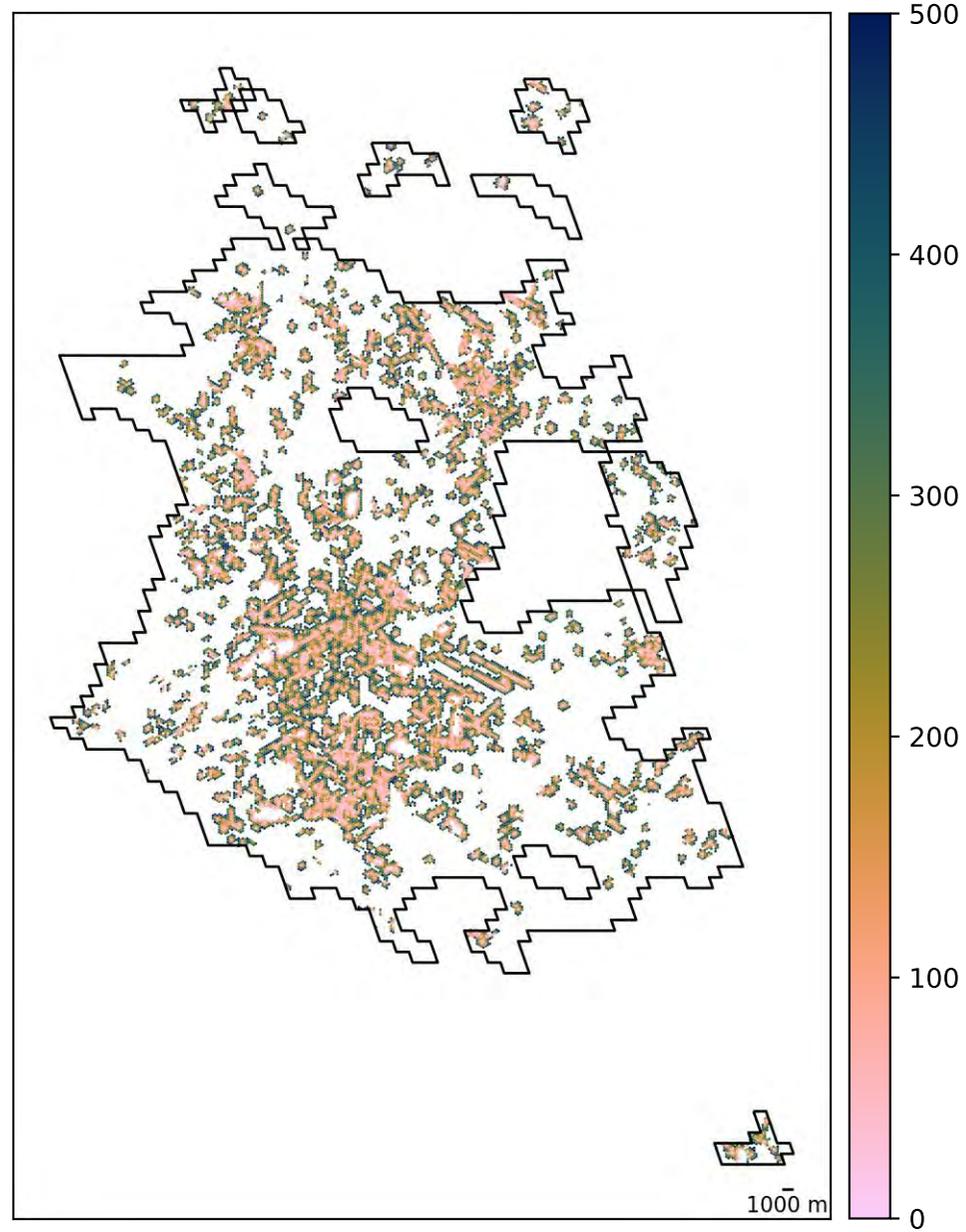

Distance to nearest park (m; up to 500m)

distances: Estimated Distance to nearest park (m; up to 500m) requirement for distances to destinations, measured up to a maximum distance target threshold of 500 metres

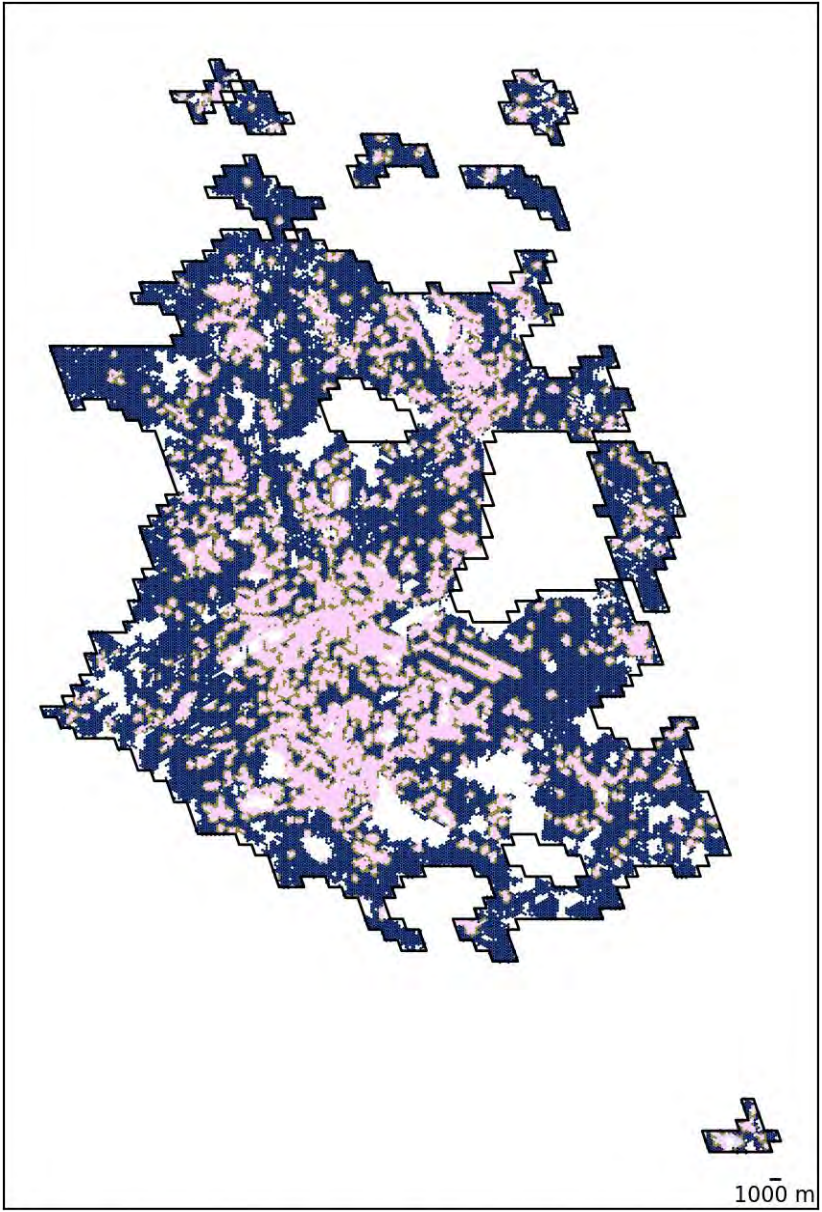

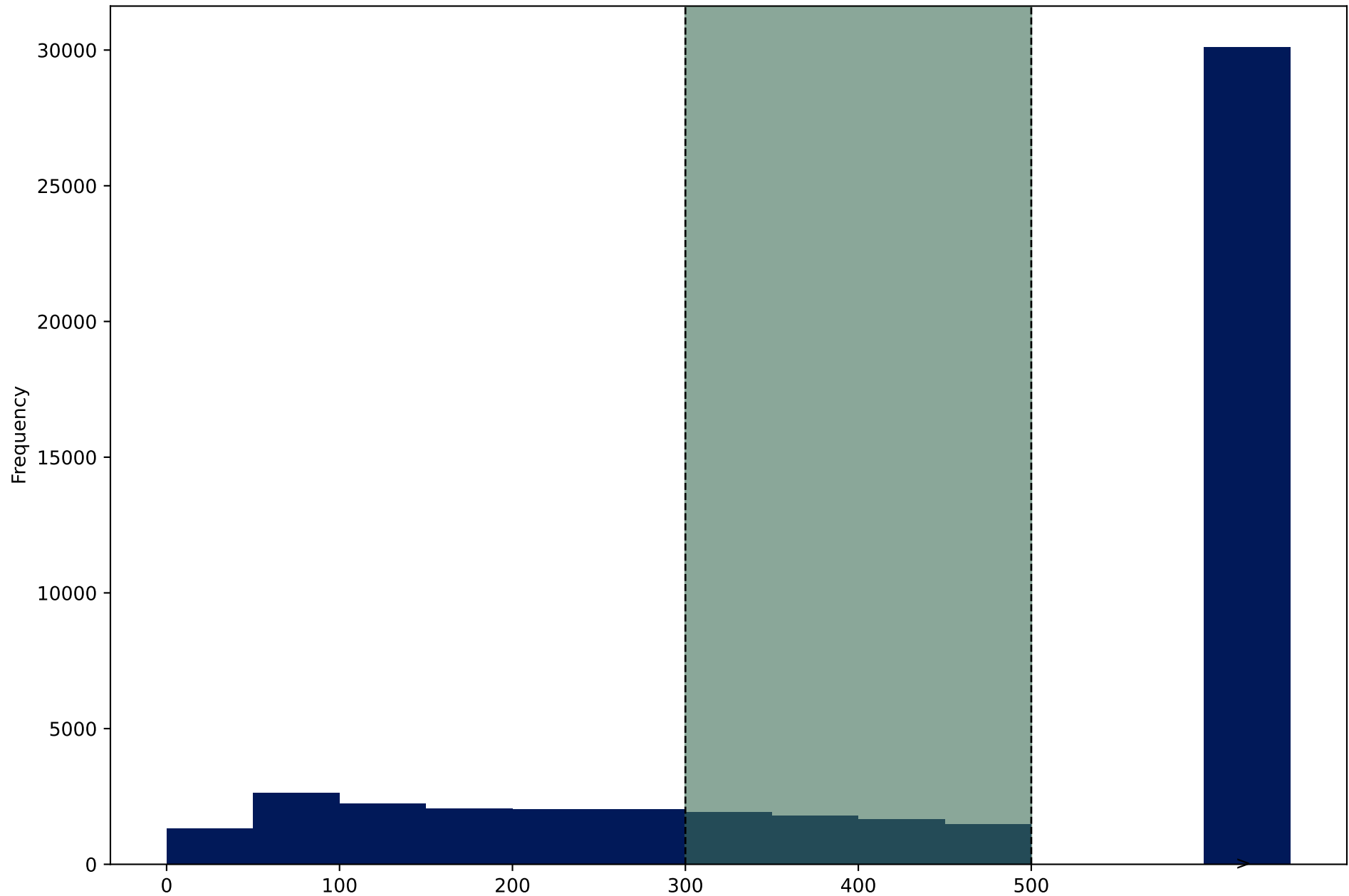

# America, North, United States, Baltimore

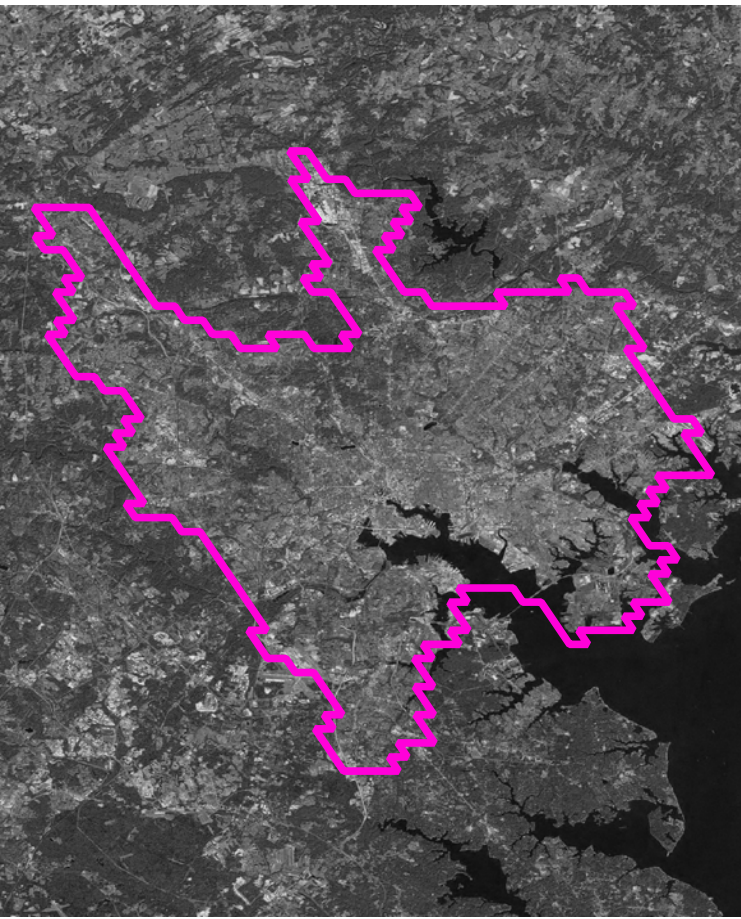
Satellite imagery of urban study region (Bing)

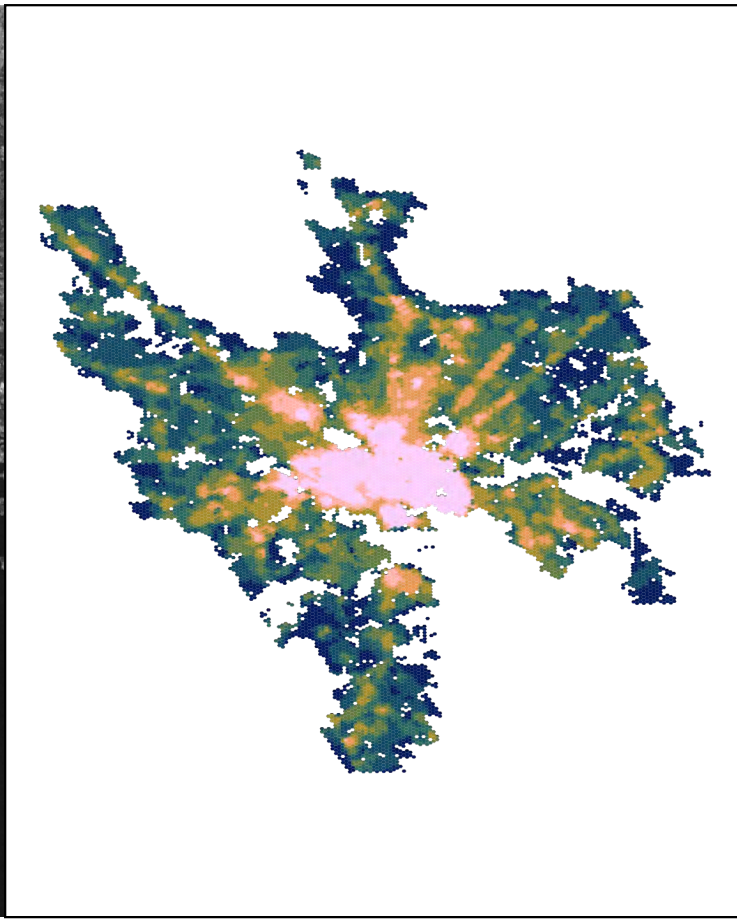
Walkability, relative to city

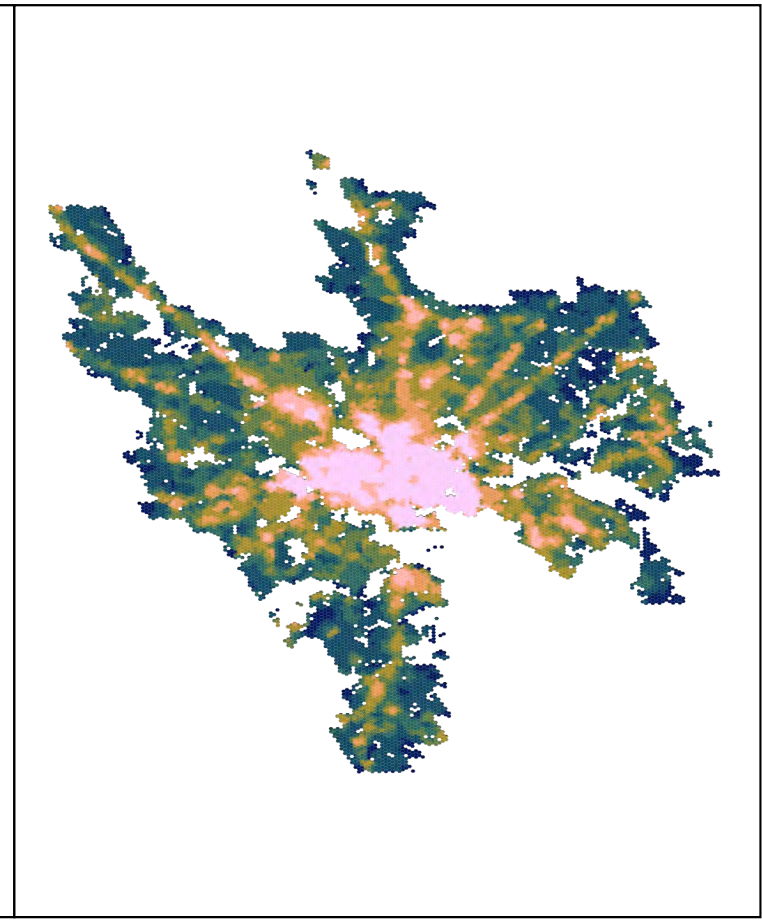
Walkability, relative to 25 global cities

Urban boundary

0 — 18 — 36 km

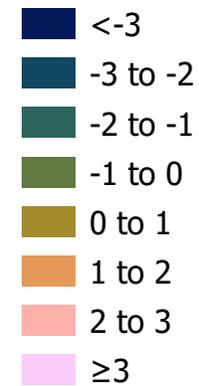
Walkability score
- <-3
- -3 to -2
- -2 to -1
- -1 to 0
- 0 to 1
- 1 to 2
- 2 to 3
- ≥3

Walkability relative to all cities by component variables (2D histograms), and overall (histogram)

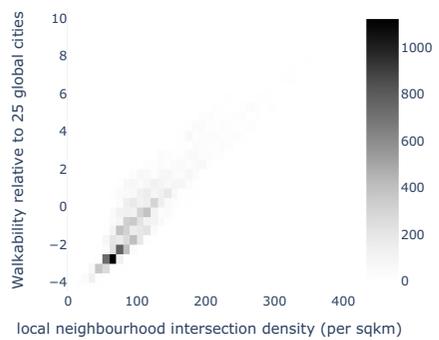
local neighbourhood intersection density (per sqkm)

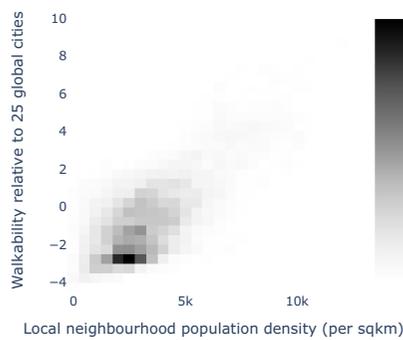
Local neighbourhood population density (per sqkm)

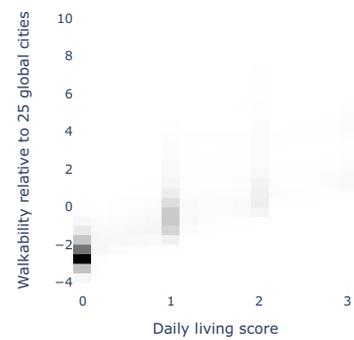
Daily living score

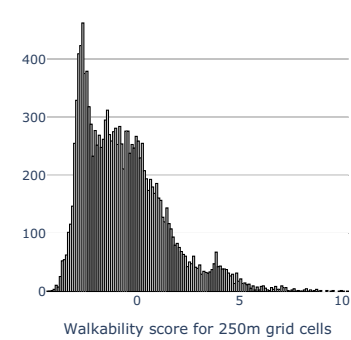
Walkability score for 250m grid cells

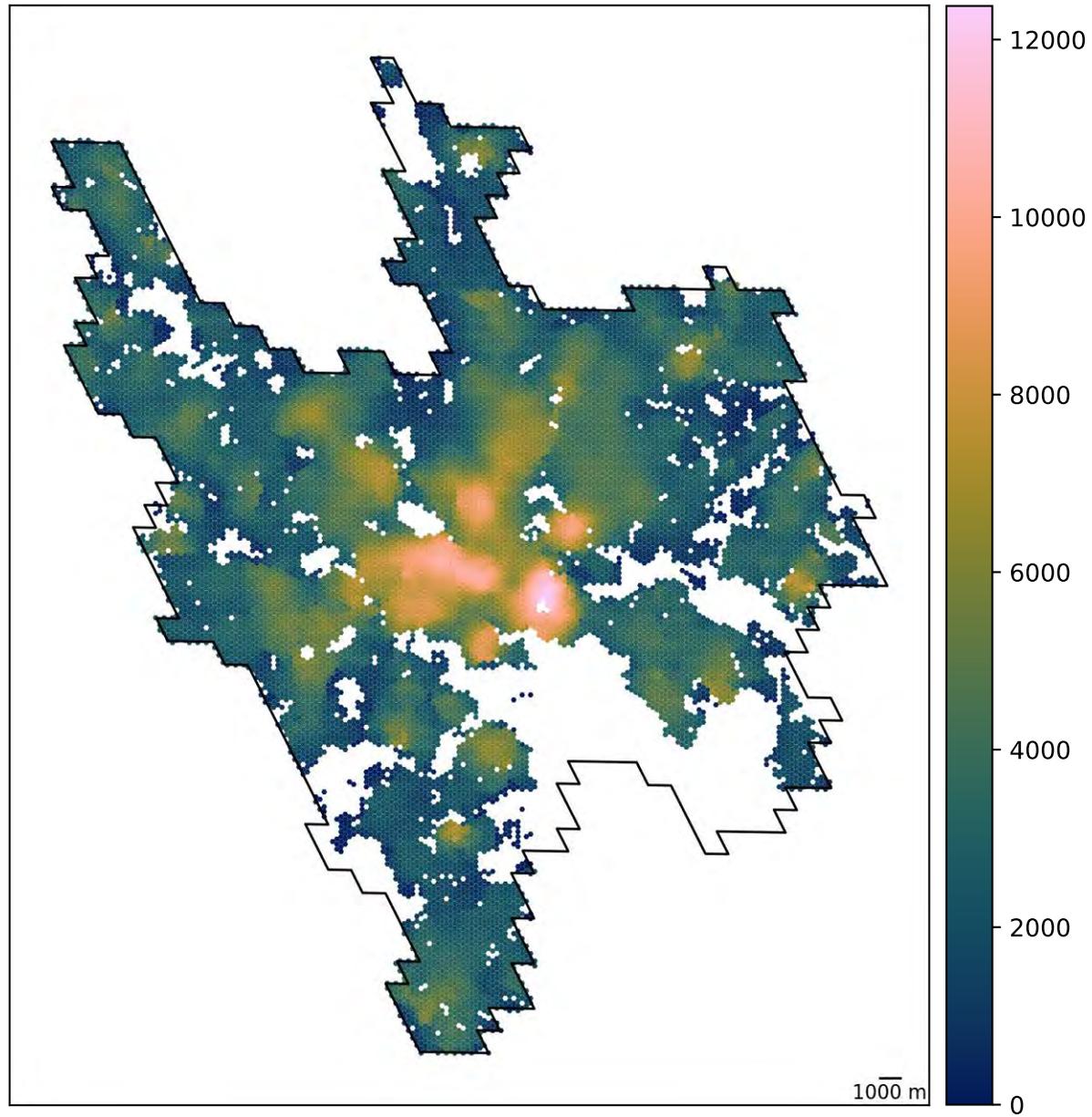

Mean 1000 m neighbourhood population per km²

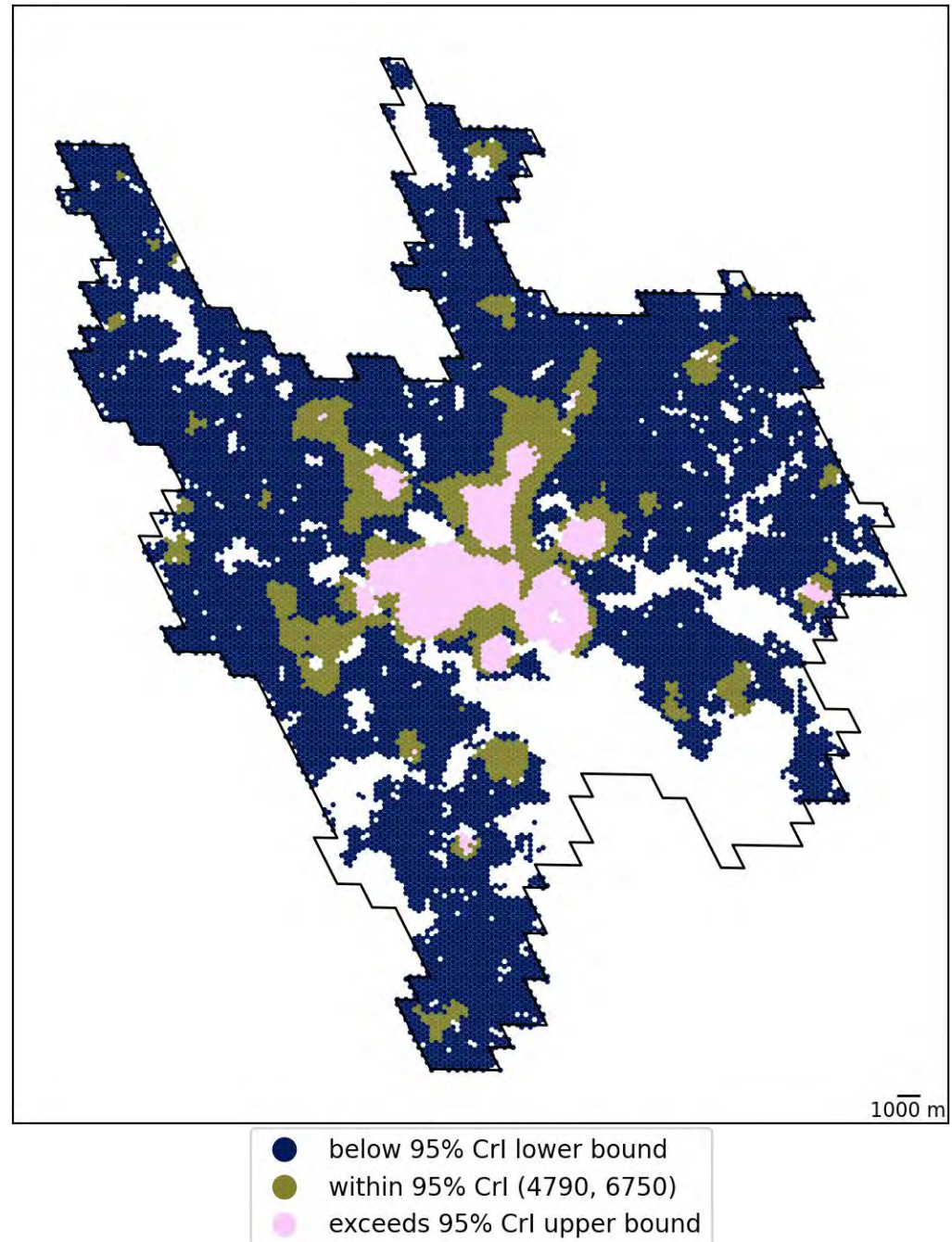

A: Estimated Mean 1000 m neighbourhood population per km² requirement for ≥80% probability of engaging in walking for transport

- below 95% CrI lower bound
- within 95% CrI (4790, 6750)
- exceeds 95% CrI upper bound

B: Estimated Mean 1000 m neighbourhood population per km² requirement for reaching the WHO's target of a ≥15% relative reduction in insufficient physical activity through walking

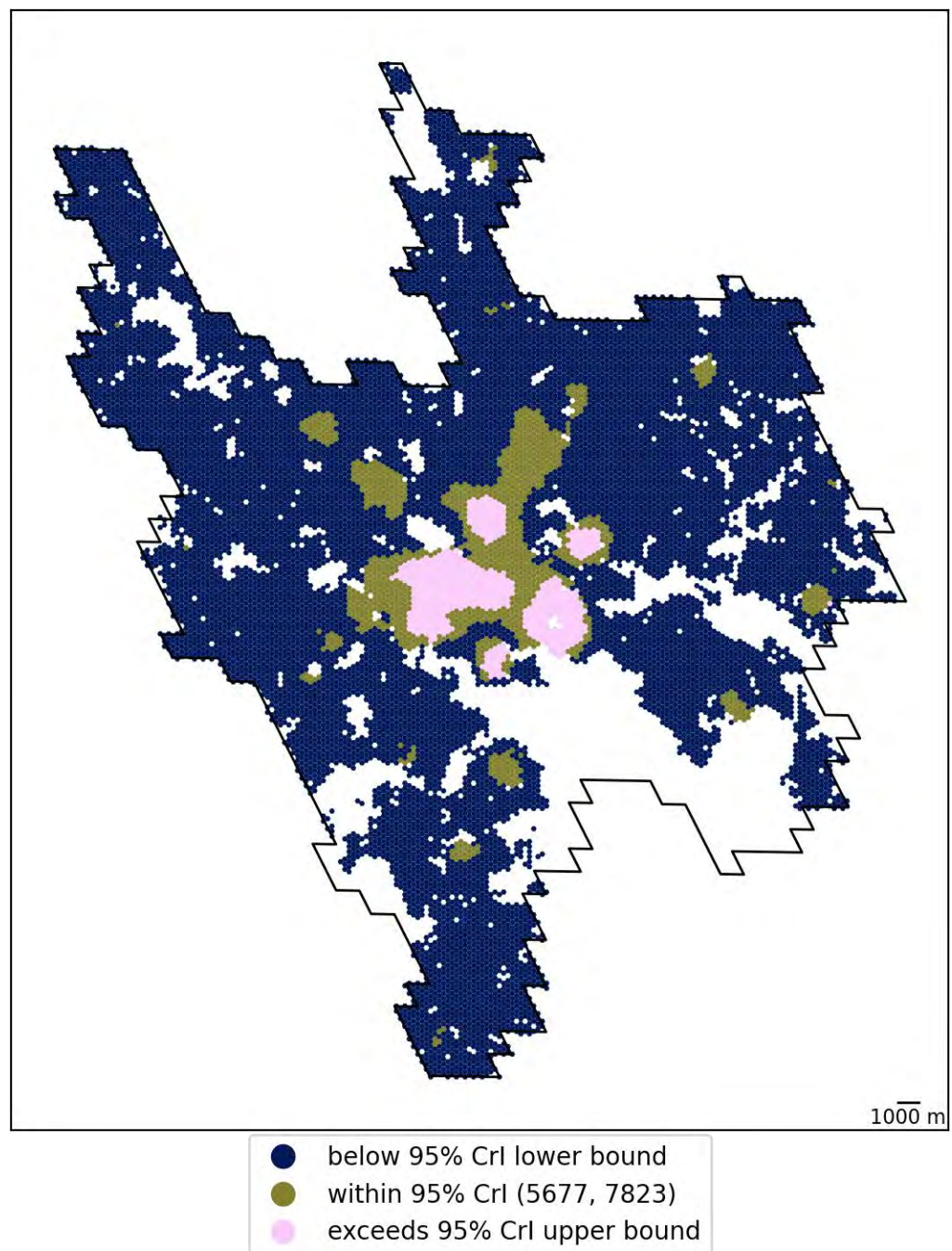

- below 95% CrI lower bound
- within 95% CrI (5677, 7823)
- exceeds 95% CrI upper bound

1000 m

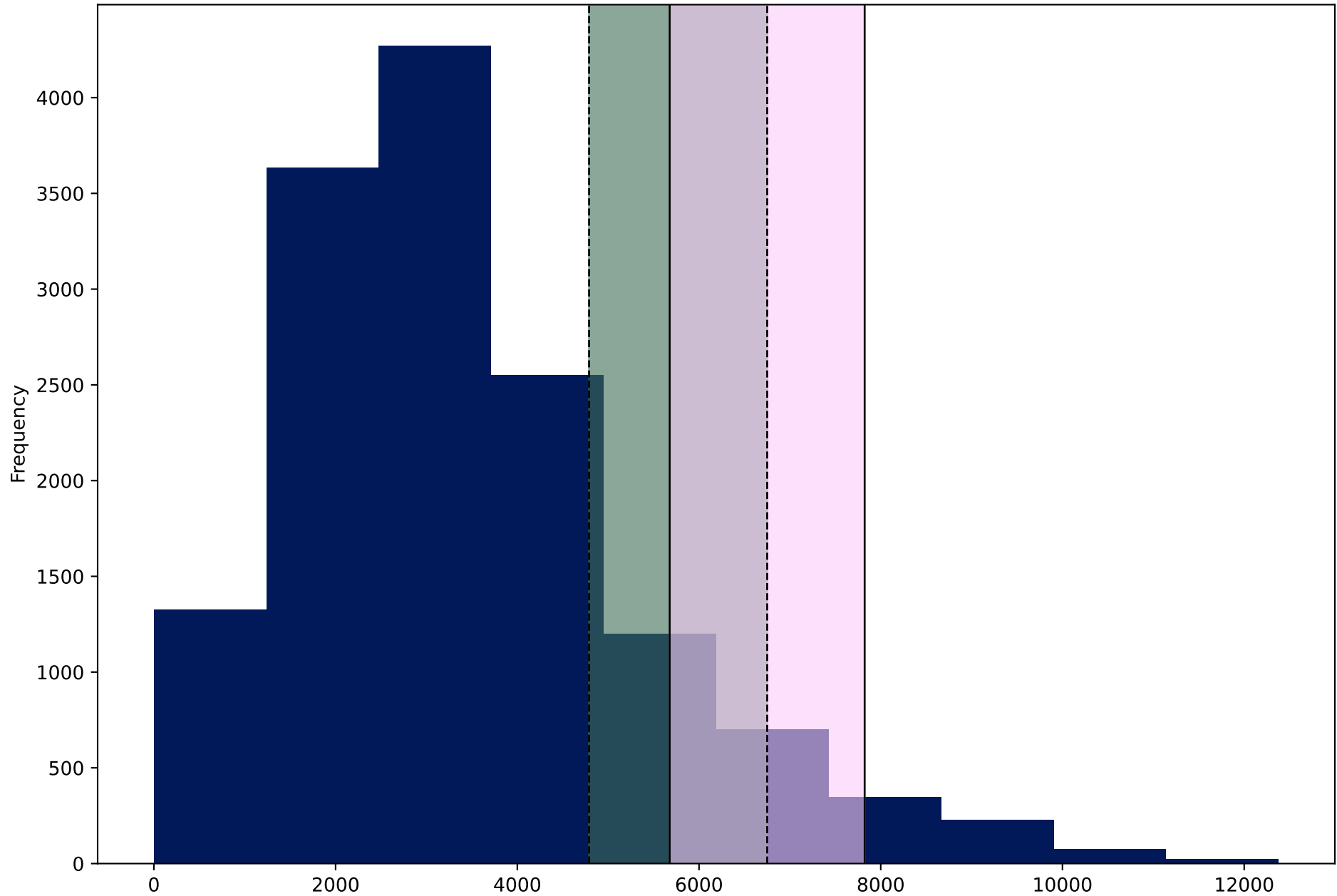

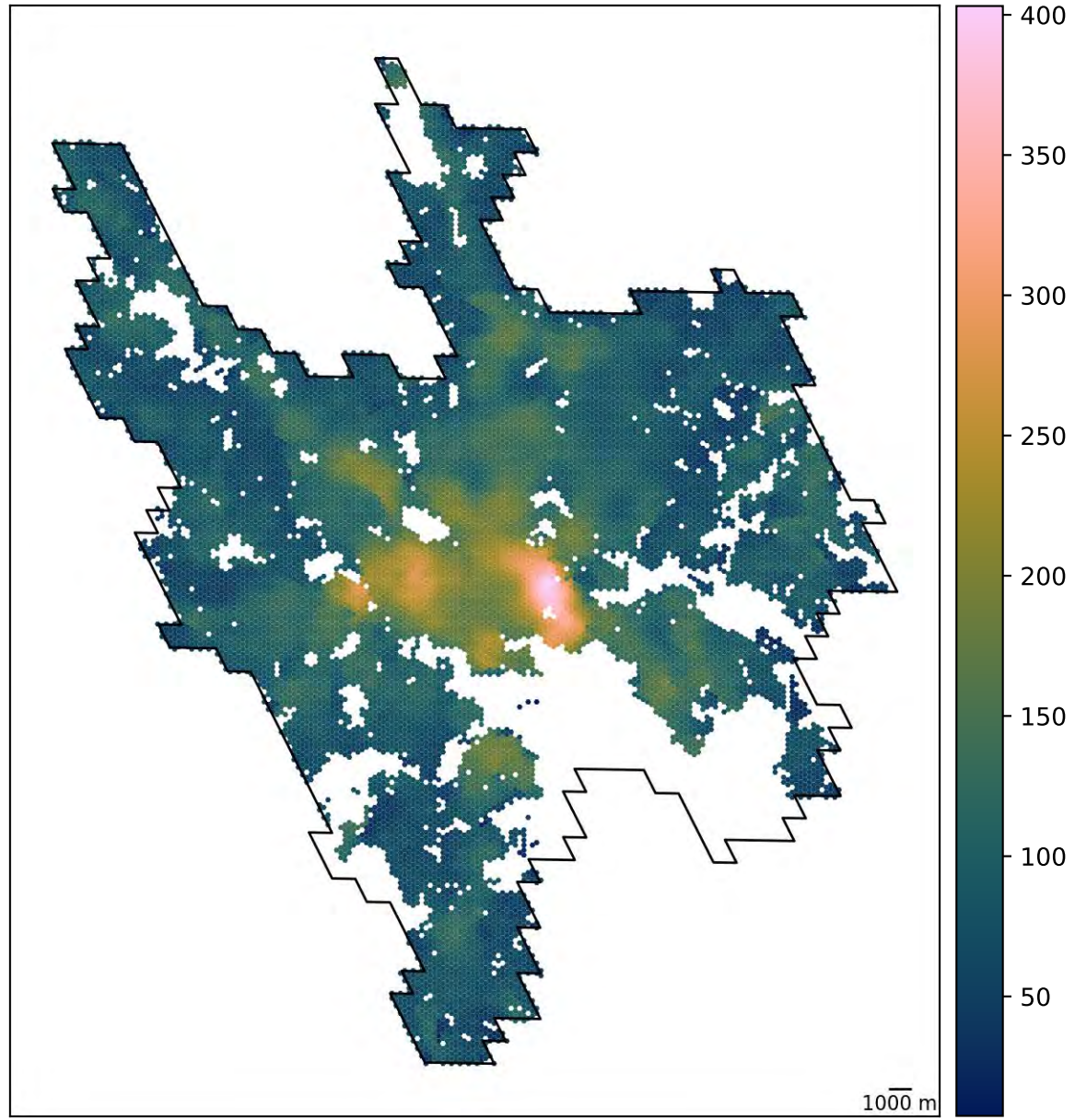

Mean 1000 m neighbourhood street intersections per km²

A: Estimated Mean 1000 m neighbourhood street intersections per km² requirement for ≥80% probability of engaging in walking for transport

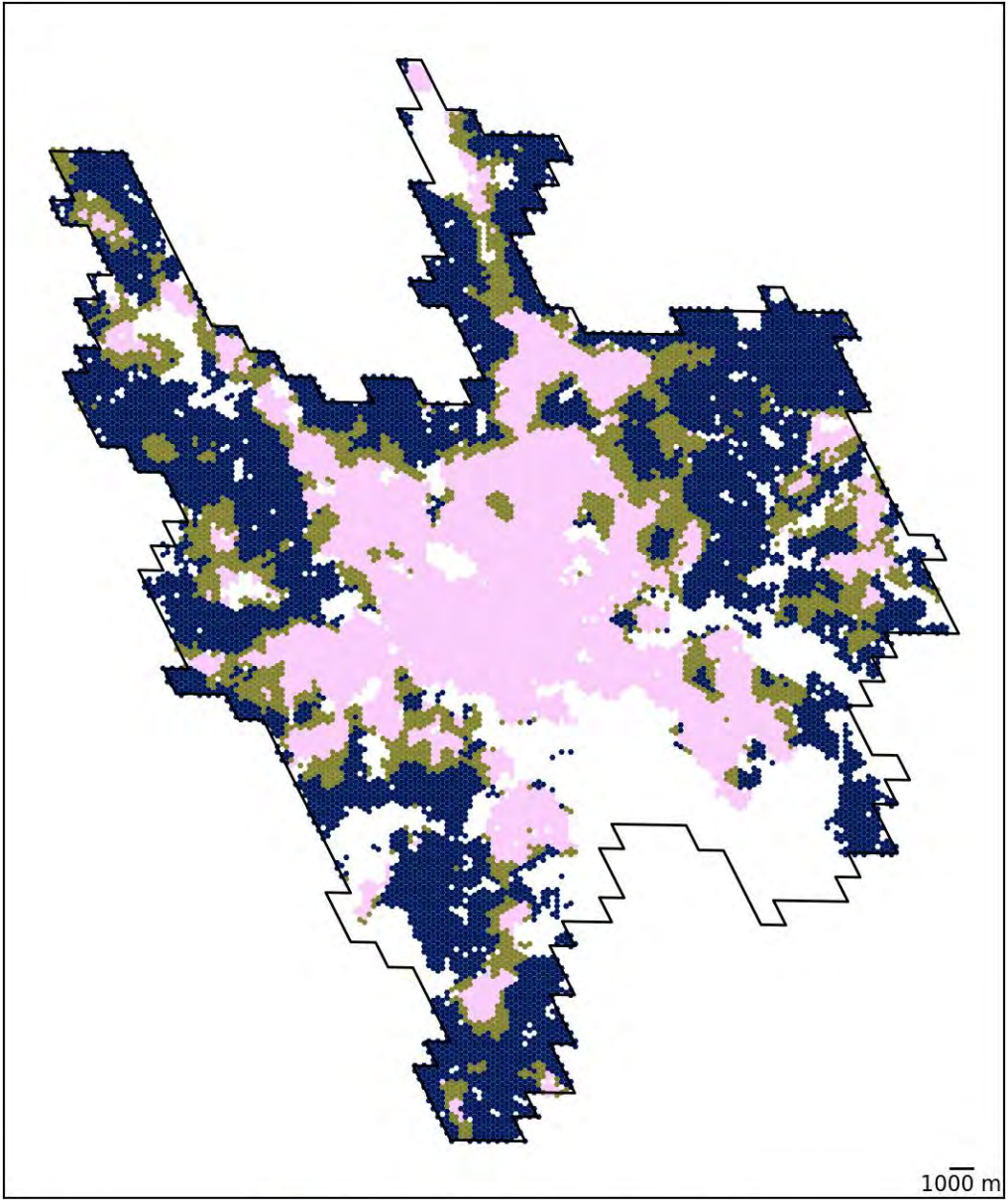

B: Estimated Mean 1000 m neighbourhood street intersections per km² requirement for reaching the WHO's target of a ≥15% relative reduction in insufficient physical activity through walking

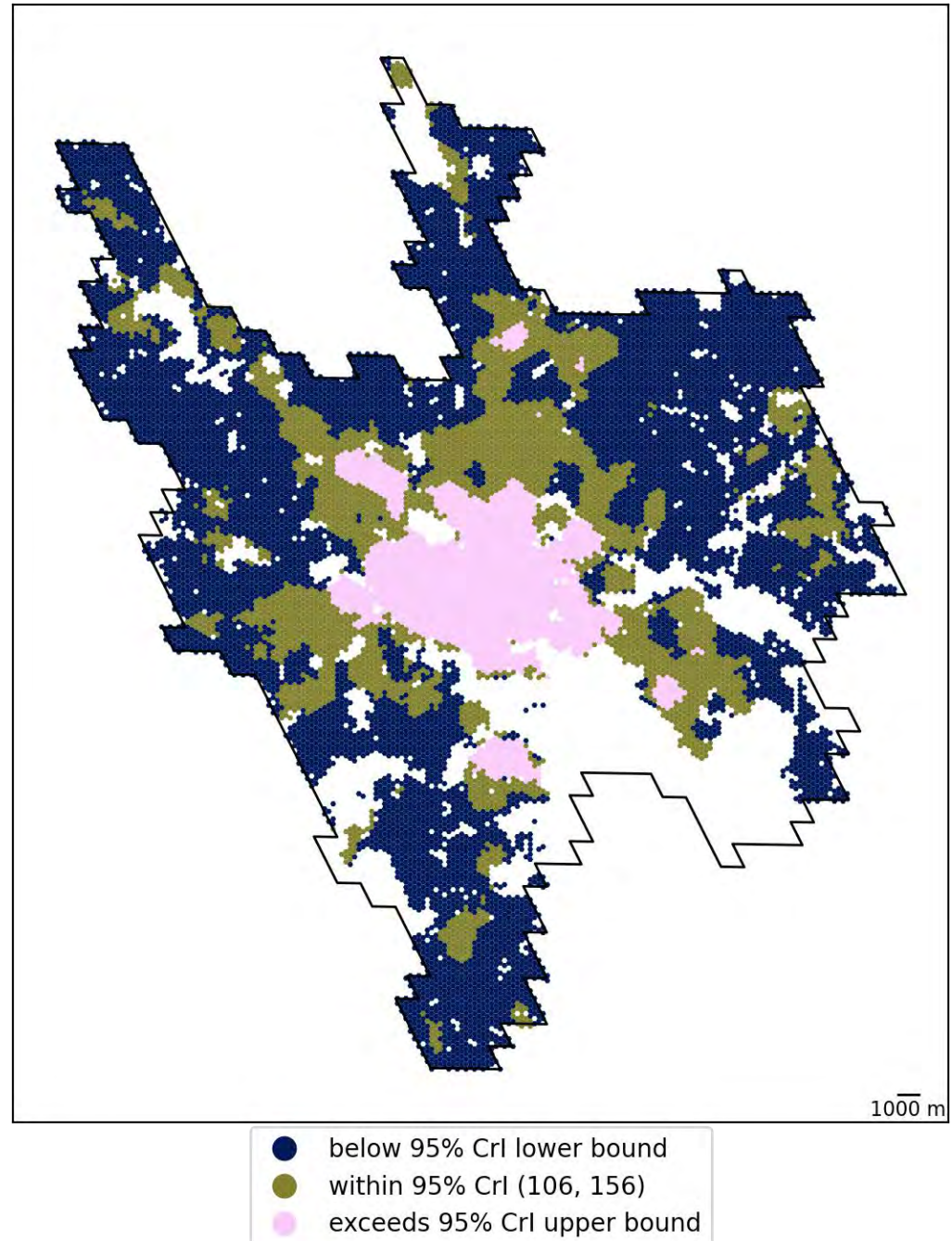

- below 95% CrI lower bound
- within 95% CrI (106, 156)
- exceeds 95% CrI upper bound

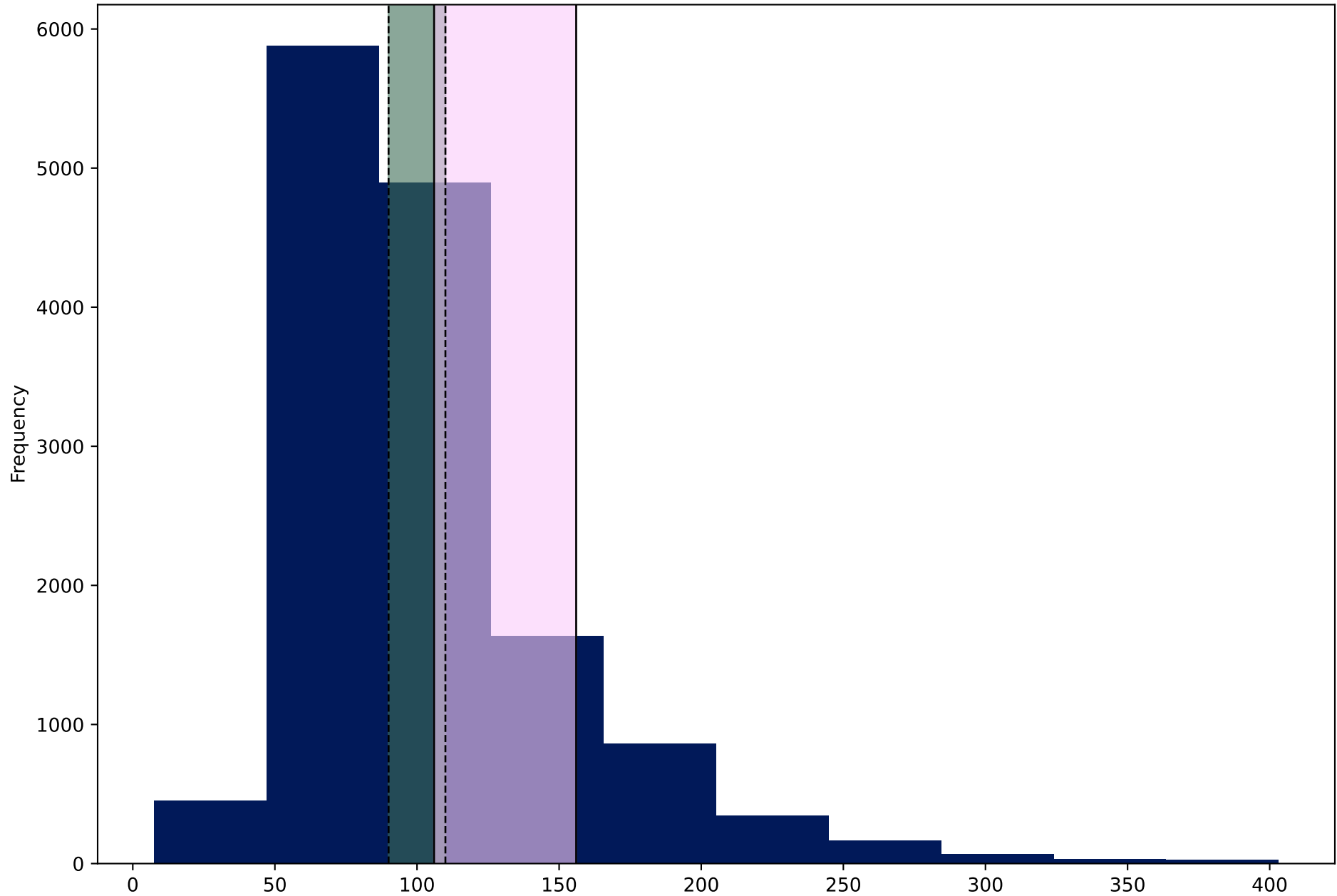

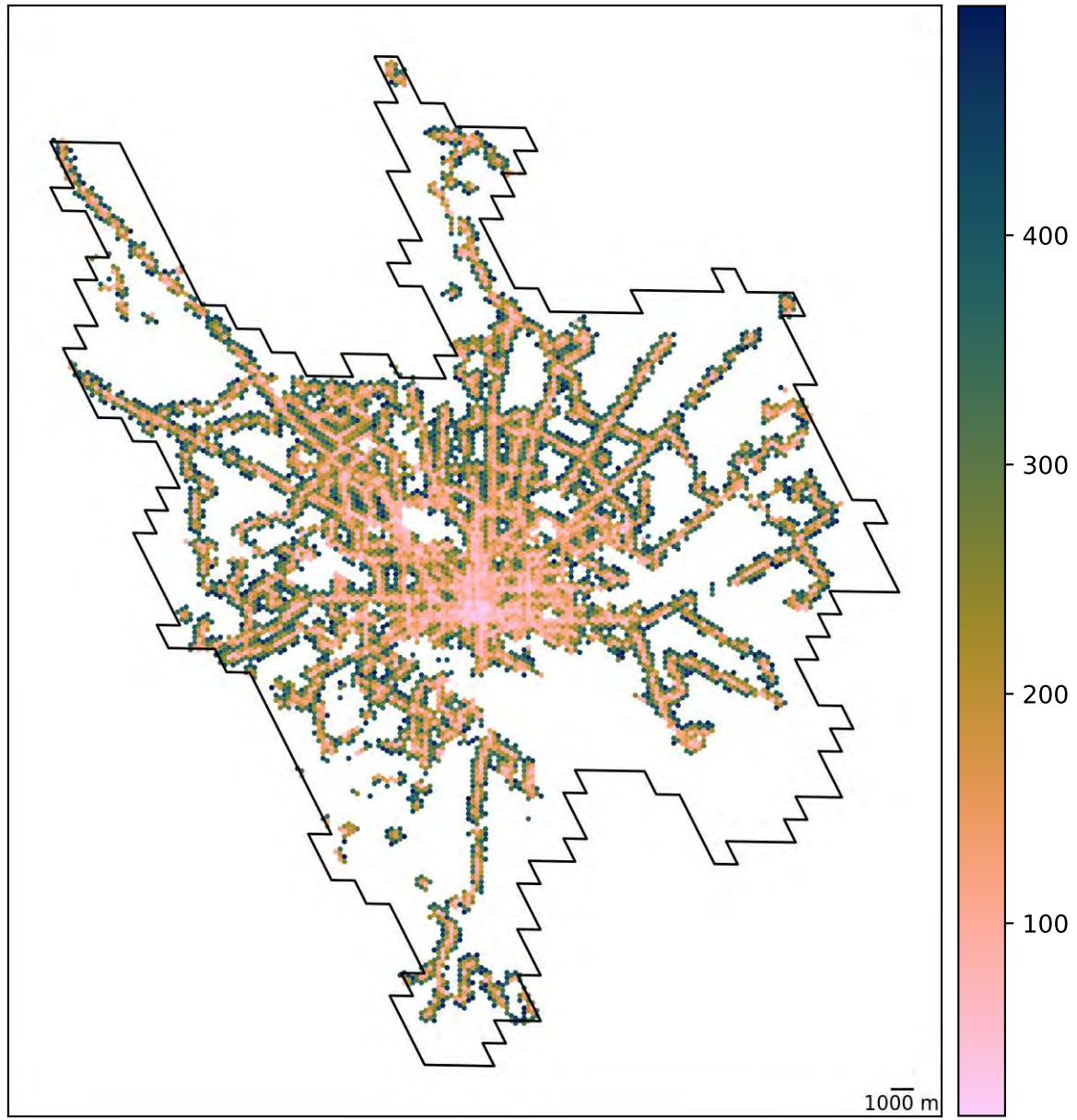

Distance to nearest public transport stops (m; up to 500m)

distances: Estimated Distance to nearest public transport stops (m; up to 500m) requirement for distances to destinations, measured up to a maximum distance target threshold of 500 metres

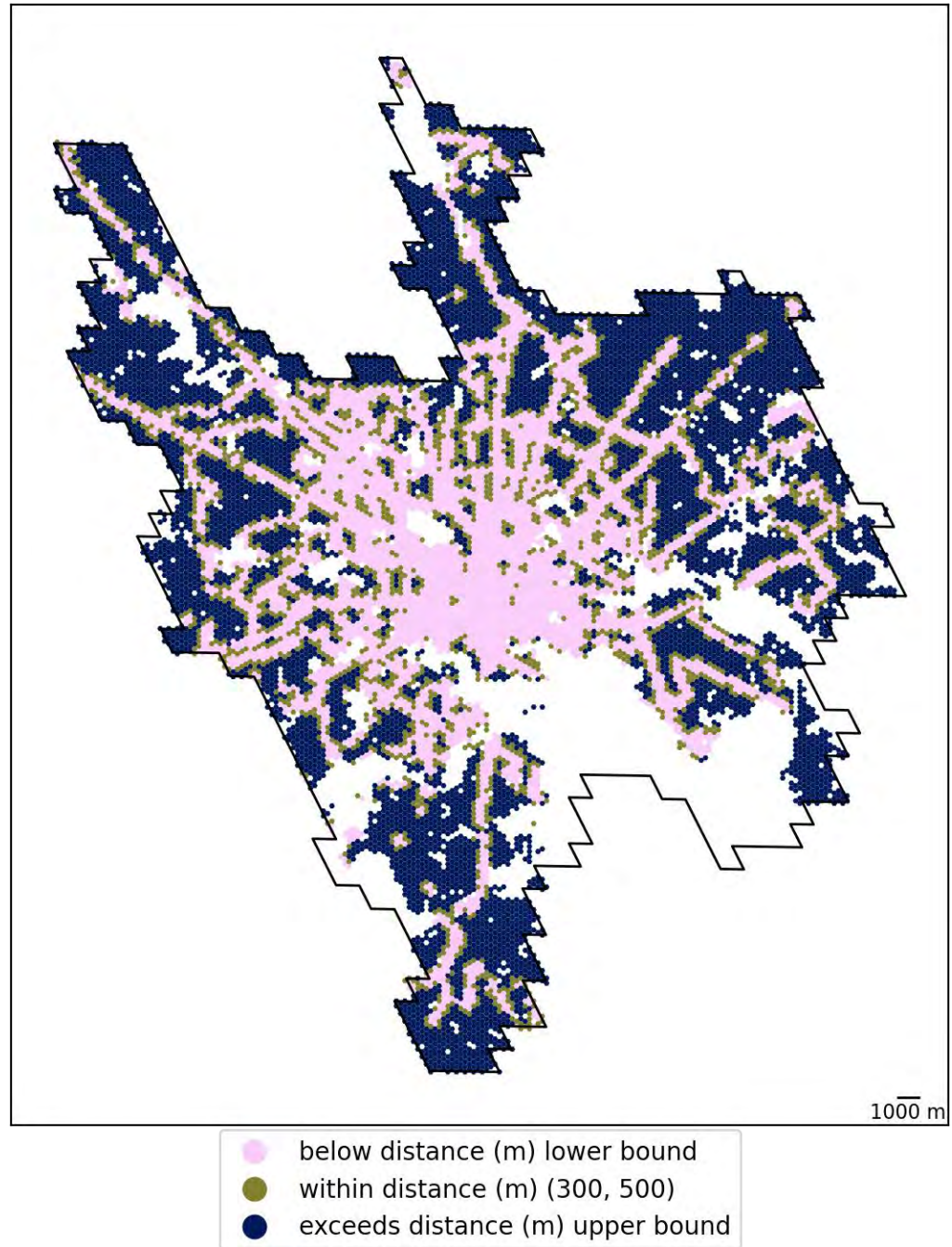

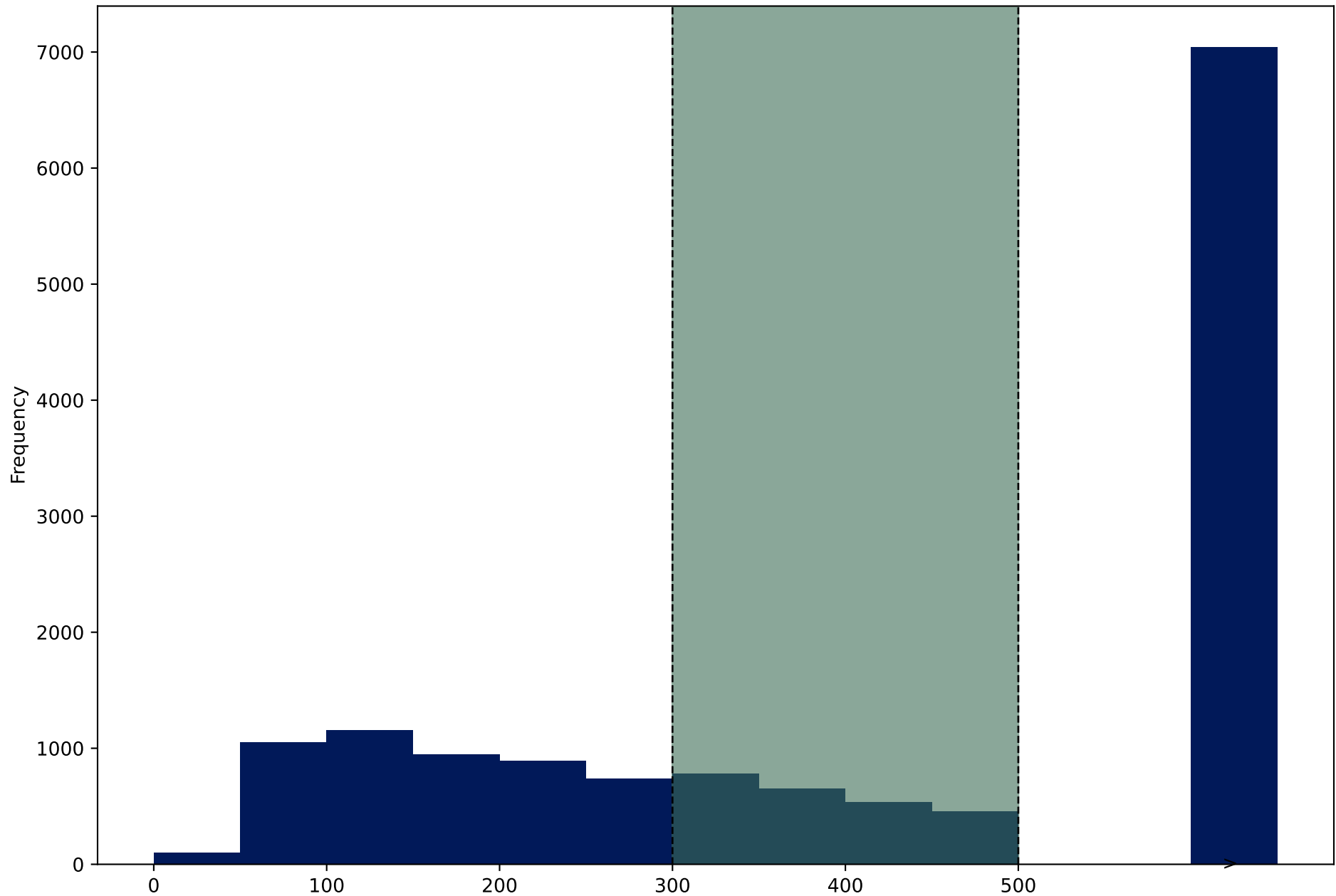

# America, North, United States, Phoenix

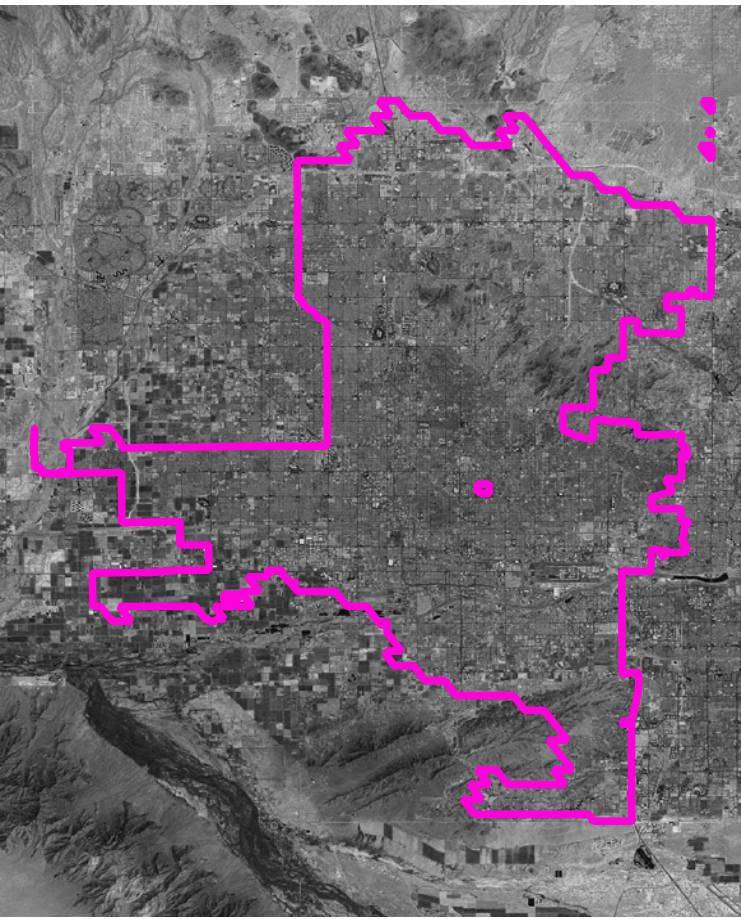
Satellite imagery of urban study region (Bing)

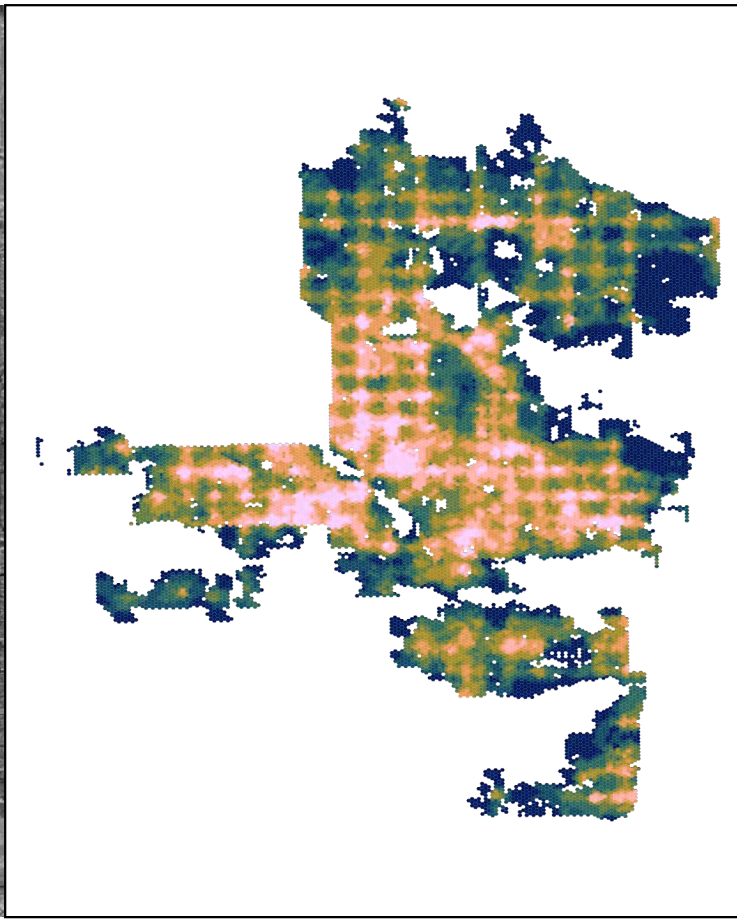
Walkability, relative to city

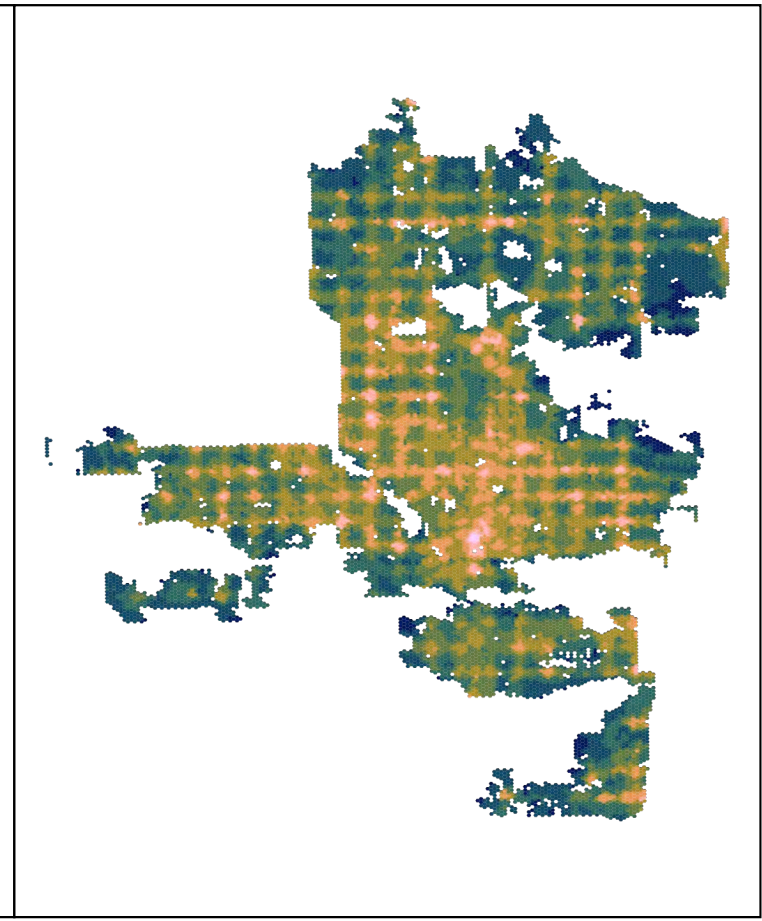
Walkability, relative to 25 global cities

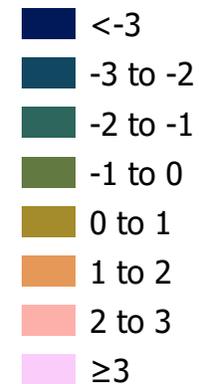

Walkability score
- <-3
- -3 to -2
- -2 to -1
- -1 to 0
- 0 to 1
- 1 to 2
- 2 to 3
- ≥3

Walkability relative to all cities by component variables (2D histograms), and overall (histogram)

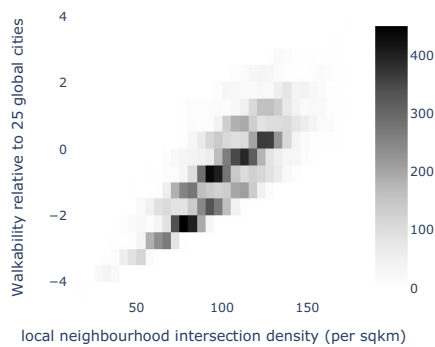
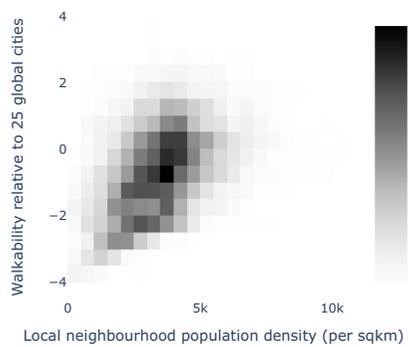
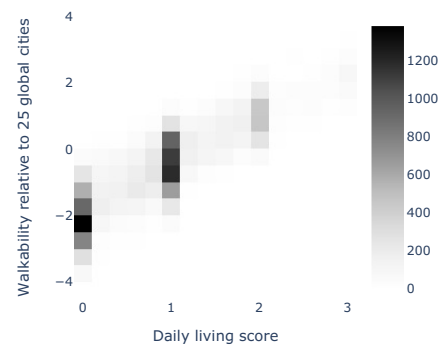
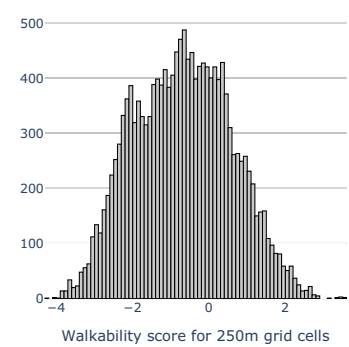

Distance to nearest park (m; up to 500m)

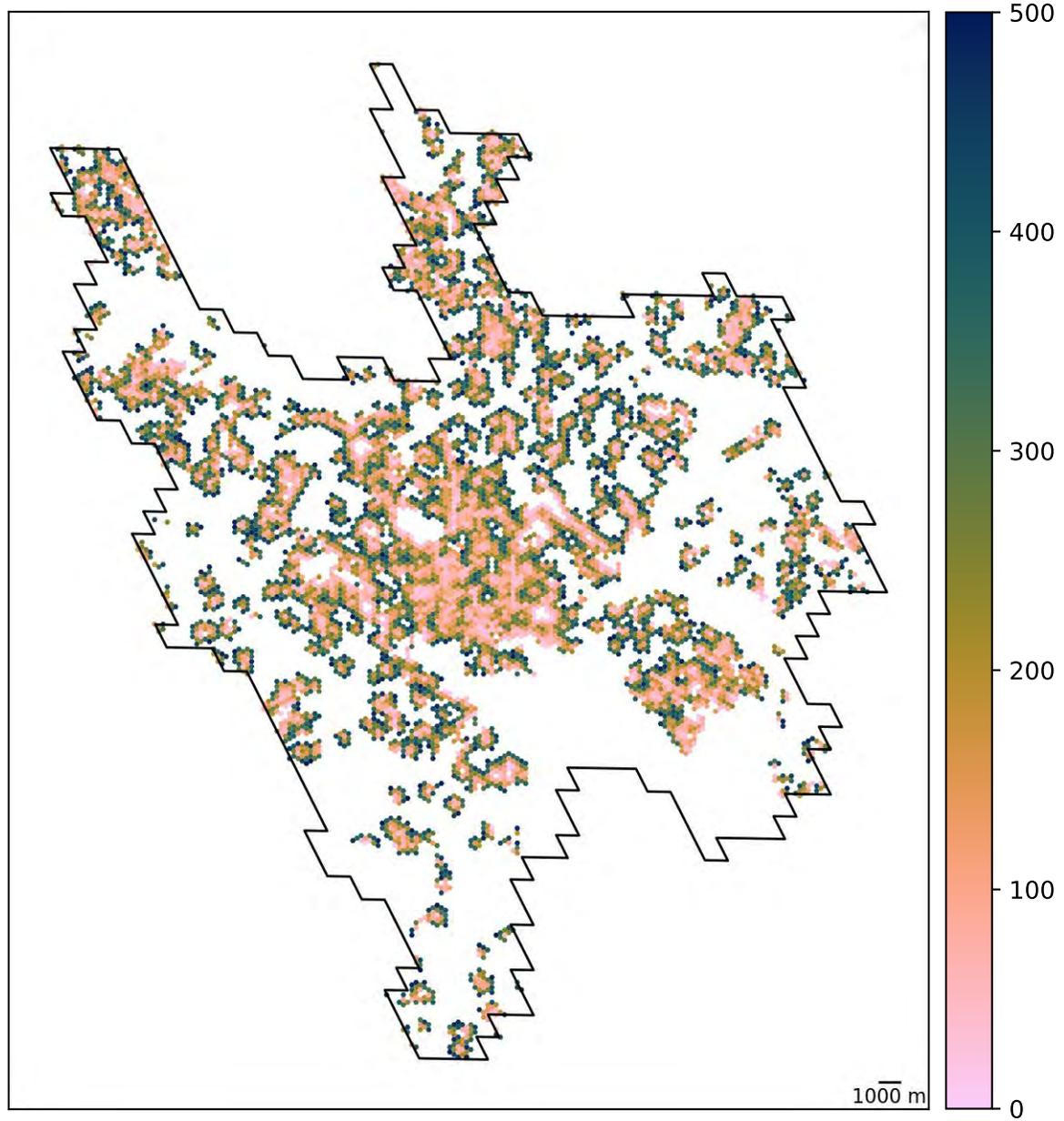

distances: Estimated Distance to nearest park (m; up to 500m) requirement for distances to destinations, measured up to a maximum distance target threshold of 500 metres

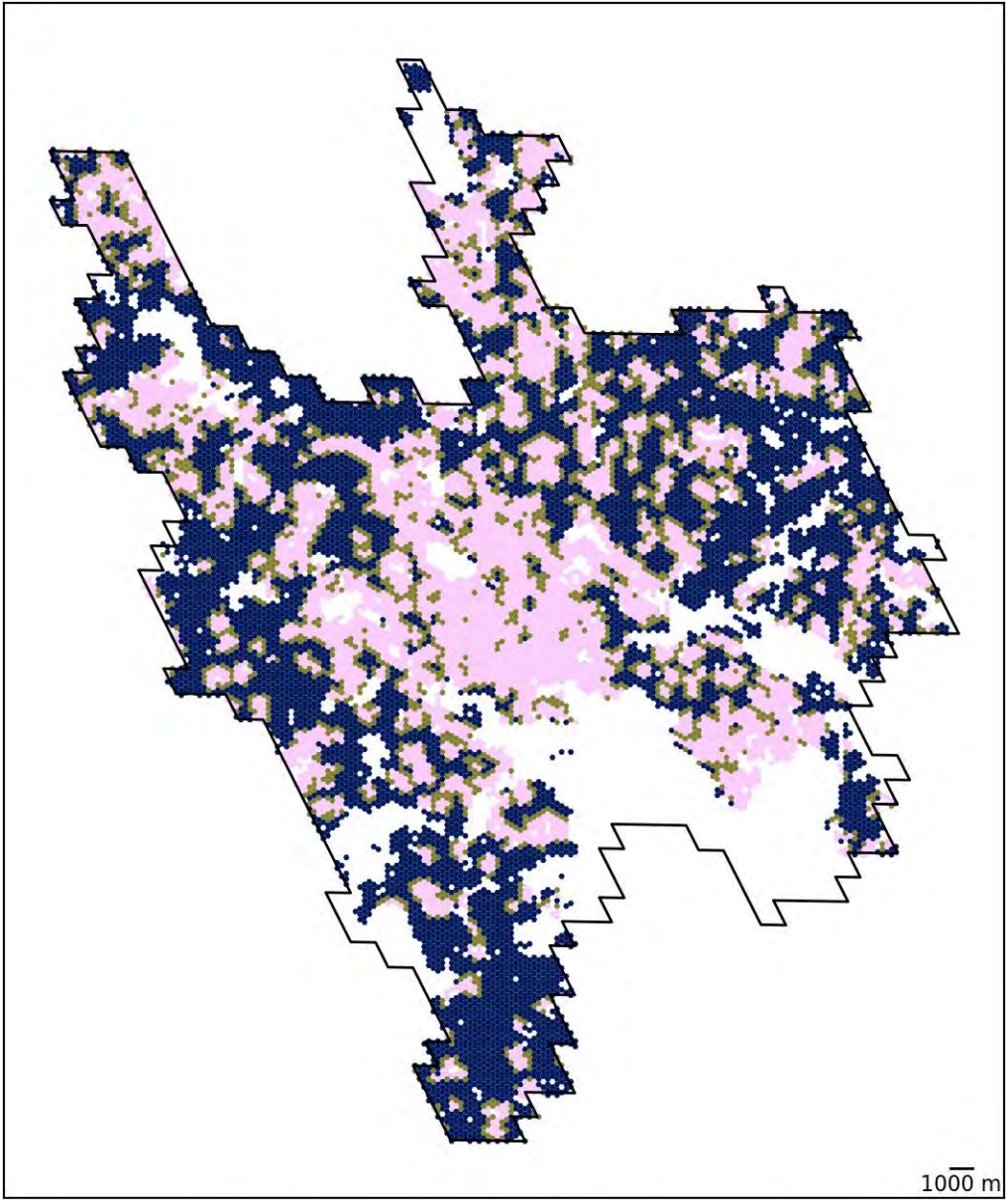

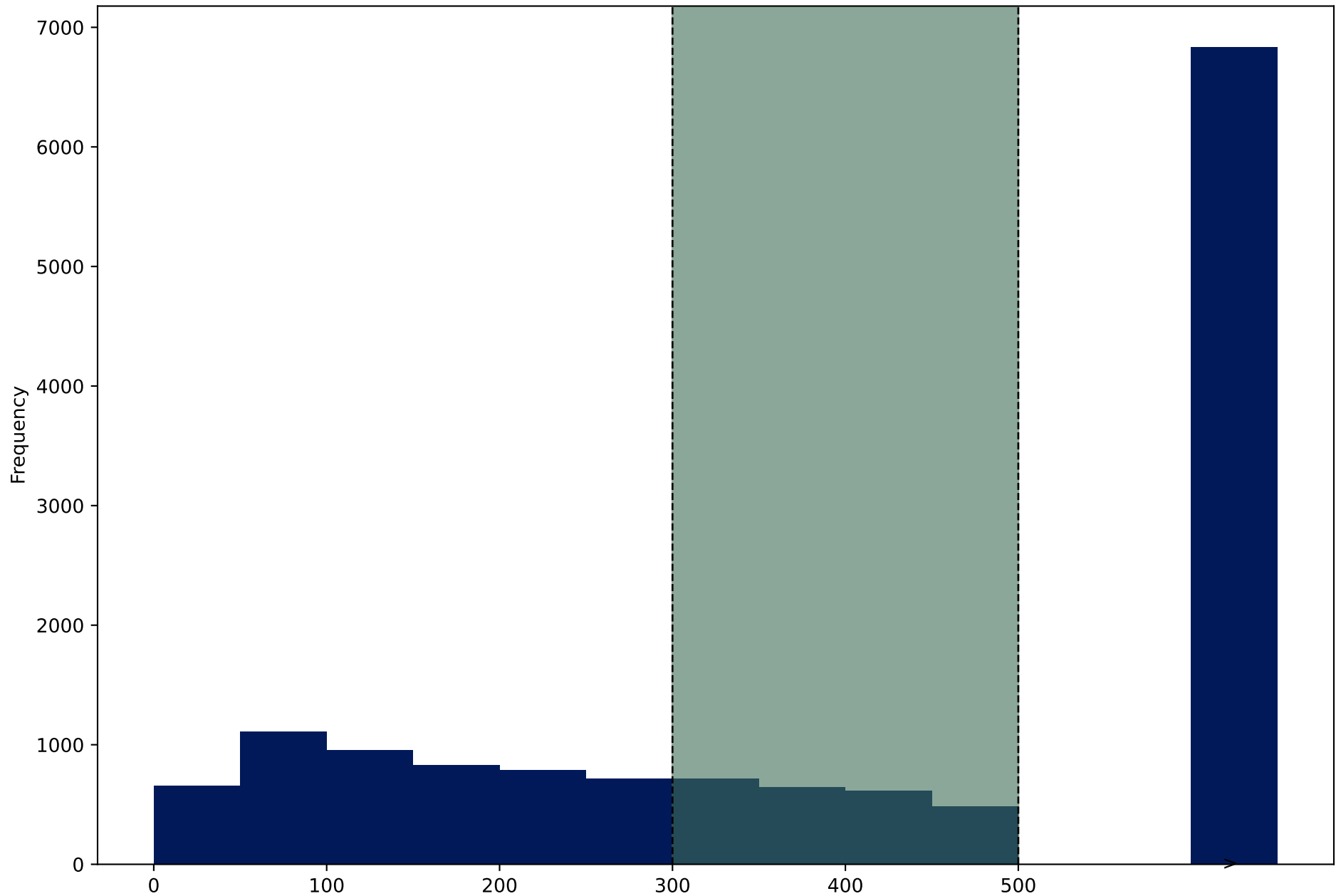

# America, North, United States, Phoenix

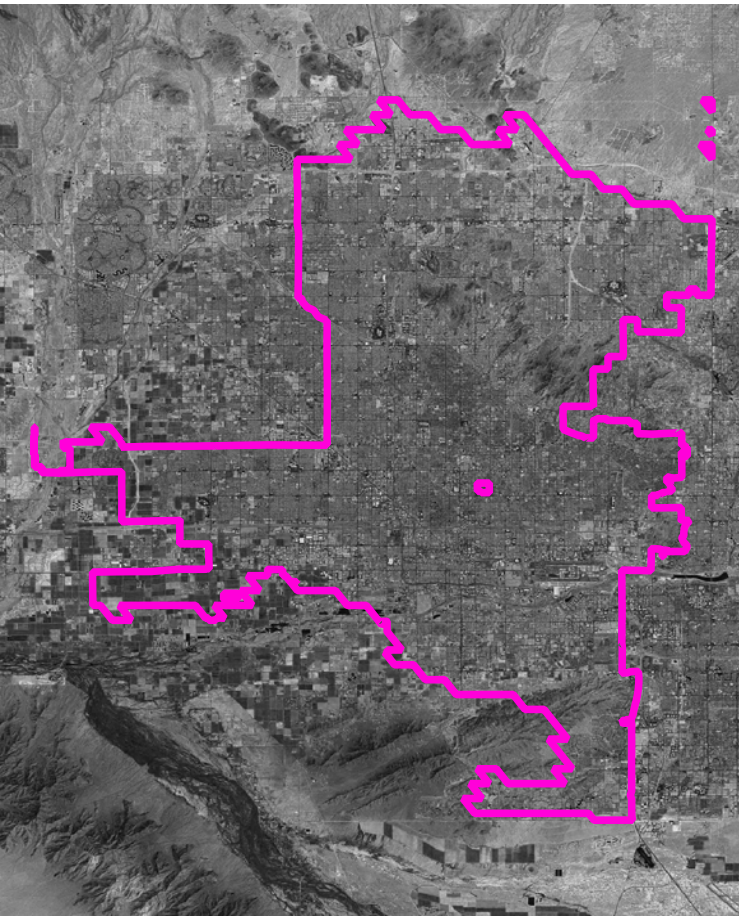
Satellite imagery of urban study region (Bing)

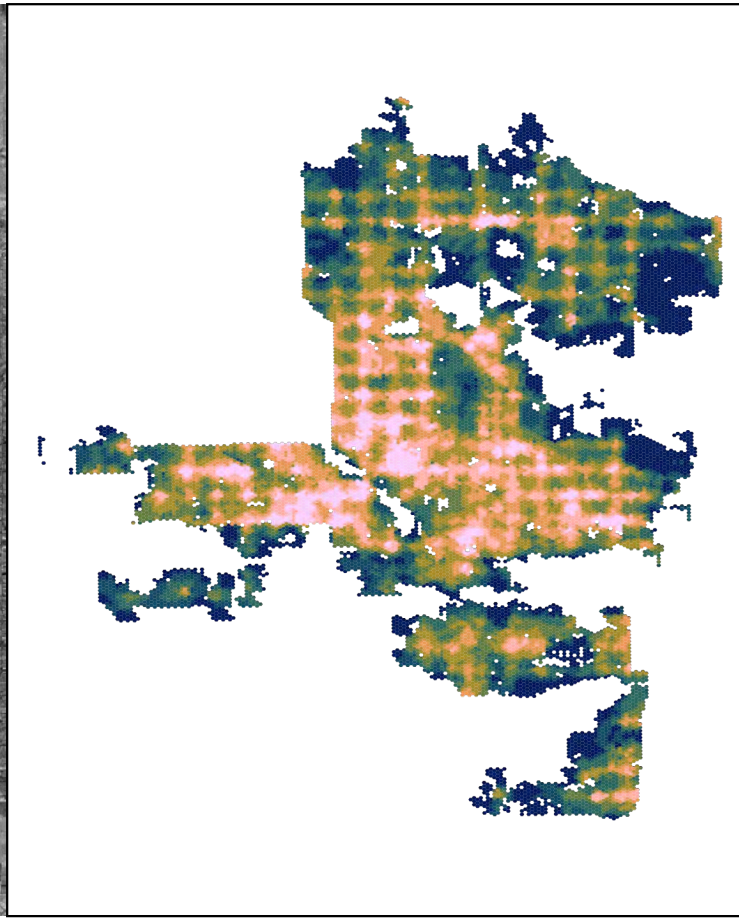
Walkability, relative to city

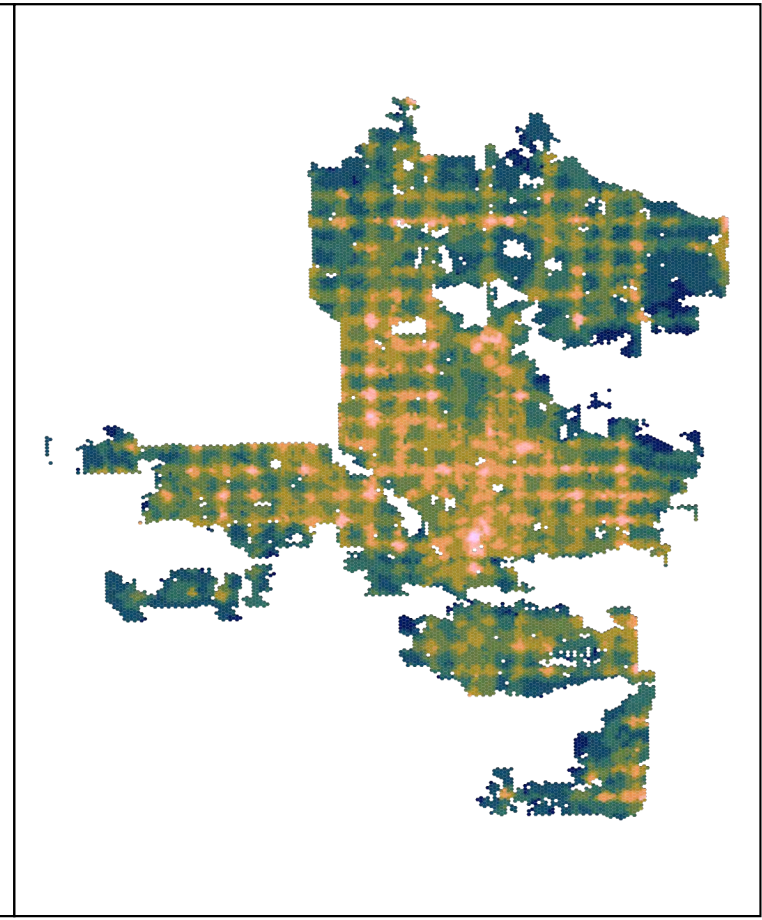
Walkability, relative to 25 global cities

Urban boundary

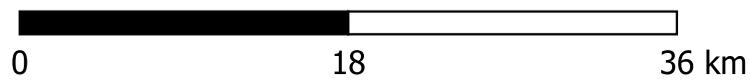
0  18  36 km

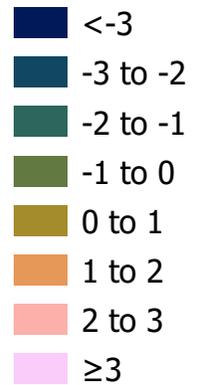
Walkability score
- <-3
- -3 to -2
- -2 to -1
- -1 to 0
- 0 to 1
- 1 to 2
- 2 to 3
- ≥3

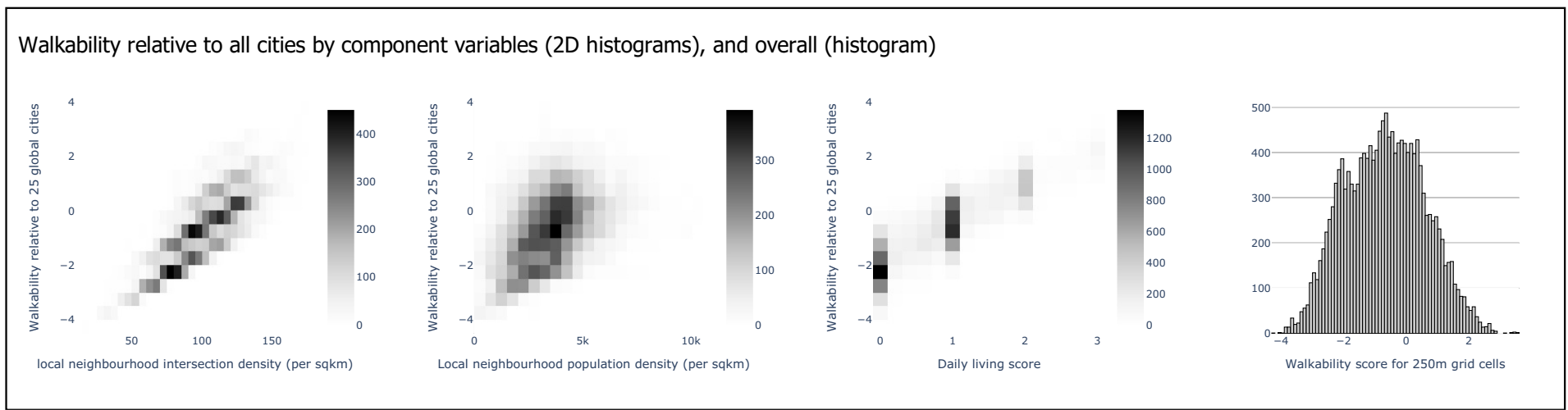
Walkability relative to all cities by component variables (2D histograms), and overall (histogram)

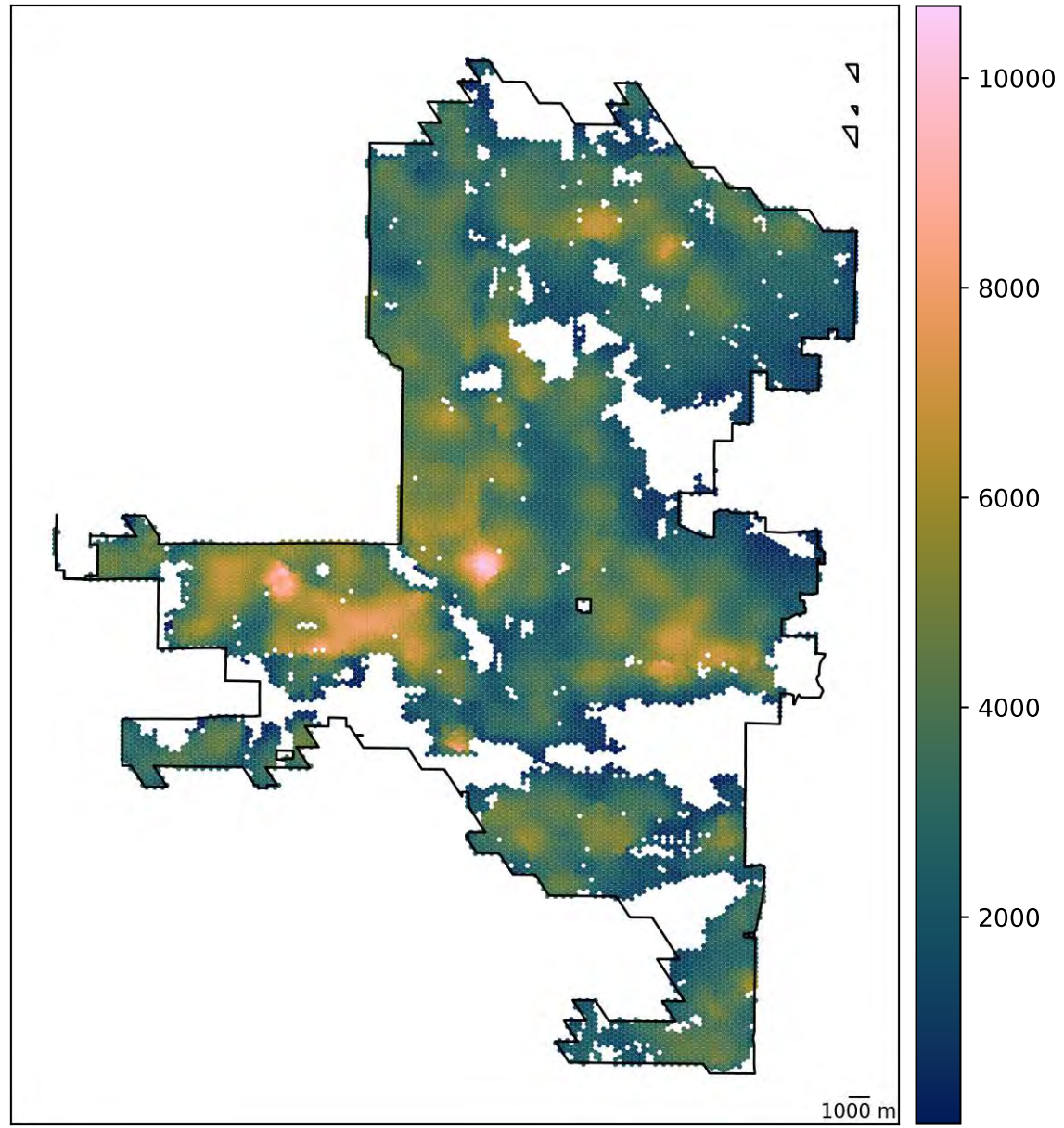

Mean 1000 m neighbourhood population per km²

A: Estimated Mean 1000 m neighbourhood population per km² requirement for ≥80% probability of engaging in walking for transport

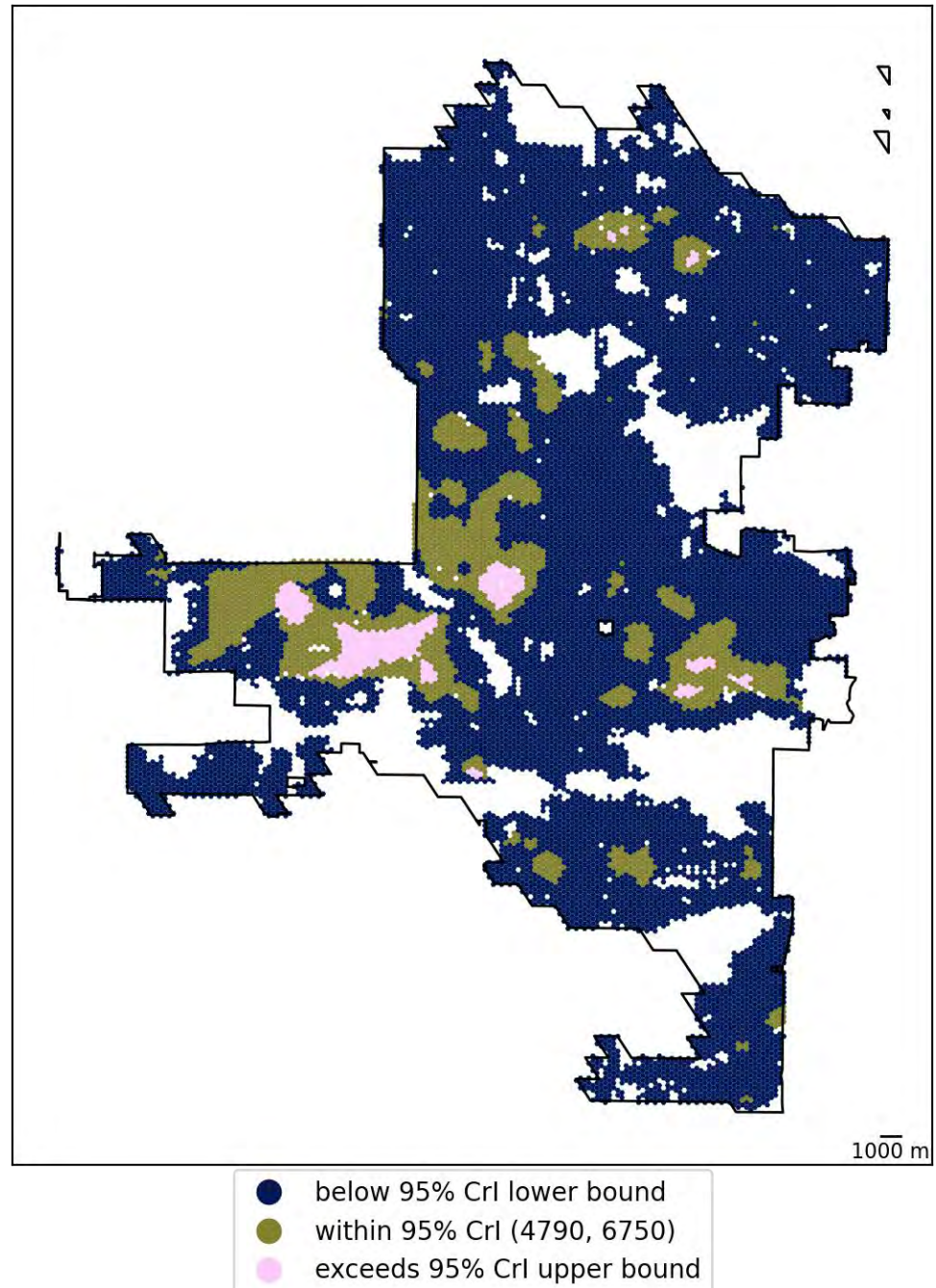

B: Estimated Mean 1000 m neighbourhood population per km² requirement for reaching the WHO's target of a ≥15% relative reduction in insufficient physical activity through walking

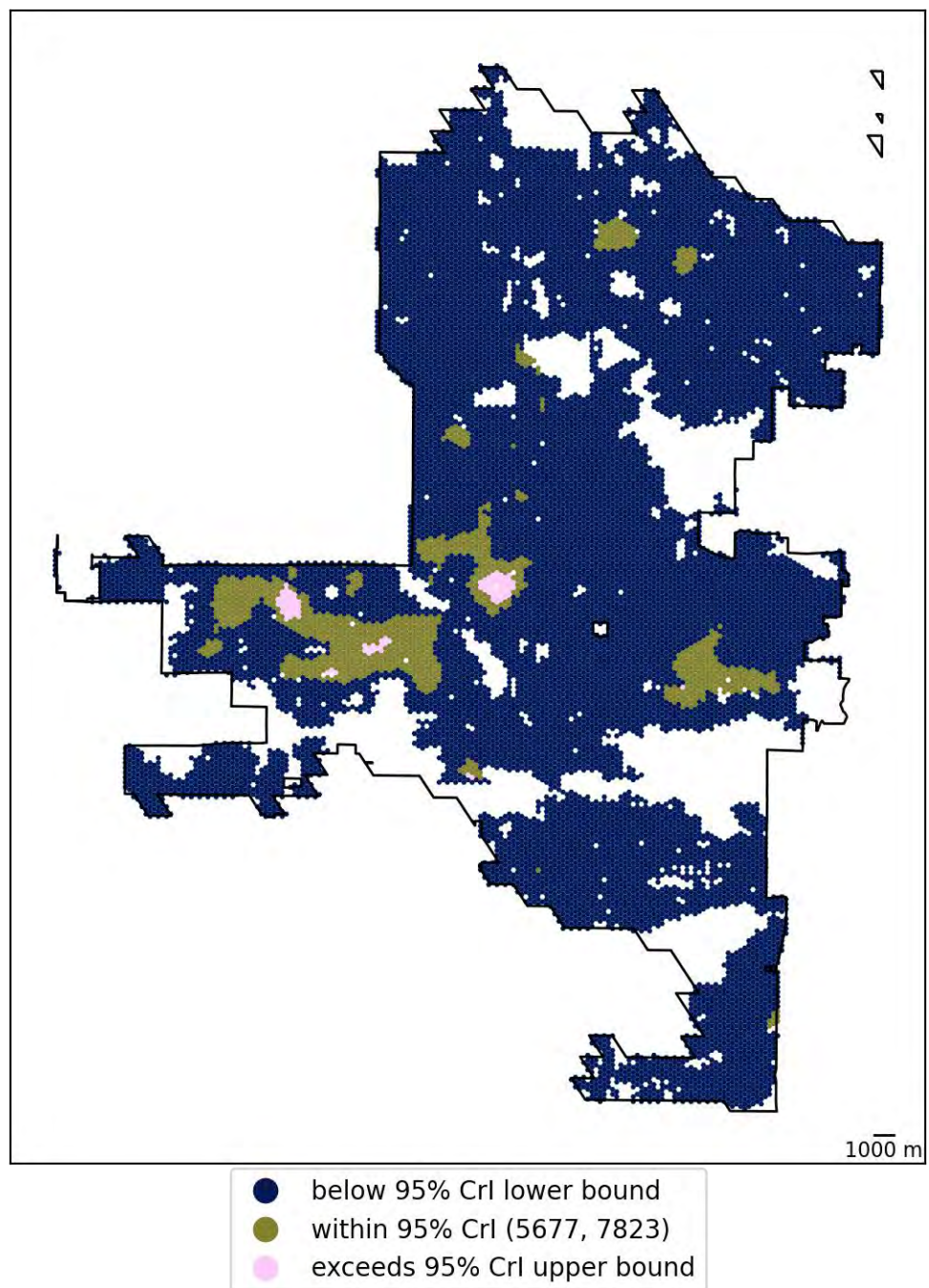

Legend:
- below 95% CrI lower bound
- within 95% CrI (5677, 7823)
- exceeds 95% CrI upper bound

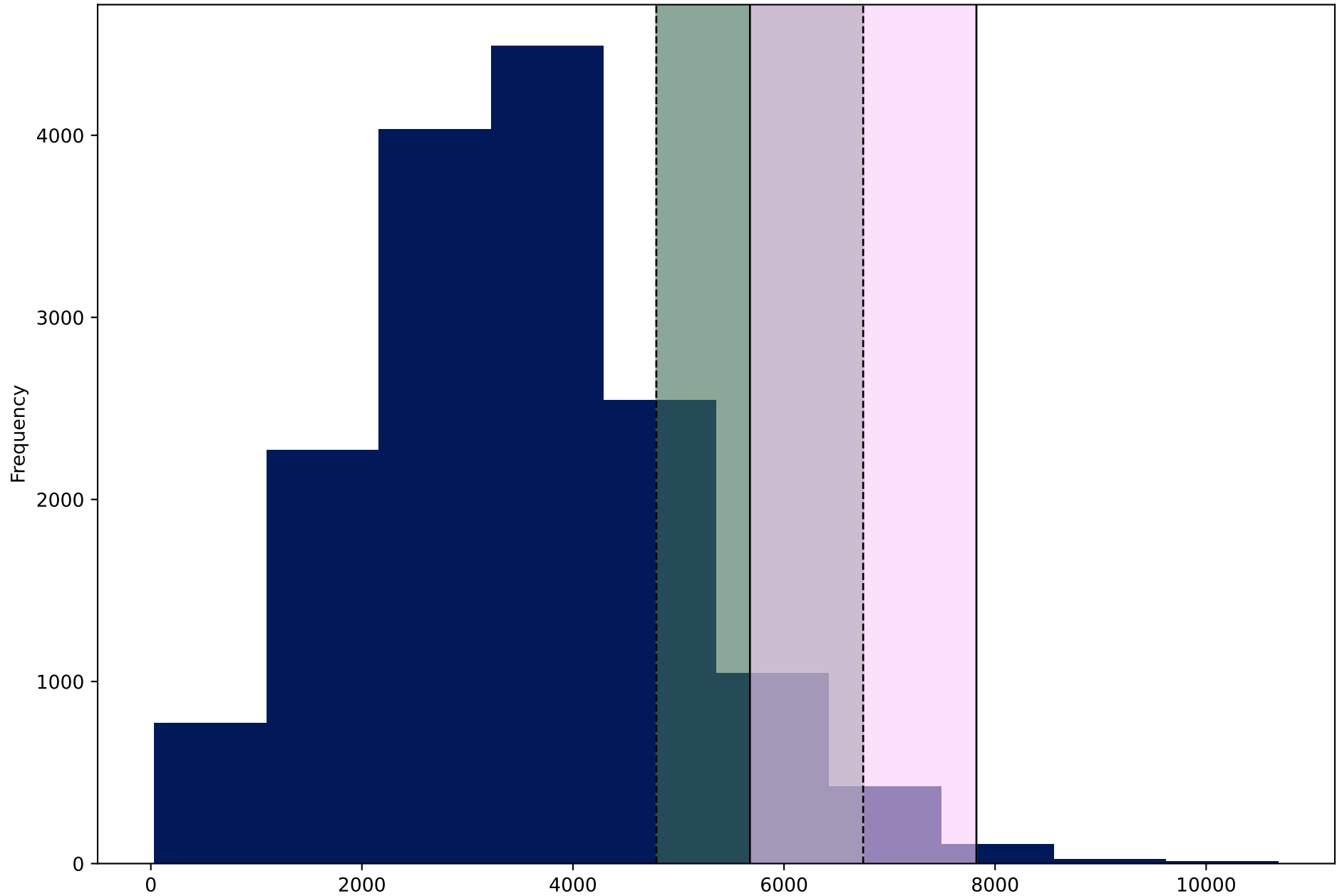
80

Mean 1000 m neighbourhood street intersections per km²

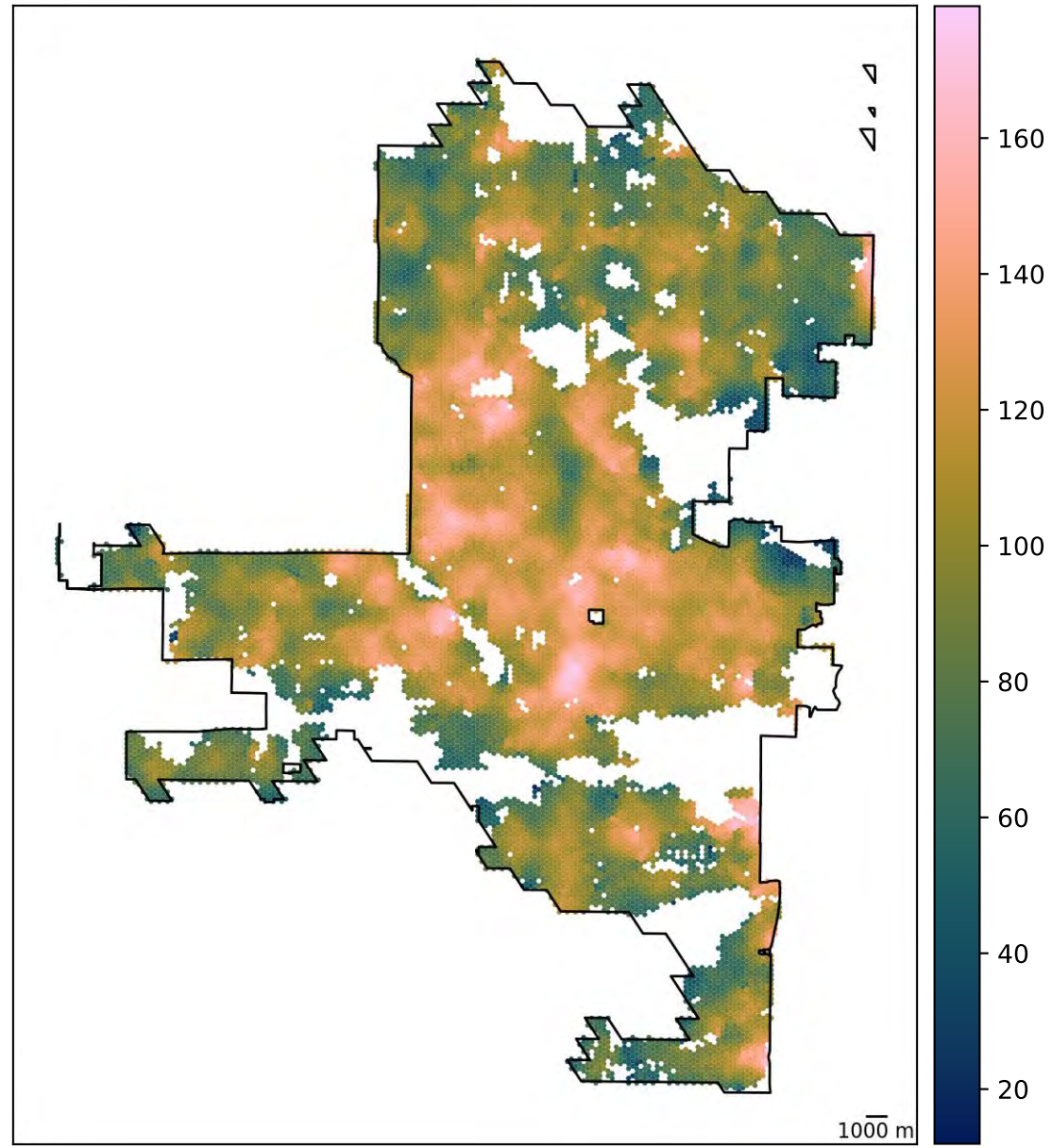

A: Estimated Mean 1000 m neighbourhood street intersections per km² requirement for ≥80% probability of engaging in walking for transport

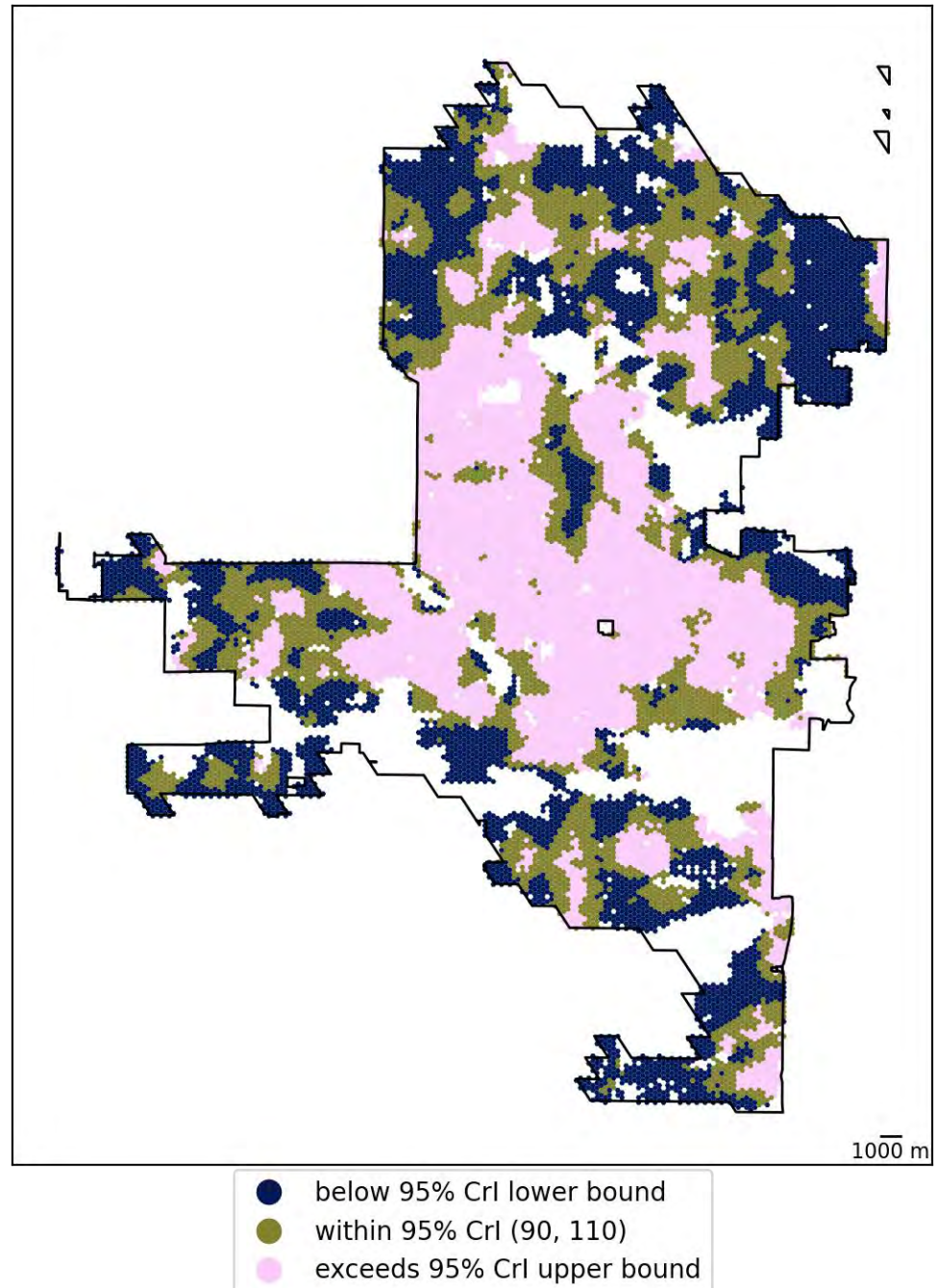

B: Estimated Mean 1000 m neighbourhood street intersections per km² requirement for reaching the WHO's target of a ≥15% relative reduction in insufficient physical activity through walking

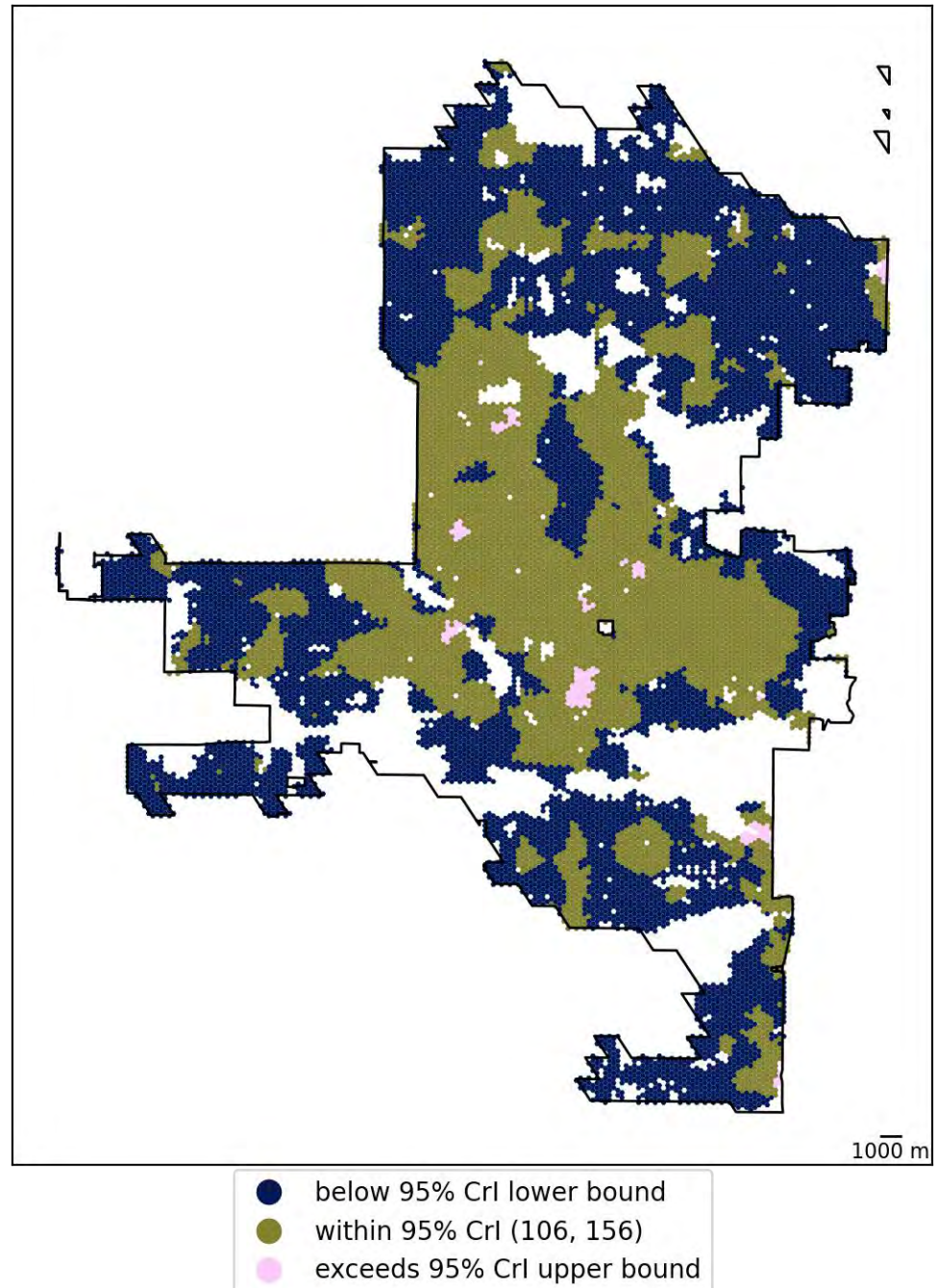

- below 95% CrI lower bound
- within 95% CrI (106, 156)
- exceeds 95% CrI upper bound

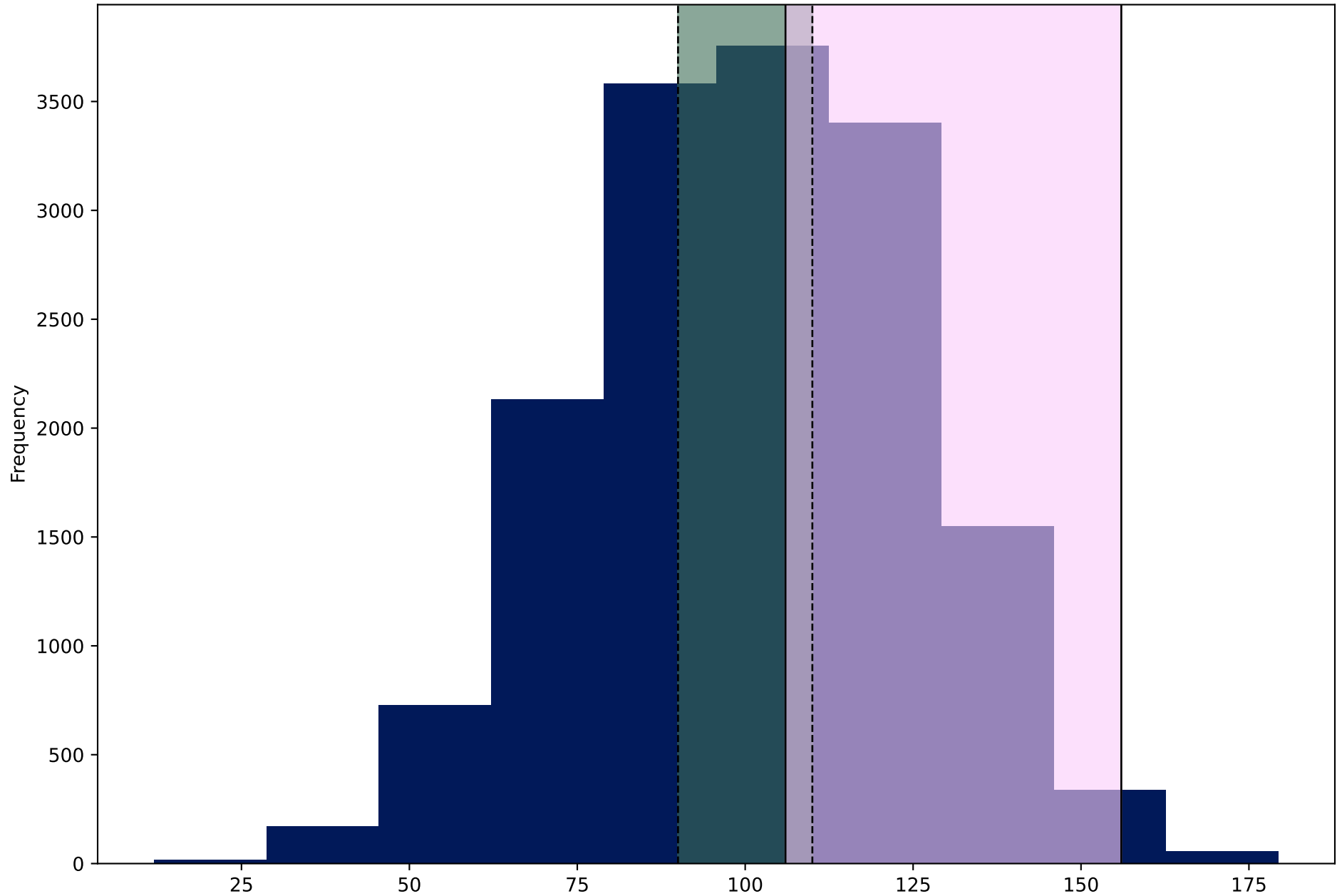

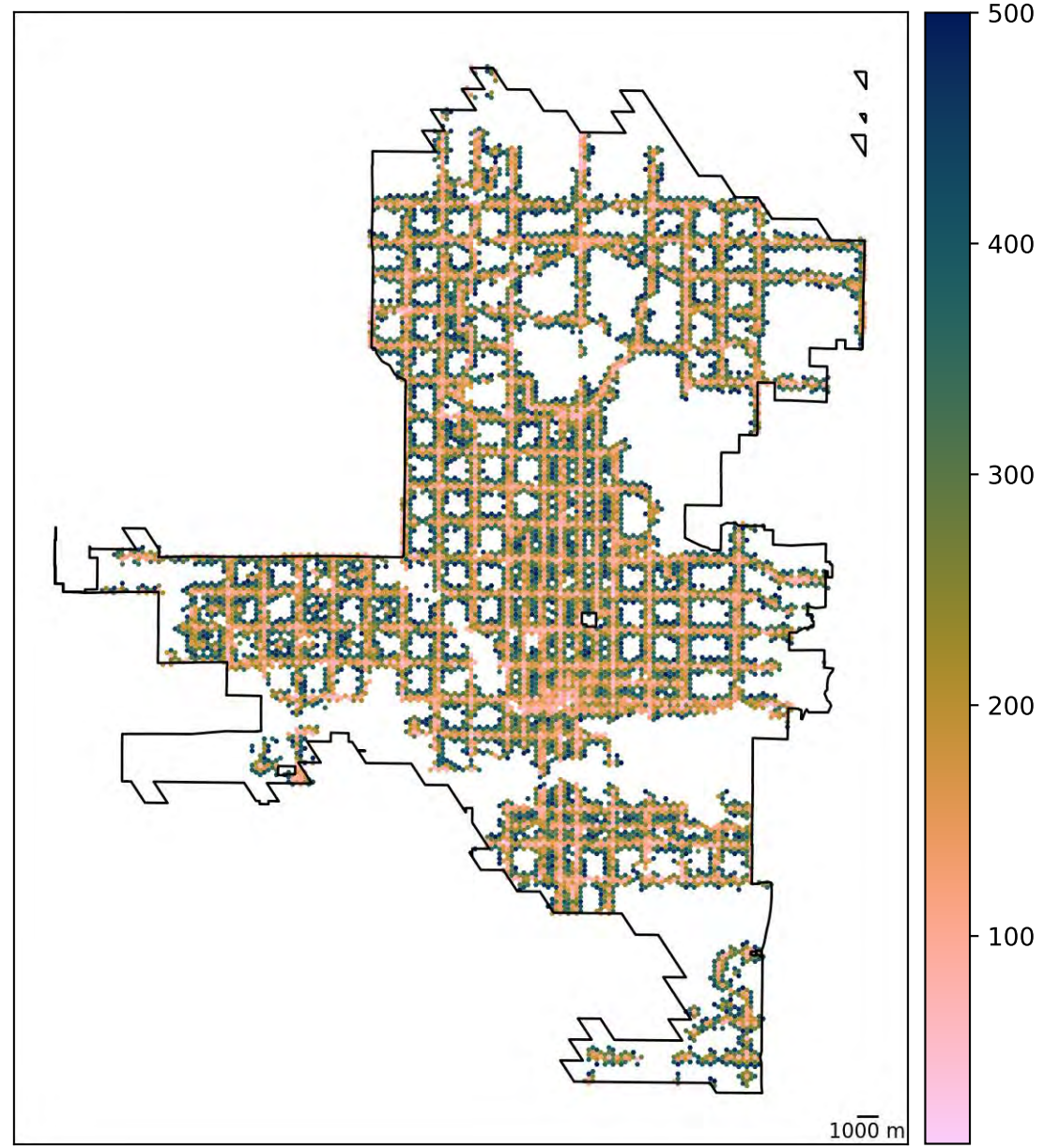

Distance to nearest public transport stops (m; up to 500m)

distances: Estimated Distance to nearest public transport stops (m; up to 500m) requirement for distances to destinations, measured up to a maximum distance target threshold of 500 metres

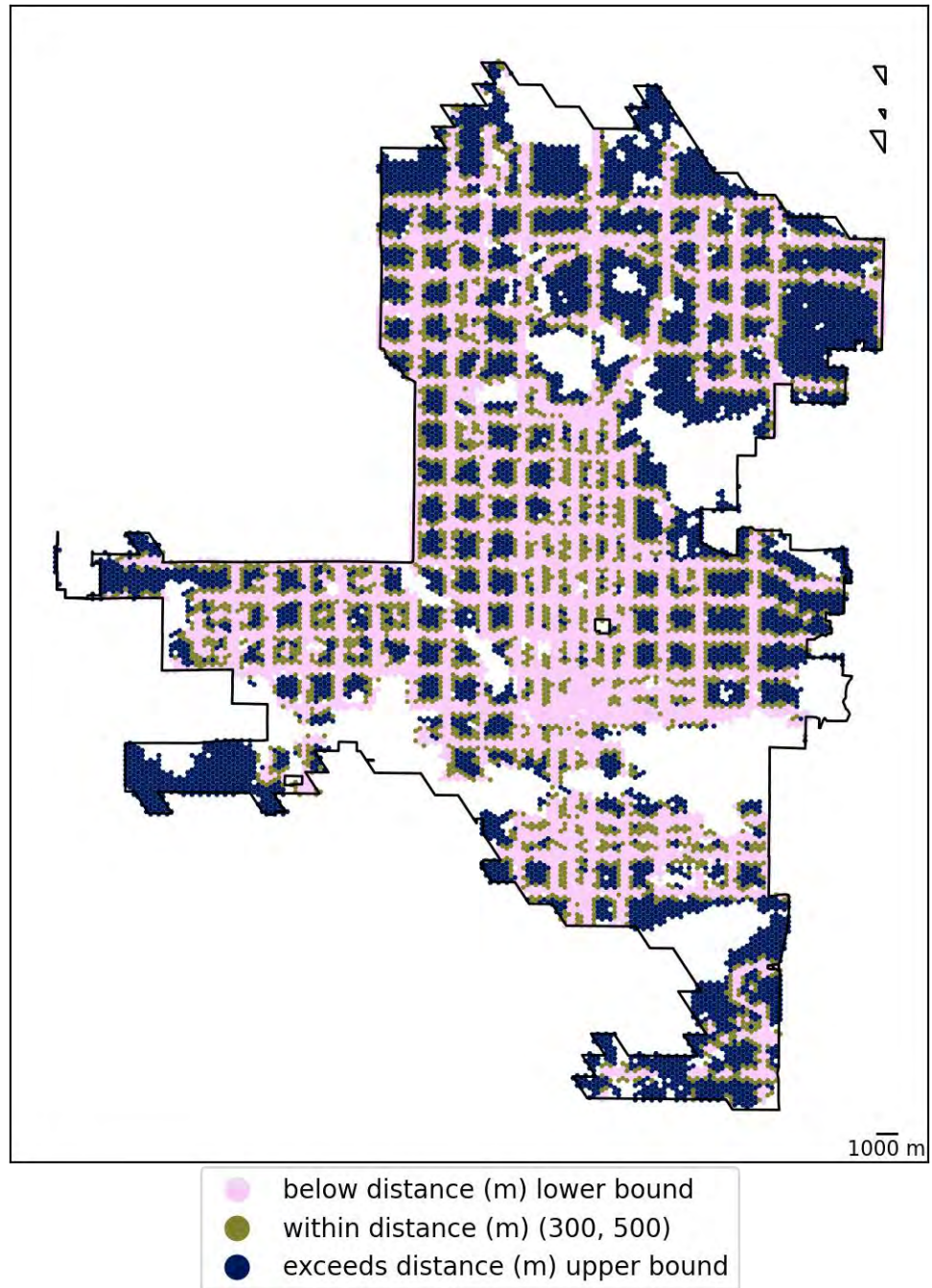

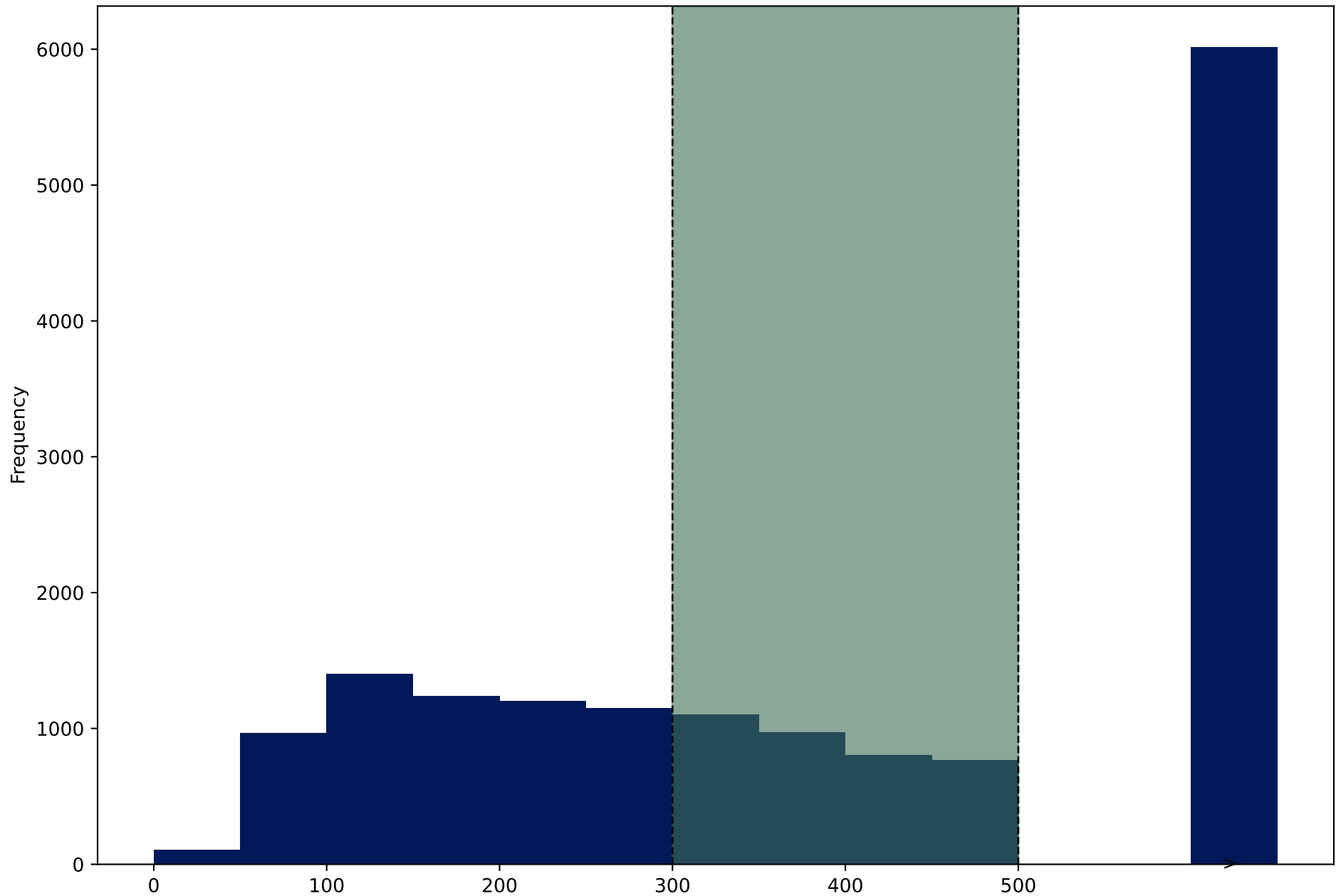

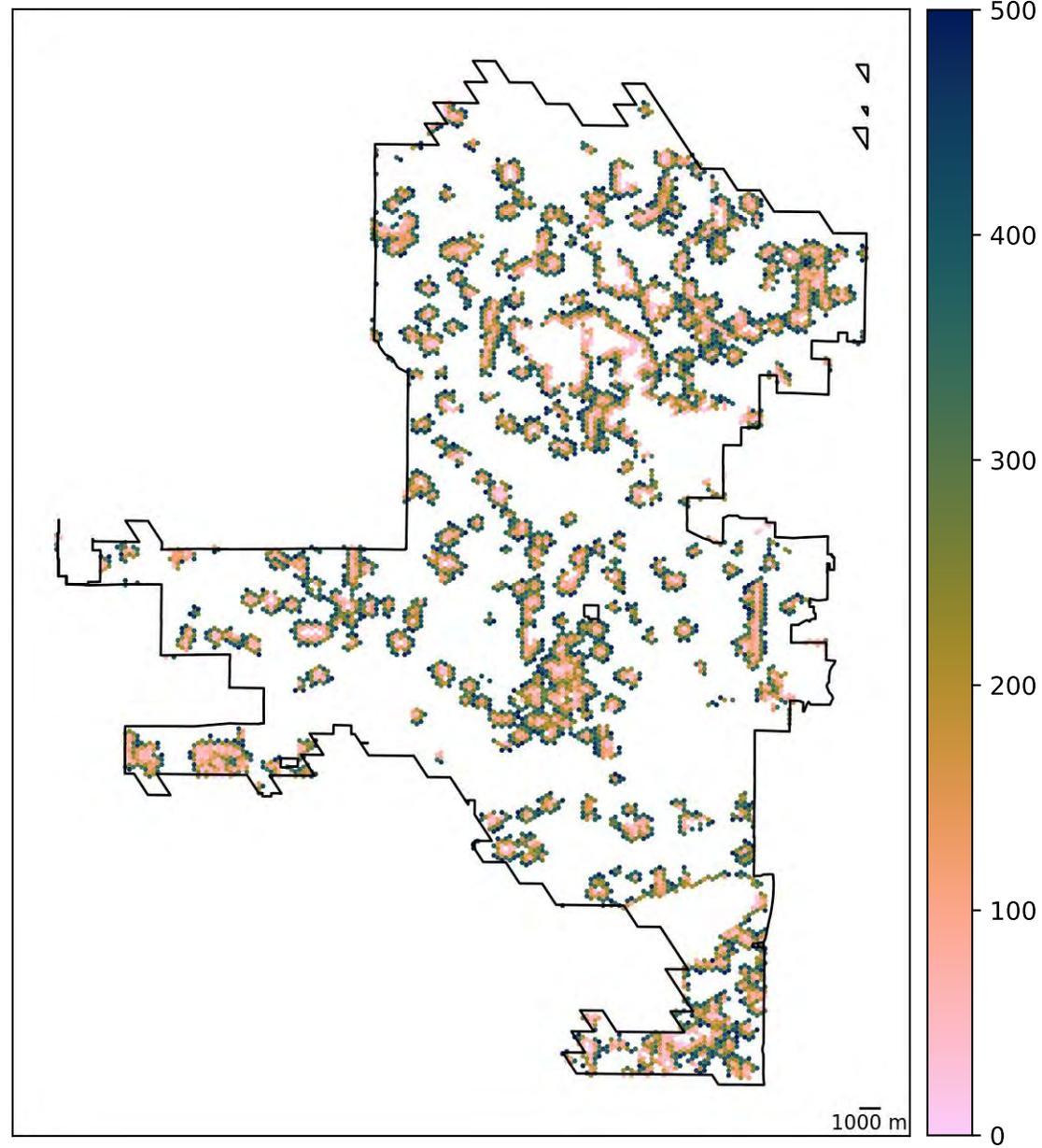

distances: Estimated Distance to nearest park (m; up to 500m) requirement for distances to destinations, measured up to a maximum distance target threshold of 500 metres

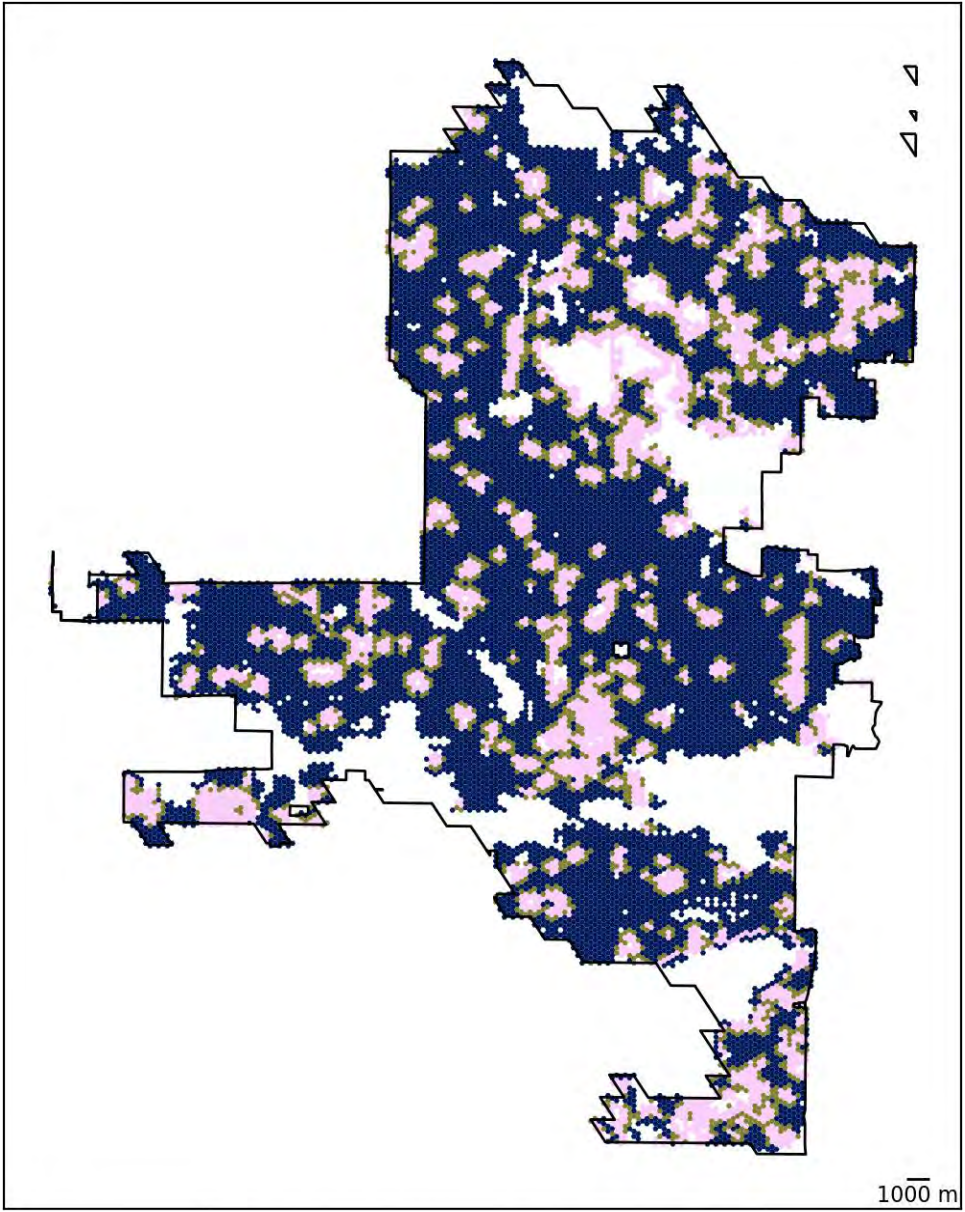

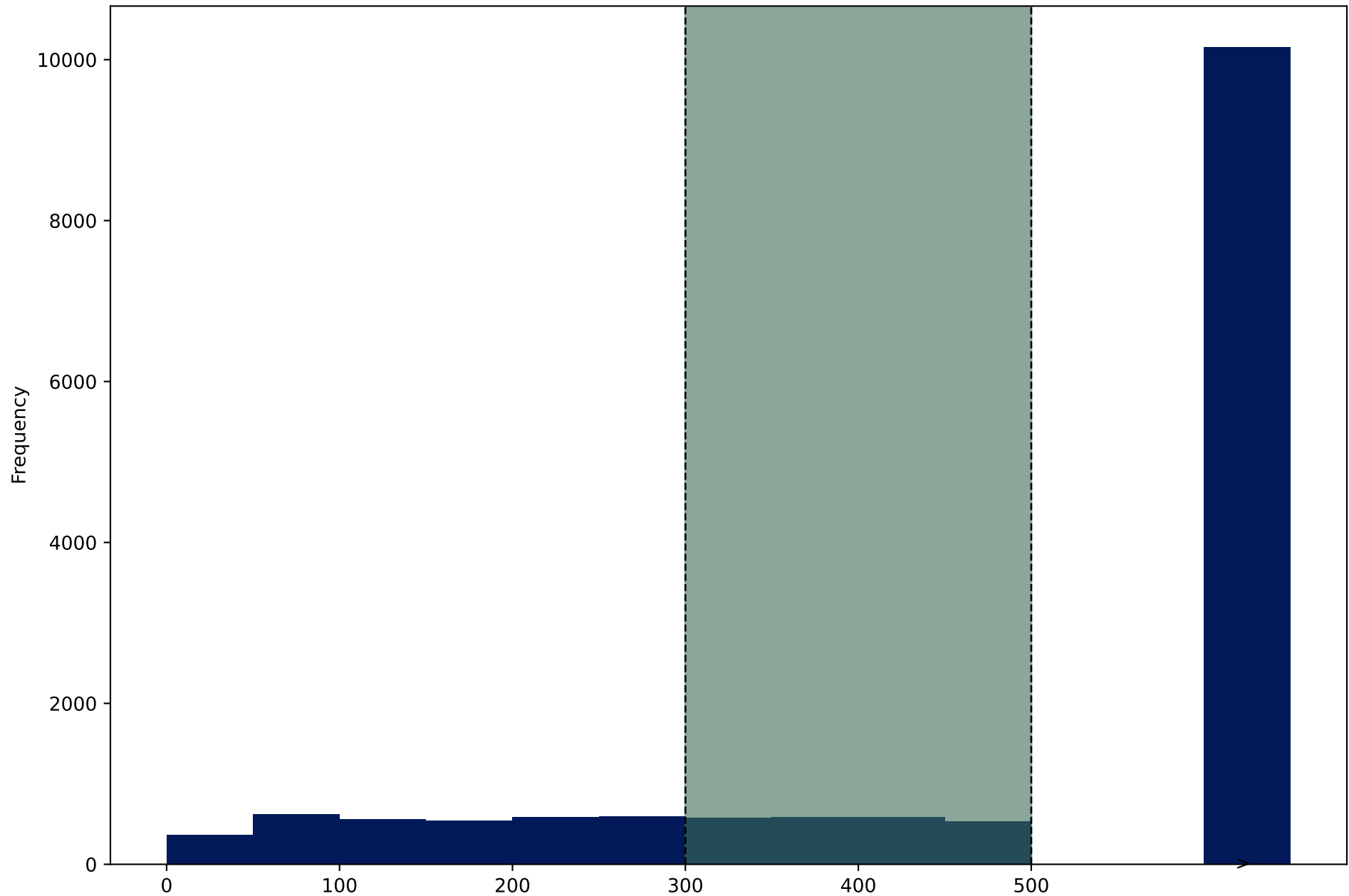

# America, North, United States, Seattle

| Satellite imagery of urban study region (Bing) | Walkability, relative to city | Walkability, relative to 25 global cities |
|---|---|---|

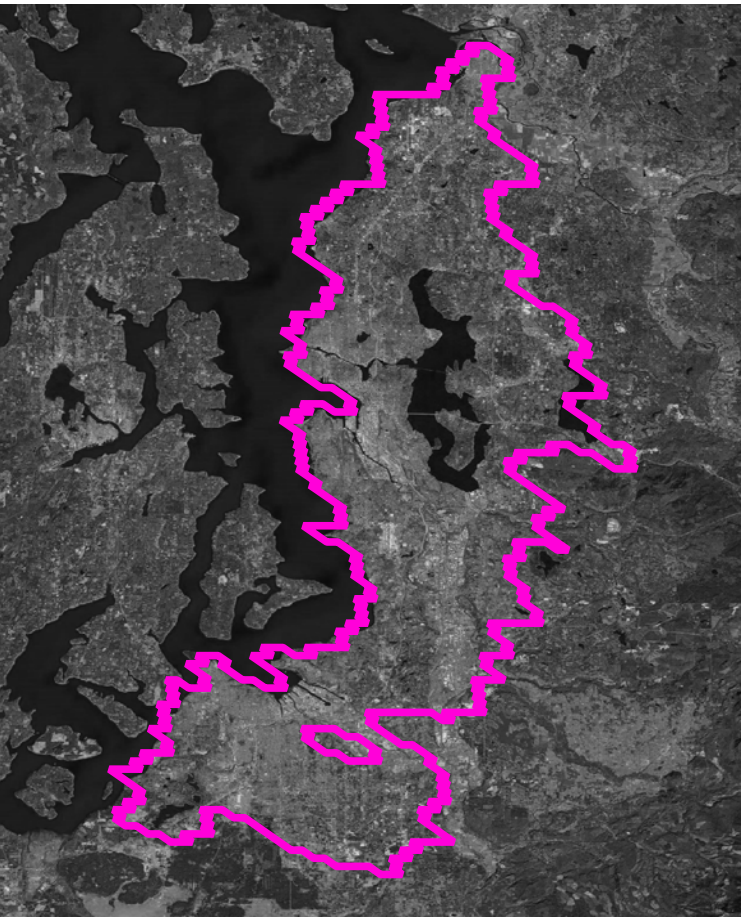
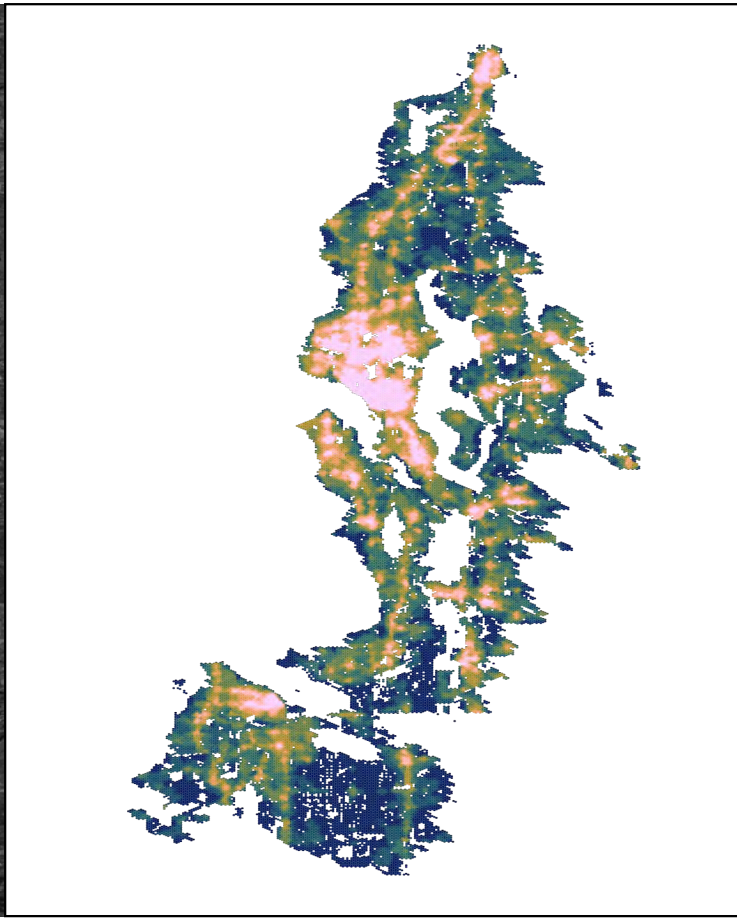
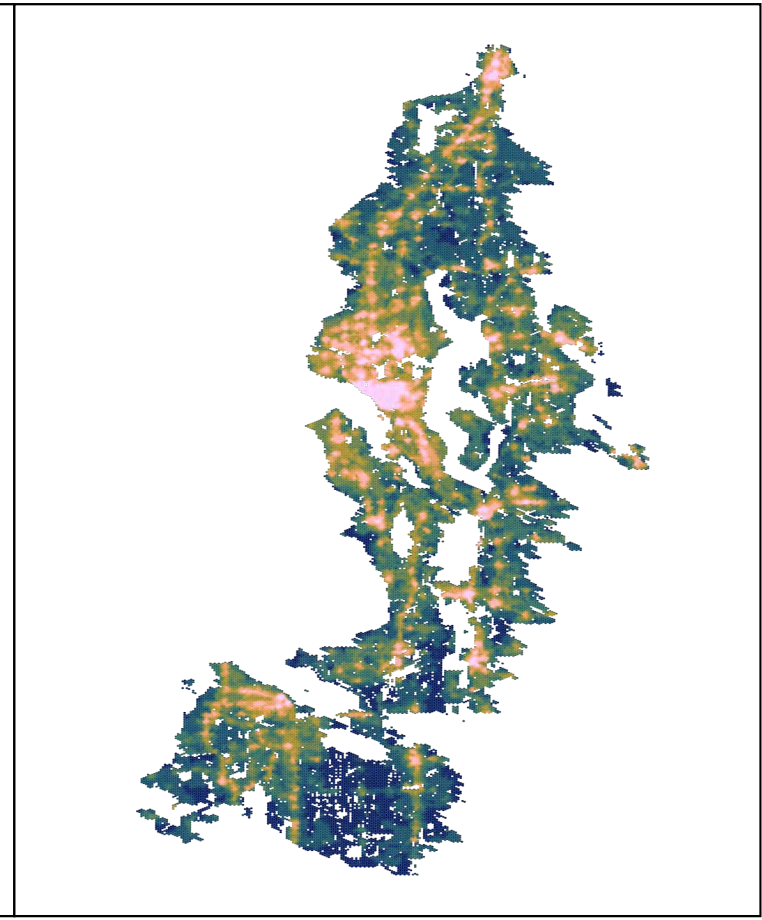

Urban boundary

0　　　　30　　　　60 km

Walkability score
- <-3
- -3 to -2
- -2 to -1
- -1 to 0
- 0 to 1
- 1 to 2
- 2 to 3
- ≥3

Walkability relative to all cities by component variables (2D histograms), and overall (histogram)

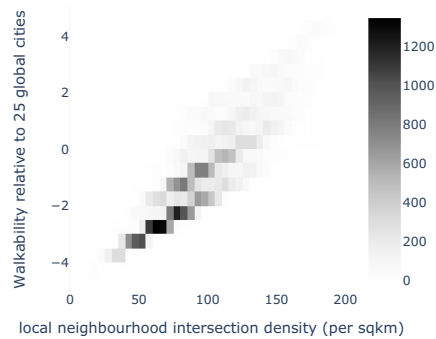
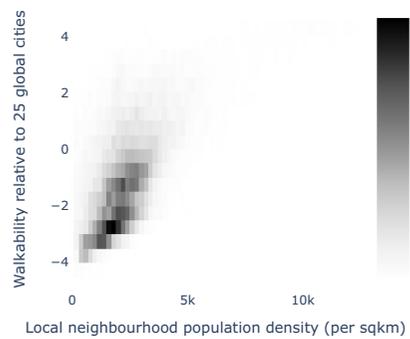
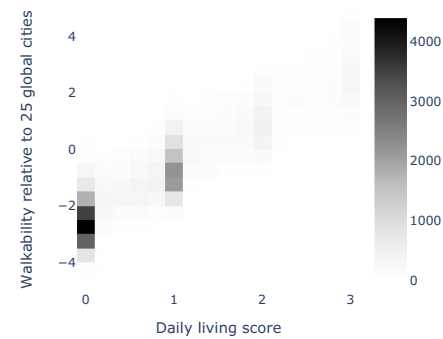
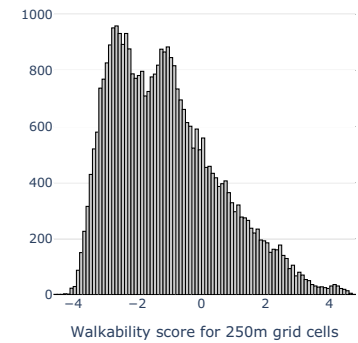

Mean 1000 m neighbourhood population per km²

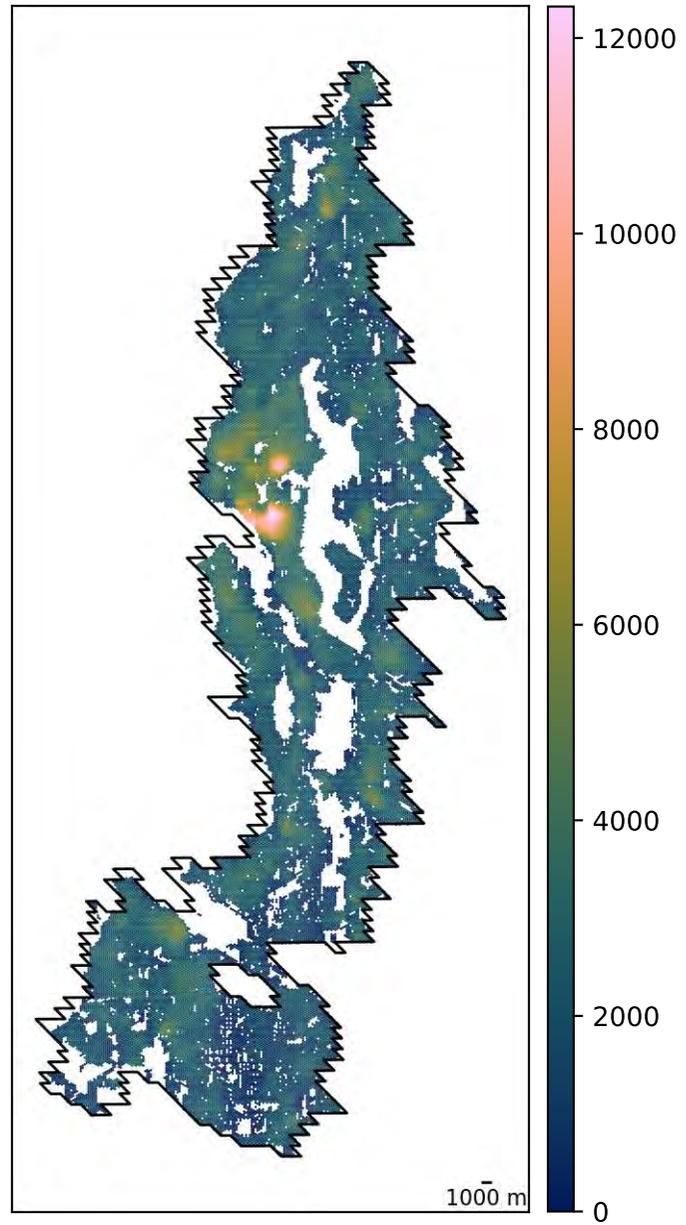

A: Estimated Mean 1000 m neighbourhood population per km² requirement for ≥80% probability of engaging in walking for transport

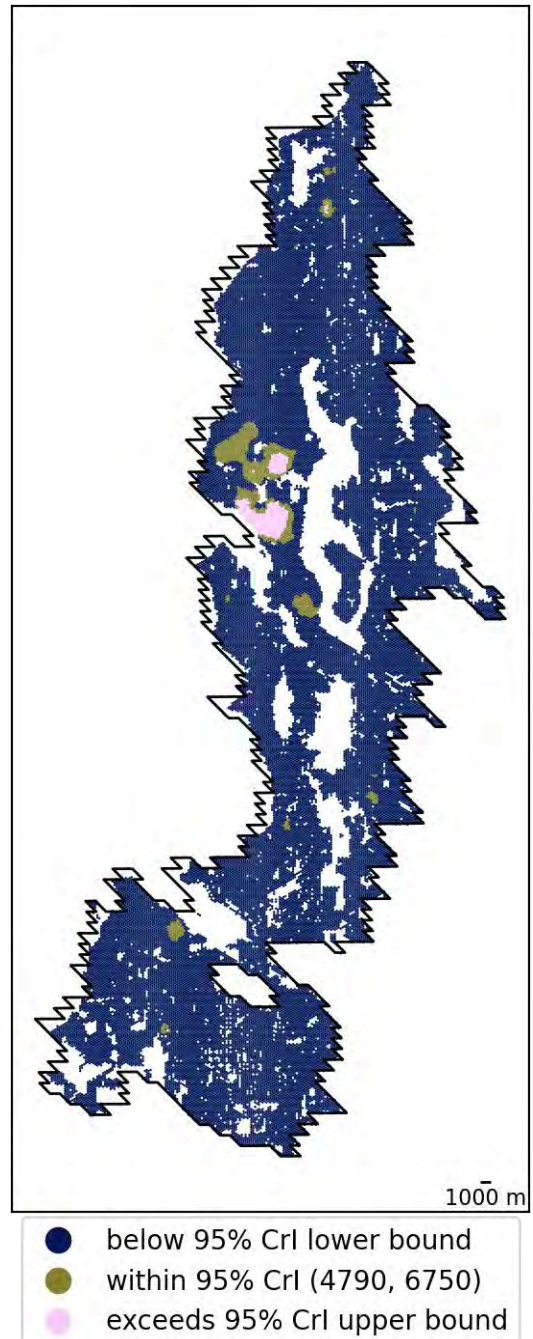

B: Estimated Mean 1000 m neighbourhood population per km² requirement for reaching the WHO's target of a ≥15% relative reduction in insufficient physical activity through walking

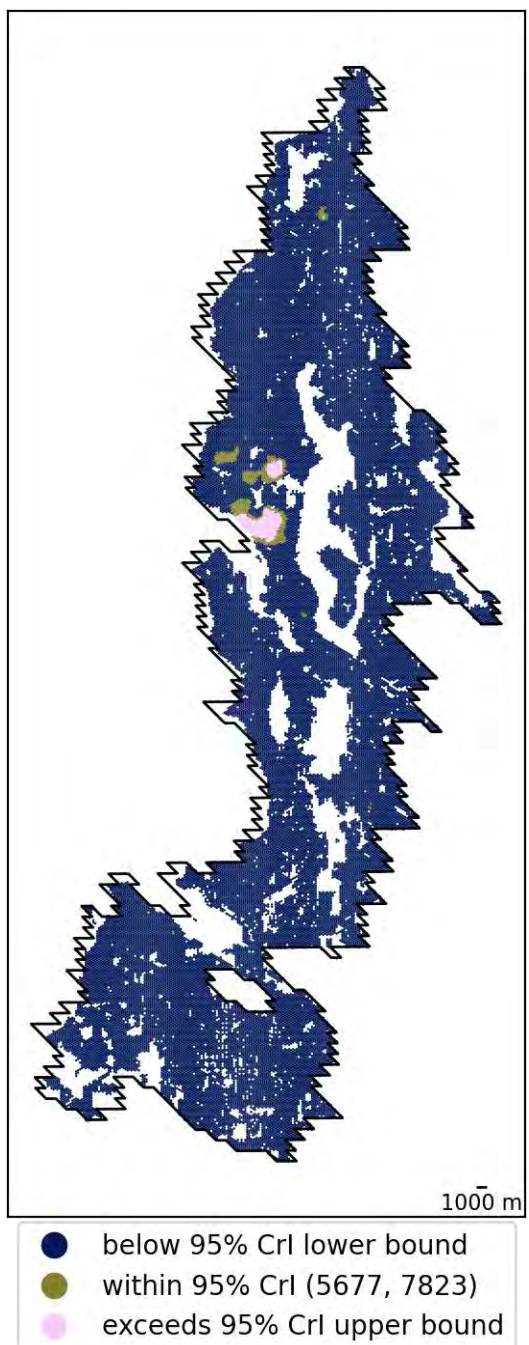

- below 95% CrI lower bound
- within 95% CrI (5677, 7823)
- exceeds 95% CrI upper bound

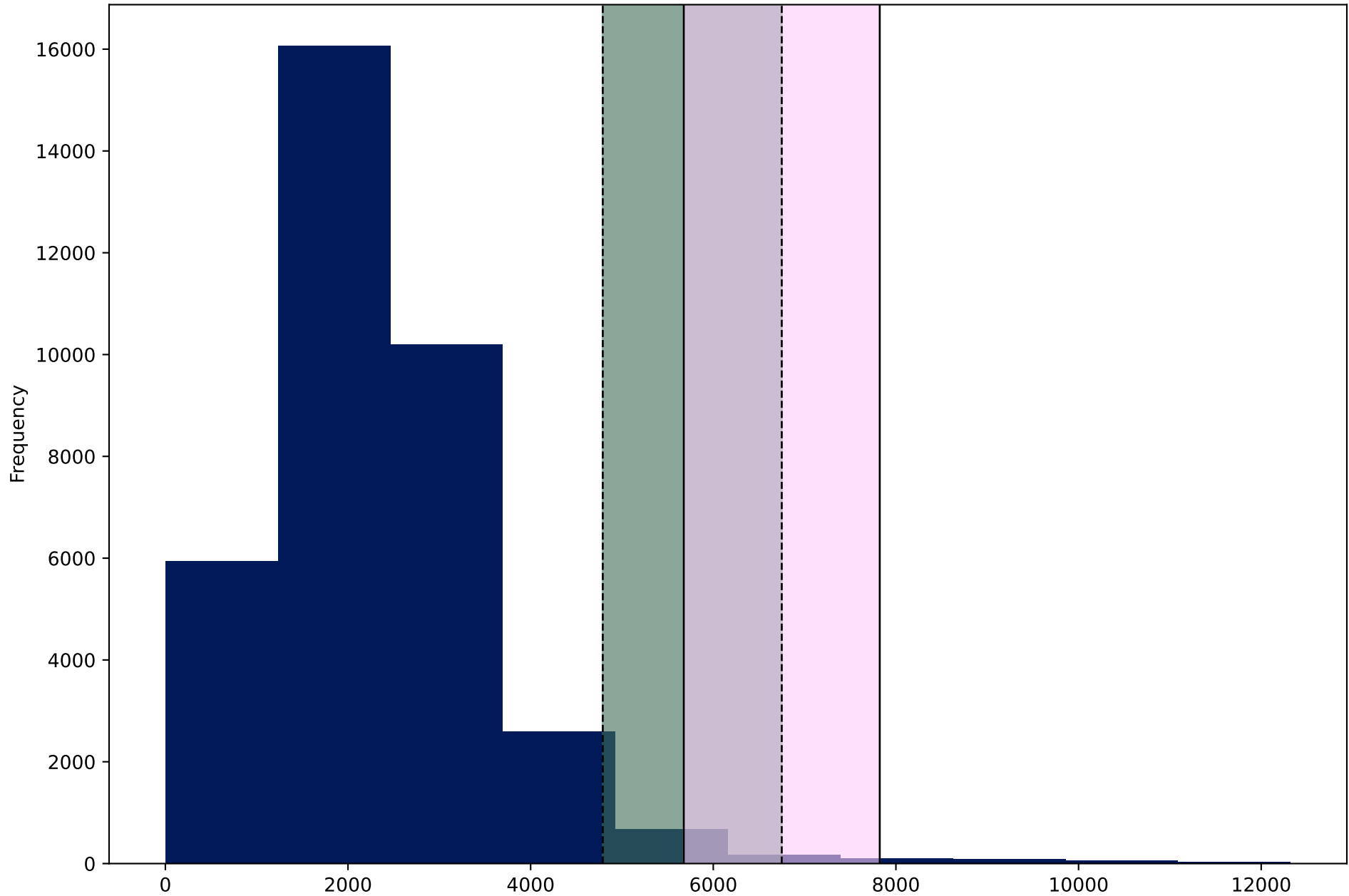

Mean 1000 m neighbourhood street intersections per km²

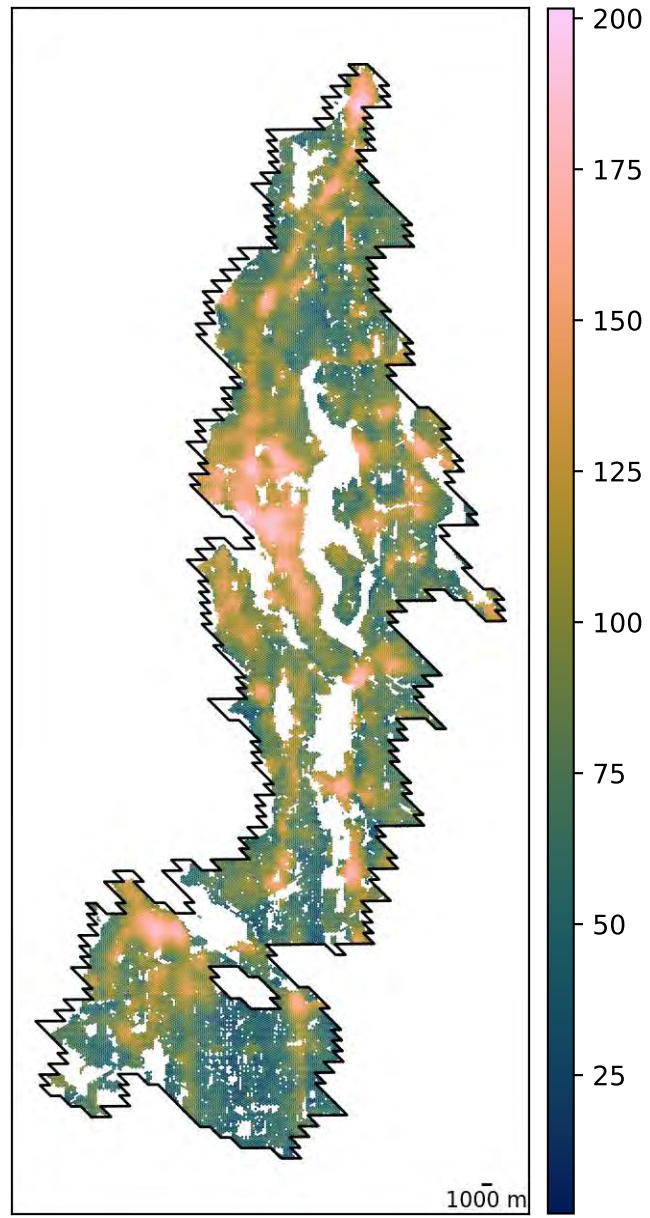

A: Estimated Mean 1000 m neighbourhood street intersections per km² requirement for ≥80% probability of engaging in walking for transport

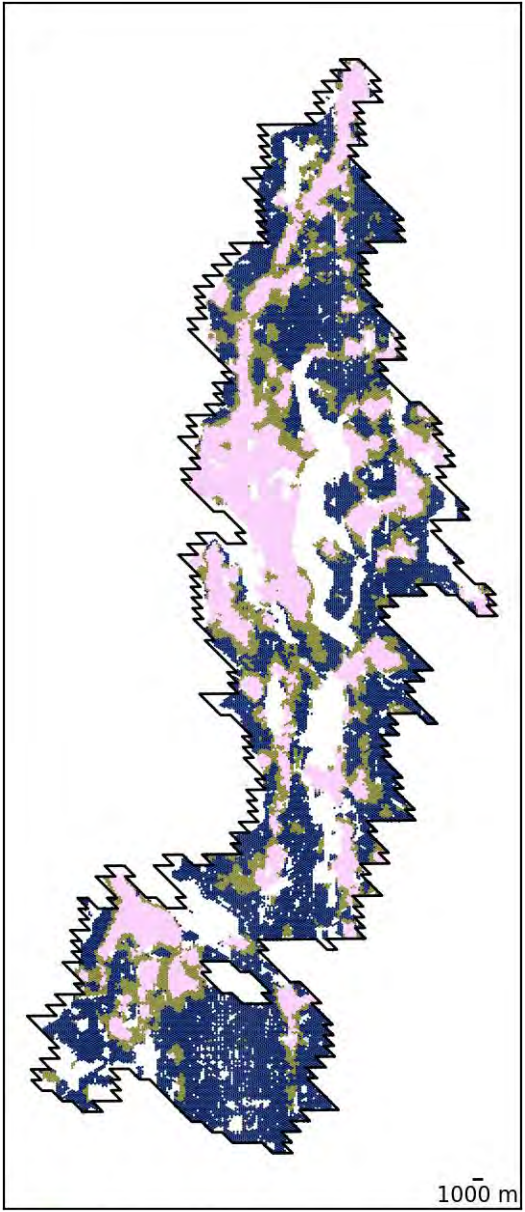

- below 95% CrI lower bound
- within 95% CrI (90, 110)
- exceeds 95% CrI upper bound

B: Estimated Mean 1000 m neighbourhood street intersections per km² requirement for reaching the WHO's target of a ≥15% relative reduction in insufficient physical activity through walking

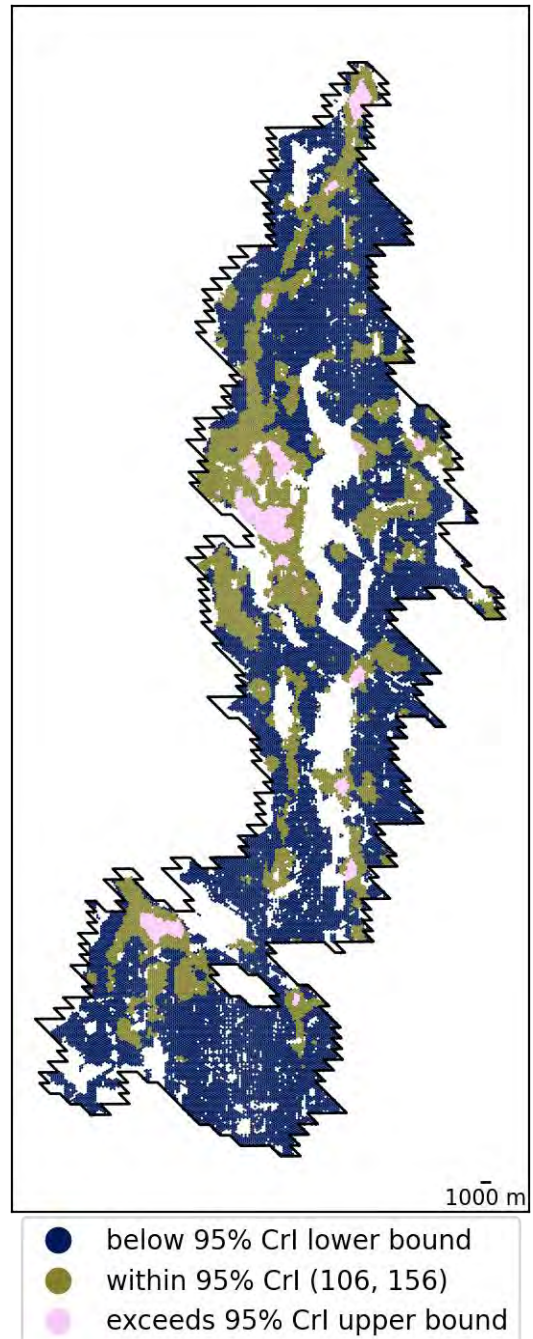

- below 95% CrI lower bound
- within 95% CrI (106, 156)
- exceeds 95% CrI upper bound

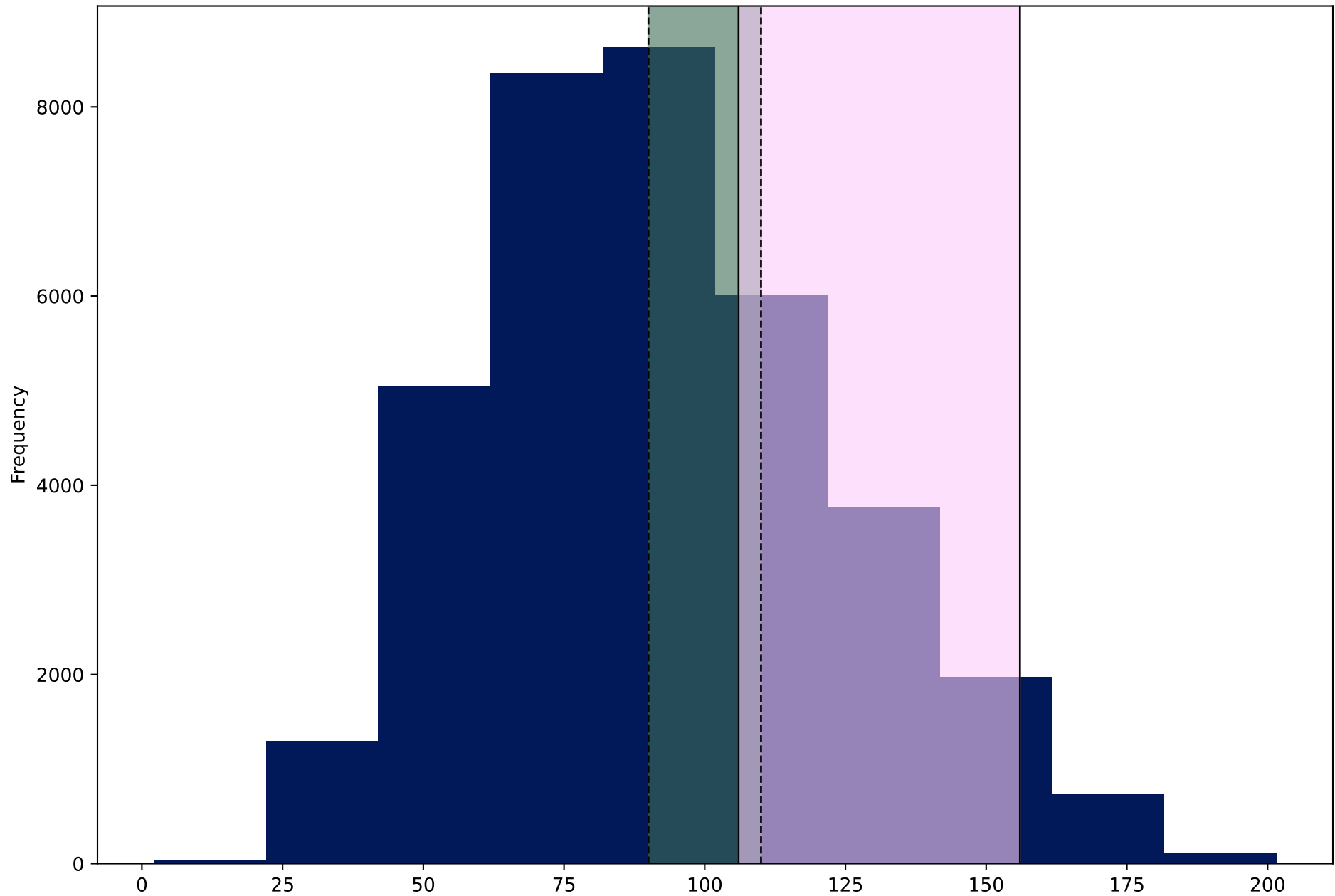

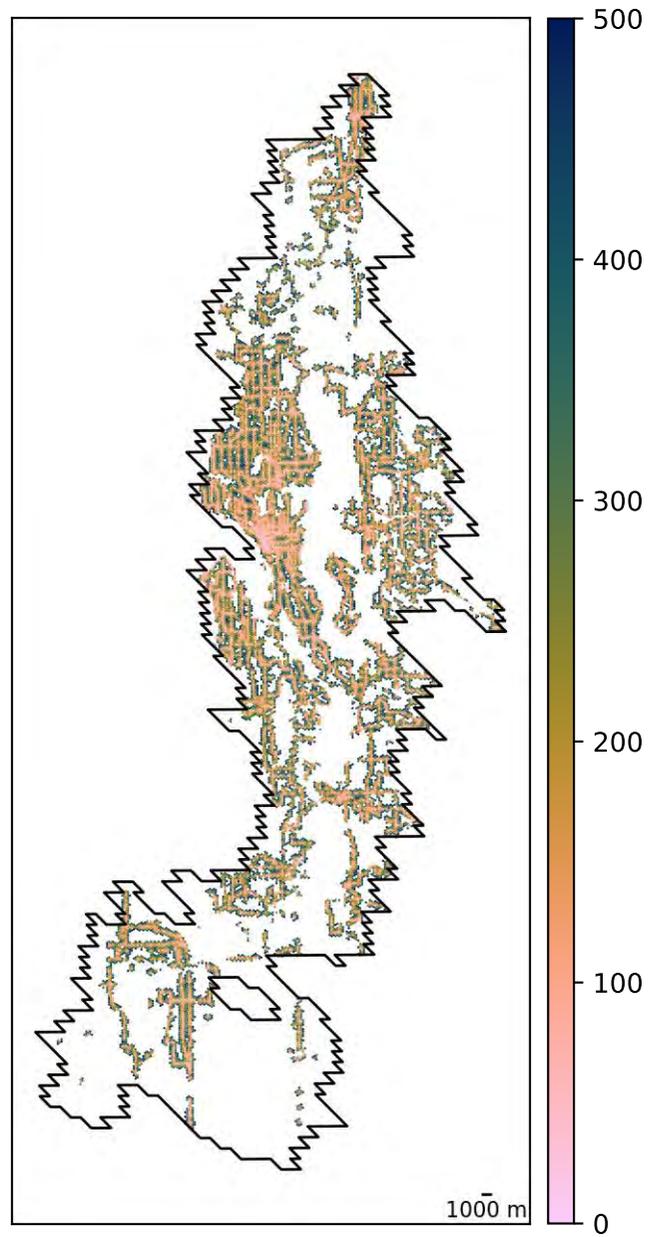

distances: Estimated Distance to nearest public transport stops (m; up to 500m) requirement for distances to destinations, measured up to a maximum distance target threshold of 500 metres

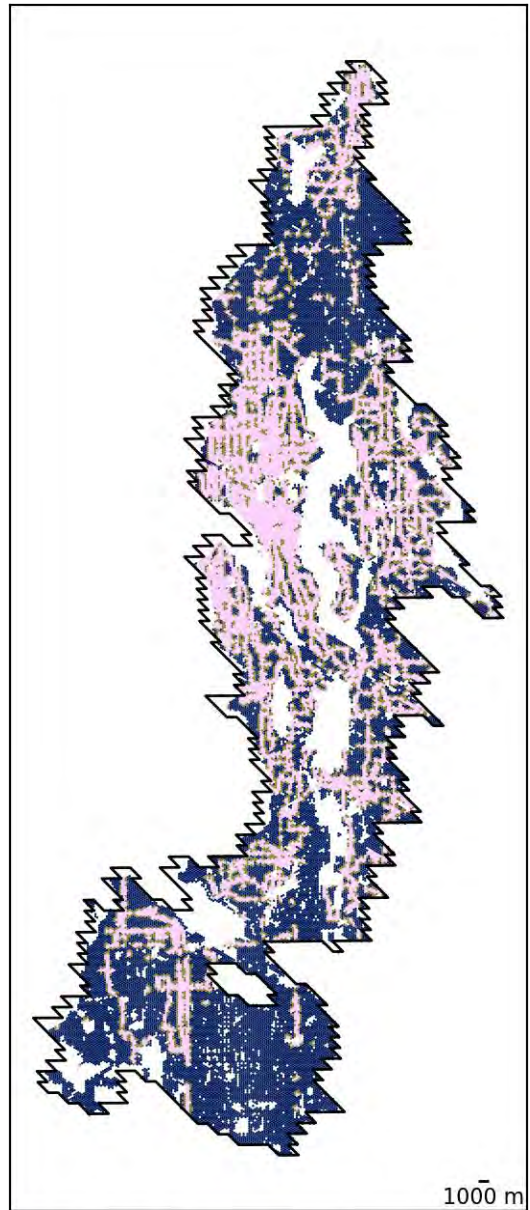

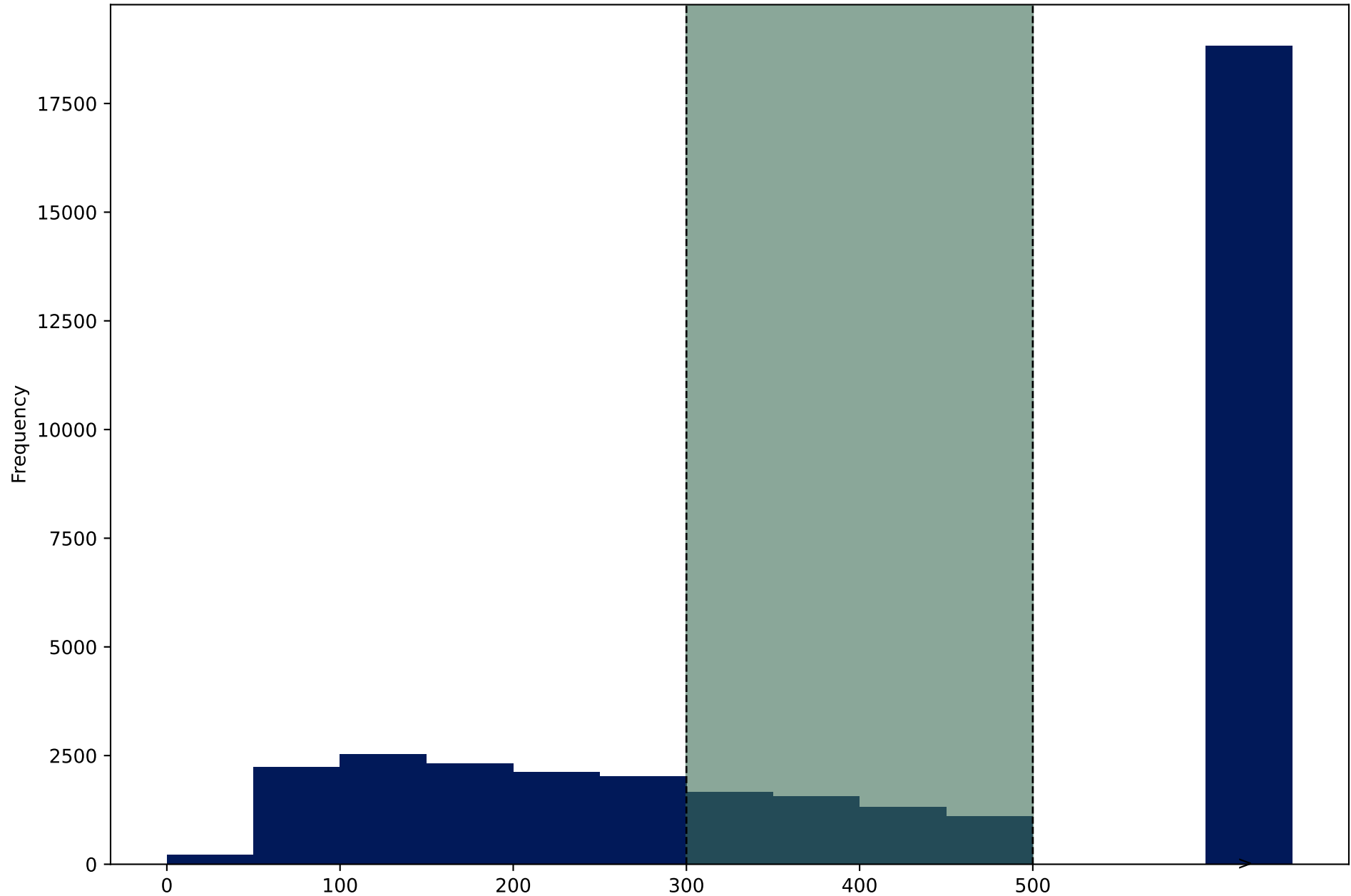
102

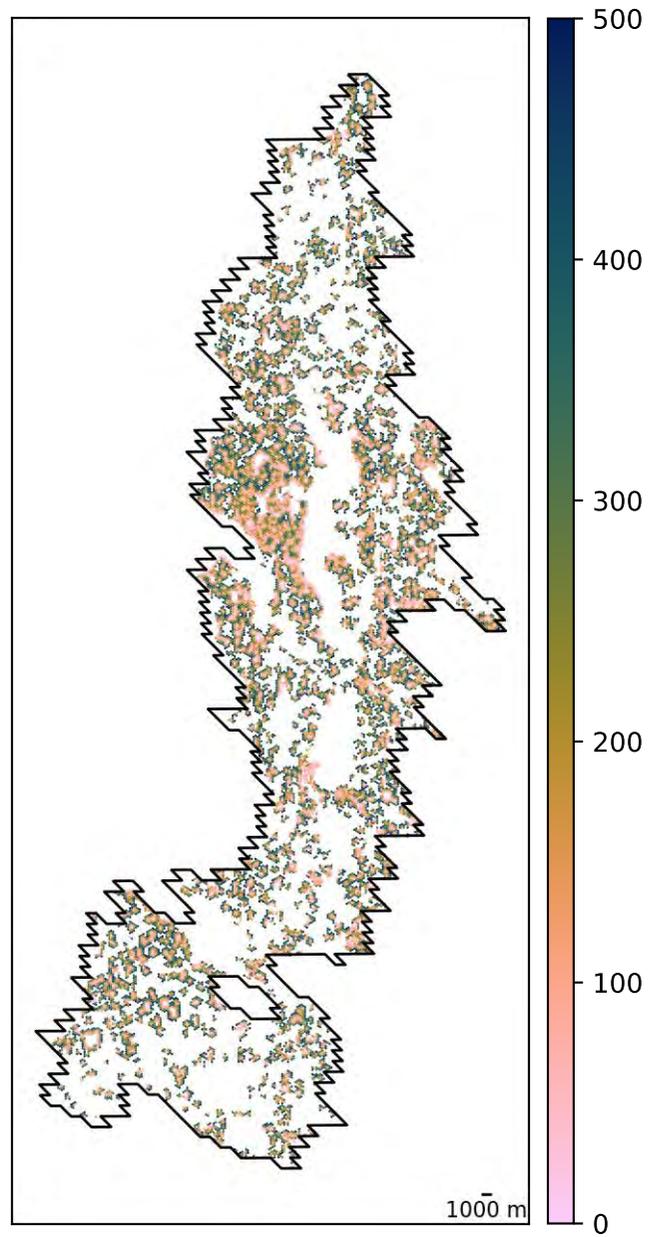

distances: Estimated Distance to nearest park (m; up to 500m) requirement for distances to destinations, measured up to a maximum distance target threshold of 500 metres

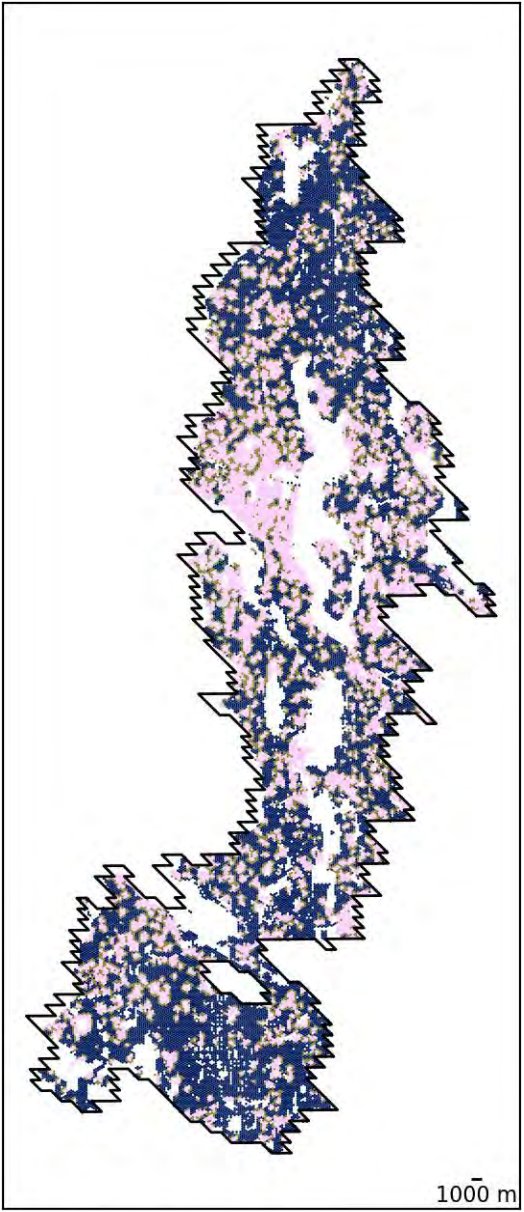

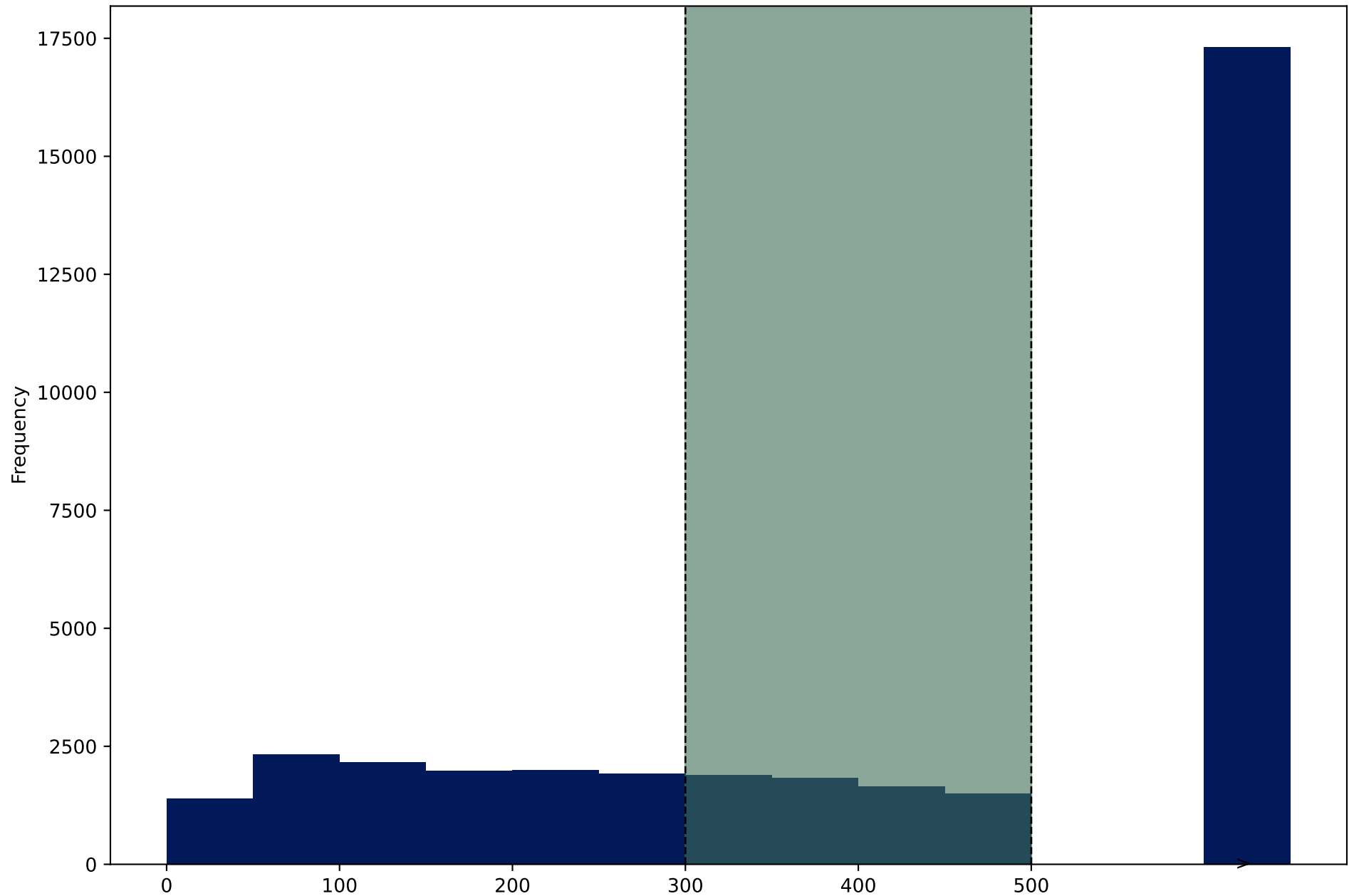

# America, South, Brazil, São Paulo

| Satellite imagery of urban study region (Bing) | Walkability, relative to city | Walkability, relative to 25 global cities |
|---|---|---|

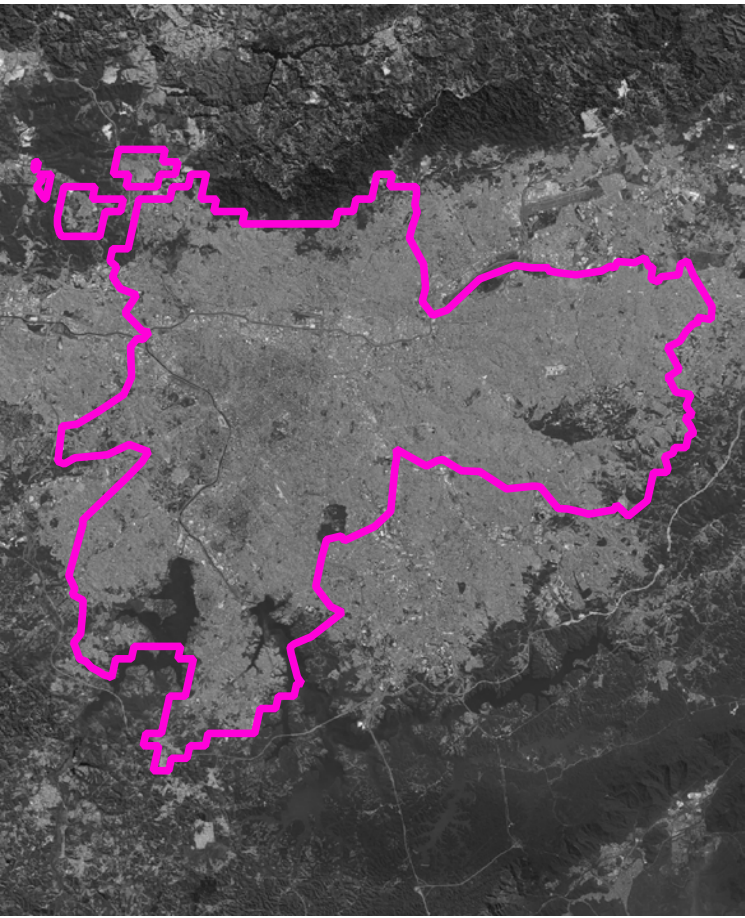
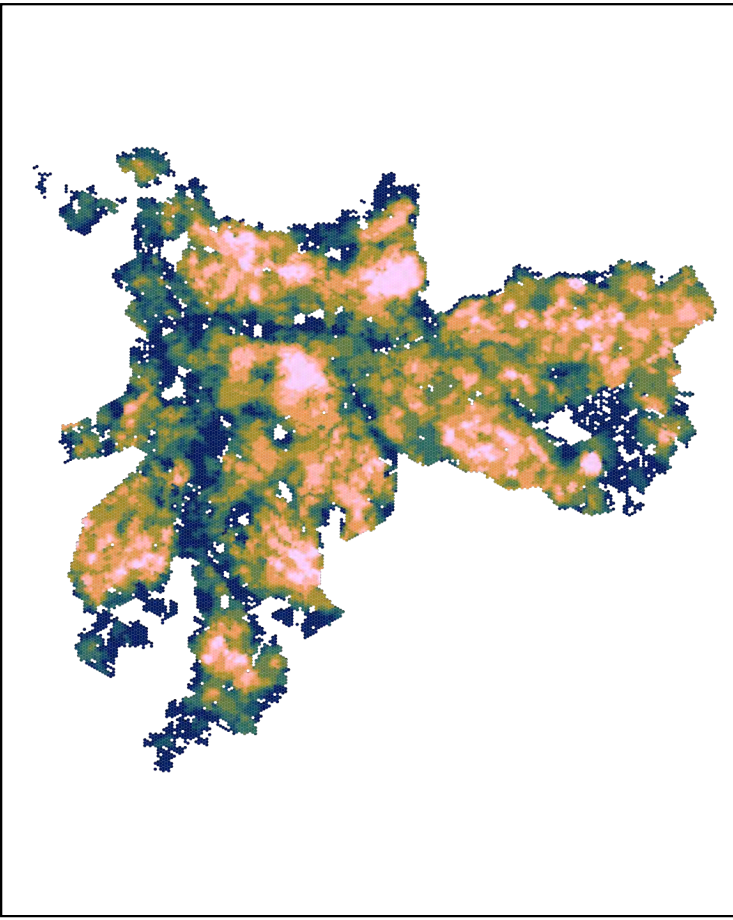
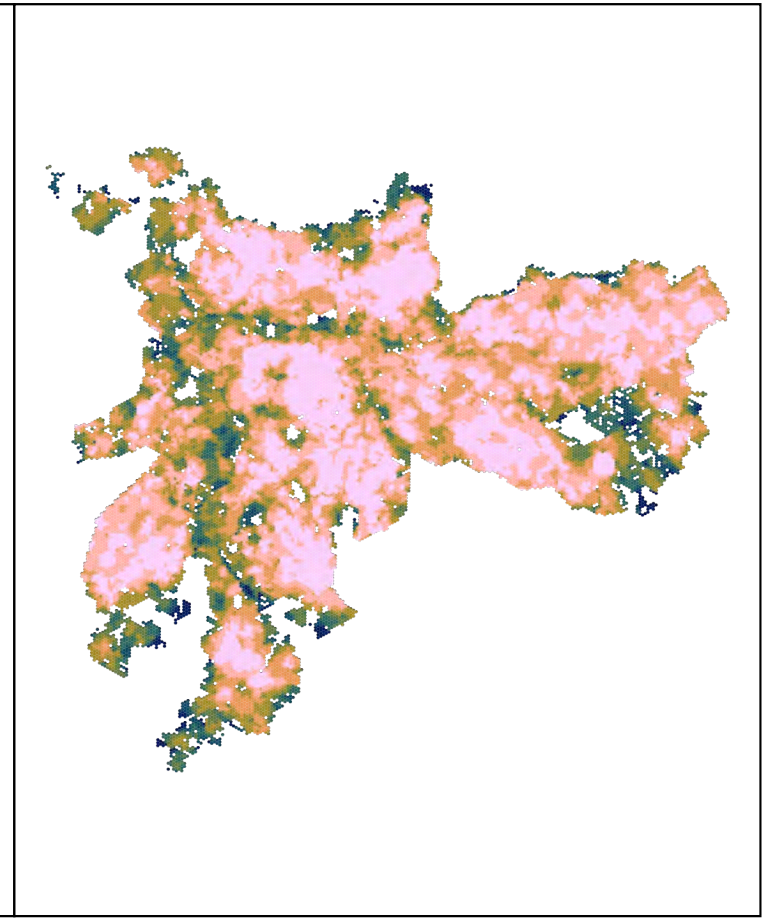

Urban boundary

Walkability score: <-3; -3 to -2; -2 to -1; -1 to 0; 0 to 1; 1 to 2; 2 to 3; ≥3

Walkability relative to all cities by component variables (2D histograms), and overall (histogram)

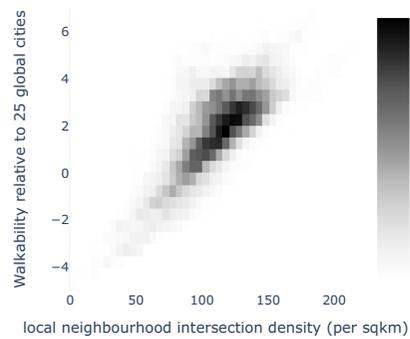
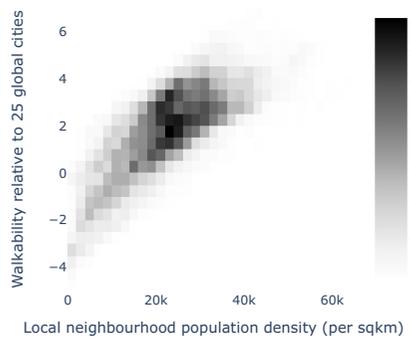
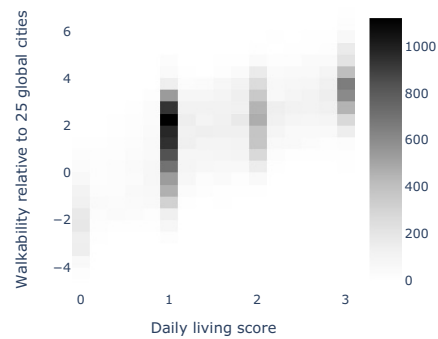
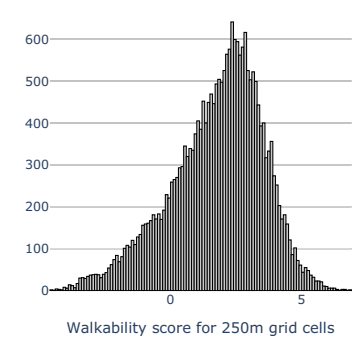

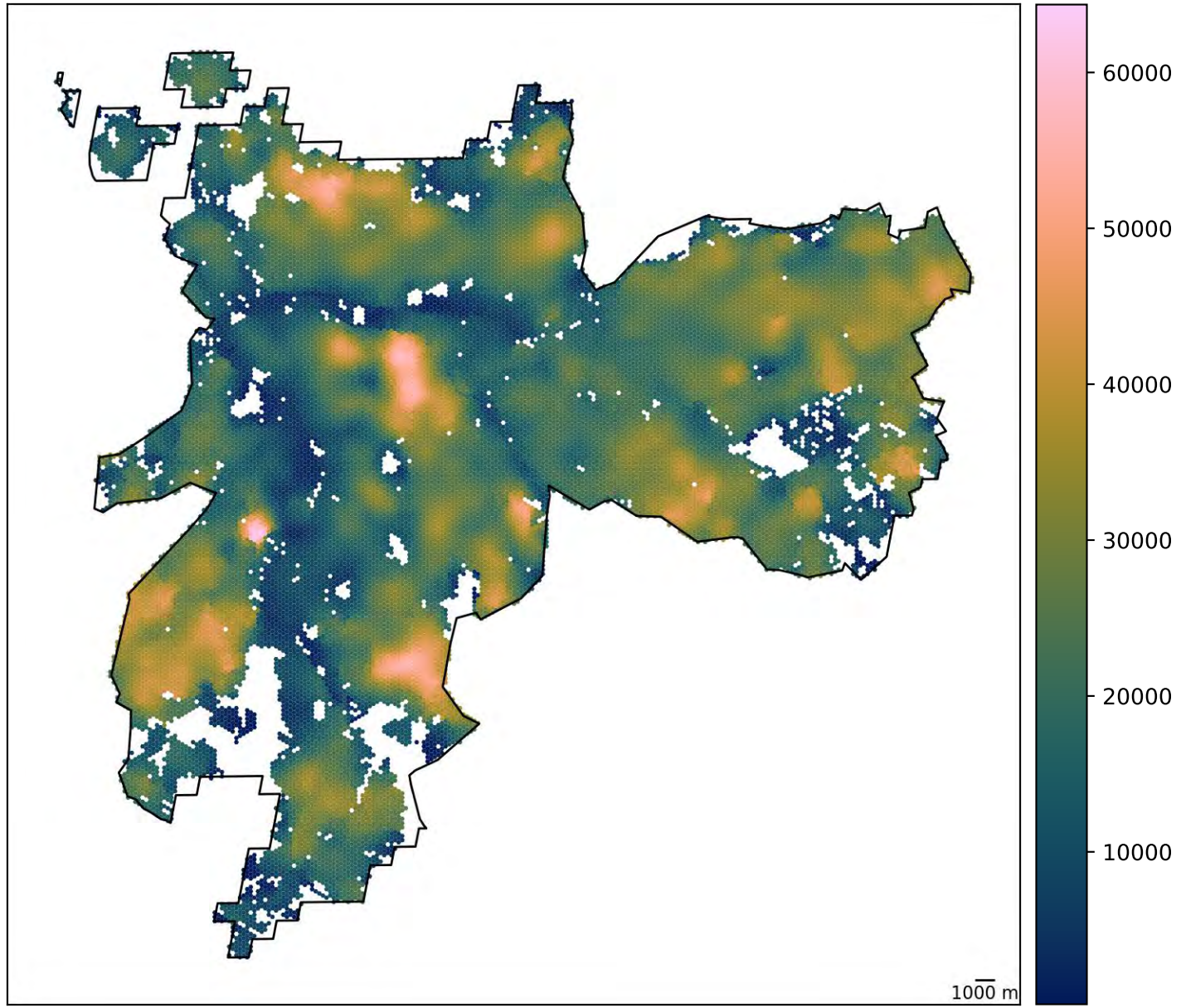

A: Estimated Mean 1000 m neighbourhood population per km² requirement for ≥80% probability of engaging in walking for transport

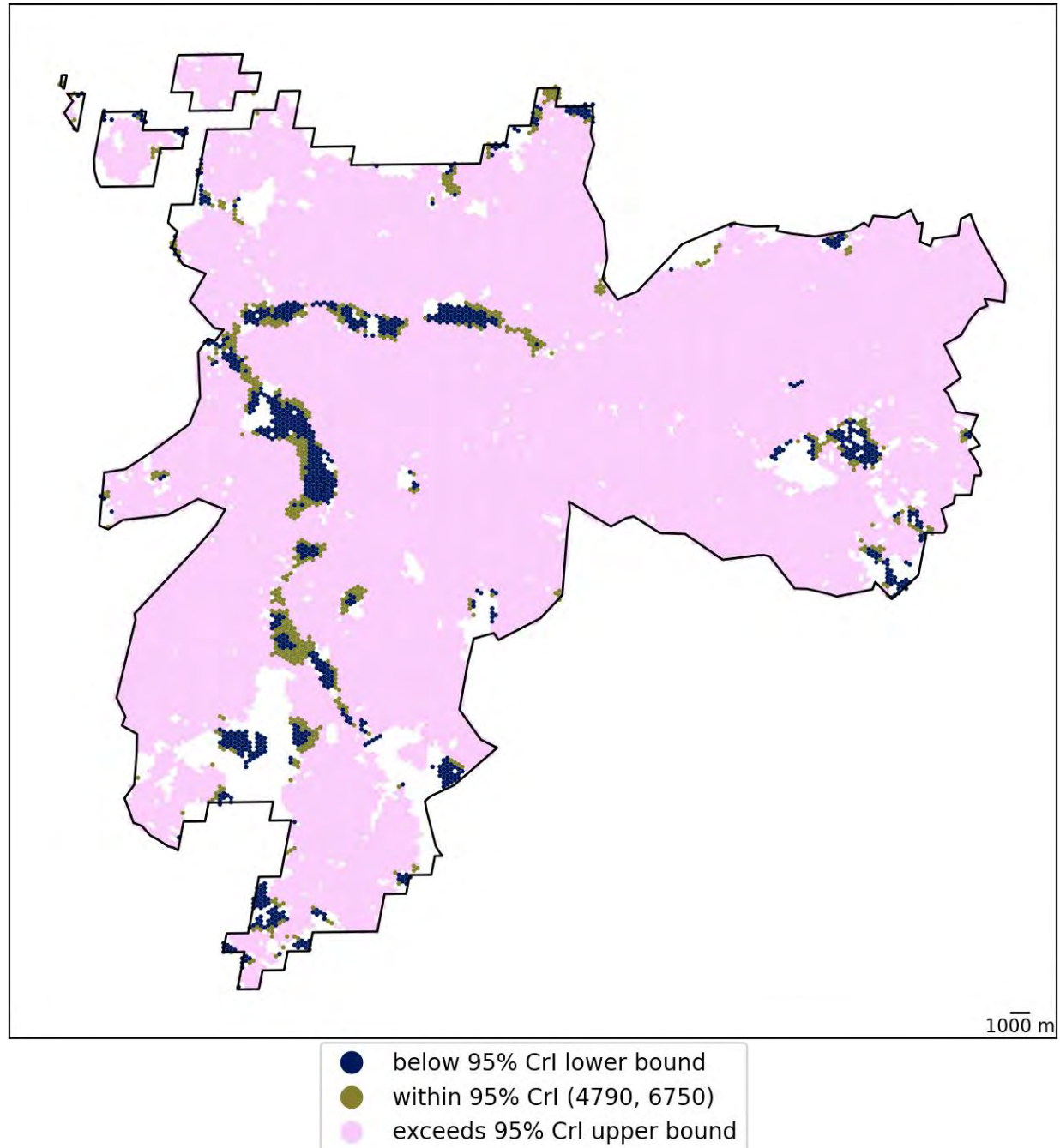

B: Estimated Mean 1000 m neighbourhood population per km² requirement for reaching the WHO's target of a ≥15% relative reduction in insufficient physical activity through walking

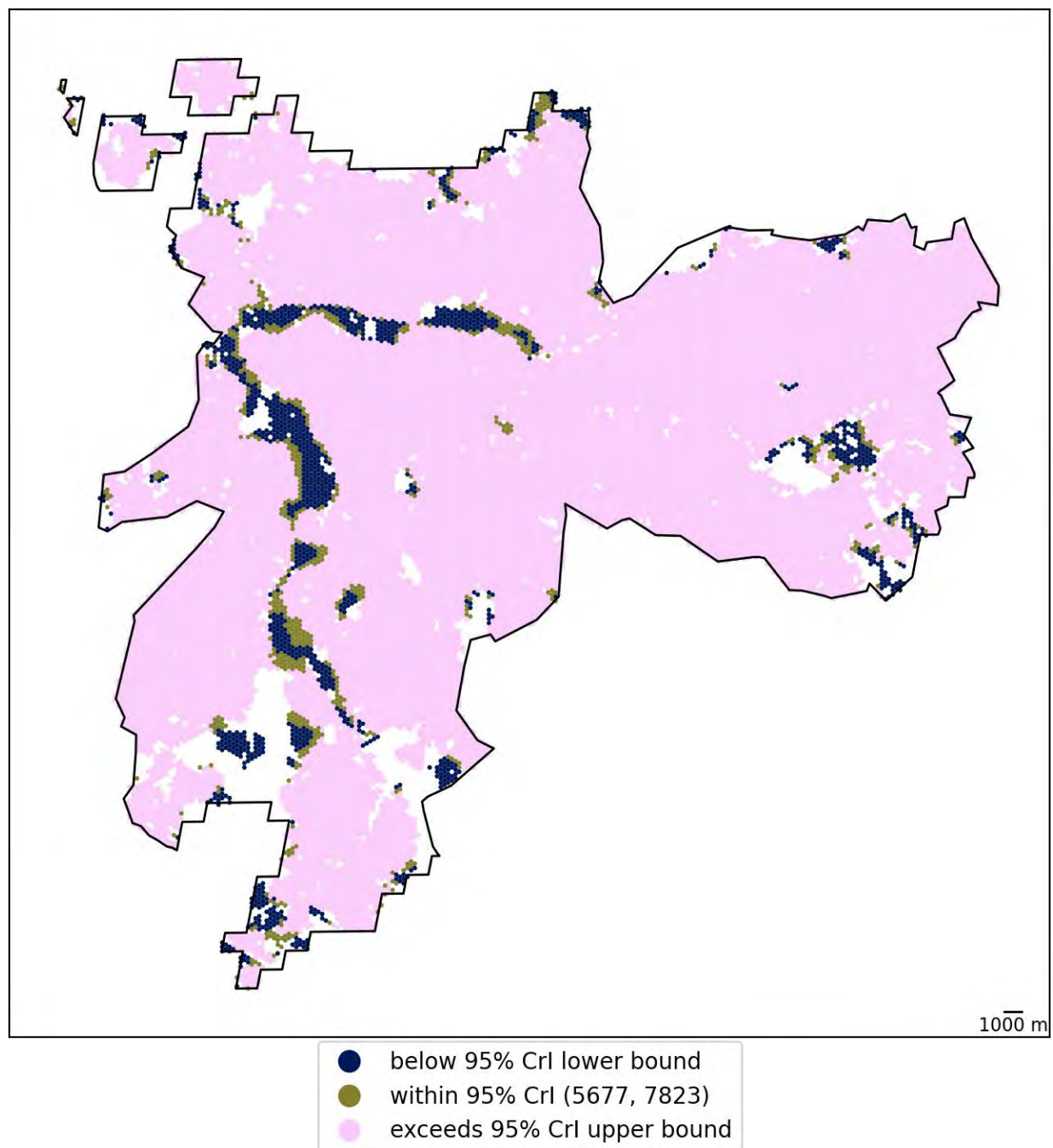

- below 95% CrI lower bound
- within 95% CrI (5677, 7823)
- exceeds 95% CrI upper bound

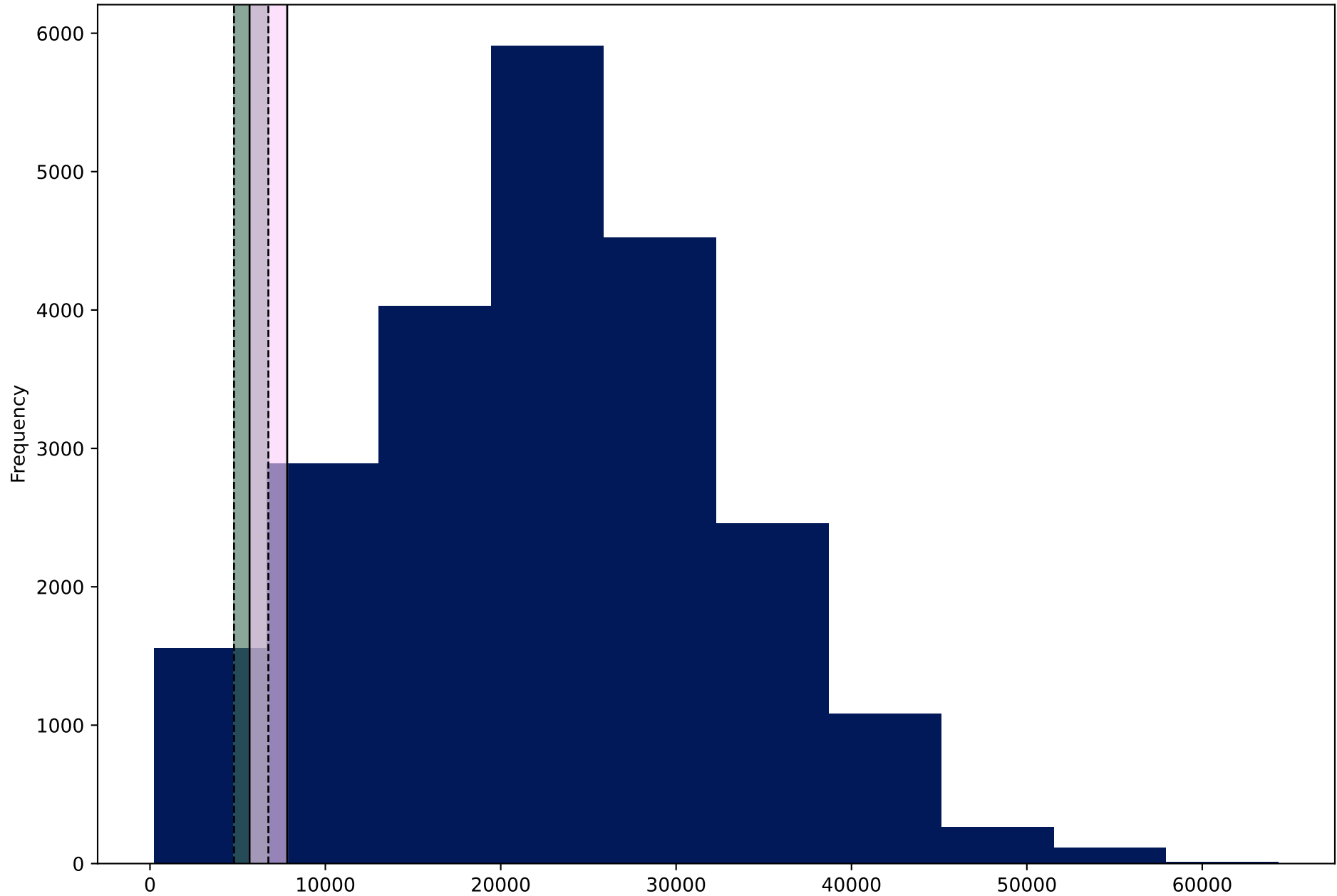

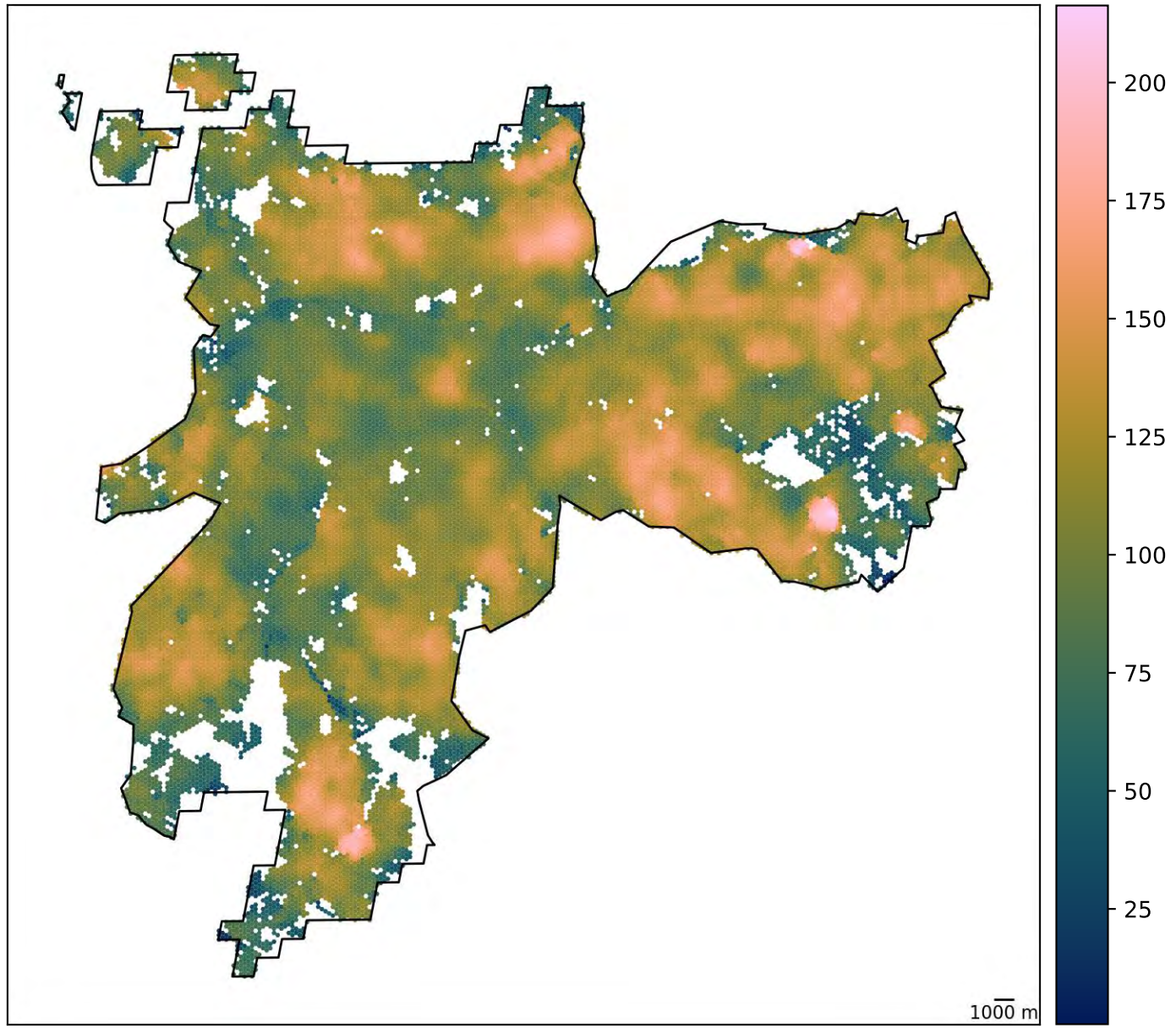

Mean 1000 m neighbourhood street intersections per km²

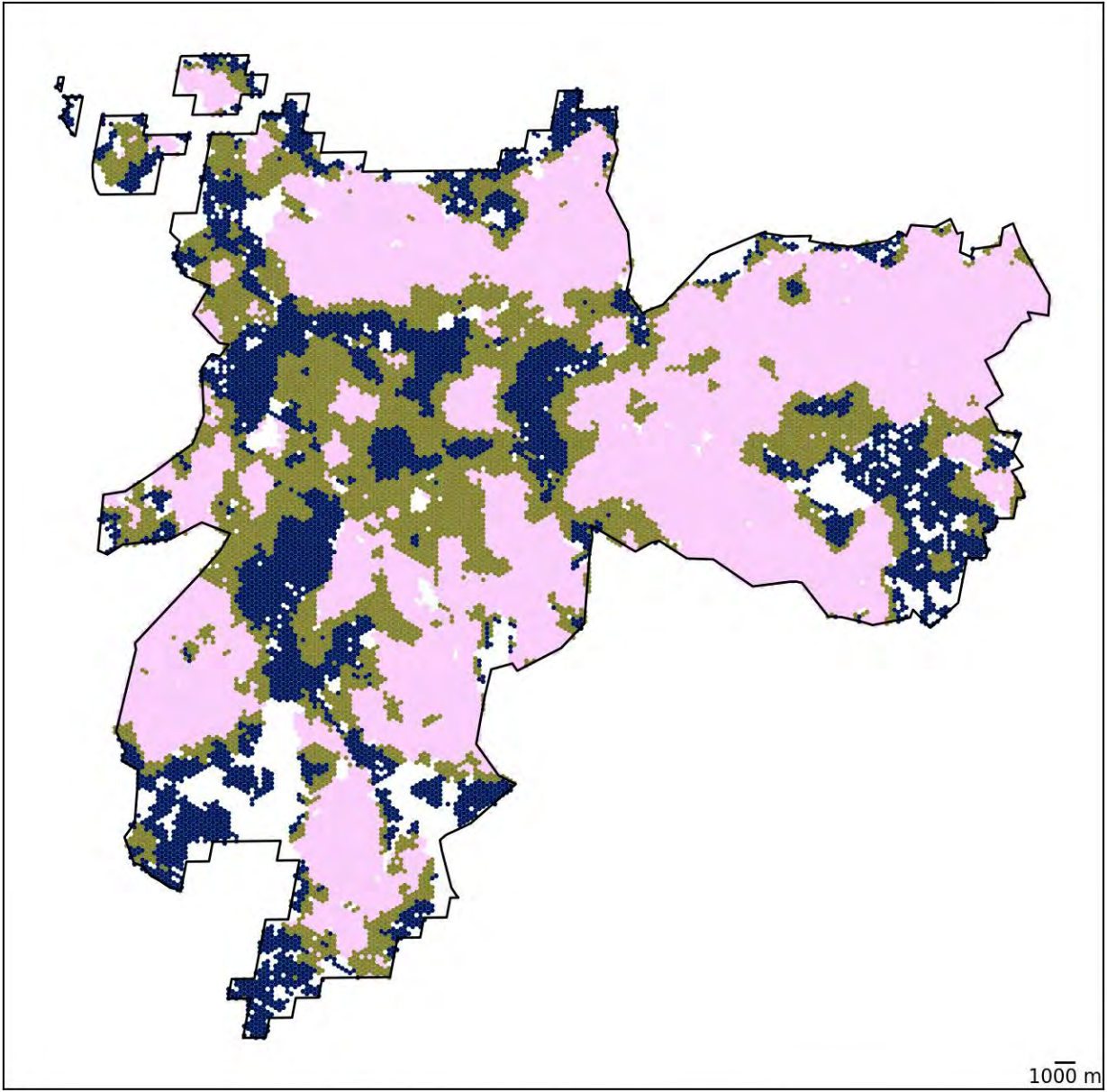

A: Estimated Mean 1000 m neighbourhood street intersections per km² requirement for ≥80% probability of engaging in walking for transport

B: Estimated Mean 1000 m neighbourhood street intersections per km² requirement for reaching the WHO's target of a ≥15% relative reduction in insufficient physical activity through walking

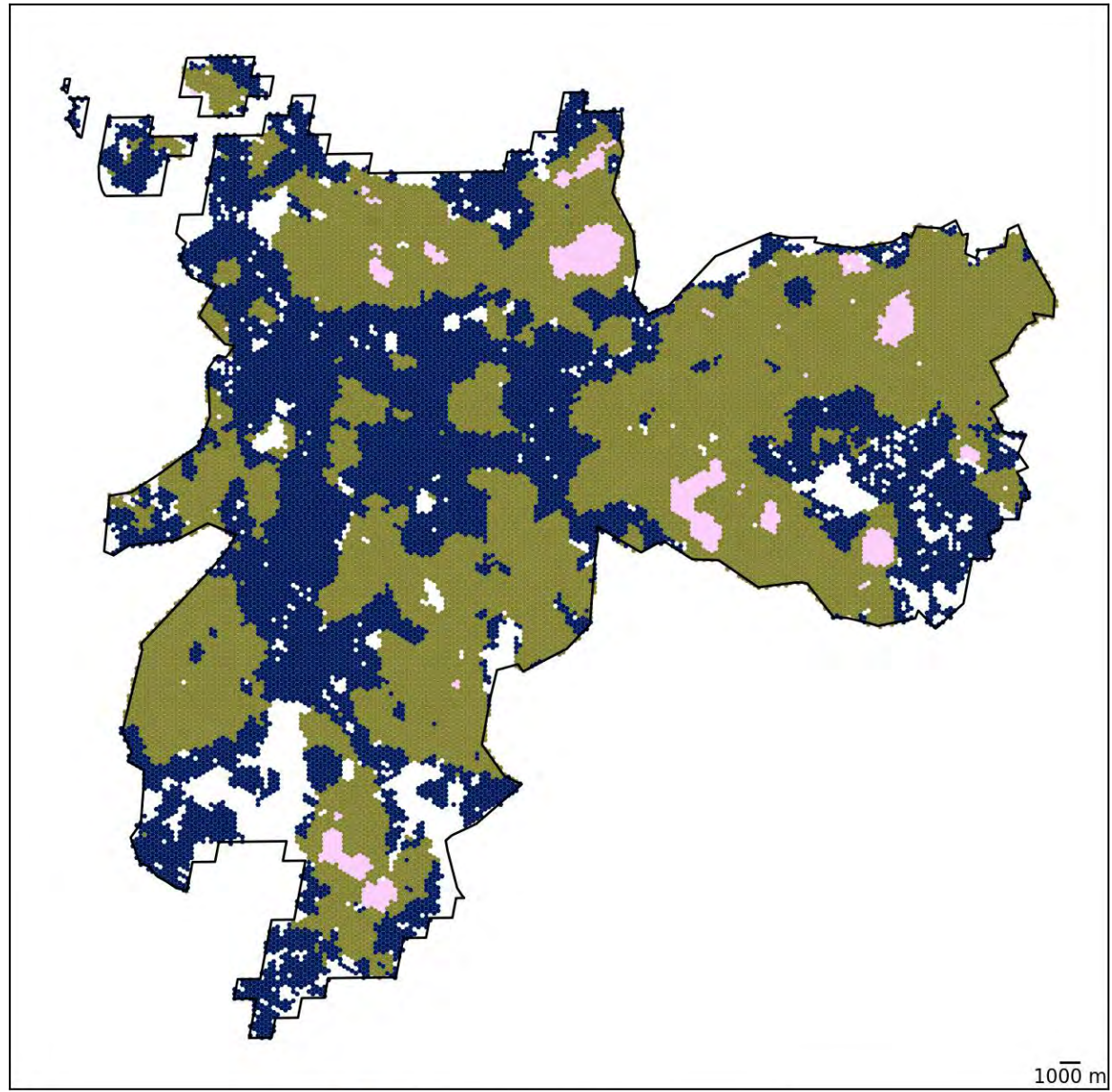

Legend:
- below 95% CrI lower bound
- within 95% CrI (106, 156)
- exceeds 95% CrI upper bound

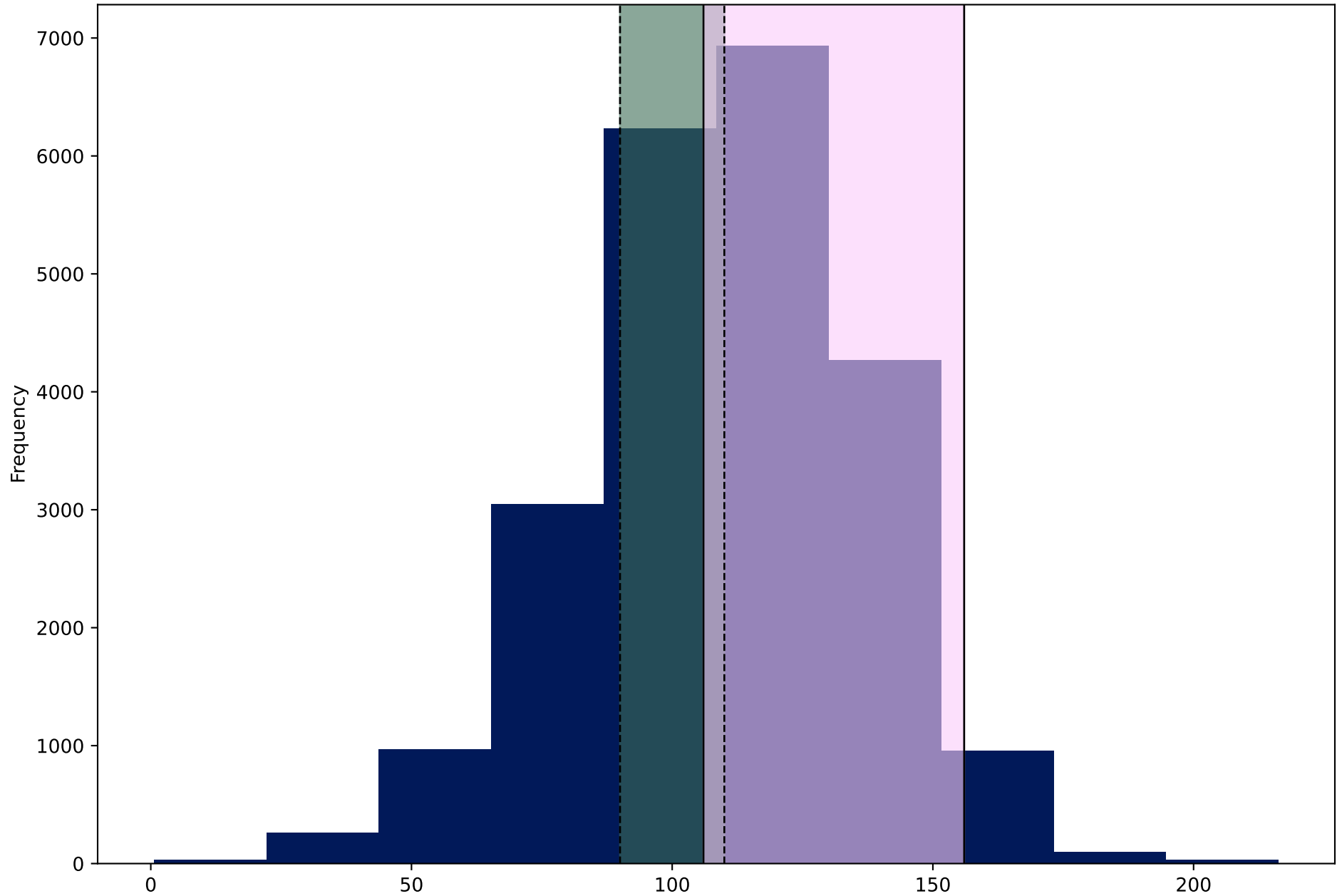

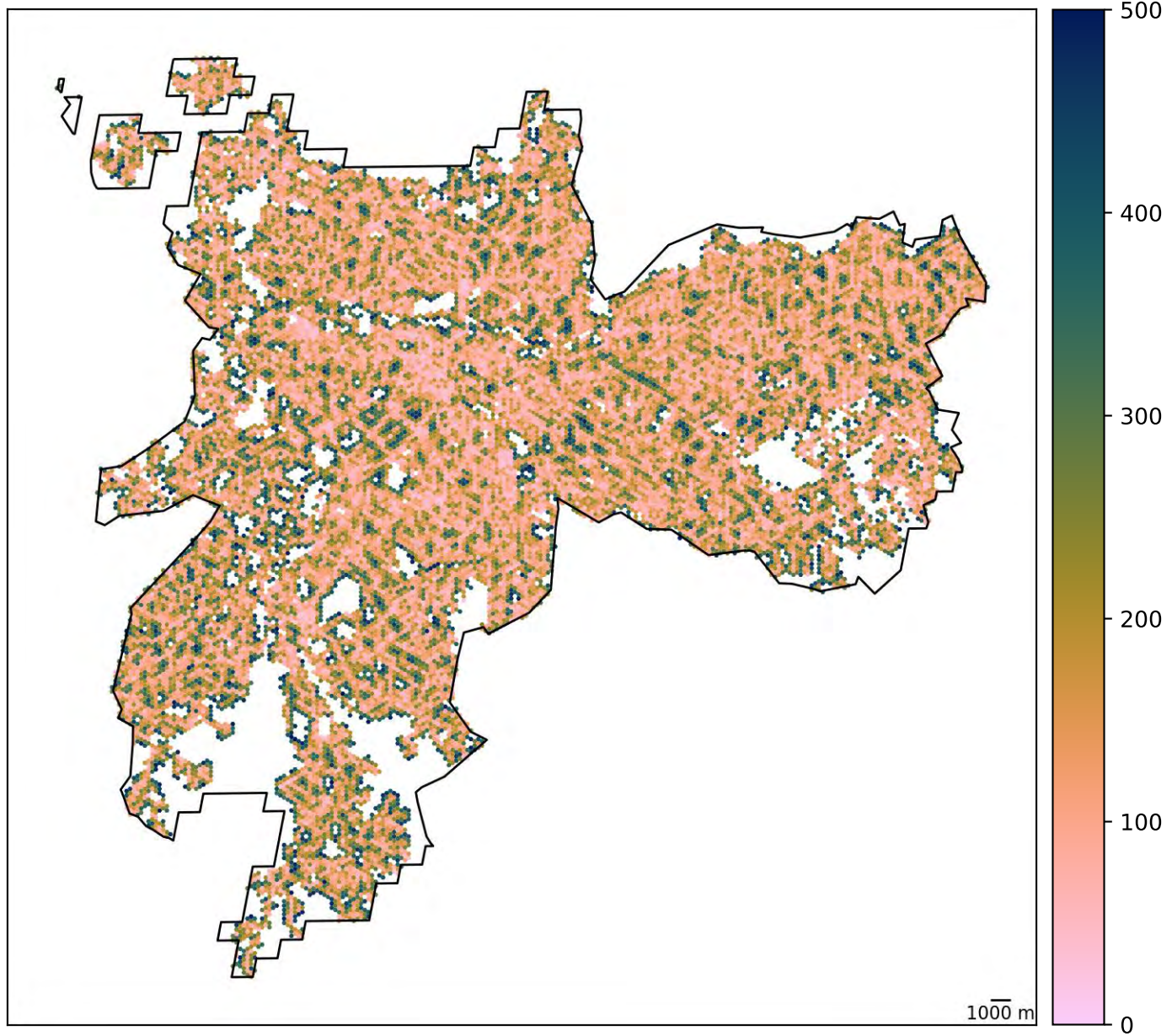

distances: Estimated Distance to nearest public transport stops (m; up to 500m) requirement for distances to destinations, measured up to a maximum distance target threshold of 500 metres

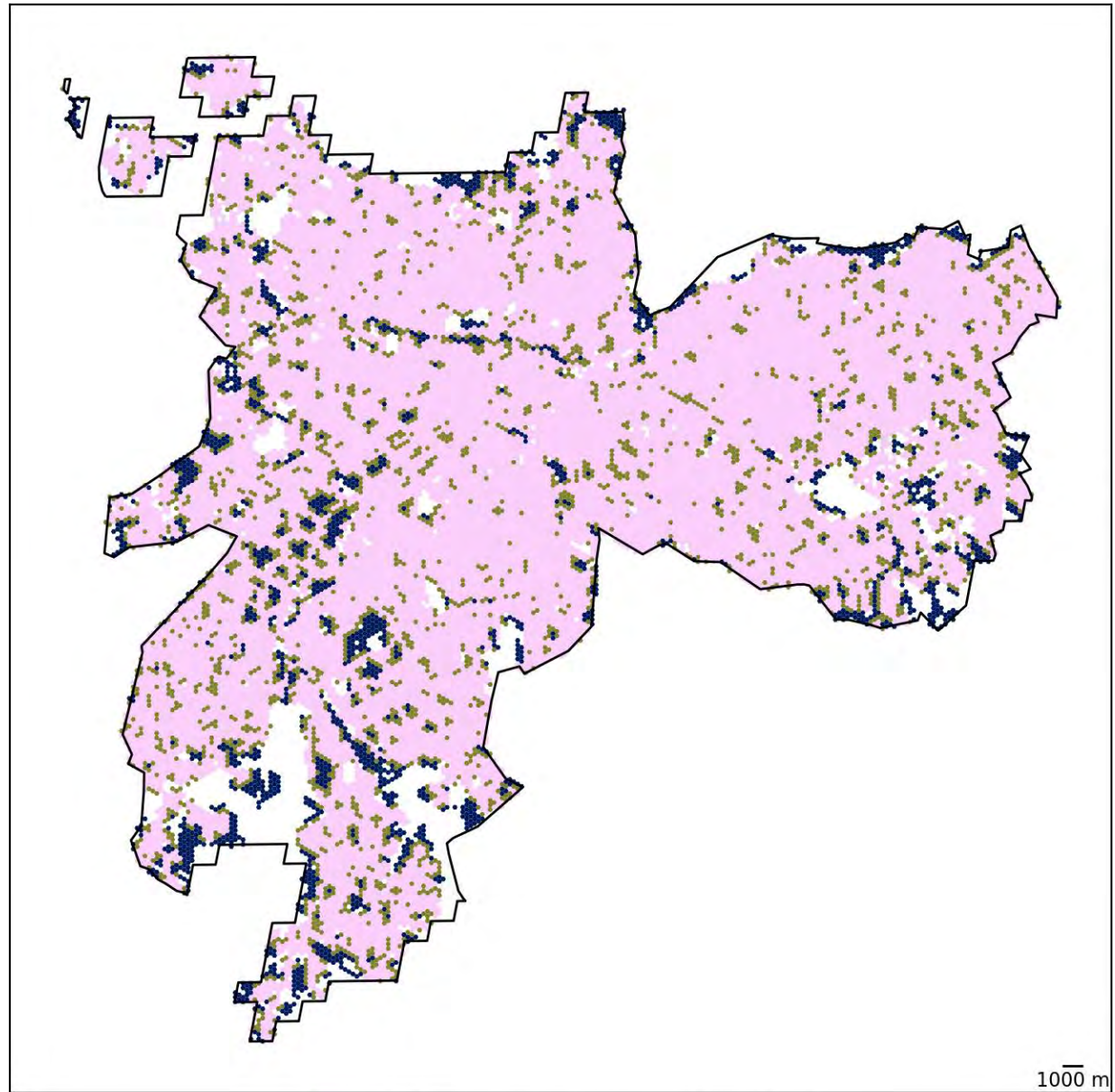

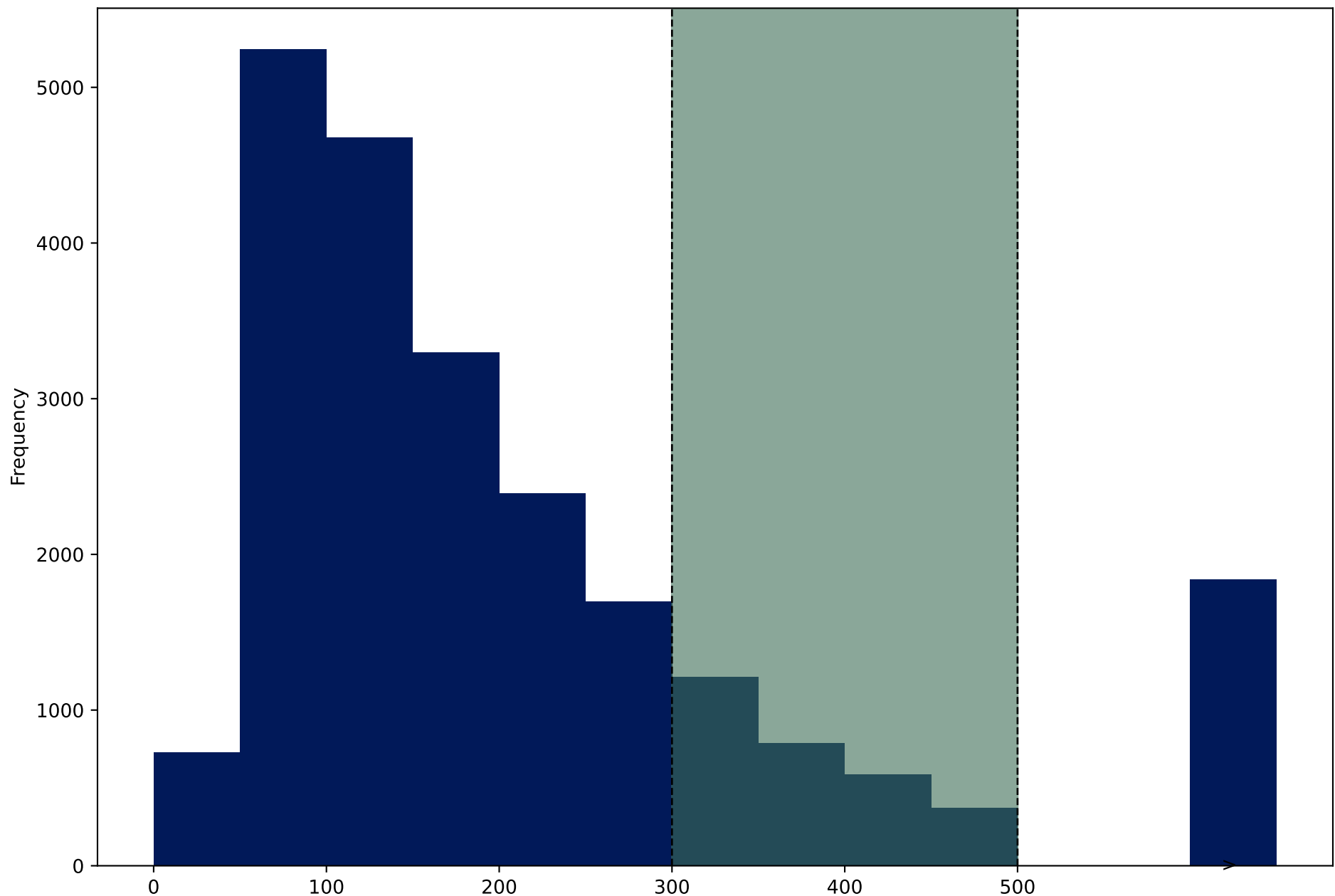

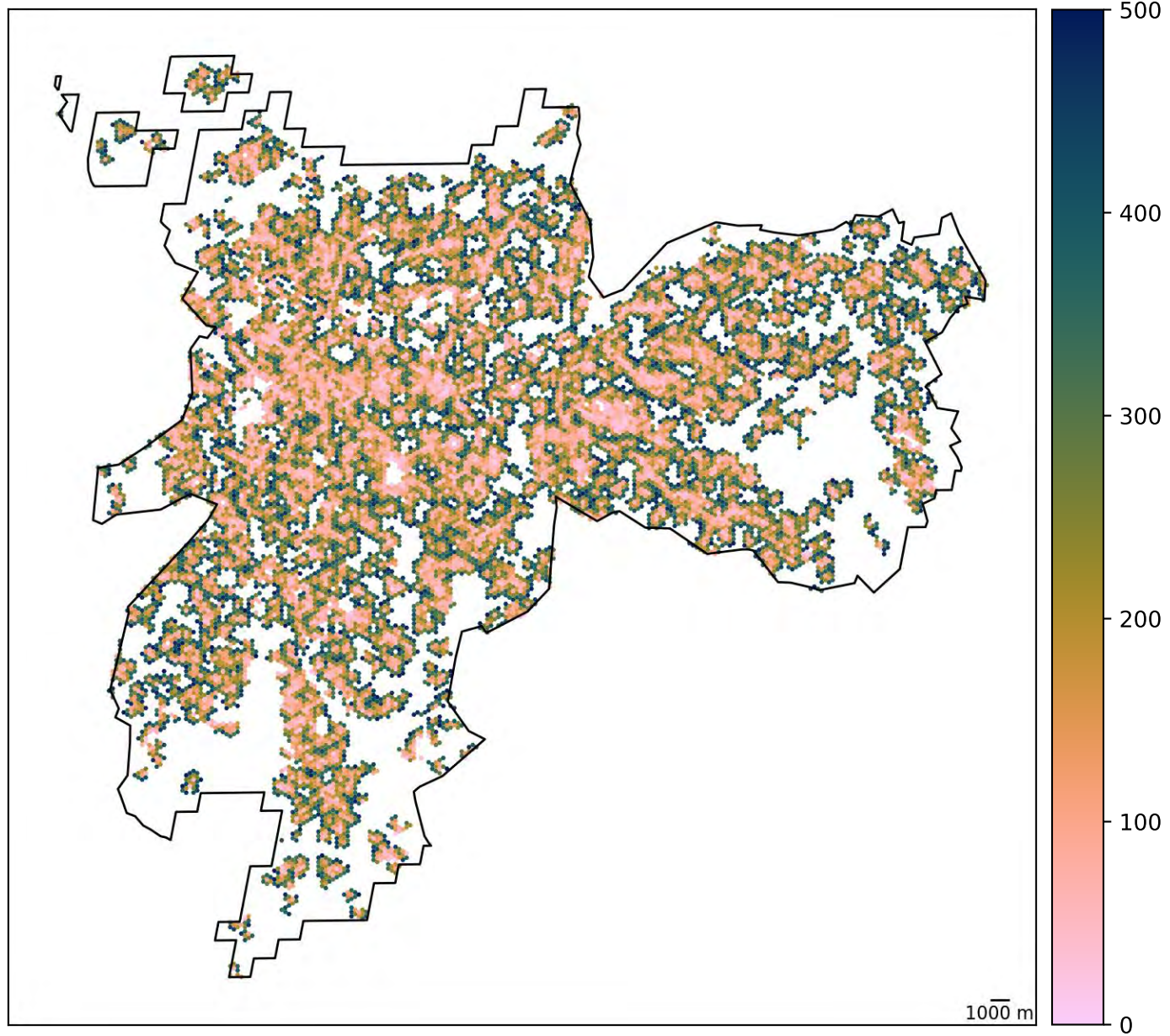

distances: Estimated Distance to nearest park (m; up to 500m) requirement for distances to destinations, measured up to a maximum distance target threshold of 500 metres

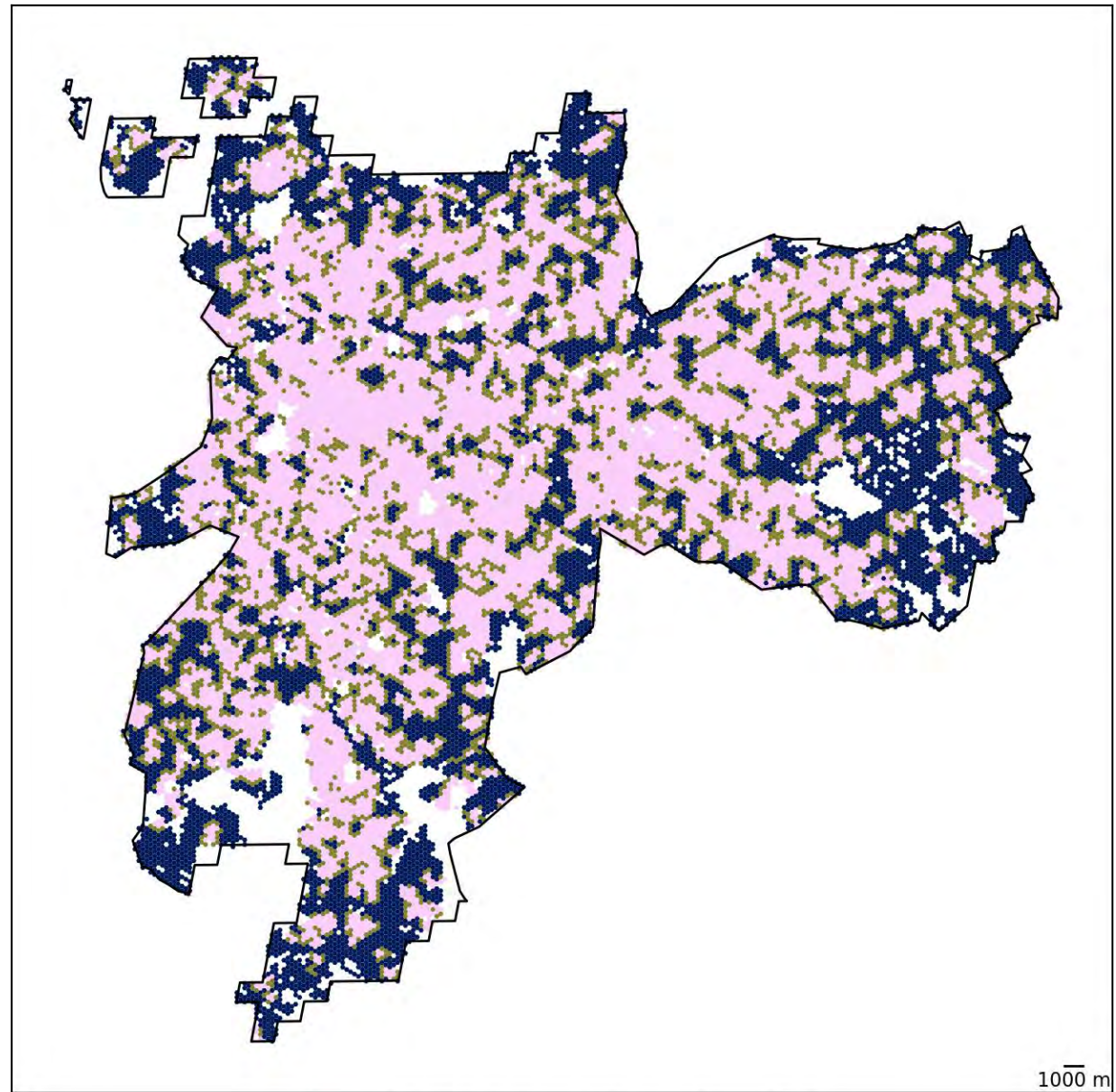

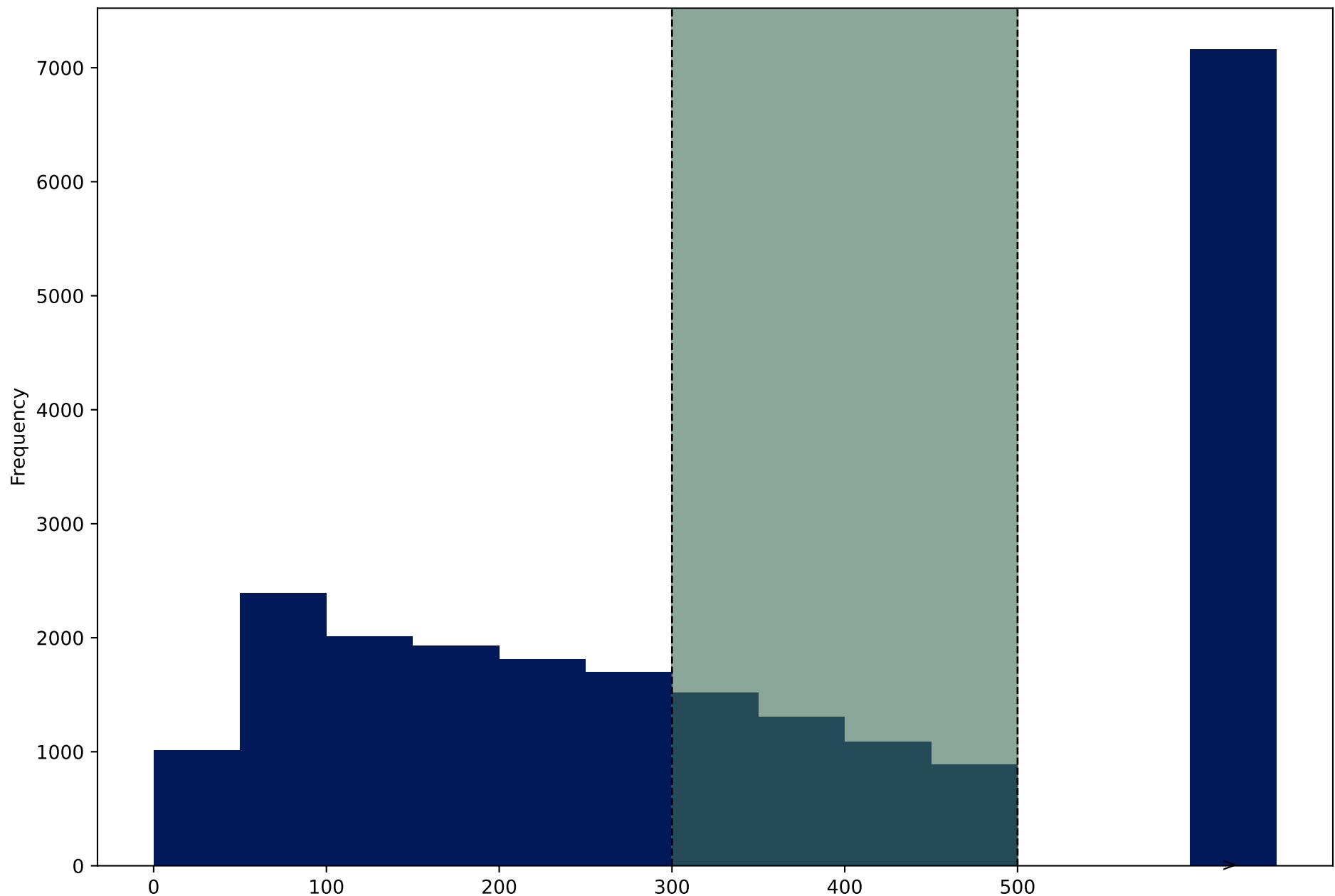

# Asia, China (SAR), Hong Kong

| Satellite imagery of urban study region (Bing) | Walkability, relative to city | Walkability, relative to 25 global cities |
|---|---|---|

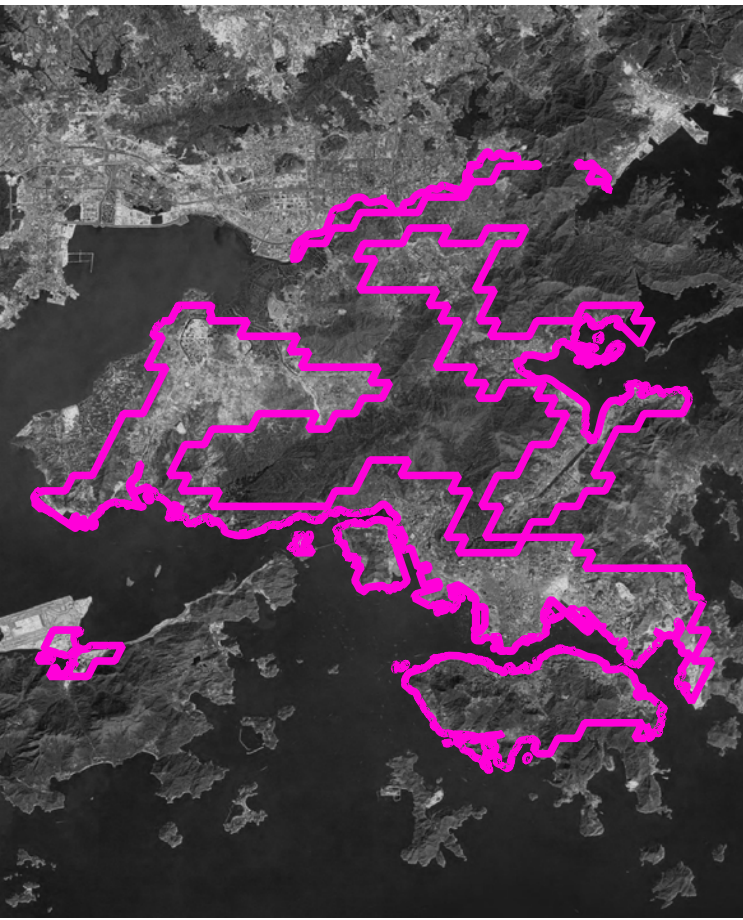
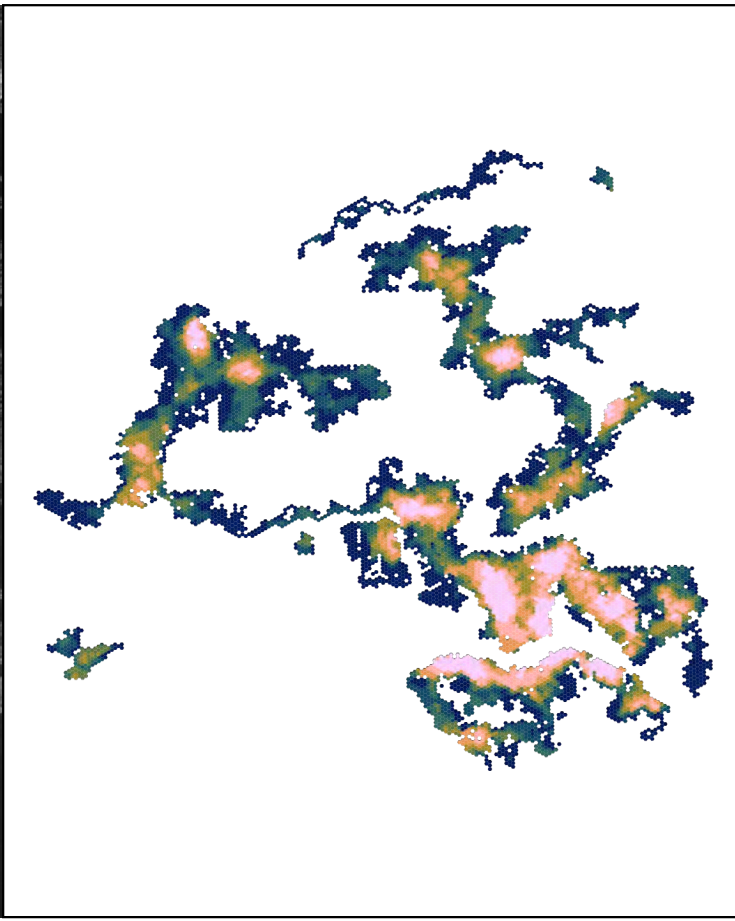
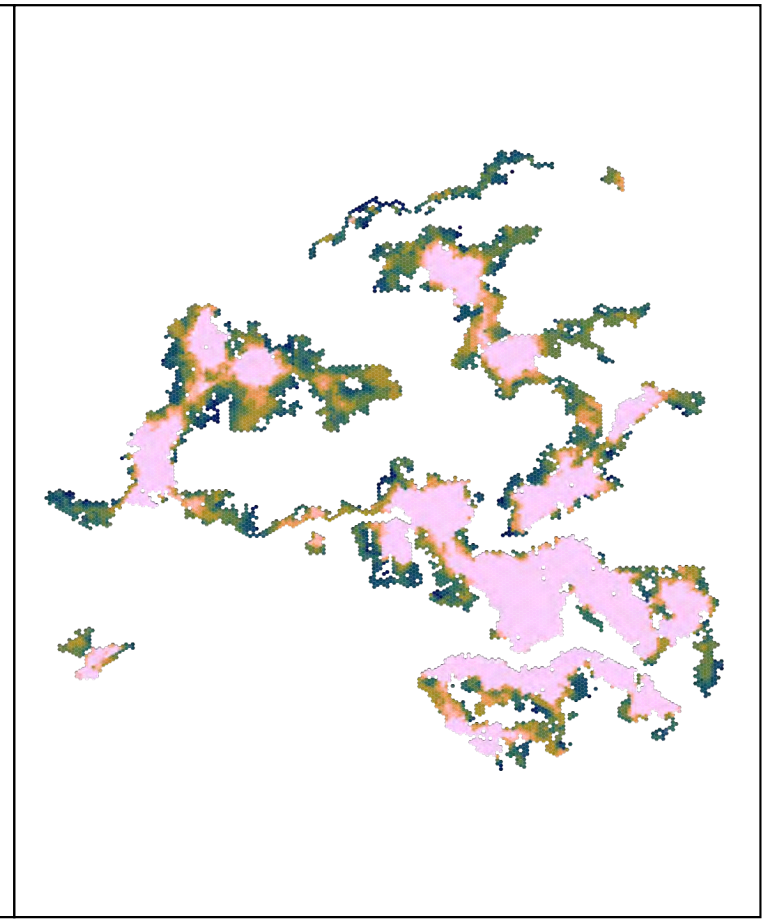

Urban boundary

0 — 19 — 38 km

**Walkability score**
- <-3
- -3 to -2
- -2 to -1
- -1 to 0
- 0 to 1
- 1 to 2
- 2 to 3
- ≥3

Walkability relative to all cities by component variables (2D histograms), and overall (histogram)

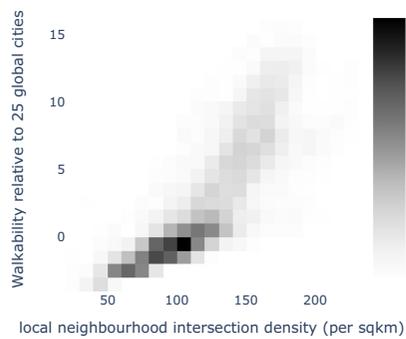
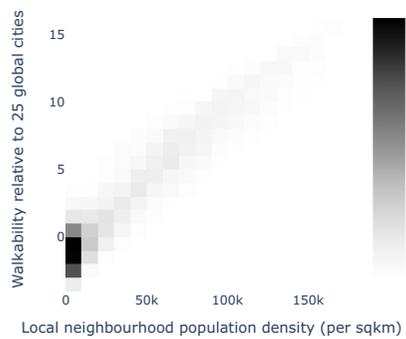
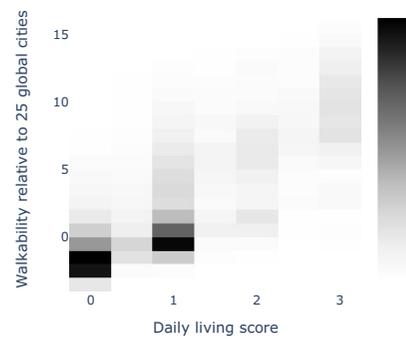
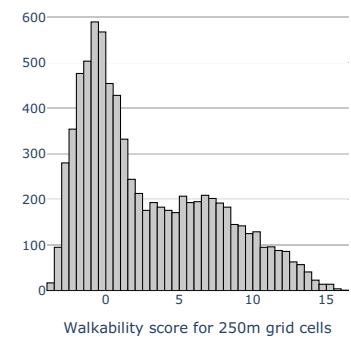

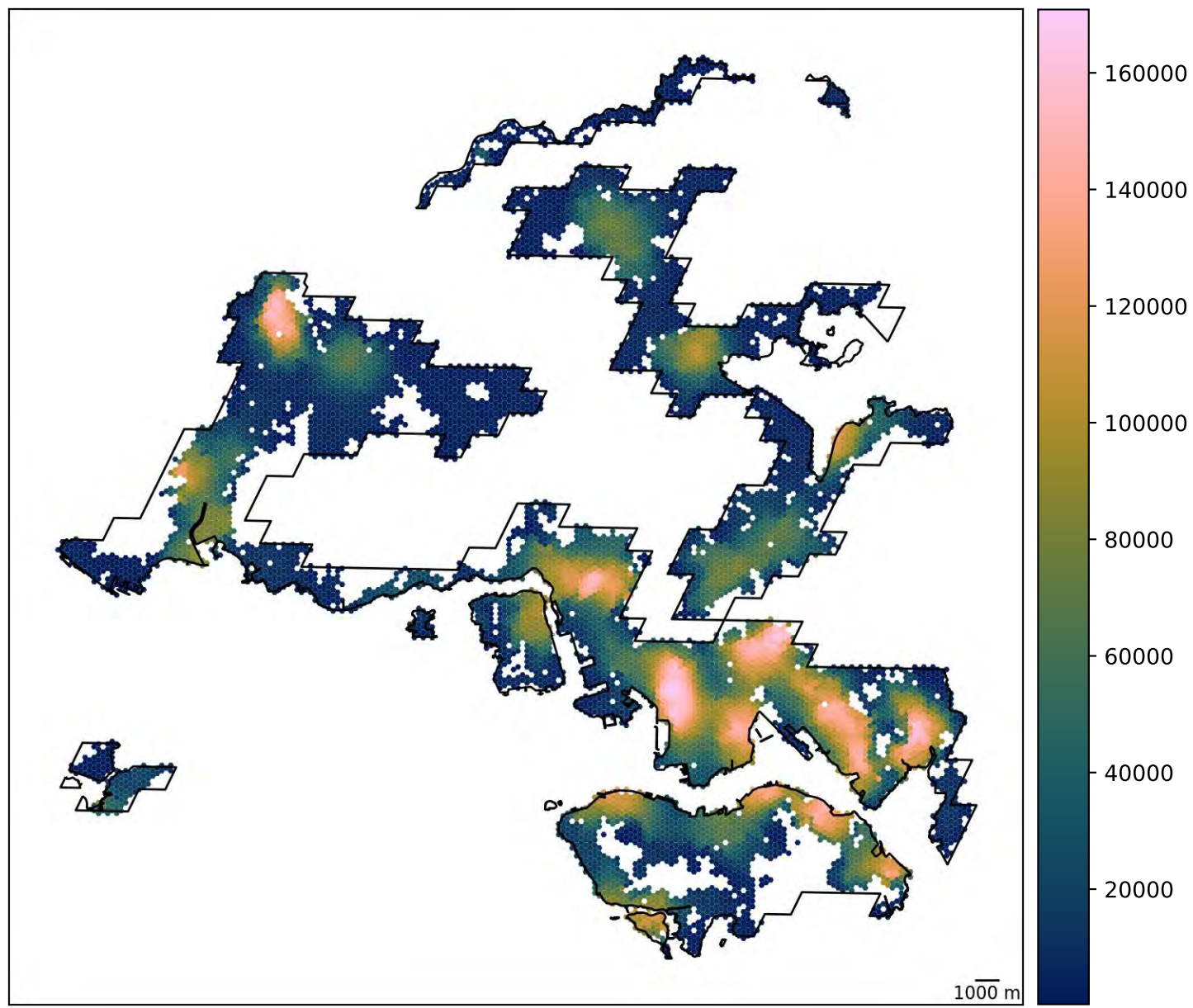

Mean 1000 m neighbourhood population per km²

# A: Estimated Mean 1000 m neighbourhood population per km² requirement for ≥80% probability of engaging in walking for transport

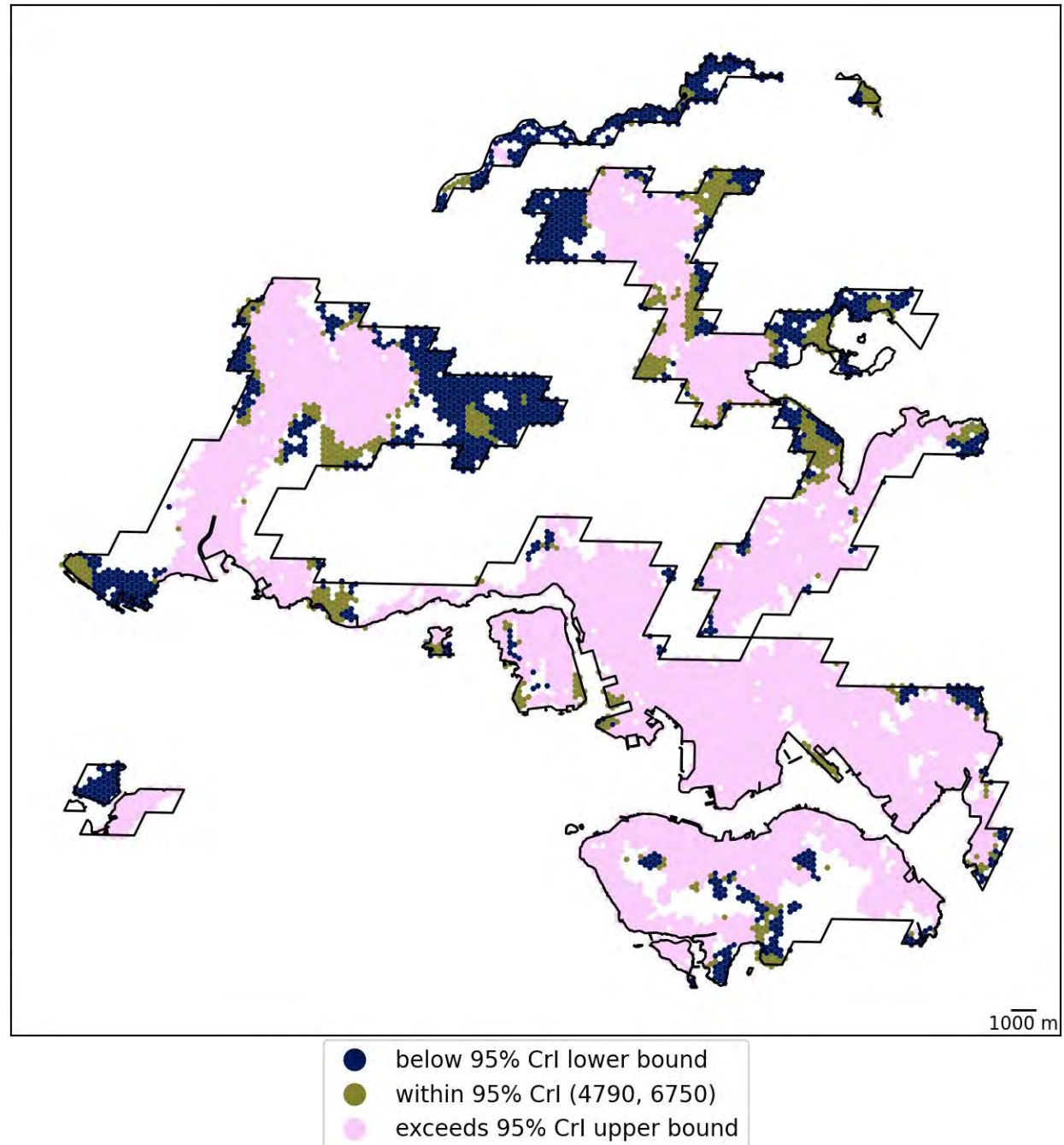

B: Estimated Mean 1000 m neighbourhood population per km² requirement for reaching the WHO's target of a ≥15% relative reduction in insufficient physical activity through walking

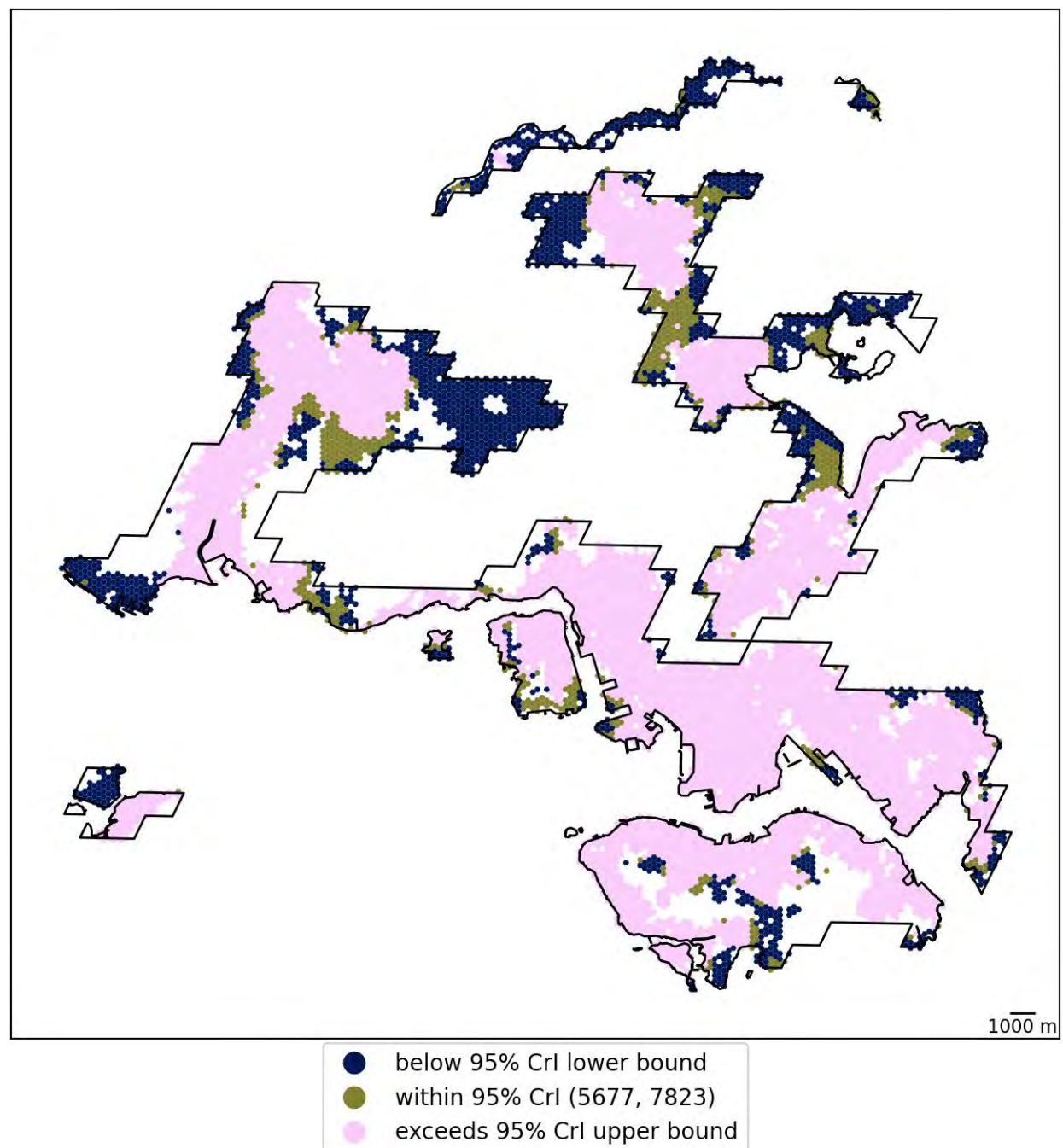

Legend:
- below 95% CrI lower bound
- within 95% CrI (5677, 7823)
- exceeds 95% CrI upper bound

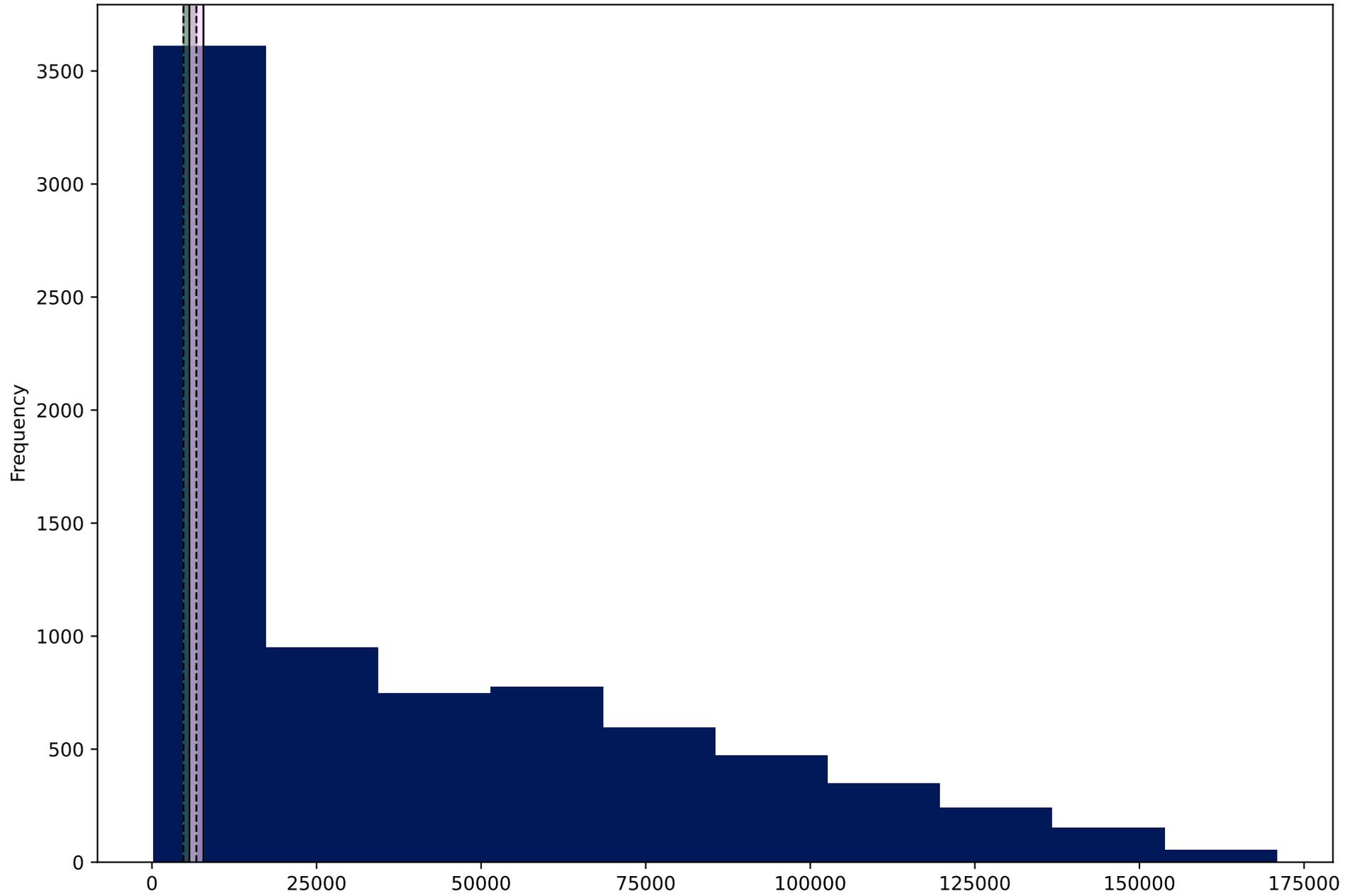

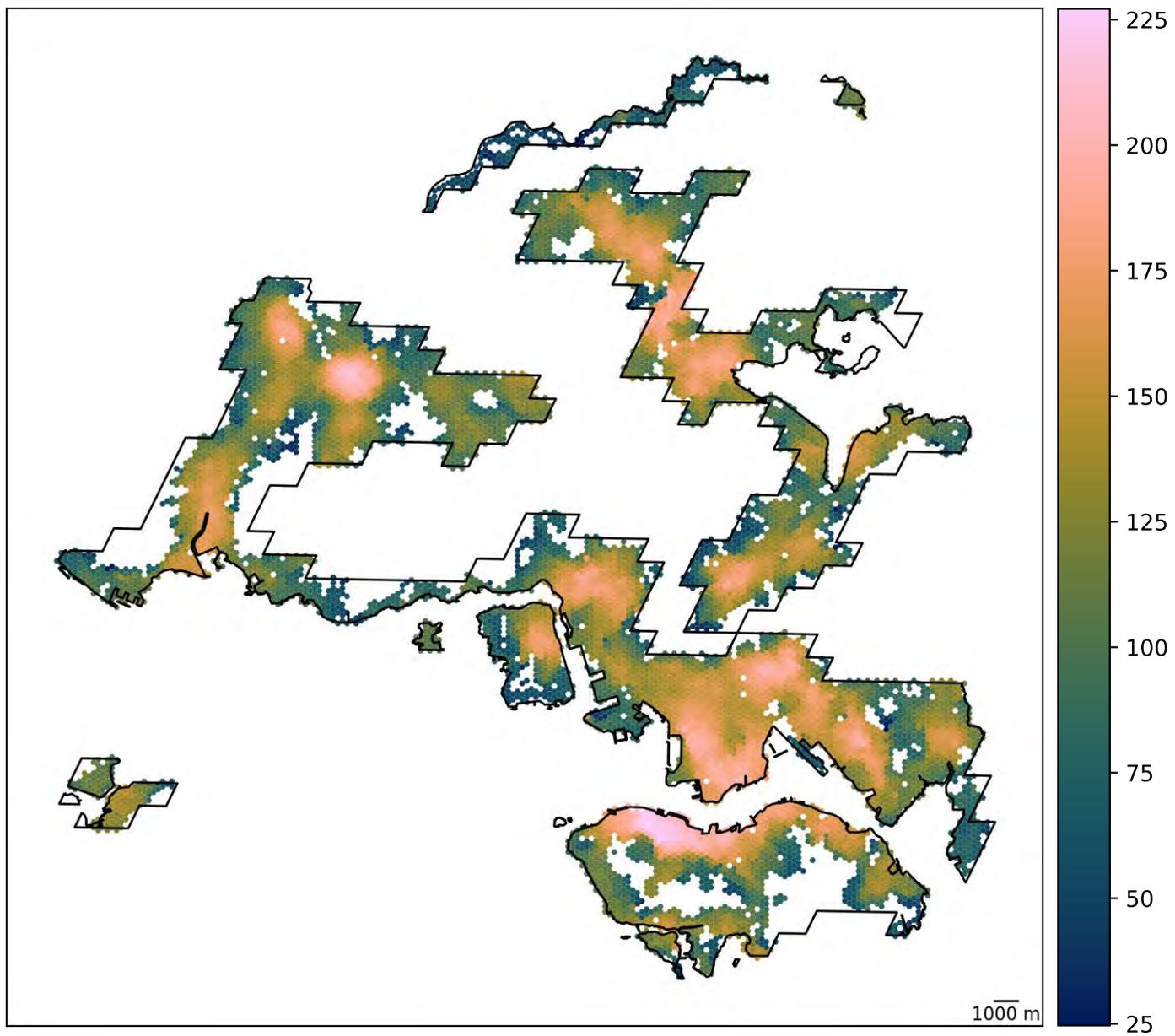

Mean 1000 m neighbourhood street intersections per km²

A: Estimated Mean 1000 m neighbourhood street intersections per km² requirement for ≥80% probability of engaging in walking for transport

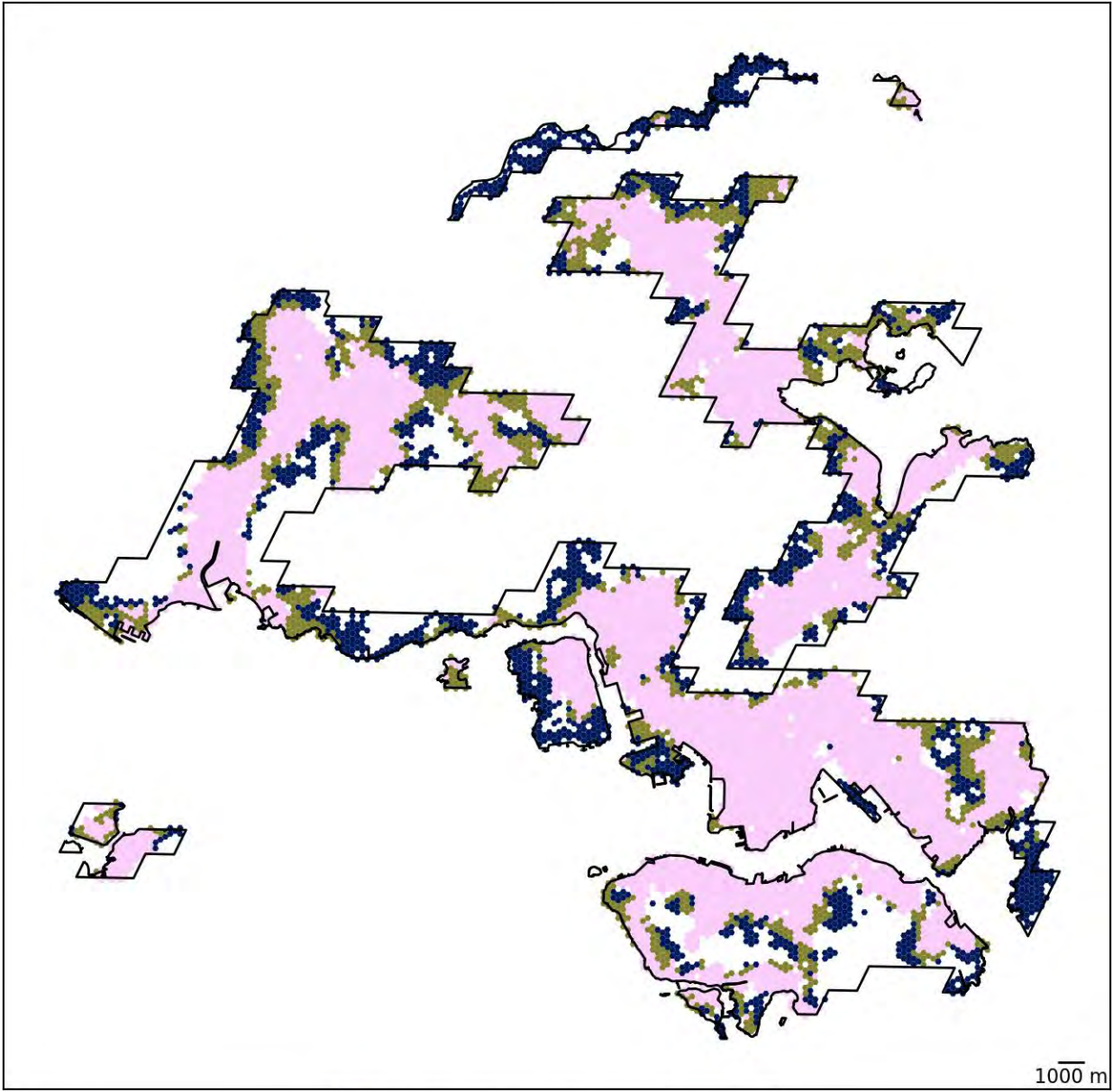

B: Estimated Mean 1000 m neighbourhood street intersections per km² requirement for reaching the WHO's target of a ≥15% relative reduction in insufficient physical activity through walking

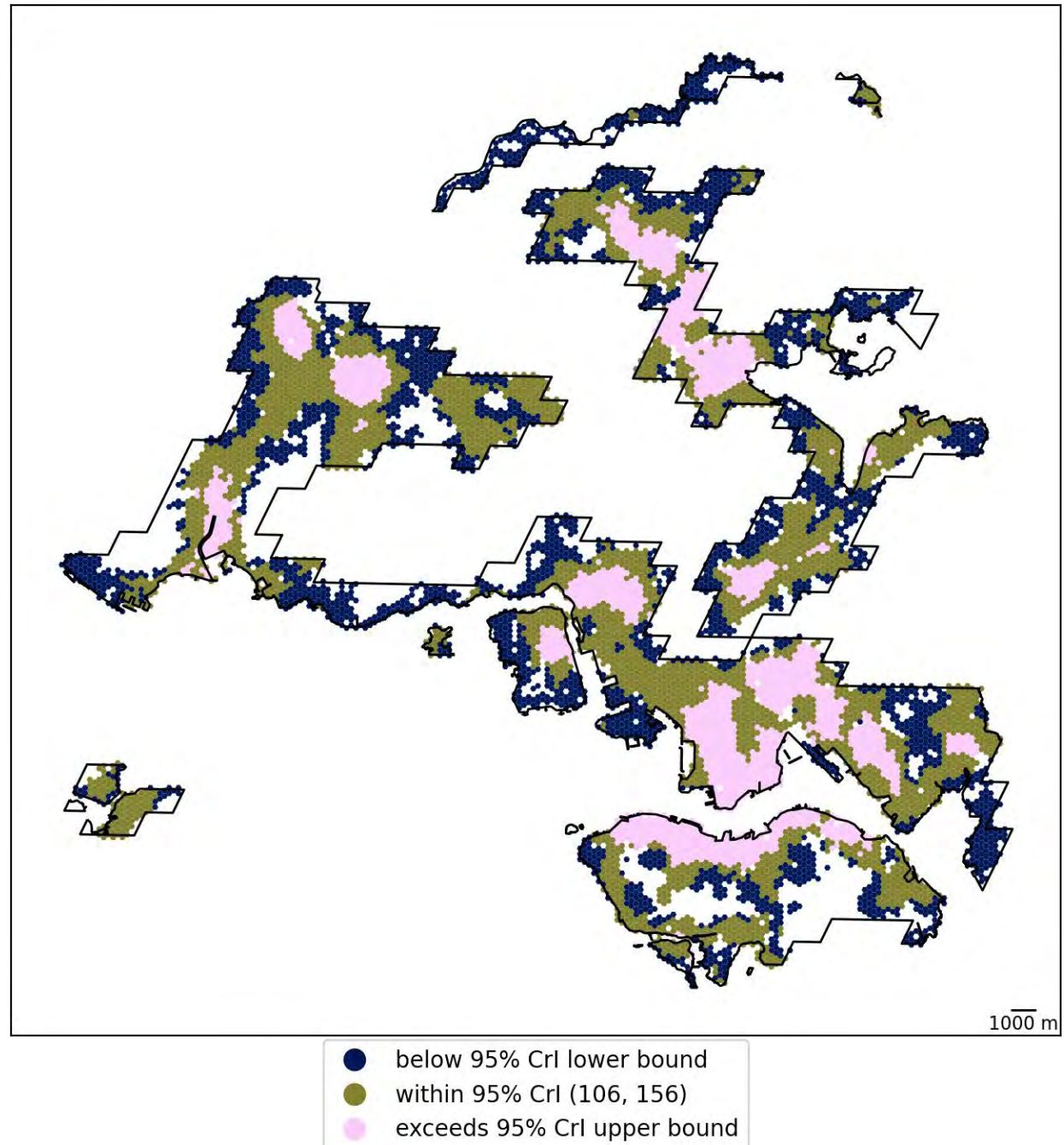

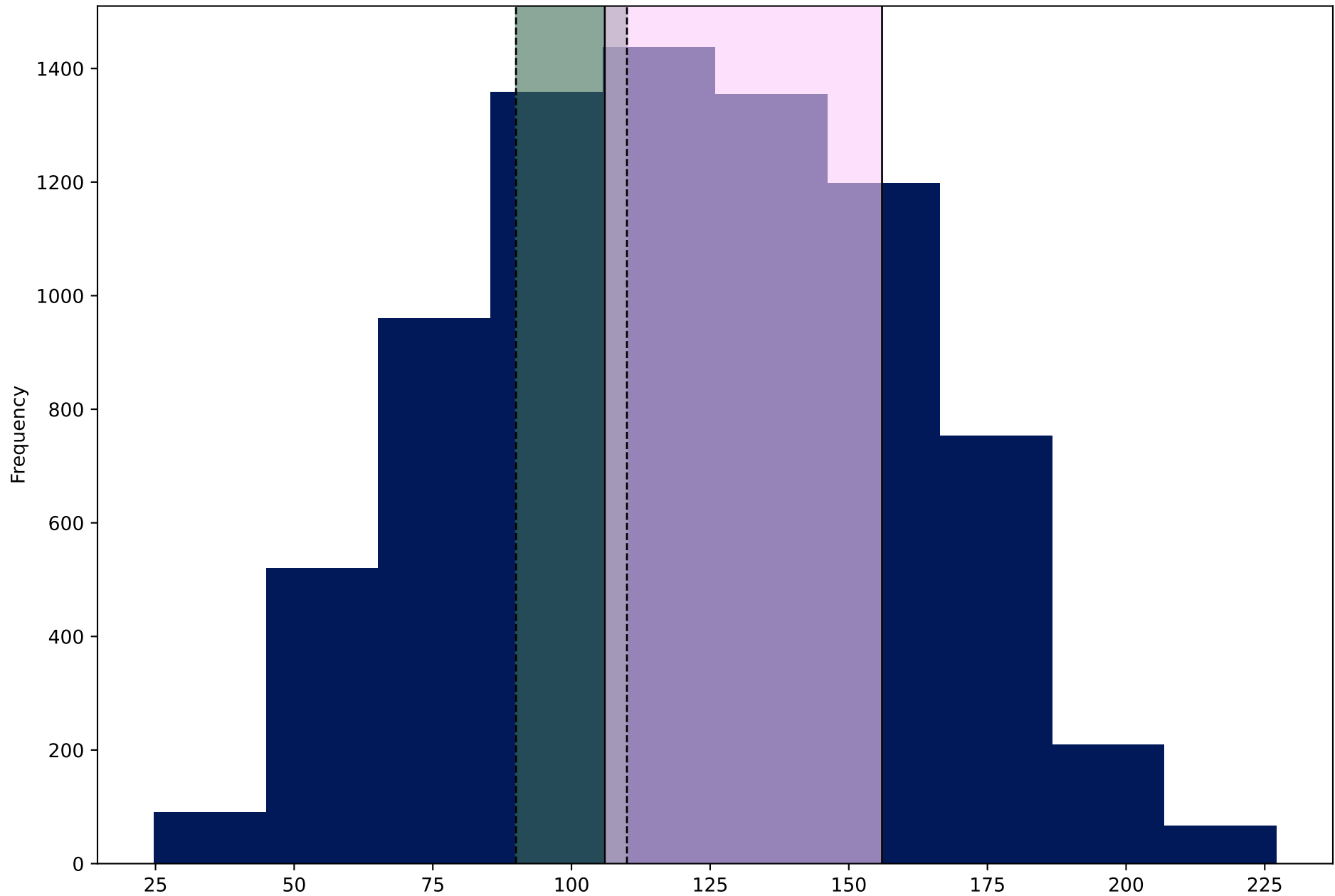

# Asia, India, Chennai

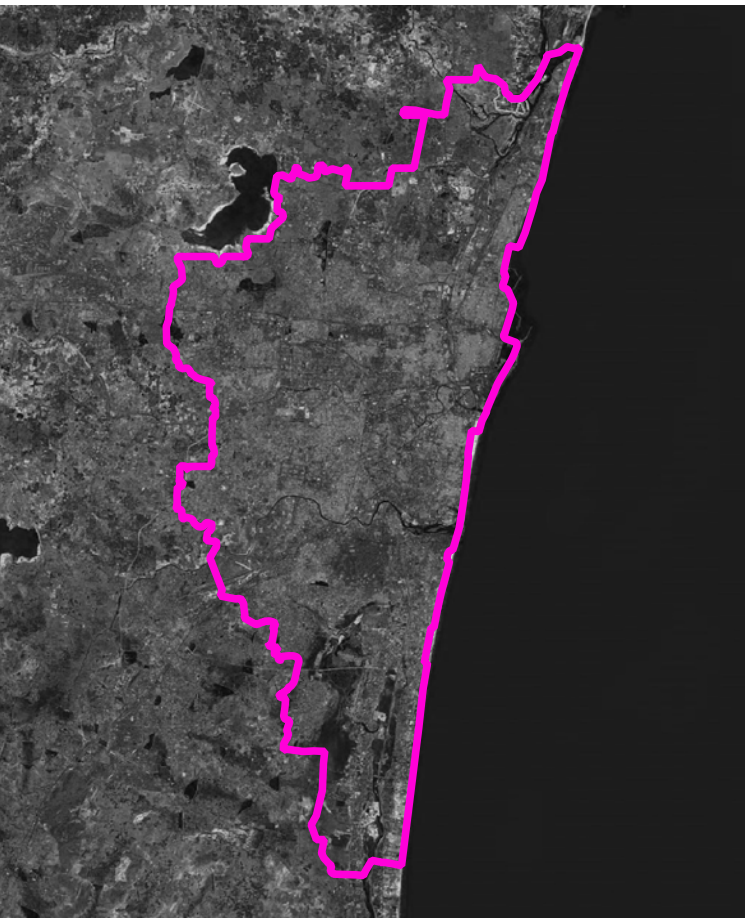
Satellite imagery of urban study region (Bing)

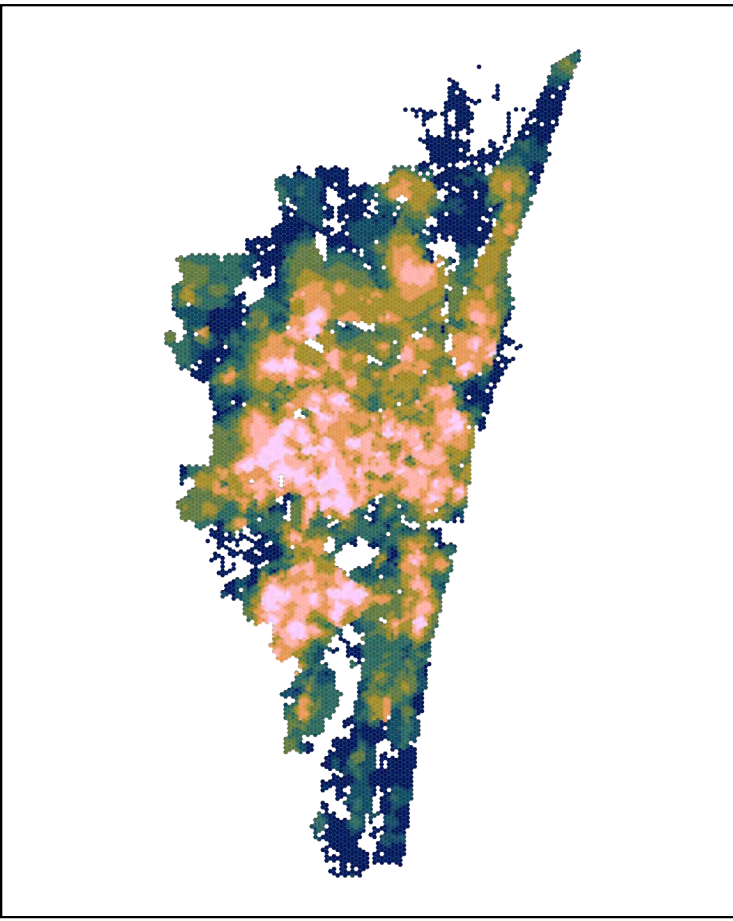
Walkability, relative to city

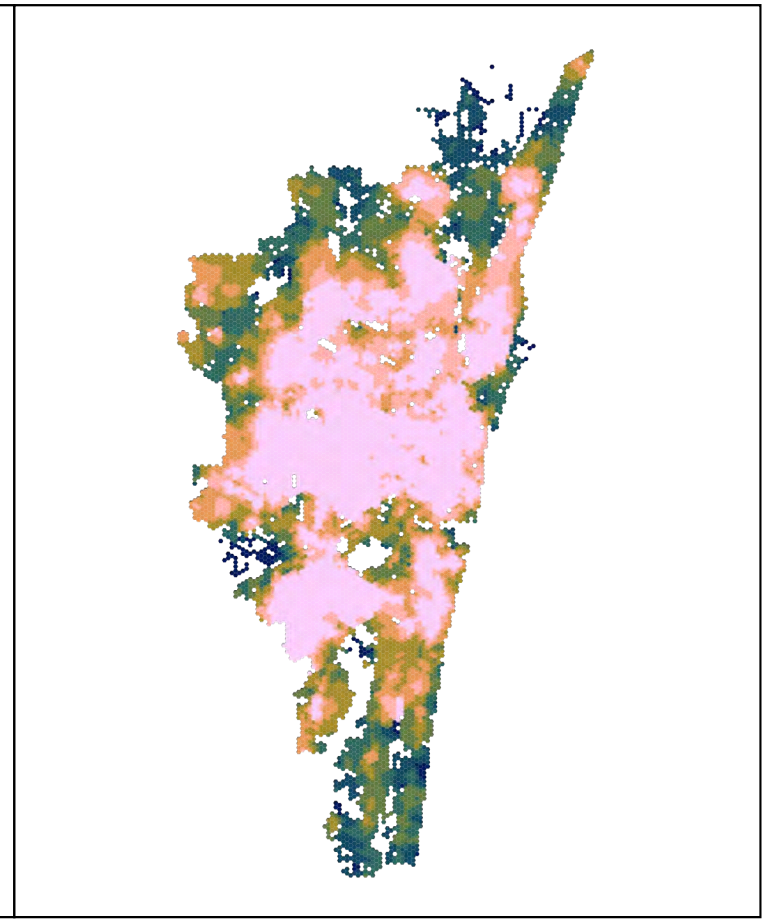
Walkability, relative to 25 global cities

Urban boundary

Walkability score
- <-3
- -3 to -2
- -2 to -1
- -1 to 0
- 0 to 1
- 1 to 2
- 2 to 3
- ≥3

Walkability relative to all cities by component variables (2D histograms), and overall (histogram)

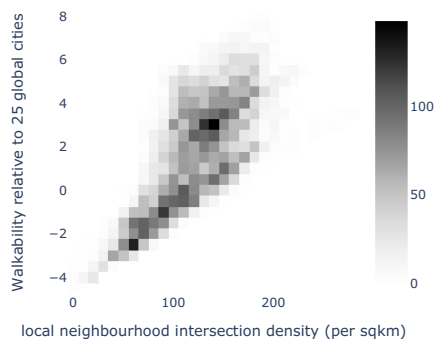
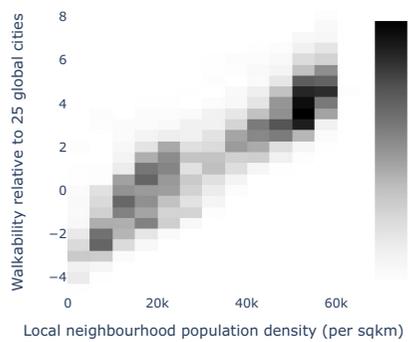
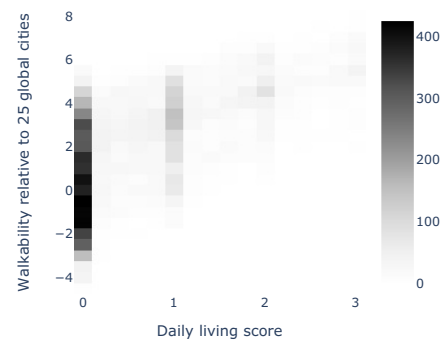
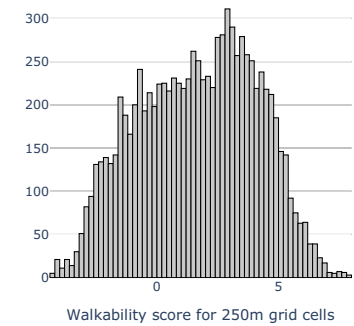

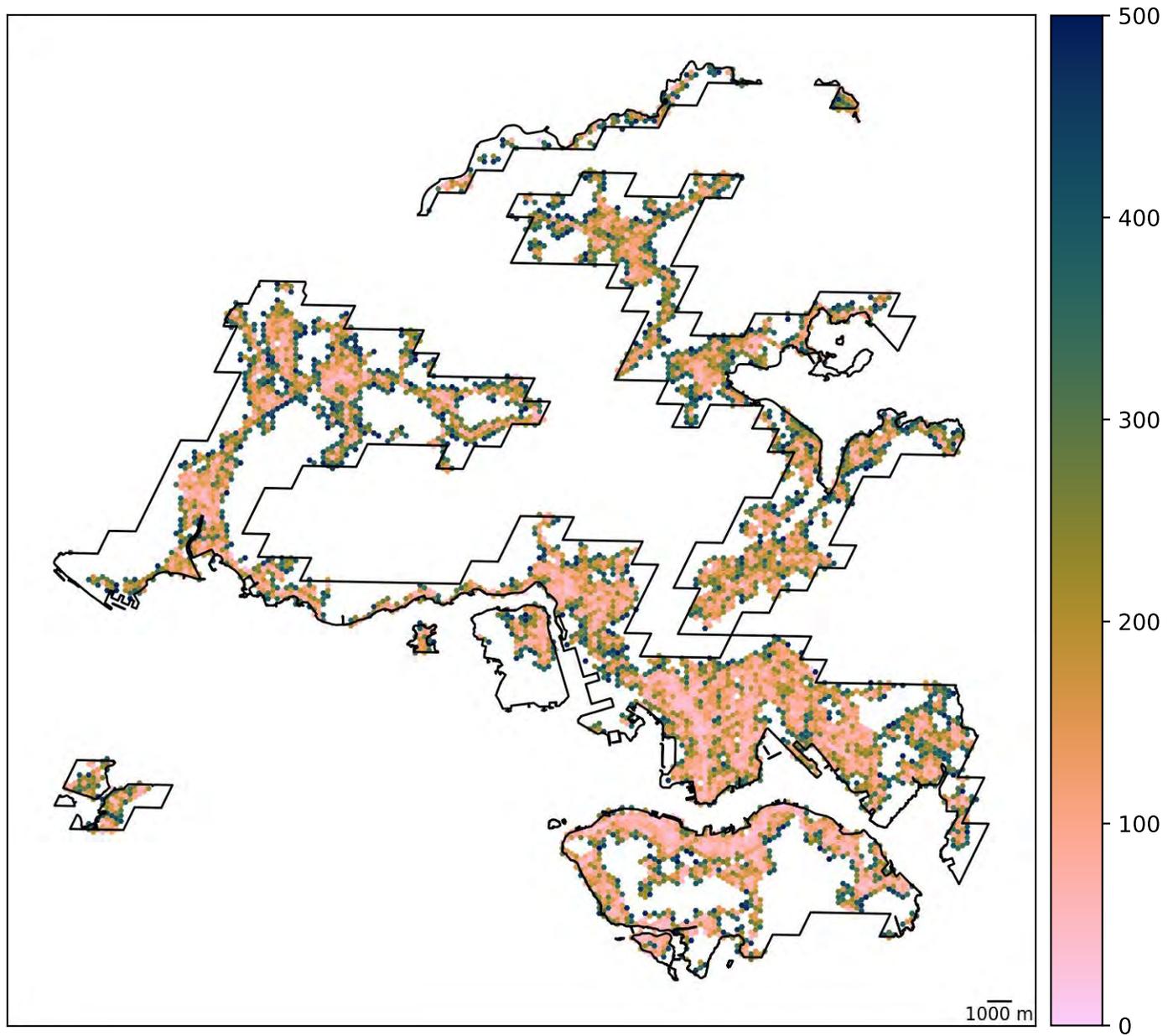
131

distances: Estimated Distance to nearest public transport stops (m; up to 500m) requirement for distances to destinations, measured up to a maximum distance target threshold of 500 metres

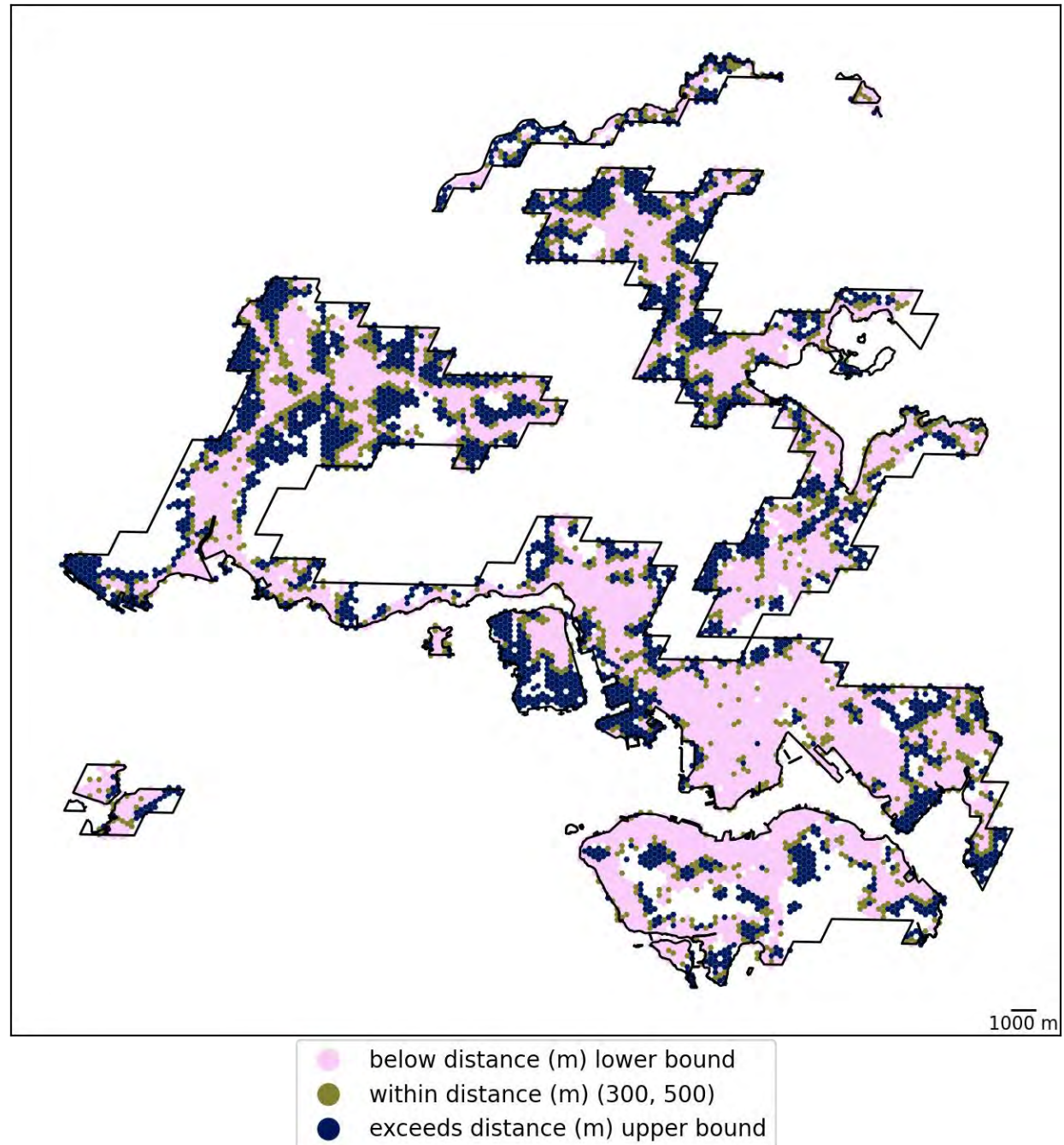

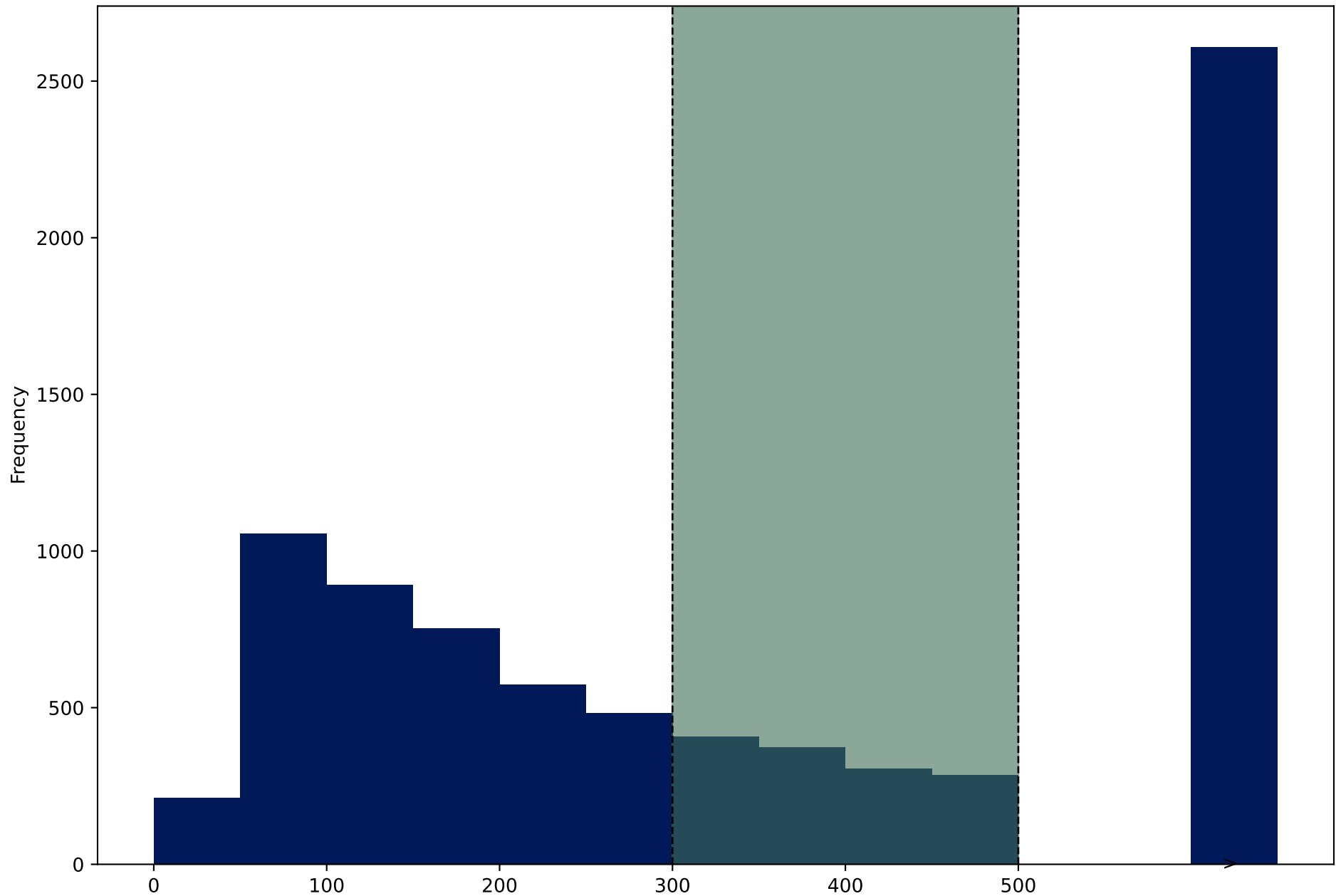

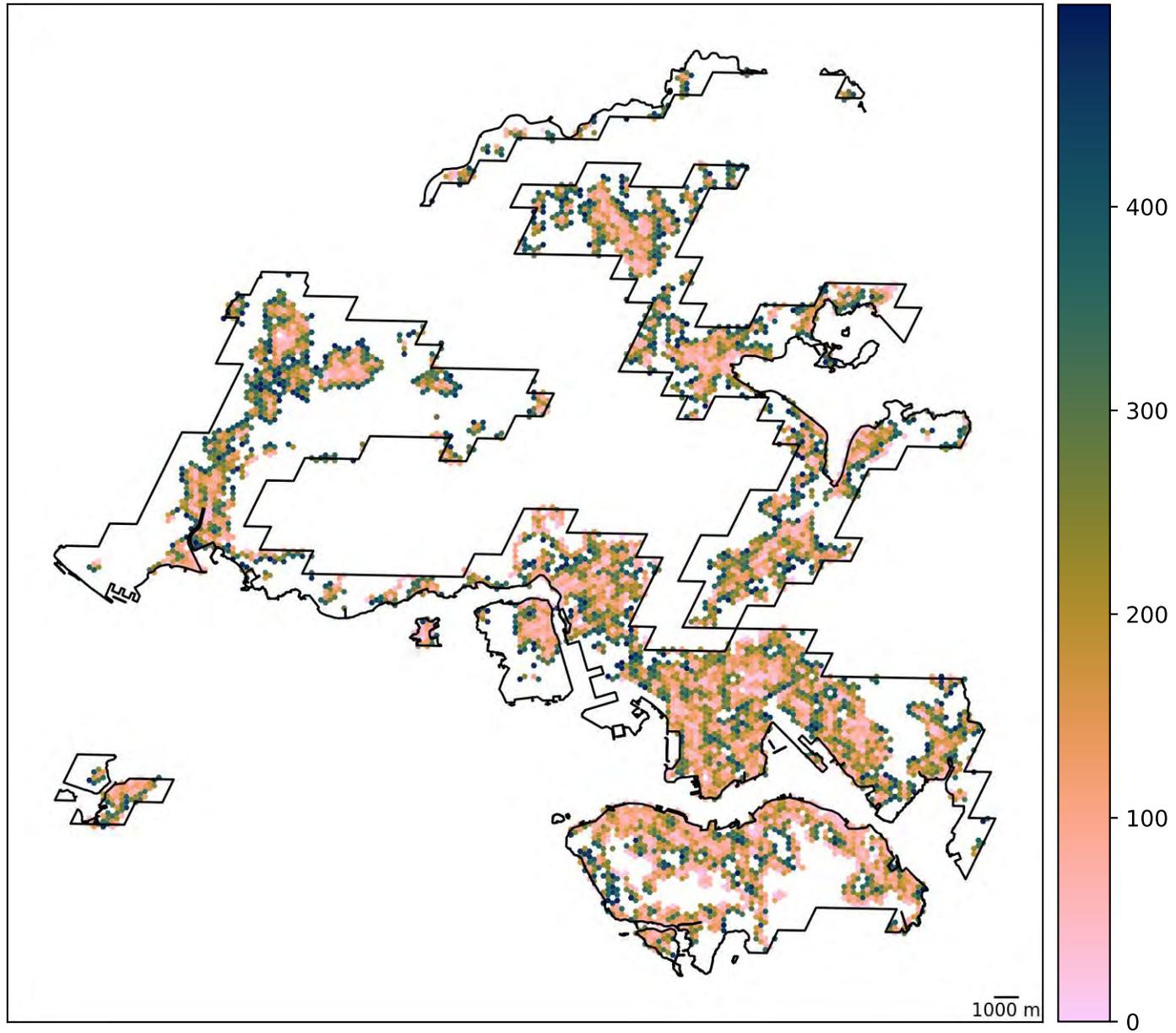

Distance to nearest park (m; up to 500m)

distances: Estimated Distance to nearest park (m; up to 500m) requirement for distances to destinations, measured up to a maximum distance target threshold of 500 metres

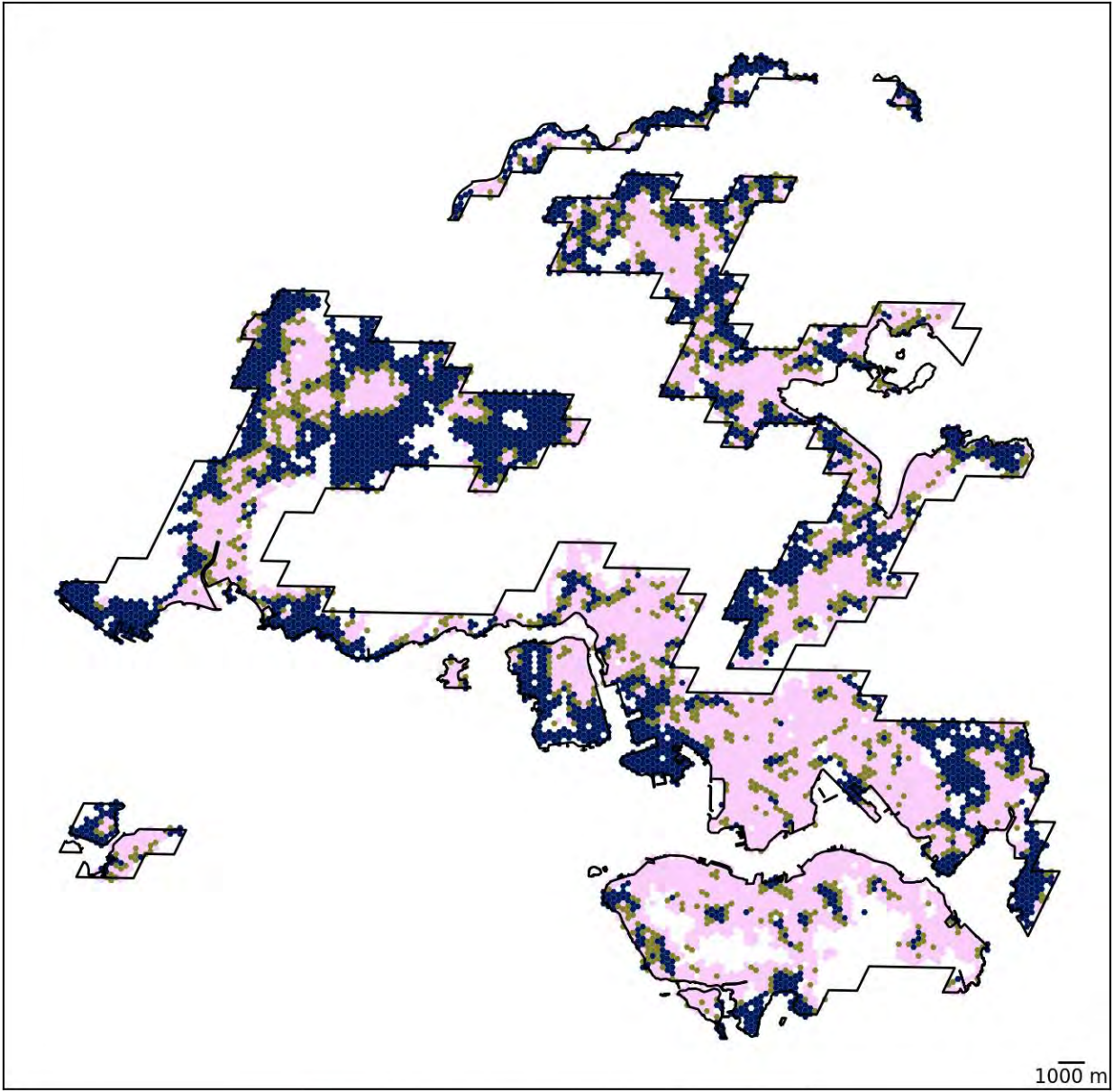

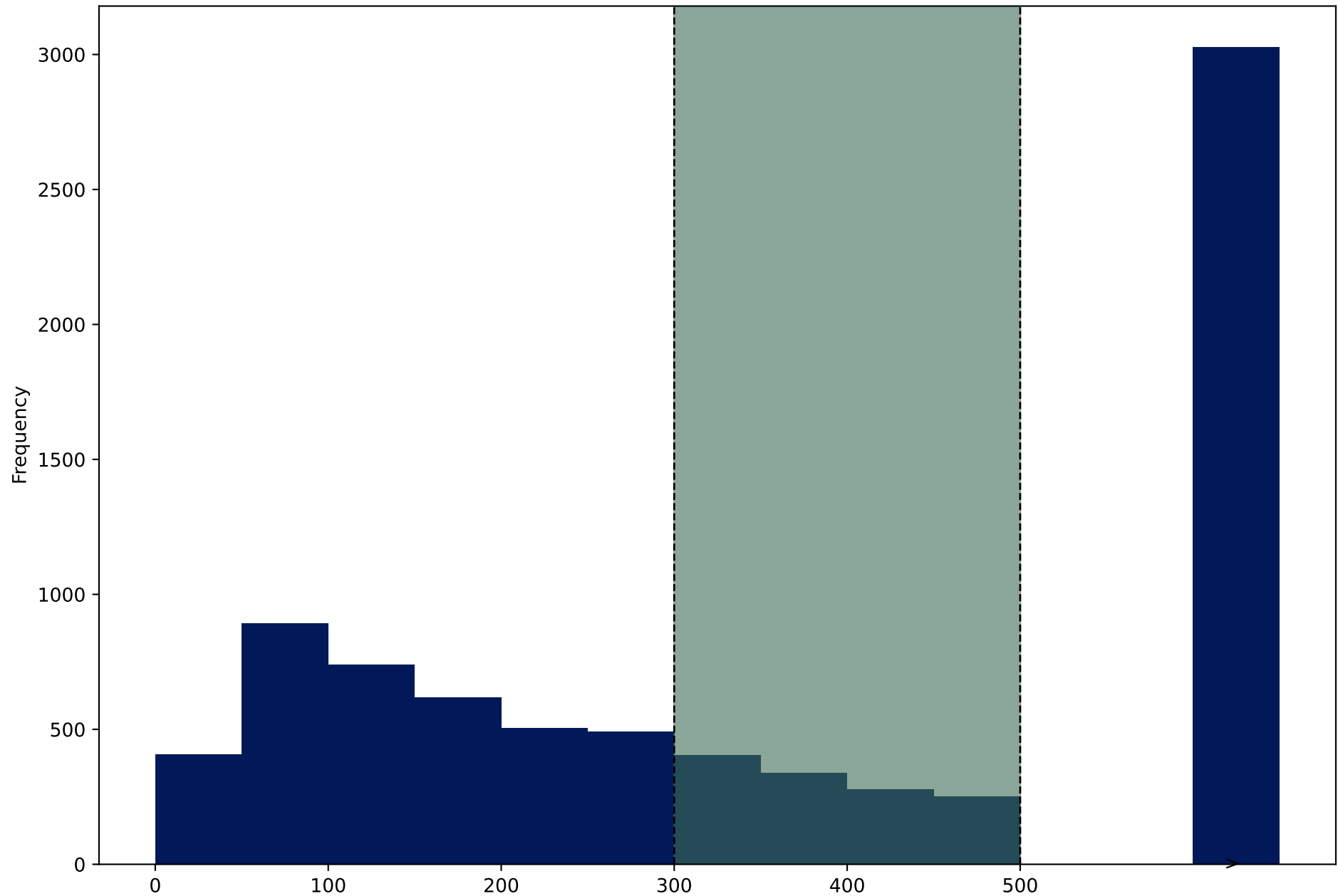

# Asia, India, Chennai

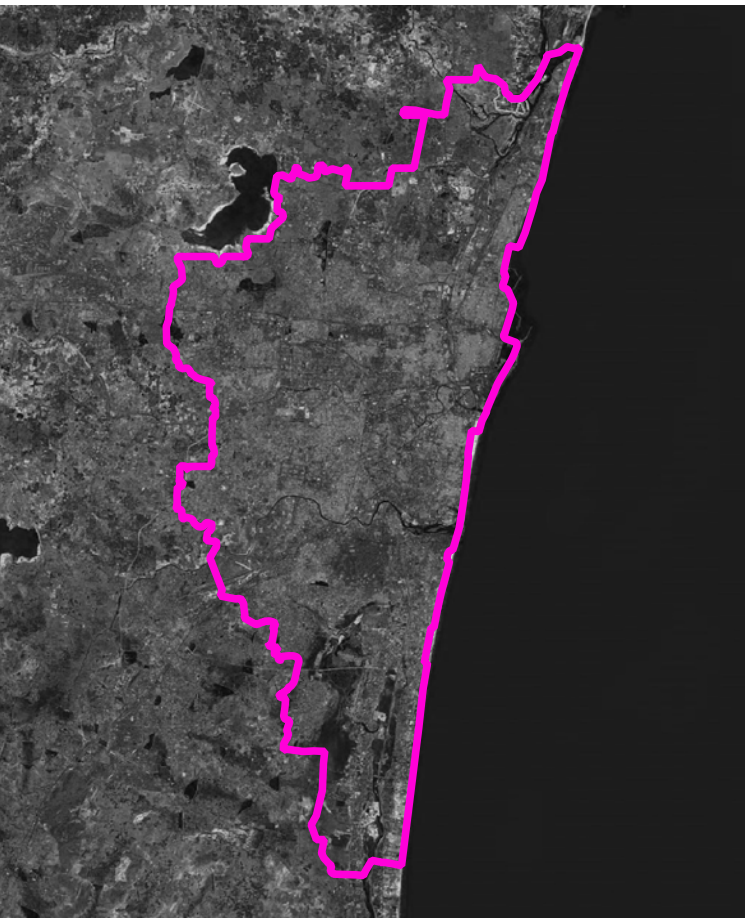
Satellite imagery of urban study region (Bing)

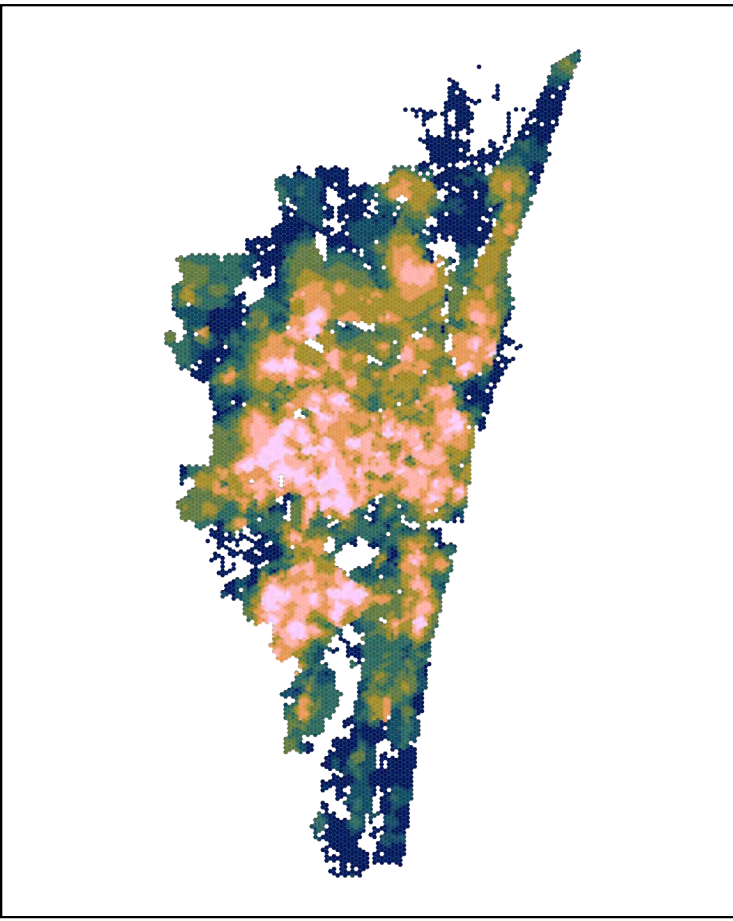
Walkability, relative to city

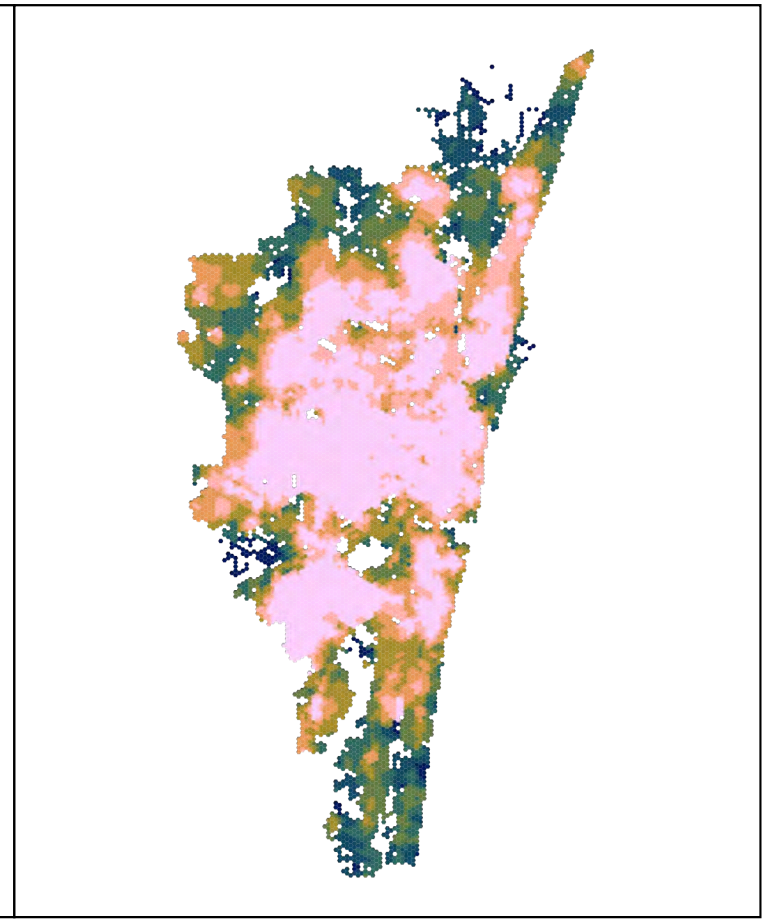
Walkability, relative to 25 global cities

▭ Urban boundary

0  10  20 km

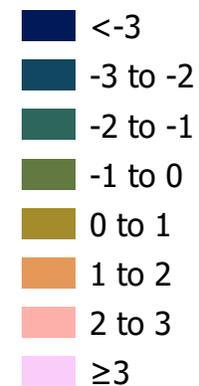

Walkability score
- <-3
- -3 to -2
- -2 to -1
- -1 to 0
- 0 to 1
- 1 to 2
- 2 to 3
- ≥3

Walkability relative to all cities by component variables (2D histograms), and overall (histogram)

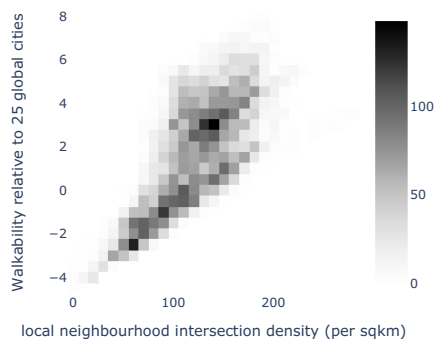
local neighbourhood intersection density (per sqkm)

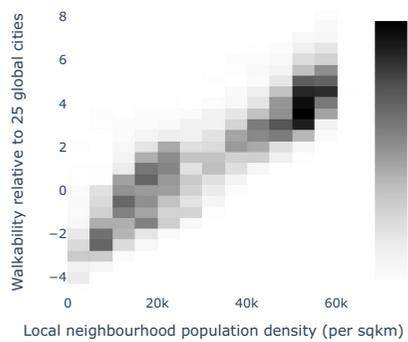
Local neighbourhood population density (per sqkm)

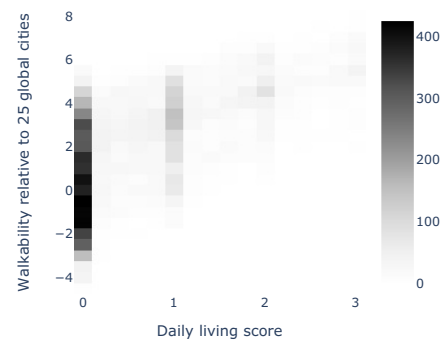
Daily living score

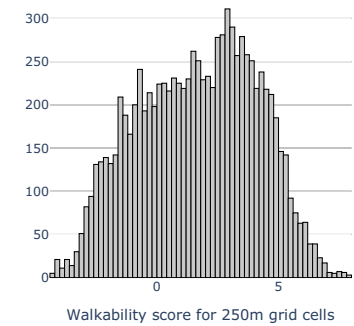
Walkability score for 250m grid cells

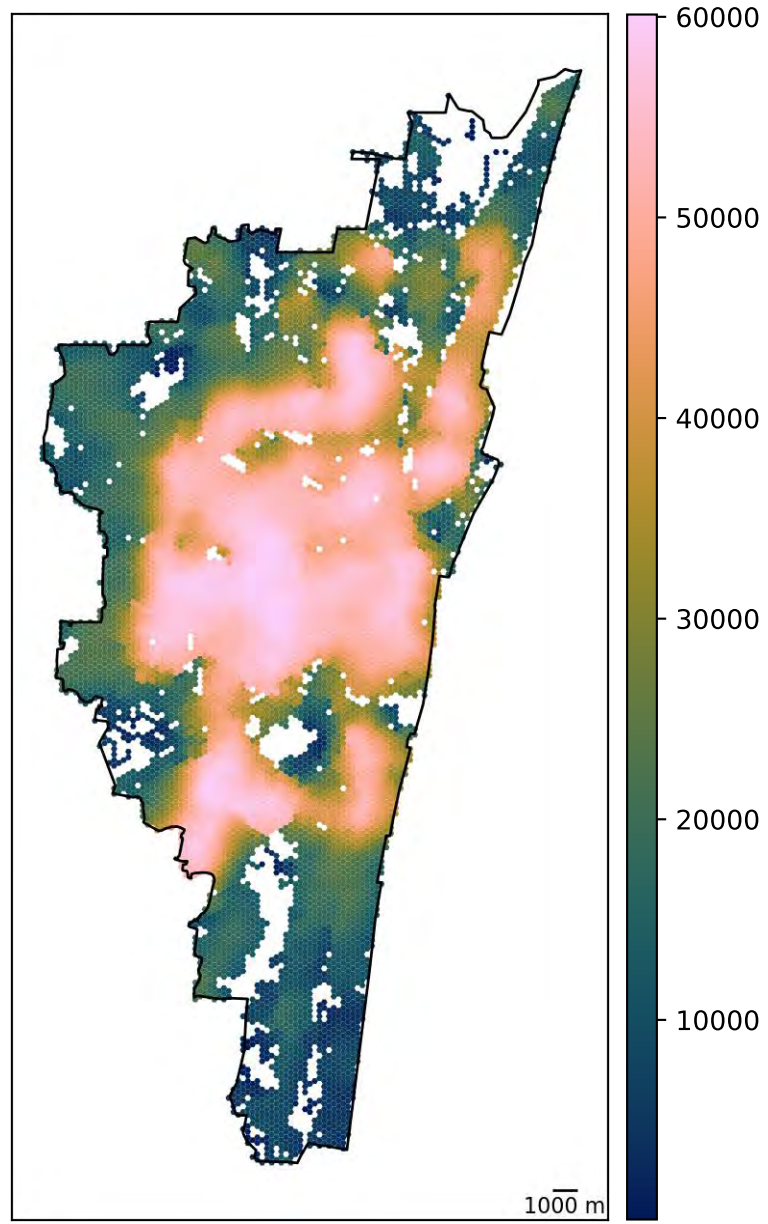

Mean 1000 m neighbourhood population per km²

A: Estimated Mean 1000 m neighbourhood population per km² requirement for ≥80% probability of engaging in walking for transport

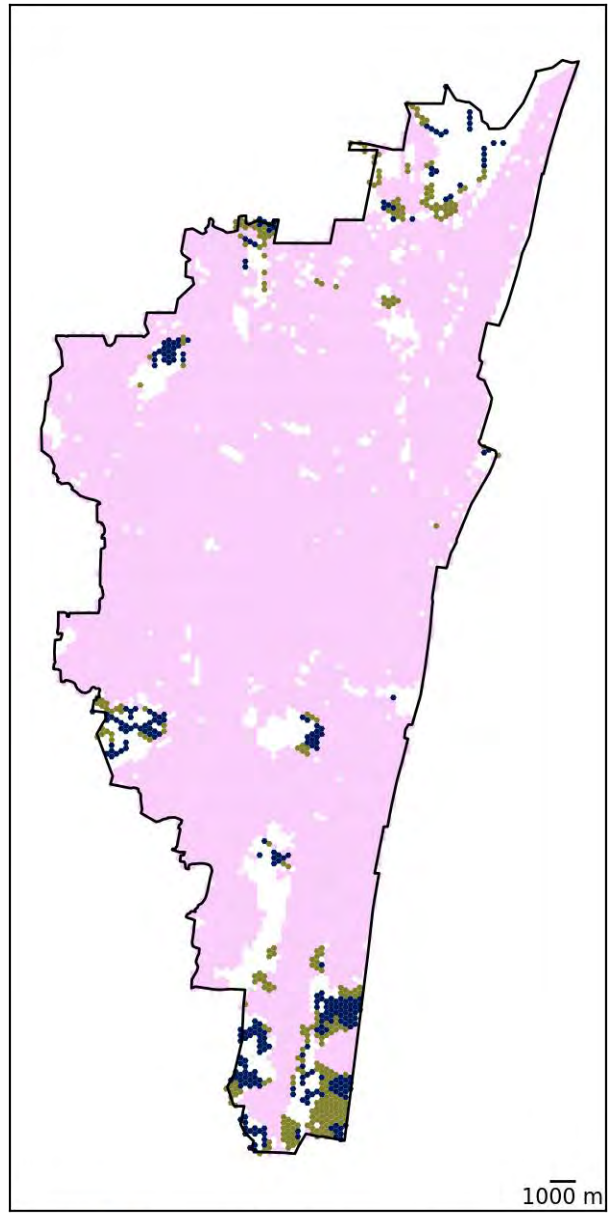

B: Estimated Mean 1000 m neighbourhood population per km² requirement for reaching the WHO's target of a ≥15% relative reduction in insufficient physical activity through walking

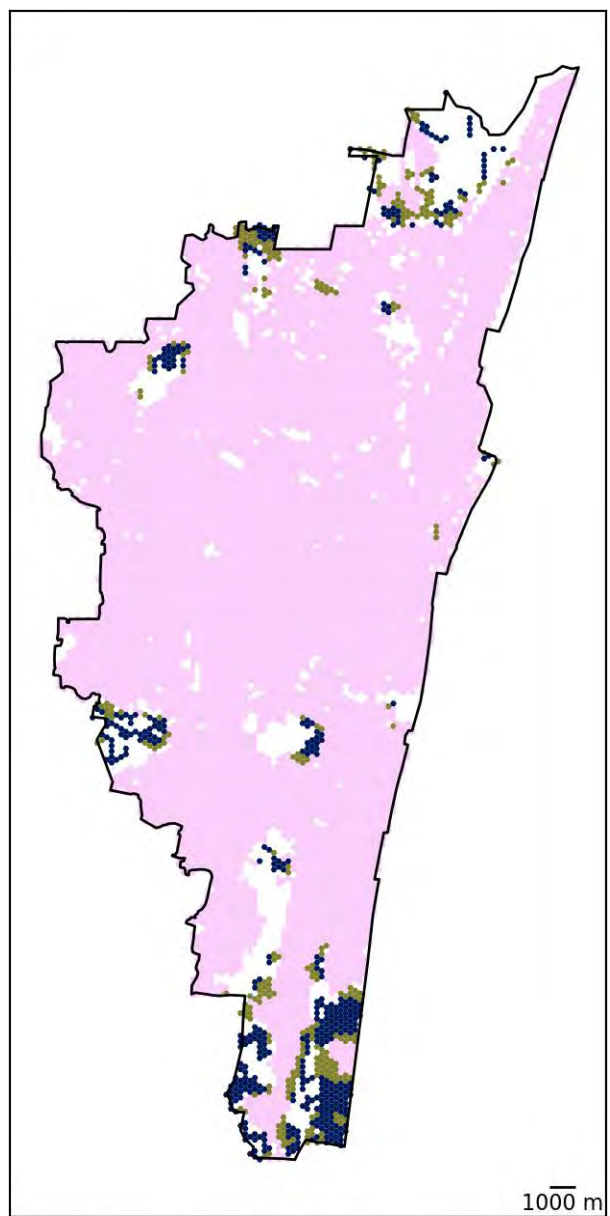

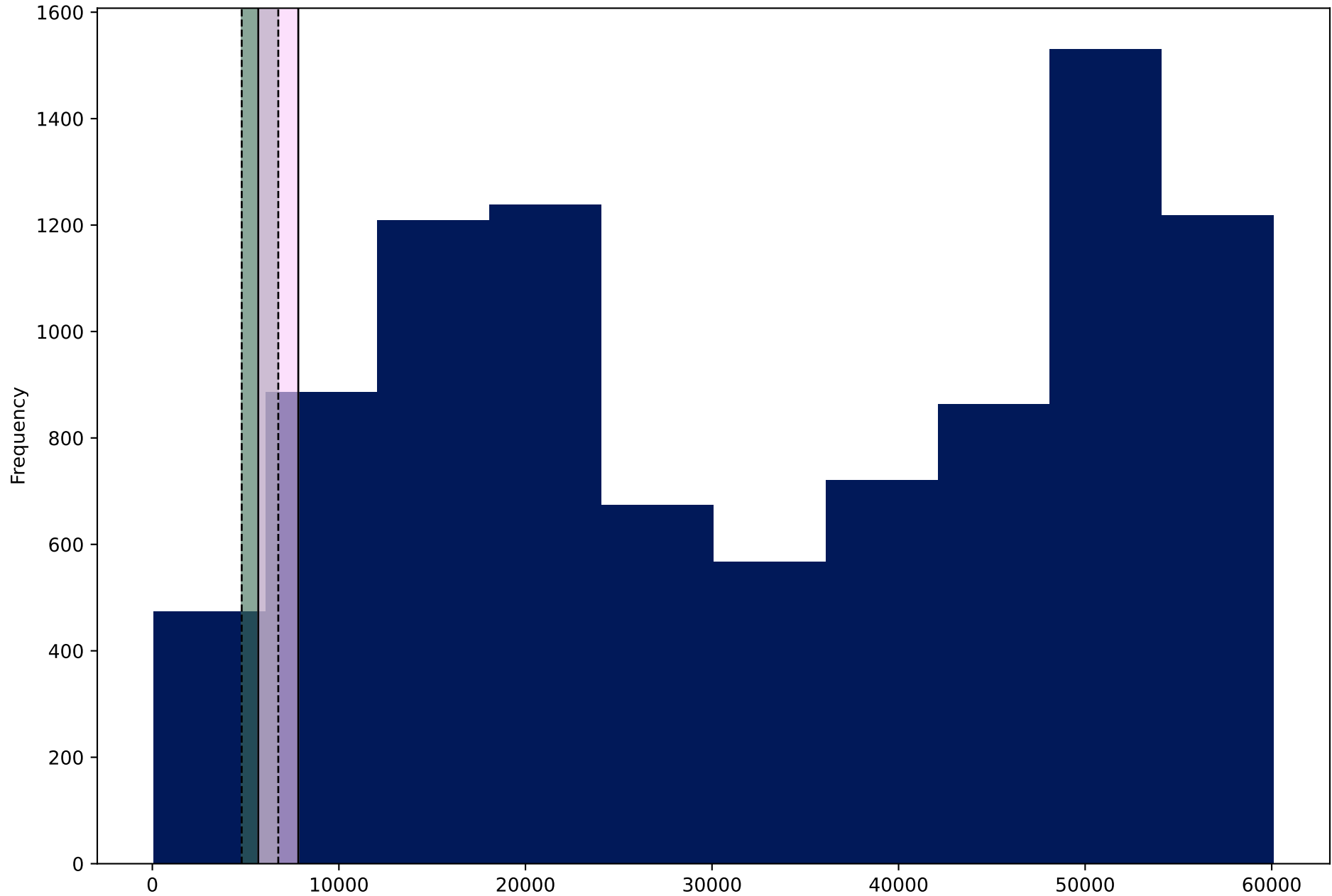

Mean 1000 m neighbourhood street intersections per km²

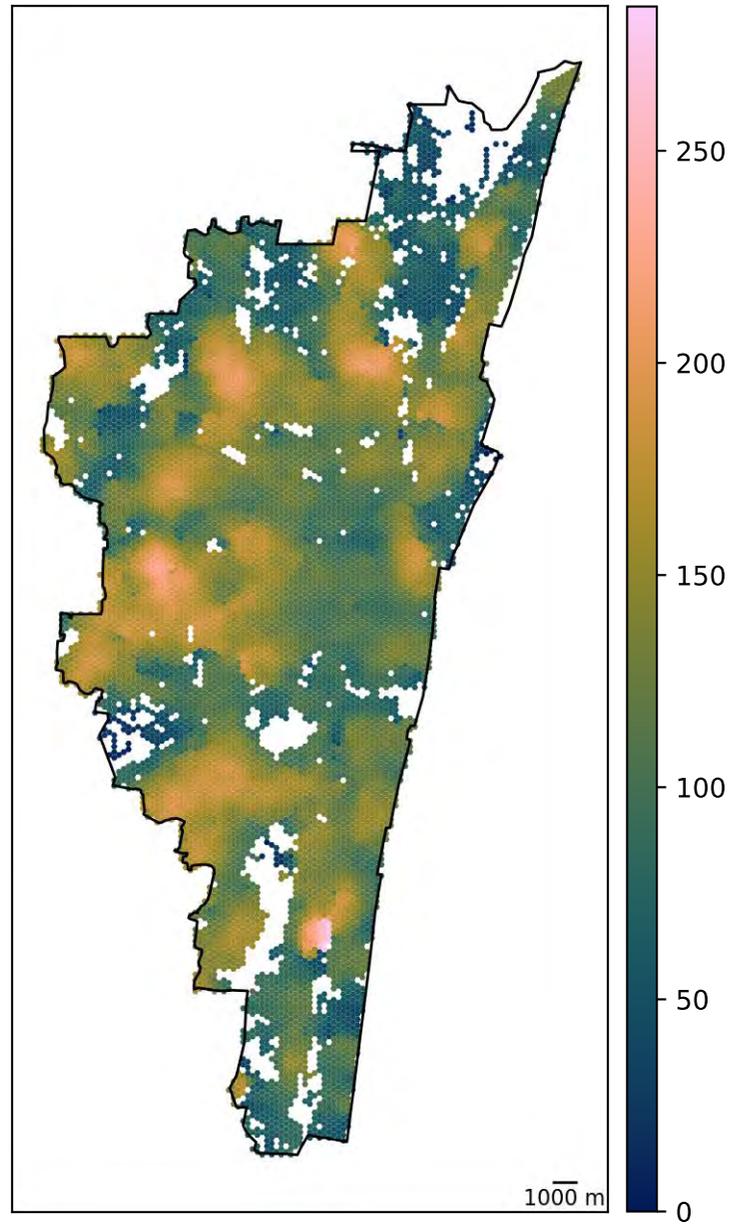

A: Estimated Mean 1000 m neighbourhood street intersections per km² requirement for ≥80% probability of engaging in walking for transport

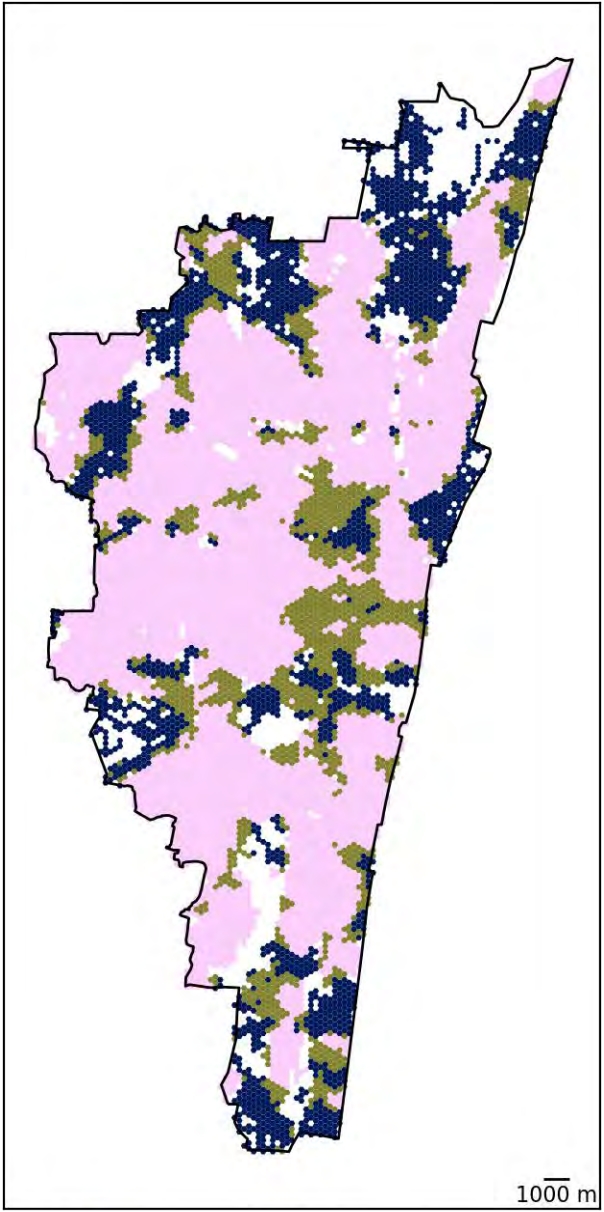

- below 95% CrI lower bound
- within 95% CrI (90, 110)
- exceeds 95% CrI upper bound

B: Estimated Mean 1000 m neighbourhood street intersections per km² requirement for reaching the WHO's target of a ≥15% relative reduction in insufficient physical activity through walking

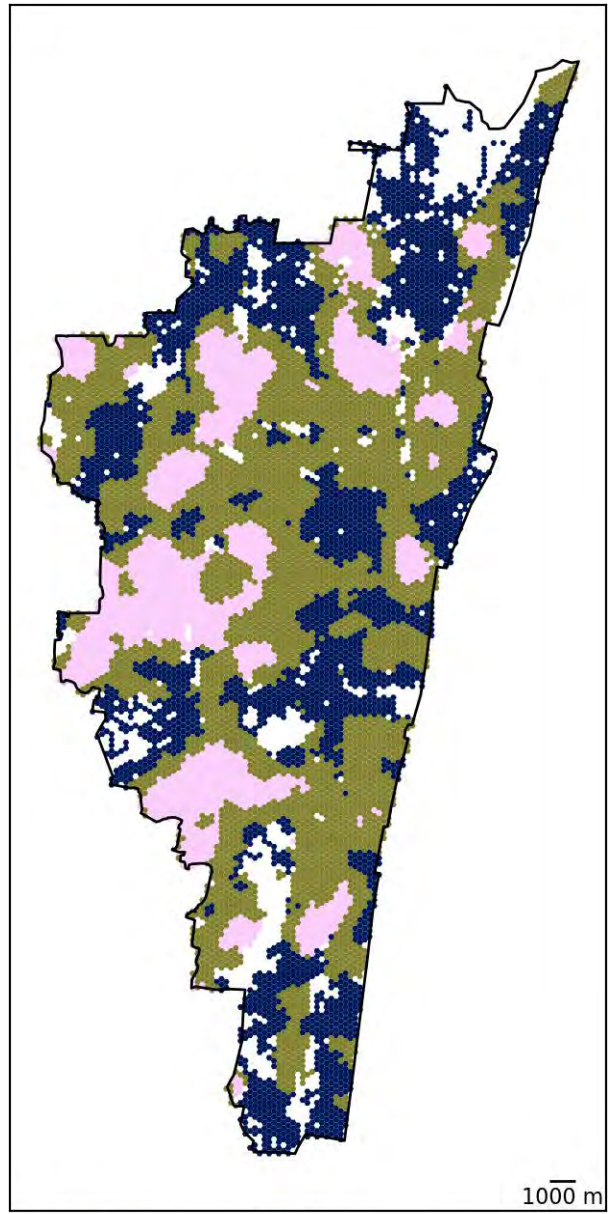

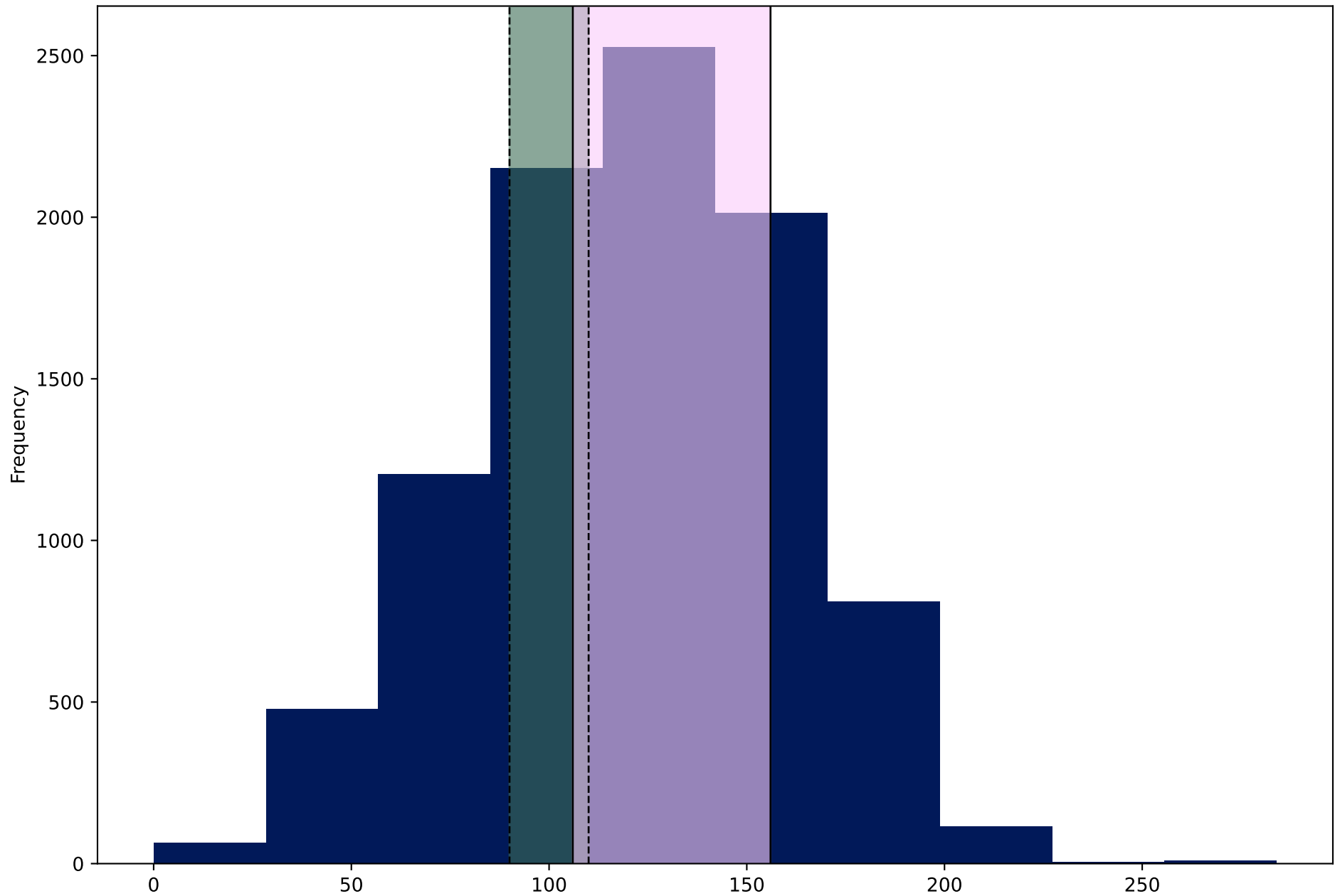

# Distance to nearest public transport stops (m; up to 500m)

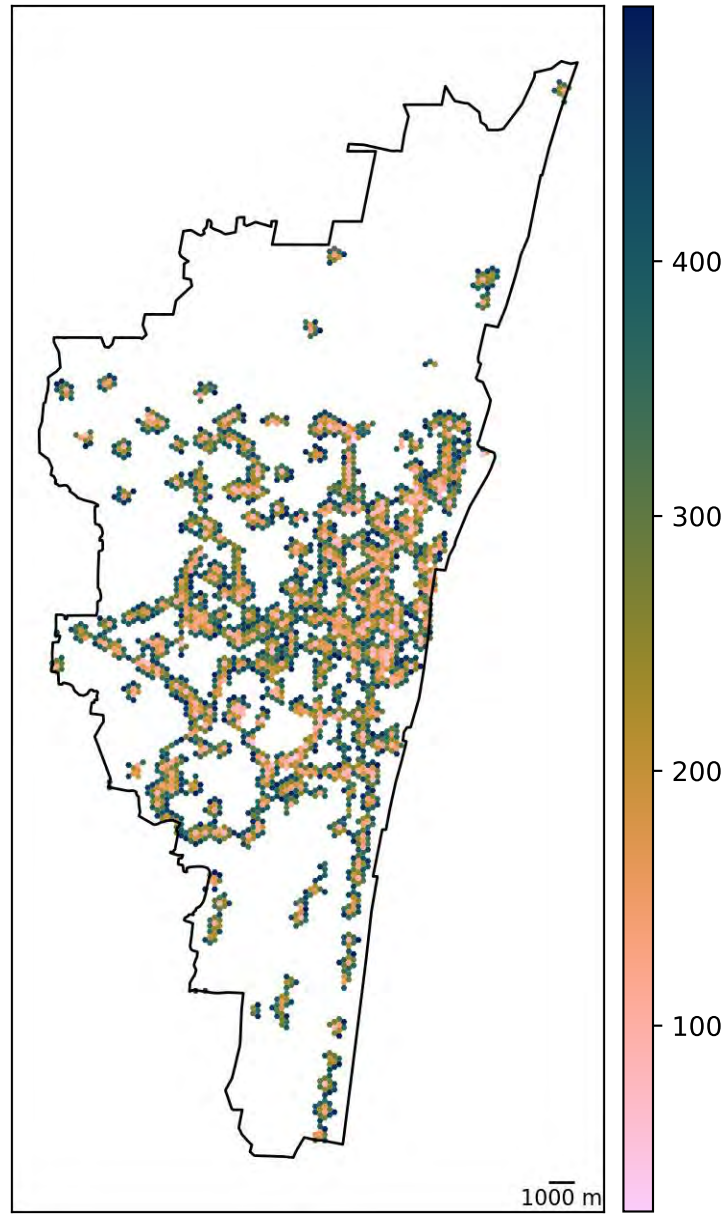

distances: Estimated Distance to nearest public transport stops (m; up to 500m) requirement for distances to destinations, measured up to a maximum distance target threshold of 500 metres

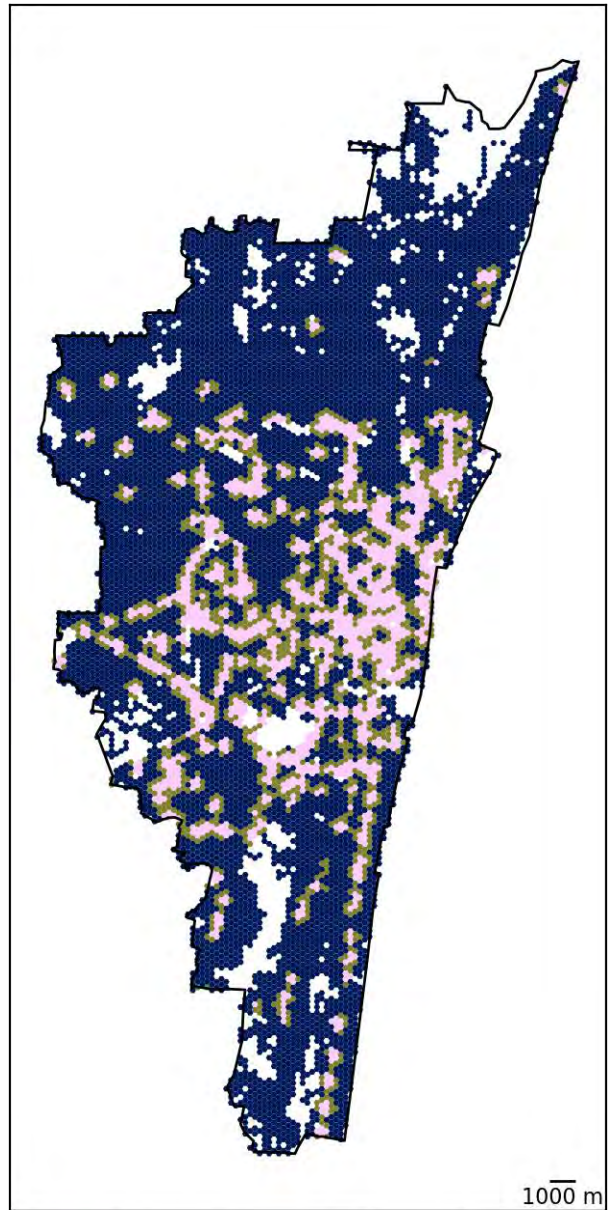

below distance (m) lower bound
within distance (m) (300, 500)
exceeds distance (m) upper bound

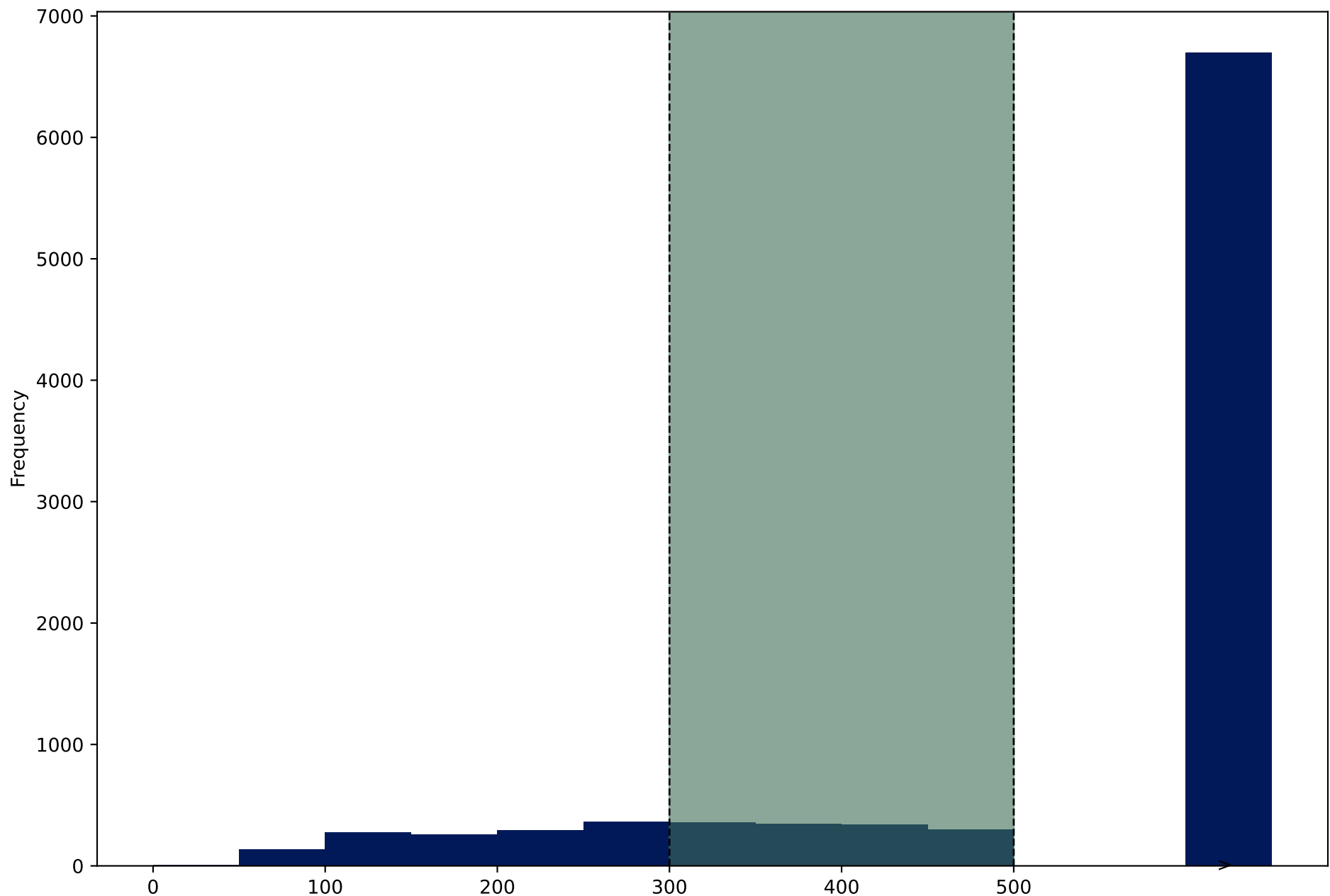

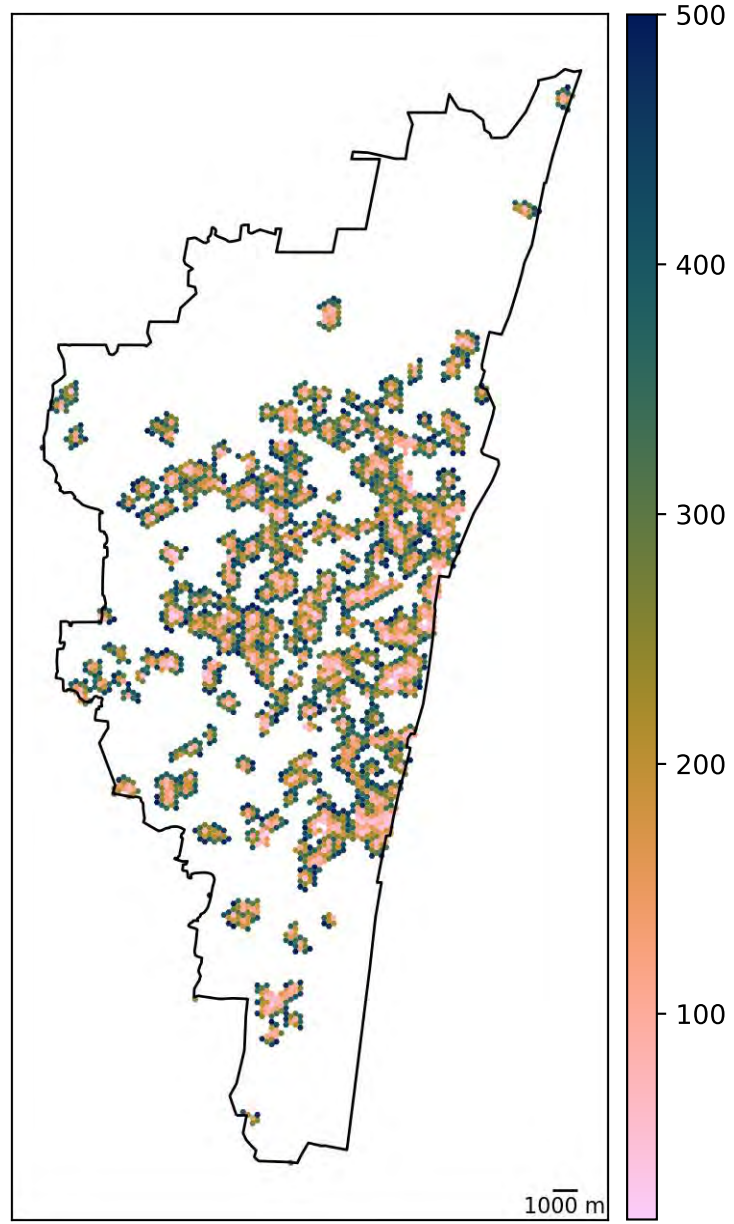

Distance to nearest park (m; up to 500m)

distances: Estimated Distance to nearest park (m; up to 500m) requirement for distances to destinations, measured up to a maximum distance target threshold of 500 metres

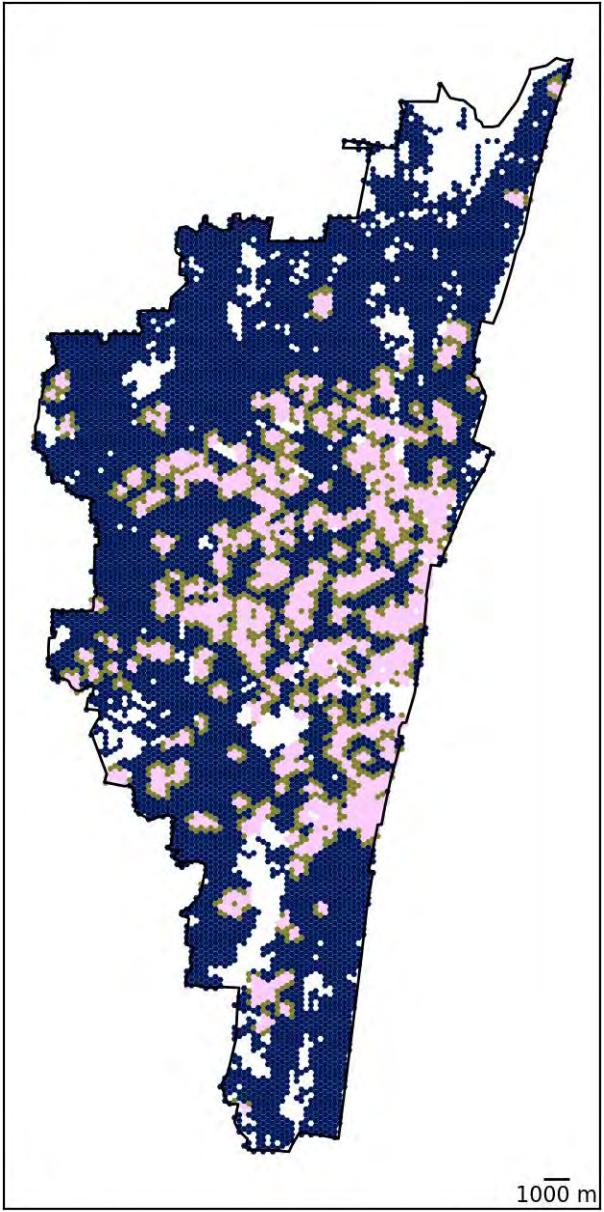

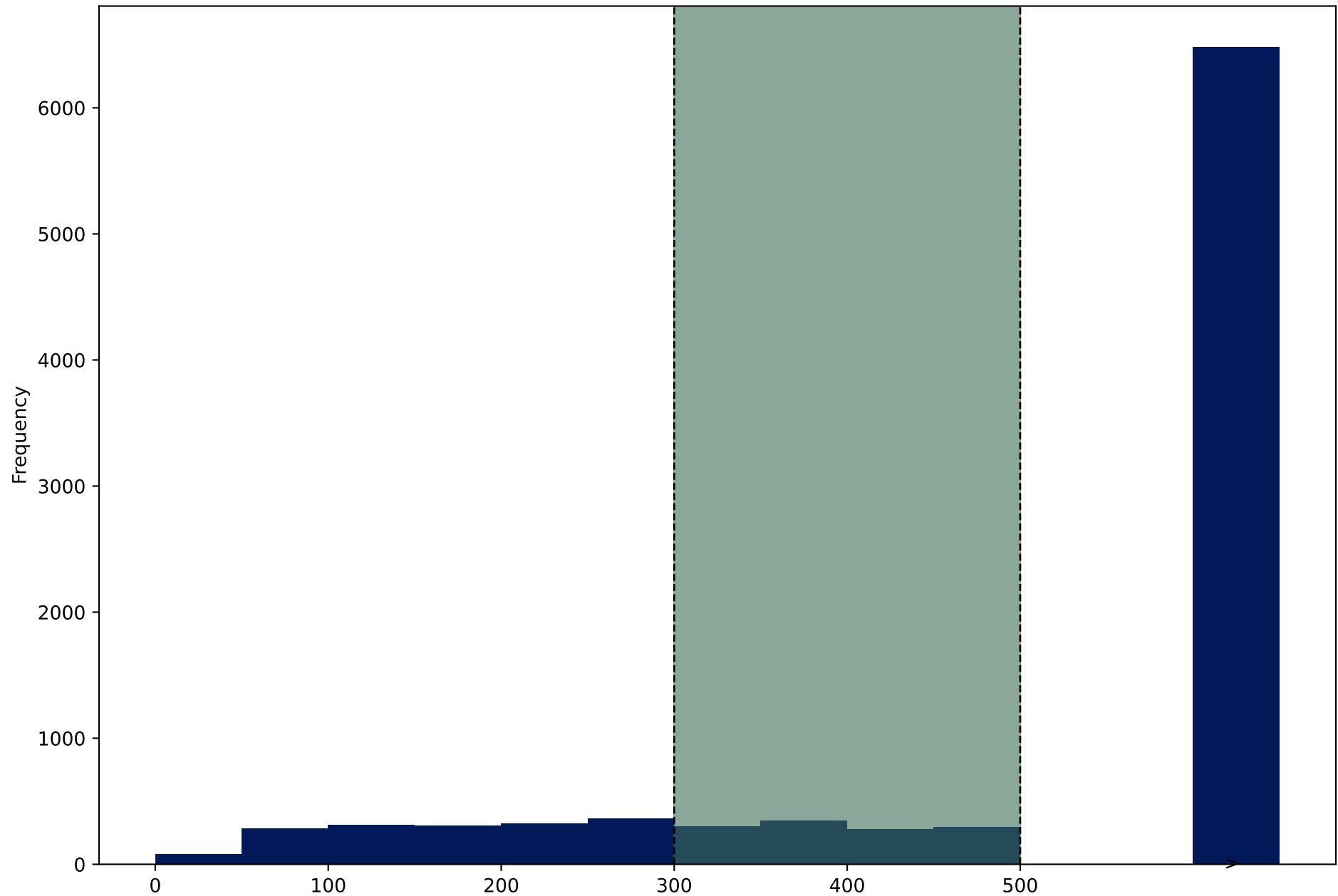

# Asia, Thailand, Bangkok

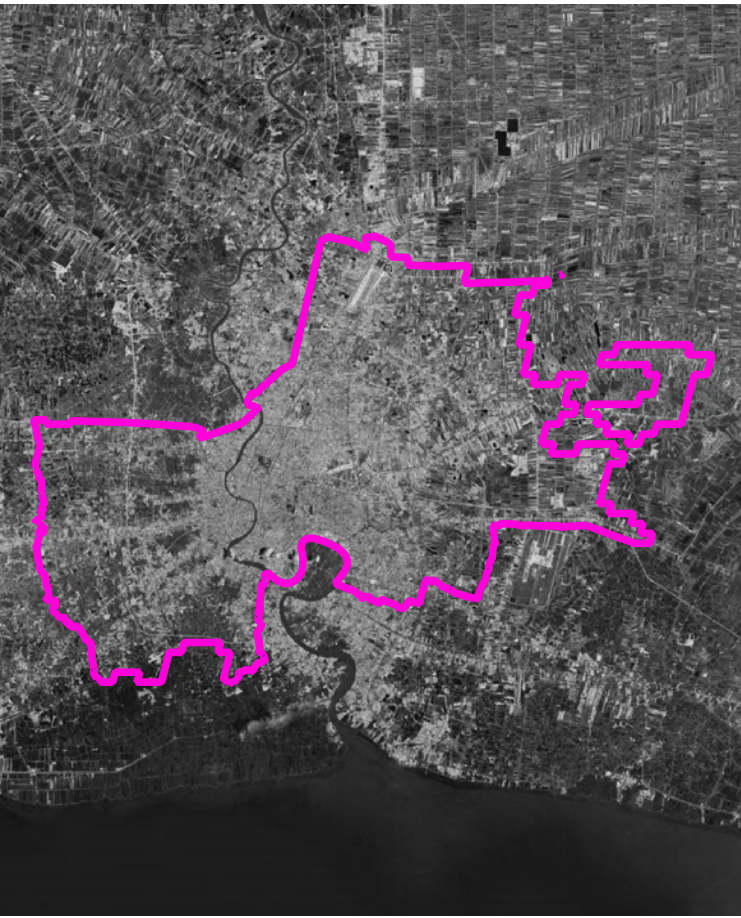
Satellite imagery of urban study region (Bing)

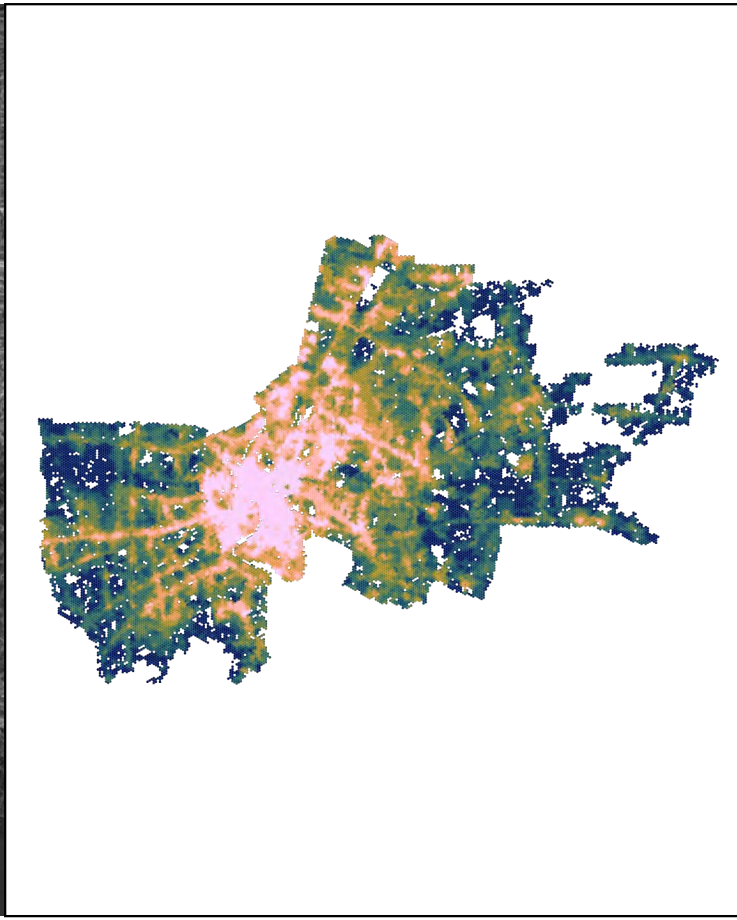
Walkability, relative to city

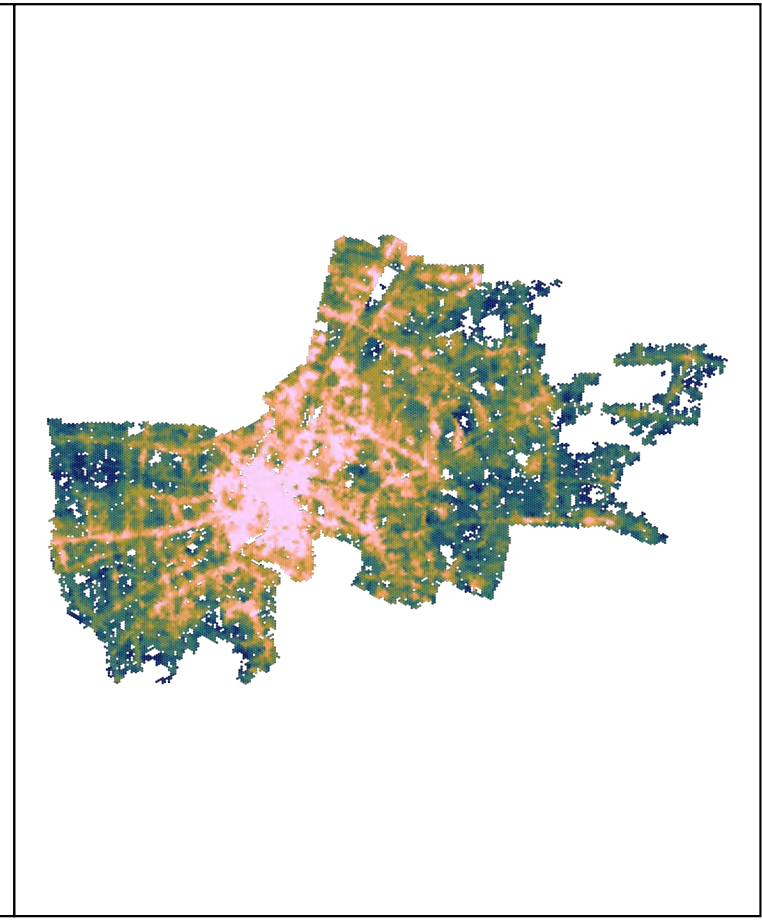
Walkability, relative to 25 global cities

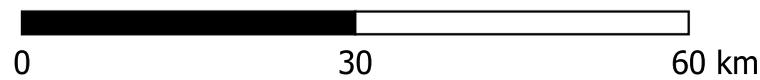

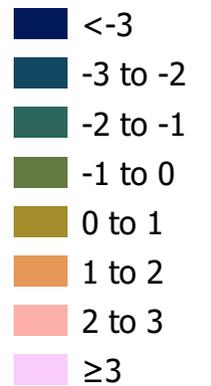
Walkability score

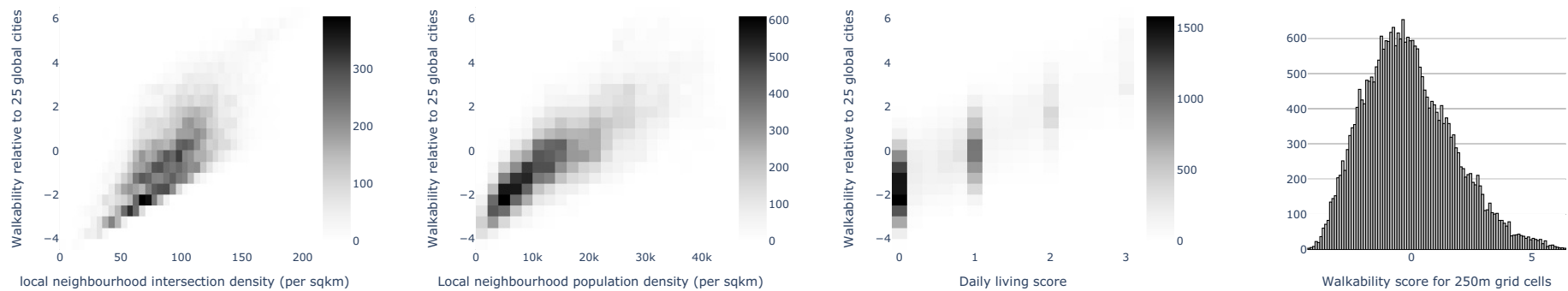
Walkability relative to all cities by component variables (2D histograms), and overall (histogram)

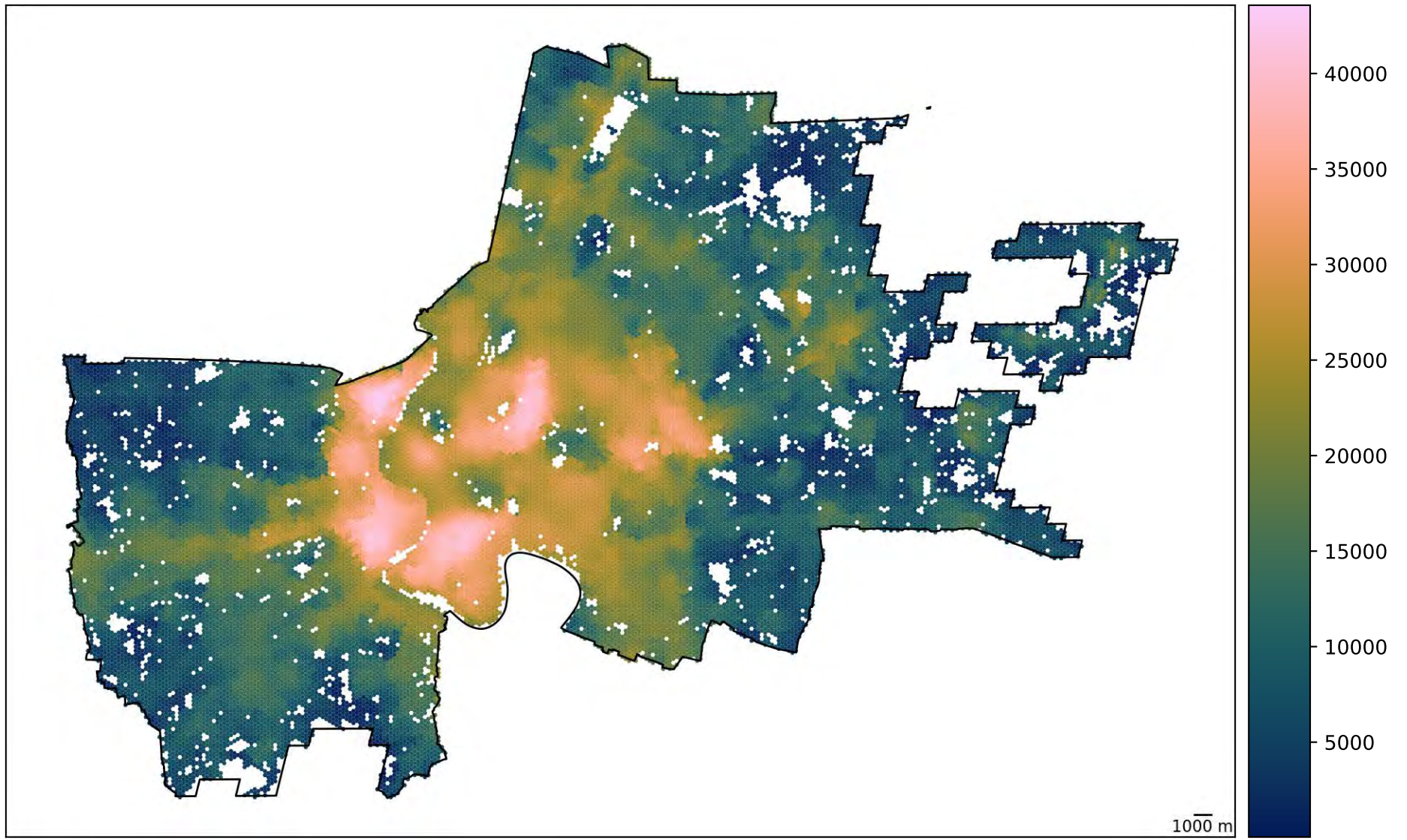

Mean 1000 m neighbourhood population per km²

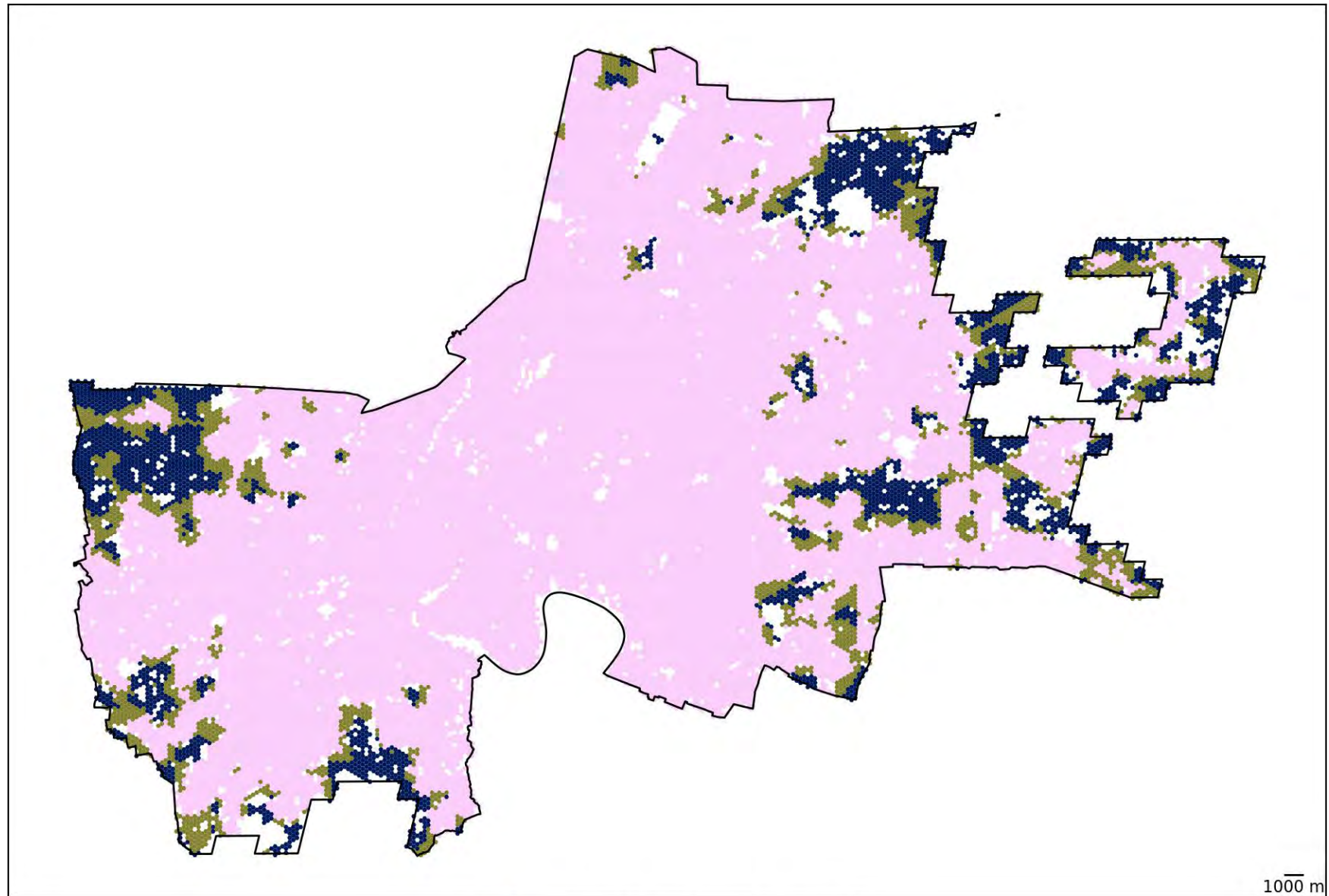

A: Estimated Mean 1000 m neighbourhood population per km² requirement for ≥80% probability of engaging in walking for transport

B: Estimated Mean 1000 m neighbourhood population per km² requirement for reaching the WHO's target of a ≥15% relative reduction in insufficient physical activity through walking

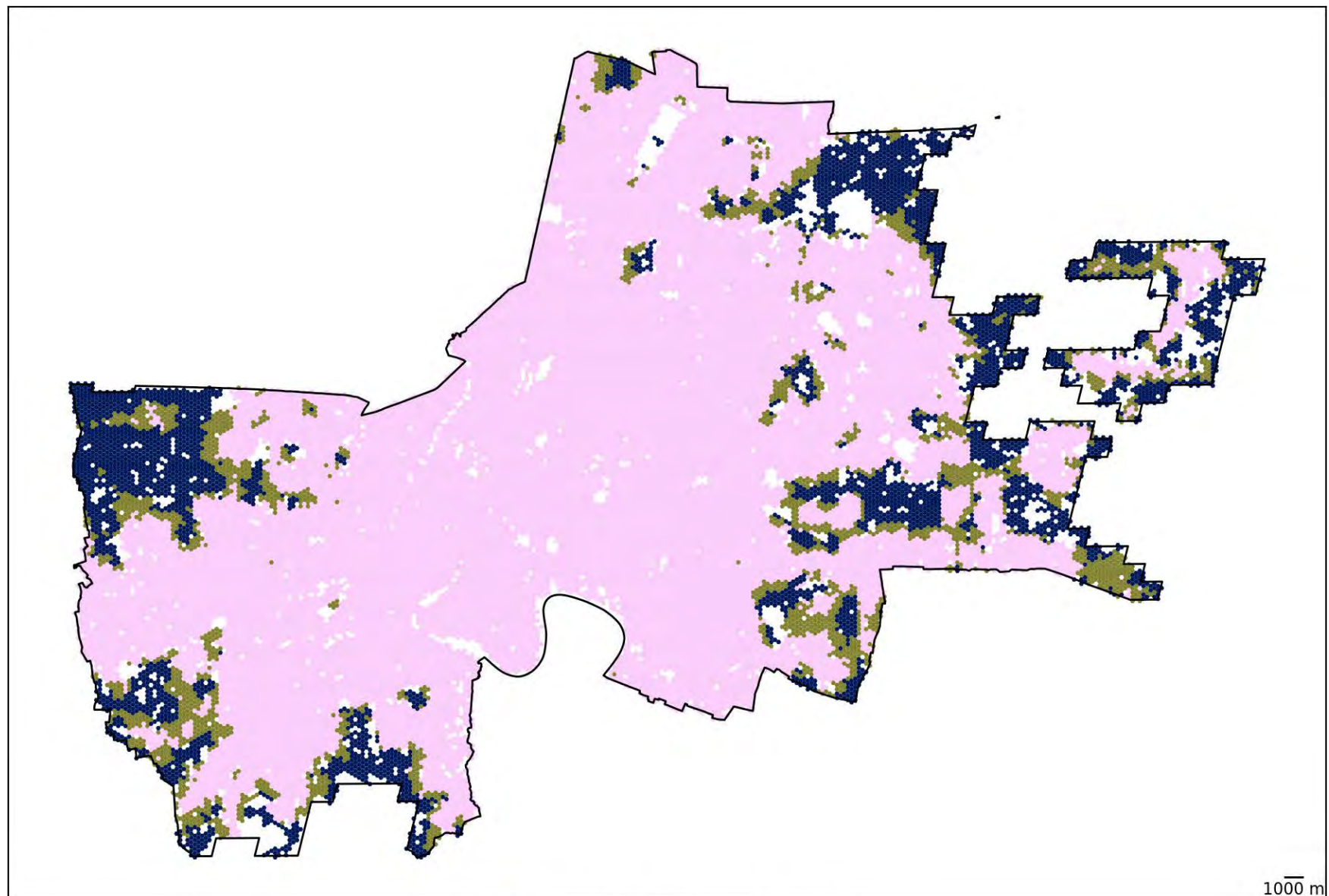

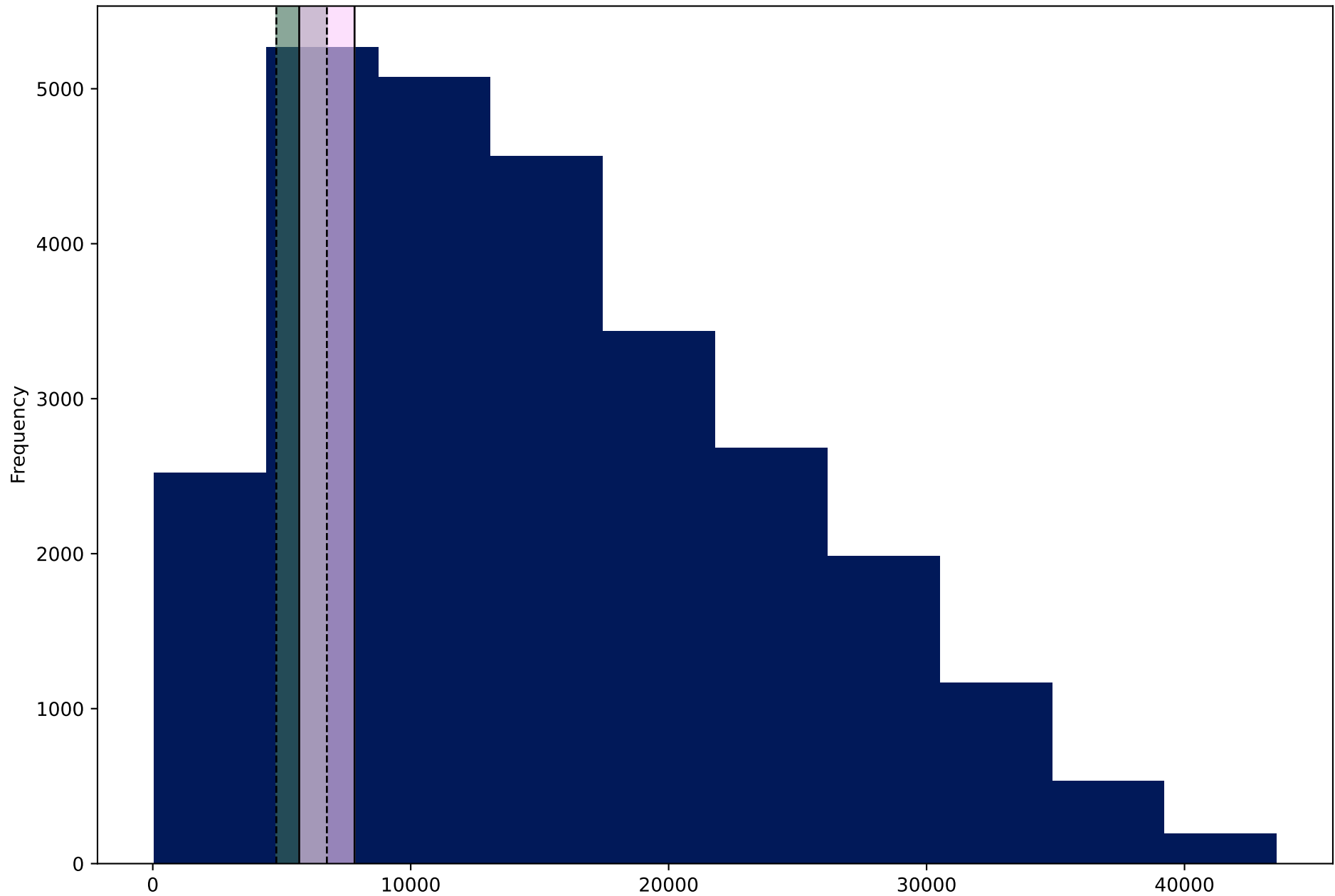

# Asia, Thailand, Bangkok

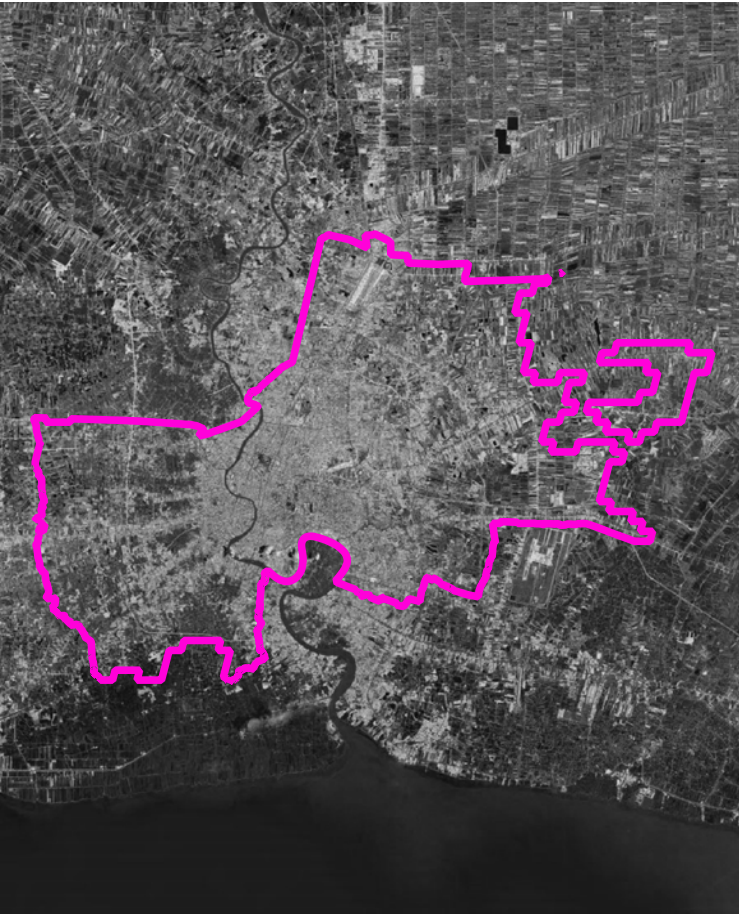
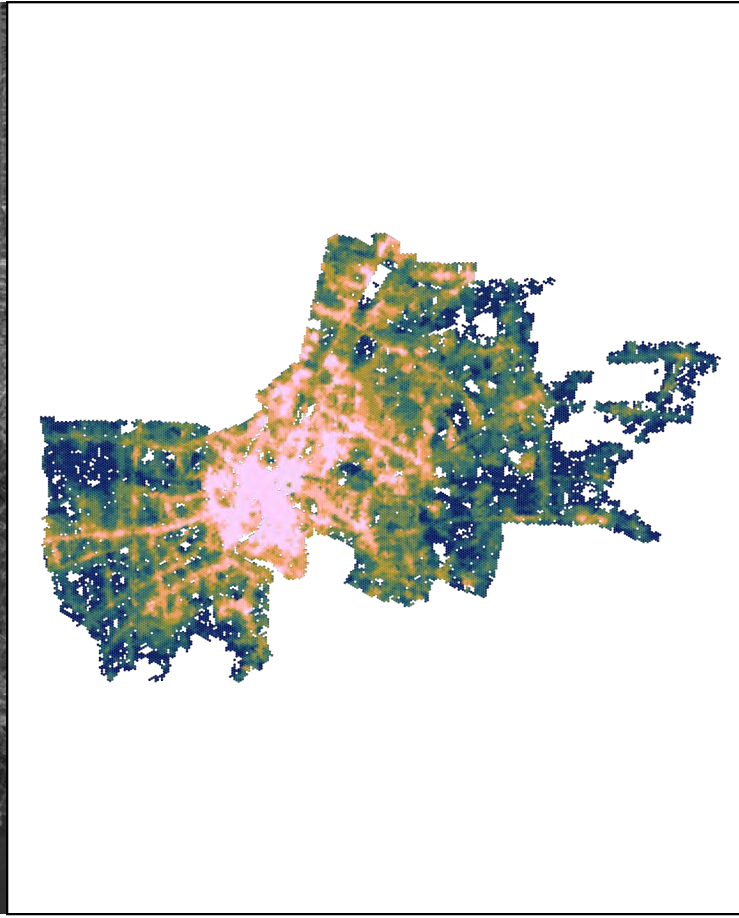
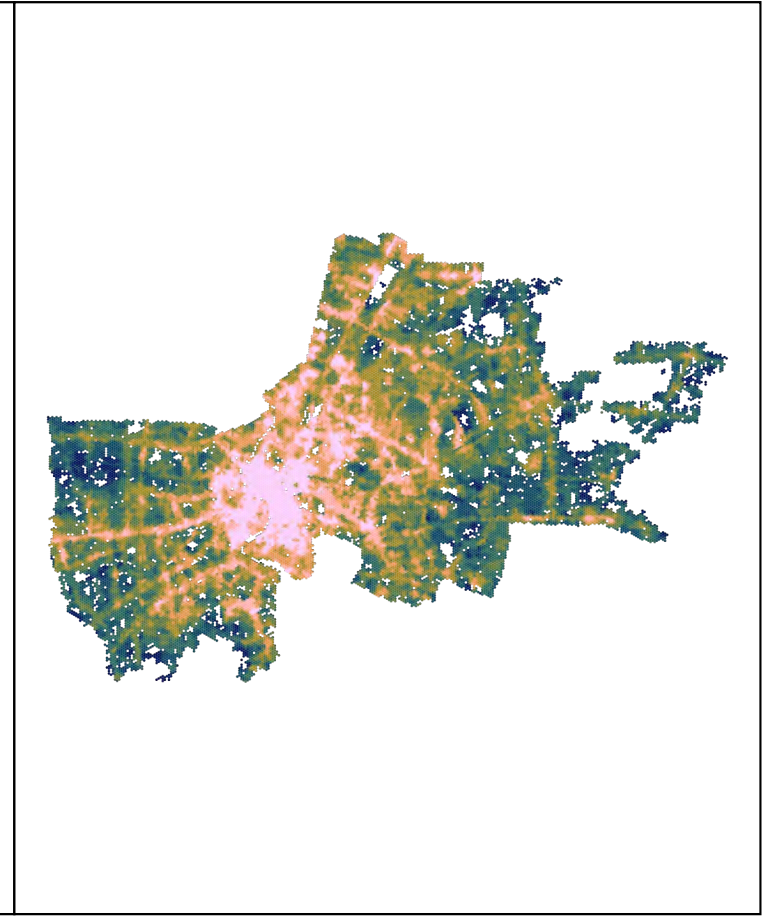
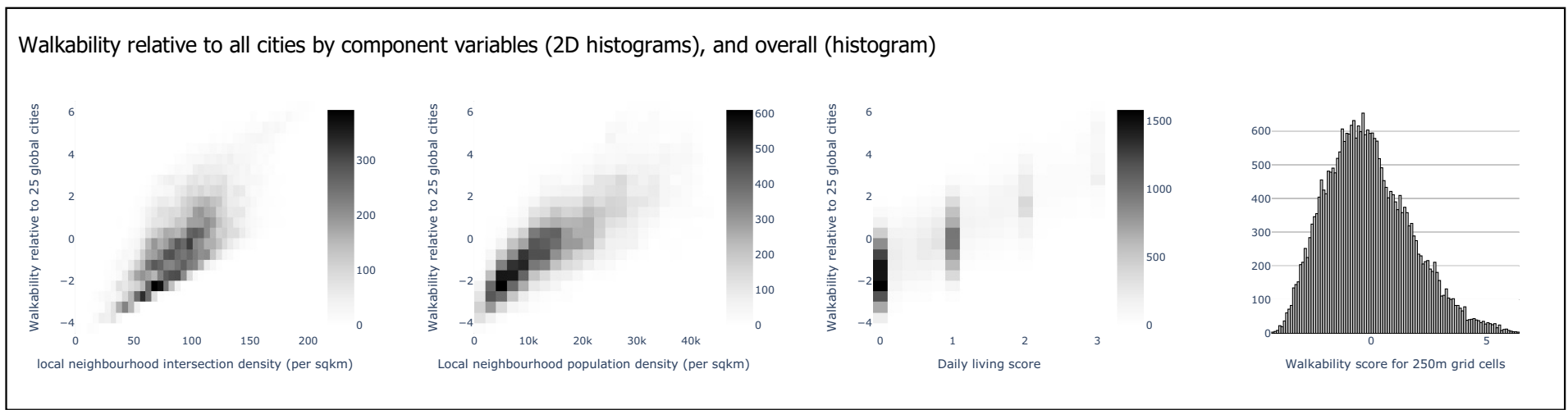

Mean 1000 m neighbourhood street intersections per km²

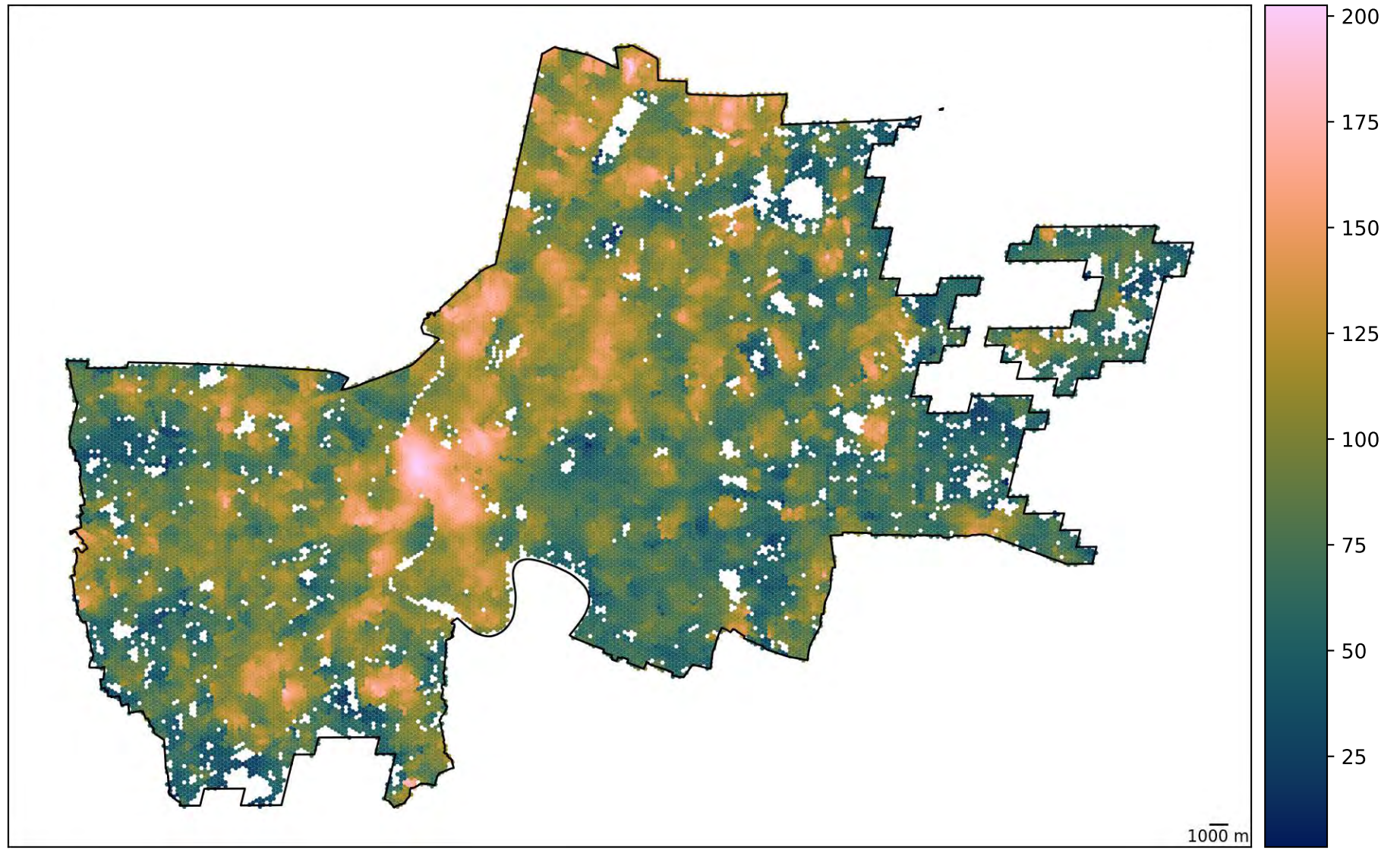

A: Estimated Mean 1000 m neighbourhood street intersections per km² requirement for ≥80% probability of engaging in walking for transport

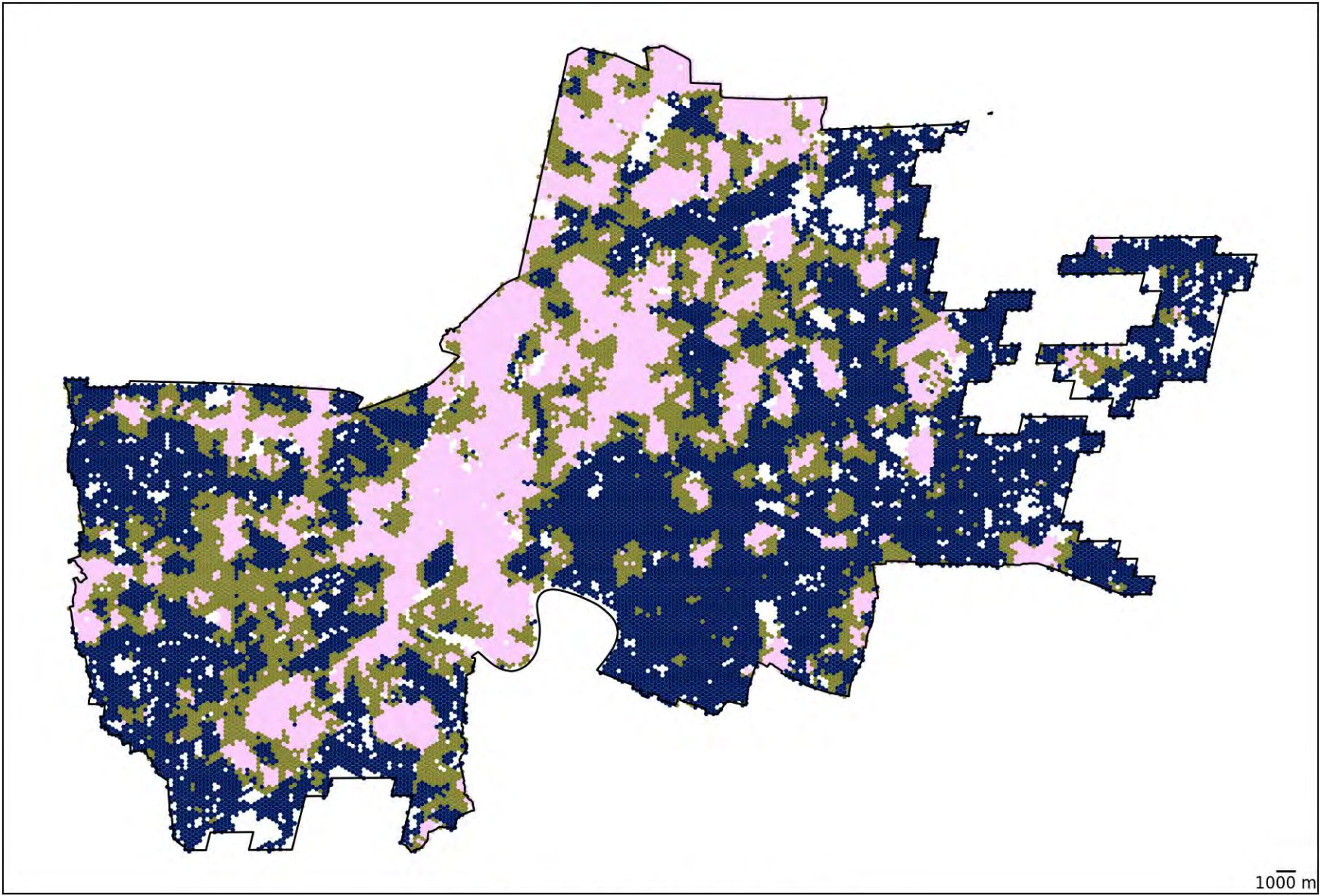

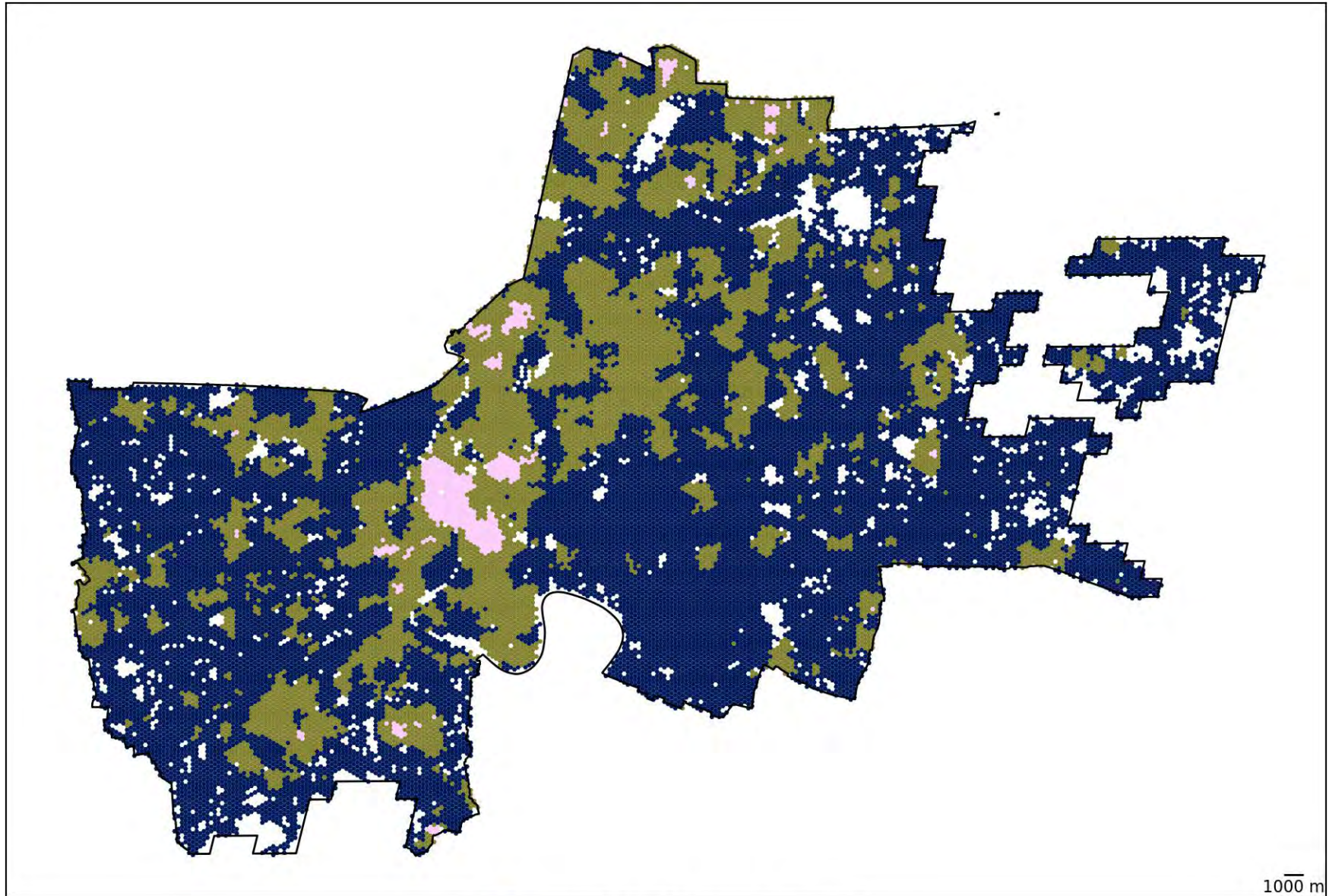

B: Estimated Mean 1000 m neighbourhood street intersections per km² requirement for reaching the WHO's target of a ≥15% relative reduction in insufficient physical activity through walking

- below 95% CrI lower bound
- within 95% CrI (106, 156)
- exceeds 95% CrI upper bound

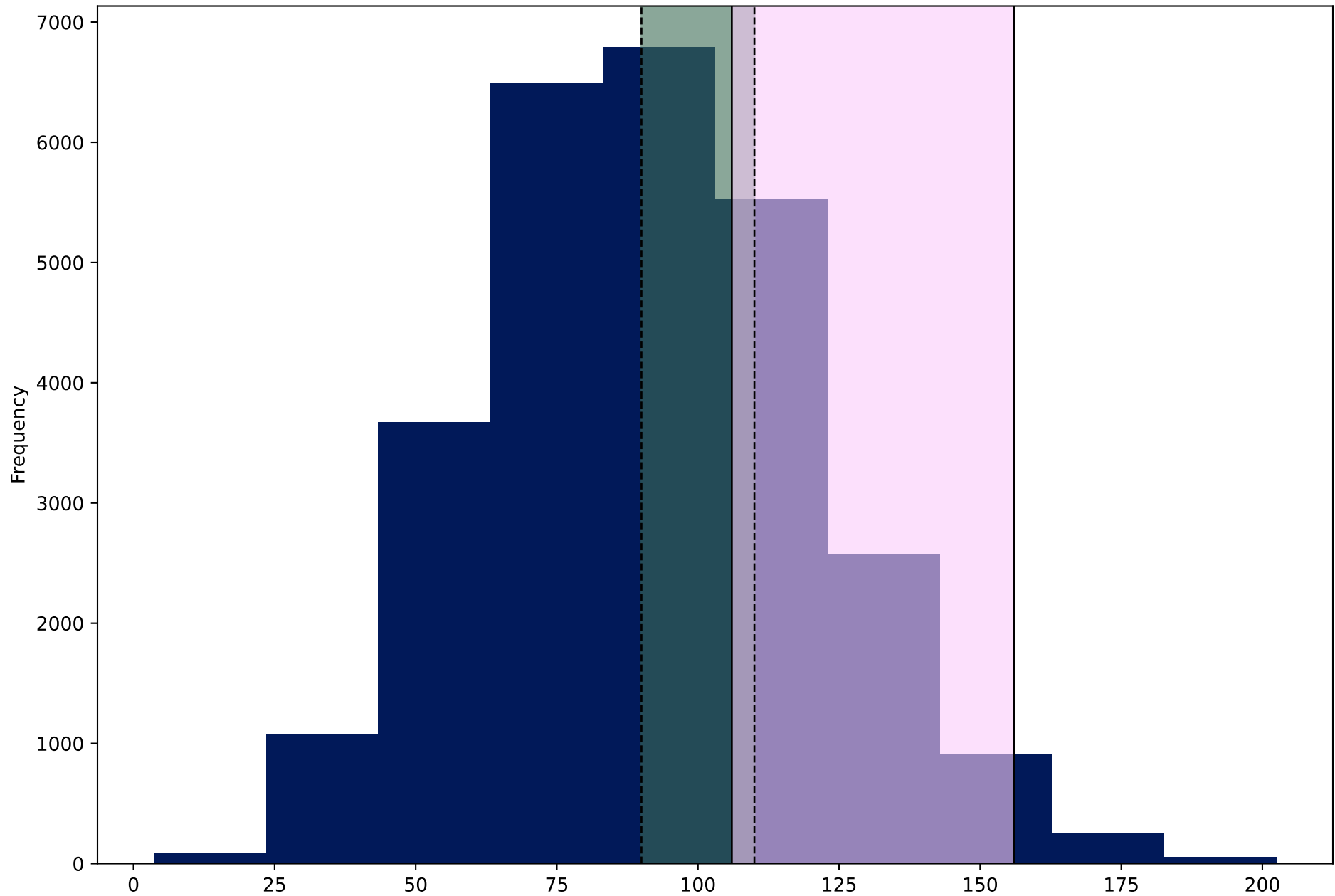

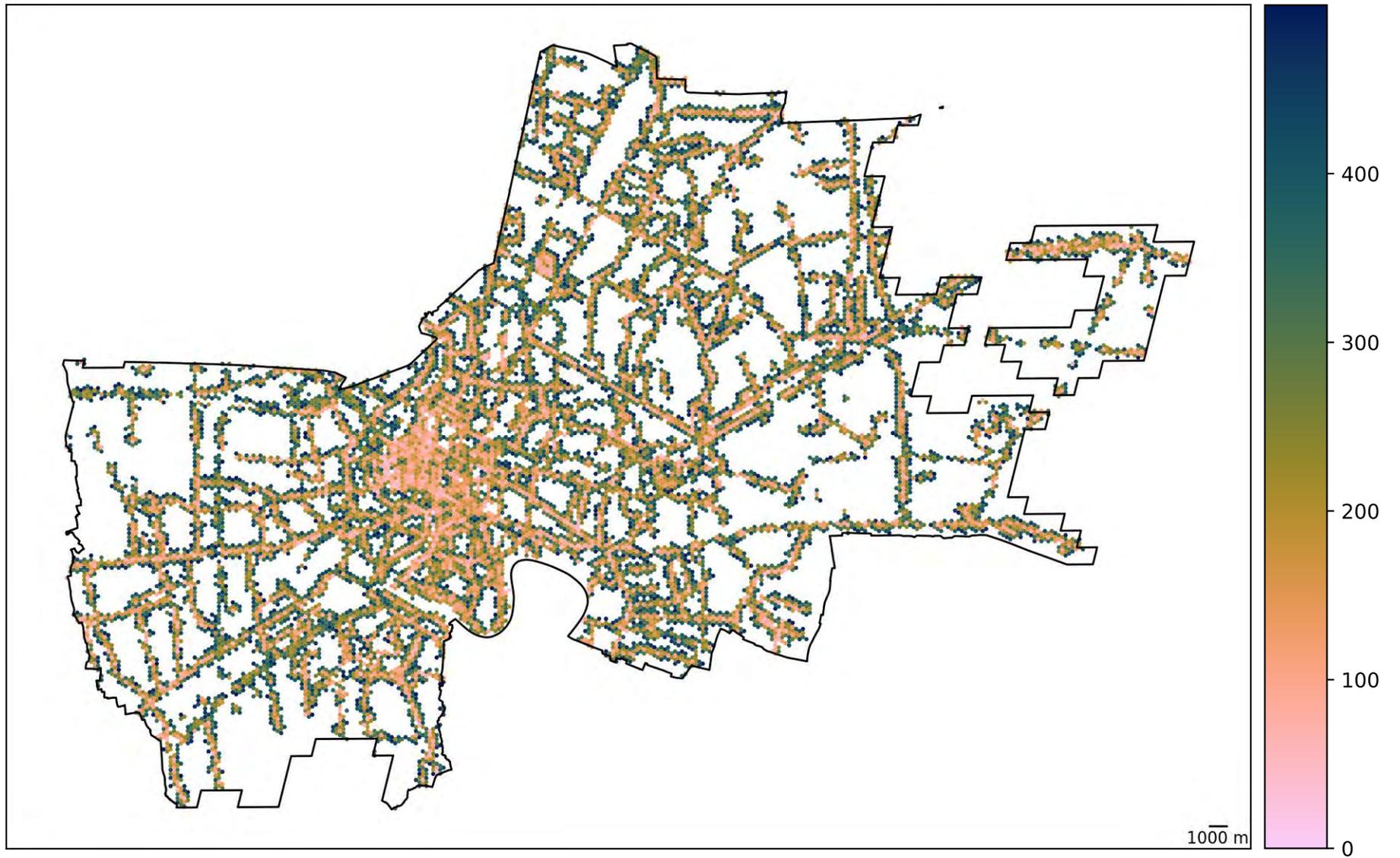

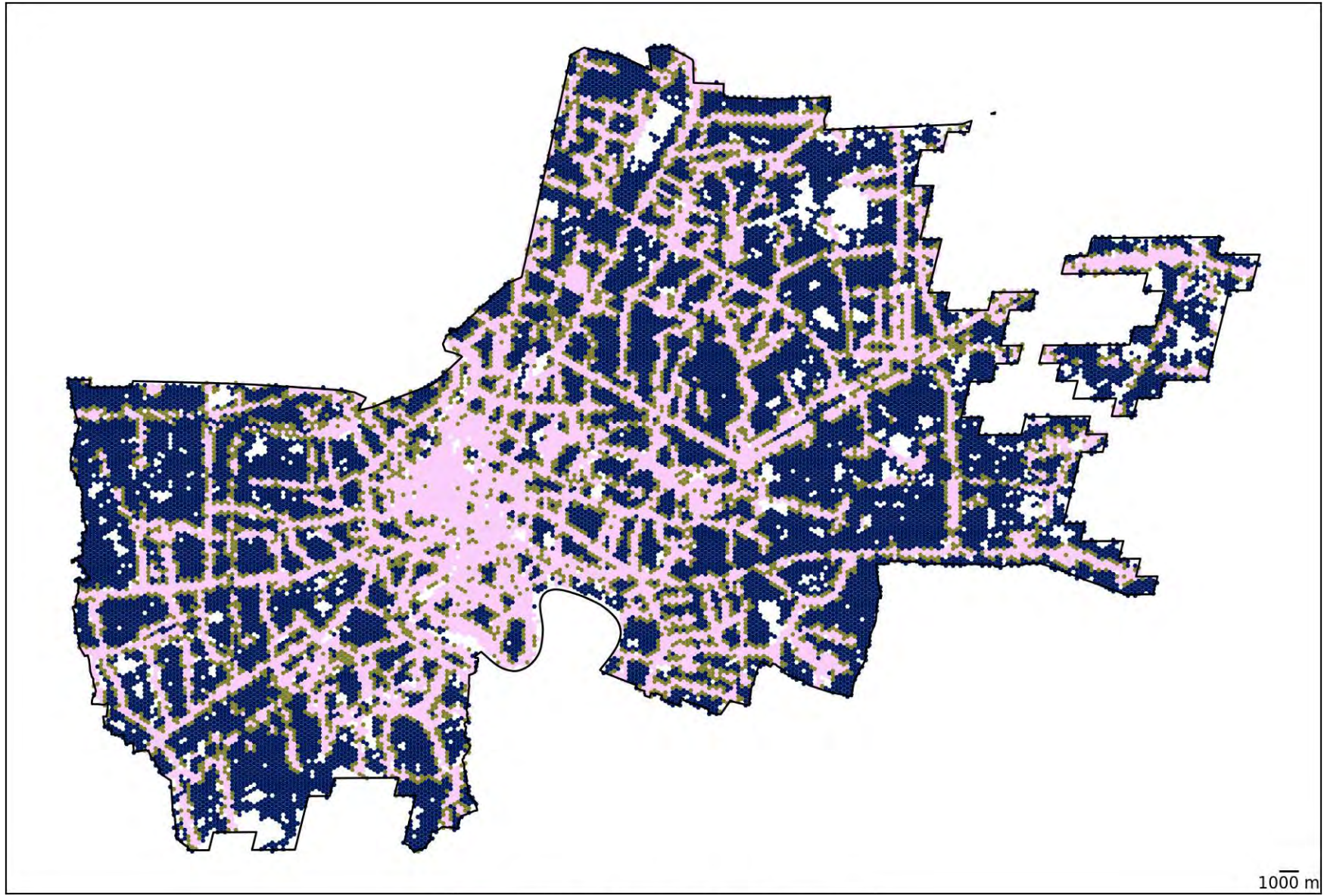

distances: Estimated Distance to nearest public transport stops (m; up to 500m) requirement for distances to destinations, measured up to a maximum distance target threshold of 500 metres

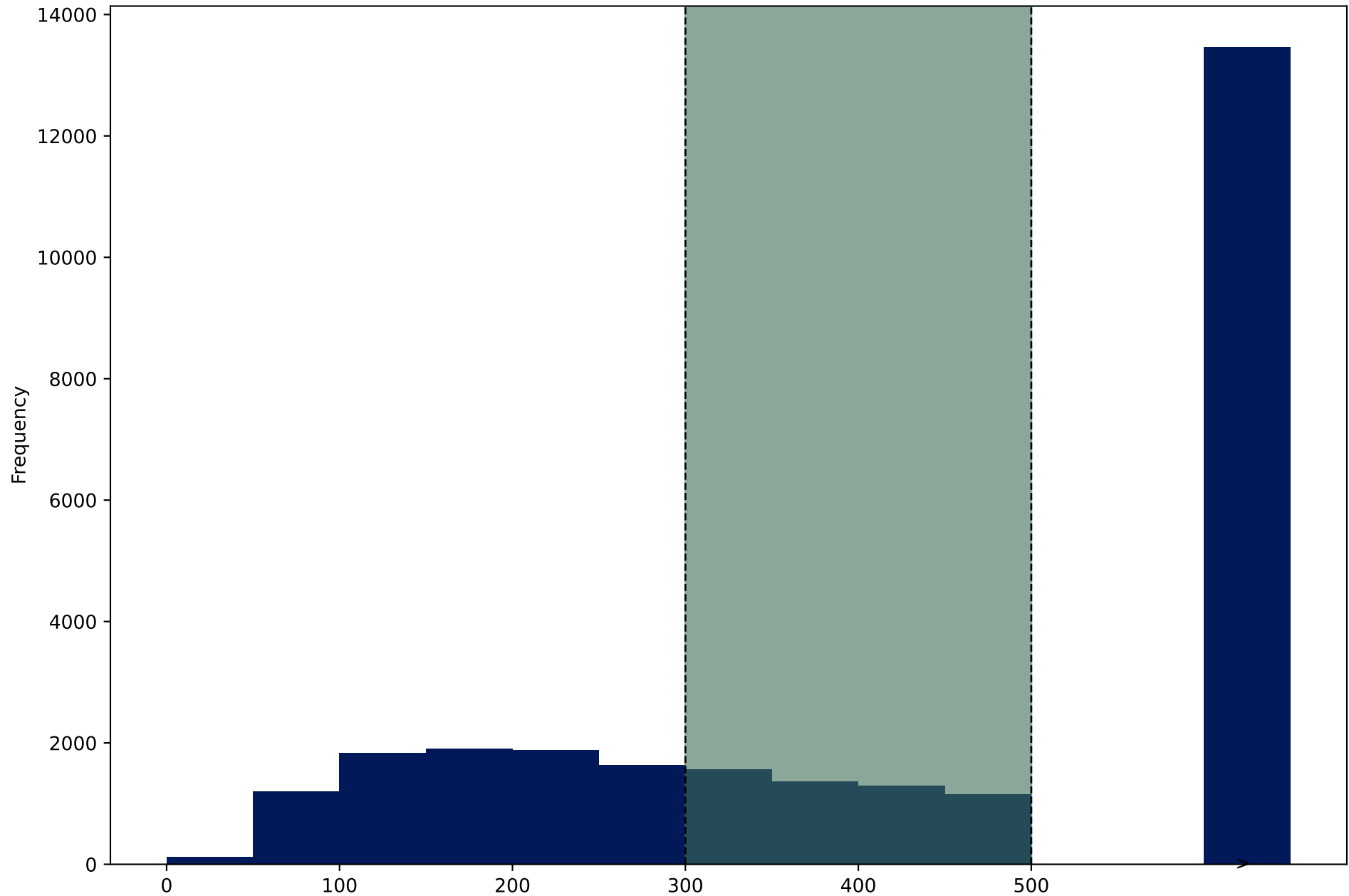

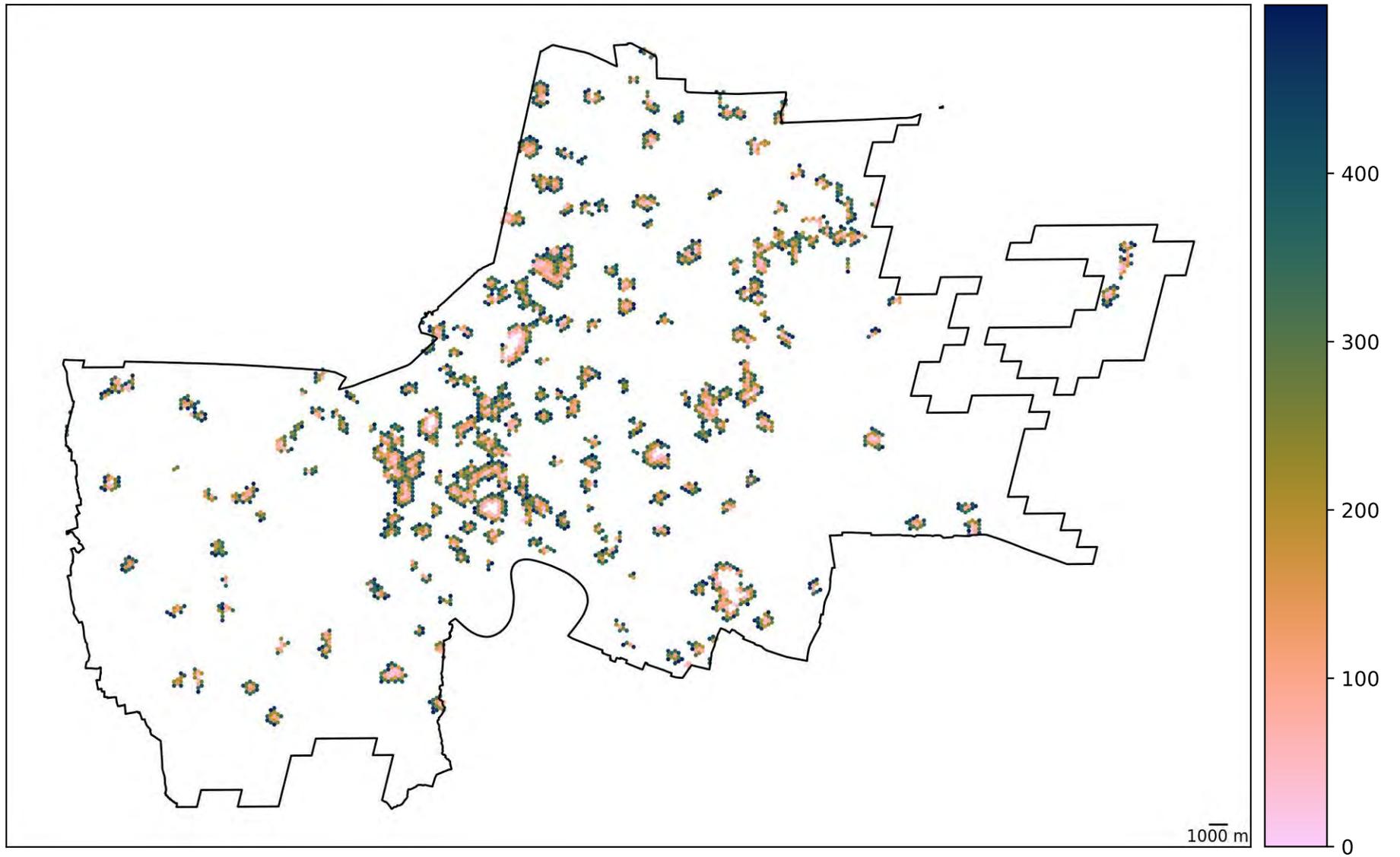

distances: Estimated Distance to nearest park (m; up to 500m) requirement for distances to destinations, measured up to a maximum distance target threshold of 500 metres

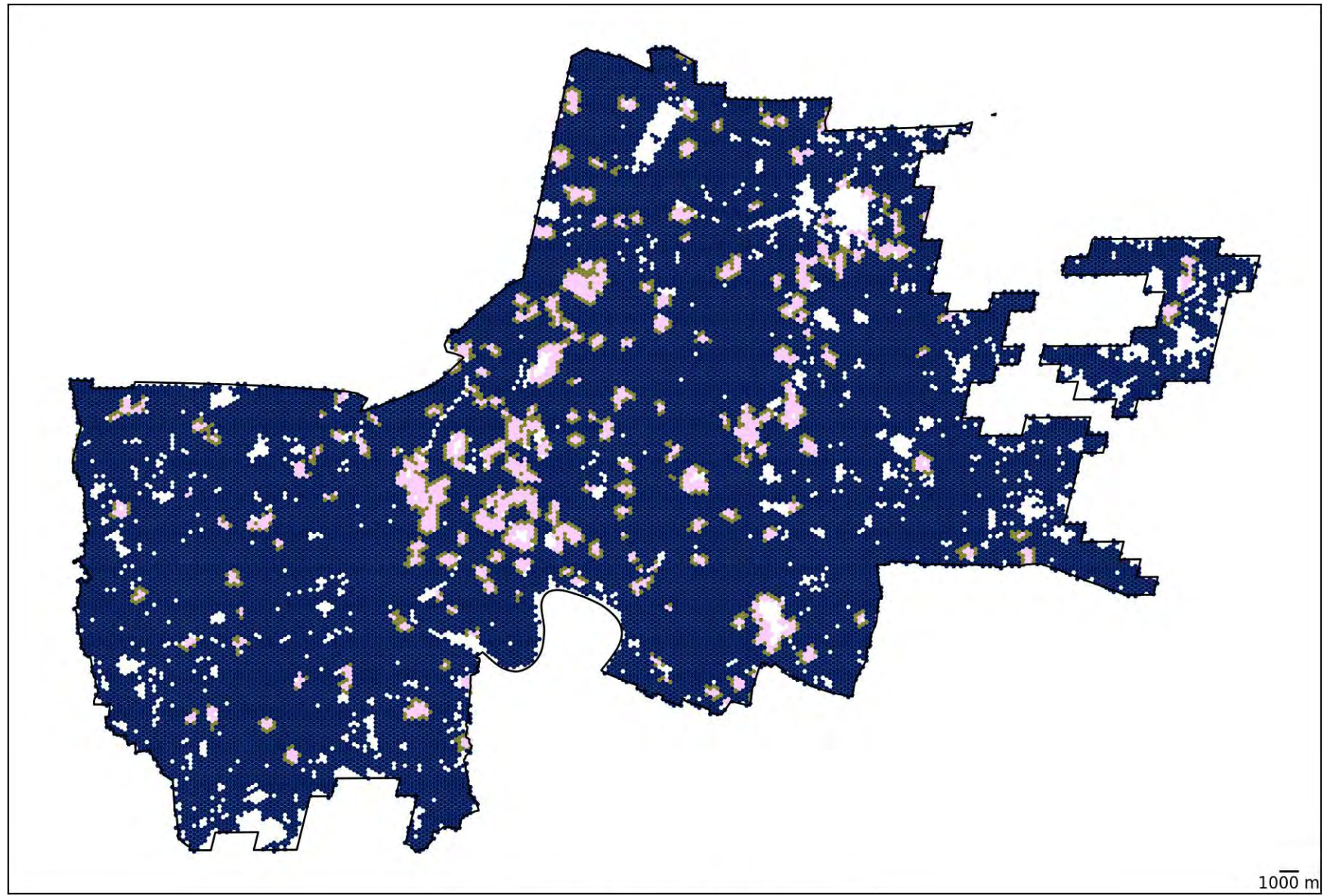

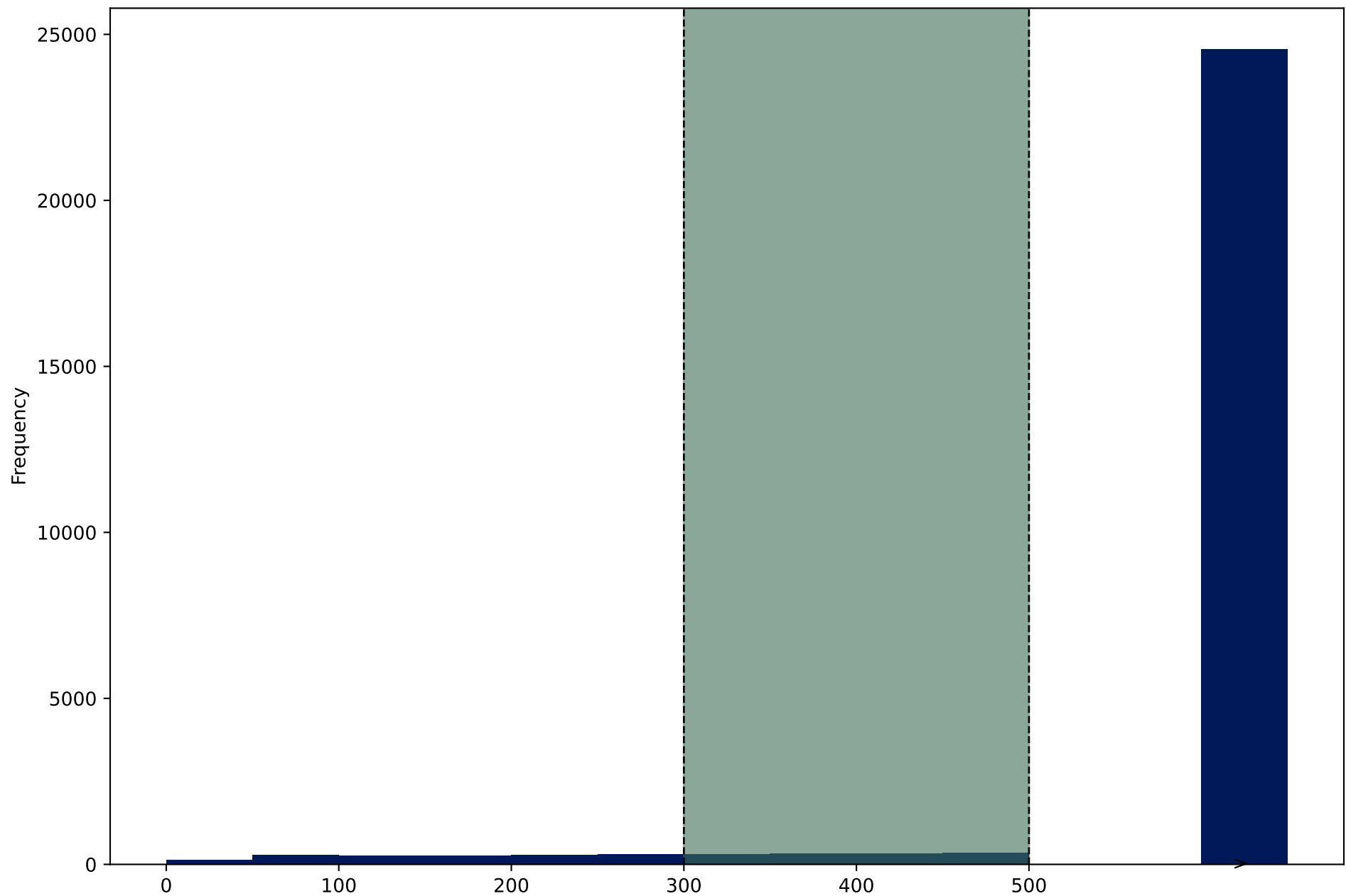

# Asia, Vietnam, Hanoi

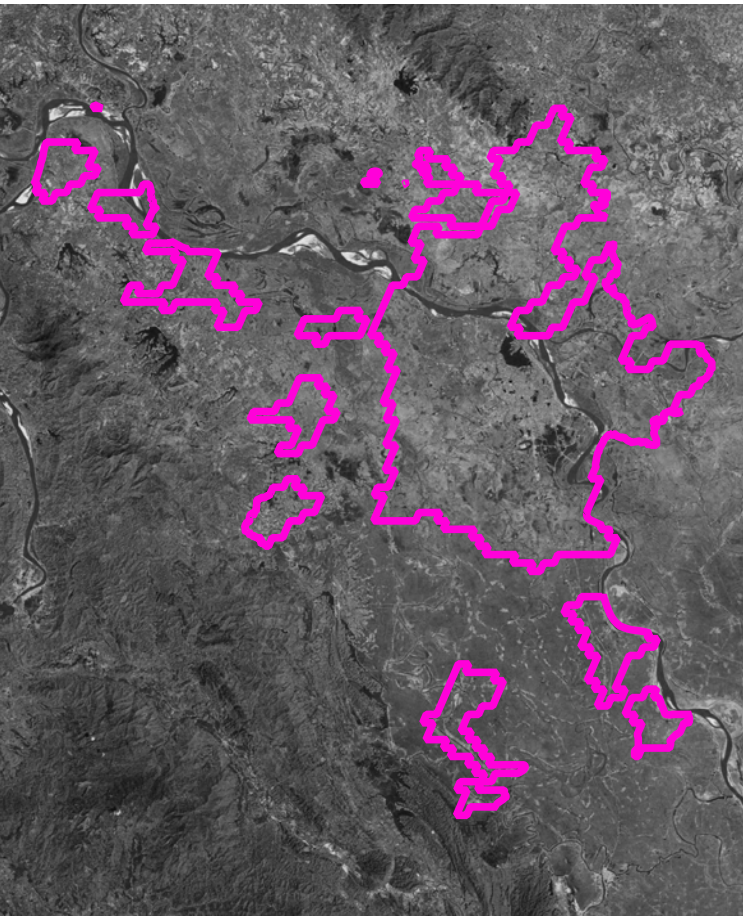
Satellite imagery of urban study region (Bing)

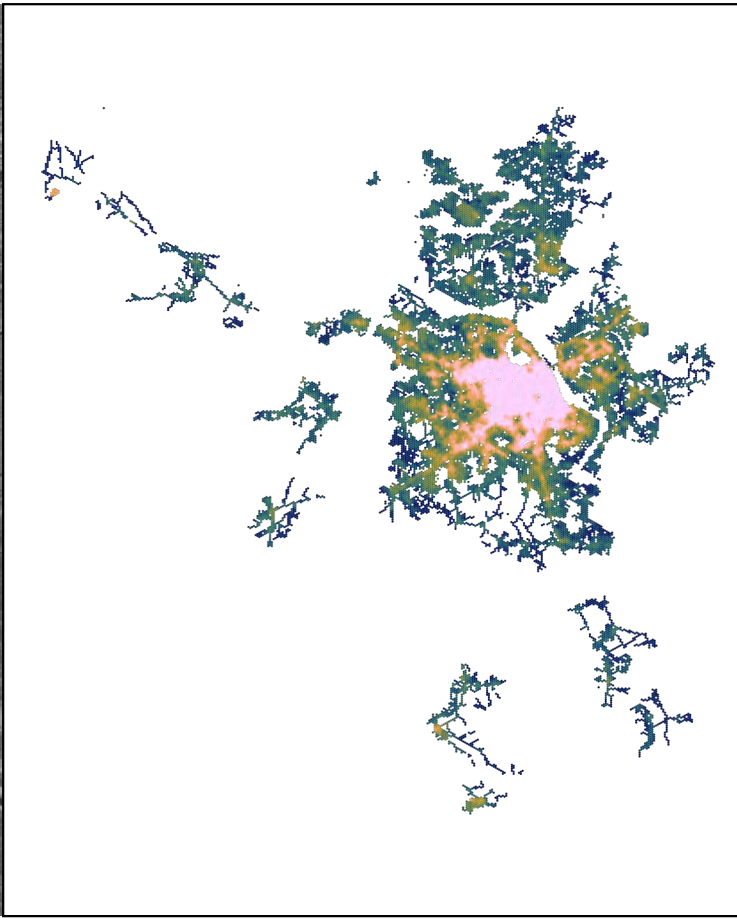
Walkability, relative to city

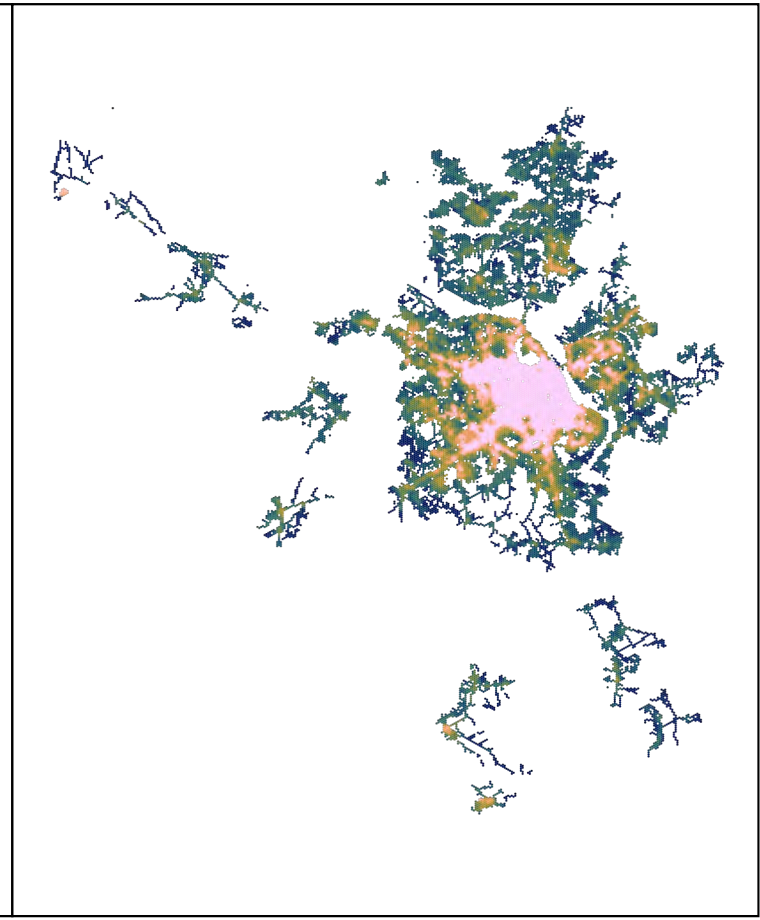
Walkability, relative to 25 global cities

Urban boundary

0  30  60 km

Walkability score
- <-3
- -3 to -2
- -2 to -1
- -1 to 0
- 0 to 1
- 1 to 2
- 2 to 3
- ≥3

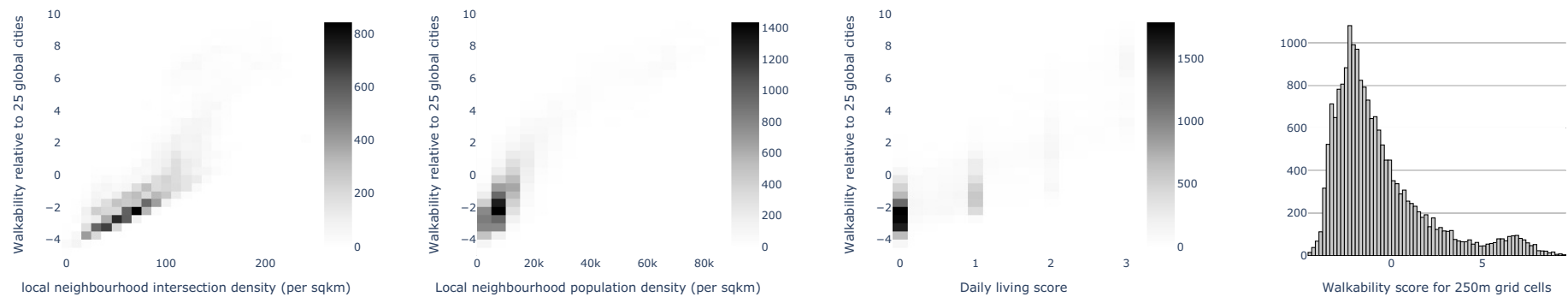
Walkability relative to all cities by component variables (2D histograms), and overall (histogram)

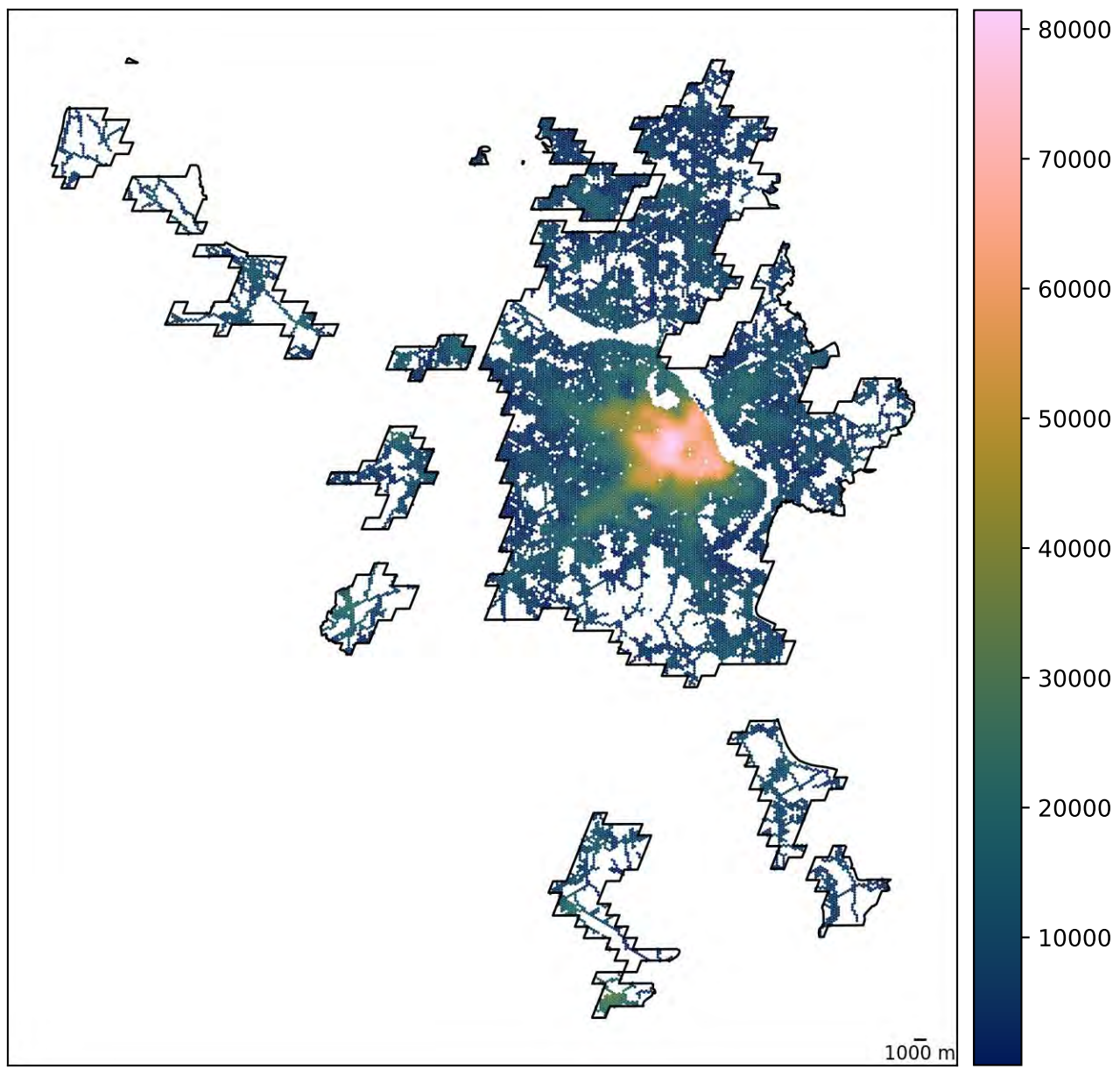

Mean 1000 m neighbourhood population per km²

A: Estimated Mean 1000 m neighbourhood population per km² requirement for ≥80% probability of engaging in walking for transport

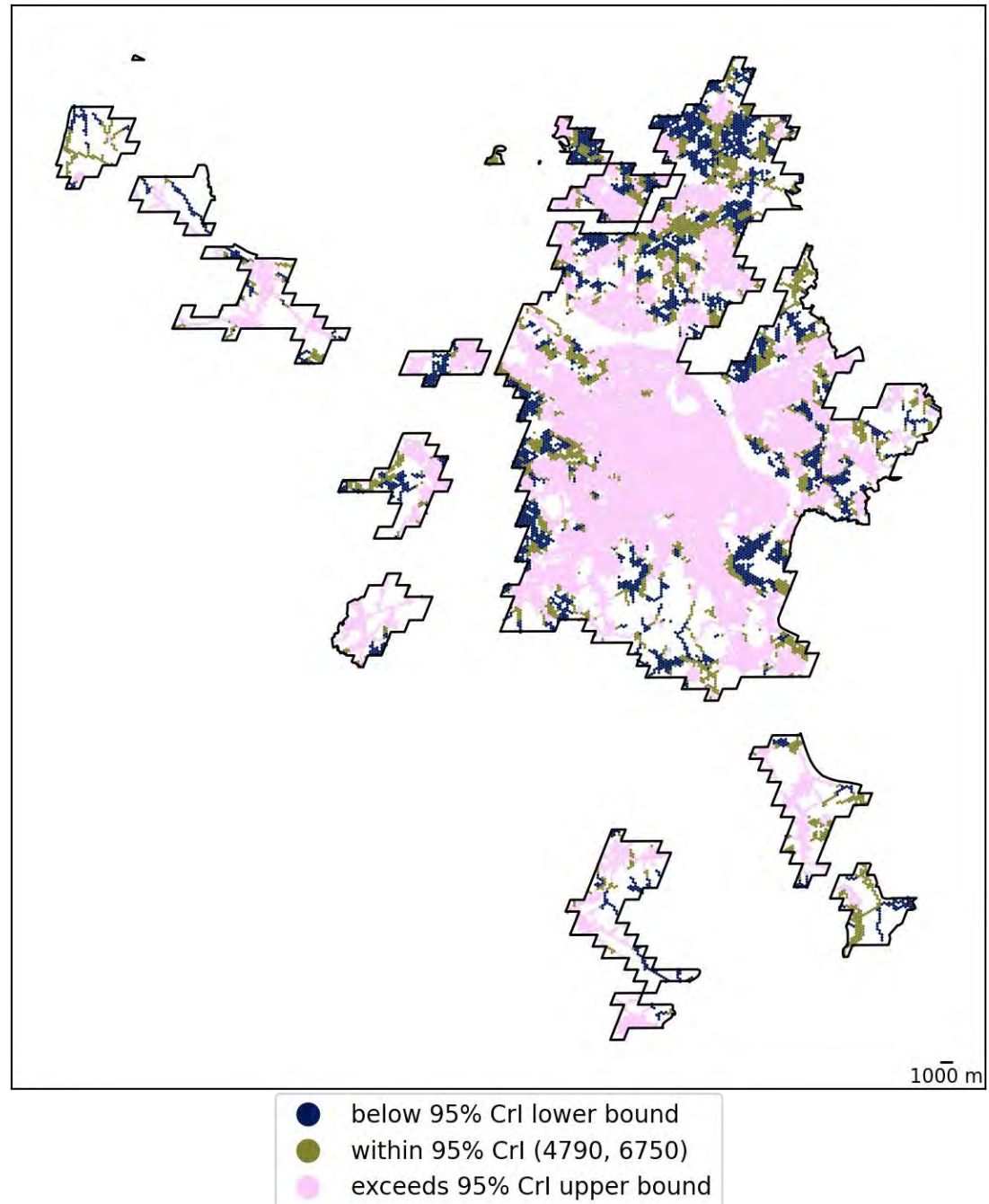

B: Estimated Mean 1000 m neighbourhood population per km² requirement for reaching the WHO's target of a ≥15% relative reduction in insufficient physical activity through walking

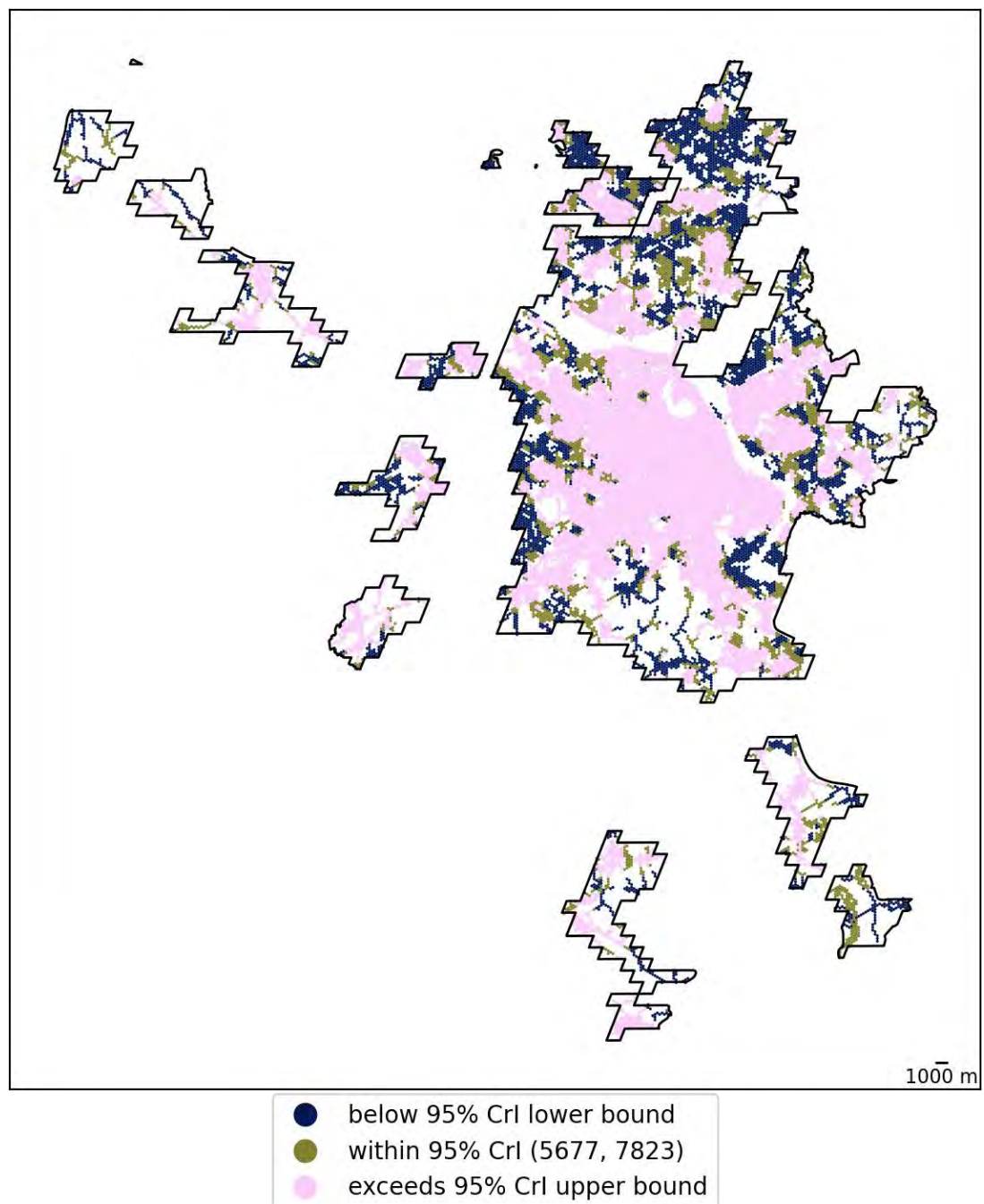

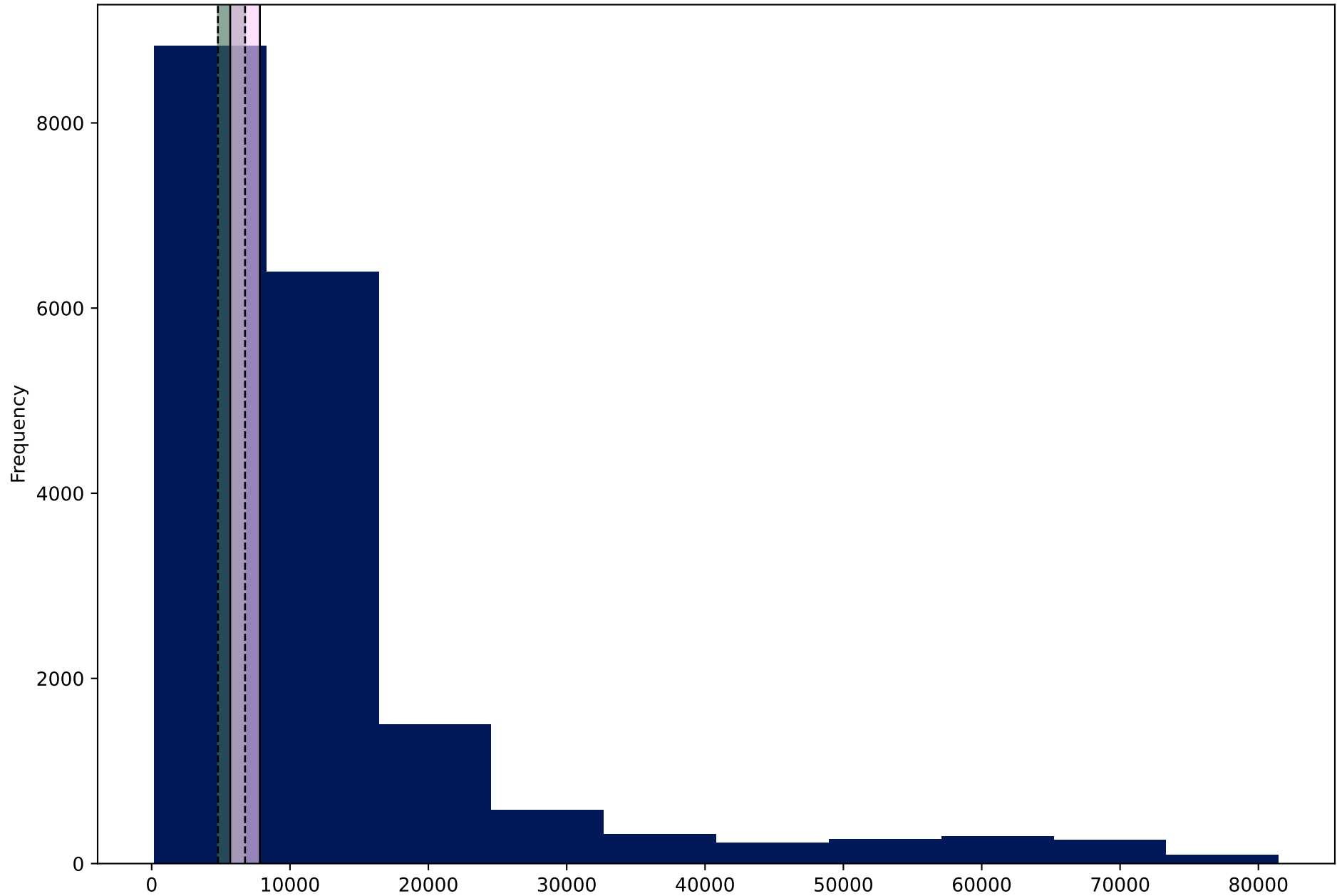

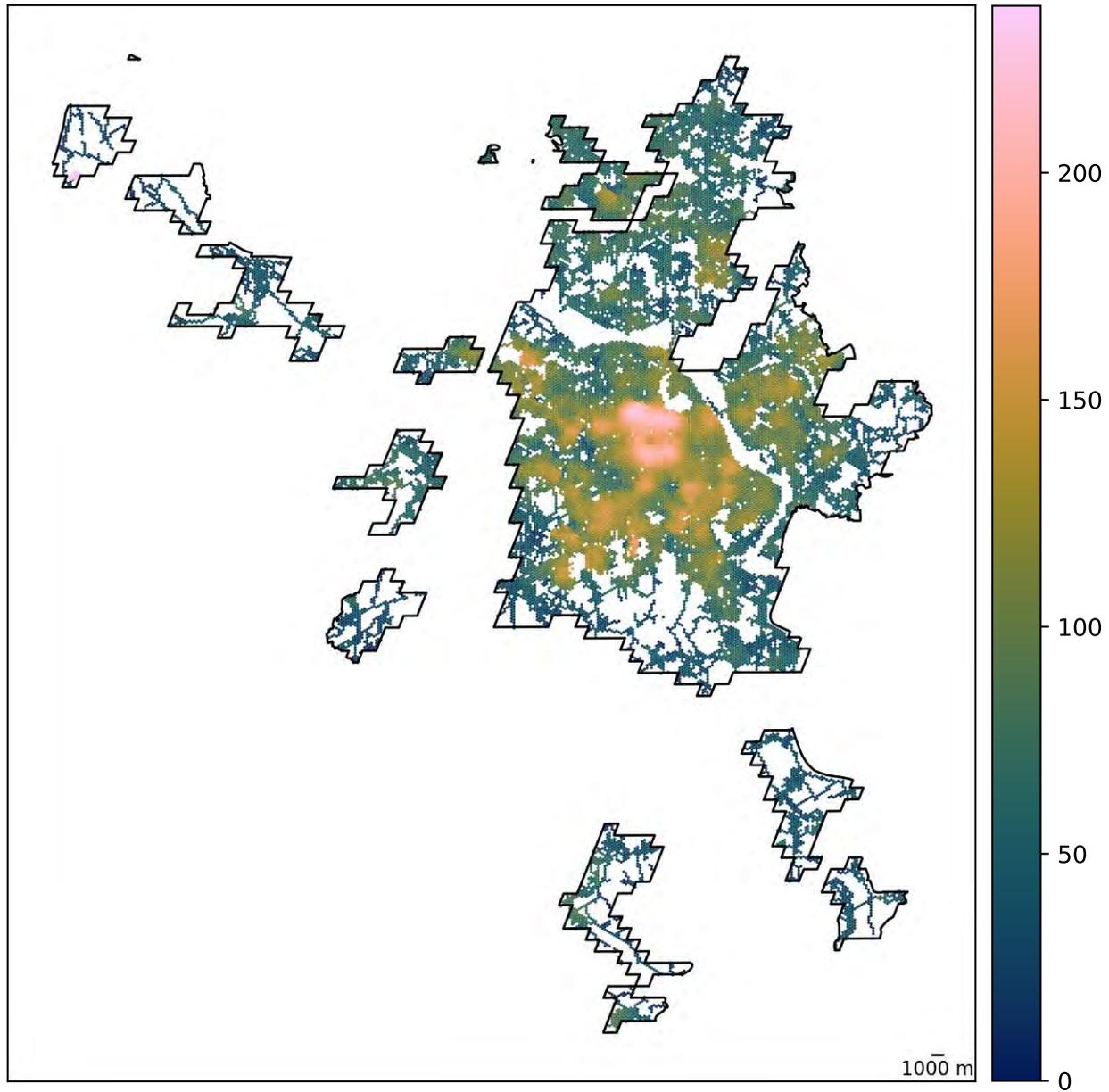

Mean 1000 m neighbourhood street intersections per km²

A: Estimated Mean 1000 m neighbourhood street intersections per km² requirement for ≥80% probability of engaging in walking for transport

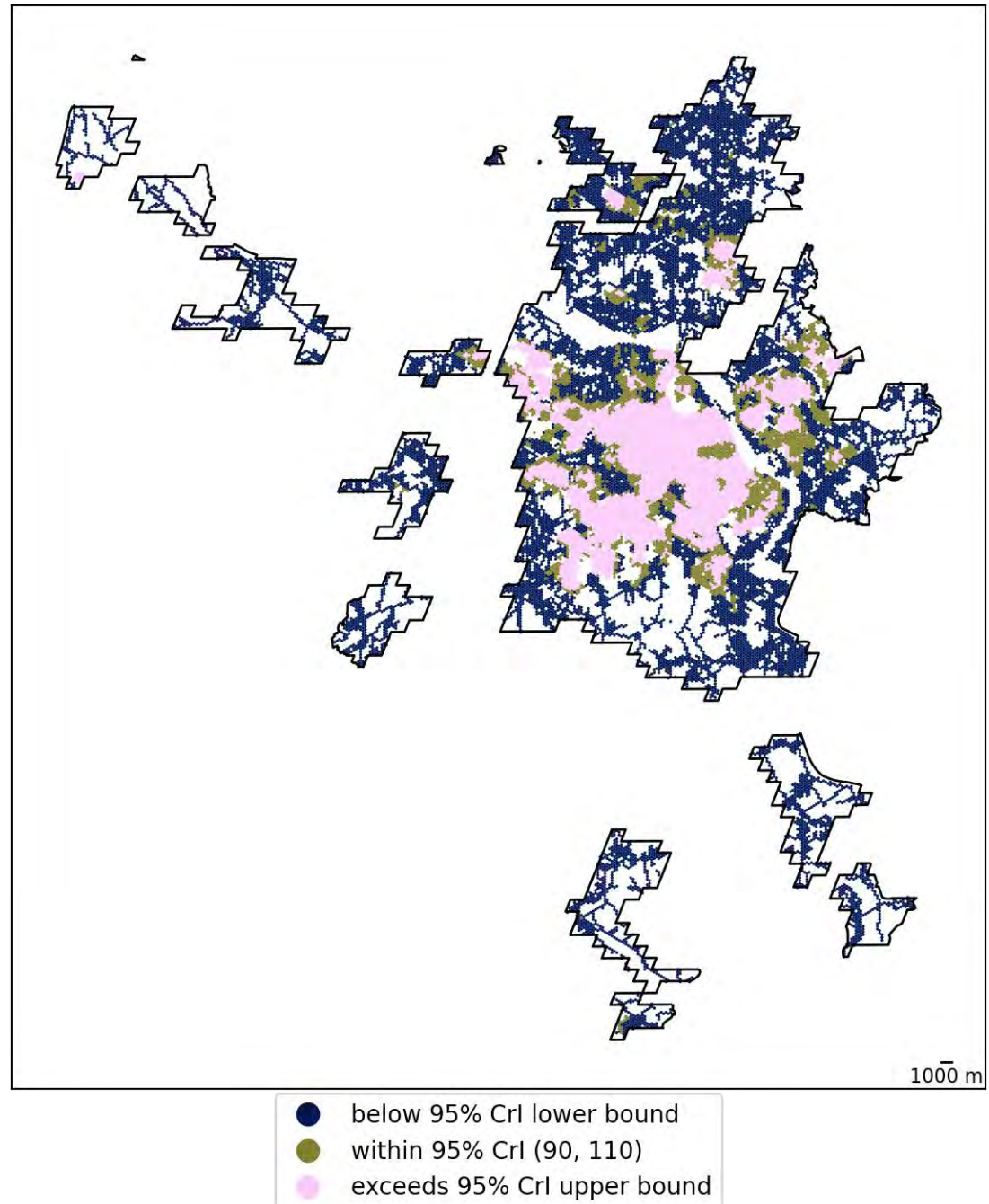

B: Estimated Mean 1000 m neighbourhood street intersections per km² requirement for reaching the WHO's target of a ≥15% relative reduction in insufficient physical activity through walking

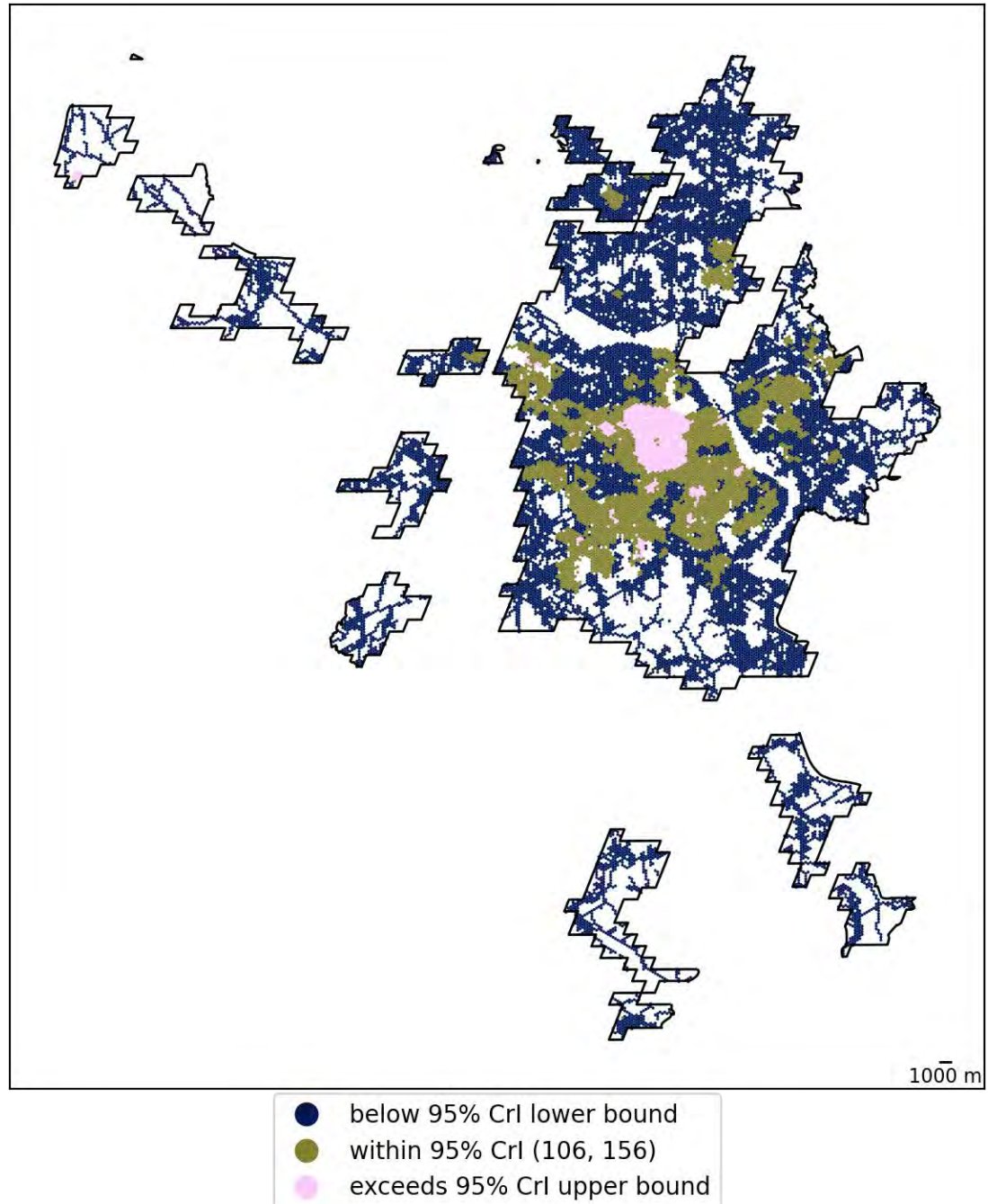

- below 95% CrI lower bound
- within 95% CrI (106, 156)
- exceeds 95% CrI upper bound

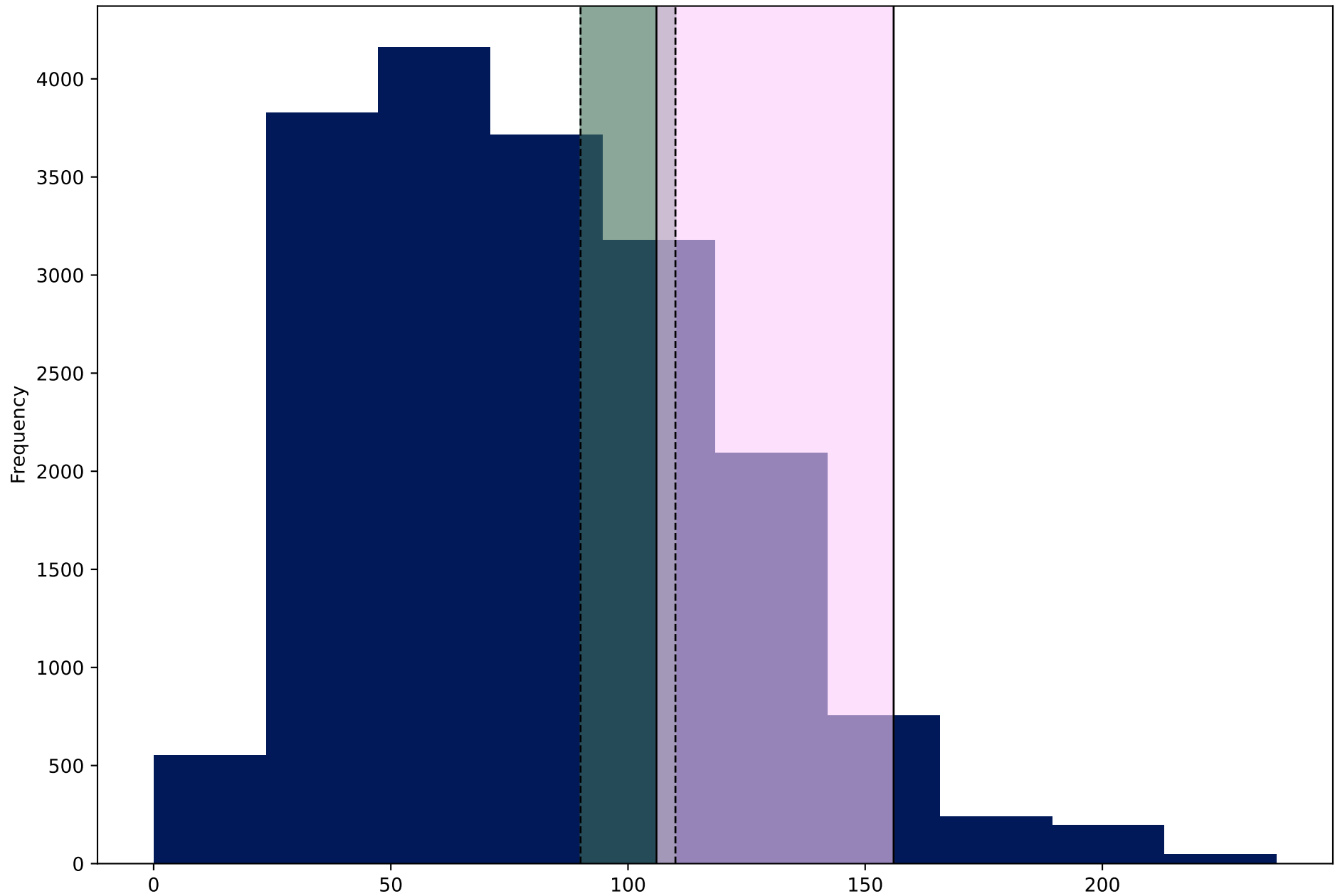

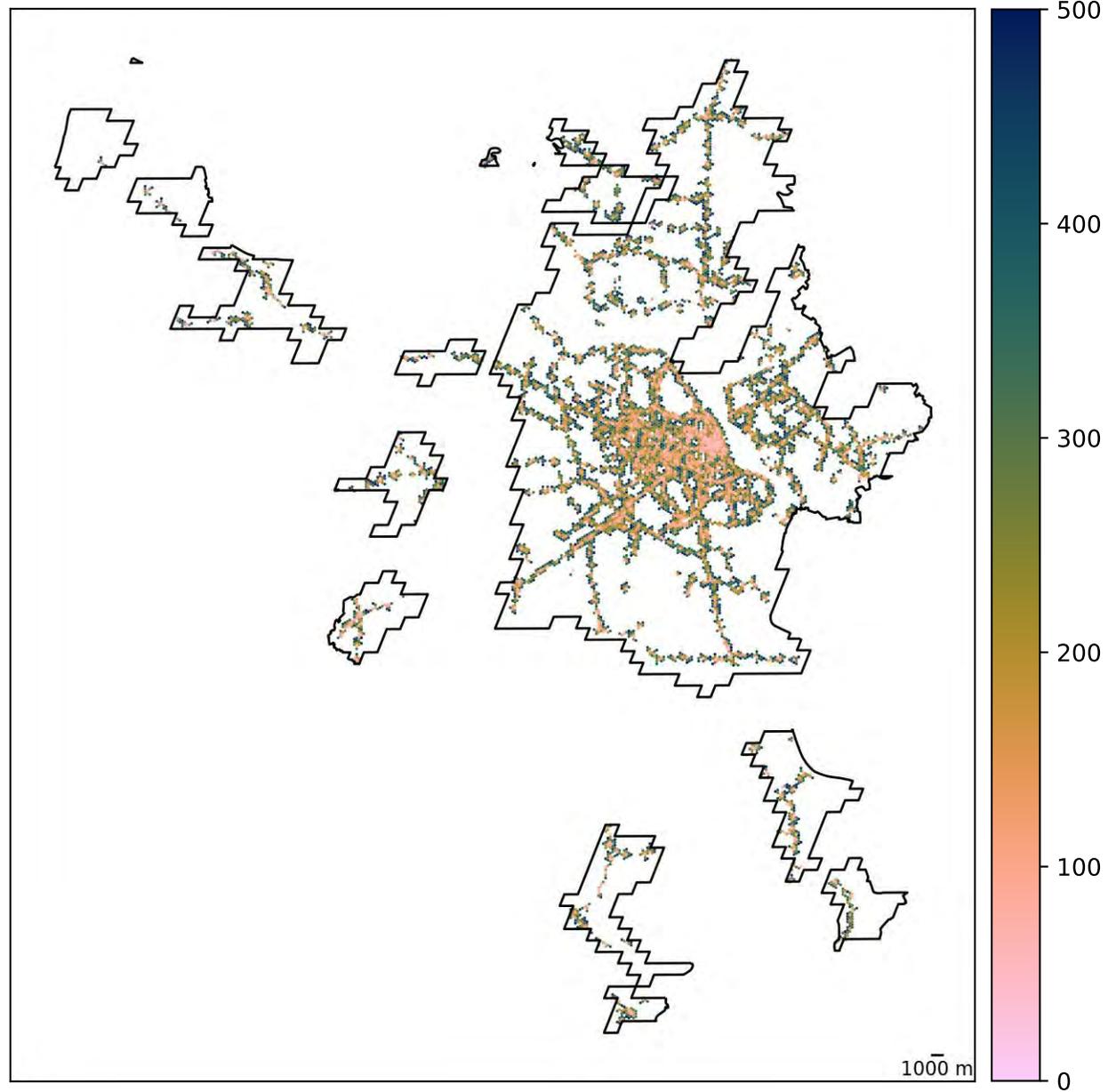

Distance to nearest public transport stops (m; up to 500m)

distances: Estimated Distance to nearest public transport stops (m; up to 500m) requirement for distances to destinations, measured up to a maximum distance target threshold of 500 metres

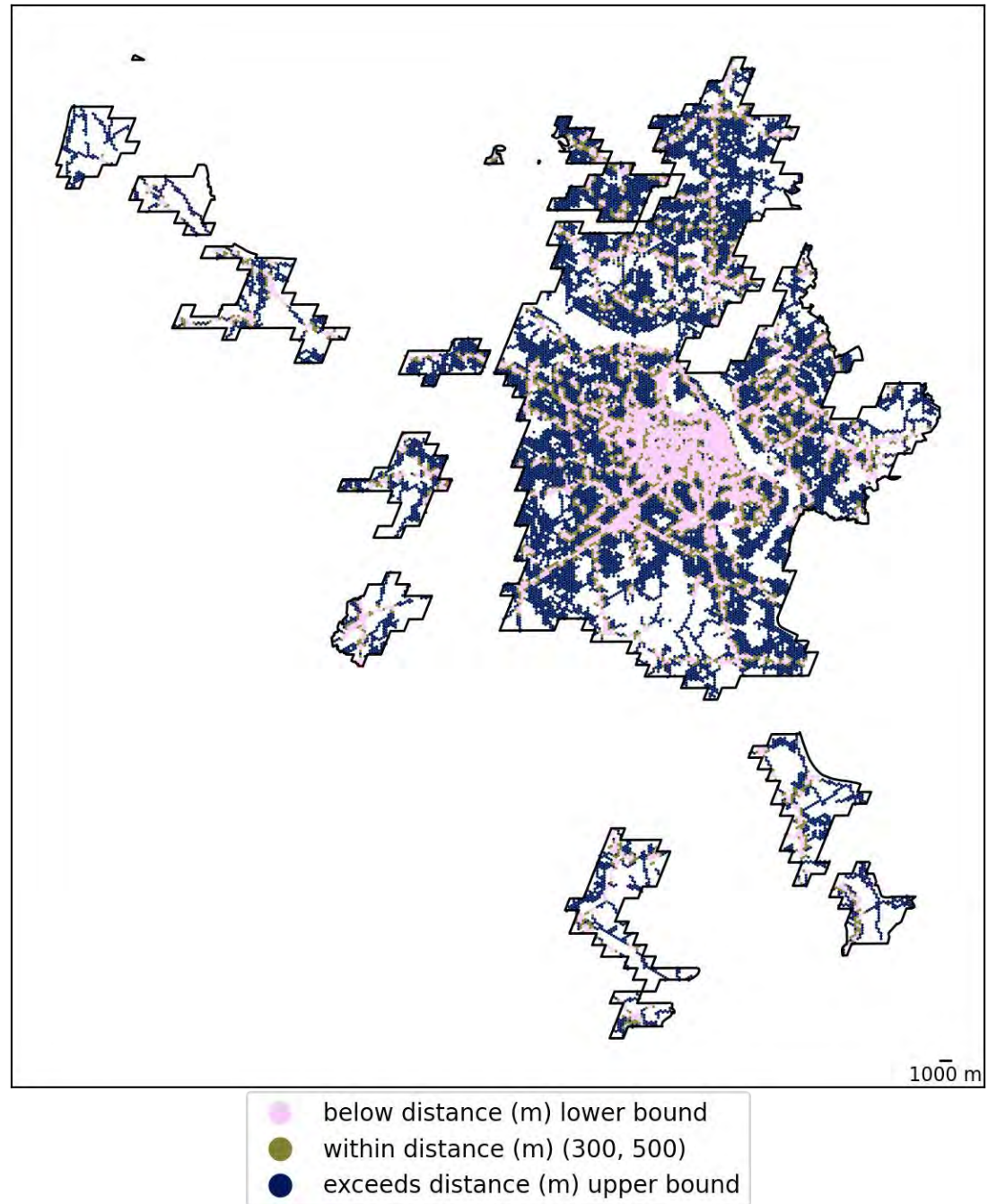

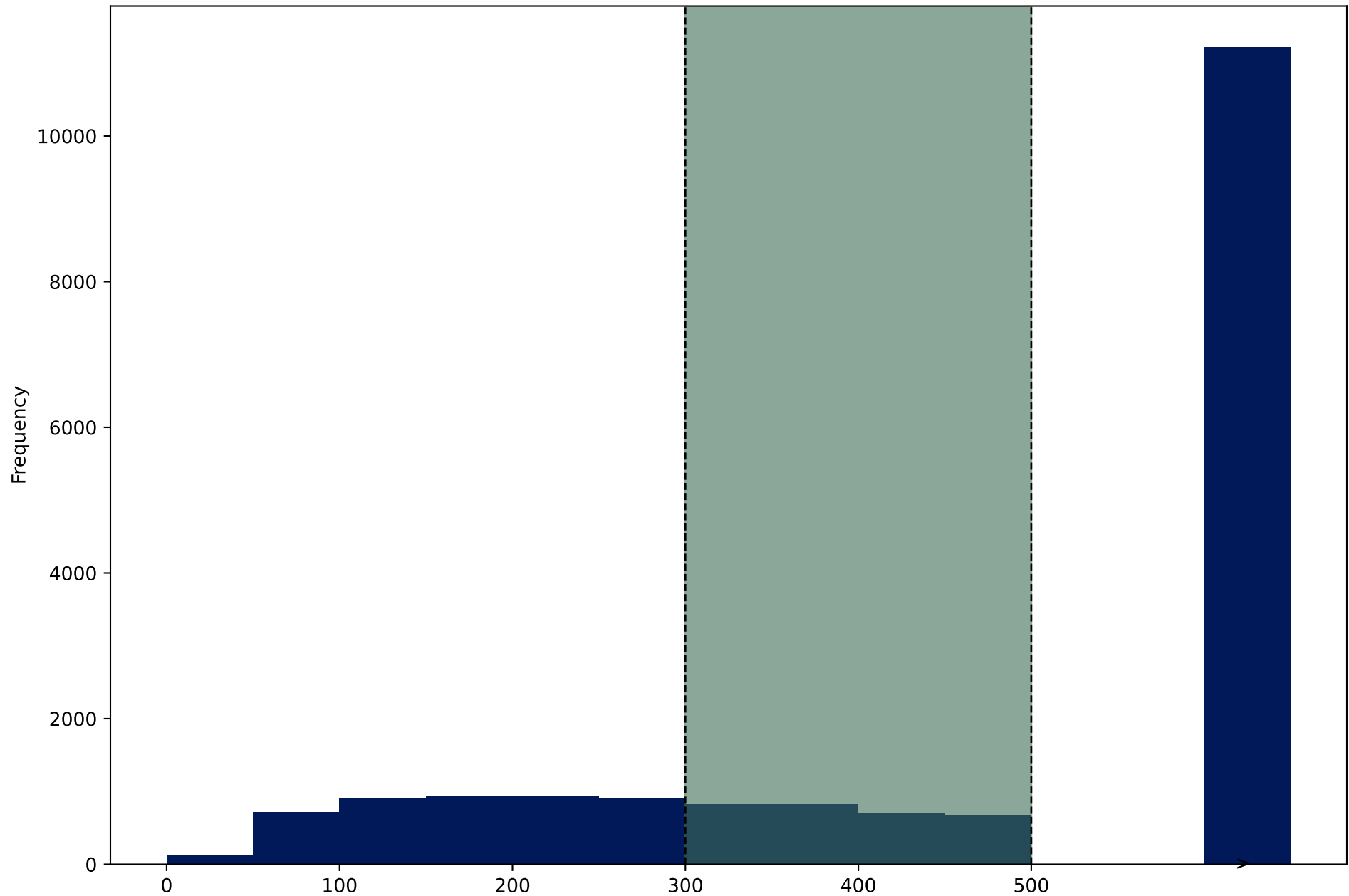

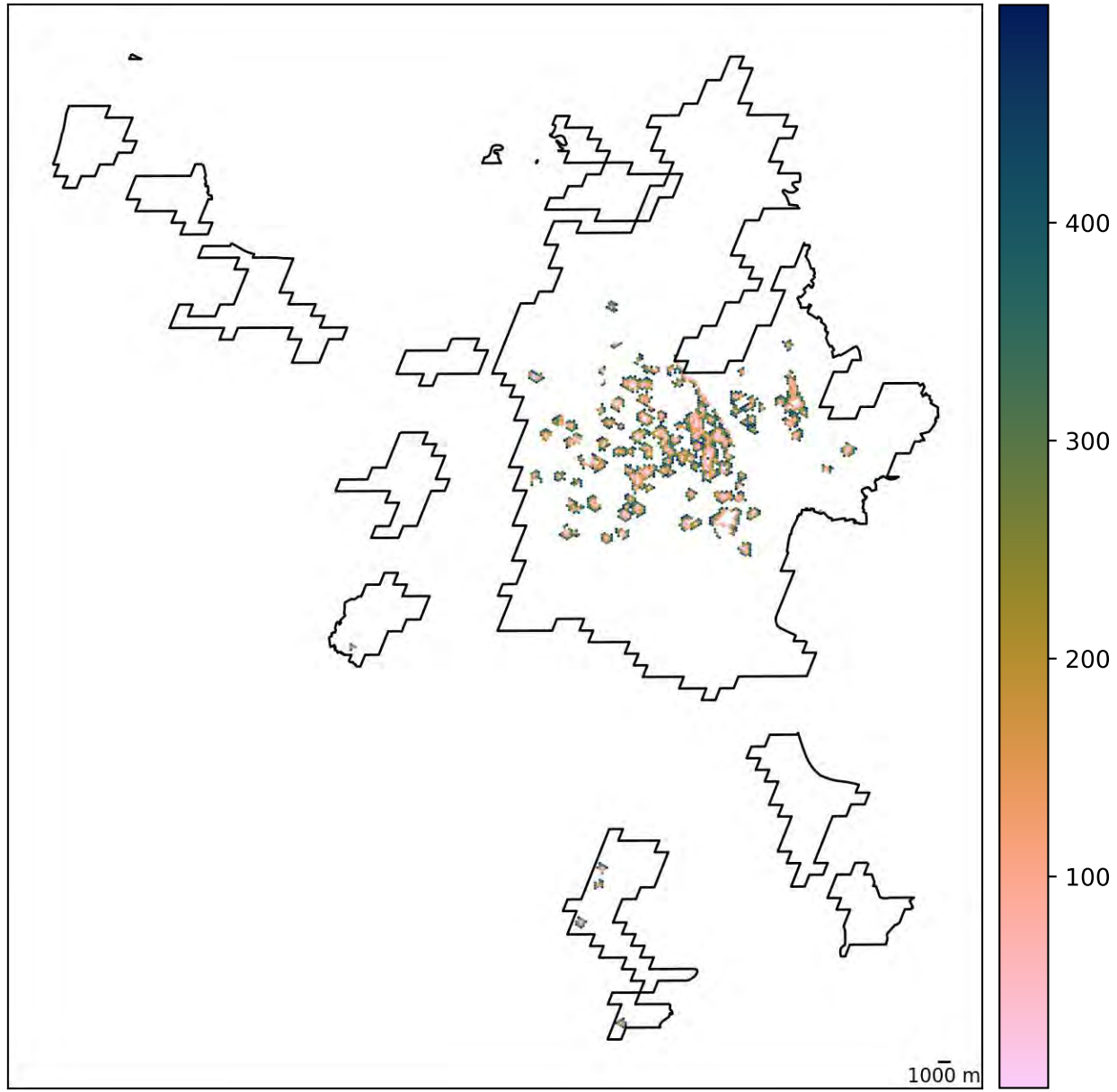

distances: Estimated Distance to nearest park (m; up to 500m) requirement for distances to destinations, measured up to a maximum distance target threshold of 500 metres

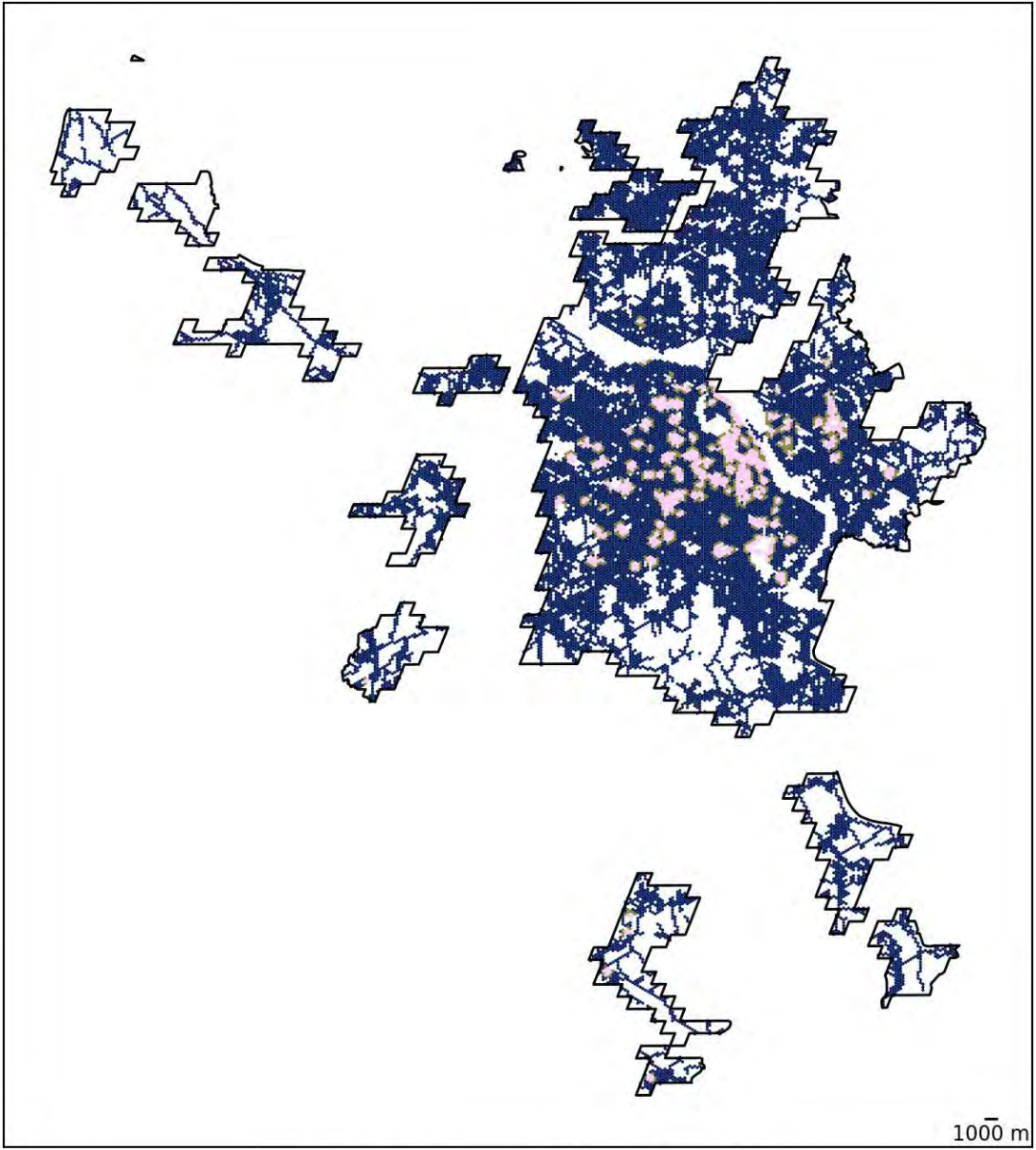

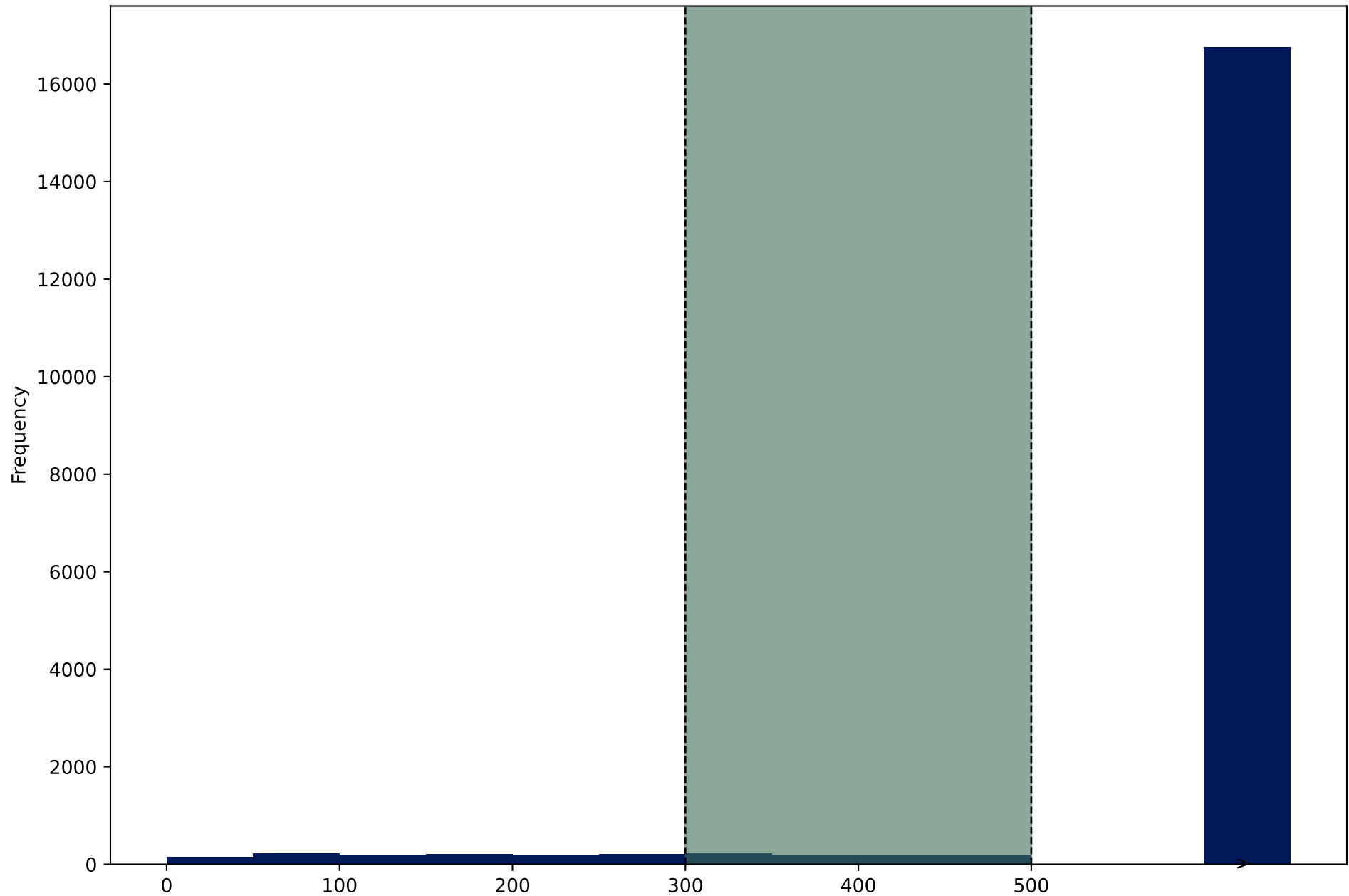

# Australasia, Australia, Adelaide

| Satellite imagery of urban study region (Bing) | Walkability, relative to city | Walkability, relative to 25 global cities |
|---|---|---|

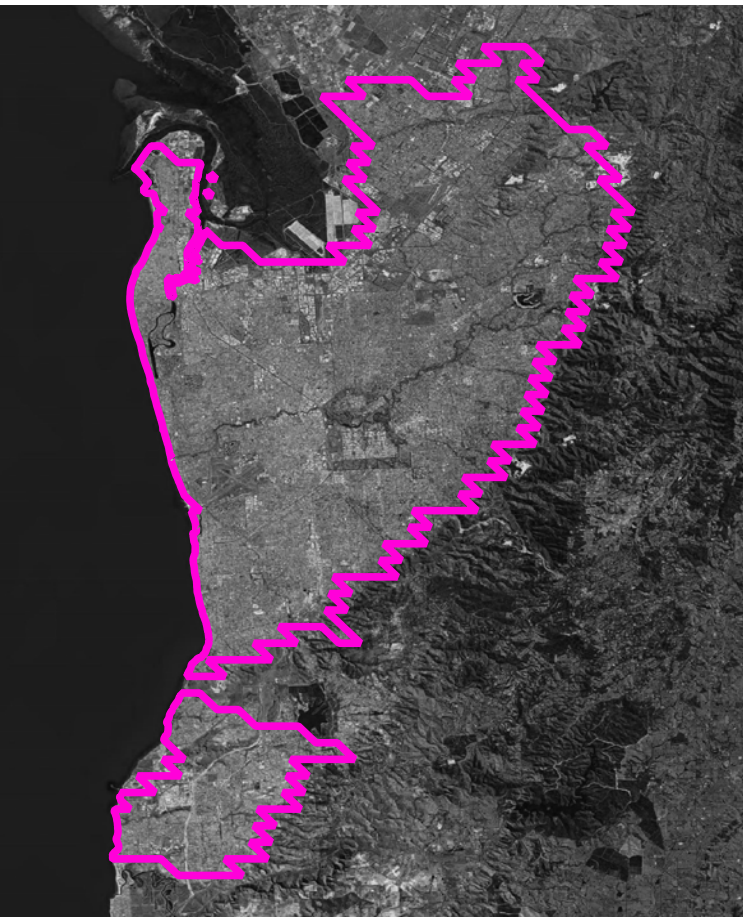
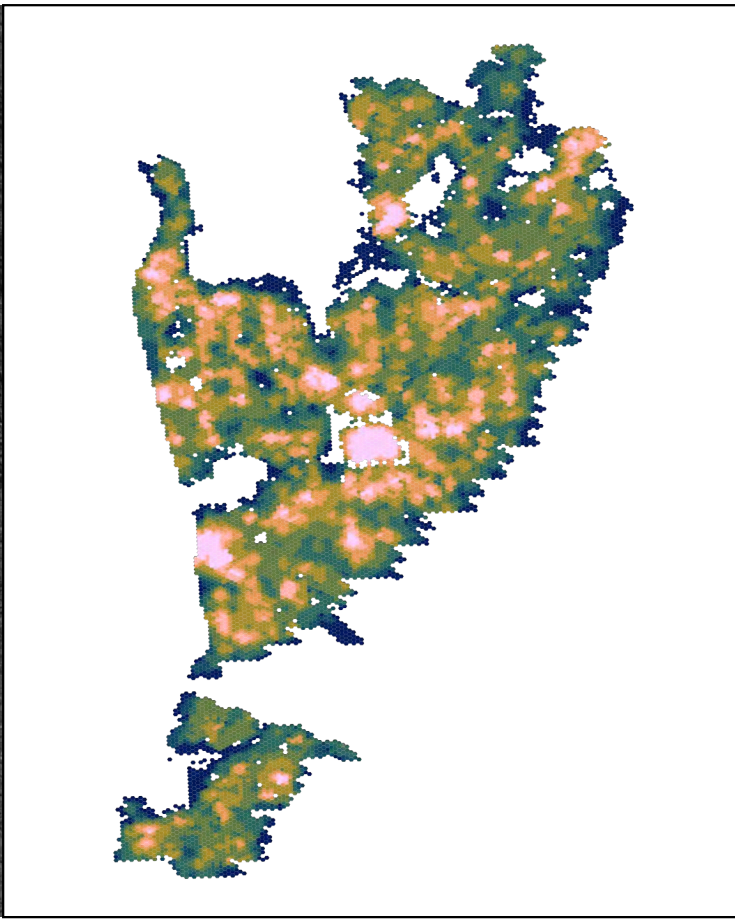
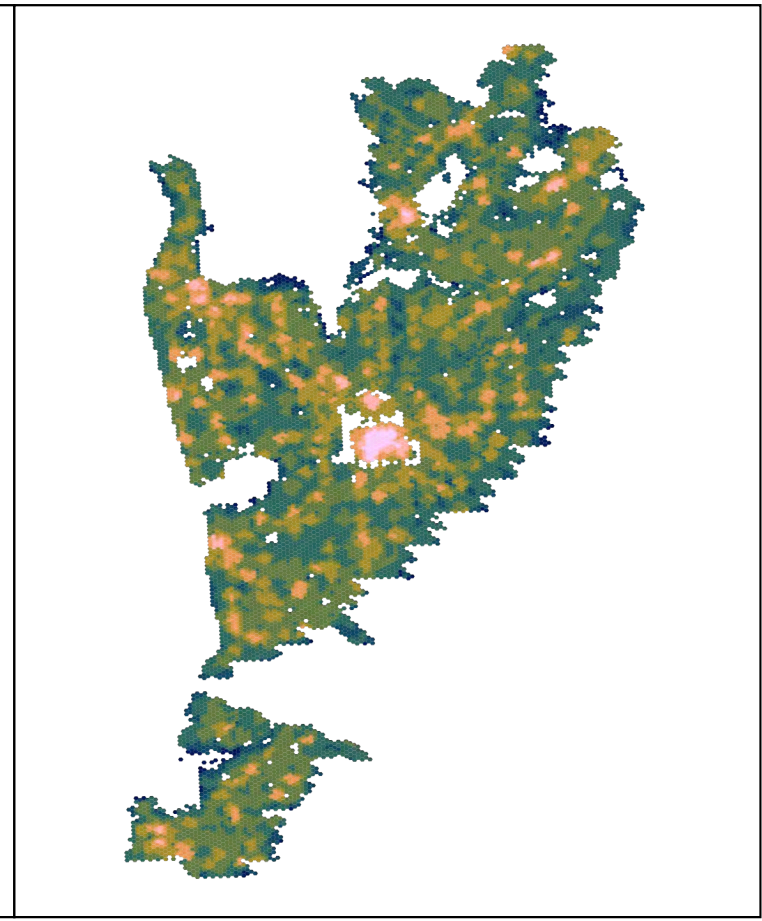

Urban boundary

0  10  20 km

Walkability relative to all cities by component variables (2D histograms), and overall (histogram)

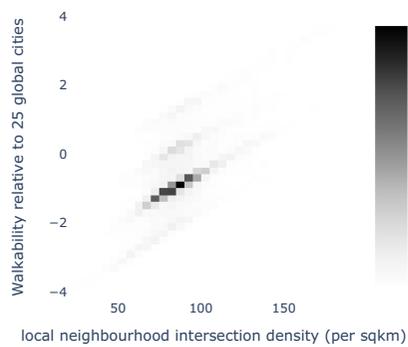
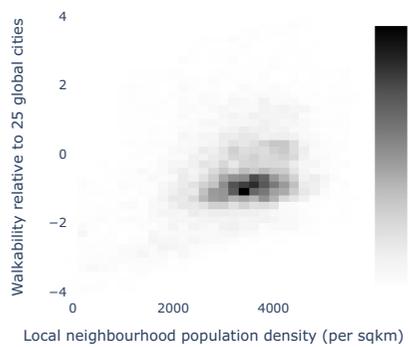
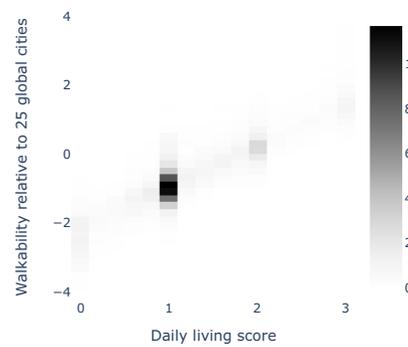
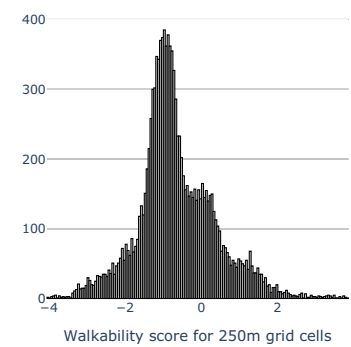

Walkability score
- <-3
- -3 to -2
- -2 to -1
- -1 to 0
- 0 to 1
- 1 to 2
- 2 to 3
- ≥3

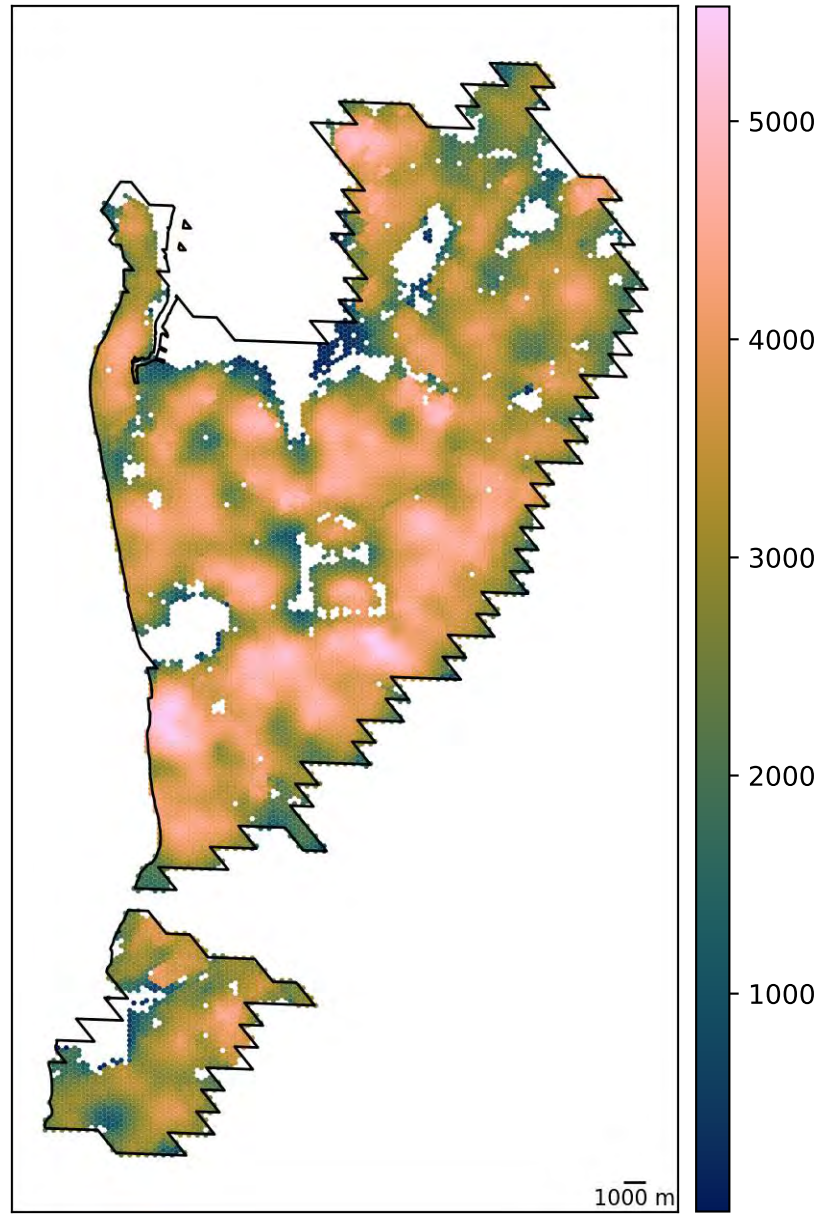

Mean 1000 m neighbourhood population per km²

A: Estimated Mean 1000 m neighbourhood population per km² requirement for ≥80% probability of engaging in walking for transport

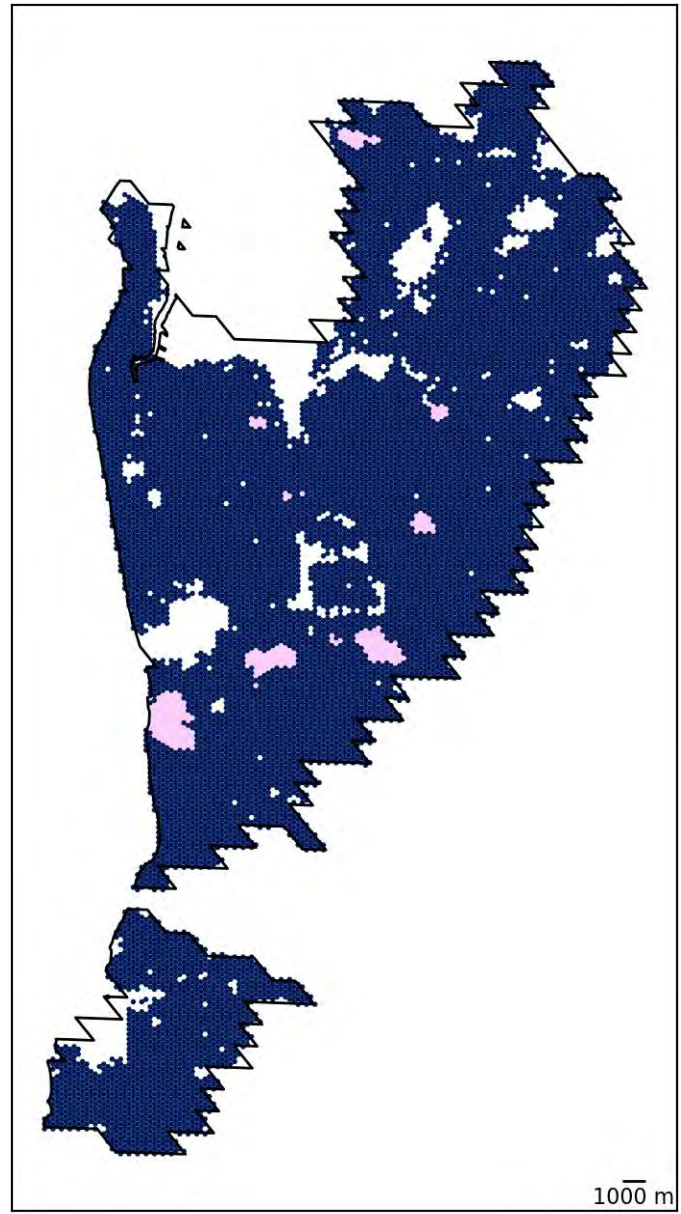

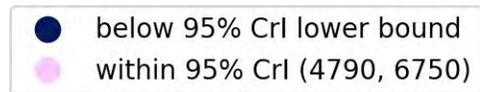

B: Estimated Mean 1000 m neighbourhood population per km² requirement for reaching the WHO's target of a ≥15% relative reduction in insufficient physical activity through walking

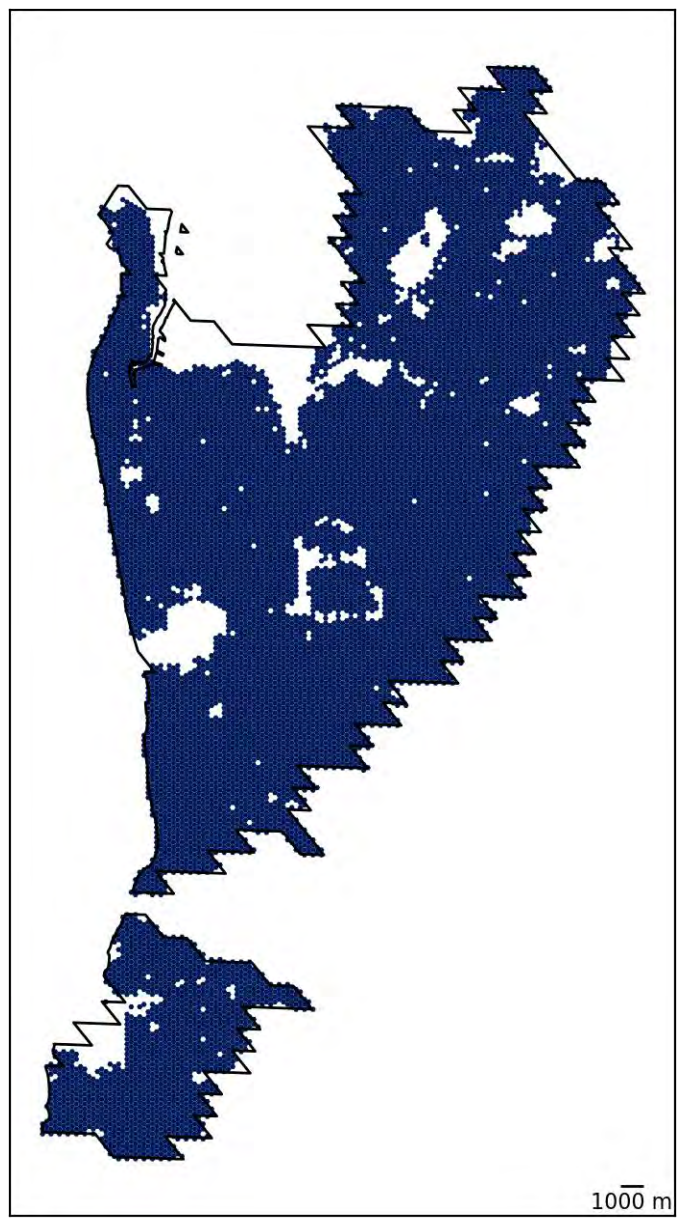

● below 95% CrI lower bound

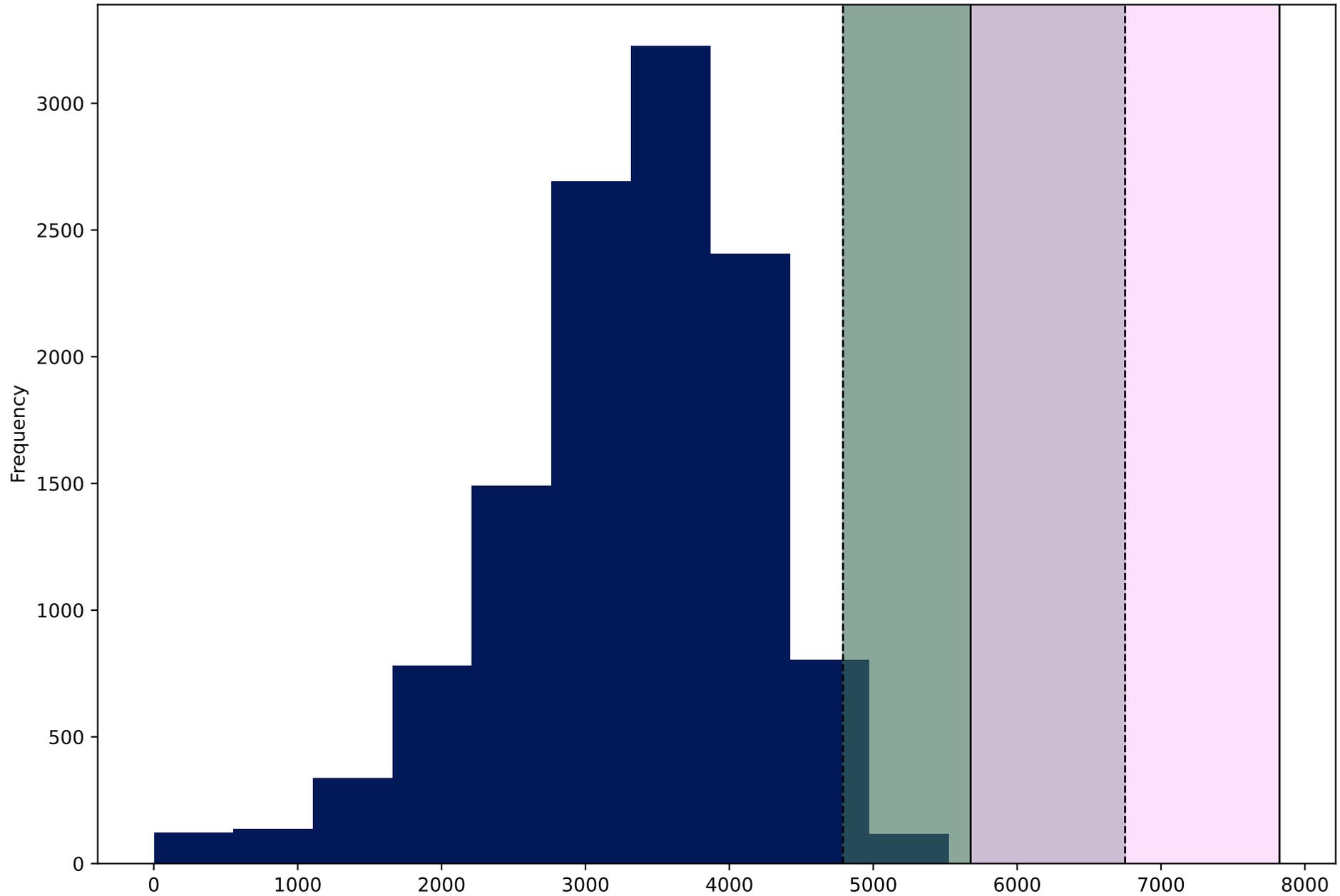

Mean 1000 m neighbourhood street intersections per km²

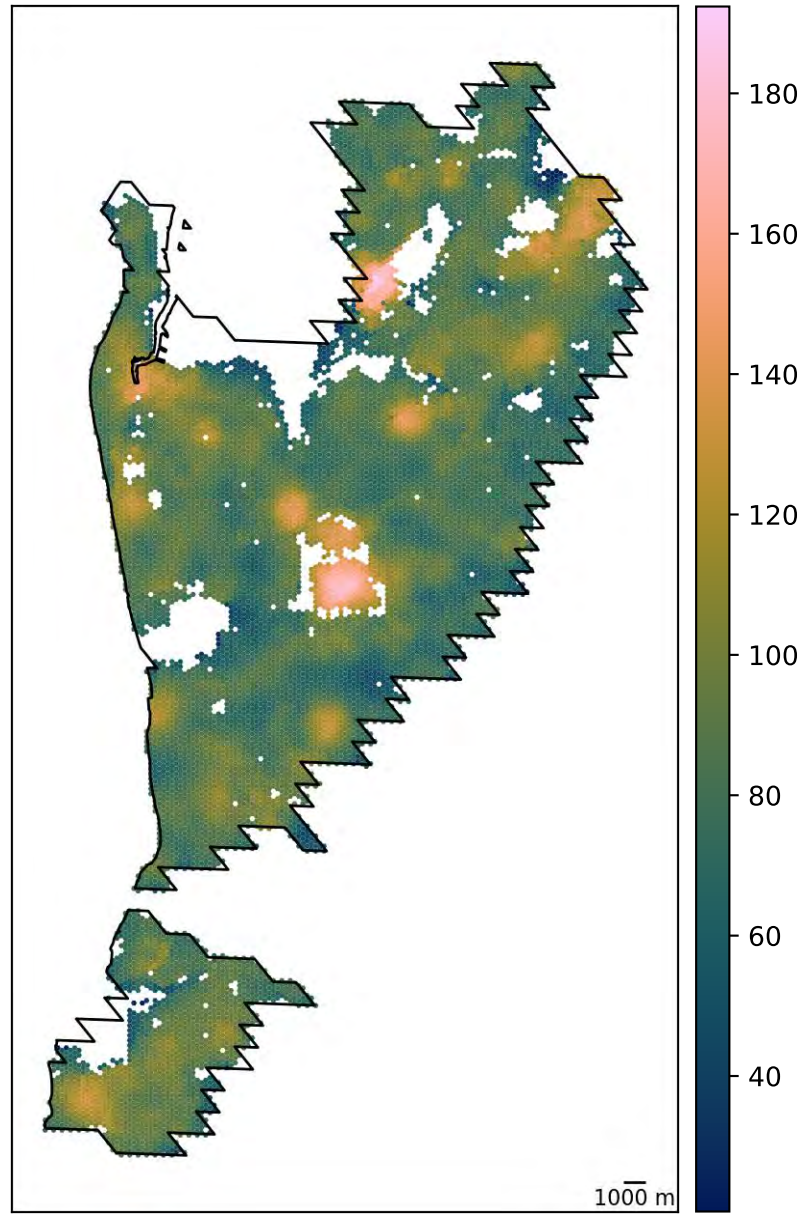

A: Estimated Mean 1000 m neighbourhood street intersections per km² requirement for ≥80% probability of engaging in walking for transport

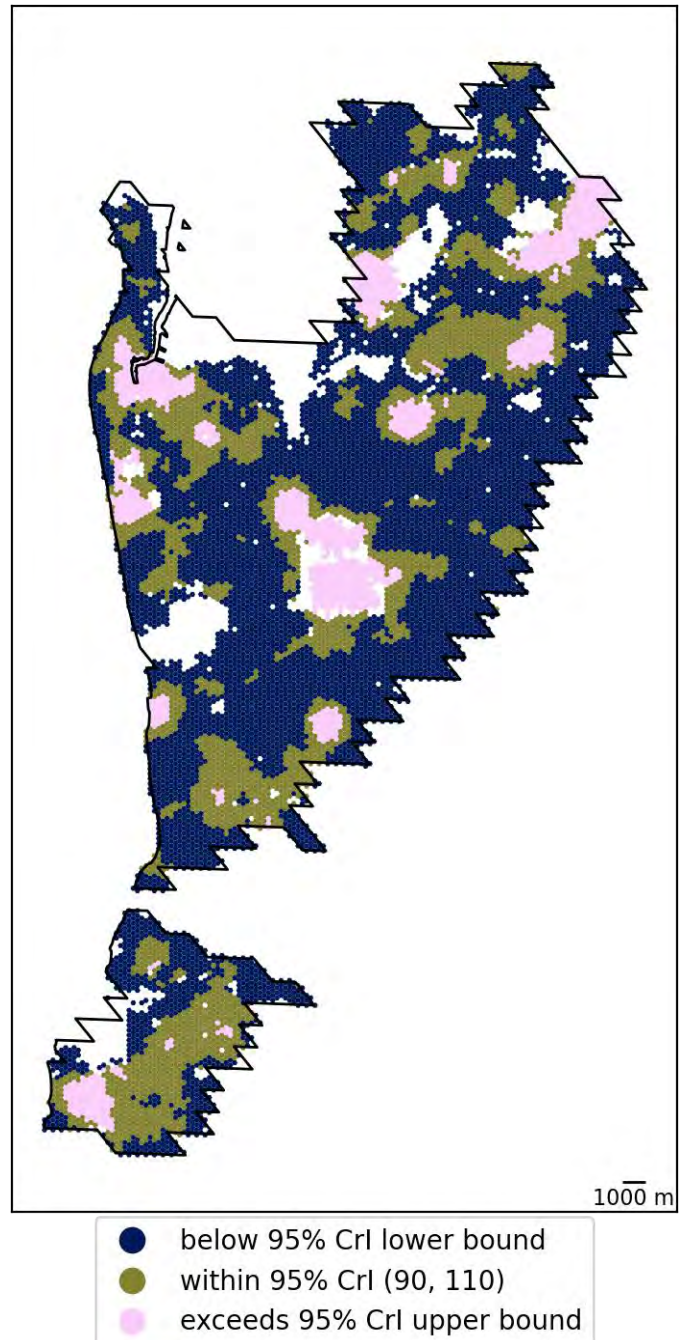

B: Estimated Mean 1000 m neighbourhood street intersections per km² requirement for reaching the WHO's target of a ≥15% relative reduction in insufficient physical activity through walking

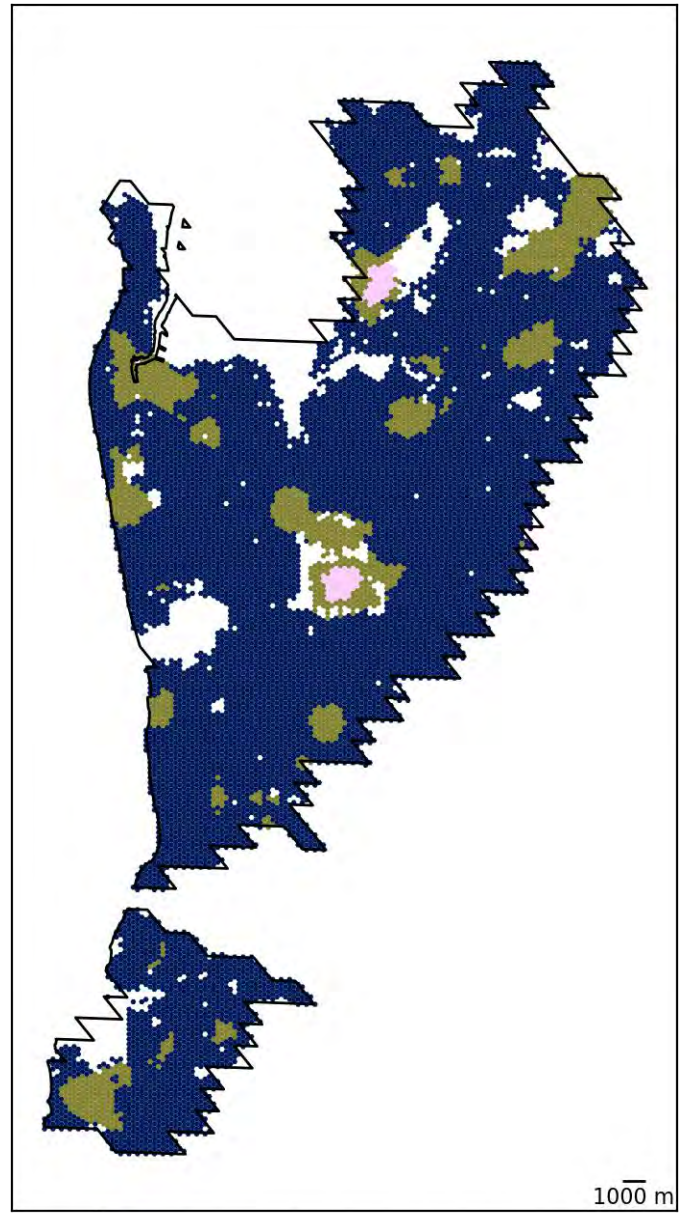

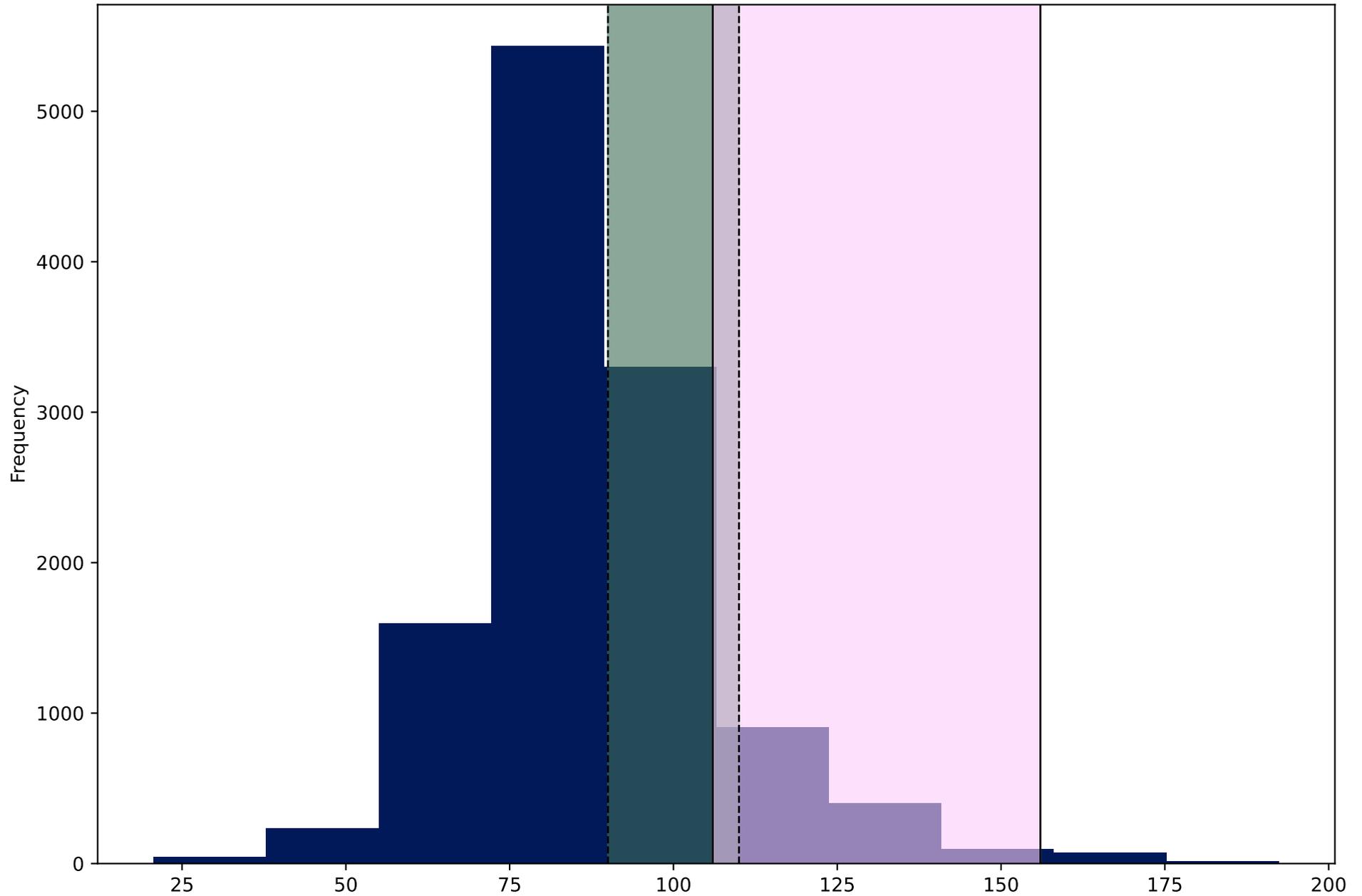

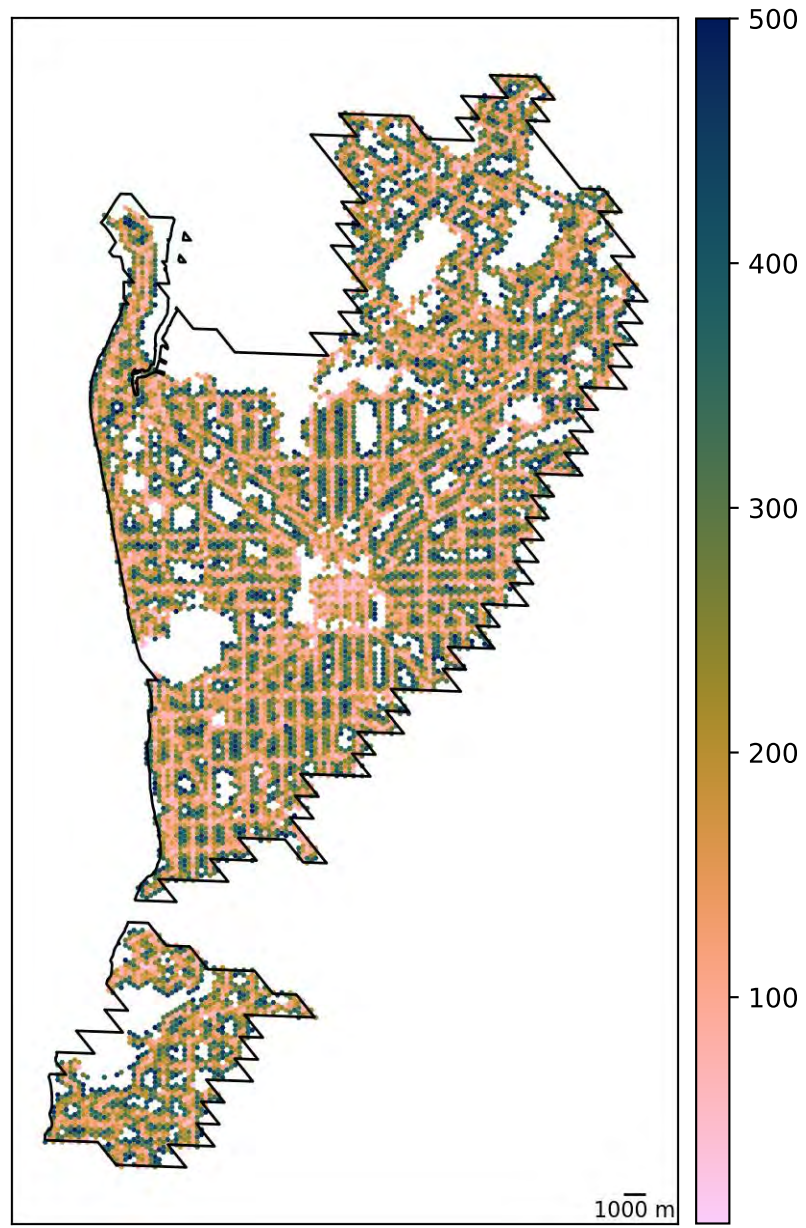

Distance to nearest public transport stops (m; up to 500m)

distances: Estimated Distance to nearest public transport stops (m; up to 500m) requirement for distances to destinations, measured up to a maximum distance target threshold of 500 metres

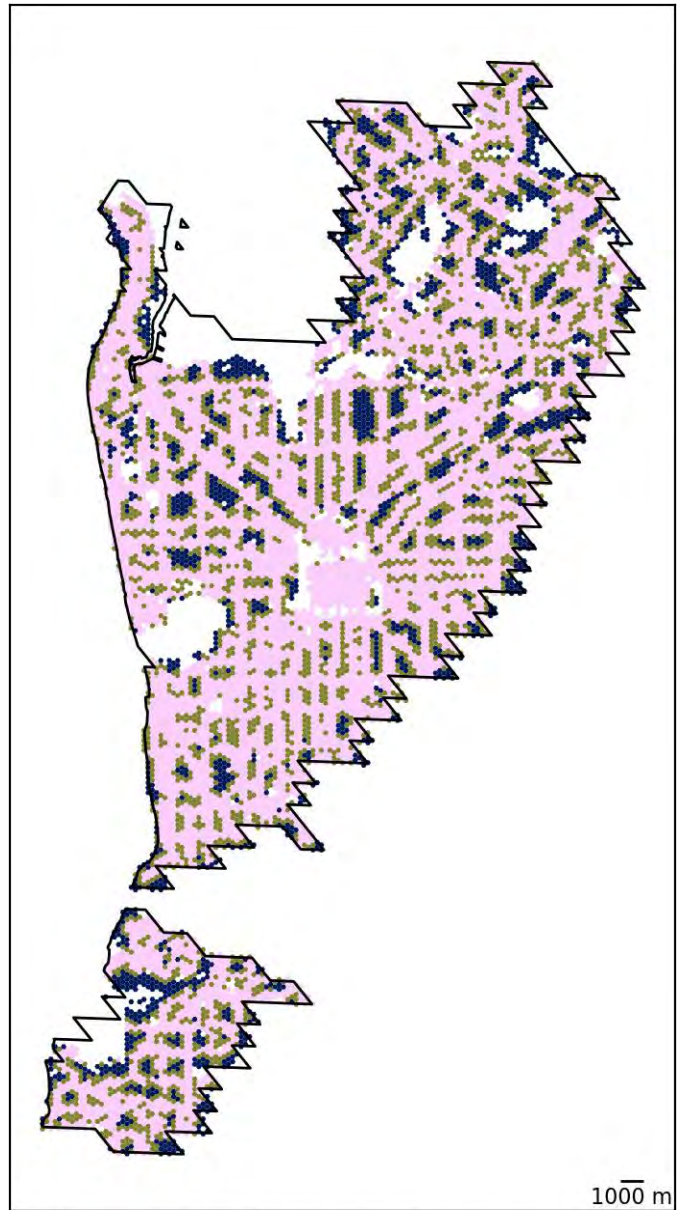

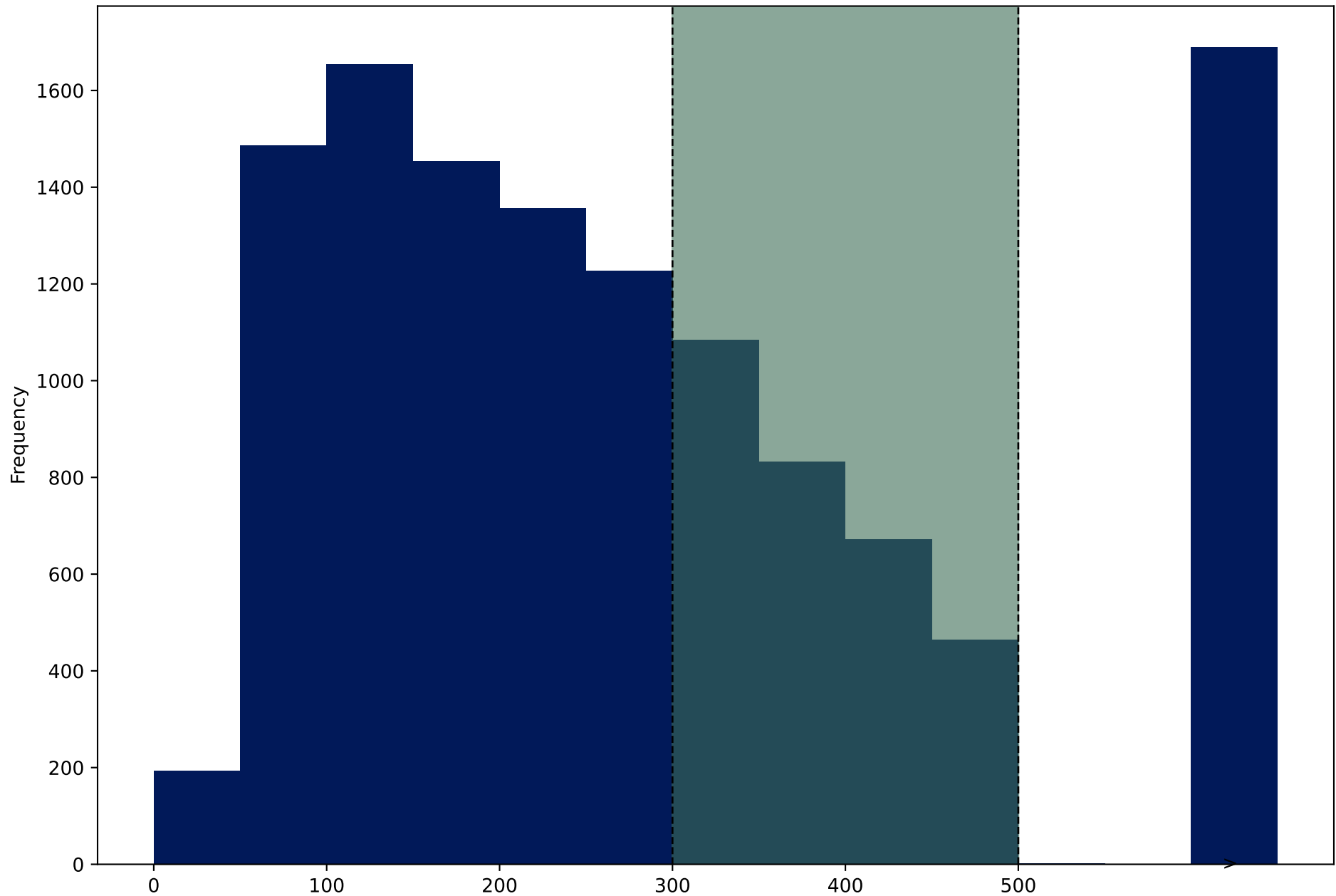

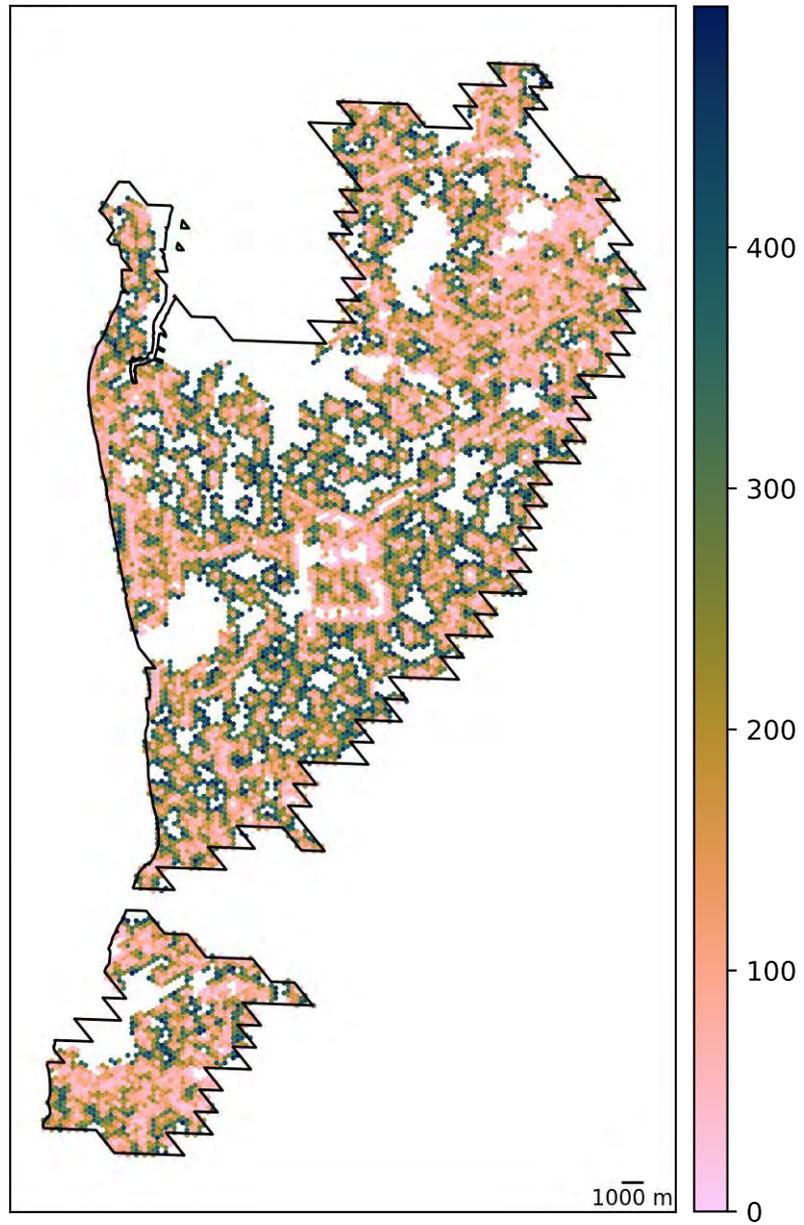

Distance to nearest park (m; up to 500m)

distances: Estimated Distance to nearest park (m; up to 500m) requirement for distances to destinations, measured up to a maximum distance target threshold of 500 metres

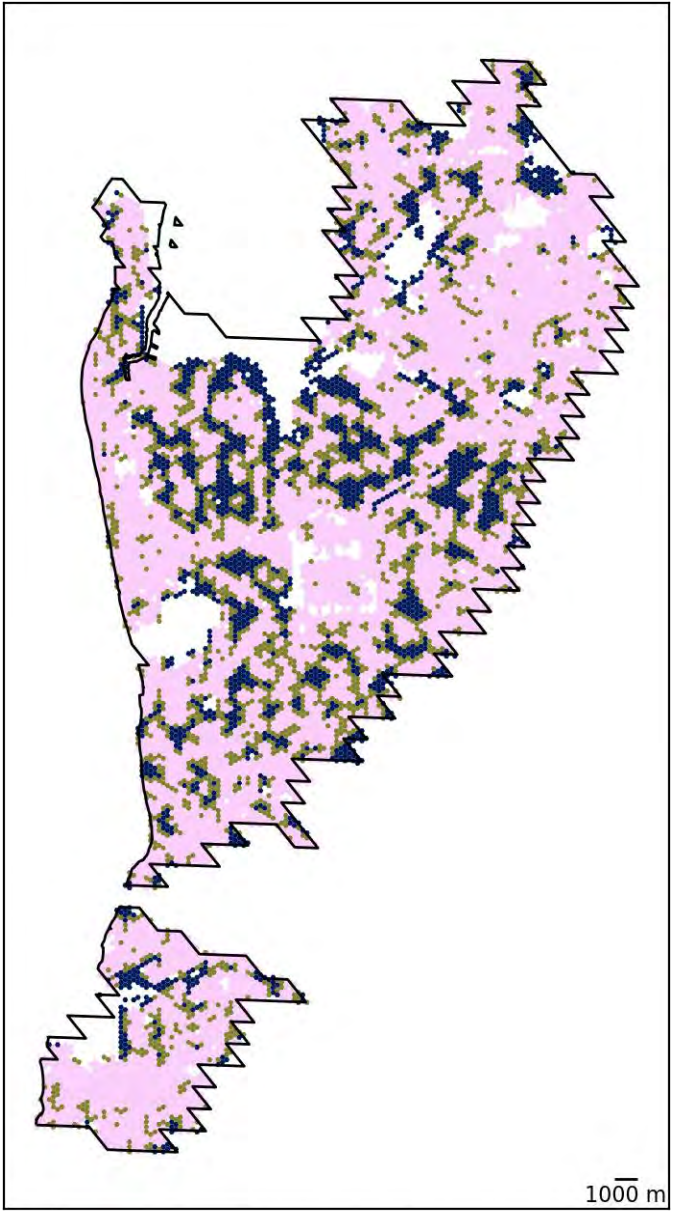

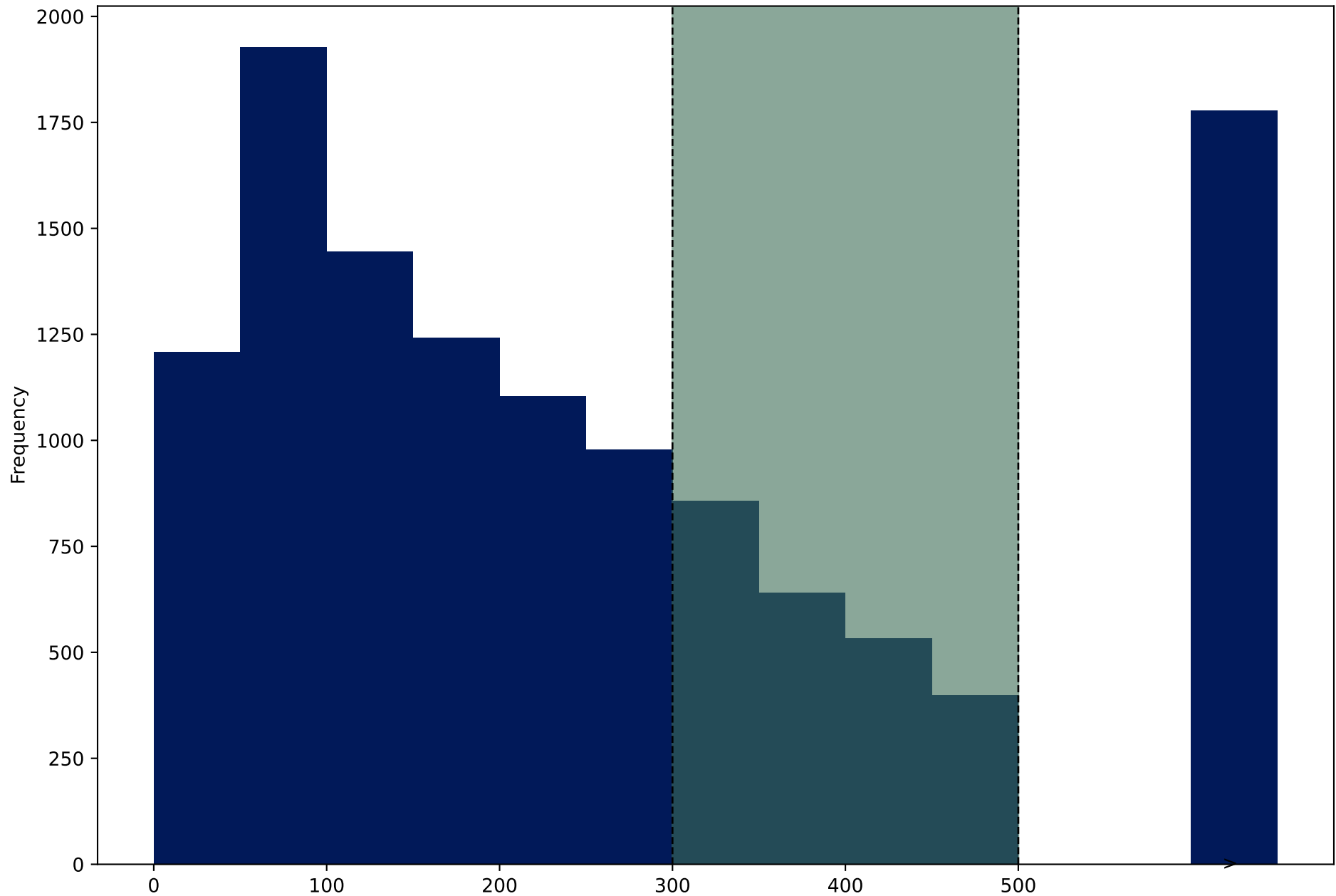

## Australasia, Australia, Melbourne

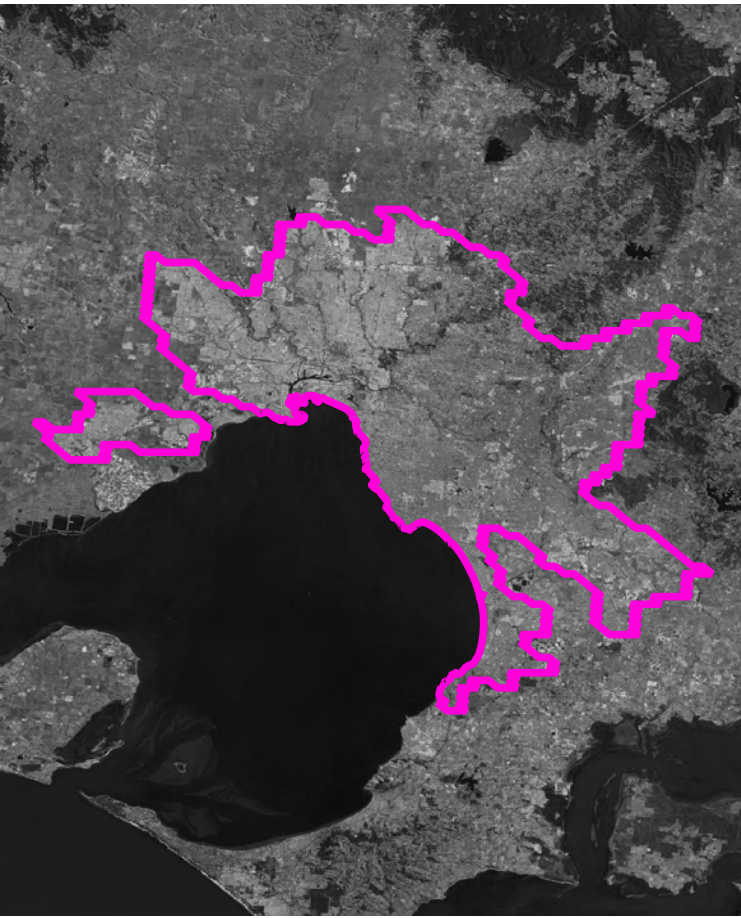
Satellite imagery of urban study region (Bing)

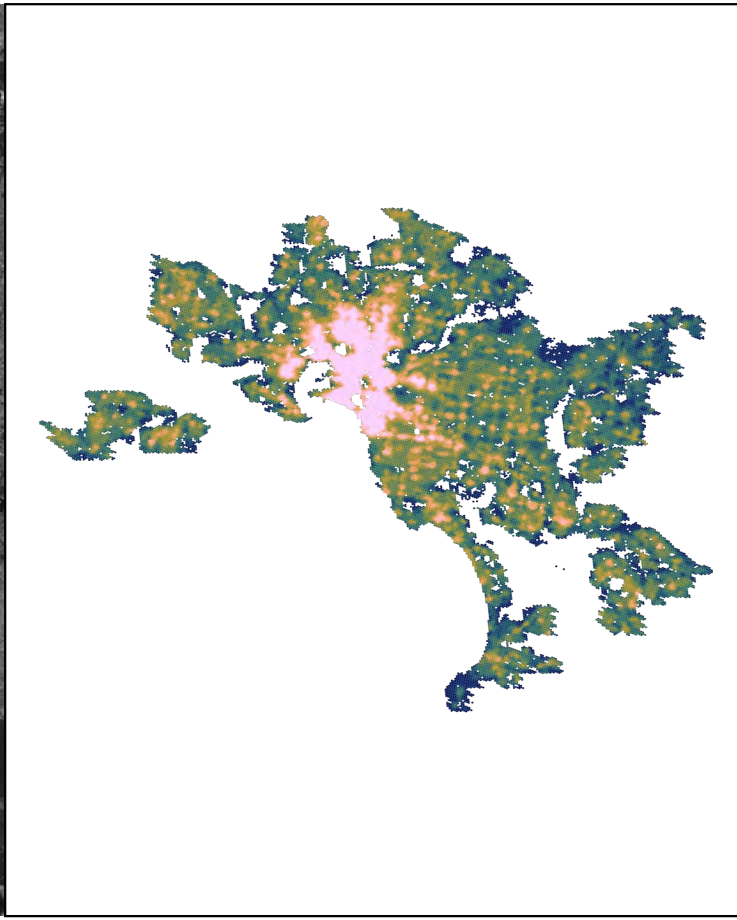
Walkability, relative to city

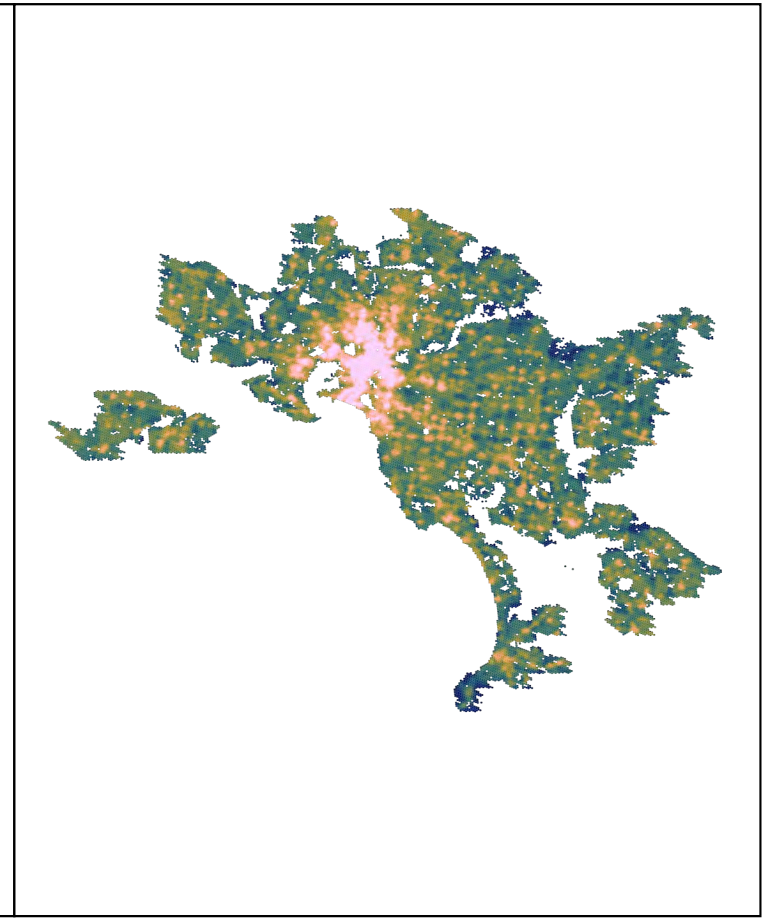
Walkability, relative to 25 global cities

Urban boundary

0 30 60 km

**Walkability score**
- <-3
- -3 to -2
- -2 to -1
- -1 to 0
- 0 to 1
- 1 to 2
- 2 to 3
- ≥3

Walkability relative to all cities by component variables (2D histograms), and overall (histogram)

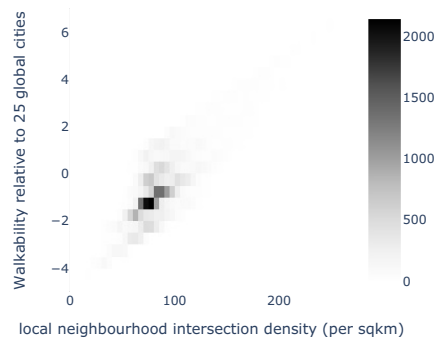
local neighbourhood intersection density (per sqkm)

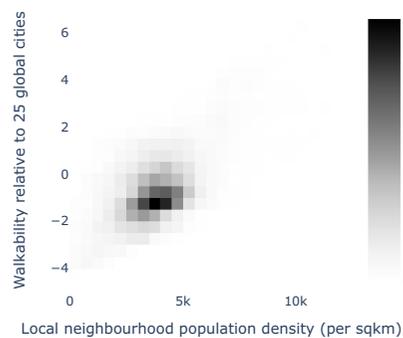
Local neighbourhood population density (per sqkm)

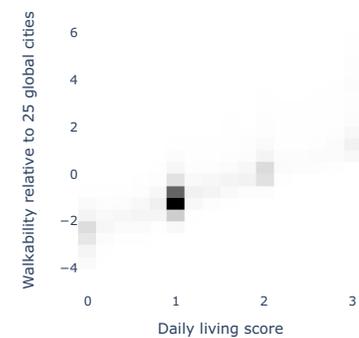
Daily living score

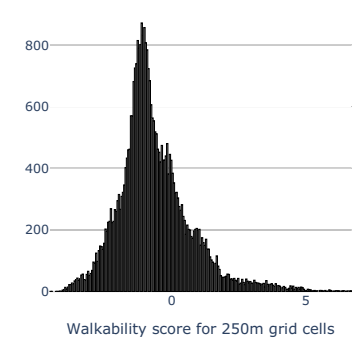
Walkability score for 250m grid cells

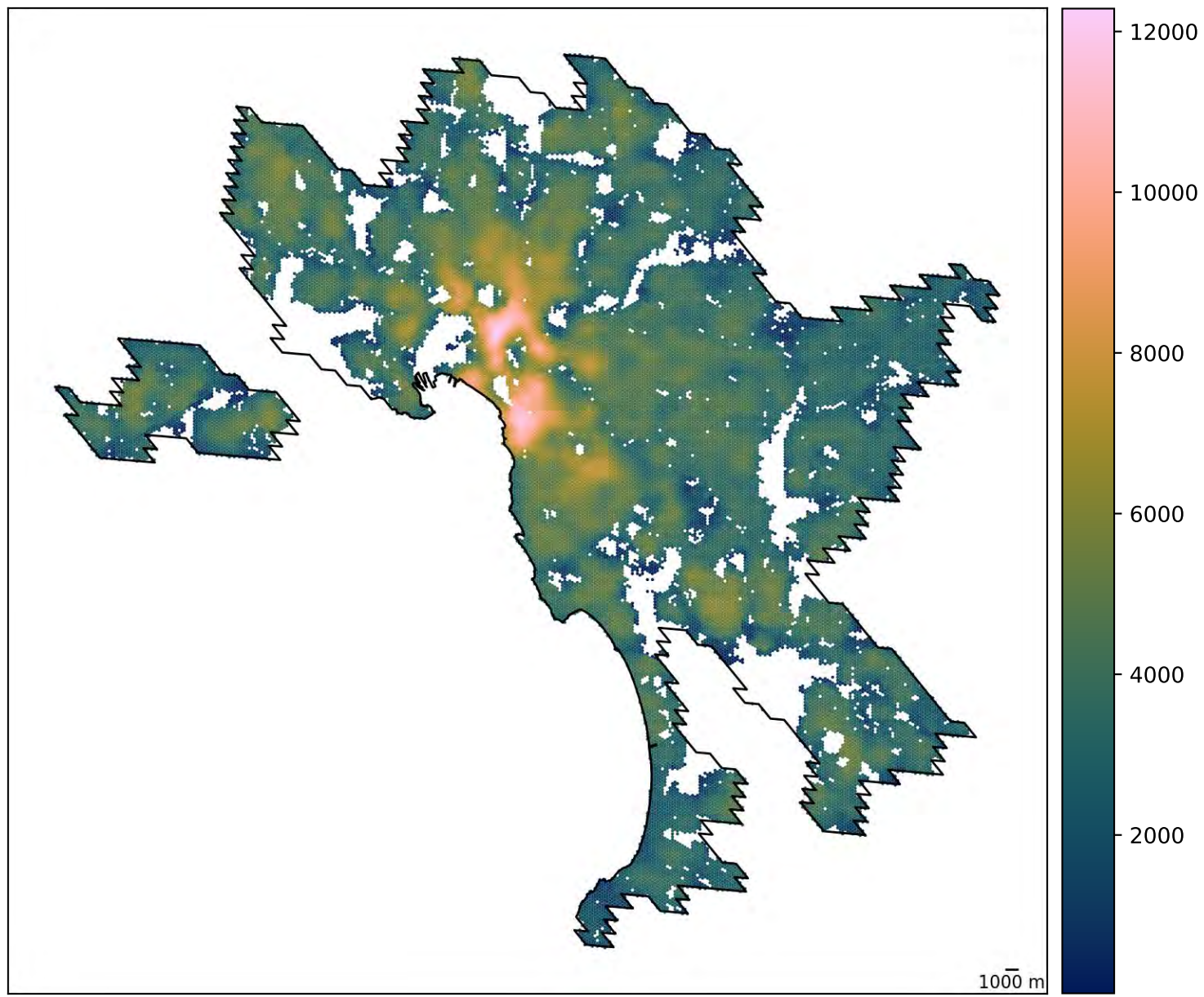

A: Estimated Mean 1000 m neighbourhood population per km² requirement for ≥80% probability of engaging in walking for transport

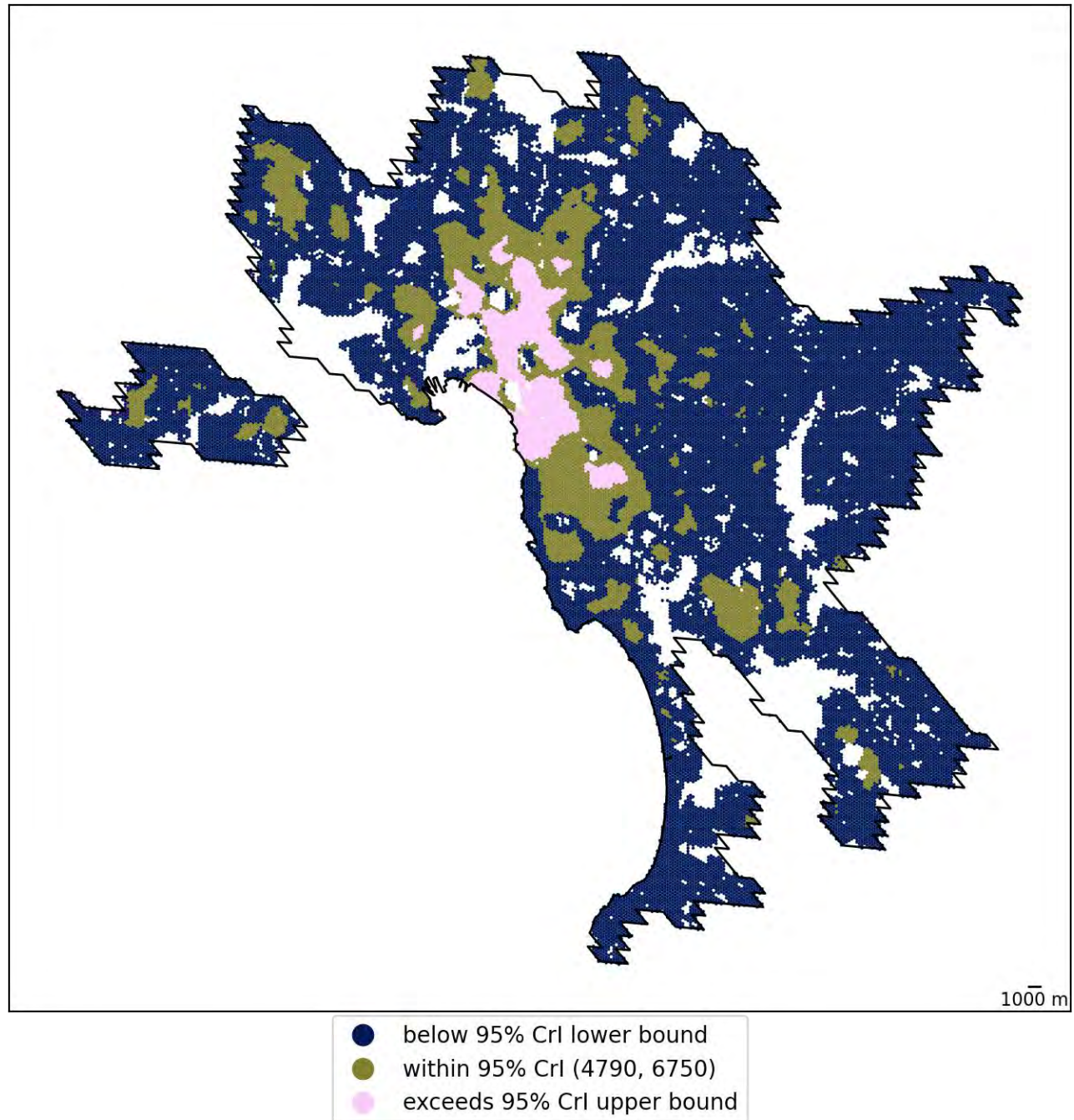

B: Estimated Mean 1000 m neighbourhood population per km² requirement for reaching the WHO's target of a ≥15% relative reduction in insufficient physical activity through walking

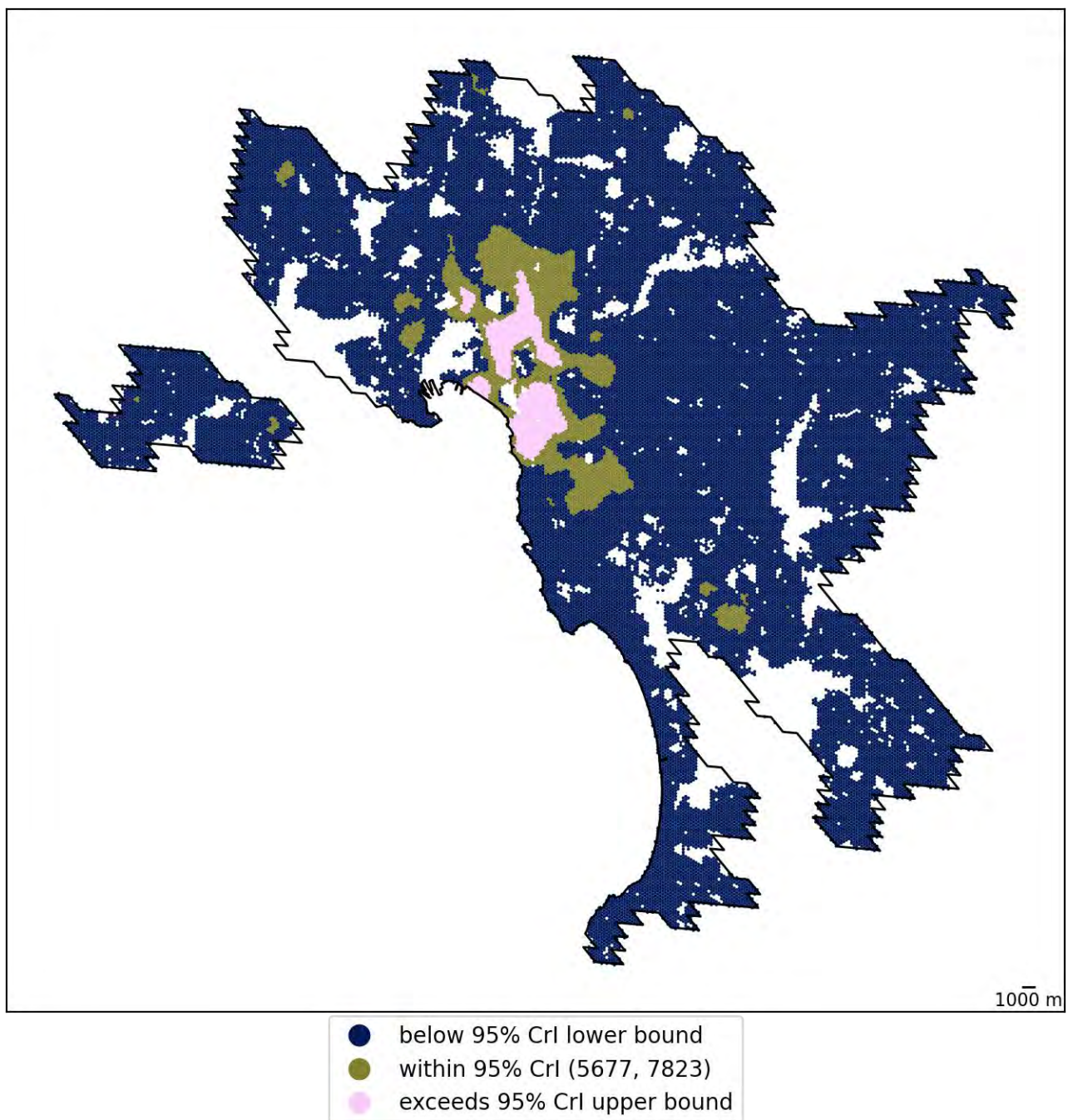

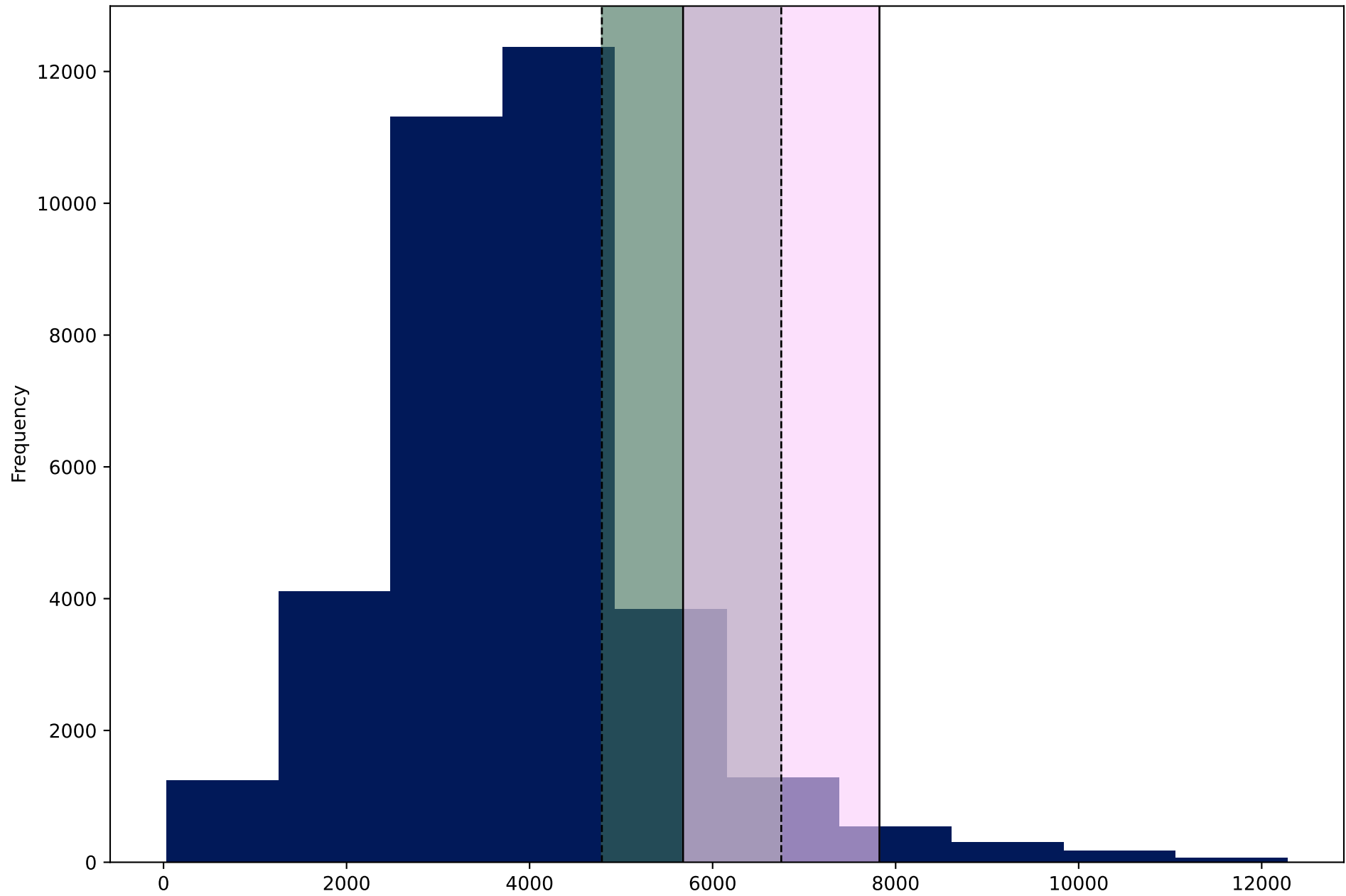

Mean 1000 m neighbourhood street intersections per km²

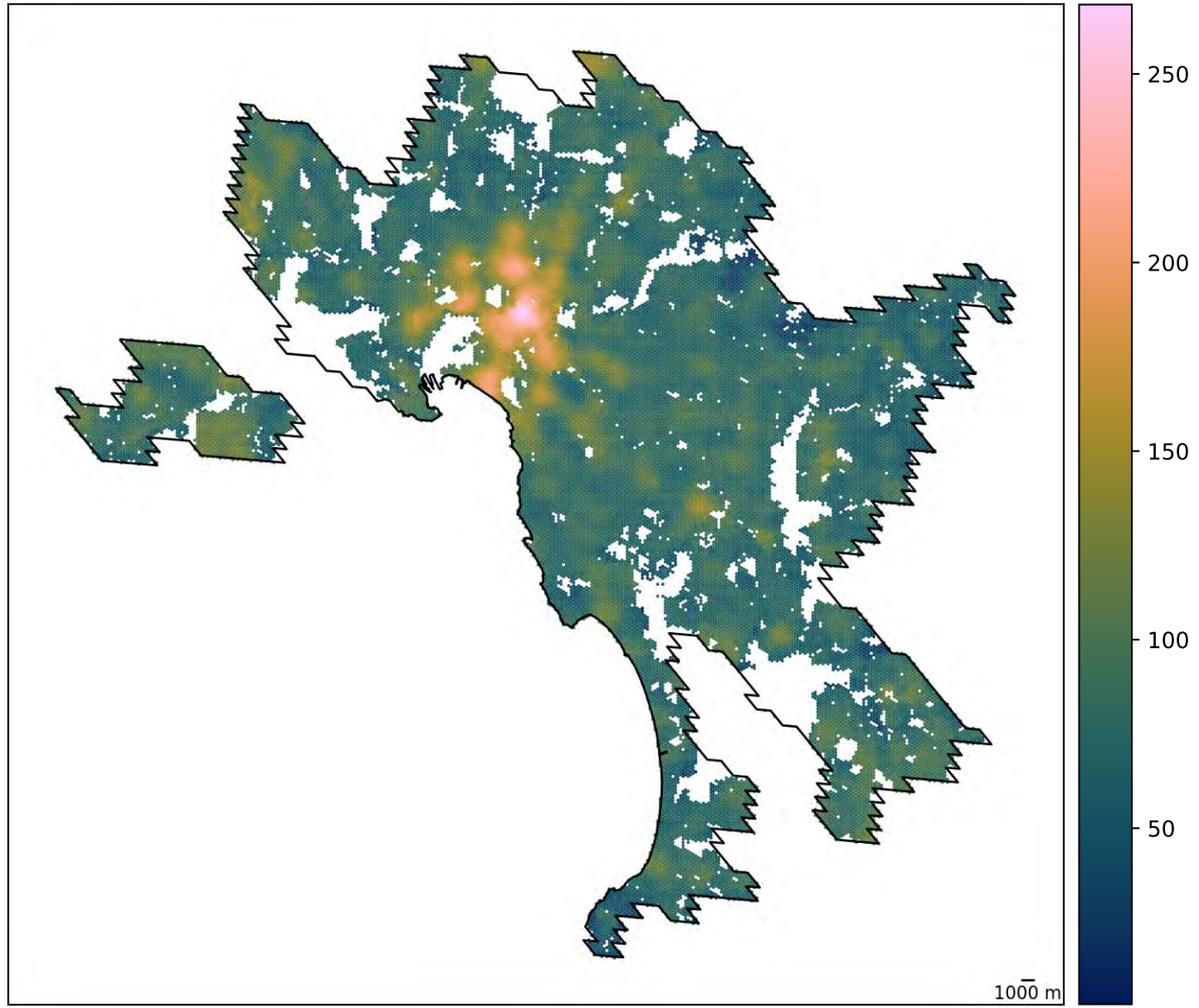

A: Estimated Mean 1000 m neighbourhood street intersections per km² requirement for ≥80% probability of engaging in walking for transport

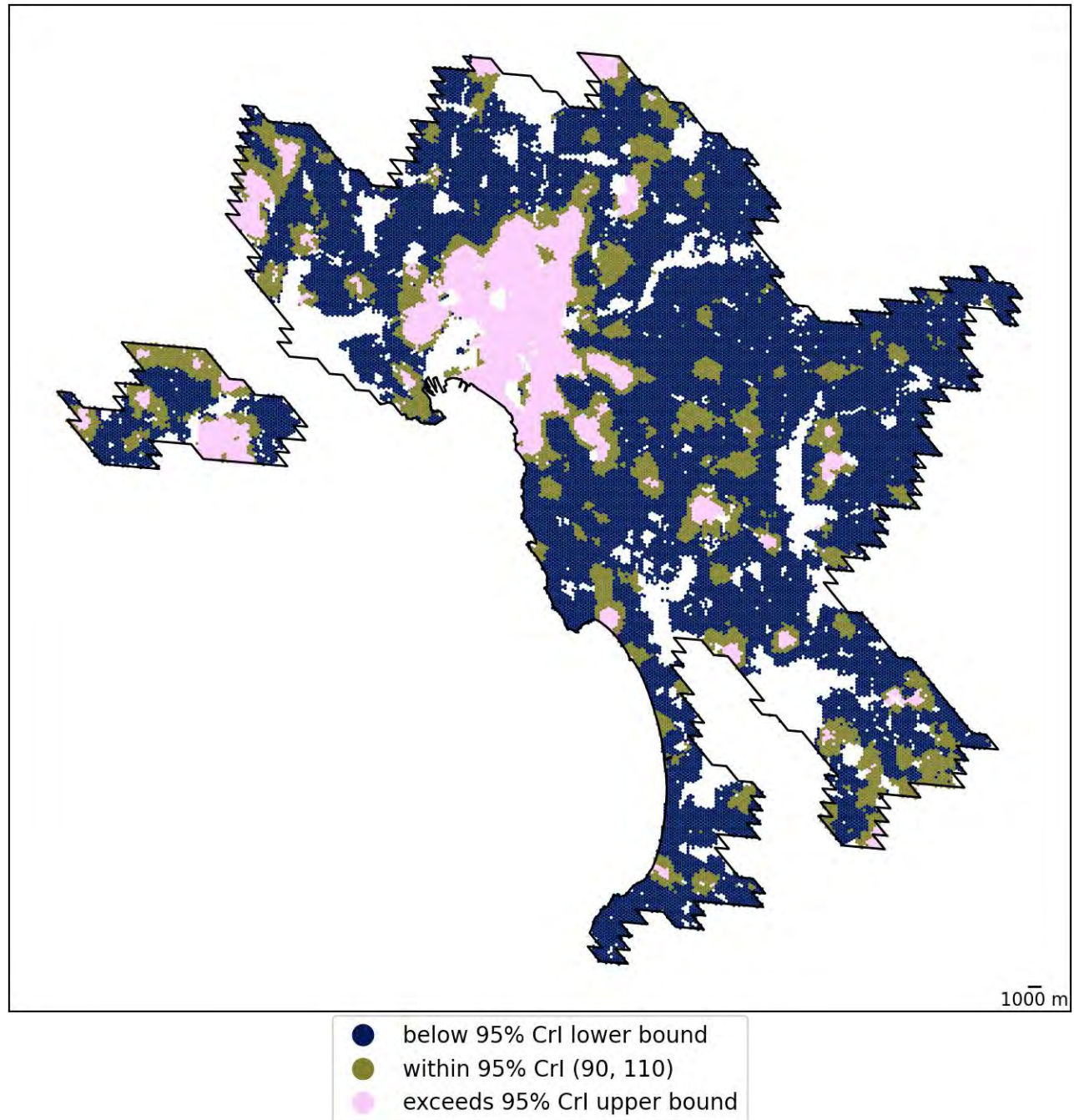

- below 95% CrI lower bound
- within 95% CrI (90, 110)
- exceeds 95% CrI upper bound

B: Estimated Mean 1000 m neighbourhood street intersections per km² requirement for reaching the WHO's target of a ≥15% relative reduction in insufficient physical activity through walking

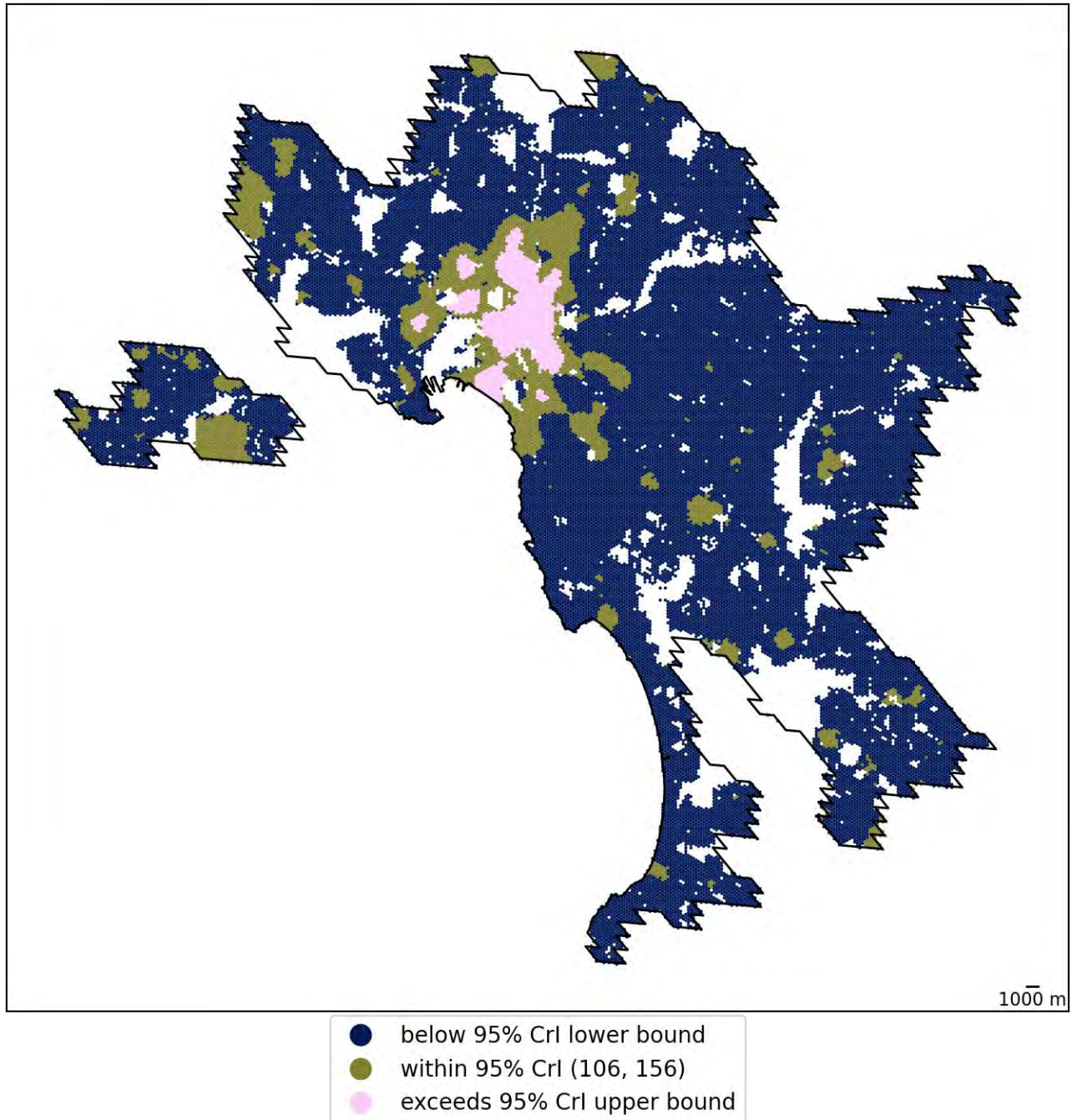

- below 95% CrI lower bound
- within 95% CrI (106, 156)
- exceeds 95% CrI upper bound

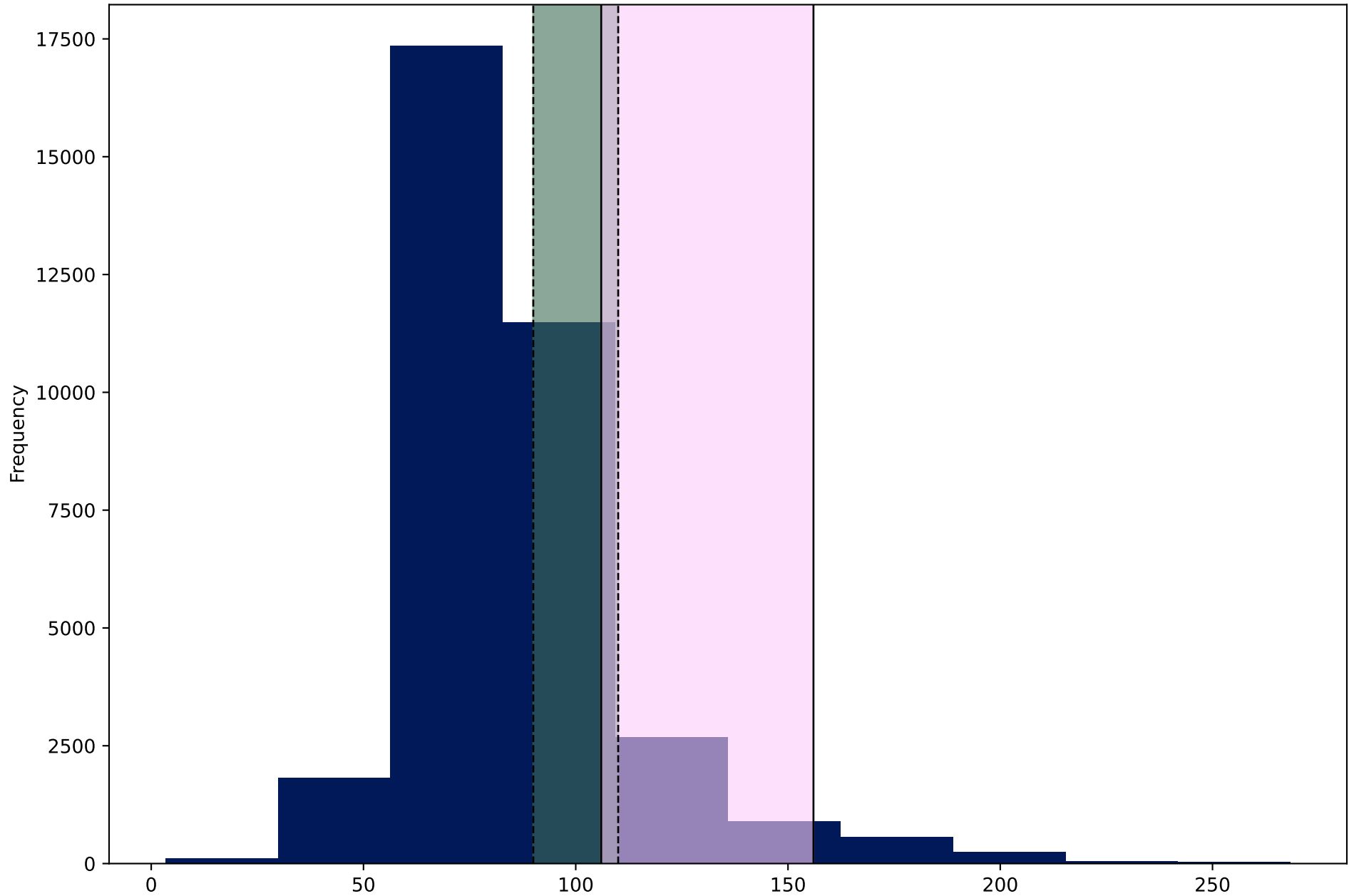

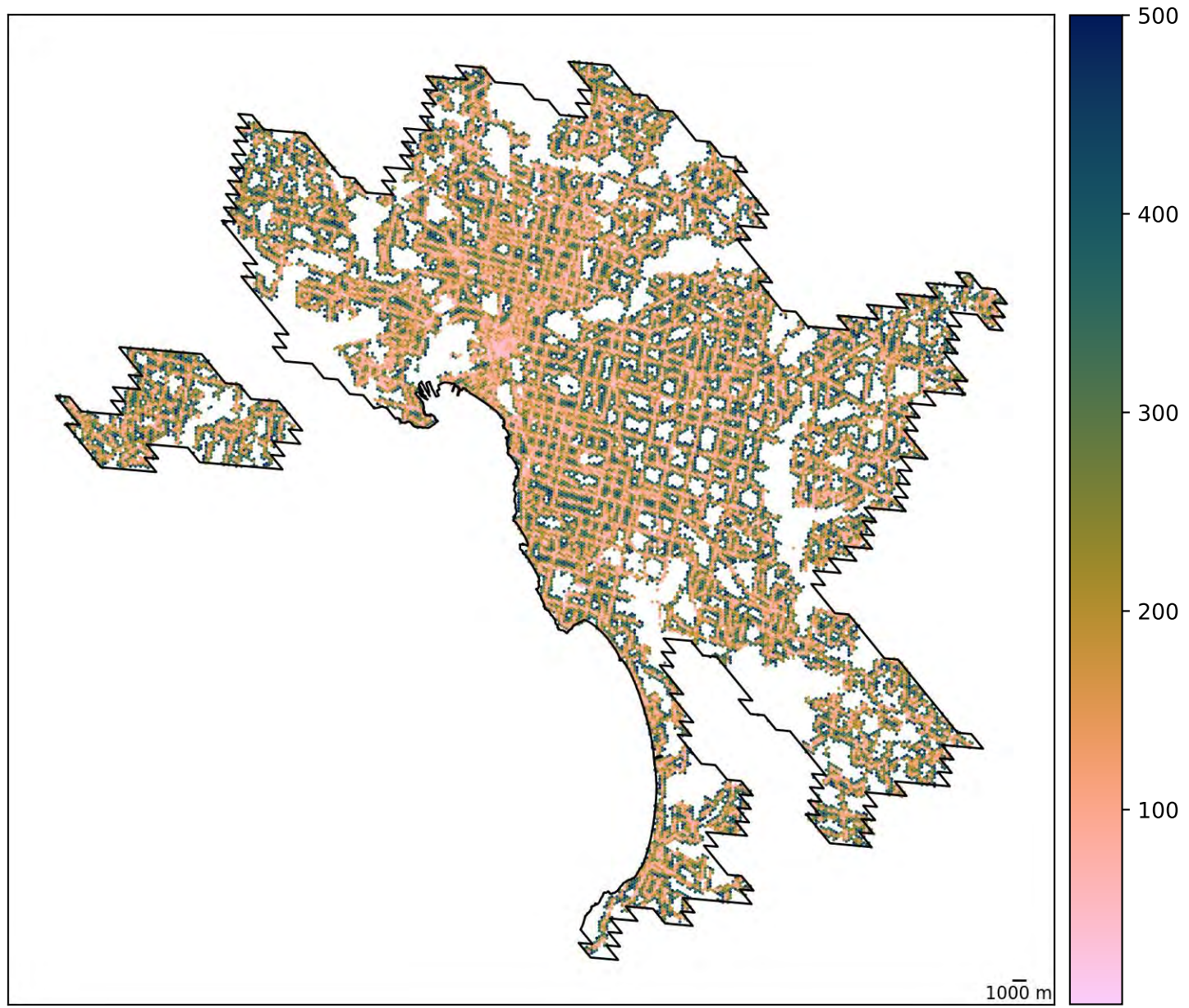

Distance to nearest public transport stops (m; up to 500m)

distances: Estimated Distance to nearest public transport stops (m; up to 500m) requirement for distances to destinations, measured up to a maximum distance target threshold of 500 metres

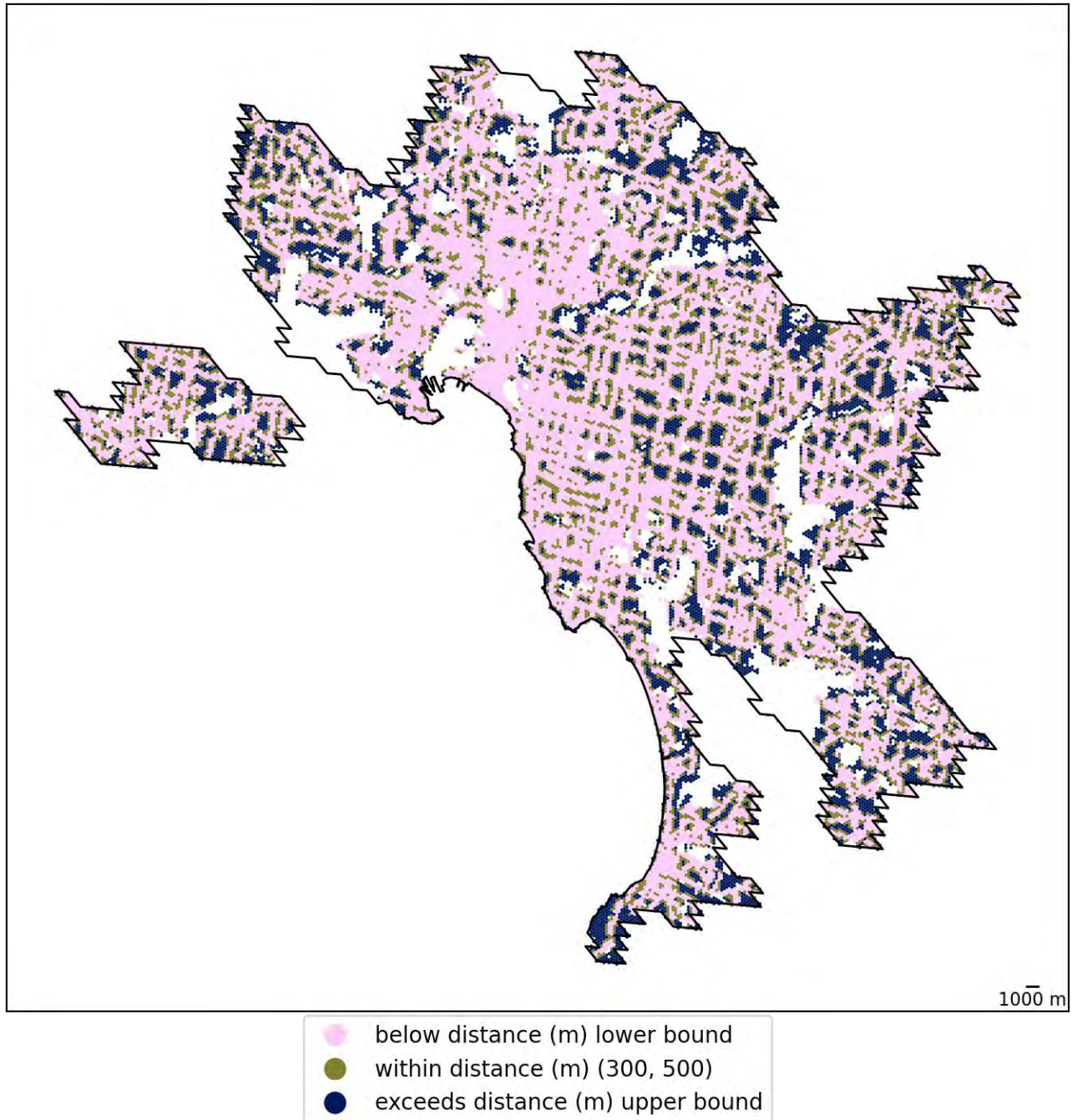

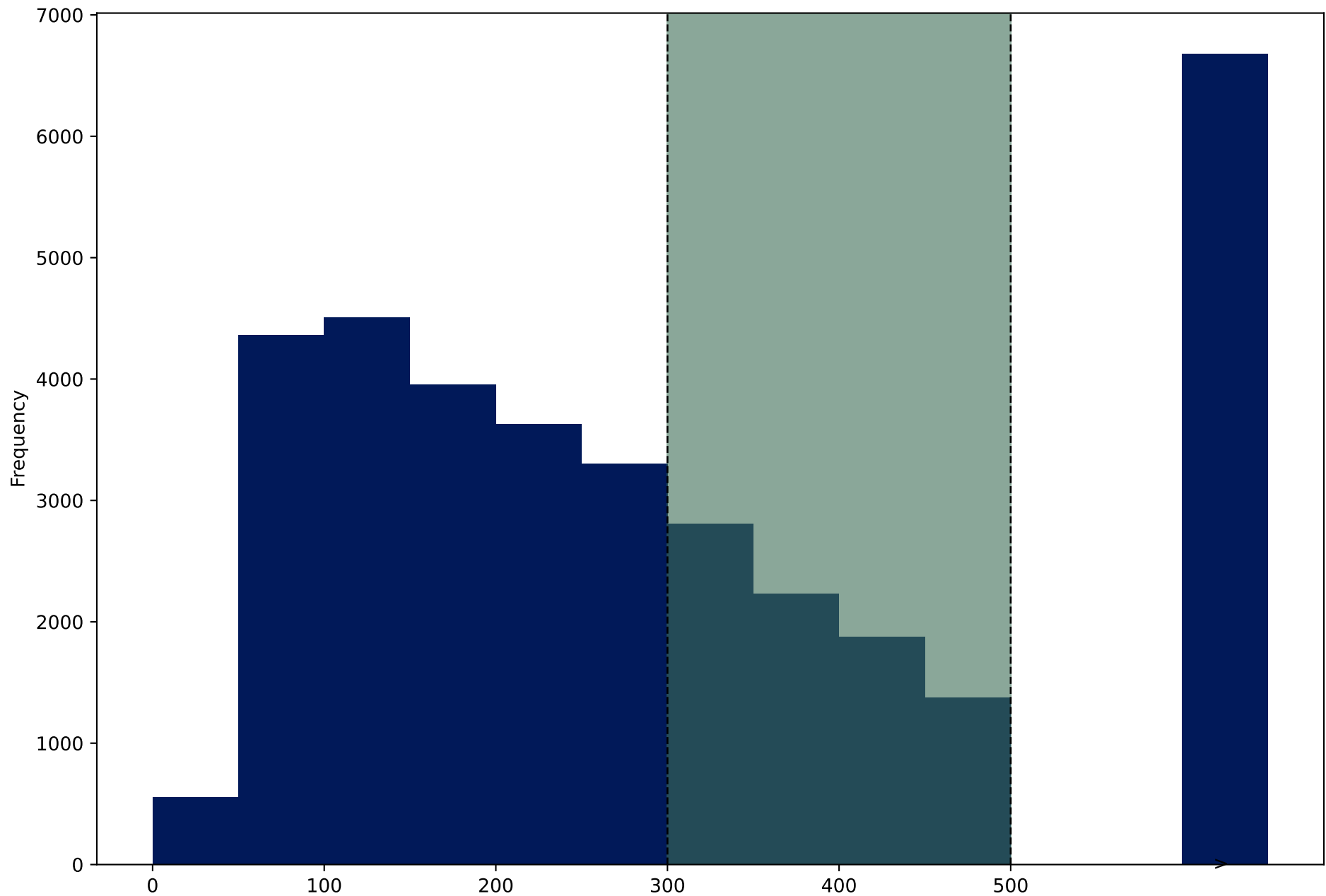

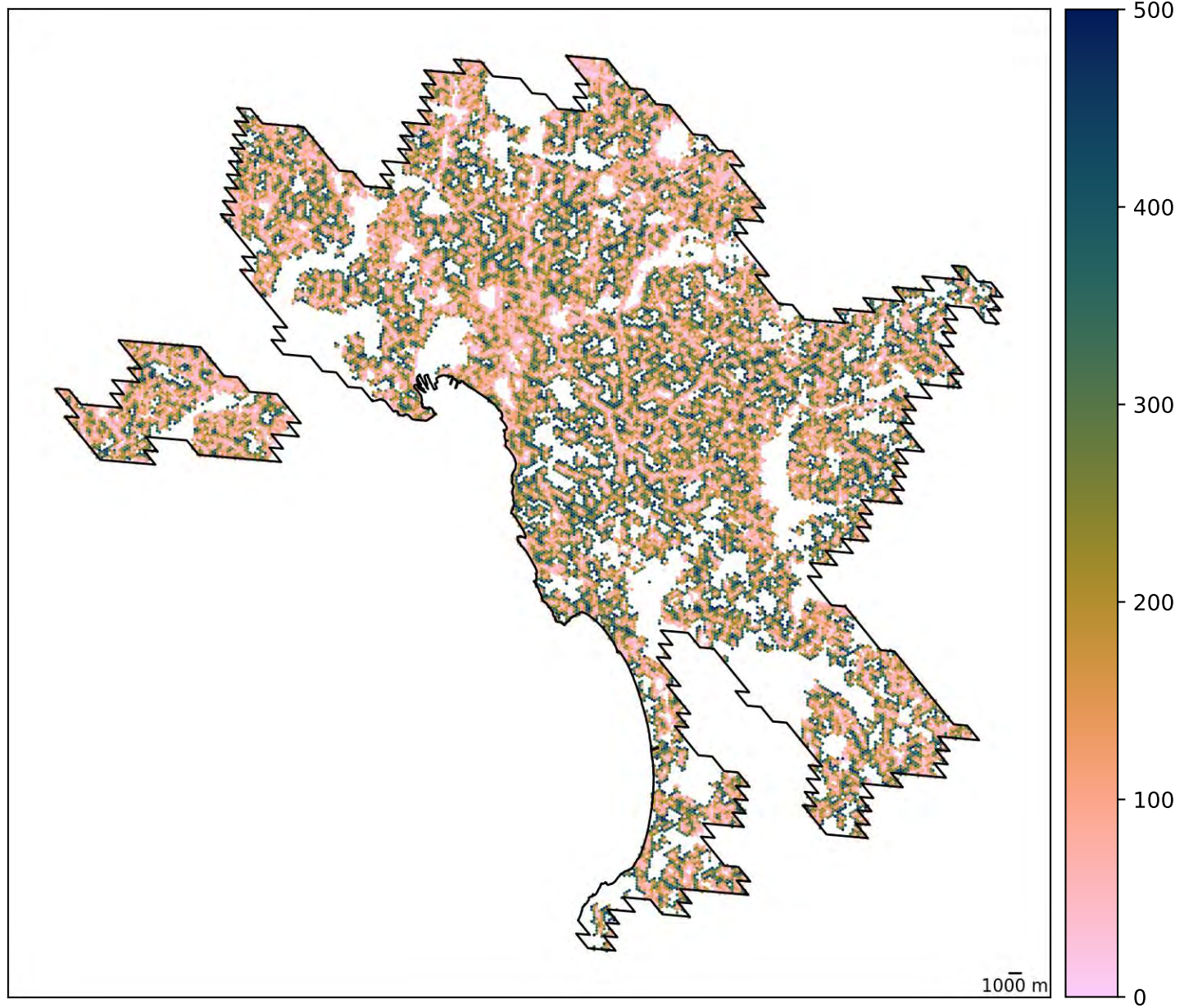

Distance to nearest park (m; up to 500m)

distances: Estimated Distance to nearest park (m; up to 500m) requirement for distances to destinations, measured up to a maximum distance target threshold of 500 metres

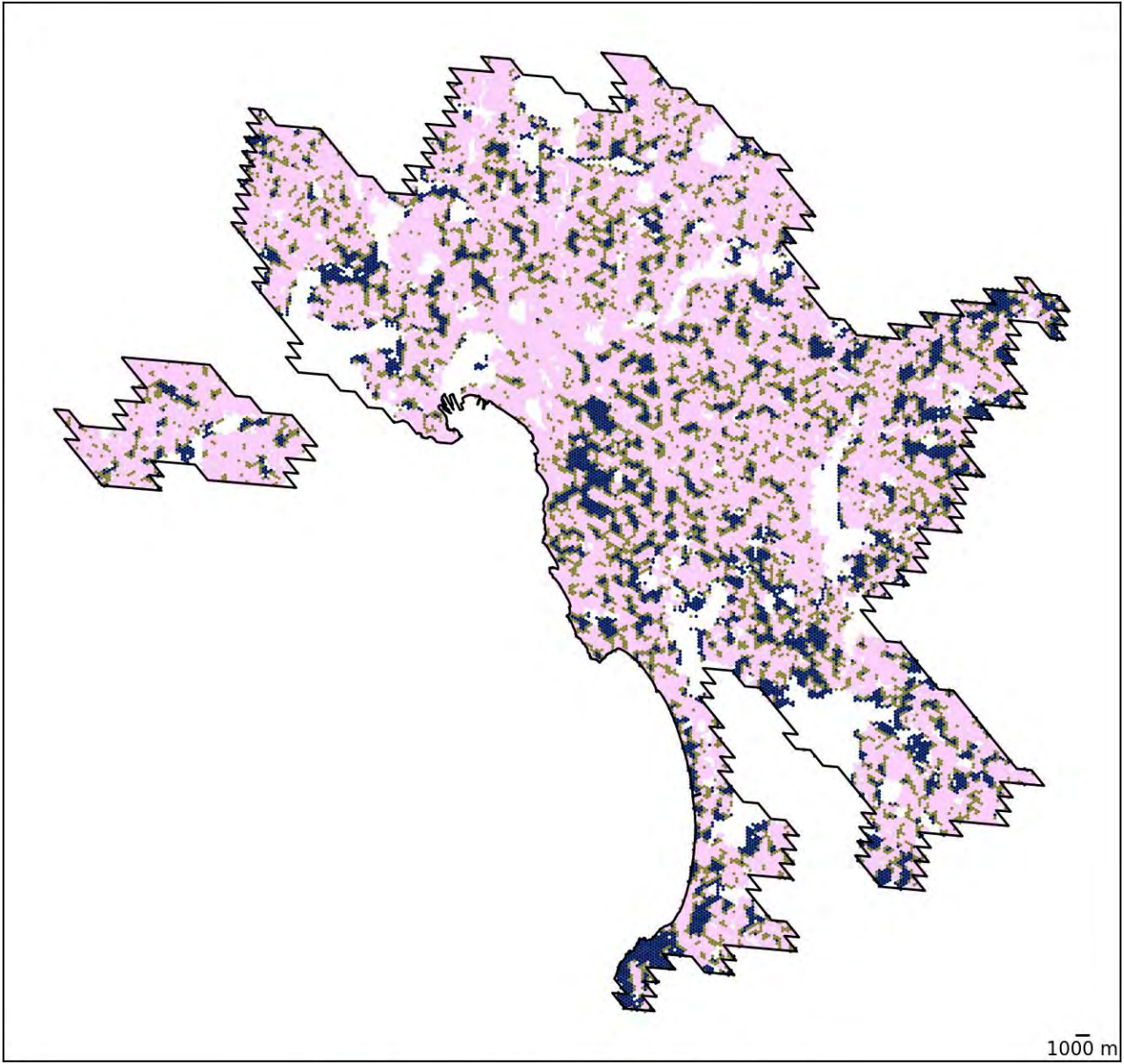

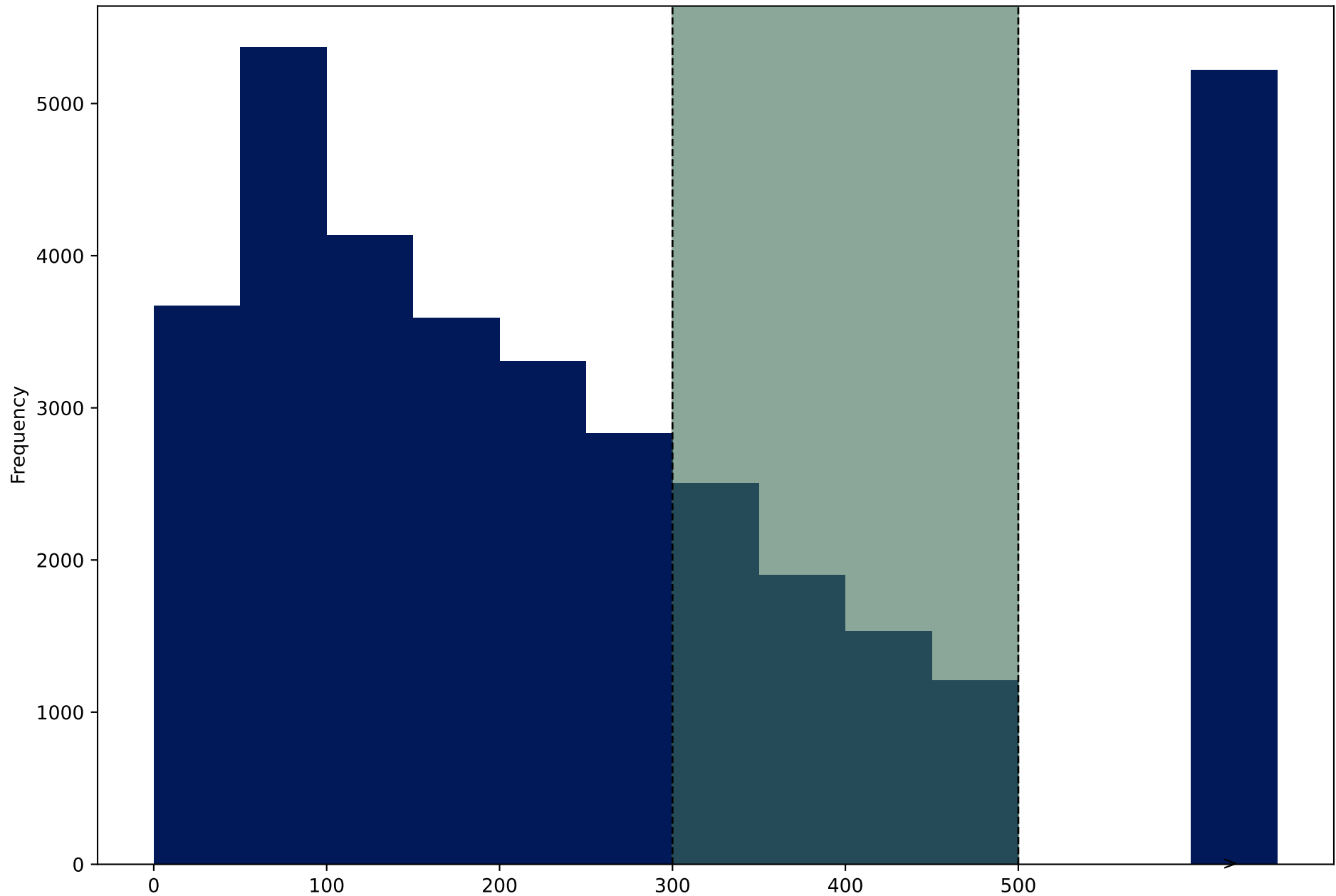

# Australasia, Australia, Sydney

| Satellite imagery of urban study region (Bing) | Walkability, relative to city | Walkability, relative to 25 global cities |
|---|---|---|
| 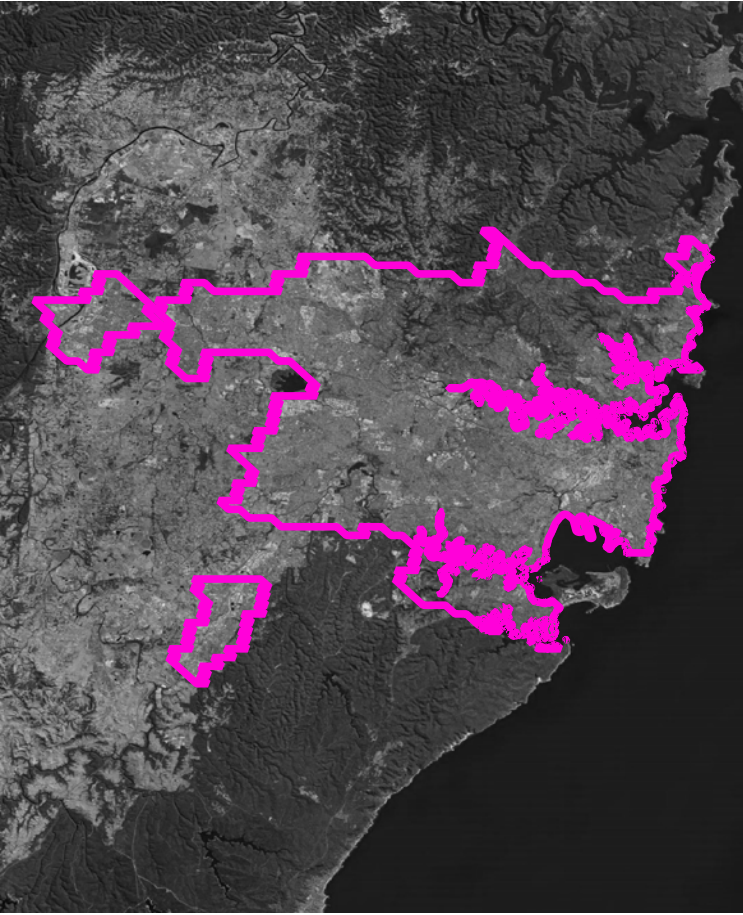 | 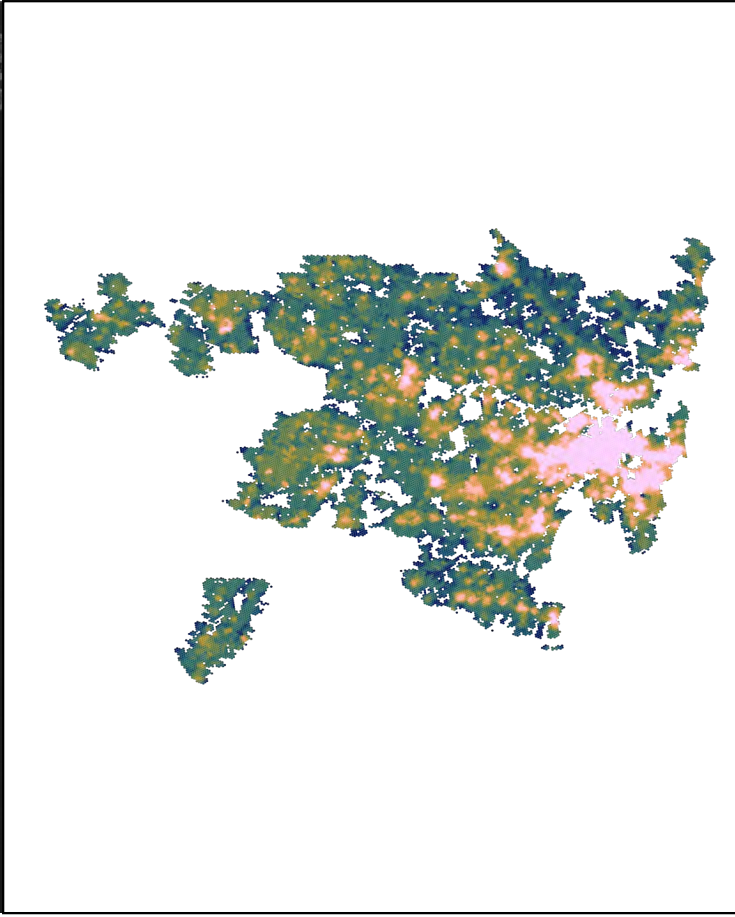 | 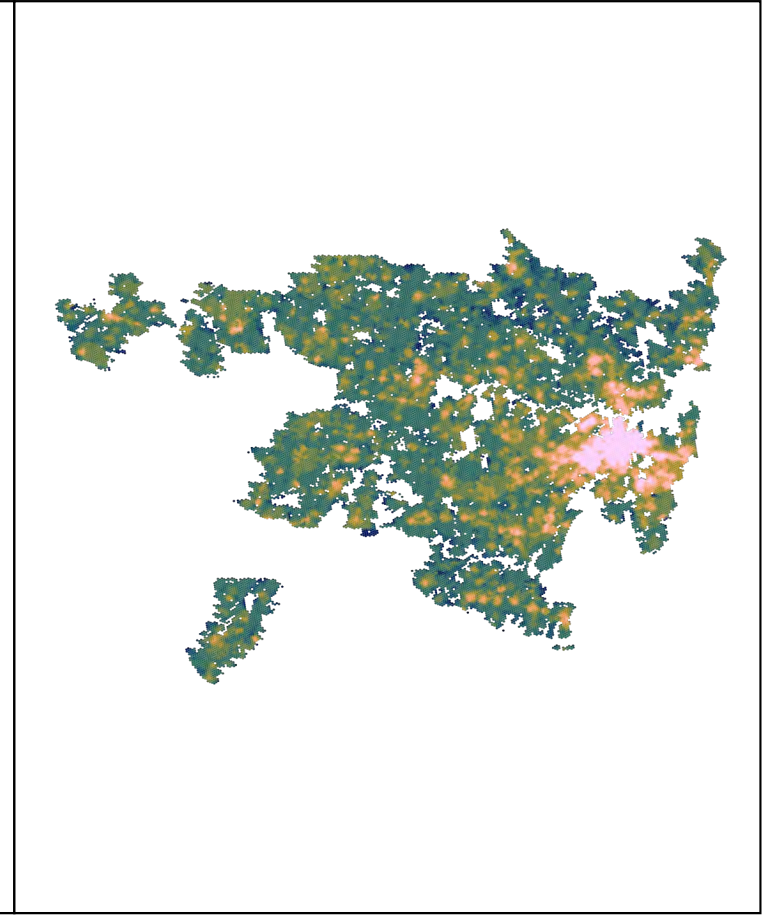 |

Walkability relative to all cities by component variables (2D histograms), and overall (histogram)

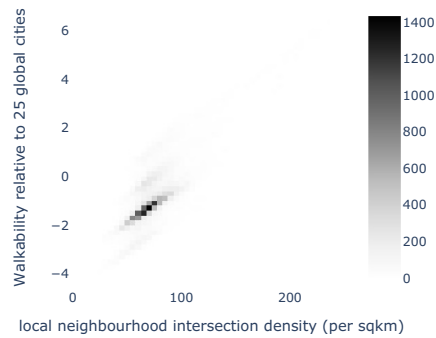 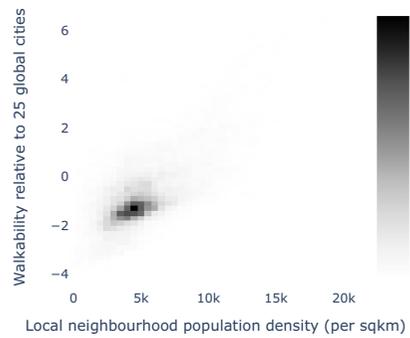 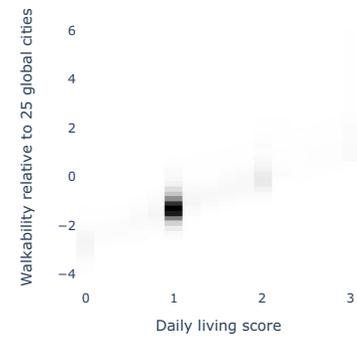 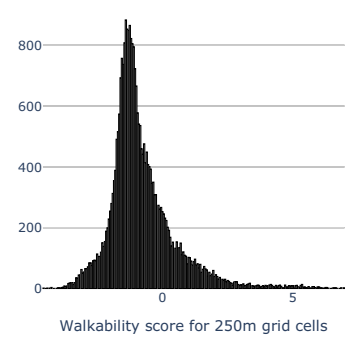

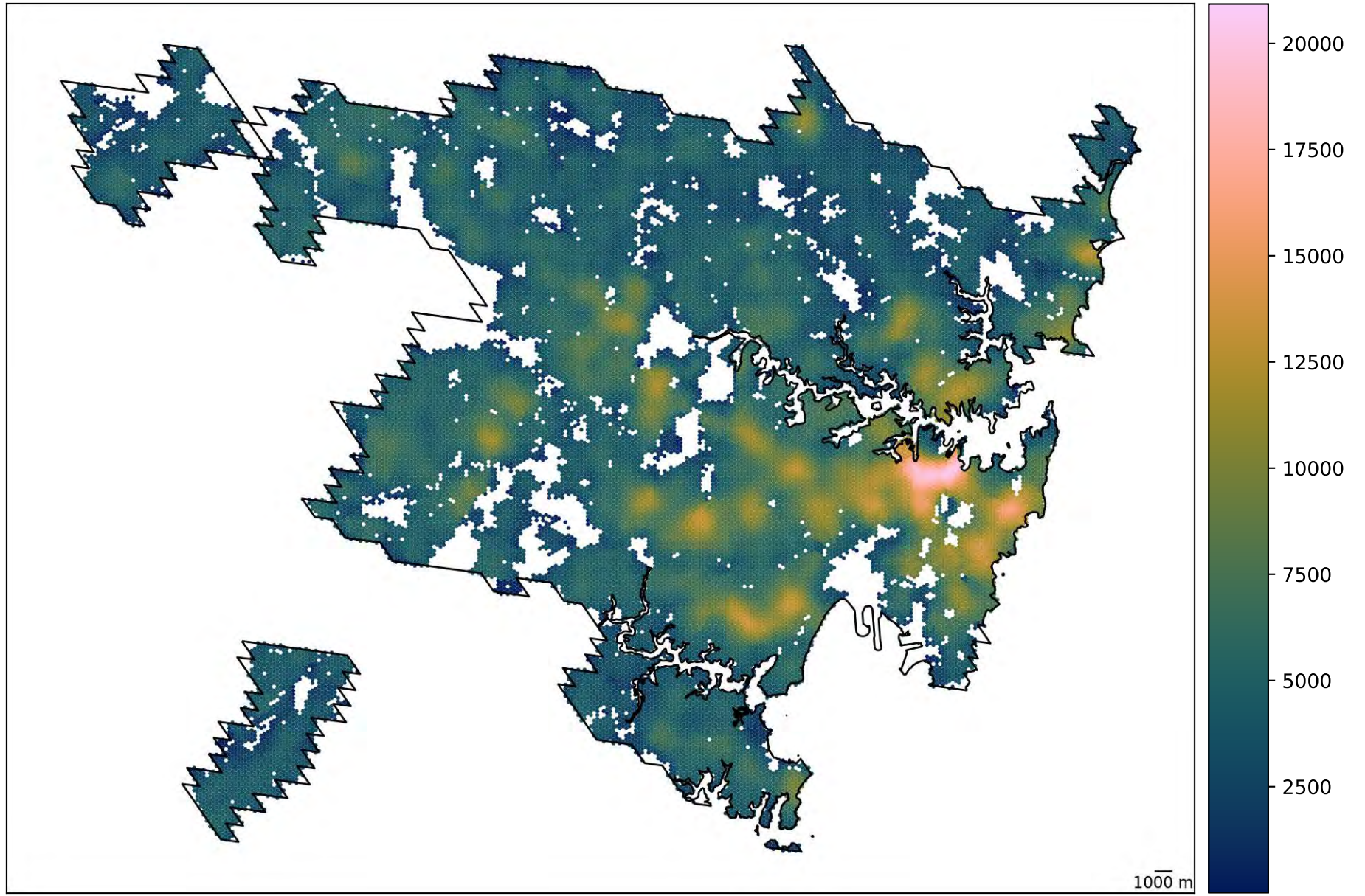

Mean 1000 m neighbourhood population per km²

A: Estimated Mean 1000 m neighbourhood population per km² requirement for ≥80% probability of engaging in walking for transport

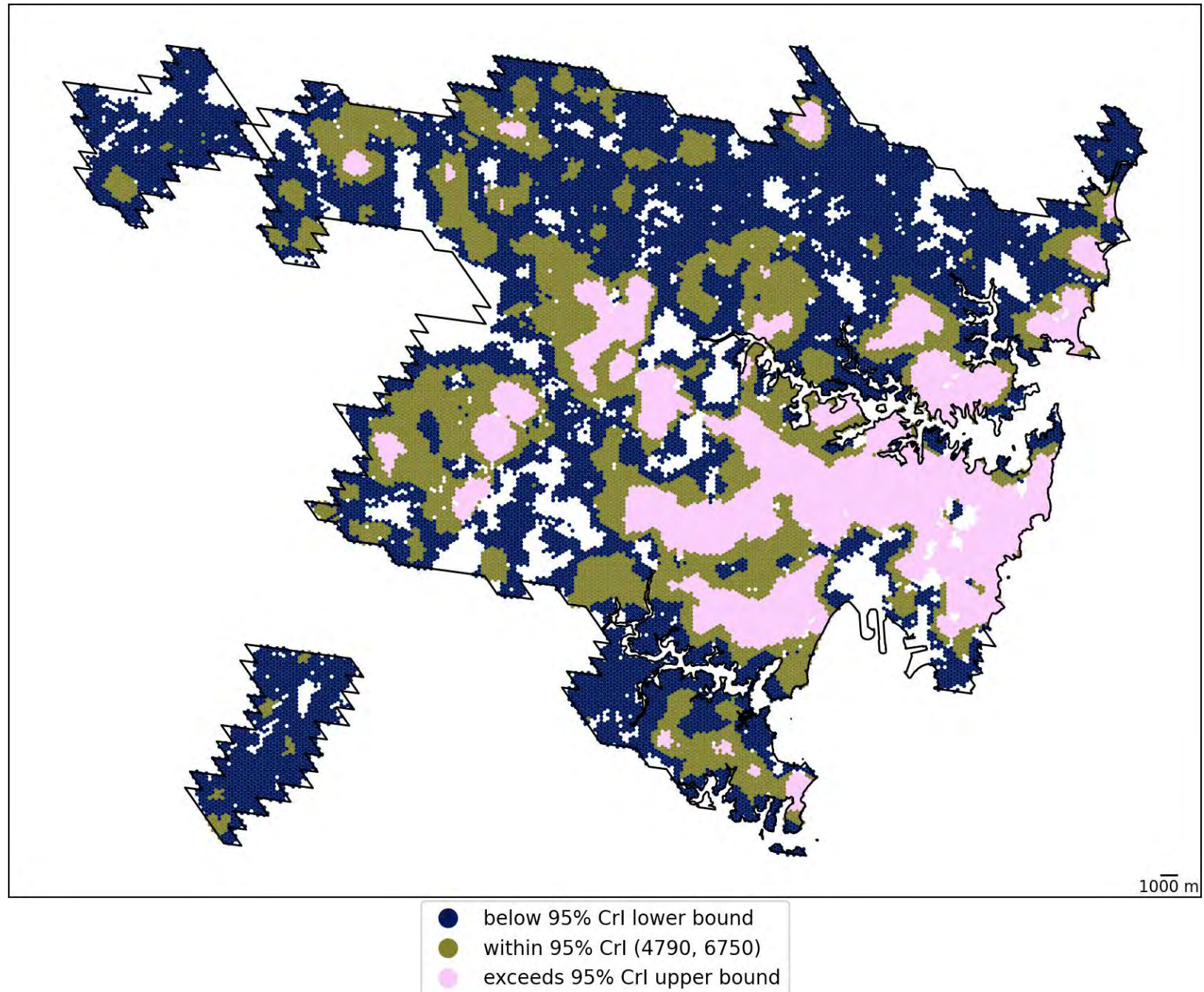

B: Estimated Mean 1000 m neighbourhood population per km² requirement for reaching the WHO's target of a ≥15% relative reduction in insufficient physical activity through walking

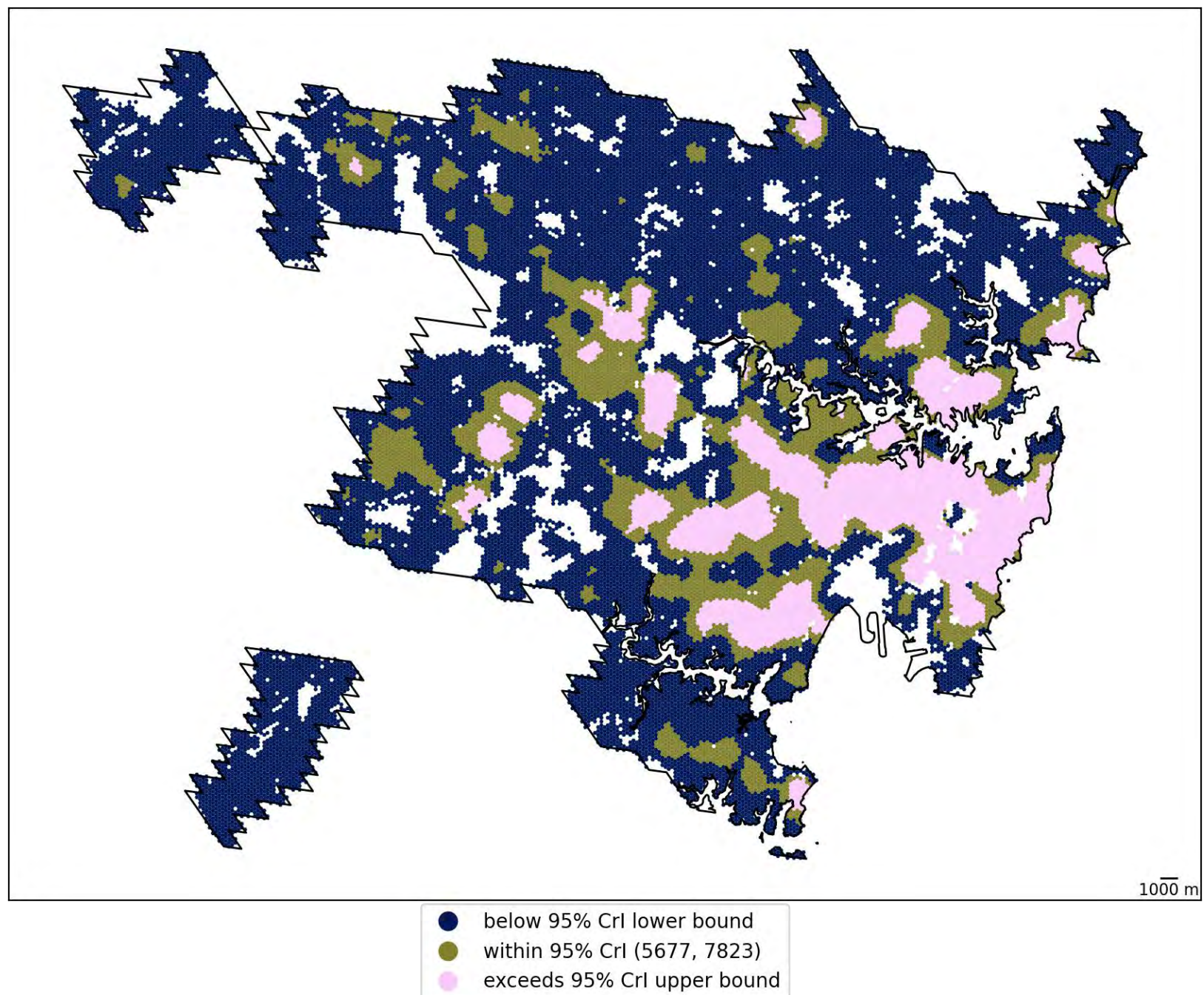

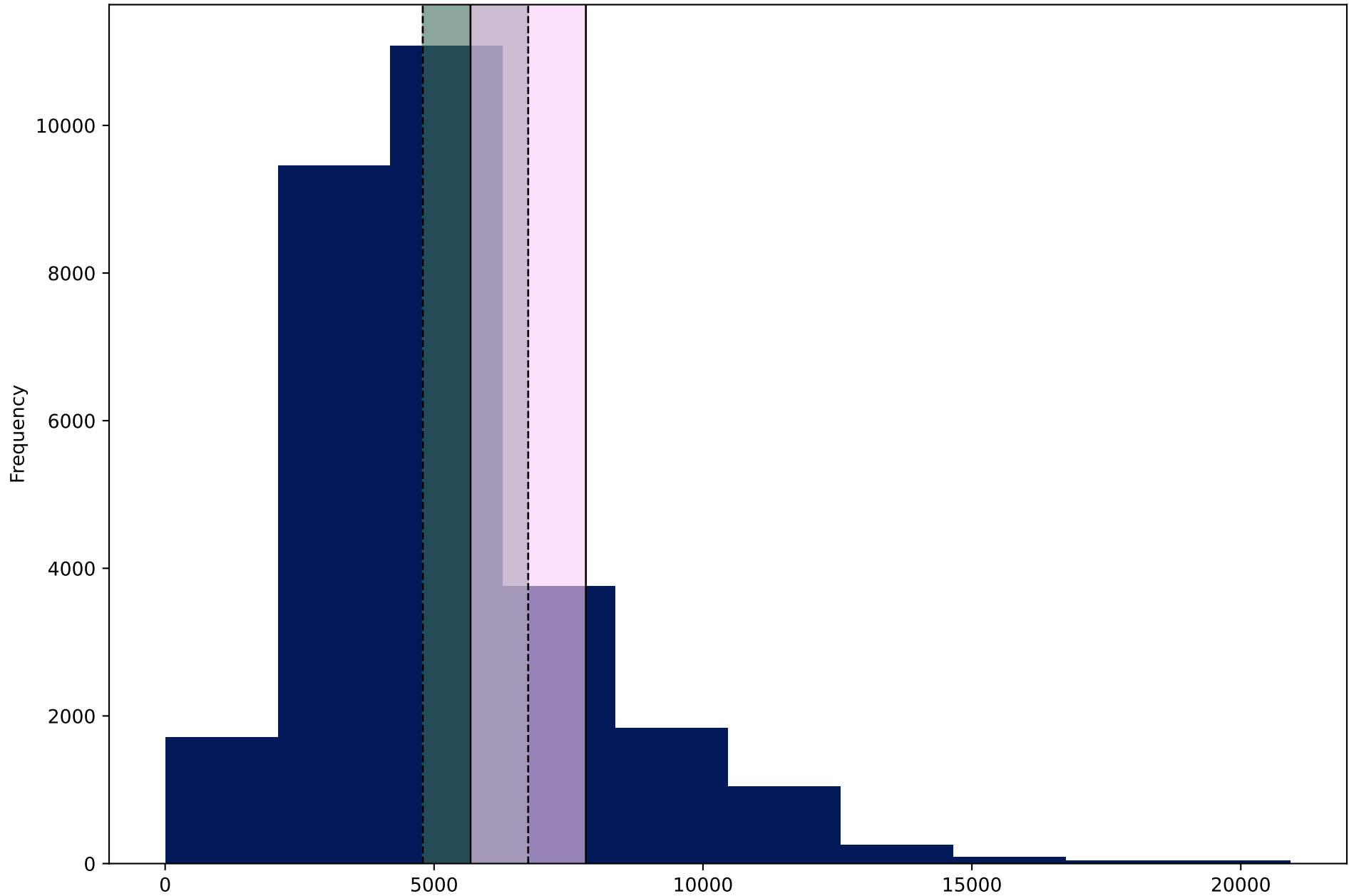

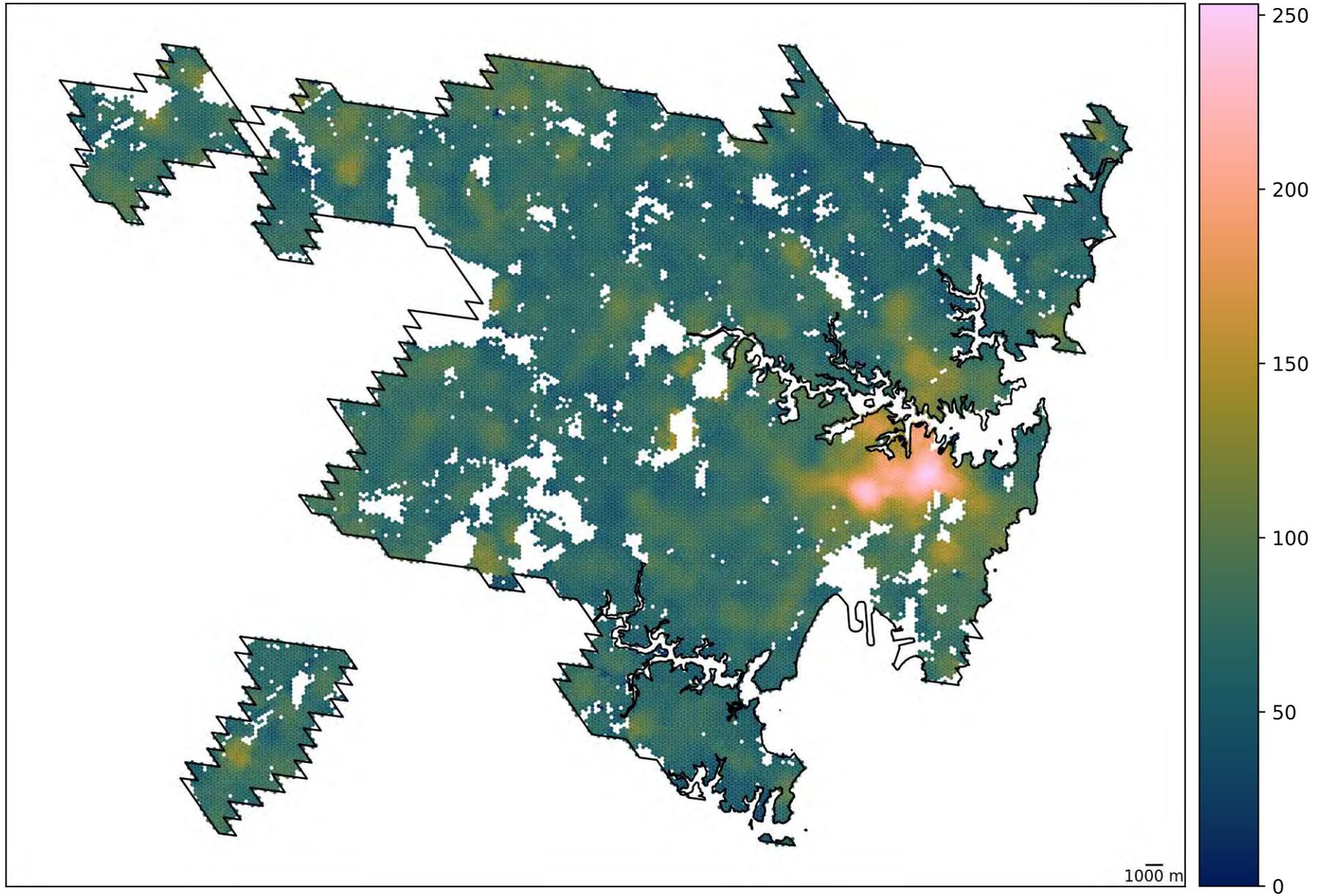

Mean 1000 m neighbourhood street intersections per km²

A: Estimated Mean 1000 m neighbourhood street intersections per km² requirement for ≥80% probability of engaging in walking for transport

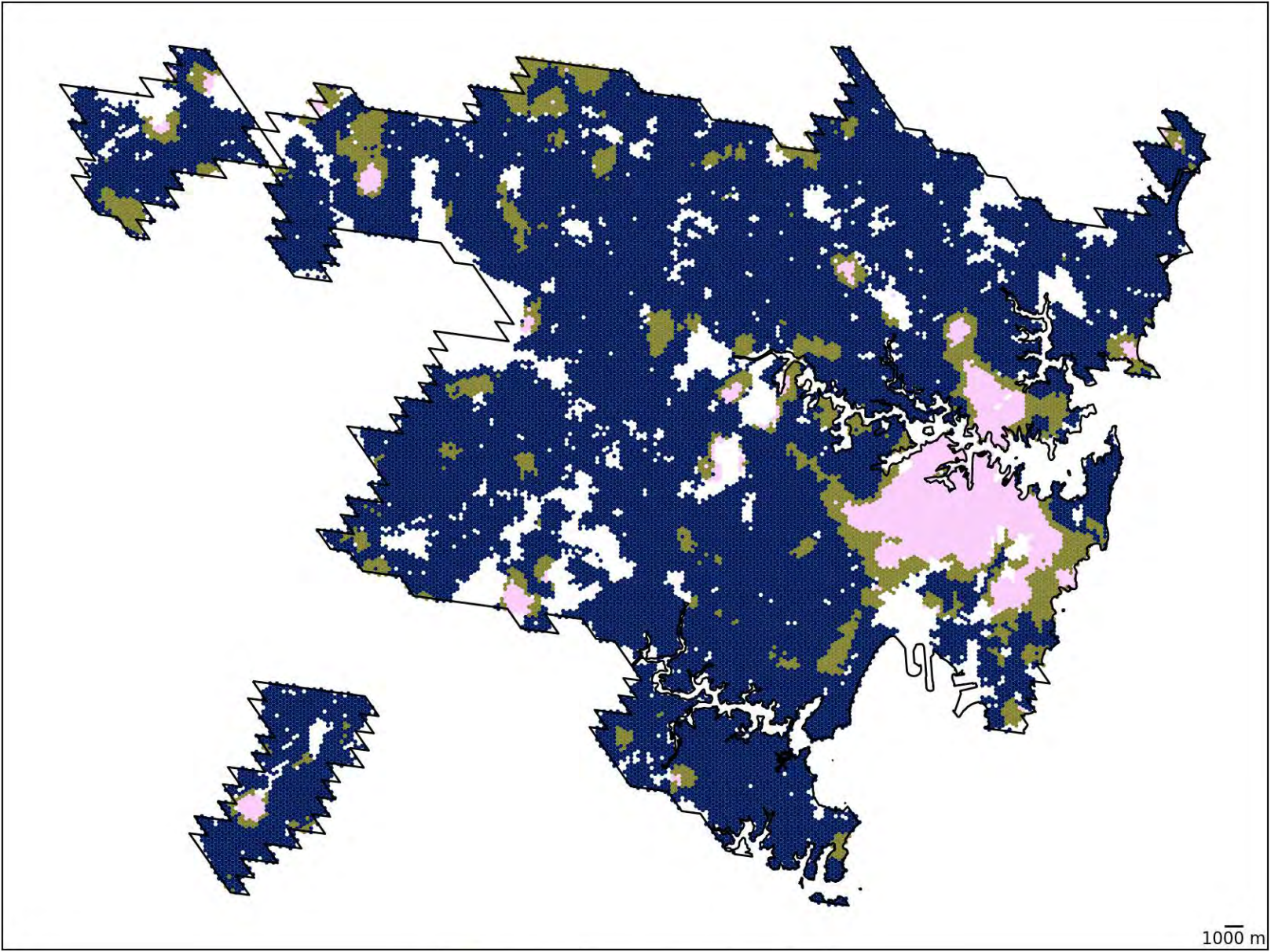

- below 95% CrI lower bound
- within 95% CrI (90, 110)
- exceeds 95% CrI upper bound

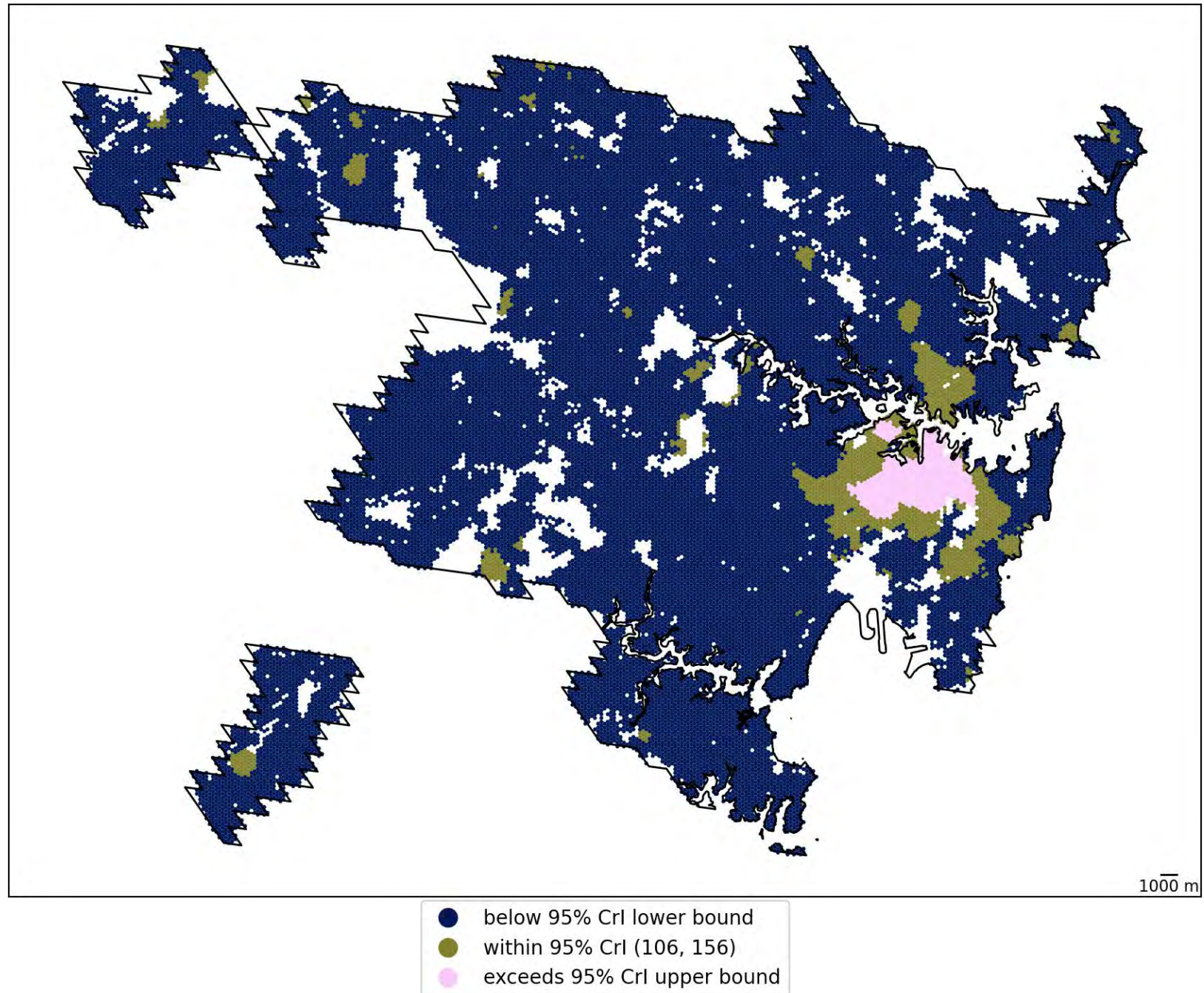

B: Estimated Mean 1000 m neighbourhood street intersections per km² requirement for reaching the WHO's target of a ≥15% relative reduction in insufficient physical activity through walking

- below 95% CrI lower bound
- within 95% CrI (106, 156)
- exceeds 95% CrI upper bound

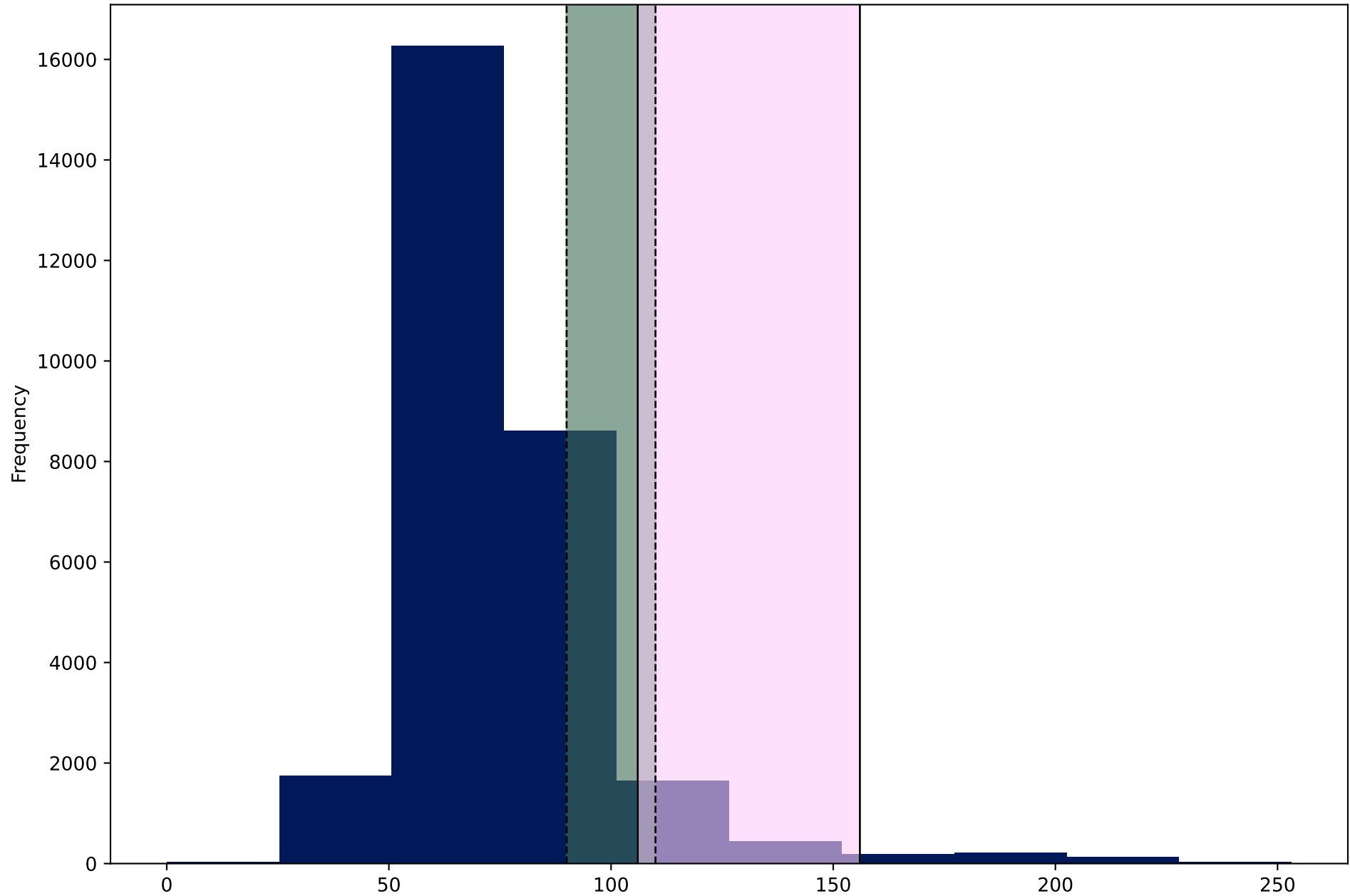

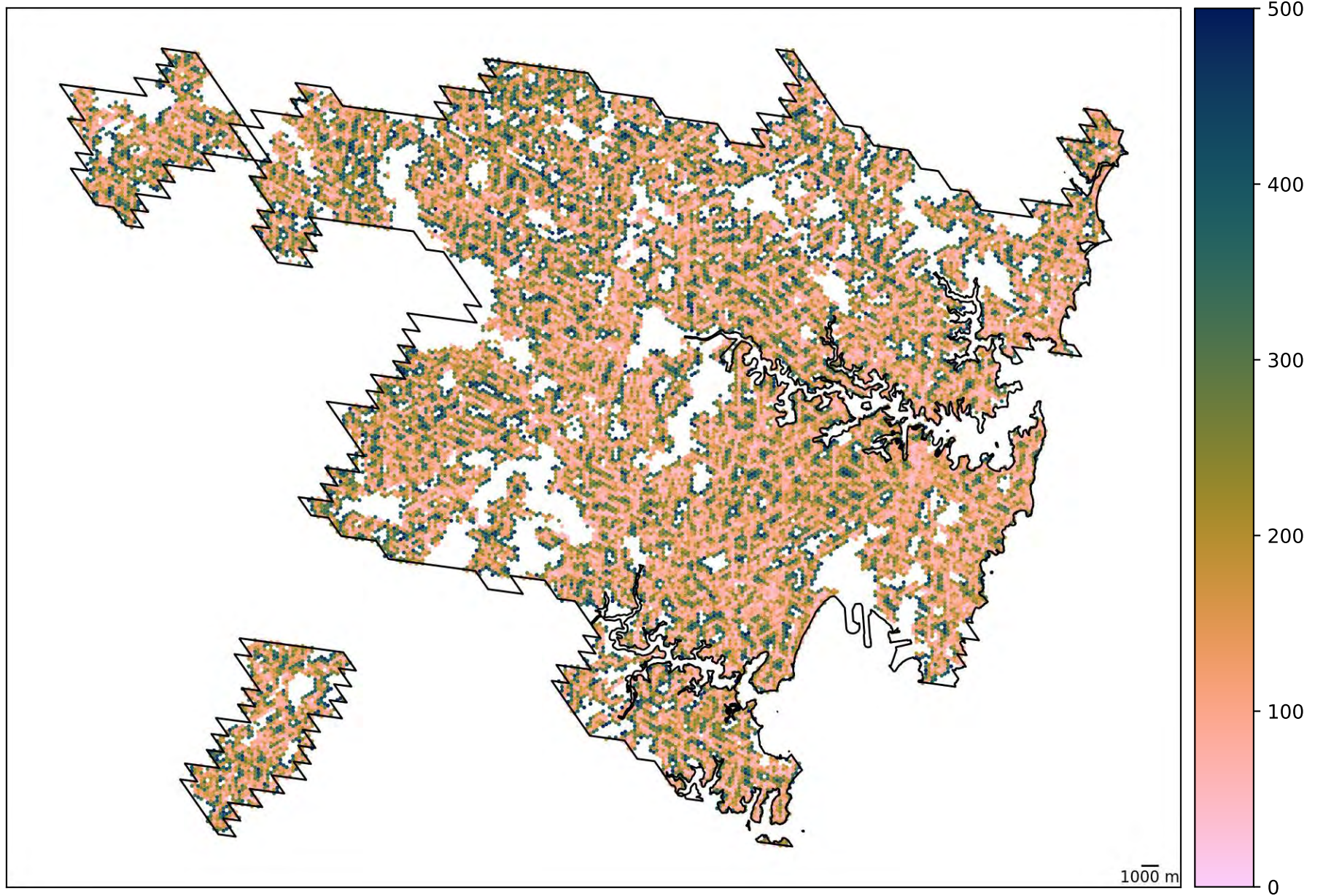

distances: Estimated Distance to nearest public transport stops (m; up to 500m) requirement for distances to destinations, measured up to a maximum distance target threshold of 500 metres

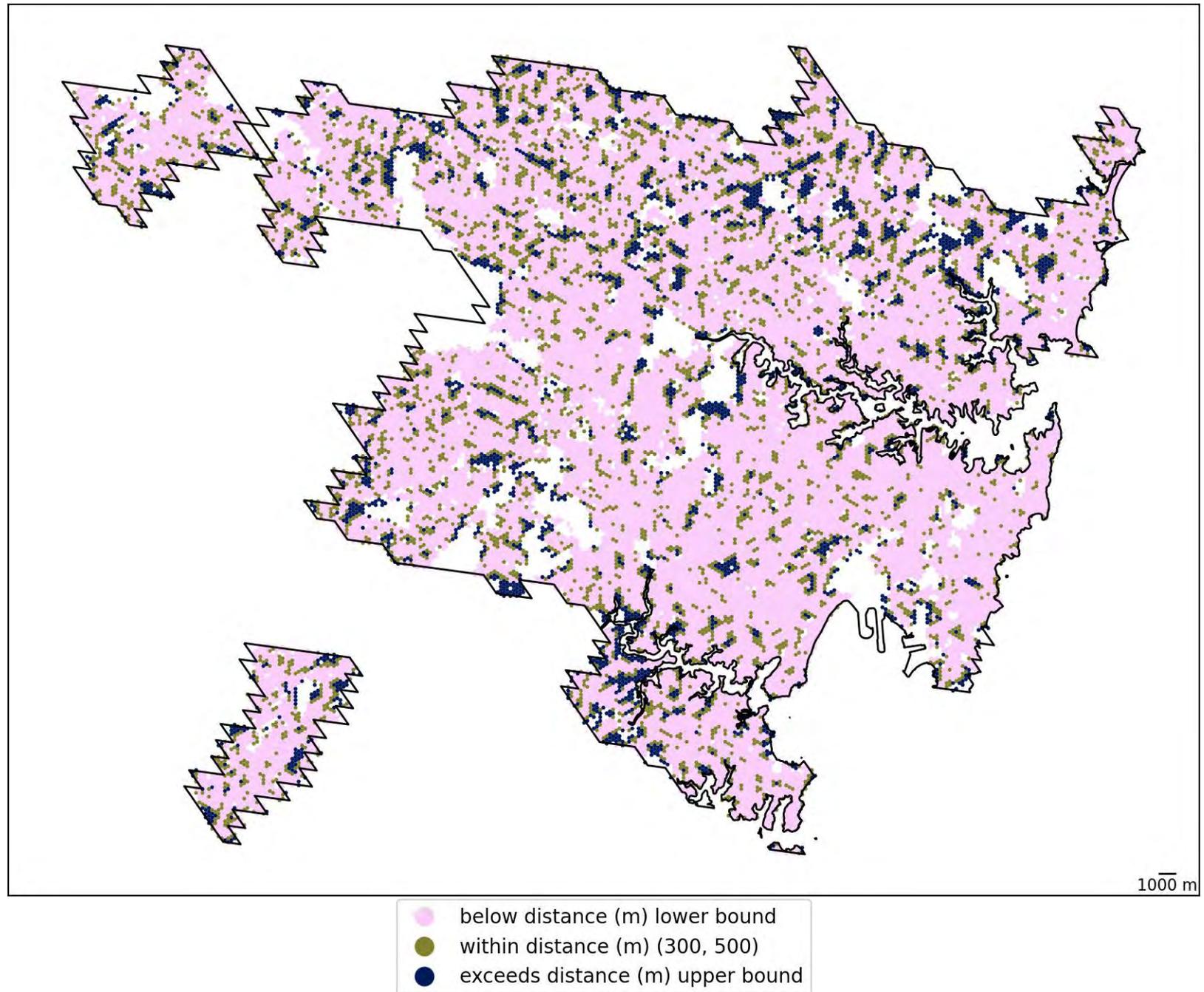

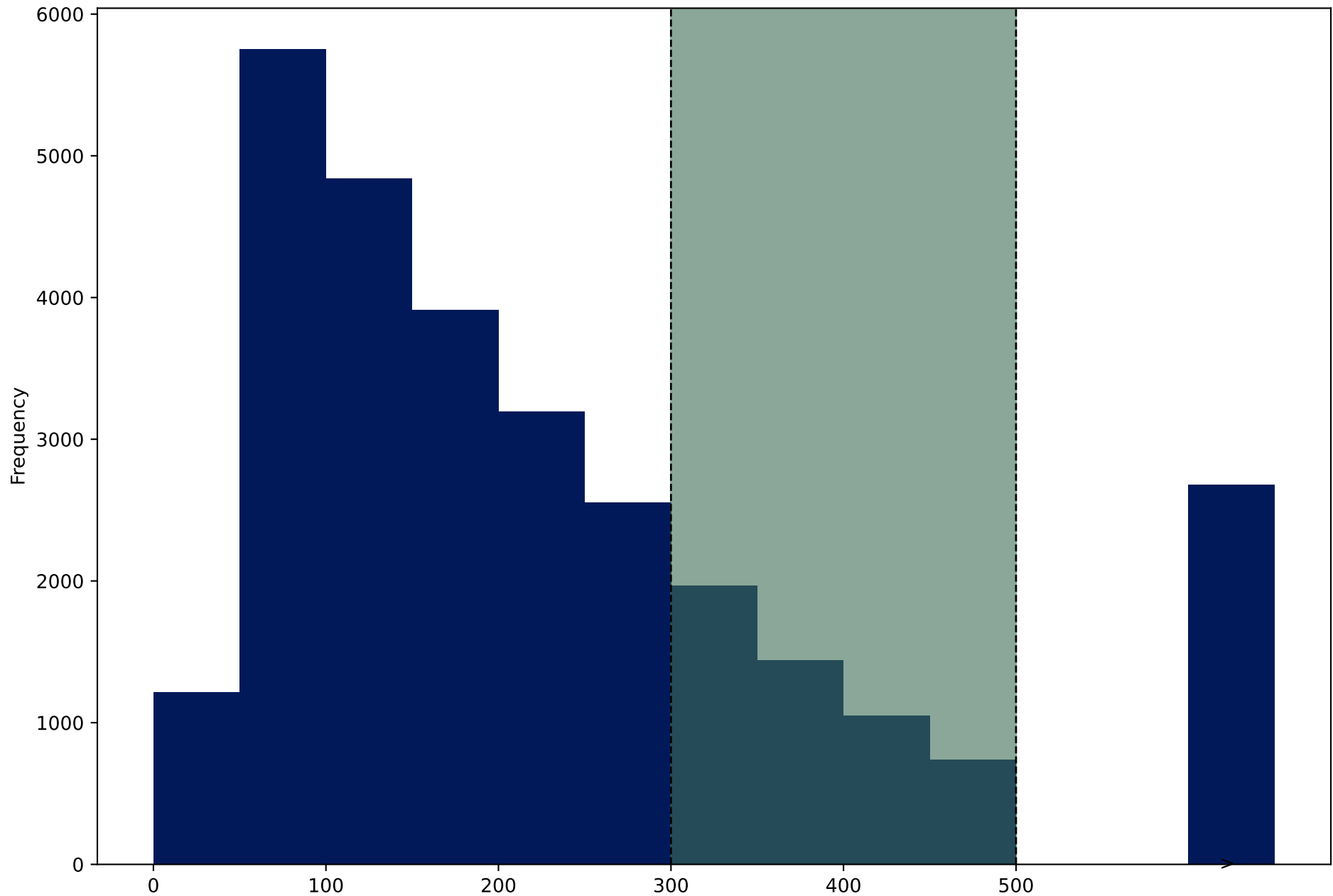

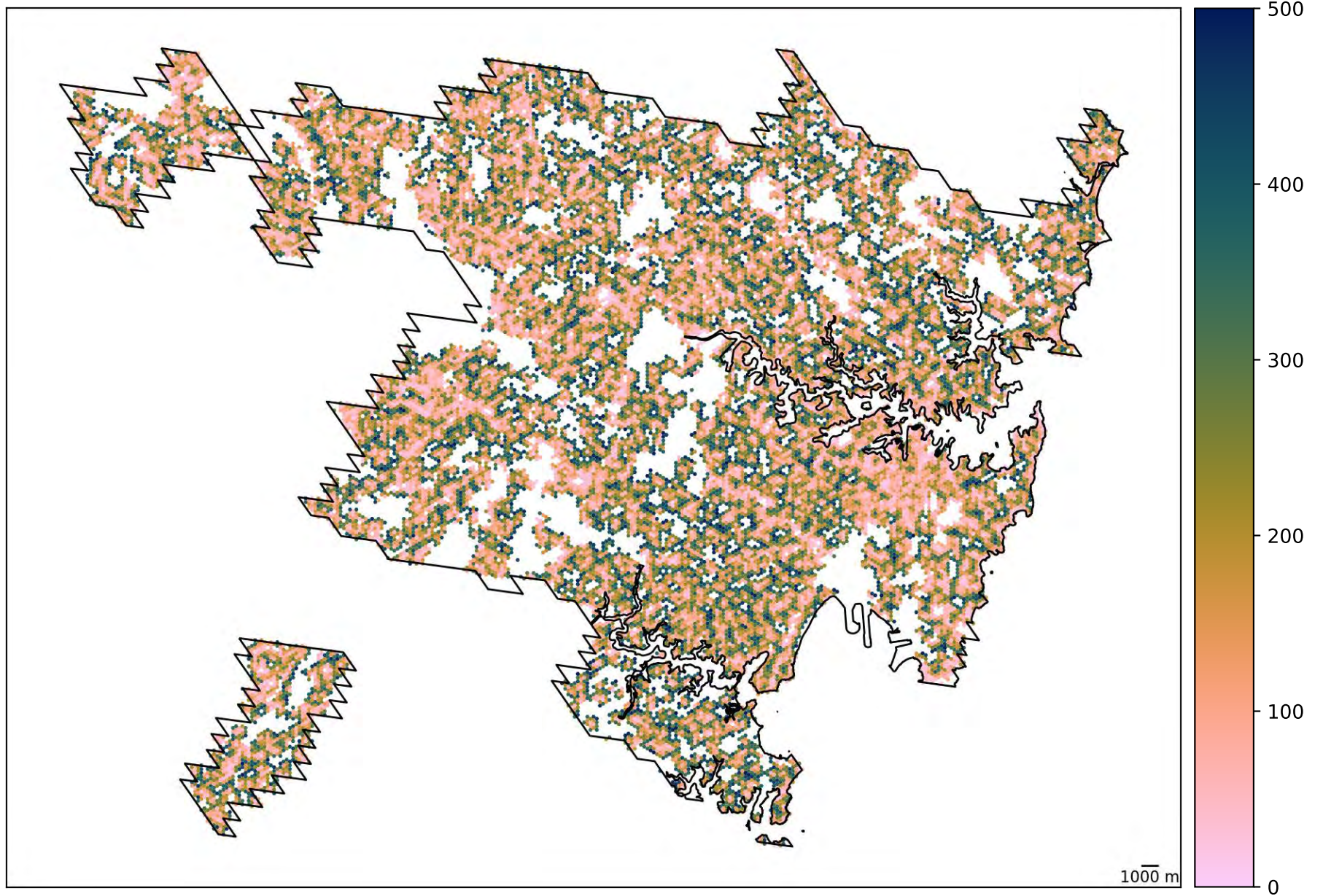

distances: Estimated Distance to nearest park (m; up to 500m) requirement for distances to destinations, measured up to a maximum distance target threshold of 500 metres

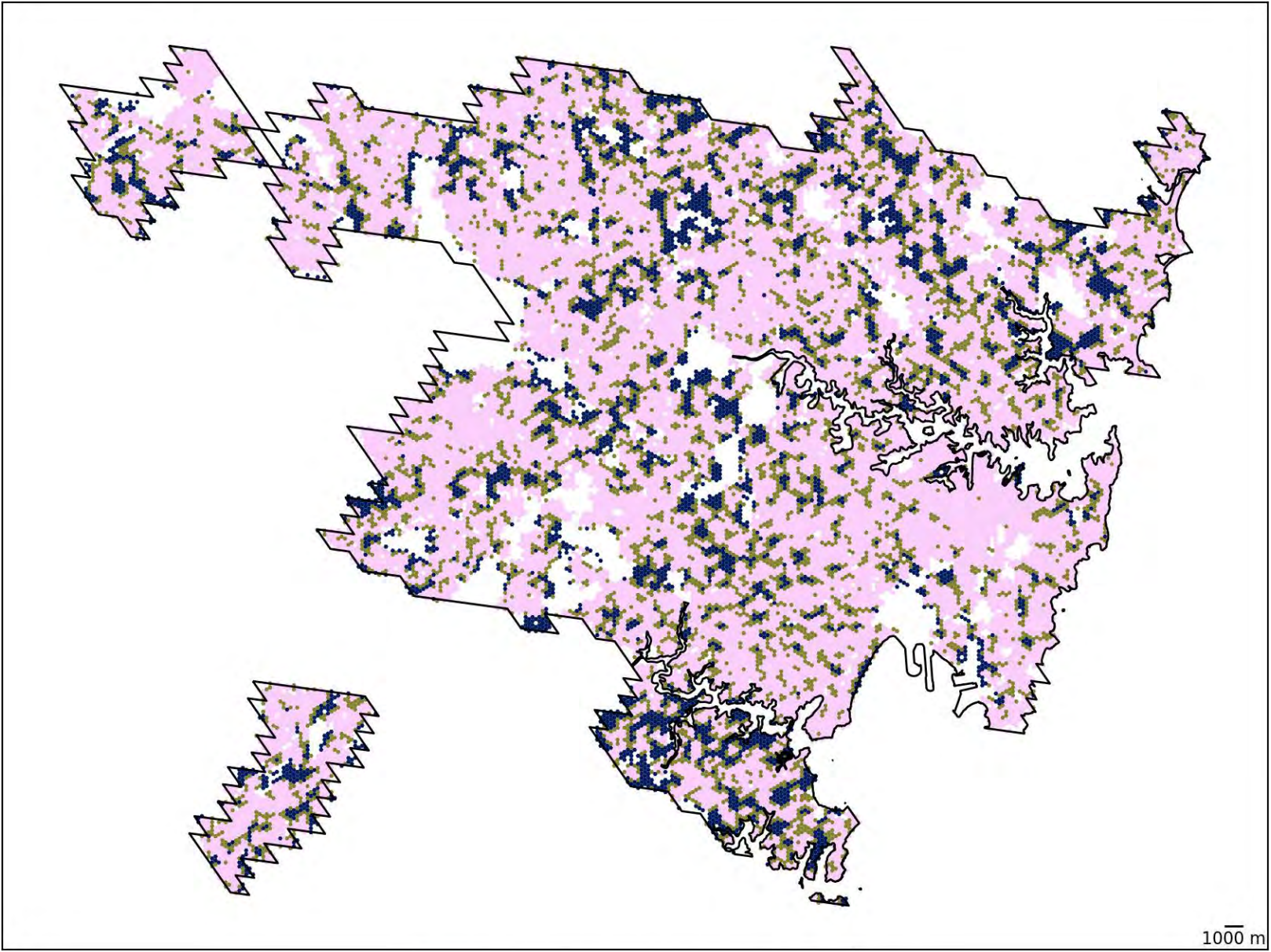

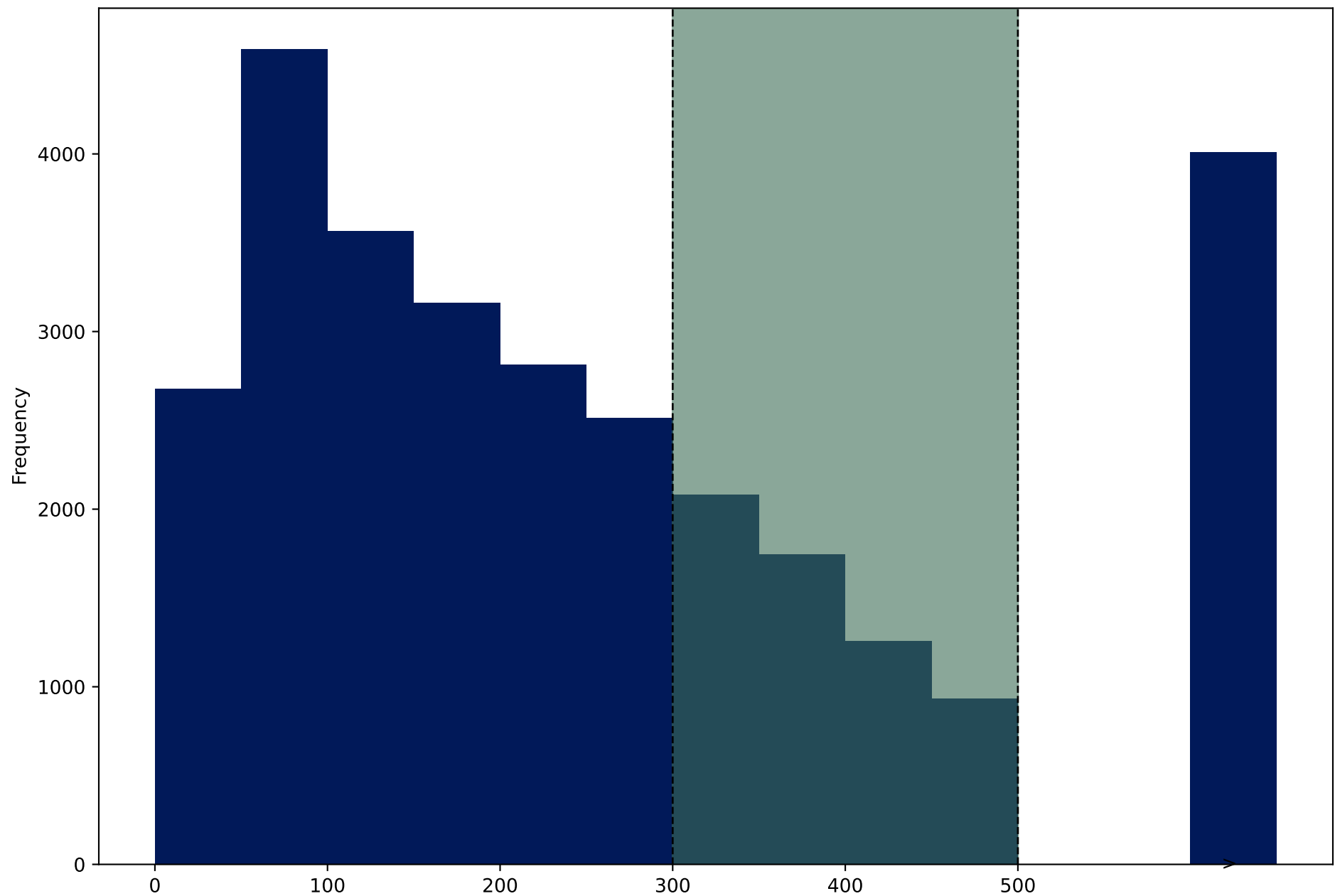

# Australasia, New Zealand, Auckland

Satellite imagery of urban study region (Bing) | Walkability, relative to city | Walkability, relative to 25 global cities

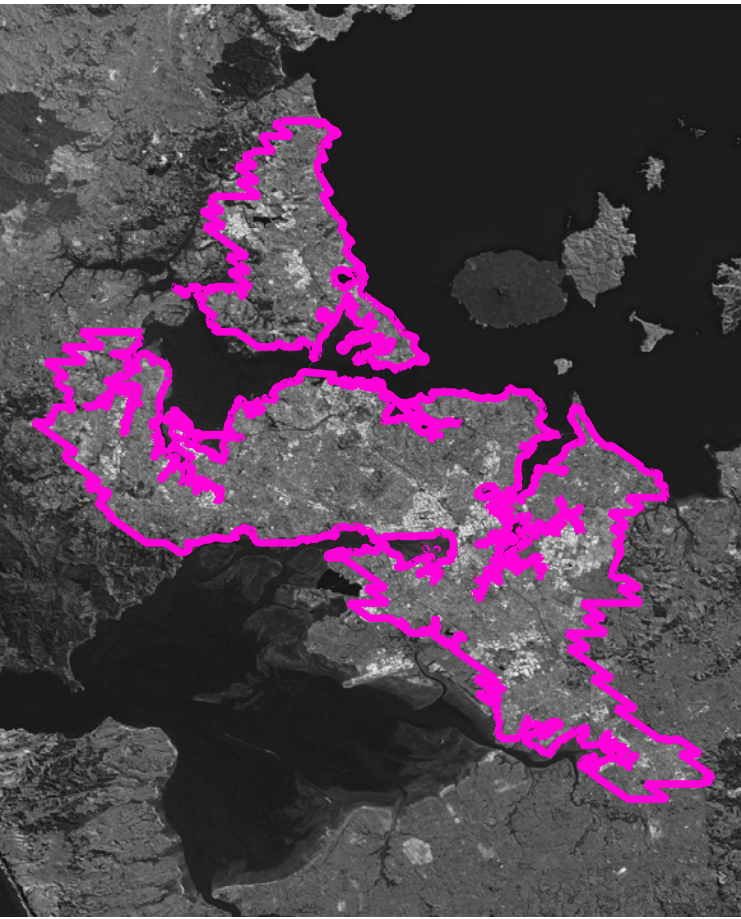
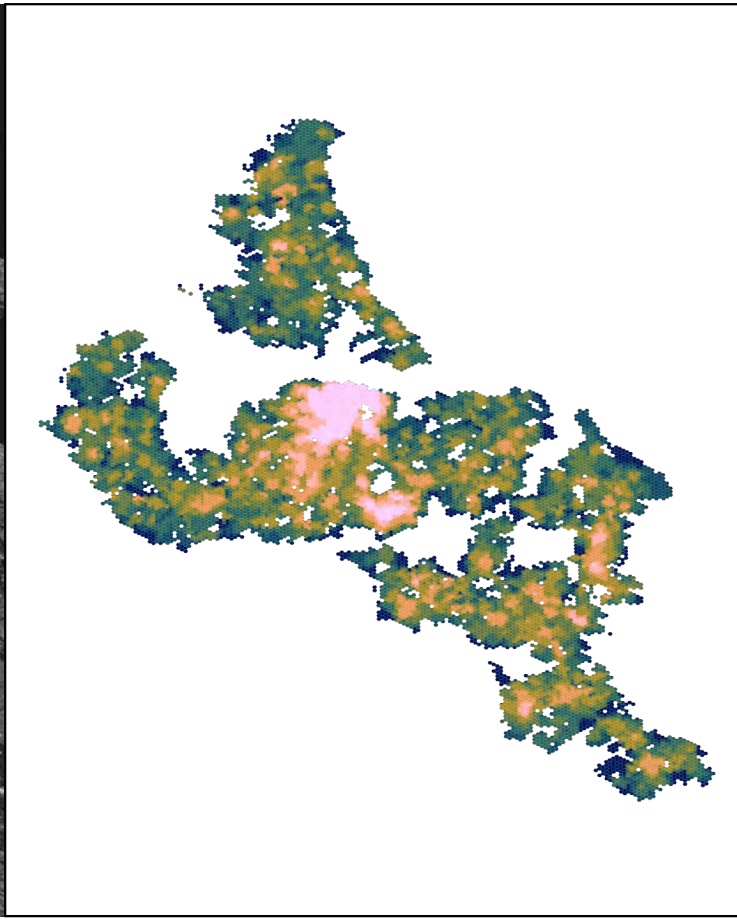
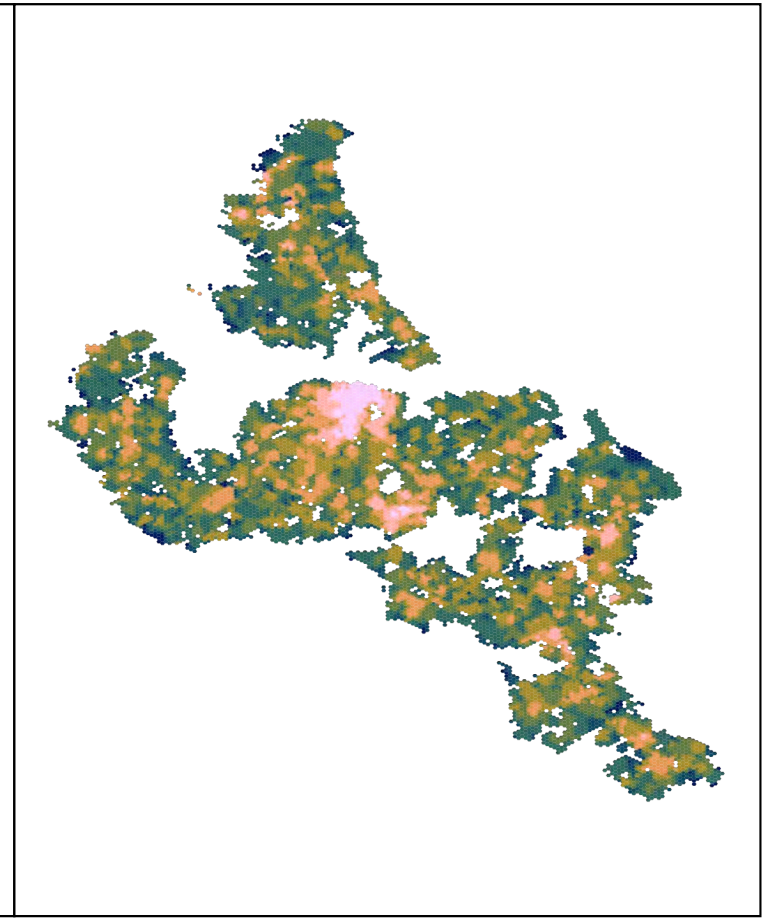

Urban boundary

0 10 20 km

Walkability score
- <-3
- -3 to -2
- -2 to -1
- -1 to 0
- 0 to 1
- 1 to 2
- 2 to 3
- ≥3

Walkability relative to all cities by component variables (2D histograms), and overall (histogram)

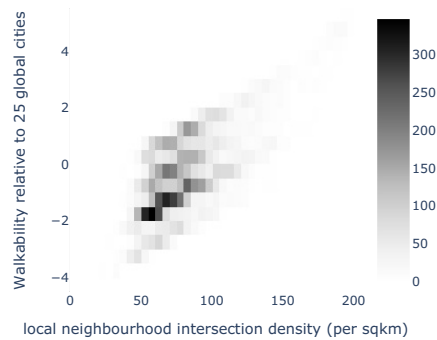
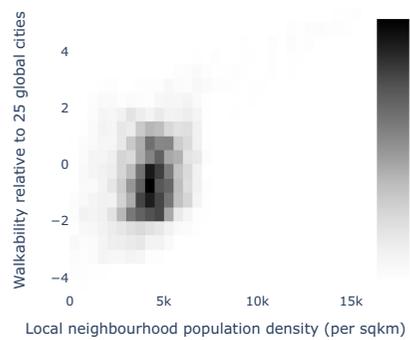
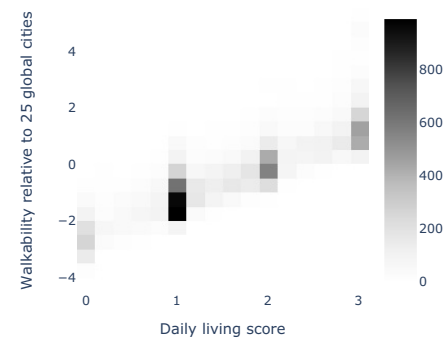
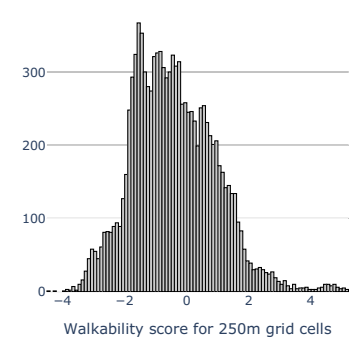

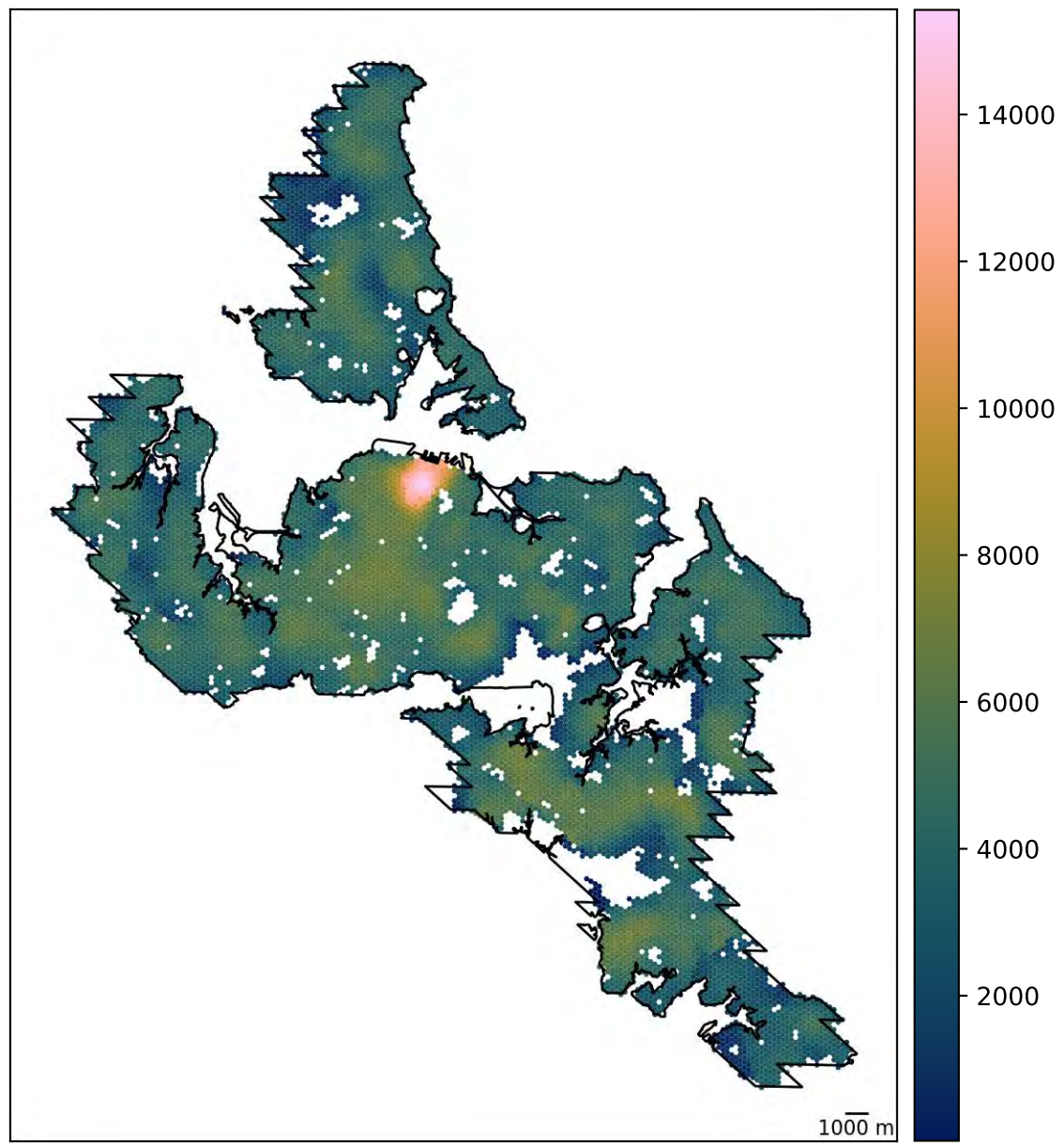

Mean 1000 m neighbourhood population per km²

A: Estimated Mean 1000 m neighbourhood population per km² requirement for ≥80% probability of engaging in walking for transport

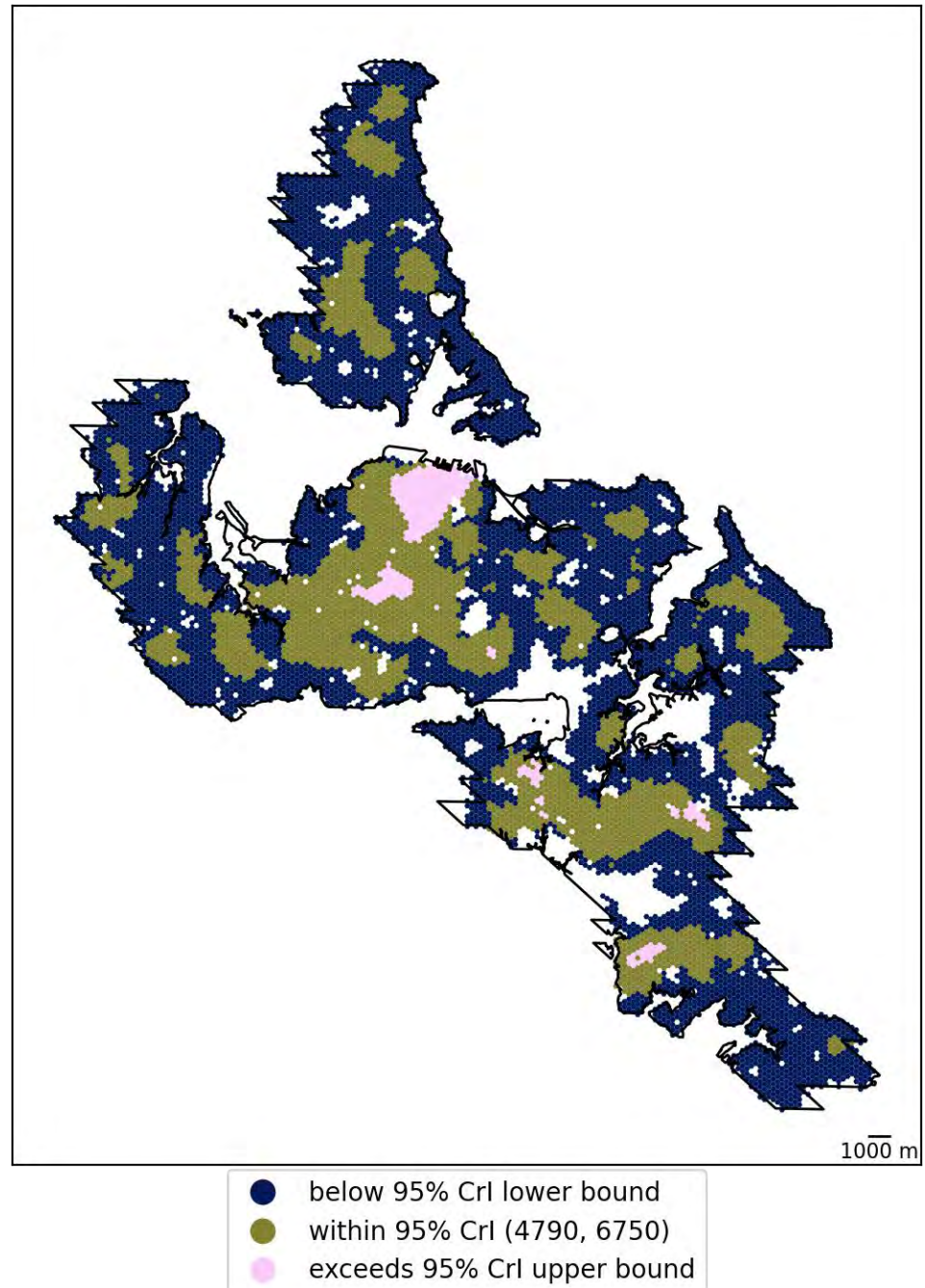

B: Estimated Mean 1000 m neighbourhood population per km² requirement for reaching the WHO's target of a ≥15% relative reduction in insufficient physical activity through walking

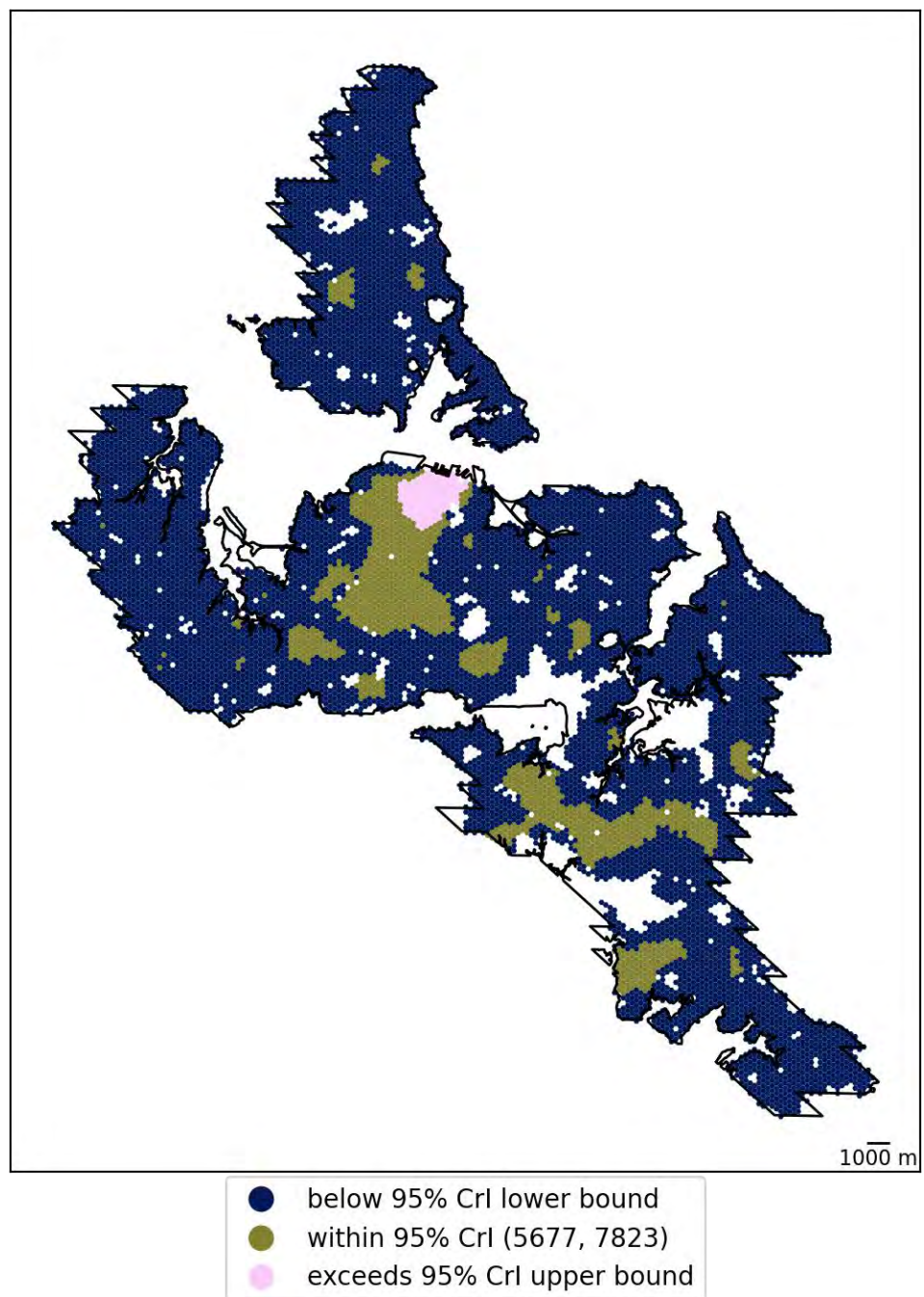

- below 95% CrI lower bound
- within 95% CrI (5677, 7823)
- exceeds 95% CrI upper bound

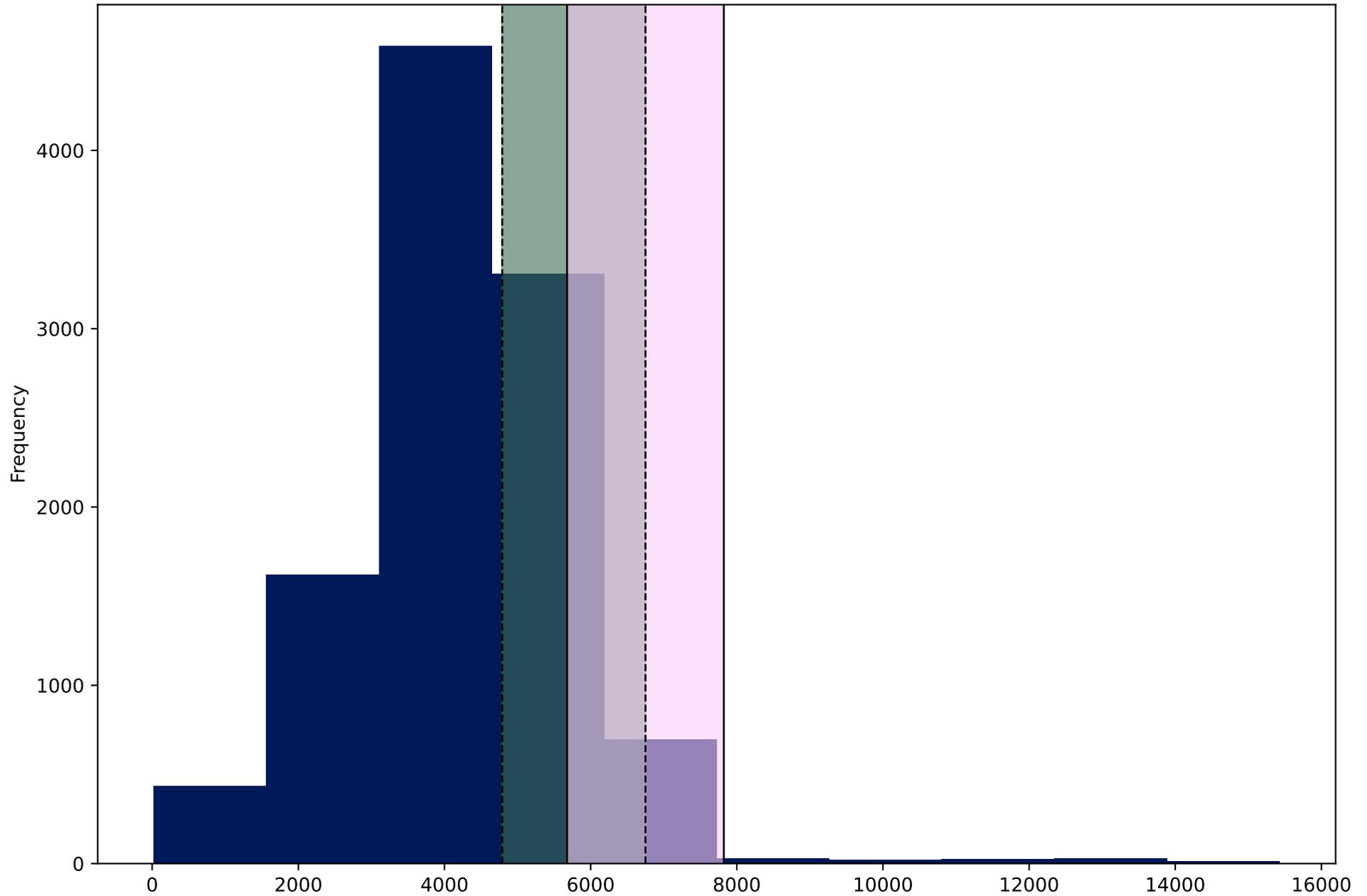

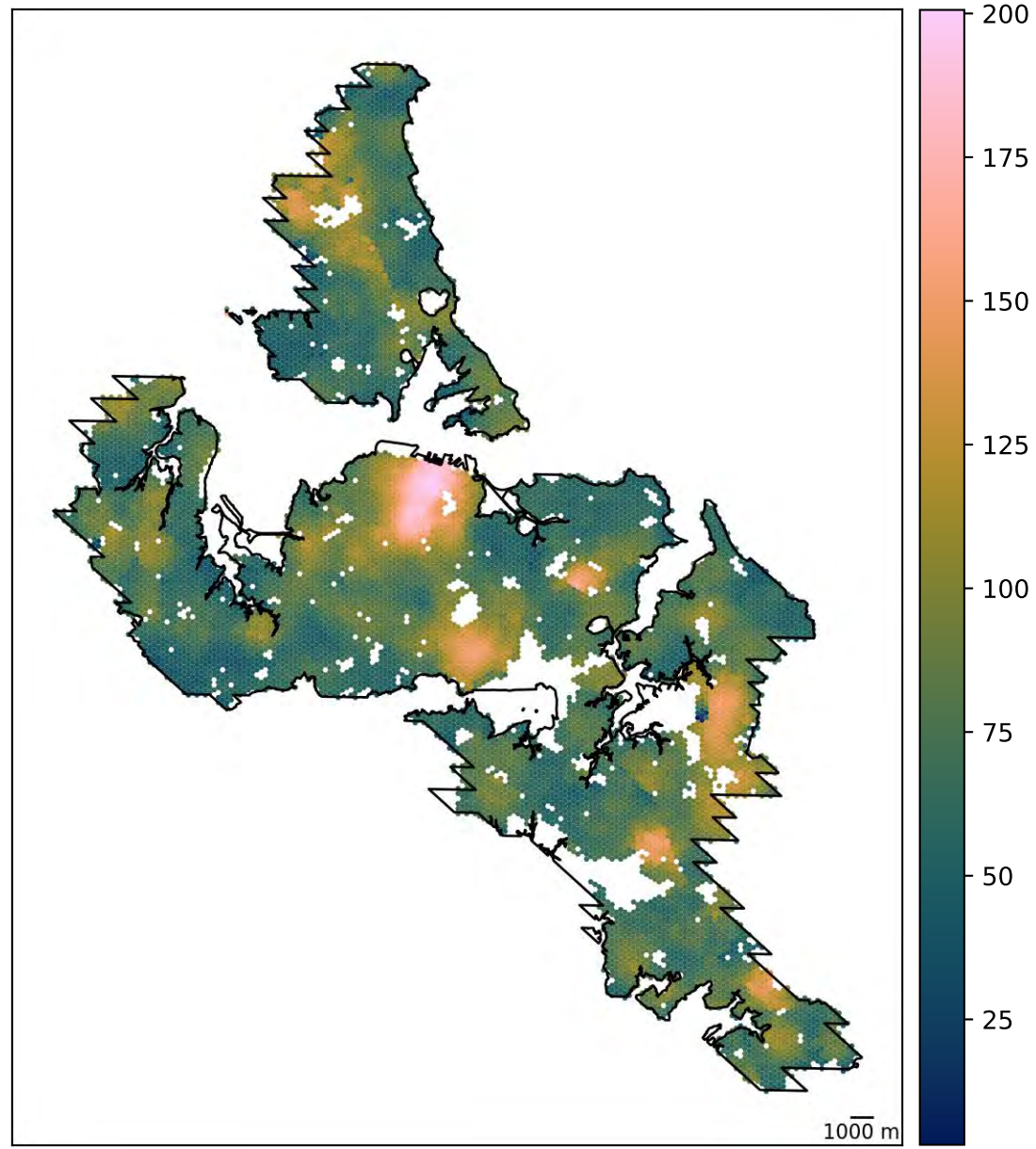

Mean 1000 m neighbourhood street intersections per km²

A: Estimated Mean 1000 m neighbourhood street intersections per km² requirement for ≥80% probability of engaging in walking for transport

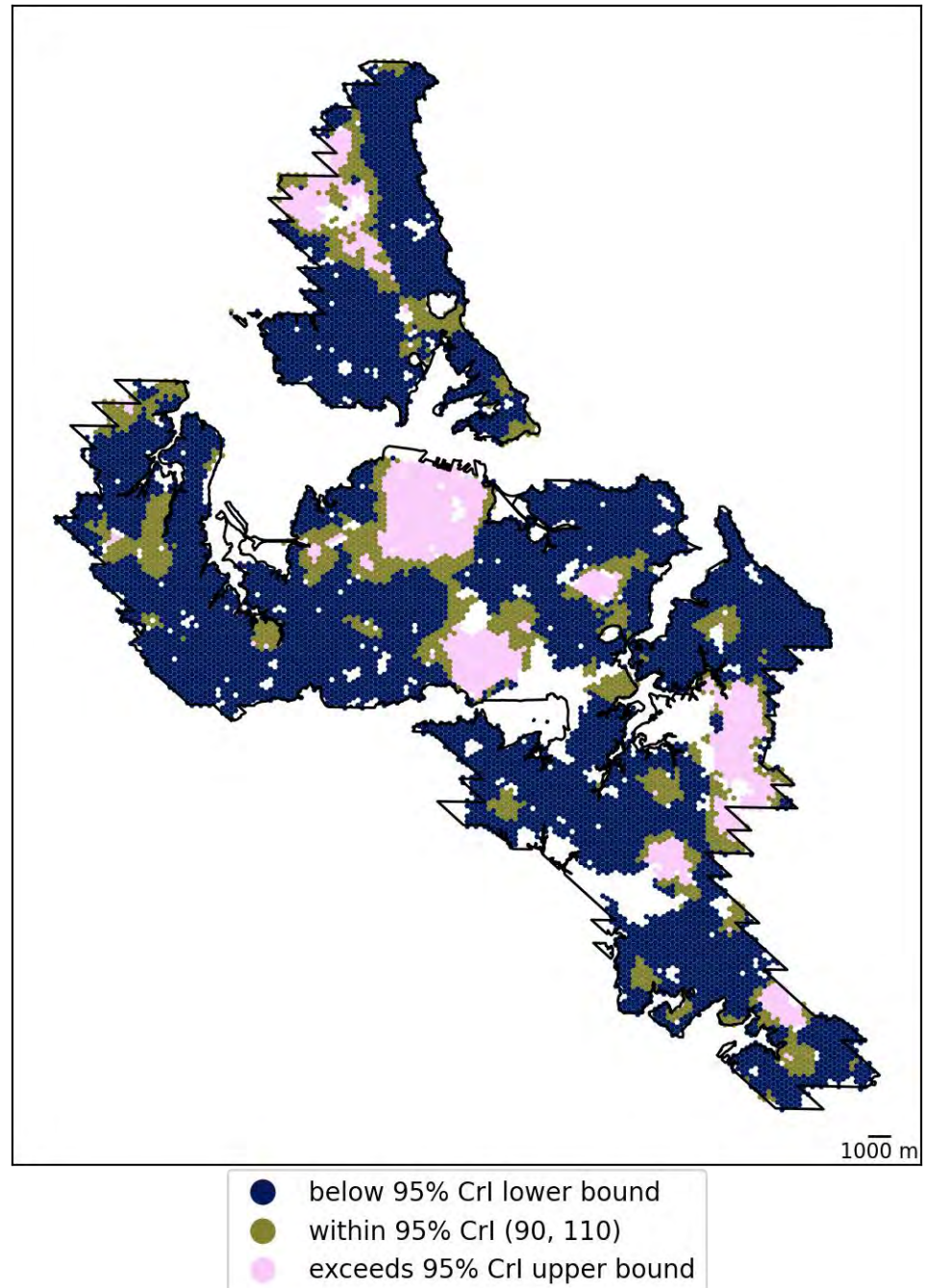

B: Estimated Mean 1000 m neighbourhood street intersections per km² requirement for reaching the WHO's target of a ≥15% relative reduction in insufficient physical activity through walking

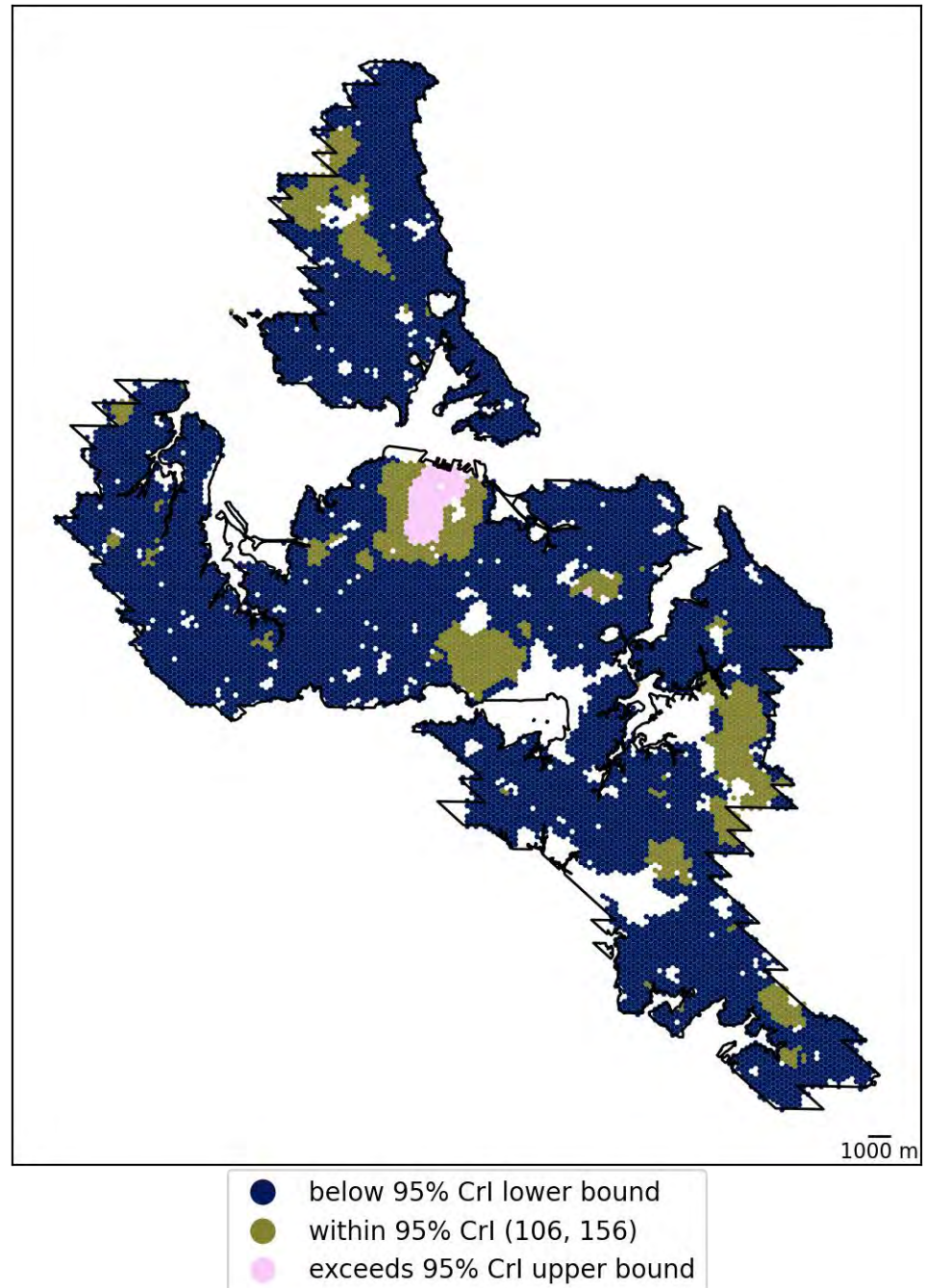

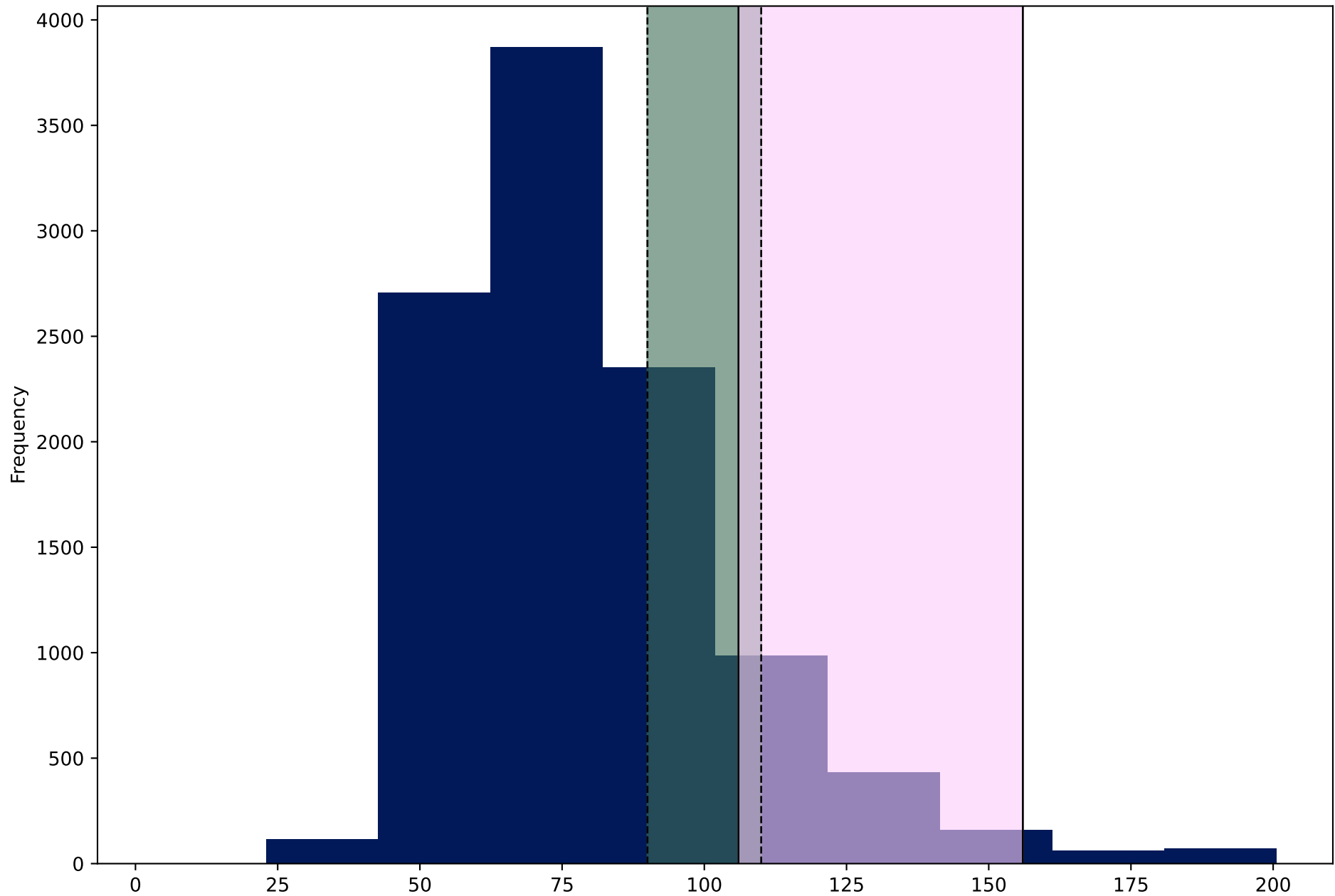
236

Distance to nearest public transport stops (m; up to 500m)

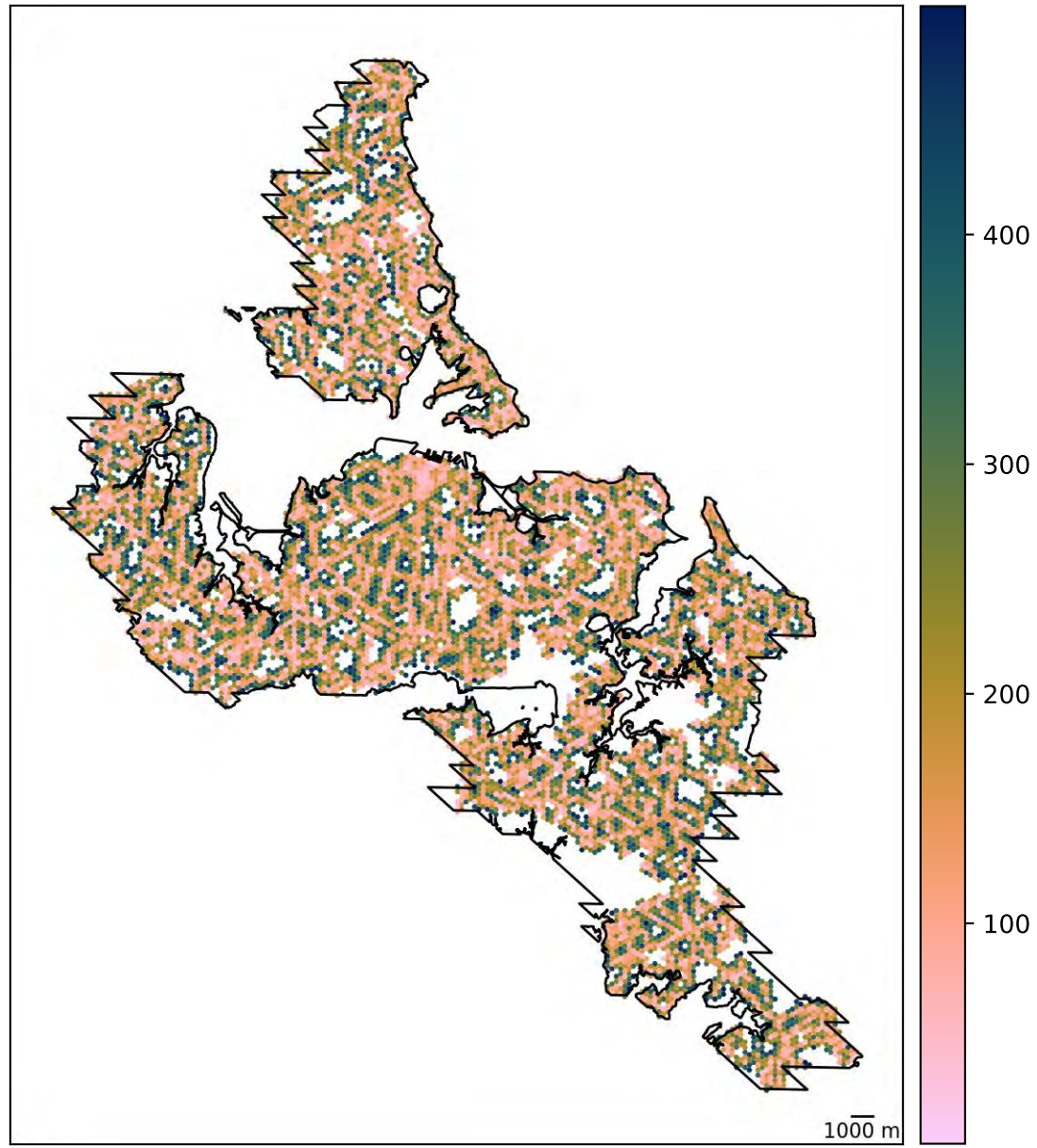

distances: Estimated Distance to nearest public transport stops (m; up to 500m) requirement for distances to destinations, measured up to a maximum distance target threshold of 500 metres

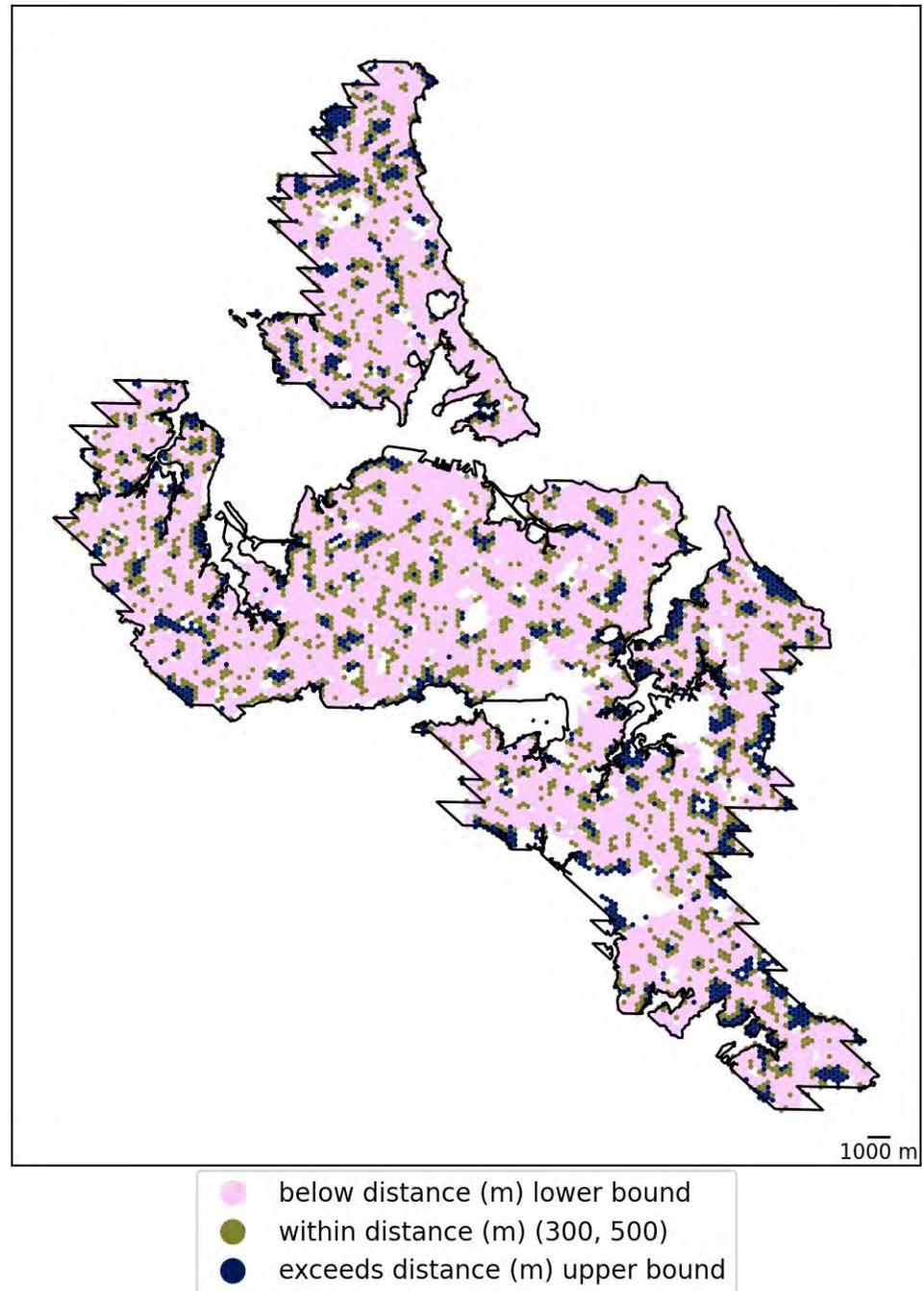

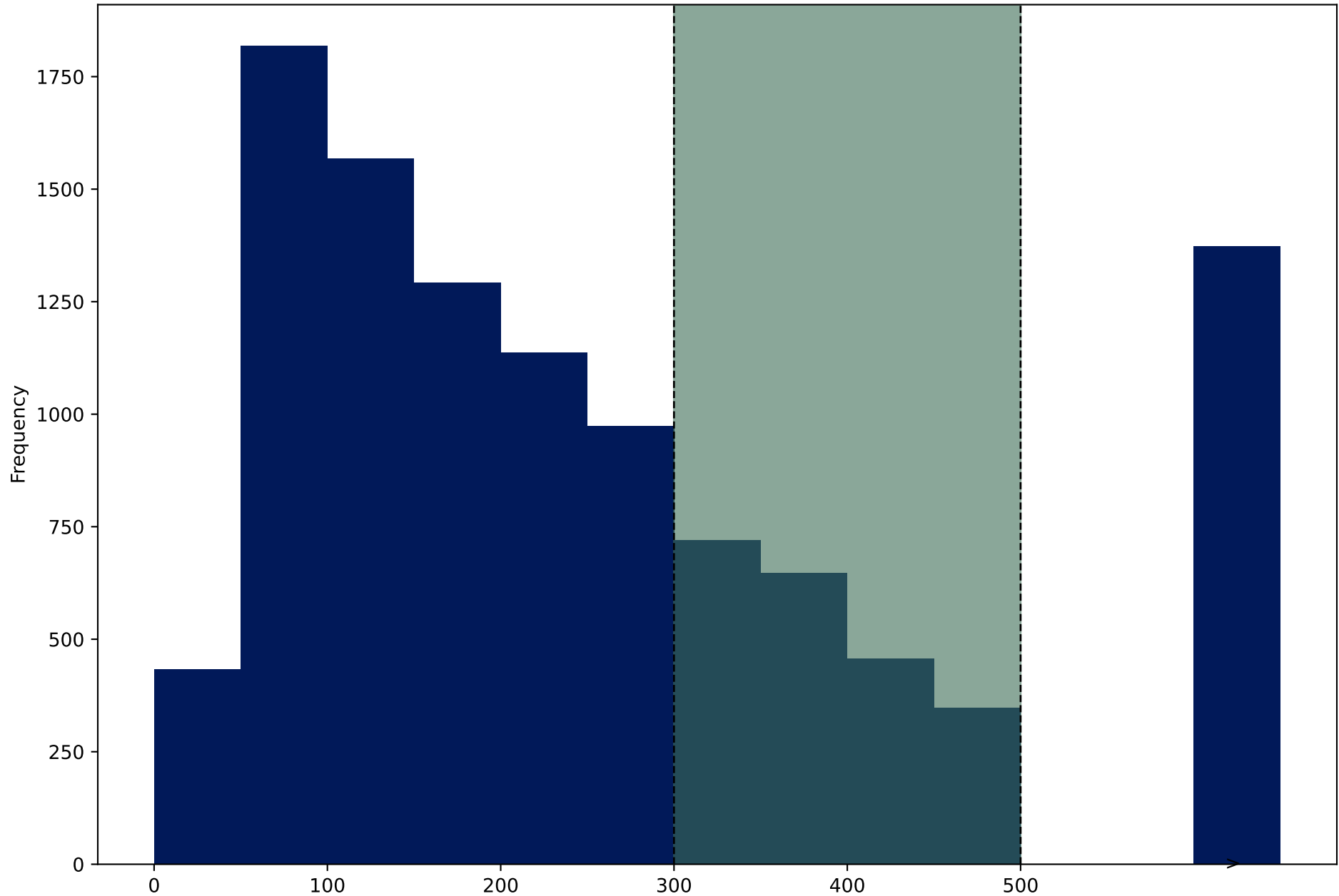

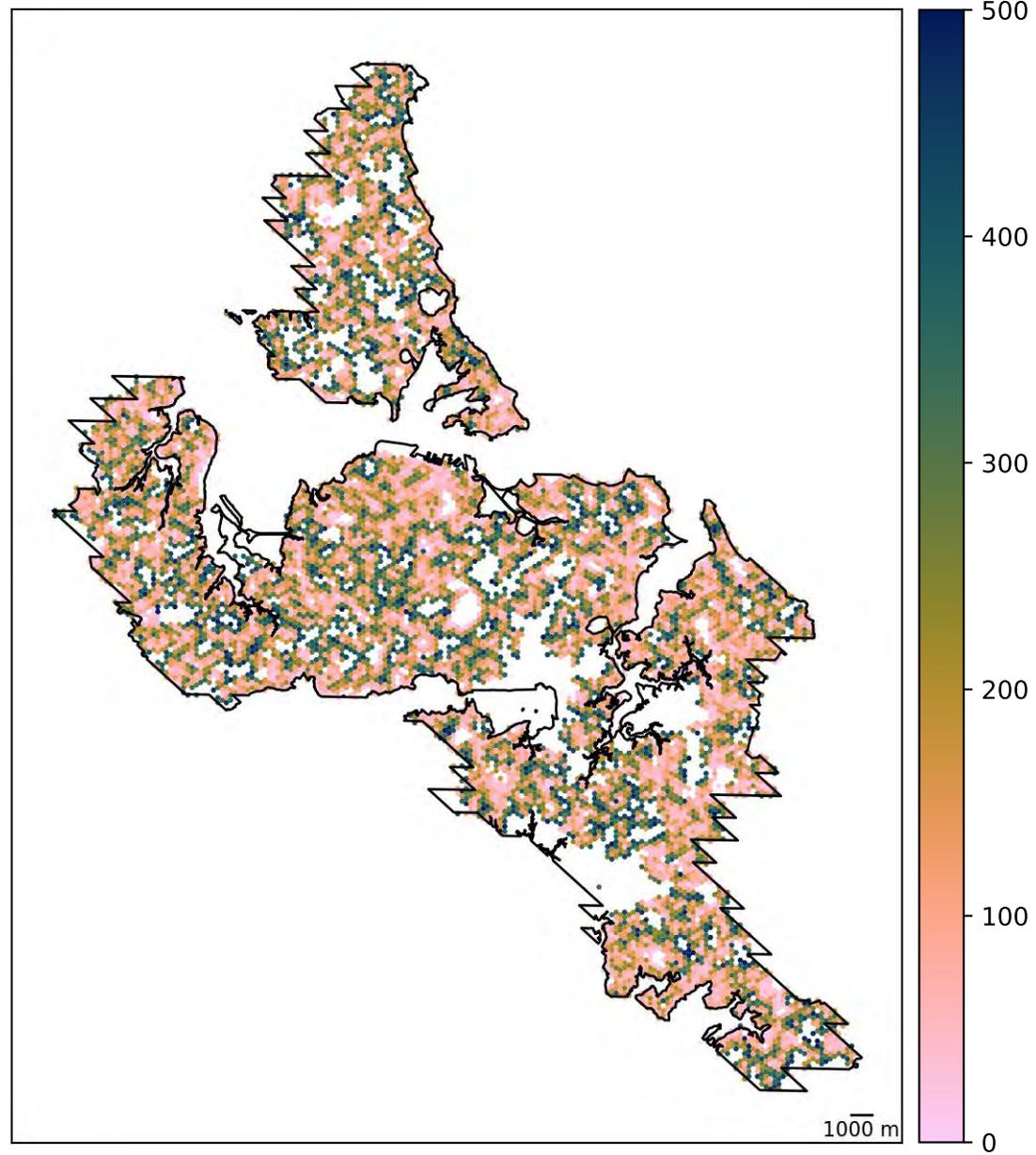

Distance to nearest park (m; up to 500m)

distances: Estimated Distance to nearest park (m; up to 500m) requirement for distances to destinations, measured up to a maximum distance target threshold of 500 metres

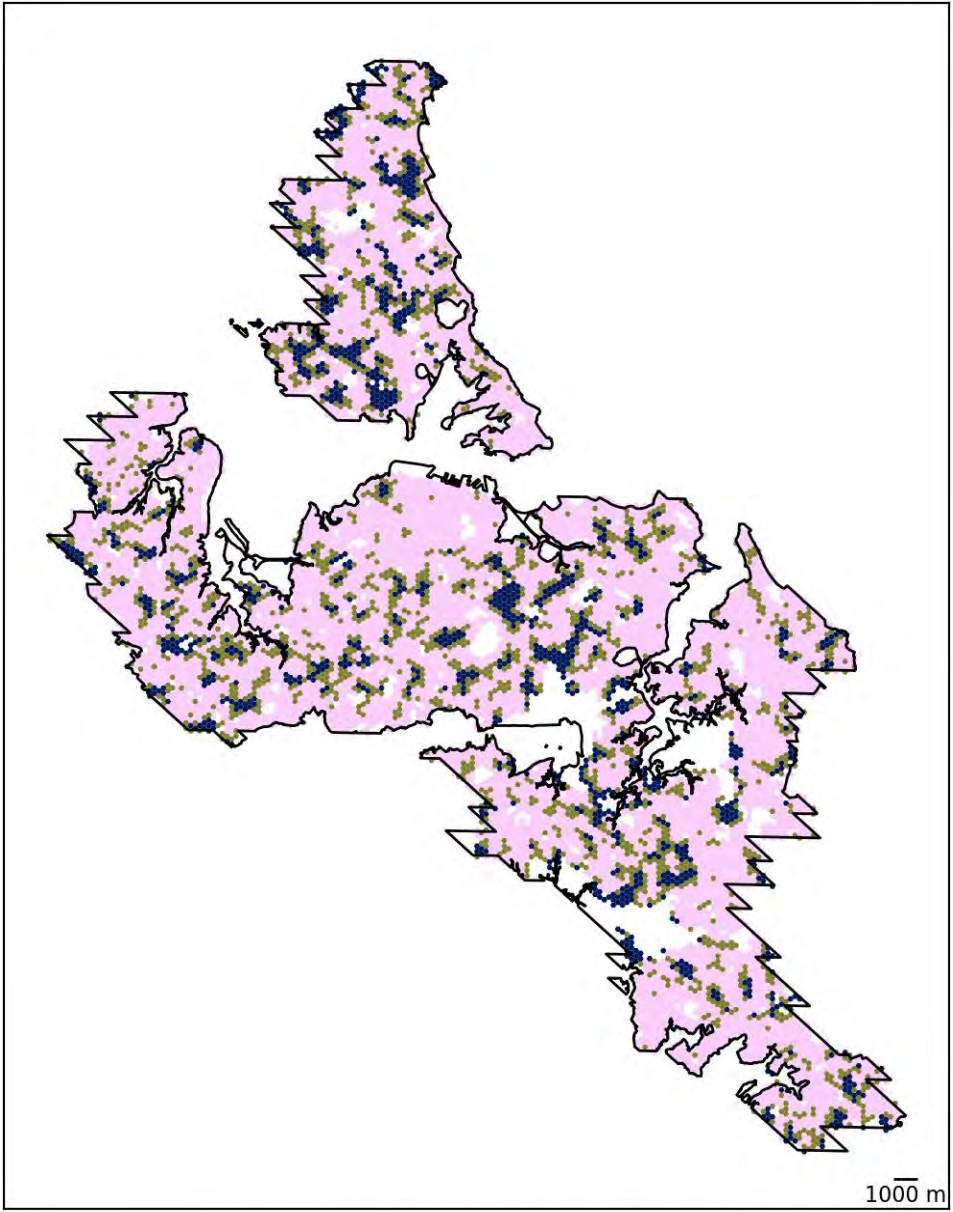

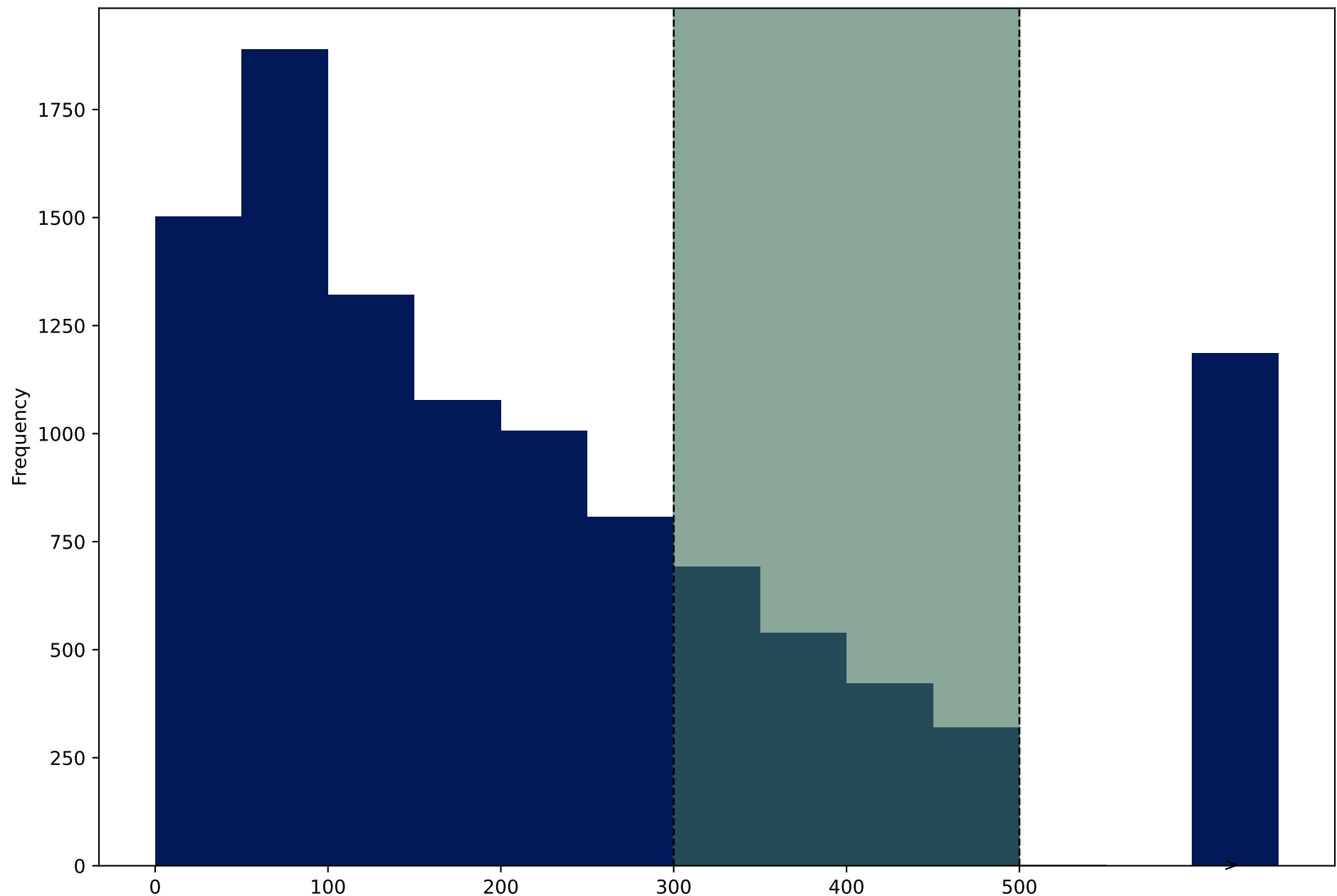

# Europe, Austria, Graz

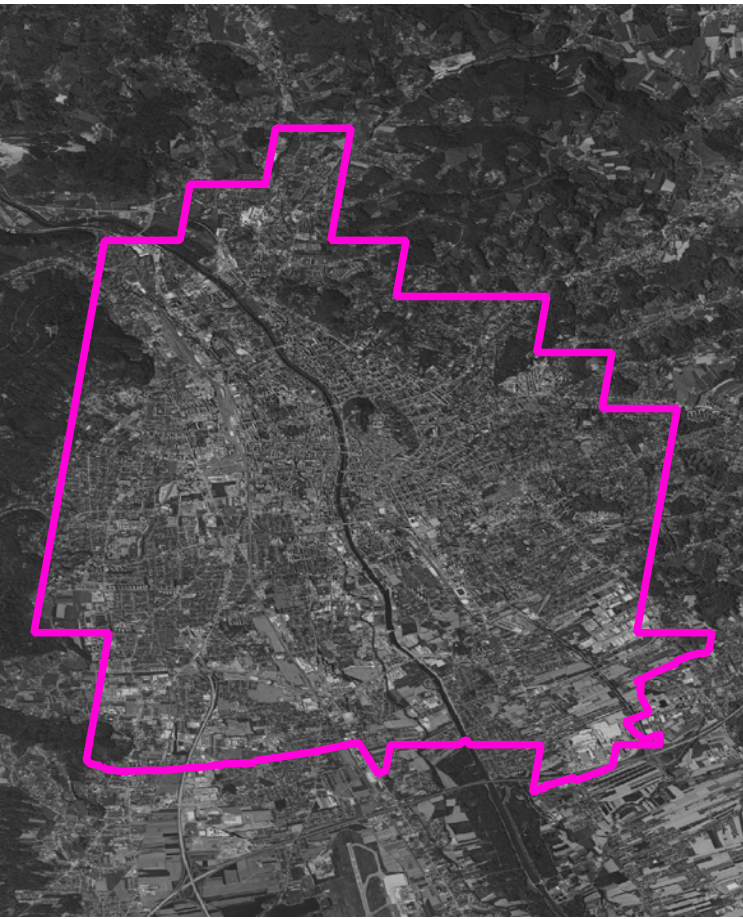
Satellite imagery of urban study region (Bing)

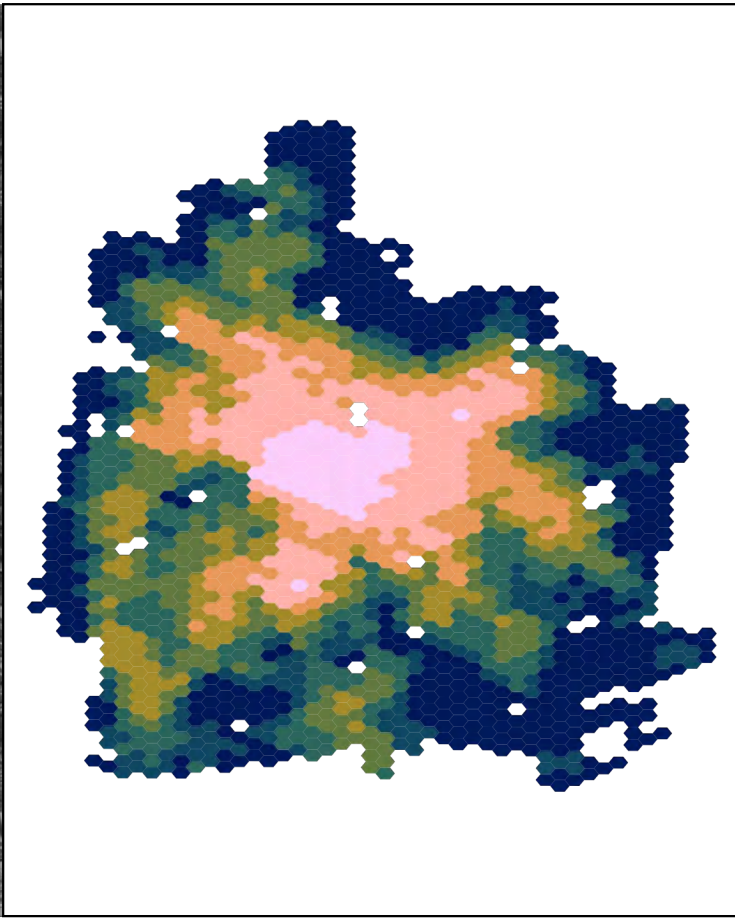
Walkability, relative to city

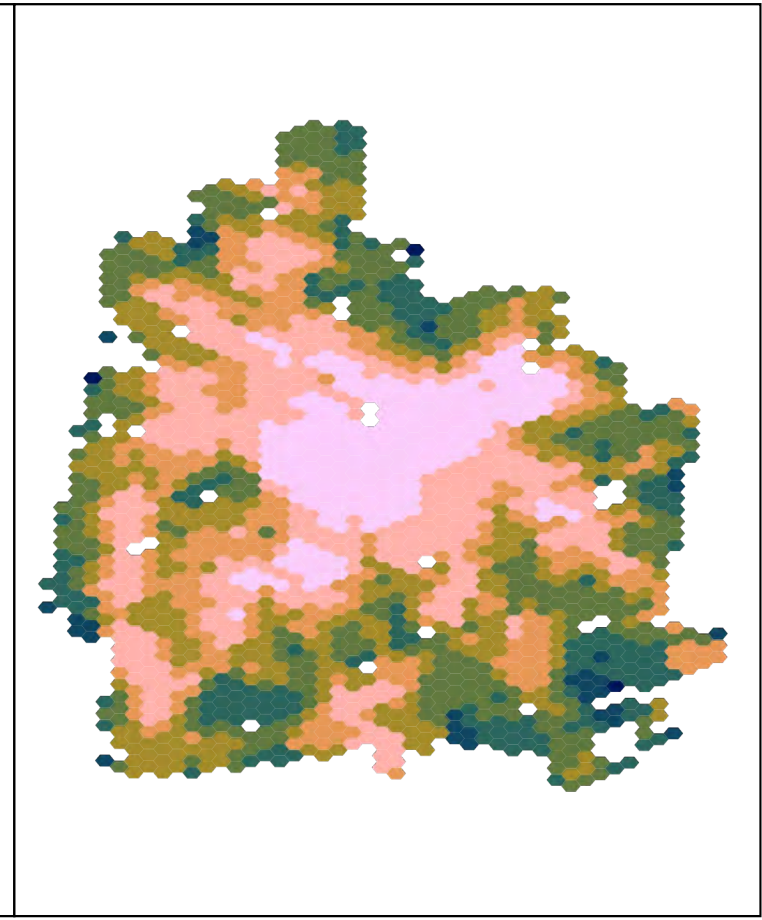
Walkability, relative to 25 global cities

— Urban boundary

0  4  8 km

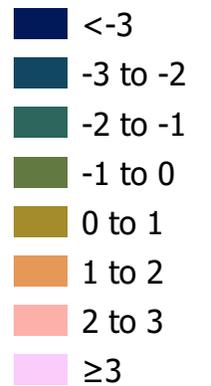

Walkability score
- <-3
- -3 to -2
- -2 to -1
- -1 to 0
- 0 to 1
- 1 to 2
- 2 to 3
- ≥3

Walkability relative to all cities by component variables (2D histograms), and overall (histogram)

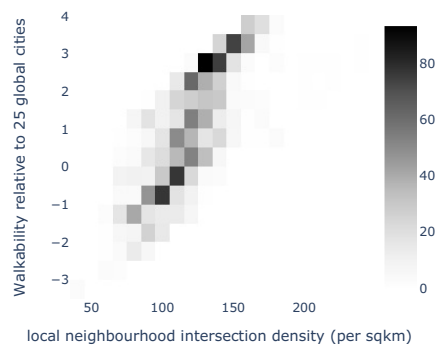
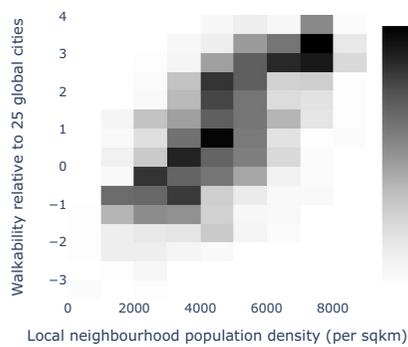
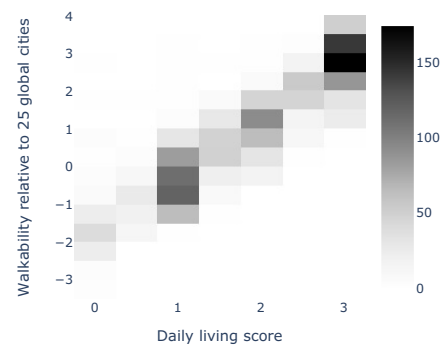
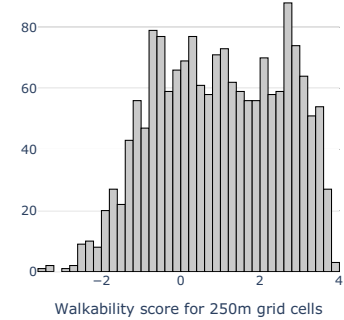

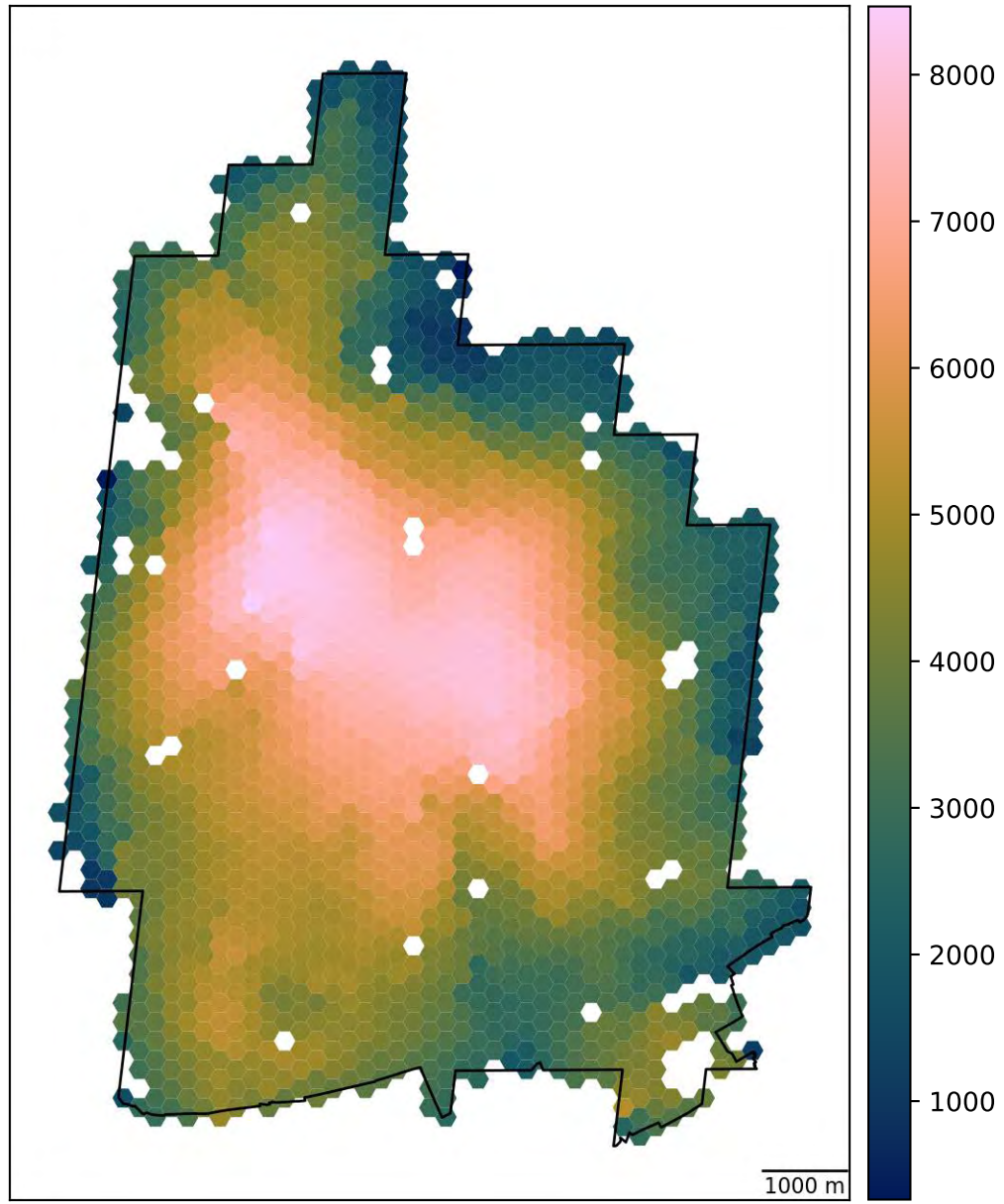

Mean 1000 m neighbourhood population per km²

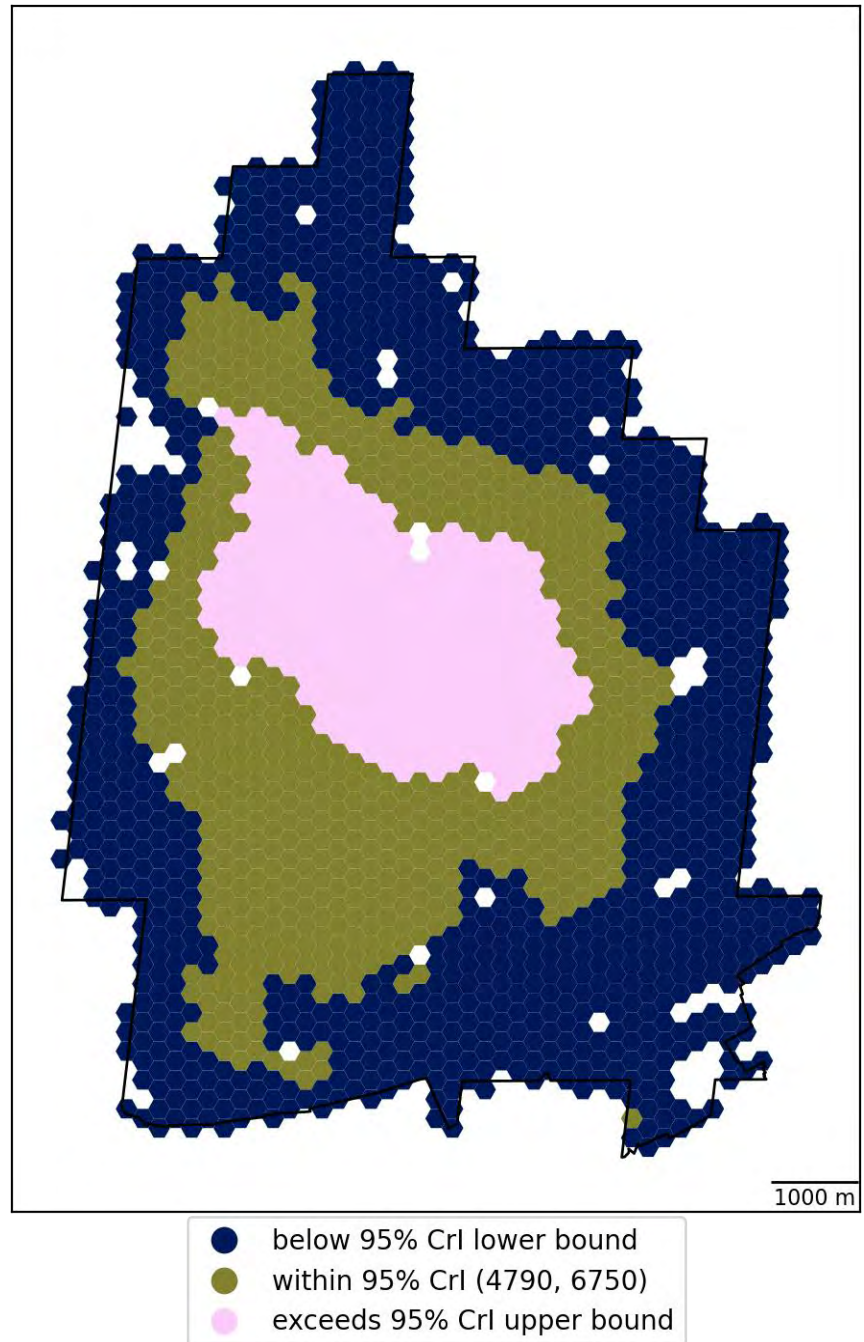

A: Estimated Mean 1000 m neighbourhood population per km² requirement for ≥80% probability of engaging in walking for transport

B: Estimated Mean 1000 m neighbourhood population per km² requirement for reaching the WHO's target of a ≥15% relative reduction in insufficient physical activity through walking

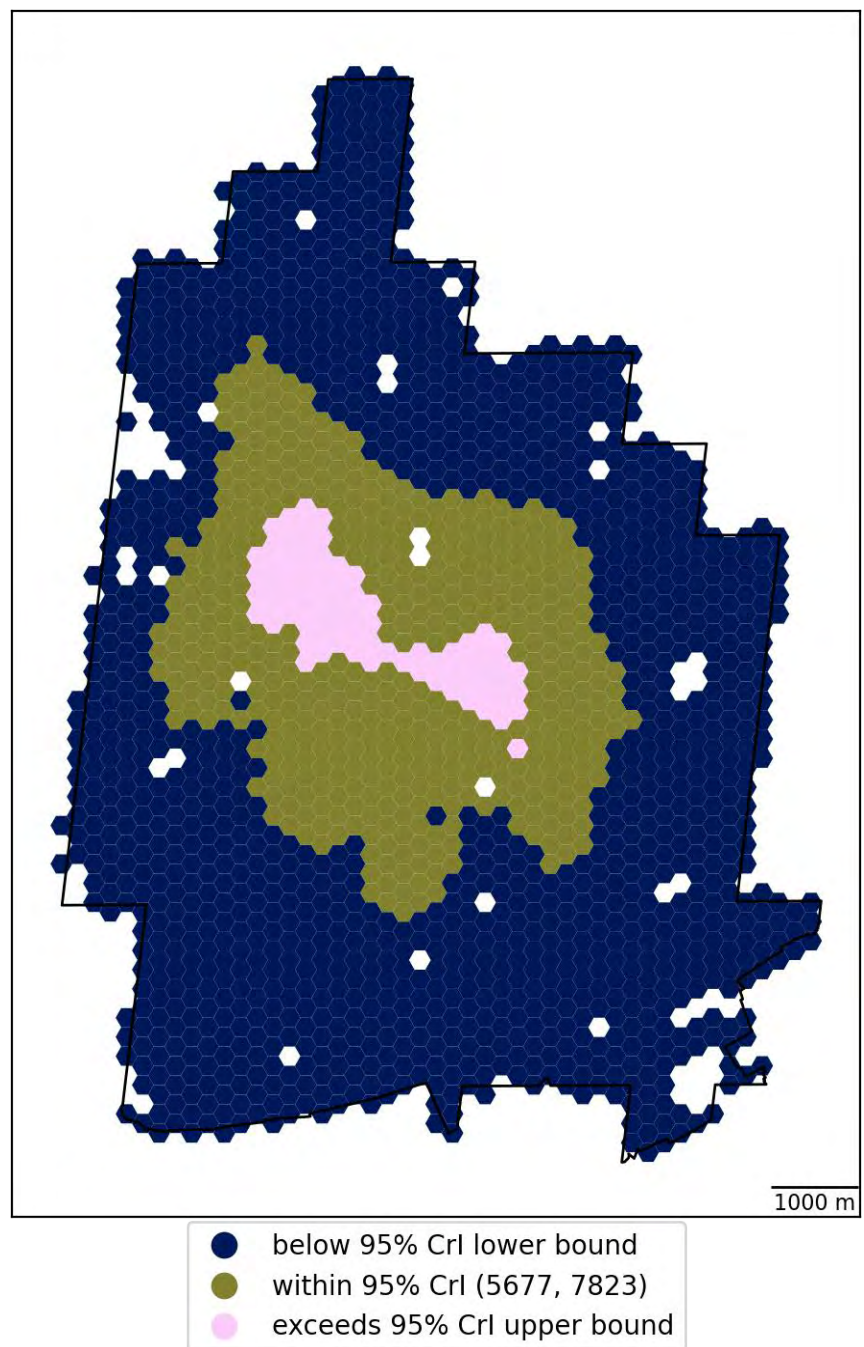

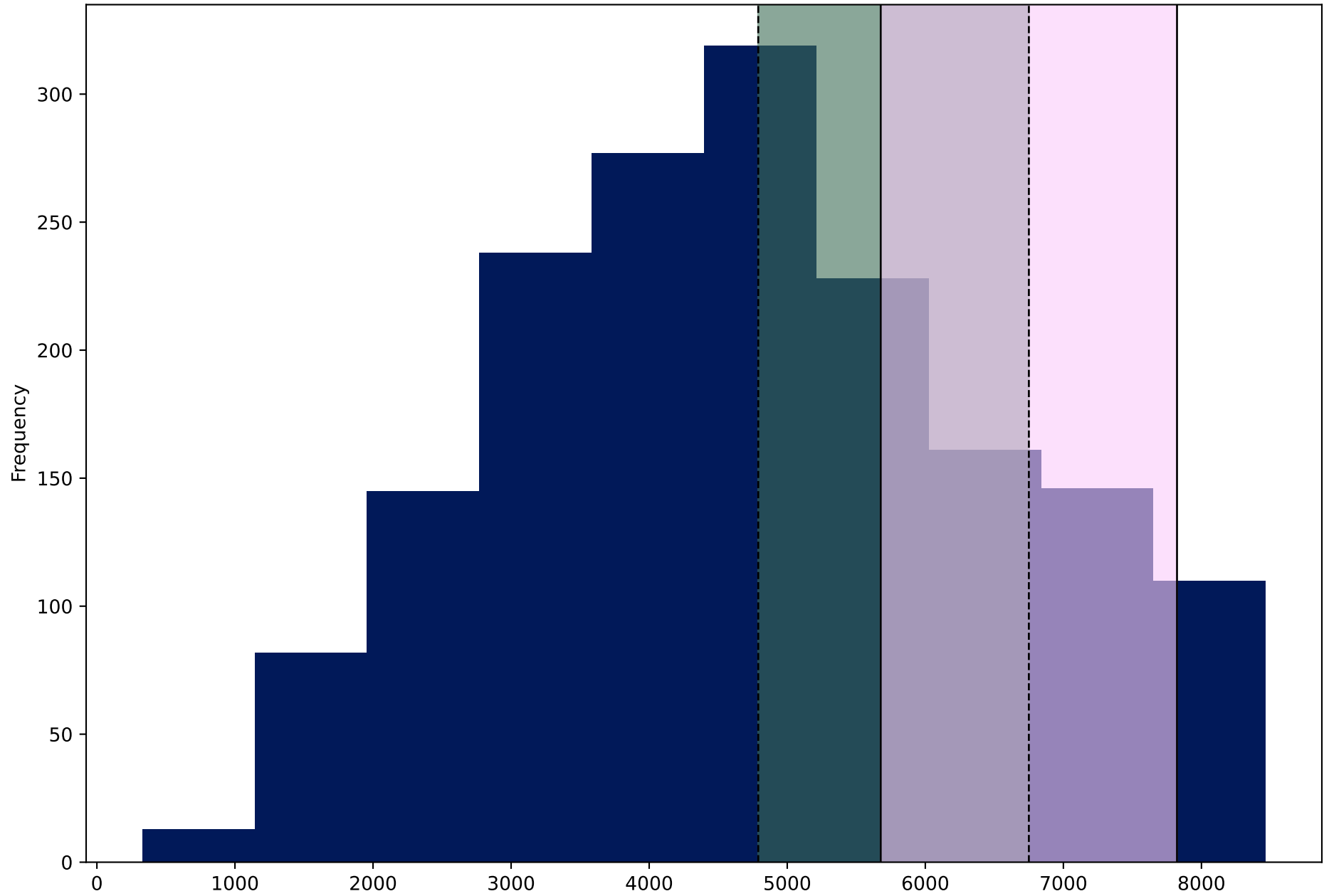

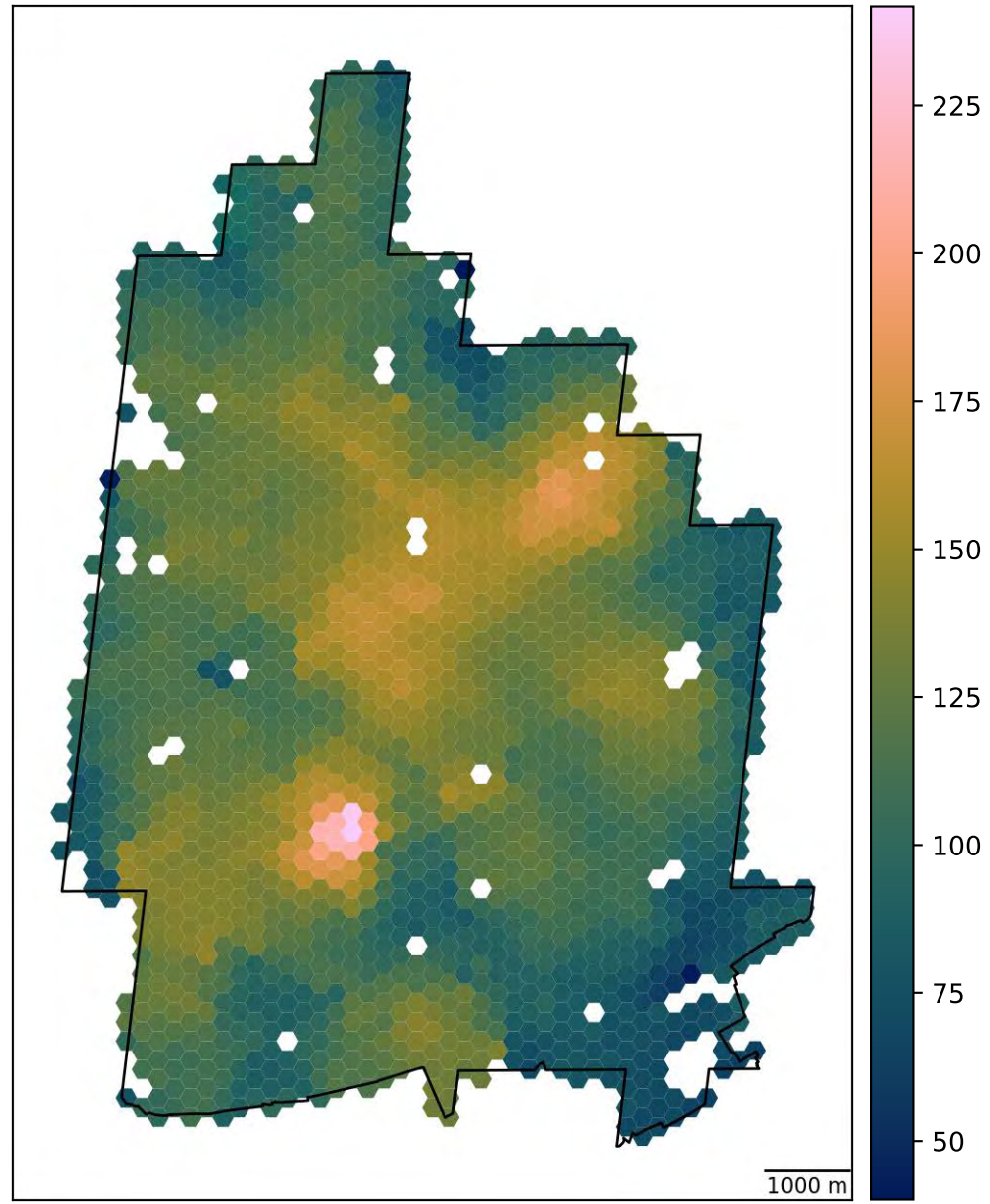

Mean 1000 m neighbourhood street intersections per km²

A: Estimated Mean 1000 m neighbourhood street intersections per km² requirement for ≥80% probability of engaging in walking for transport

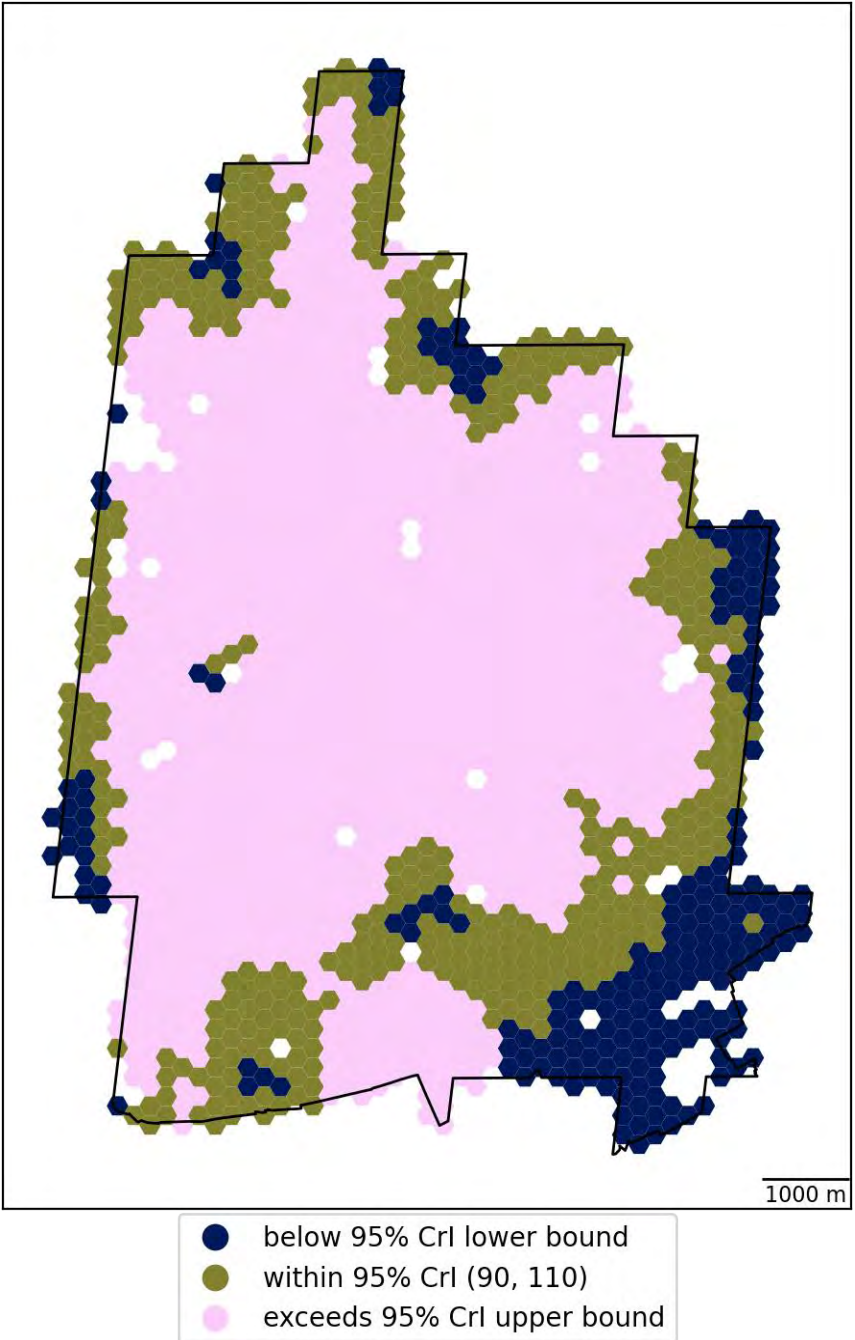

B: Estimated Mean 1000 m neighbourhood street intersections per km² requirement for reaching the WHO's target of a ≥15% relative reduction in insufficient physical activity through walking

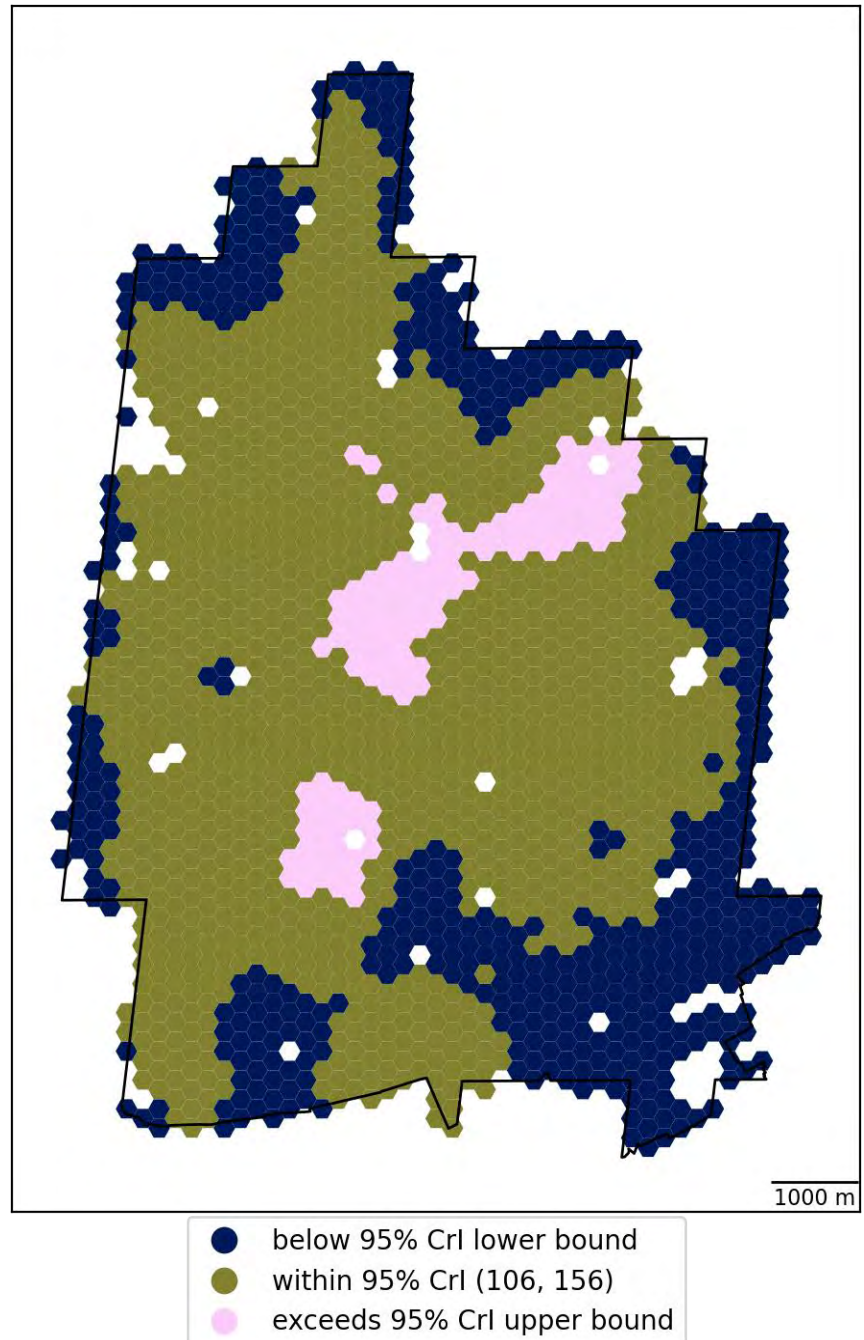

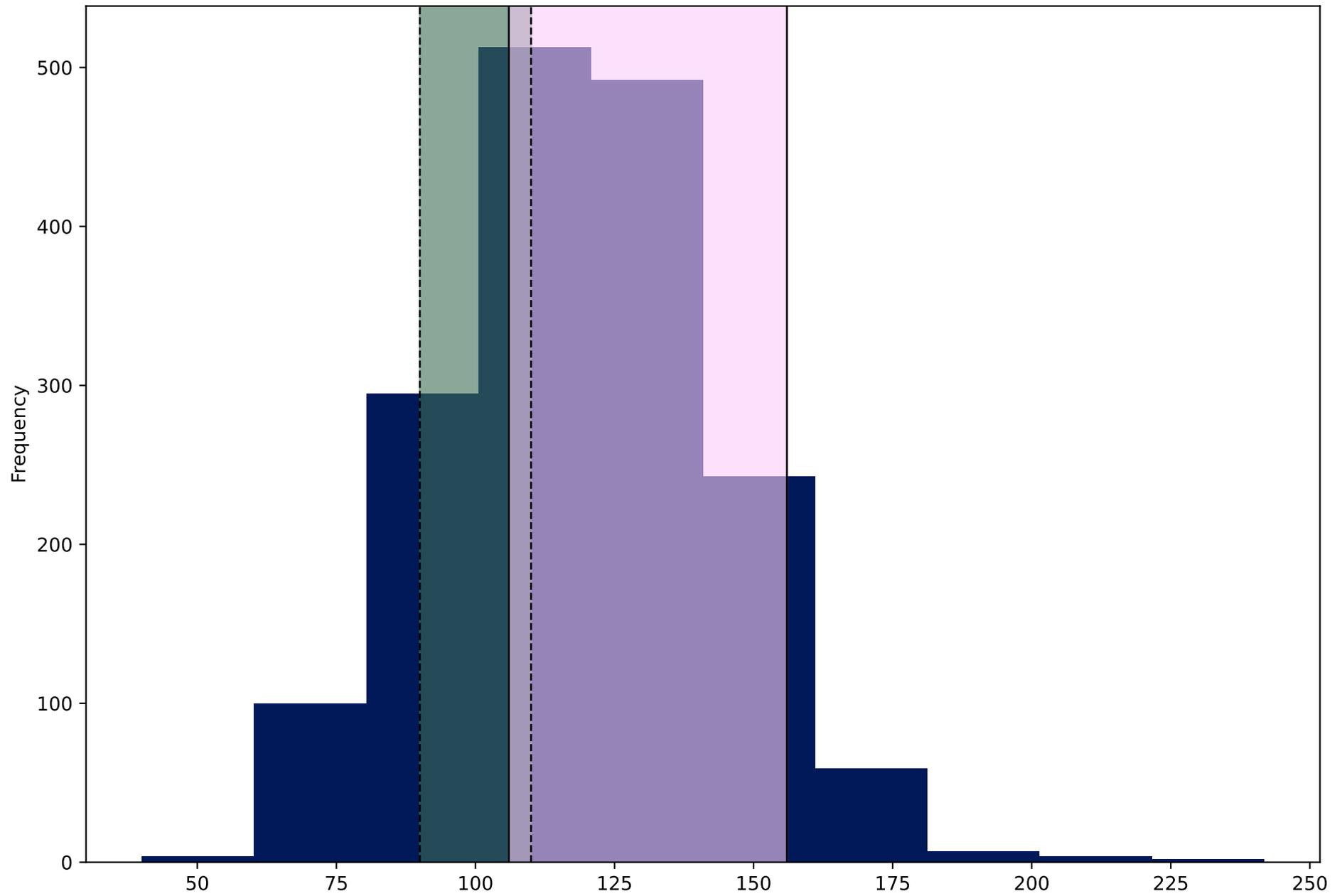

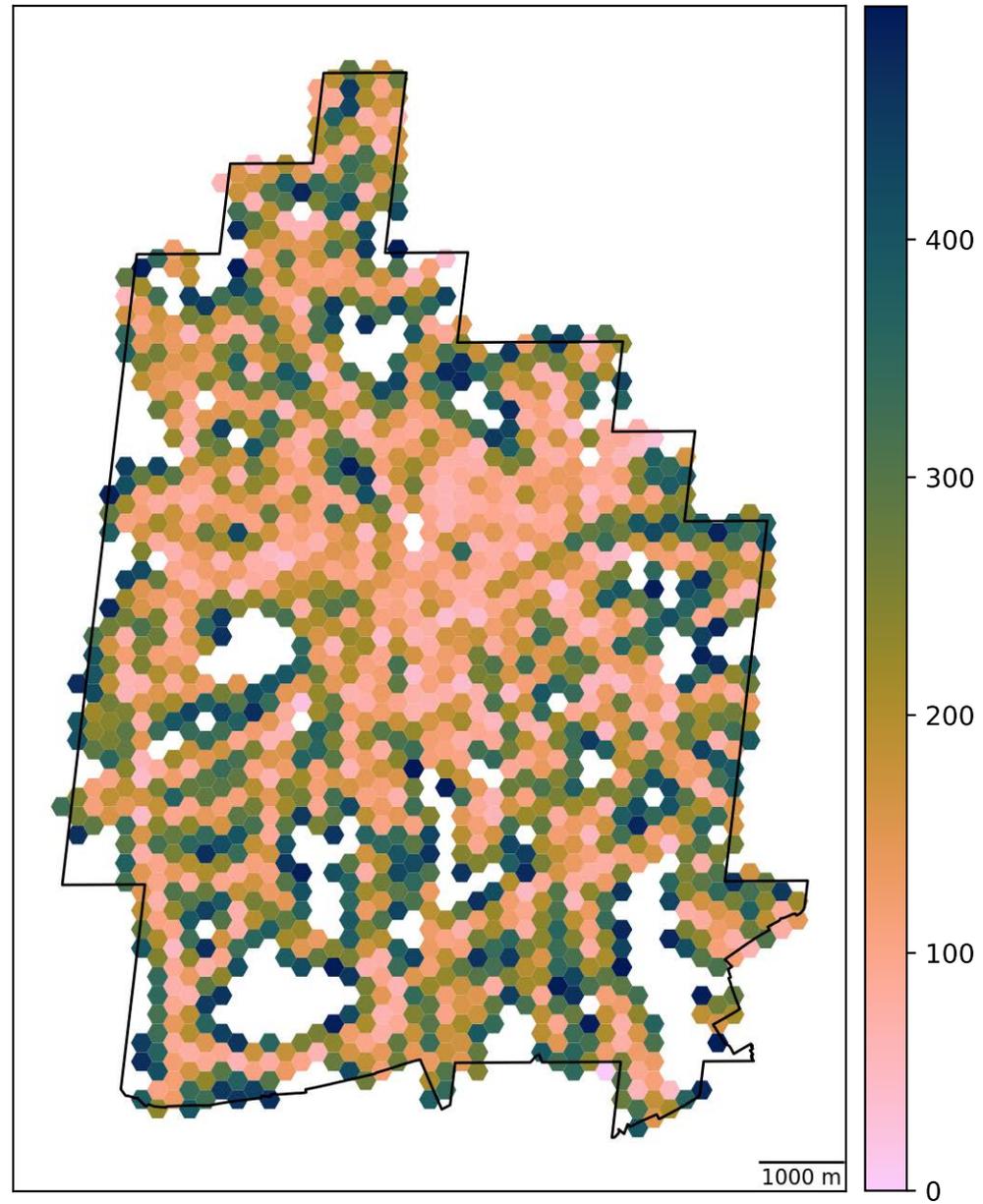

distances: Estimated Distance to nearest public transport stops (m; up to 500m) requirement for distances to destinations, measured up to a maximum distance target threshold of 500 metres

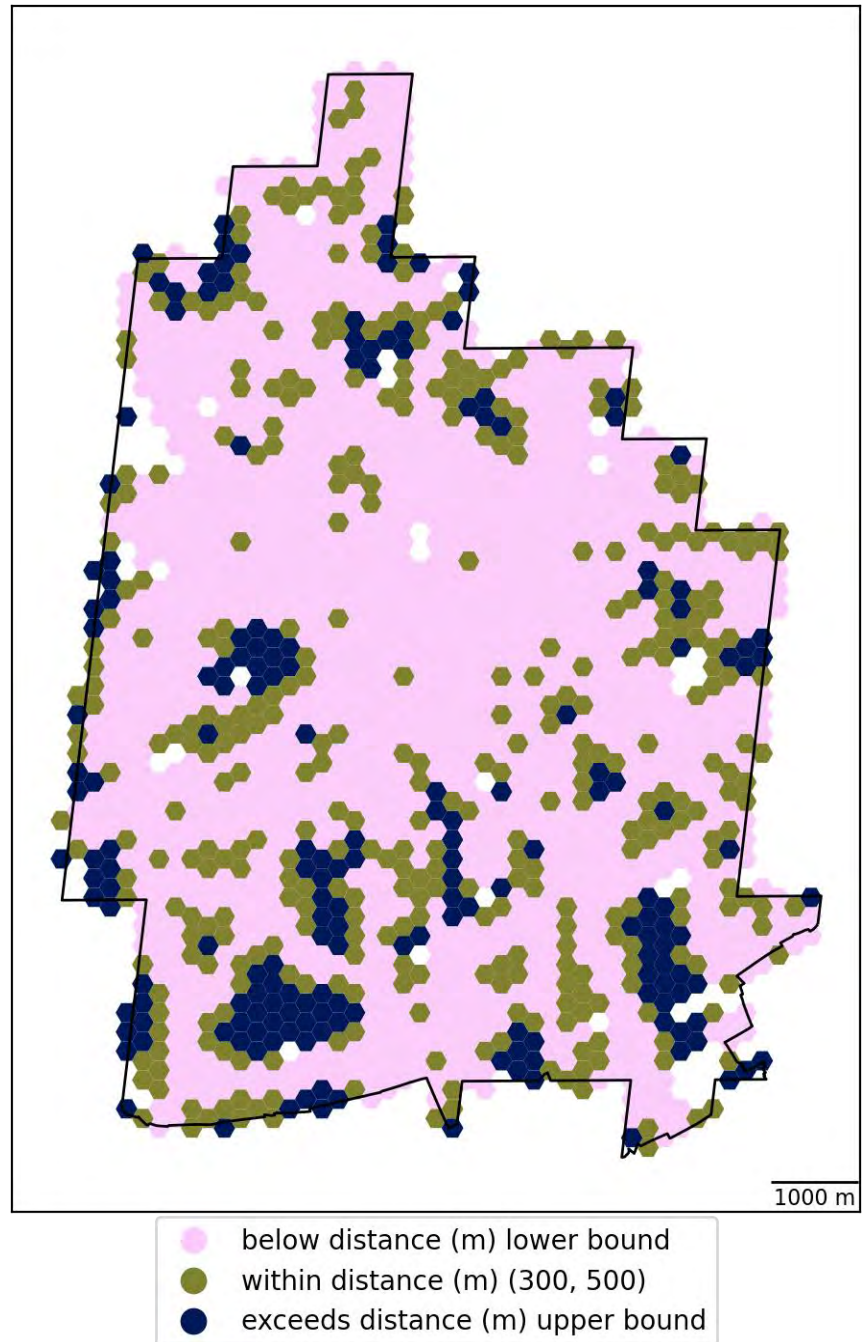

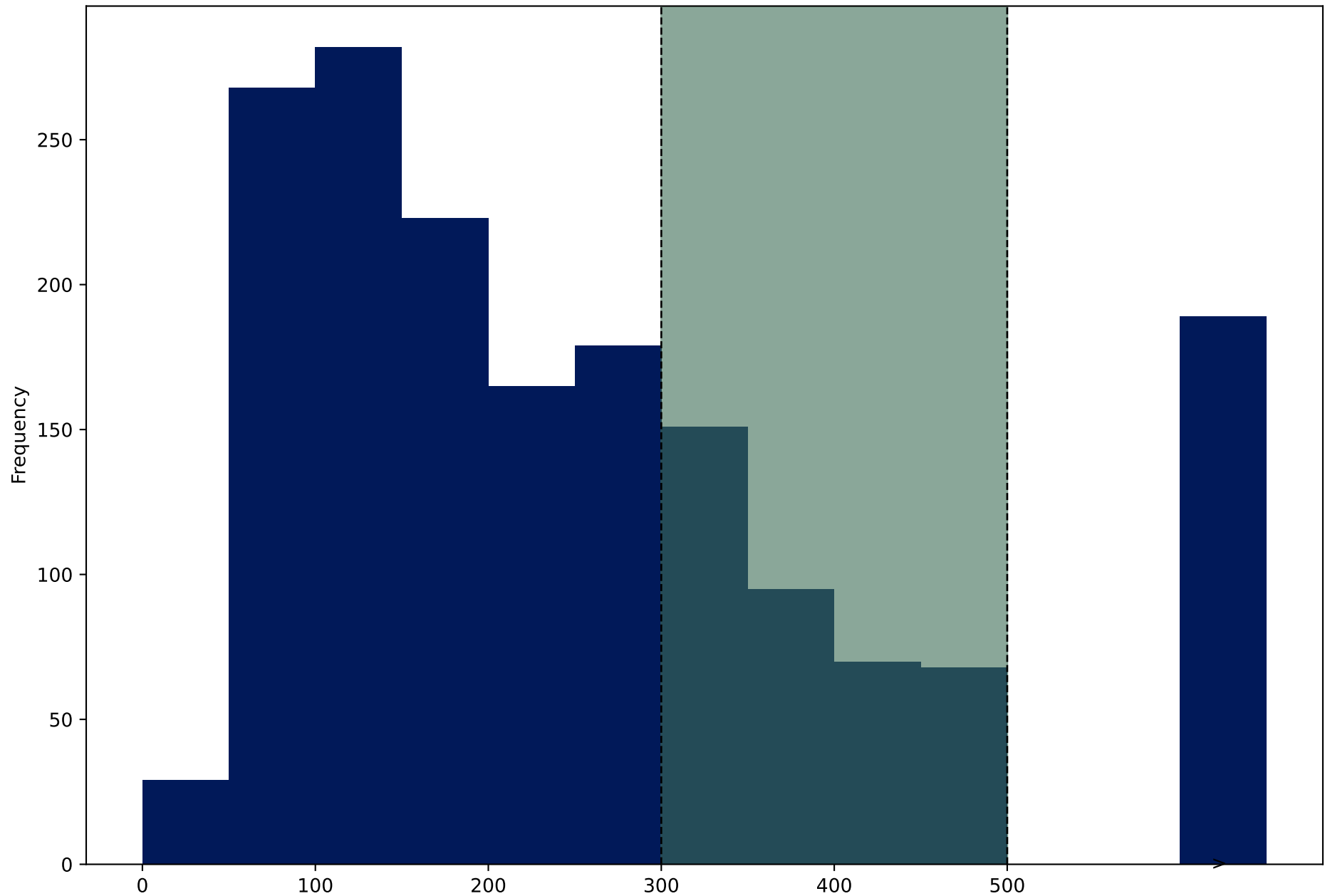

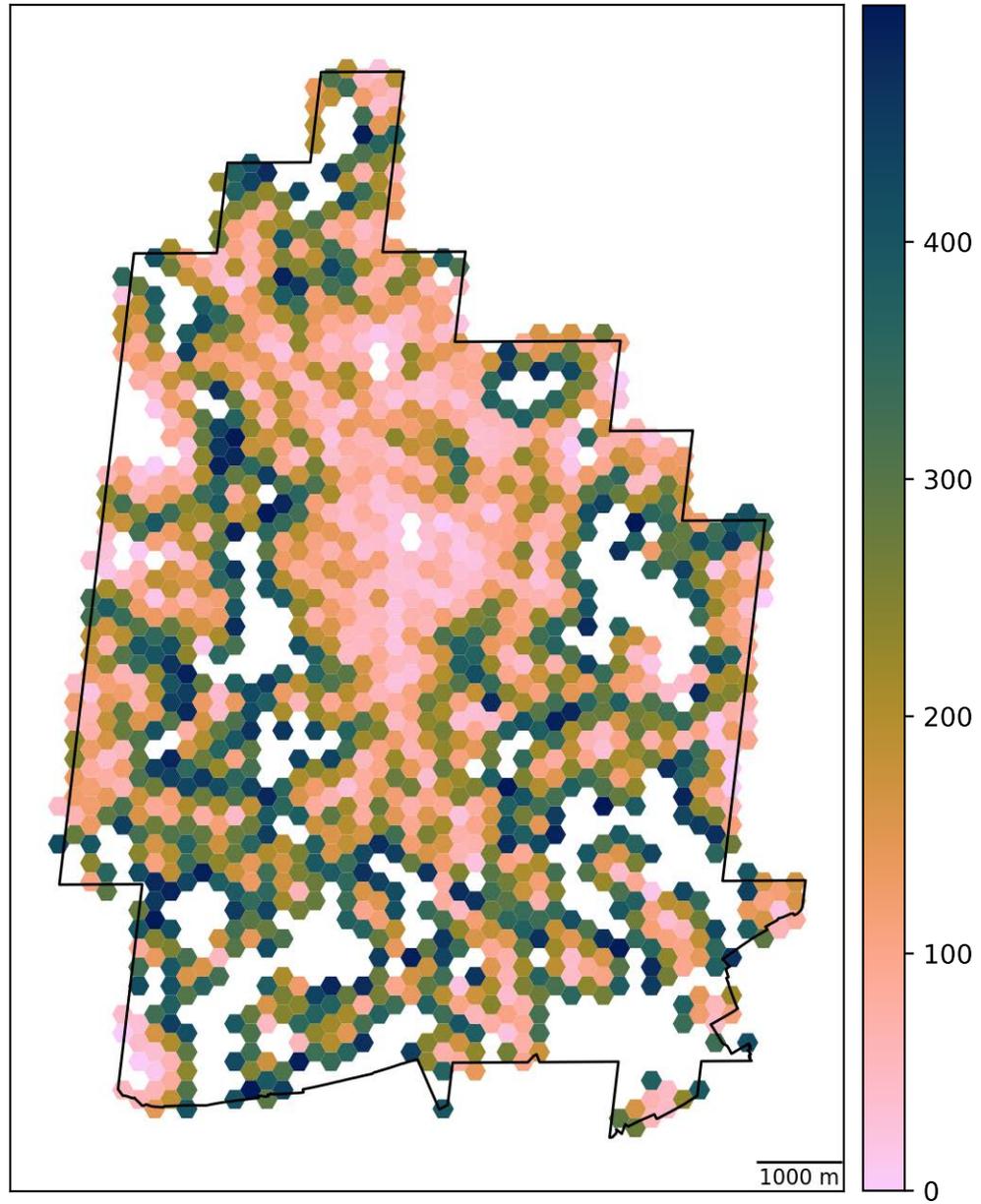
255

distances: Estimated Distance to nearest park (m; up to 500m) requirement for distances to destinations, measured up to a maximum distance target threshold of 500 metres

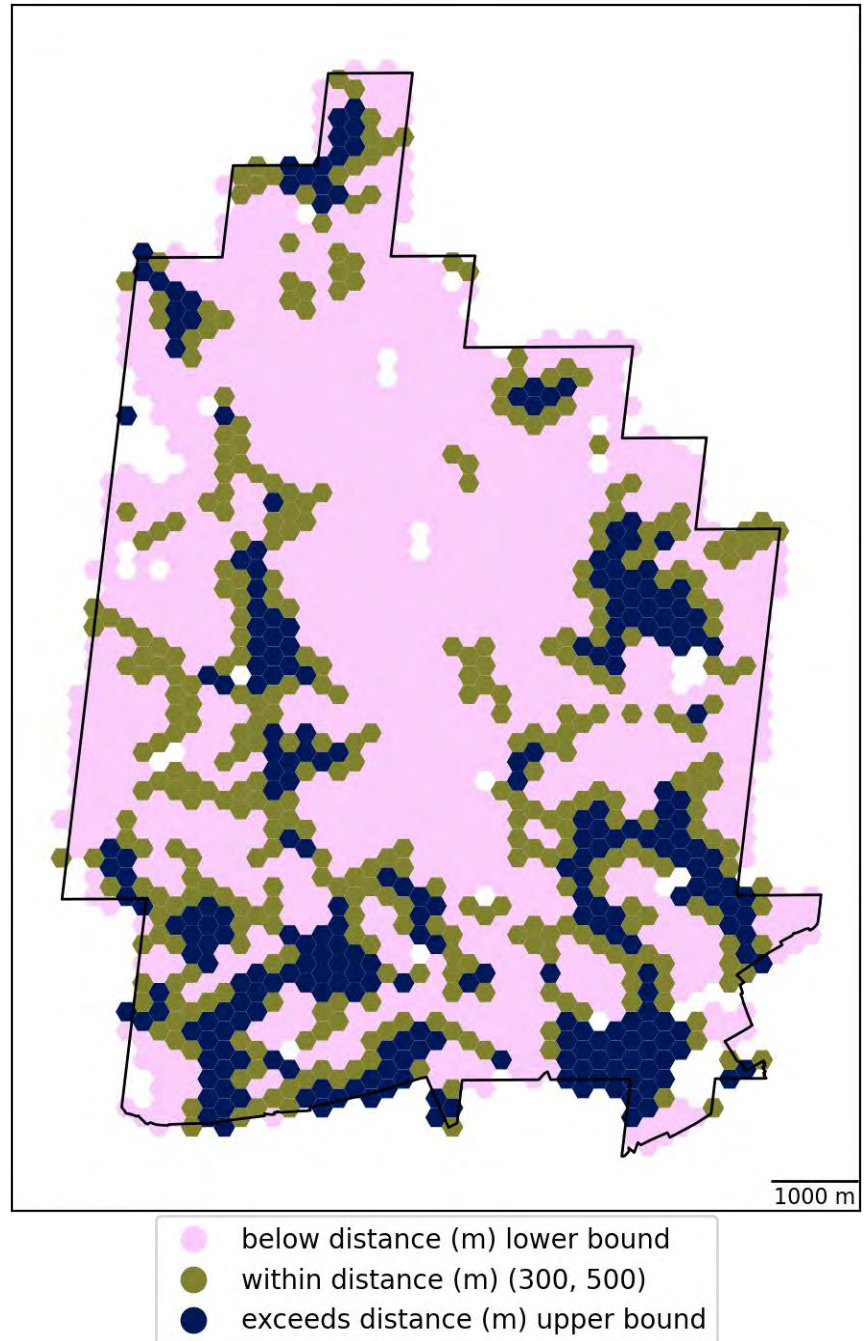

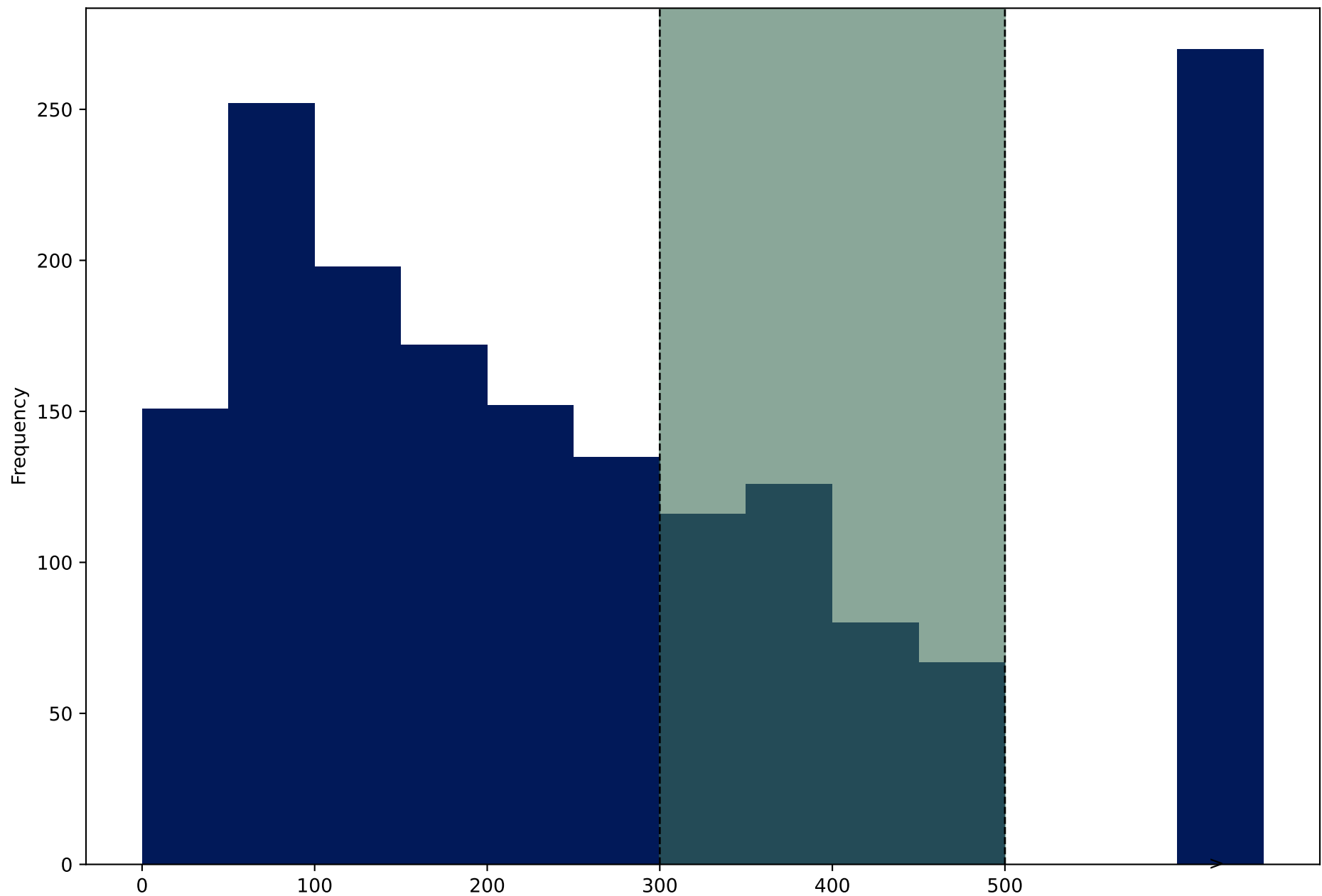

# Europe, Belgium, Ghent

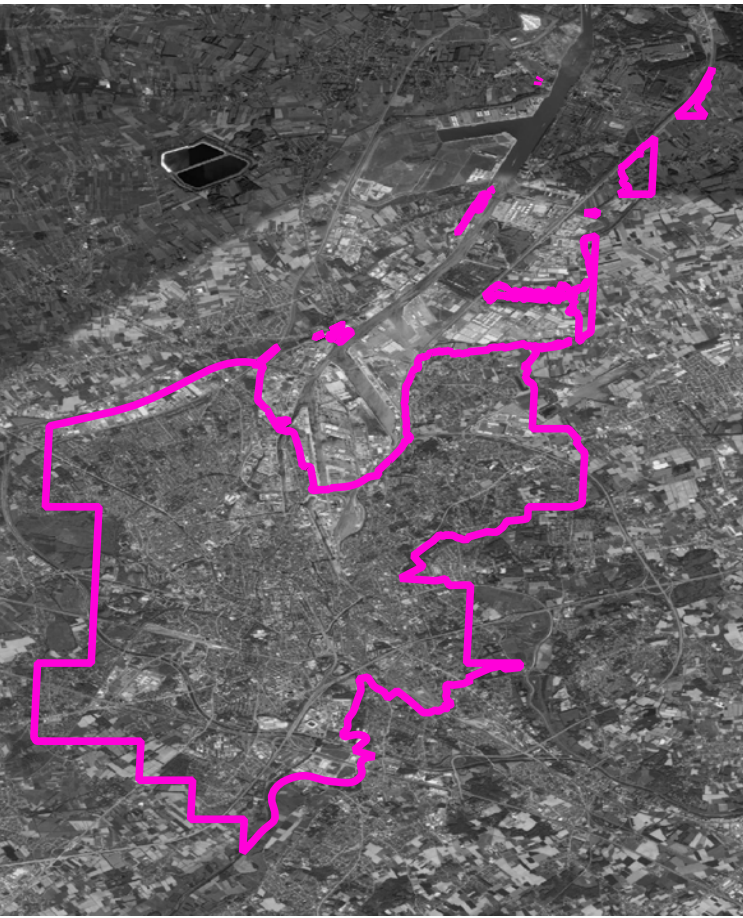
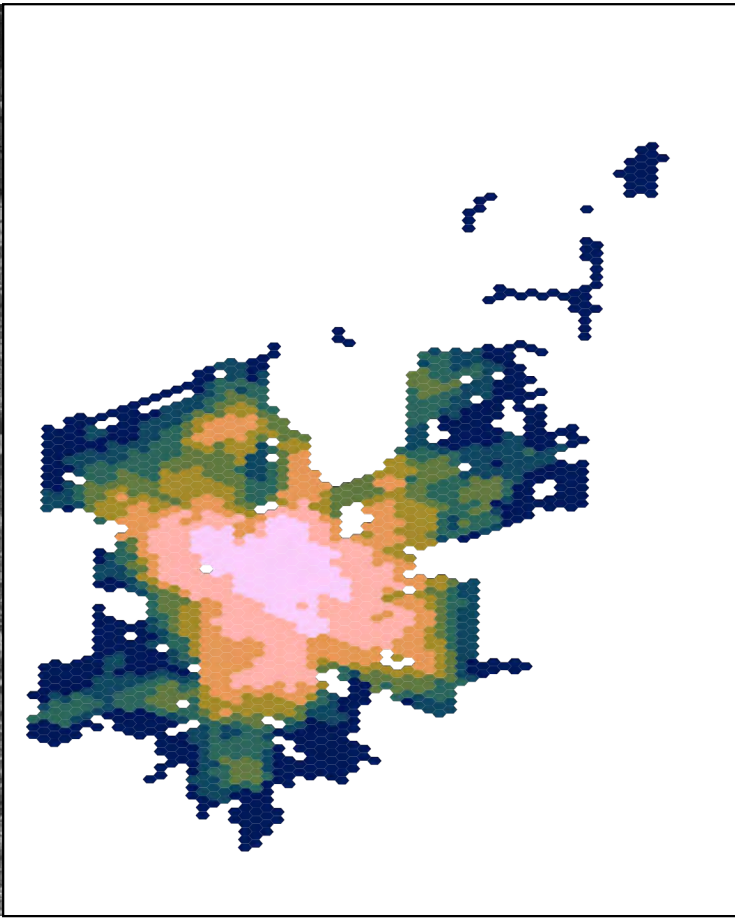
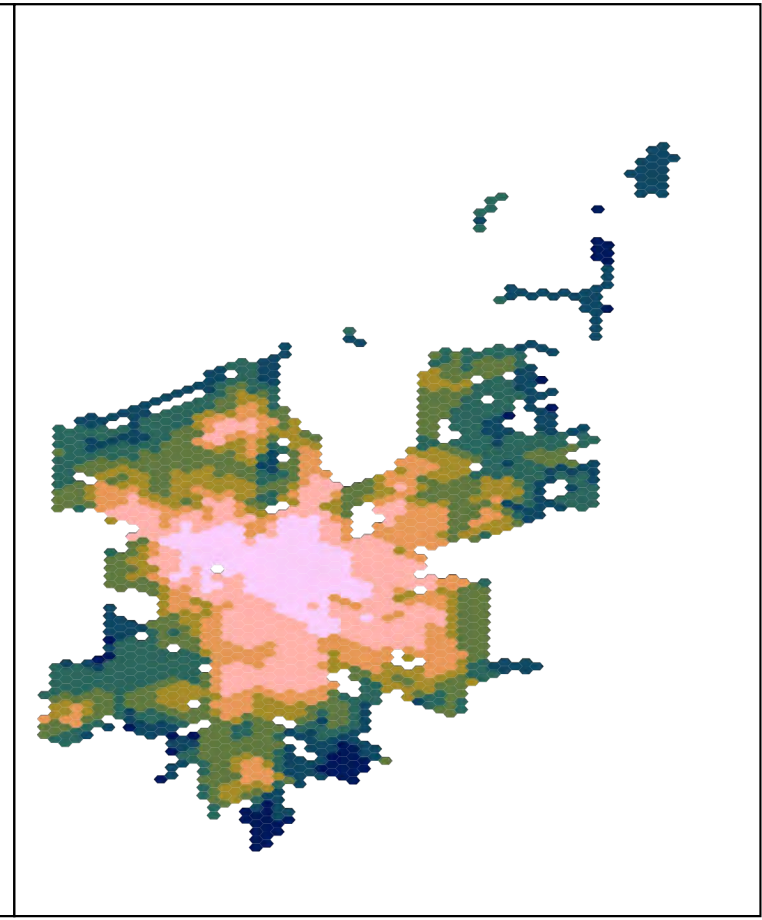
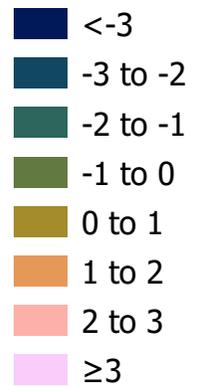
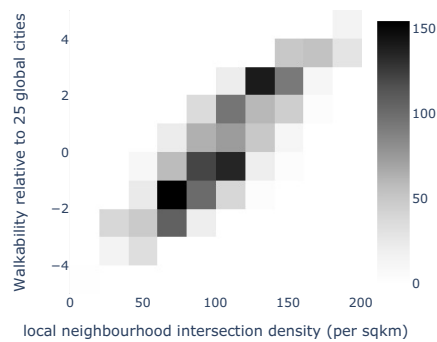
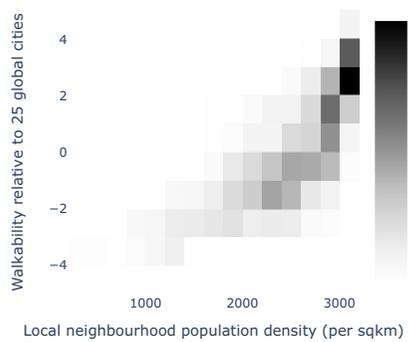
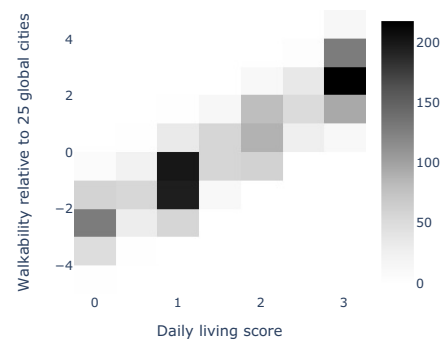
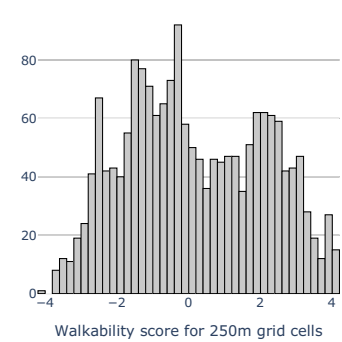

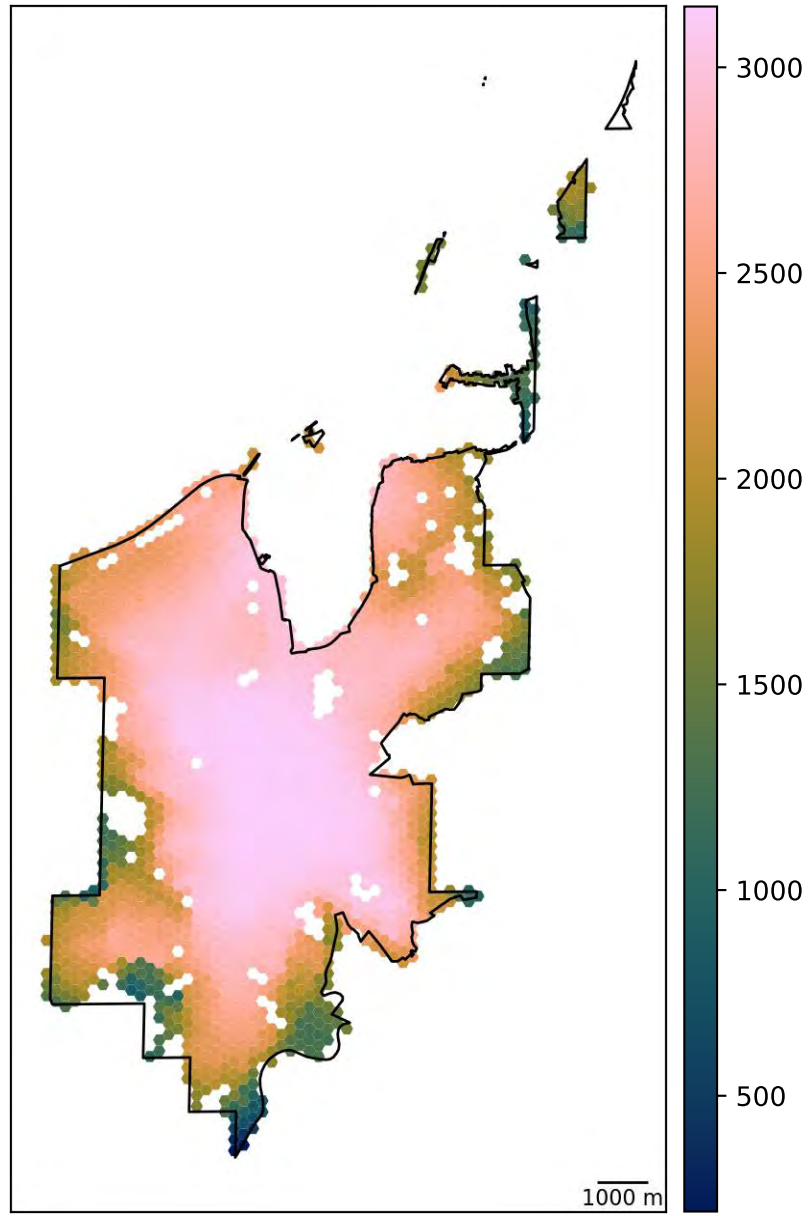

Mean 1000 m neighbourhood population per km²

A: Estimated Mean 1000 m neighbourhood population per km² requirement for ≥80% probability of engaging in walking for transport

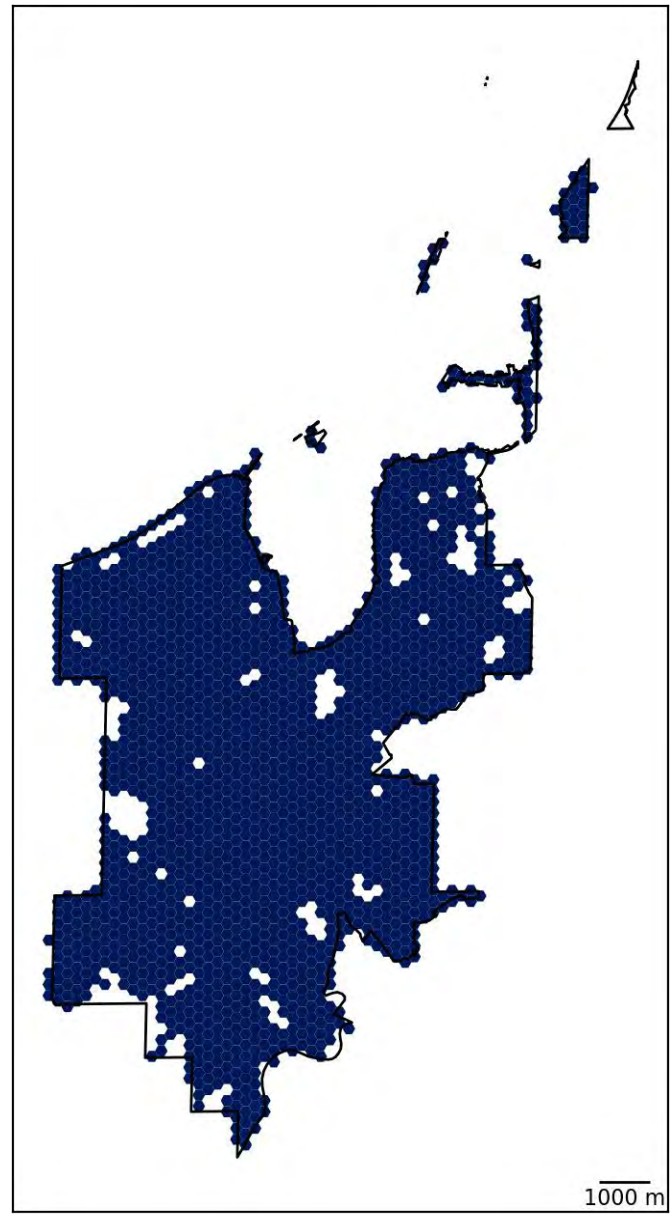

● below 95% CrI lower bound

B: Estimated Mean 1000 m neighbourhood population per km² requirement for reaching the WHO's target of a ≥15% relative reduction in insufficient physical activity through walking

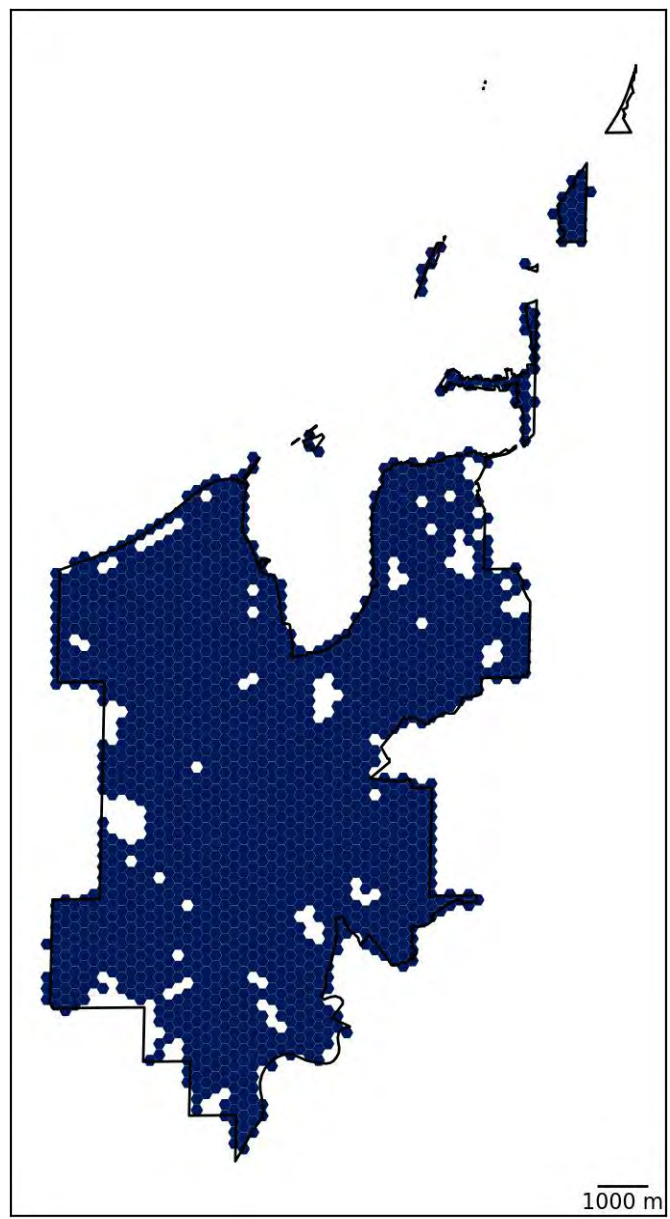

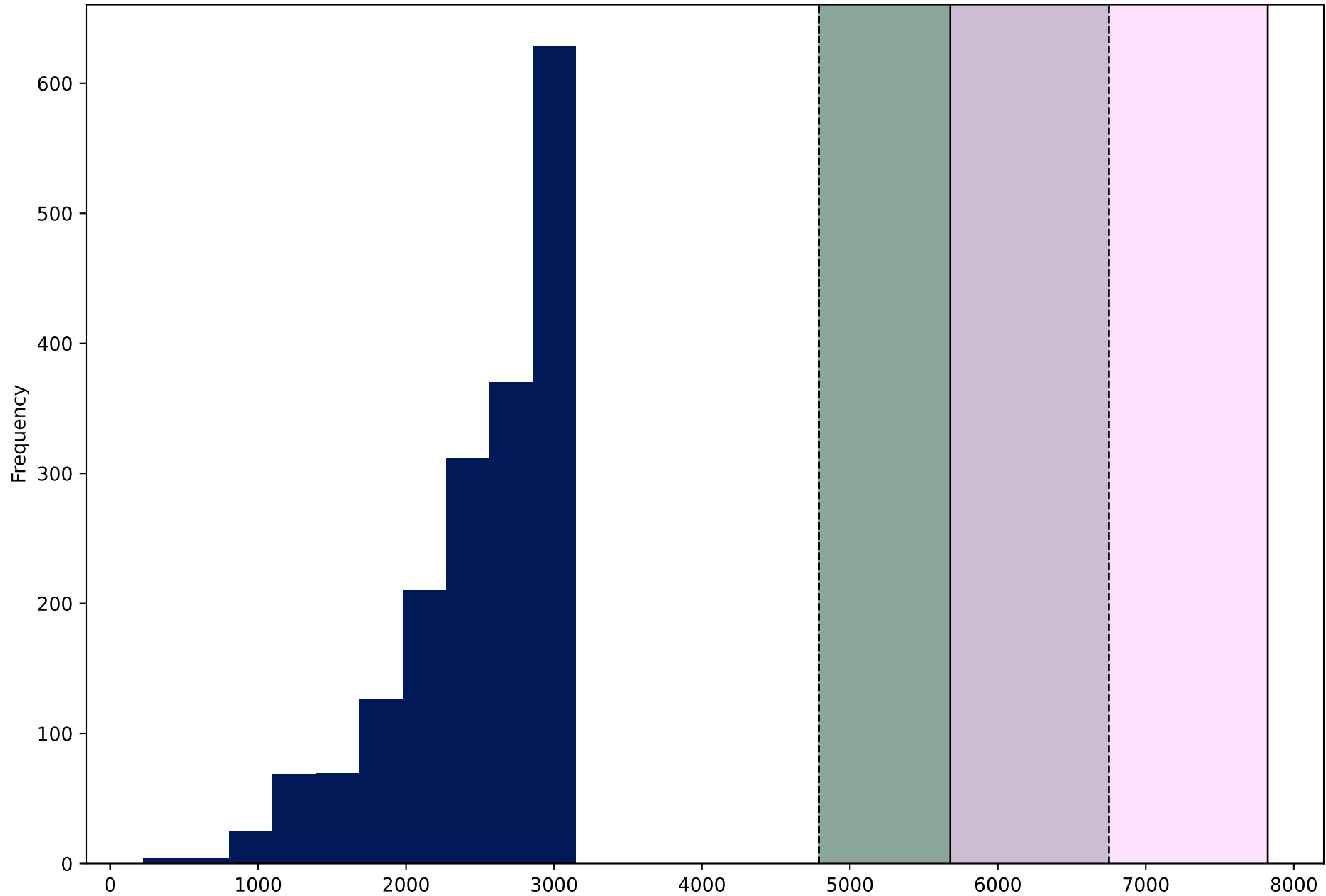
262

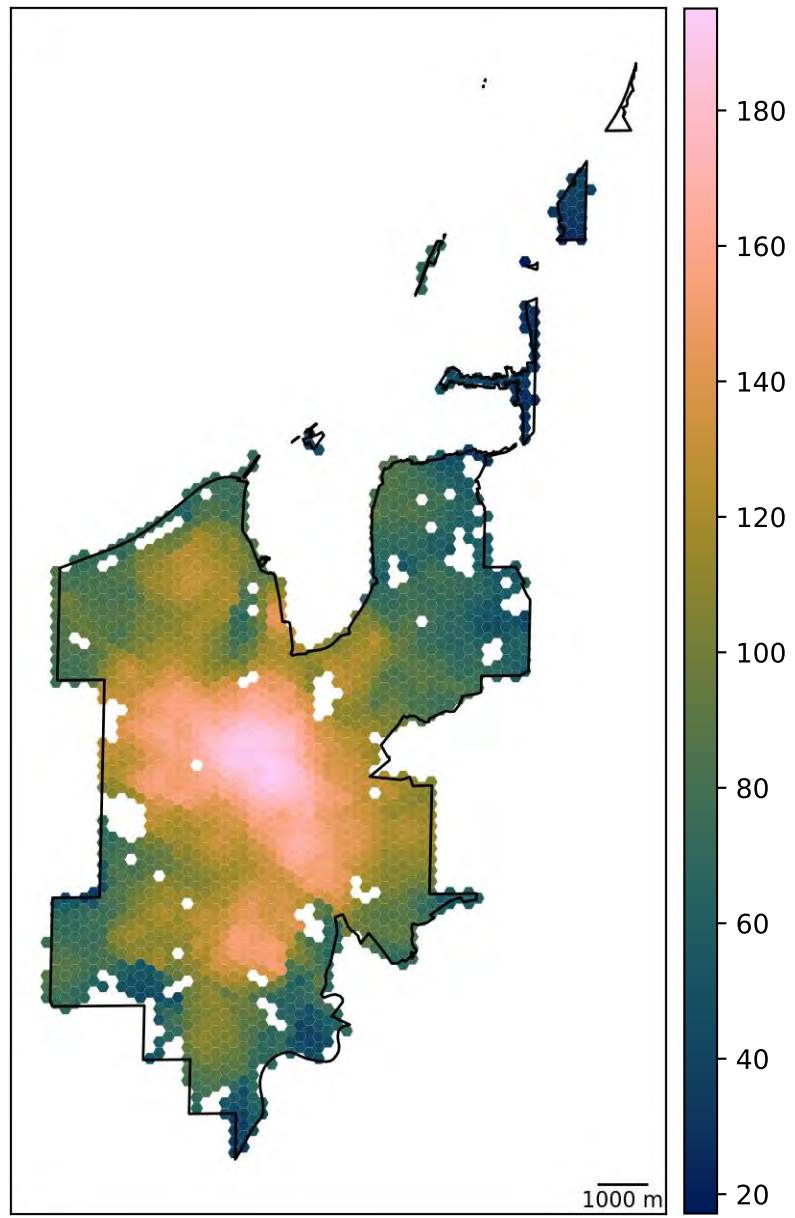

Mean 1000 m neighbourhood street intersections per km²

A: Estimated Mean 1000 m neighbourhood street intersections per km² requirement for ≥80% probability of engaging in walking for transport

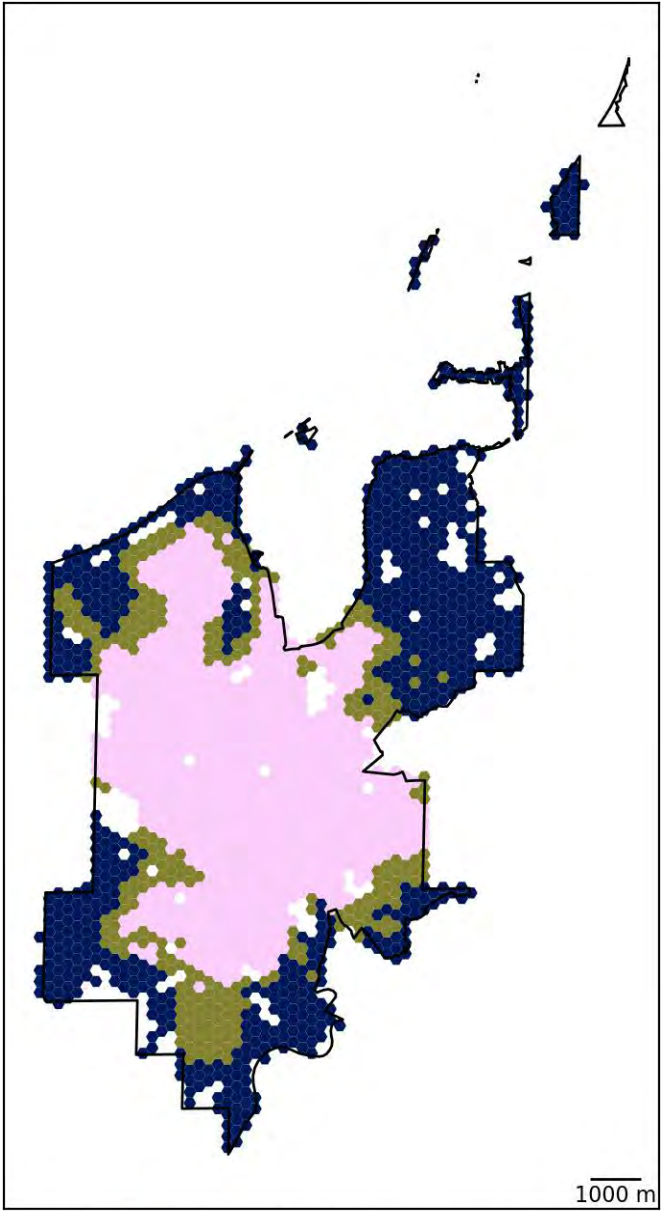

B: Estimated Mean 1000 m neighbourhood street intersections per km² requirement for reaching the WHO's target of a ≥15% relative reduction in insufficient physical activity through walking

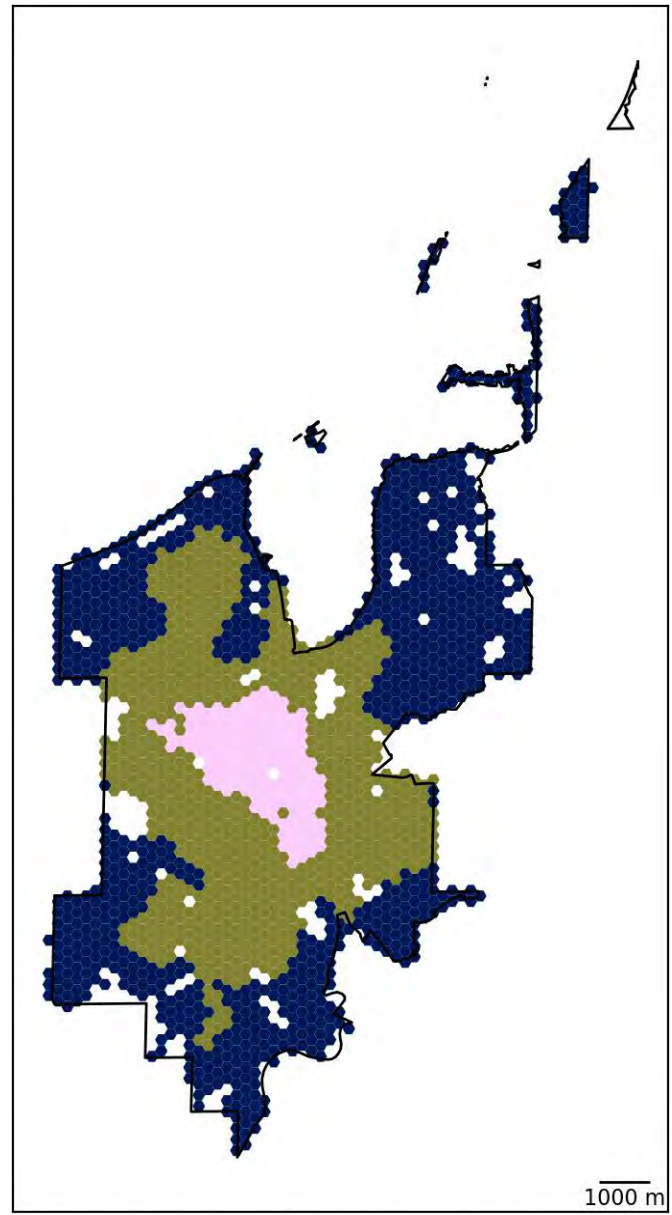

- below 95% CrI lower bound
- within 95% CrI (106, 156)
- exceeds 95% CrI upper bound

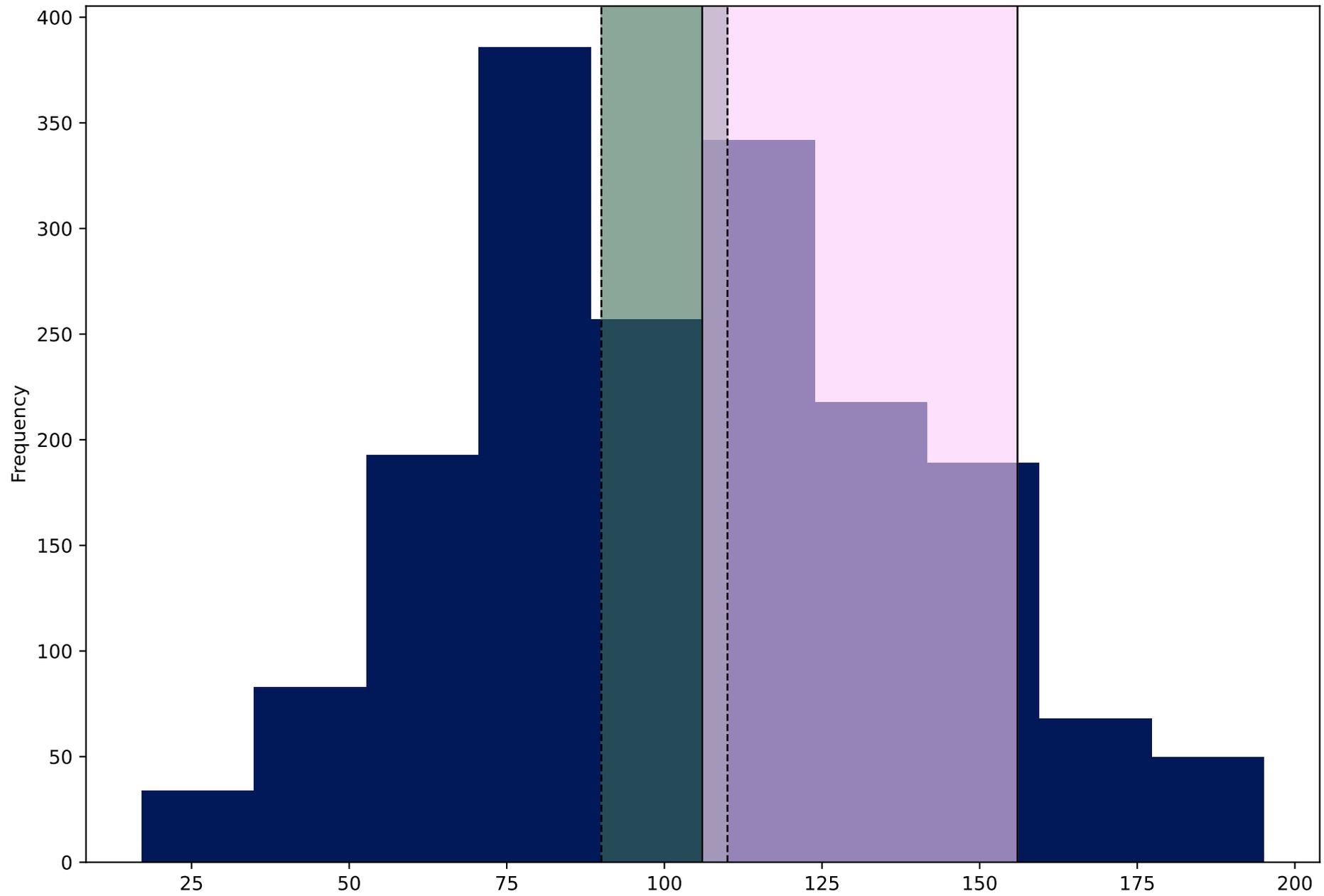

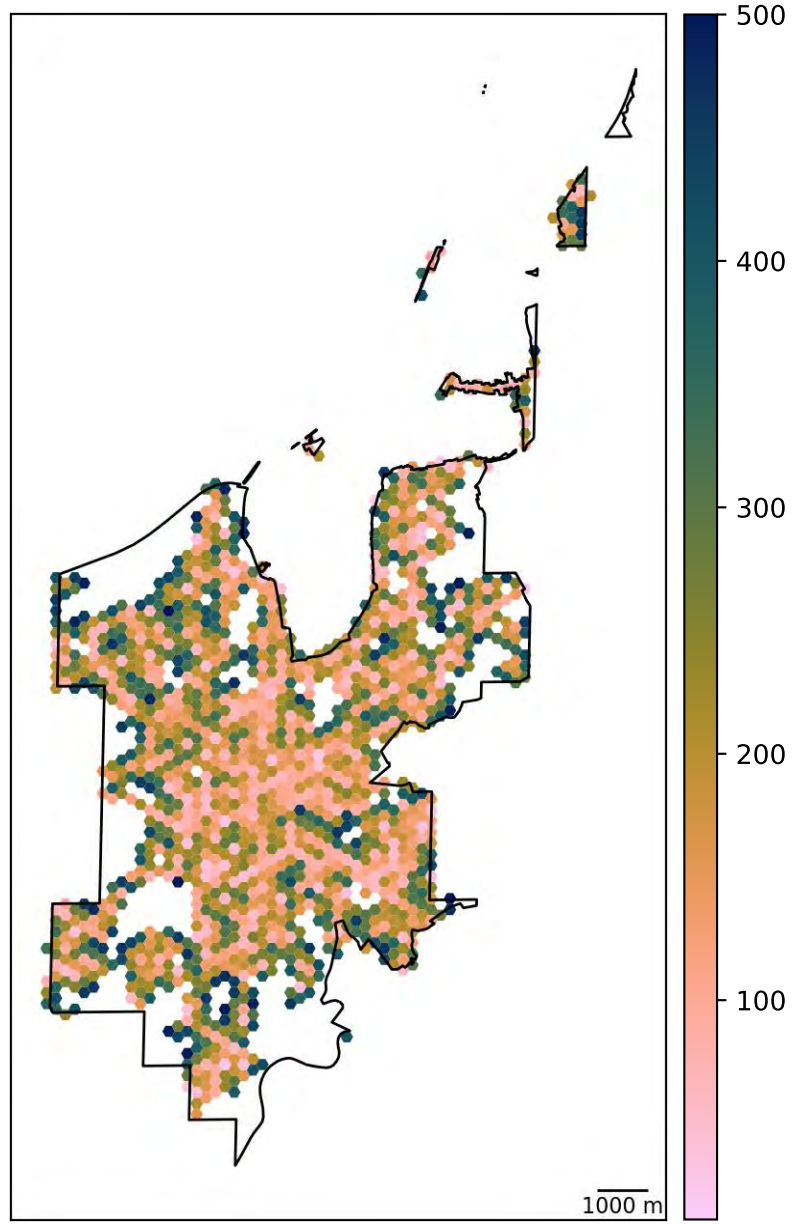

Distance to nearest public transport stops (m; up to 500m)

distances: Estimated Distance to nearest public transport stops (m; up to 500m) requirement for distances to destinations, measured up to a maximum distance target threshold of 500 metres

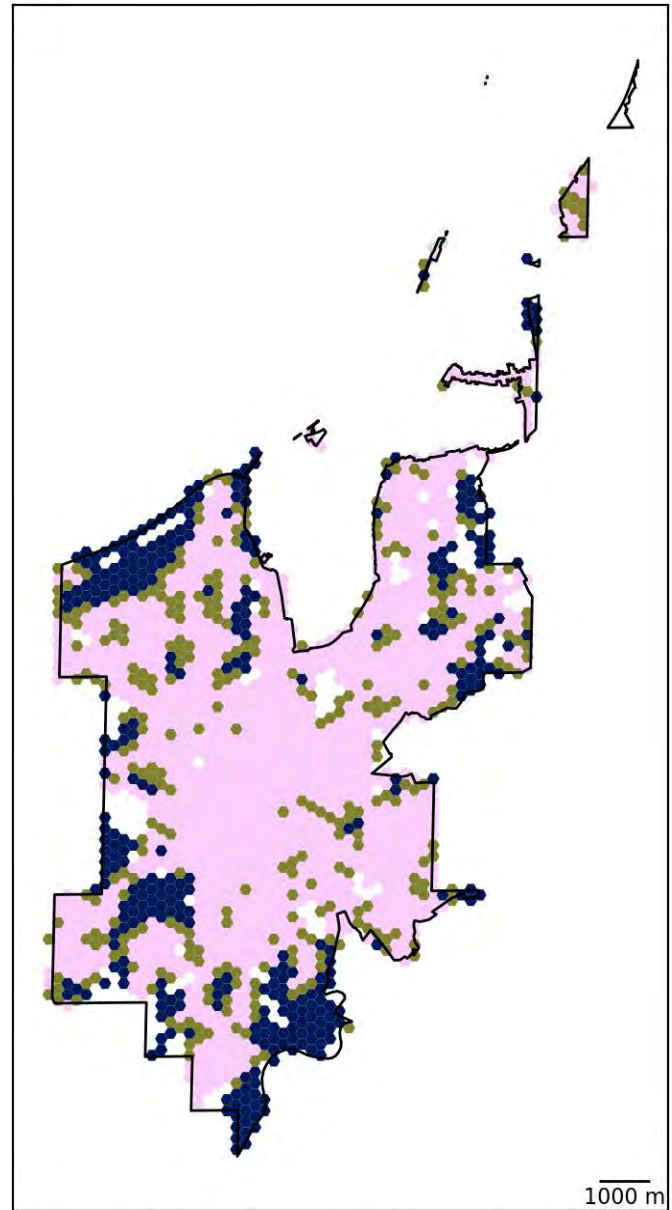

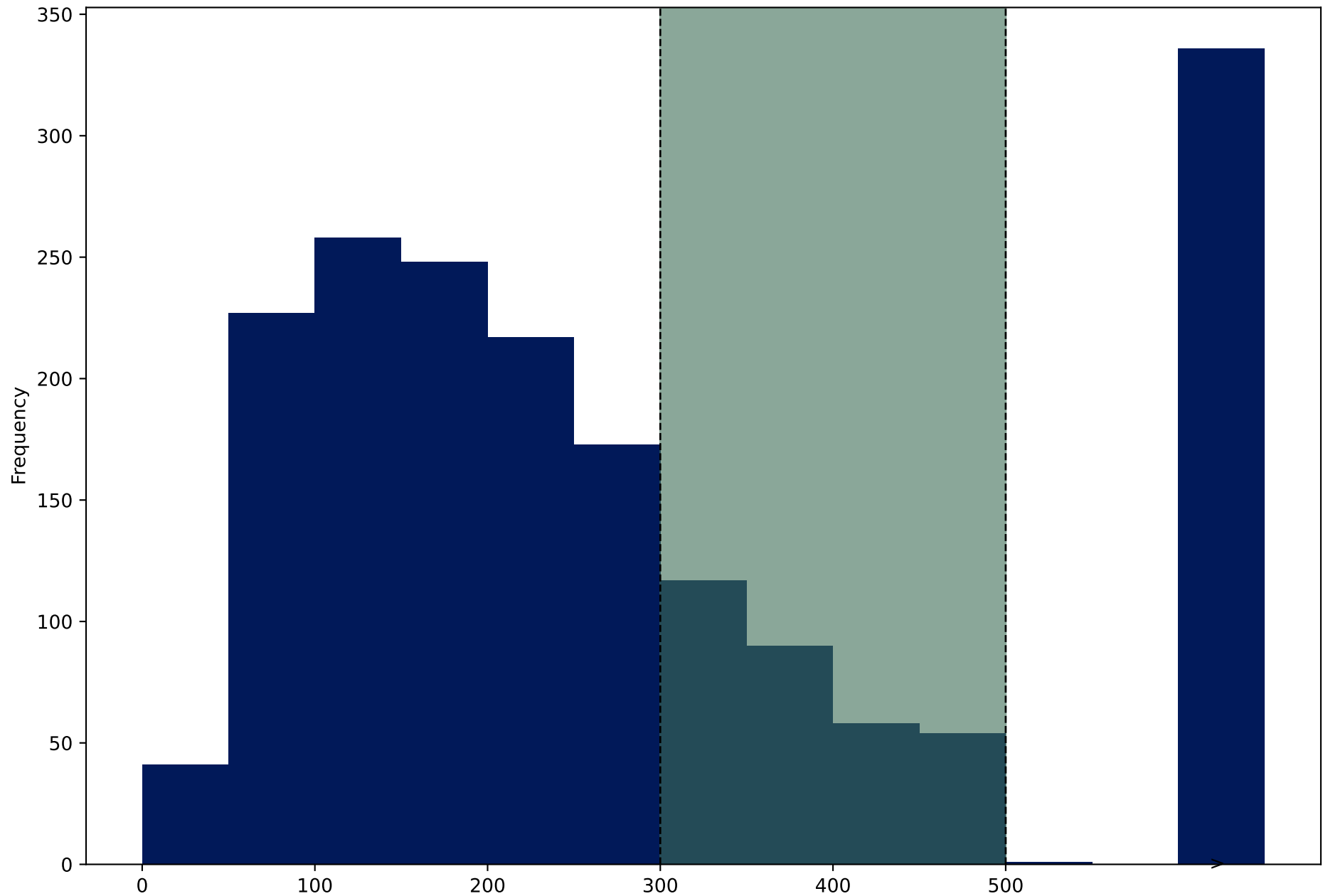

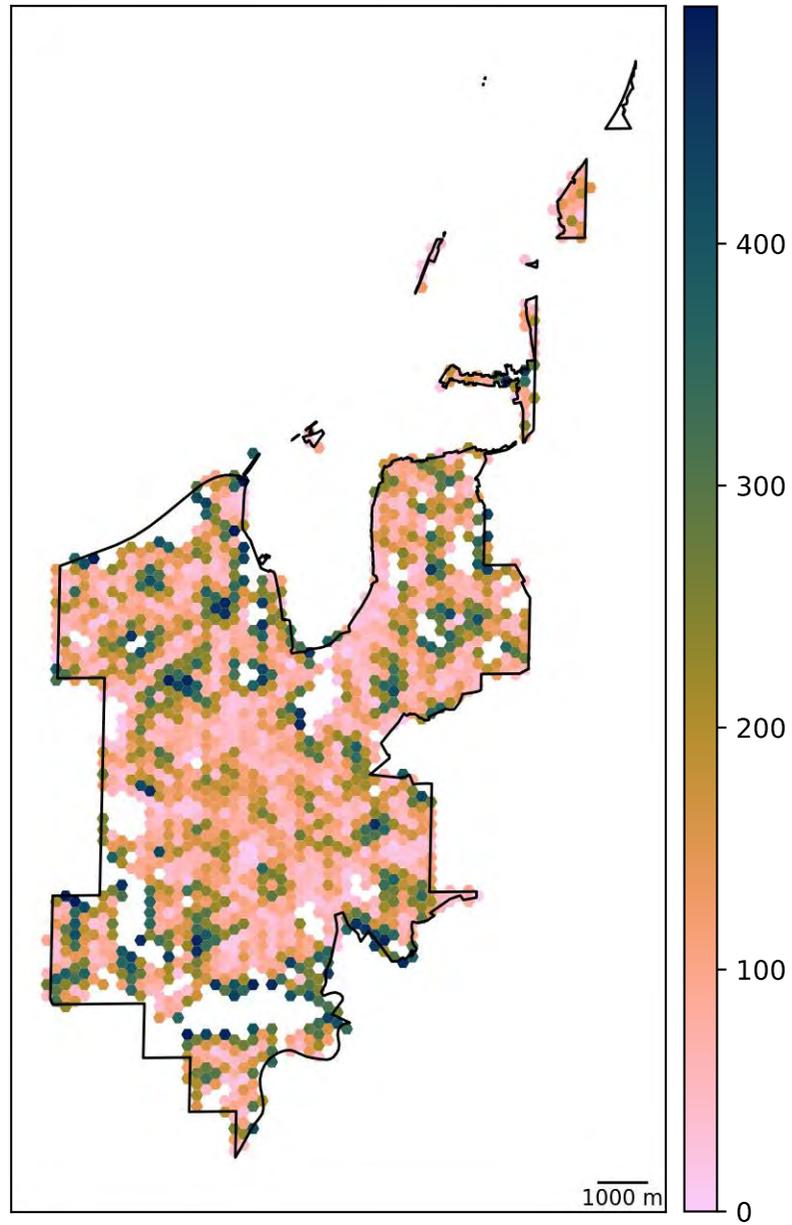

distances: Estimated Distance to nearest park (m; up to 500m) requirement for distances to destinations, measured up to a maximum distance target threshold of 500 metres

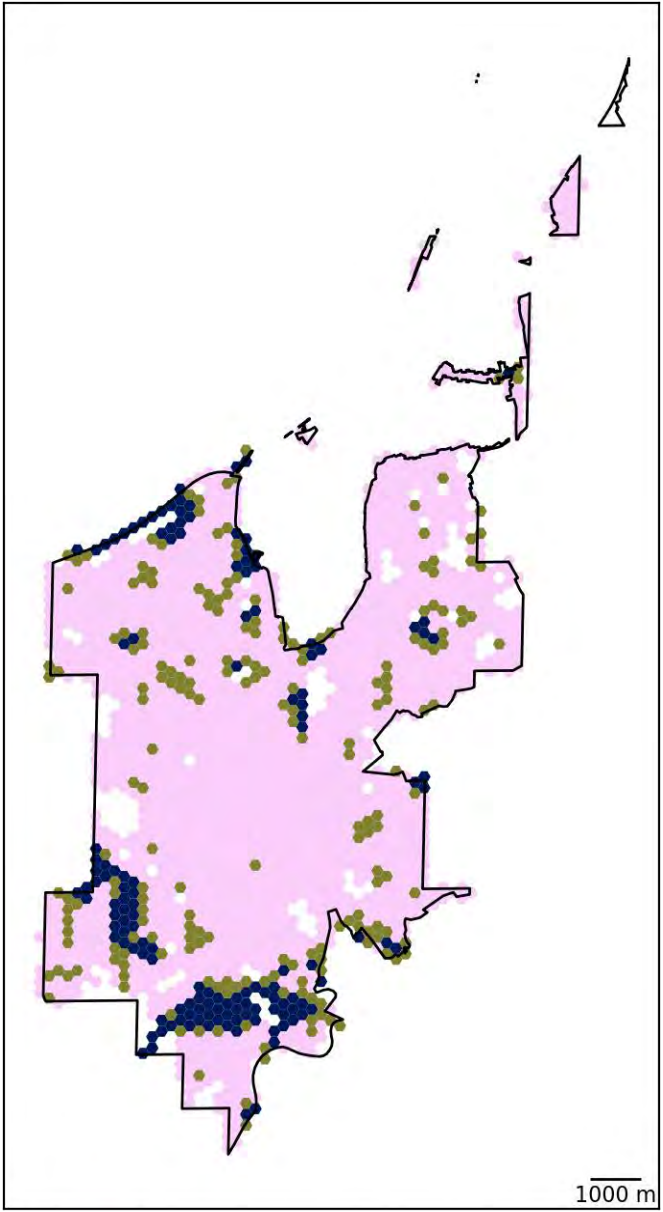

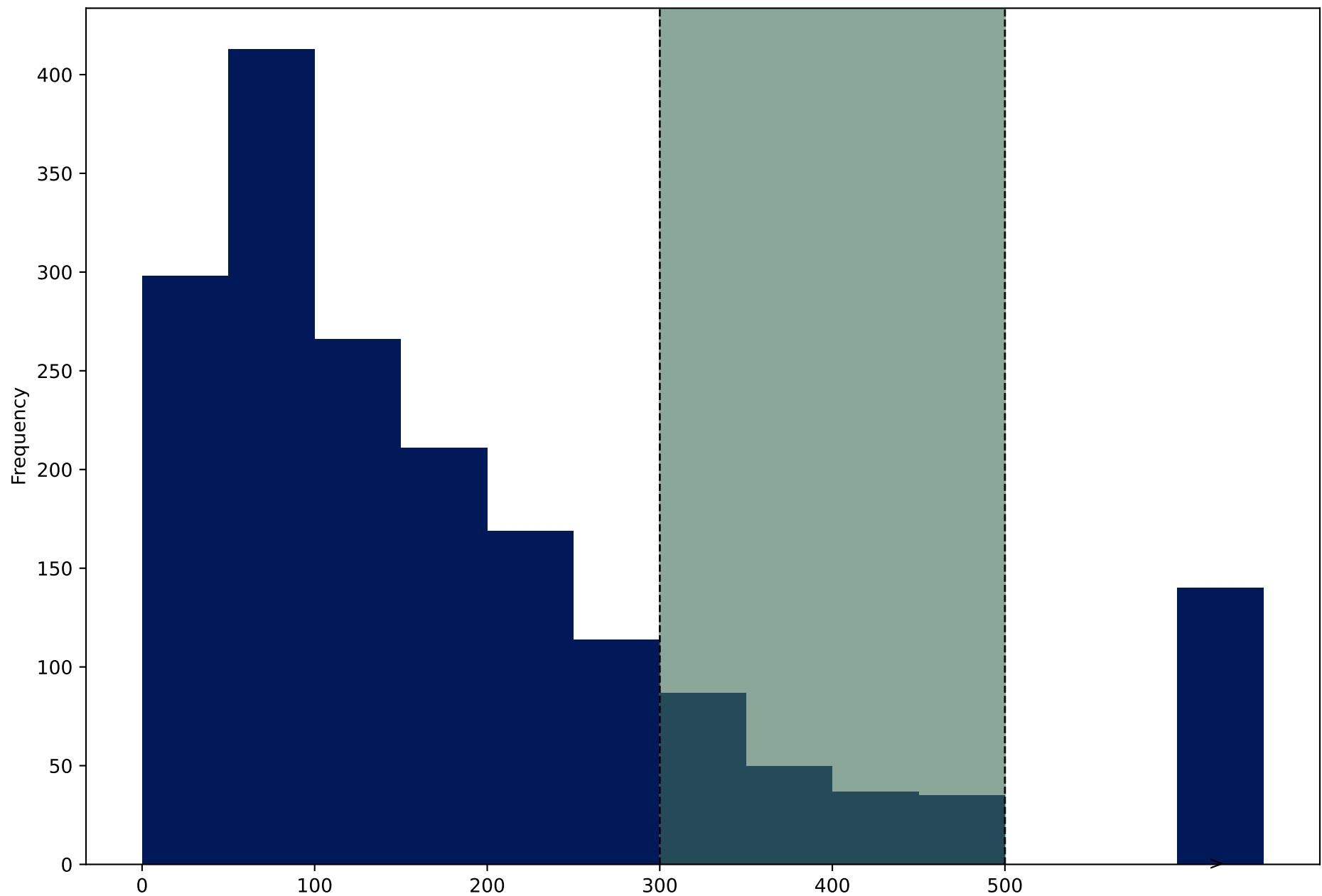

# Europe, Czech Republic, Olomouc

| Satellite imagery of urban study region (Bing) | Walkability, relative to city | Walkability, relative to 25 global cities |

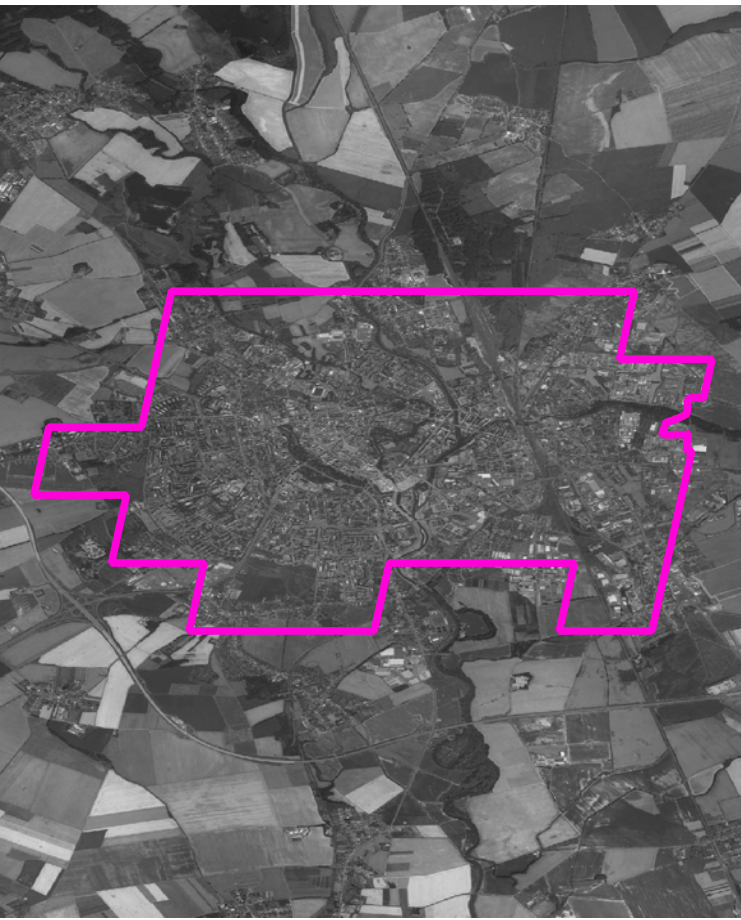
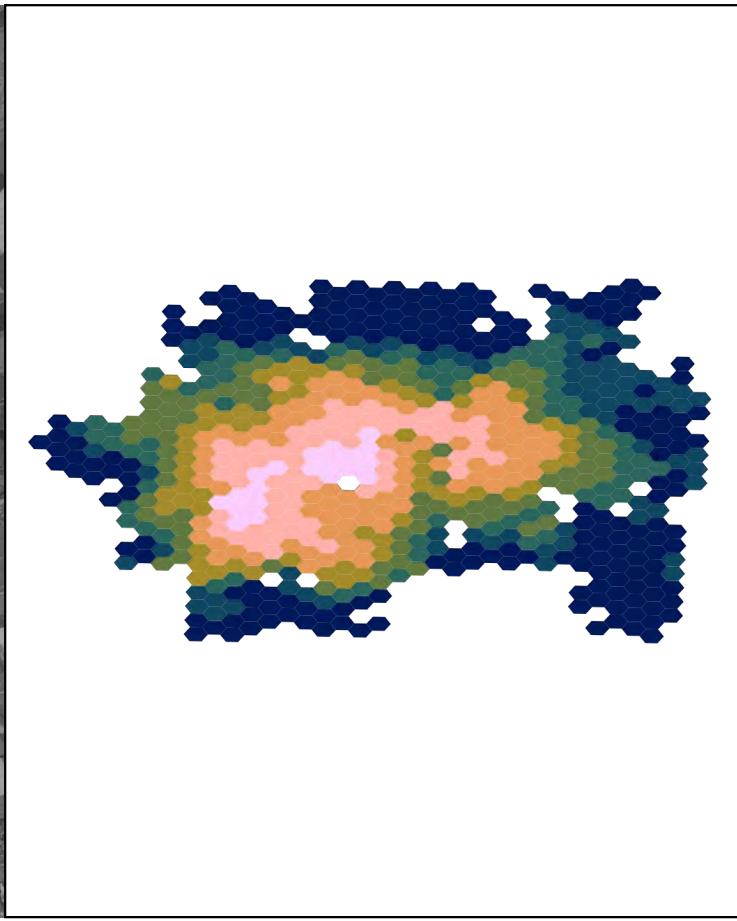
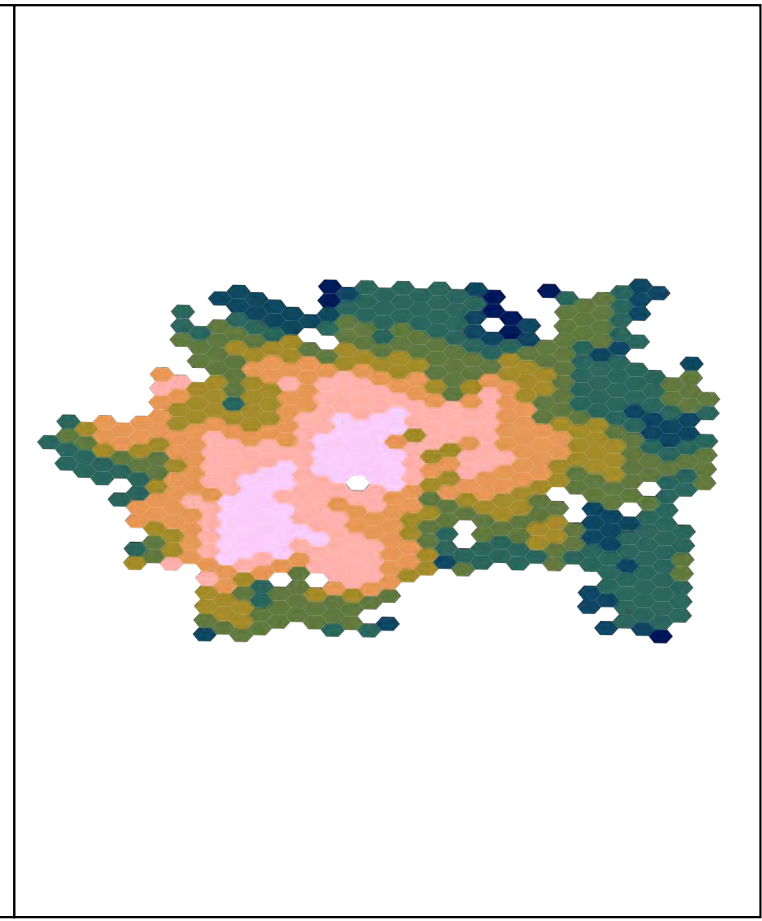

Urban boundary

0   3   6 km

Walkability score
- <-3
- -3 to -2
- -2 to -1
- -1 to 0
- 0 to 1
- 1 to 2
- 2 to 3
- ≥3

Walkability relative to all cities by component variables (2D histograms), and overall (histogram)

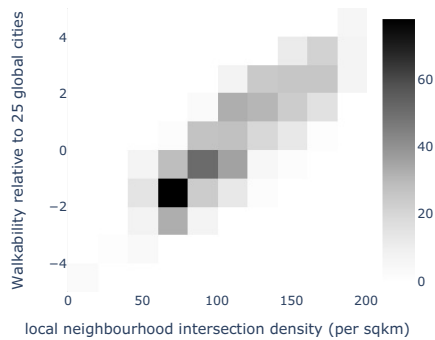
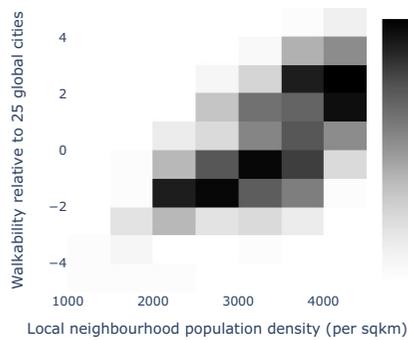
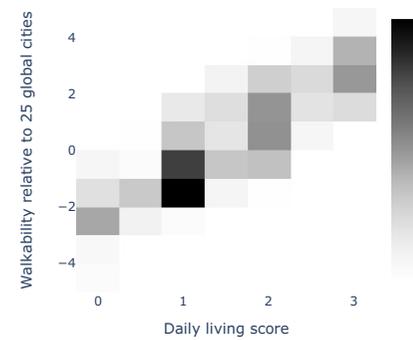
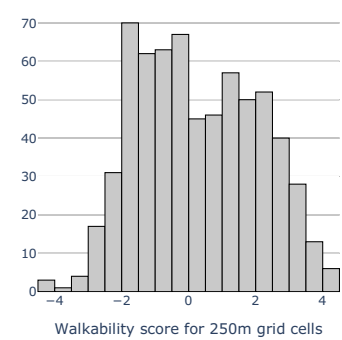

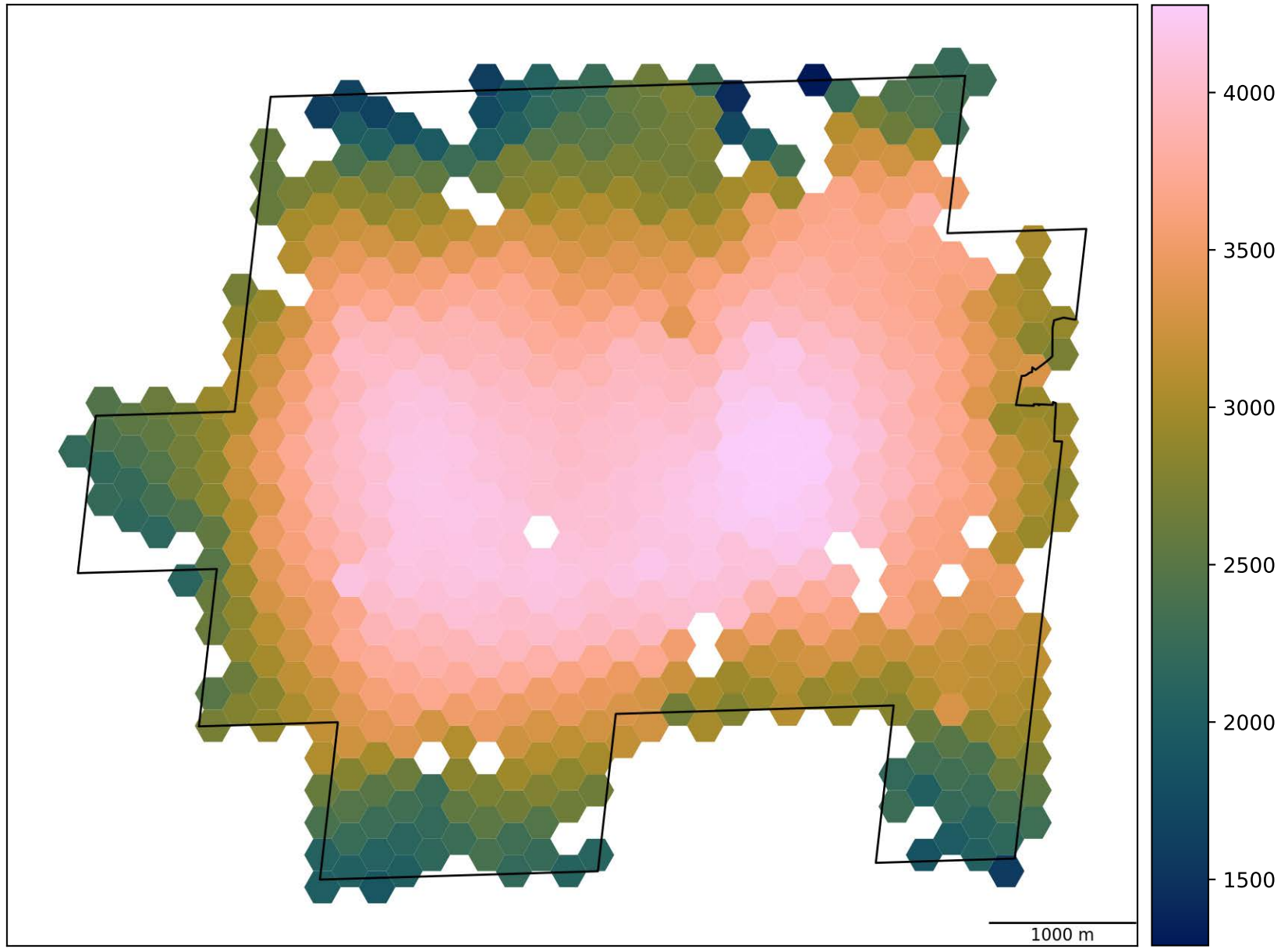

A: Estimated Mean 1000 m neighbourhood population per km² requirement for ≥80% probability of engaging in walking for transport

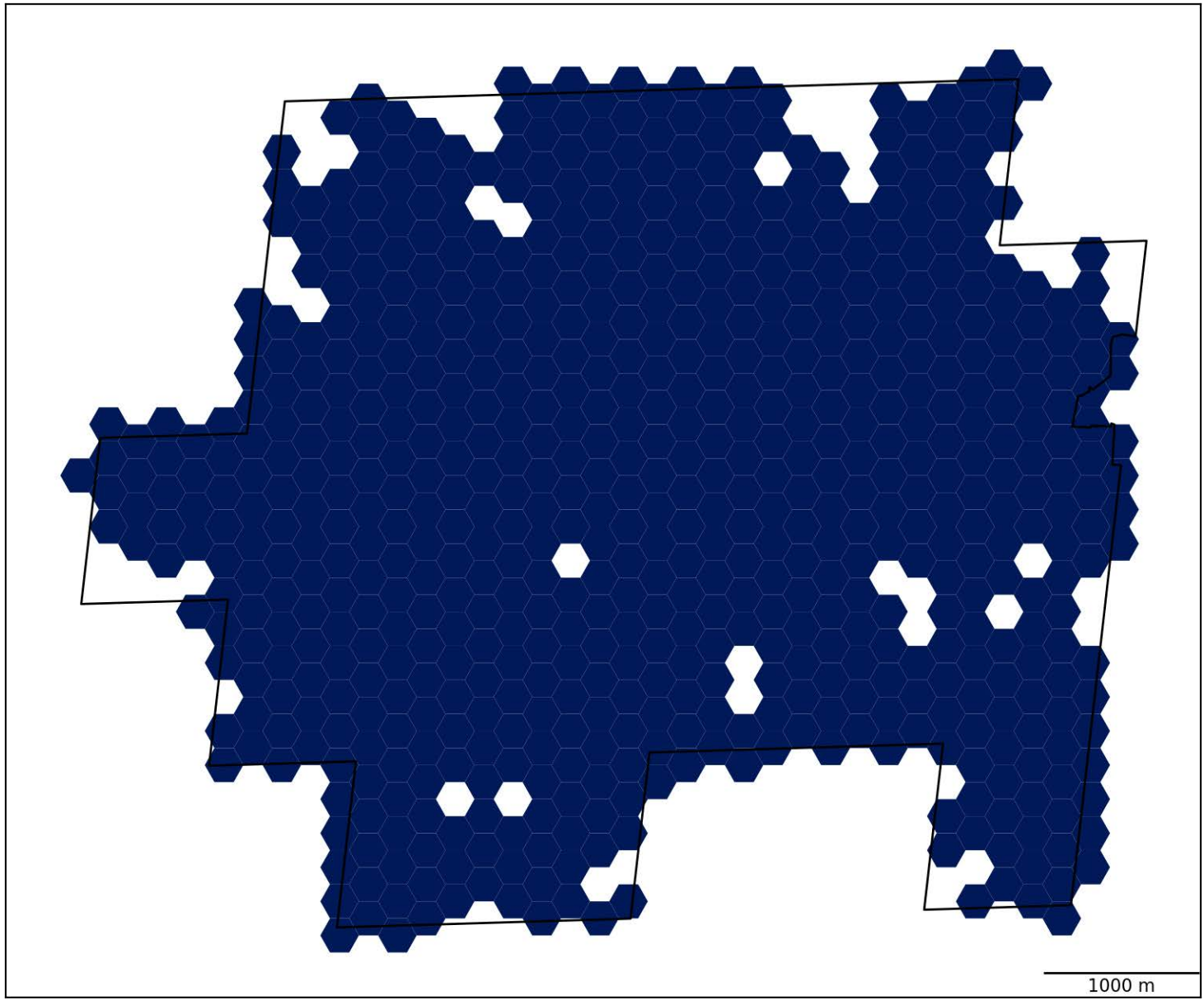

● below 95% CrI lower bound

B: Estimated Mean 1000 m neighbourhood population per km² requirement for reaching the WHO's target of a ≥15% relative reduction in insufficient physical activity through walking

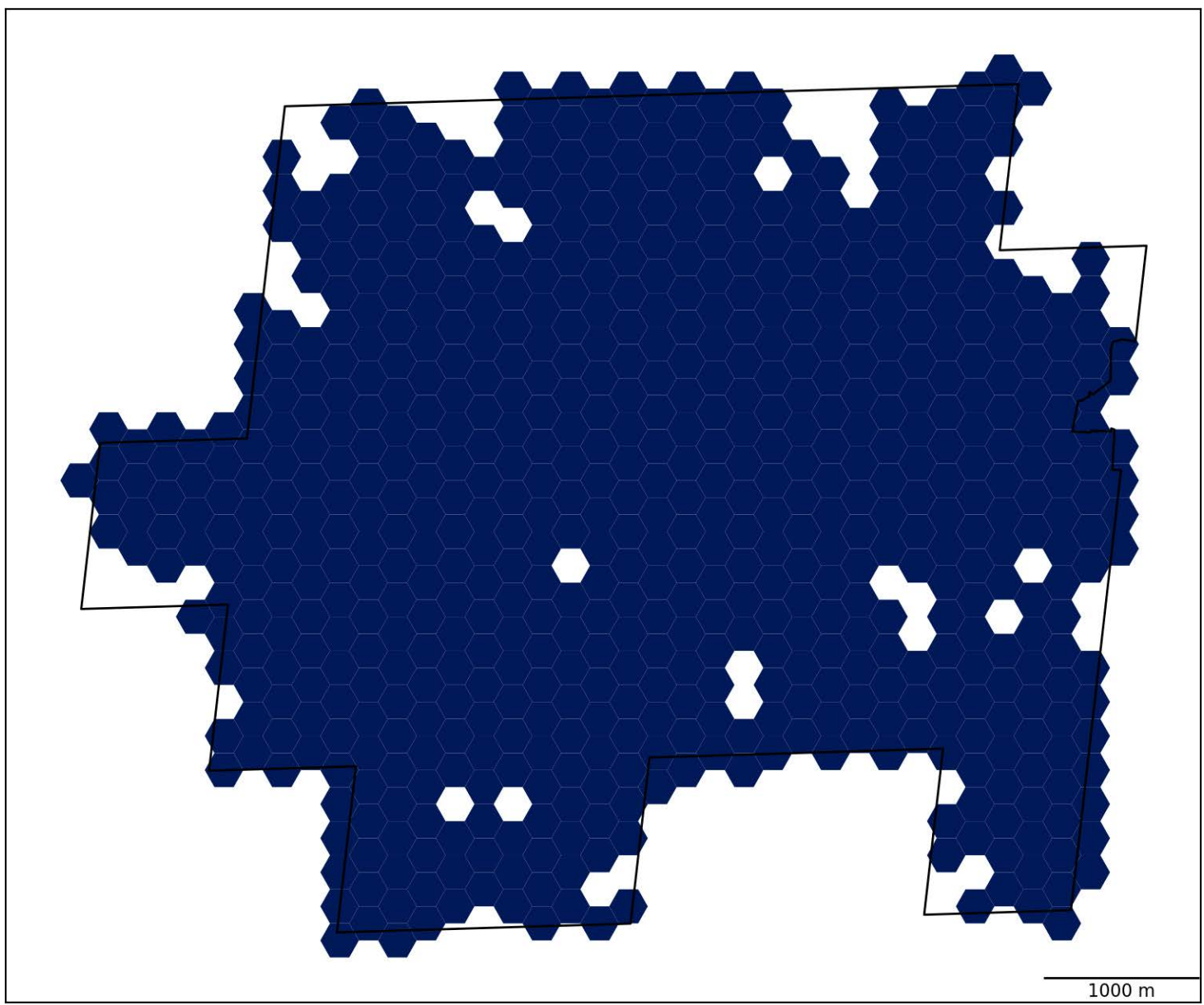

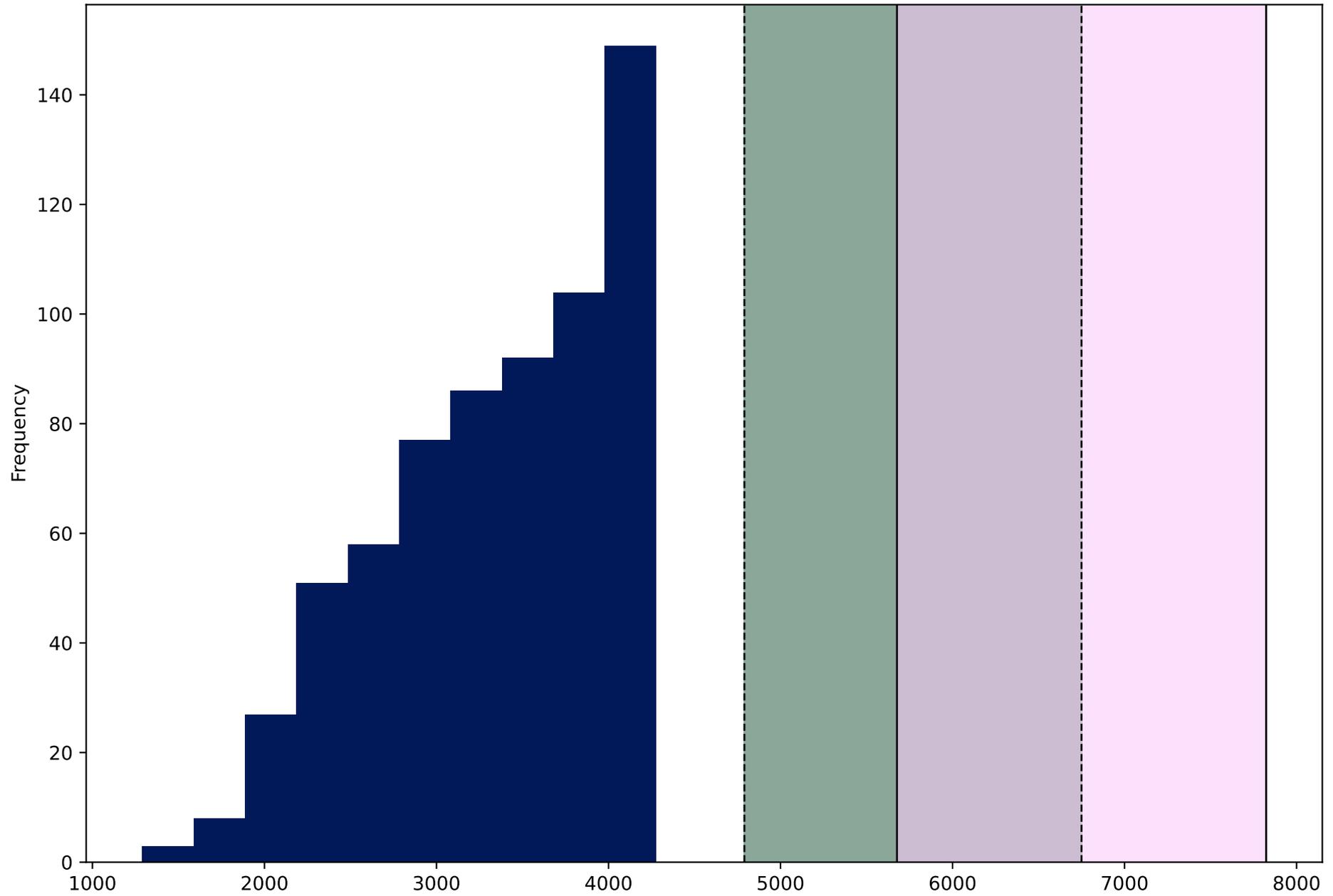

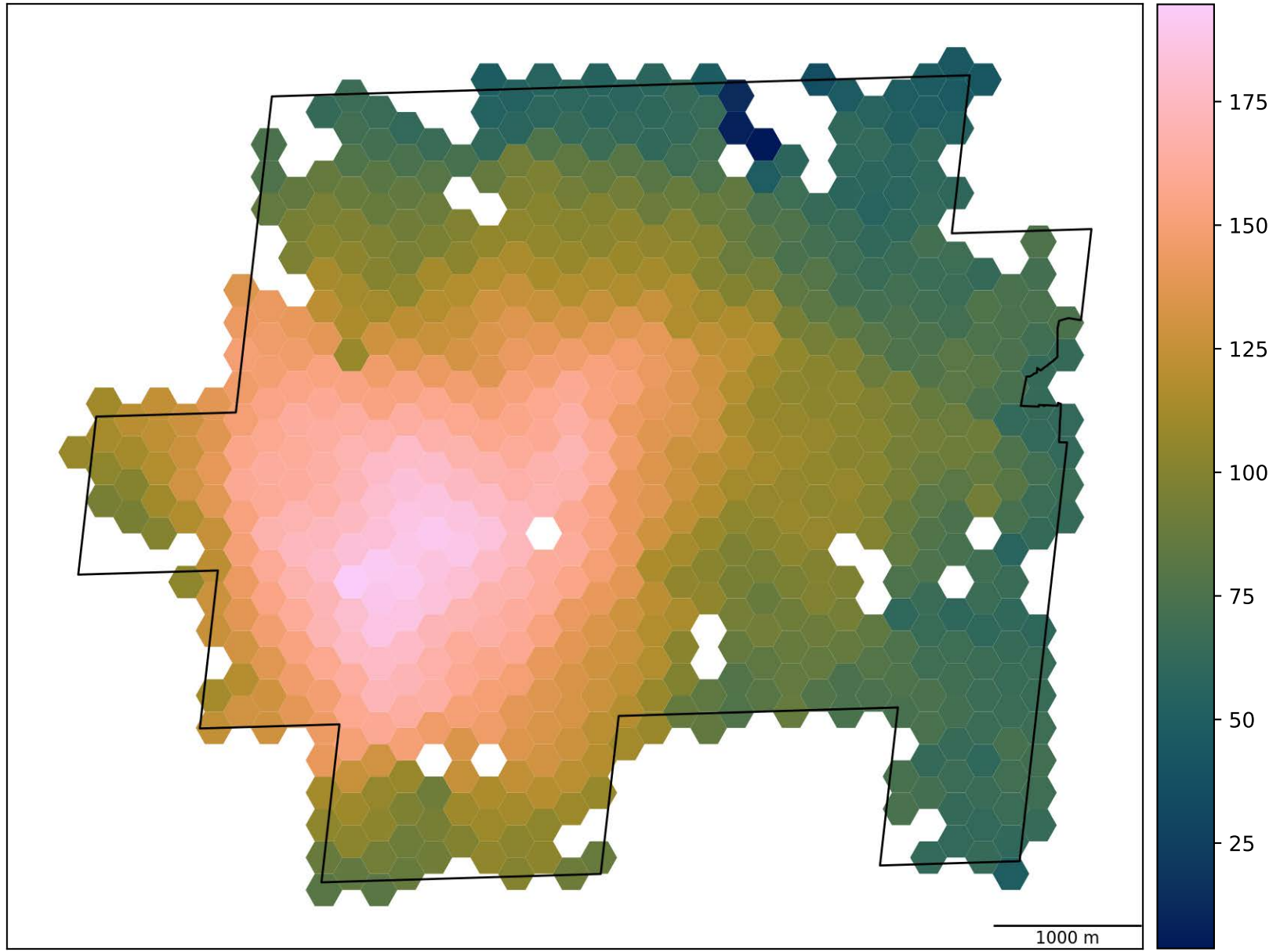

Mean 1000 m neighbourhood street intersections per km²

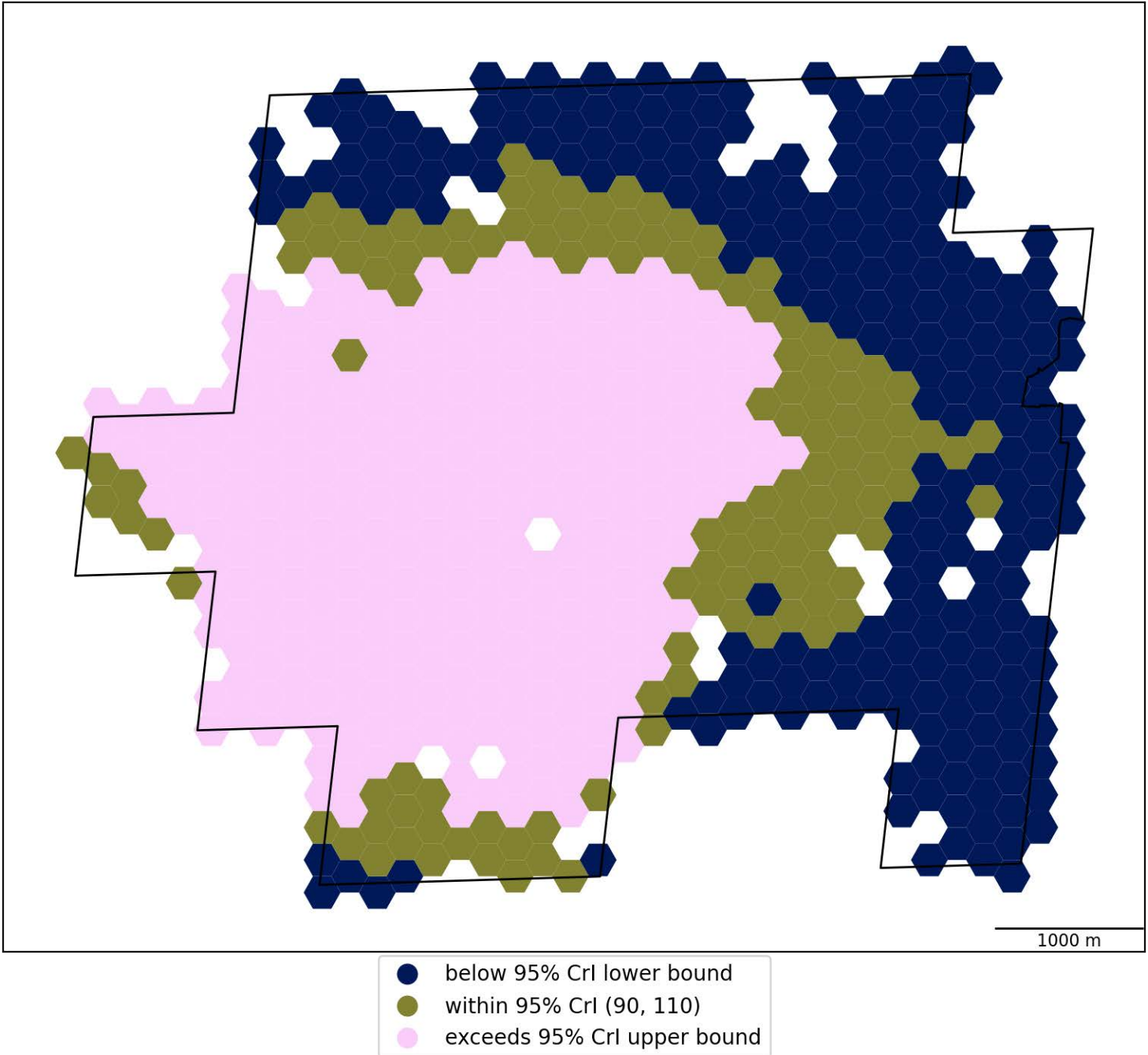

A: Estimated Mean 1000 m neighbourhood street intersections per km² requirement for ≥80% probability of engaging in walking for transport

- below 95% CrI lower bound
- within 95% CrI (90, 110)
- exceeds 95% CrI upper bound

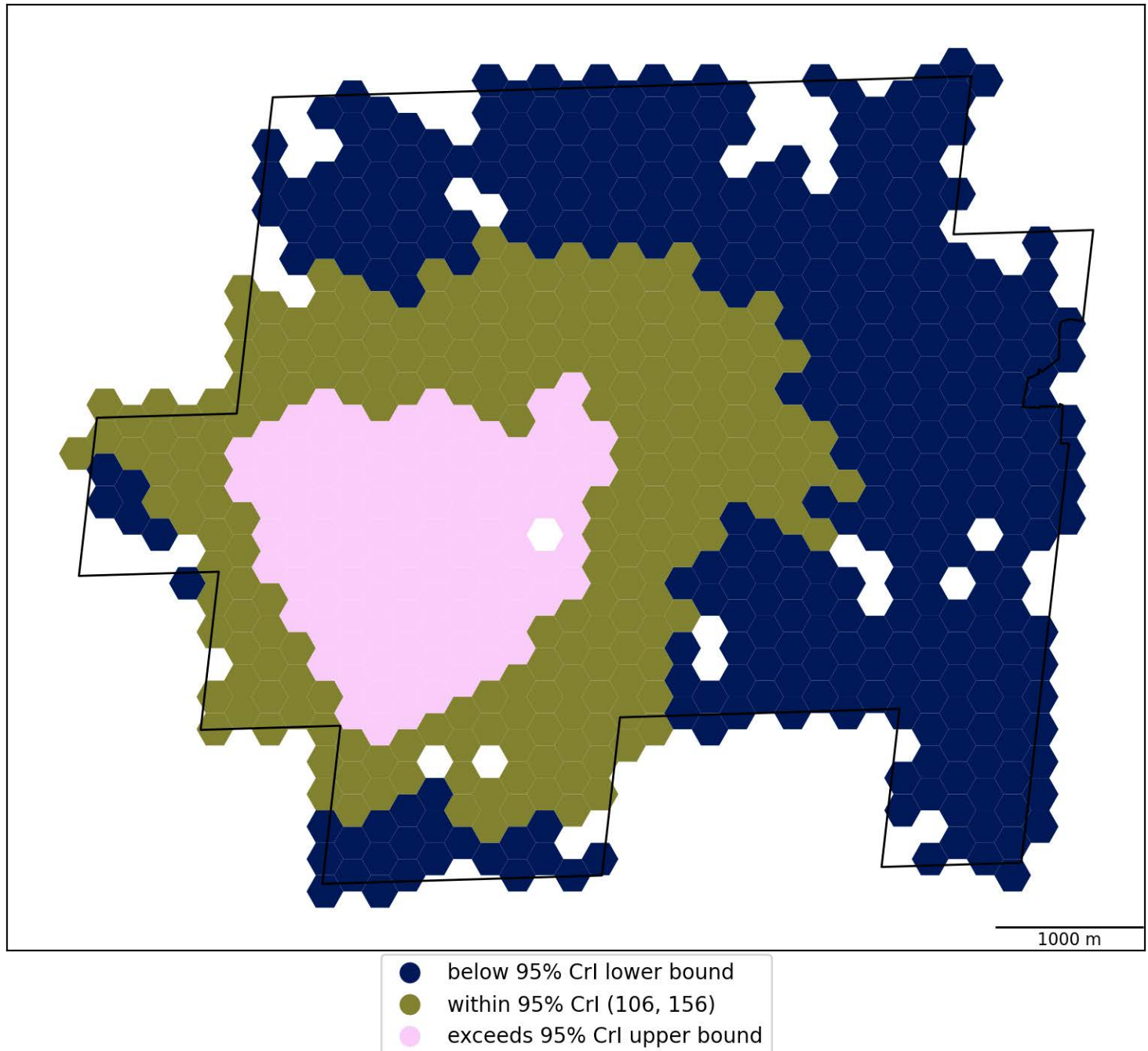

B: Estimated Mean 1000 m neighbourhood street intersections per km² requirement for reaching the WHO's target of a ≥15% relative reduction in insufficient physical activity through walking

- below 95% CrI lower bound
- within 95% CrI (106, 156)
- exceeds 95% CrI upper bound

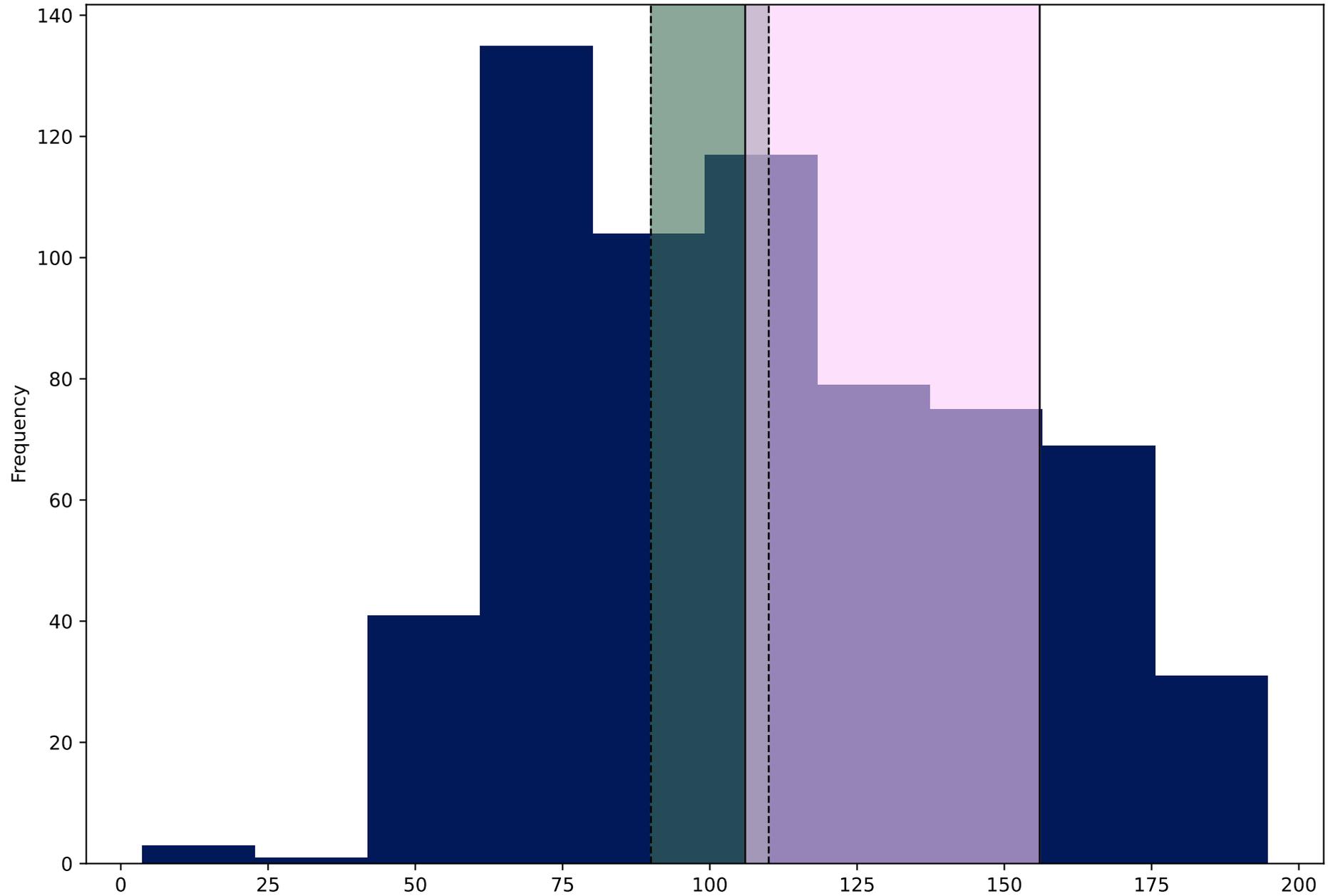

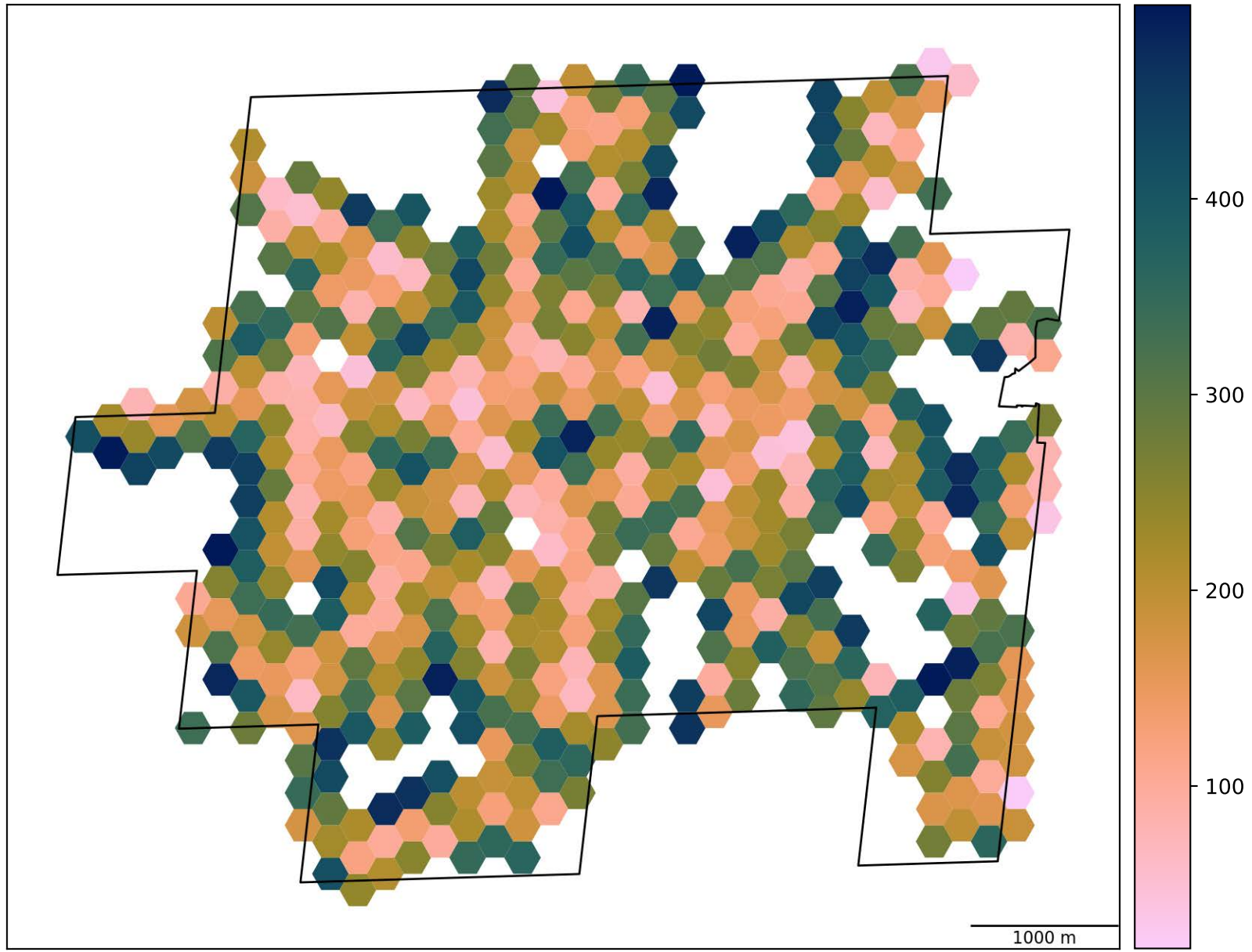

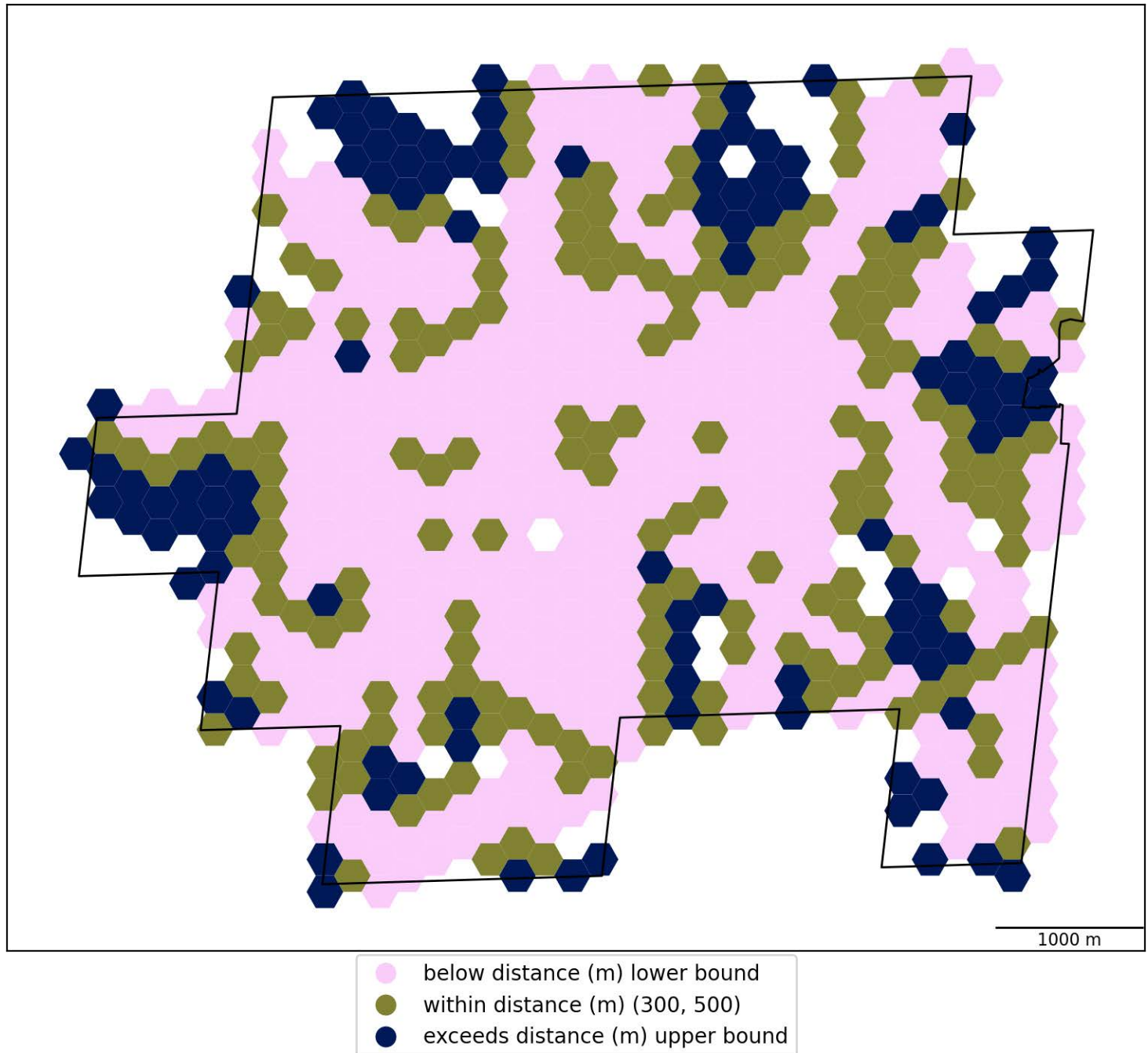

distances: Estimated Distance to nearest public transport stops (m; up to 500m) requirement for distances to destinations, measured up to a maximum distance target threshold of 500 metres

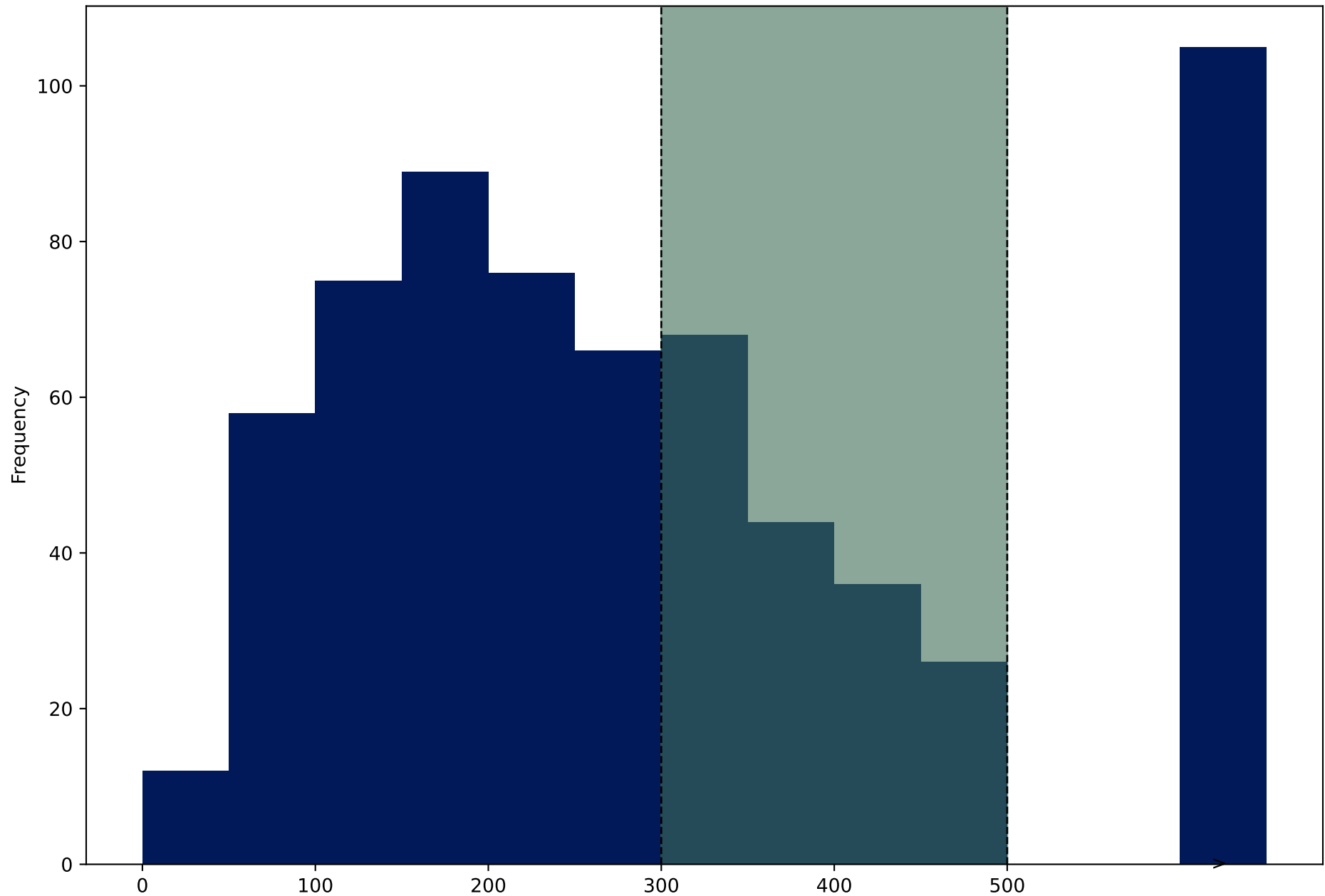

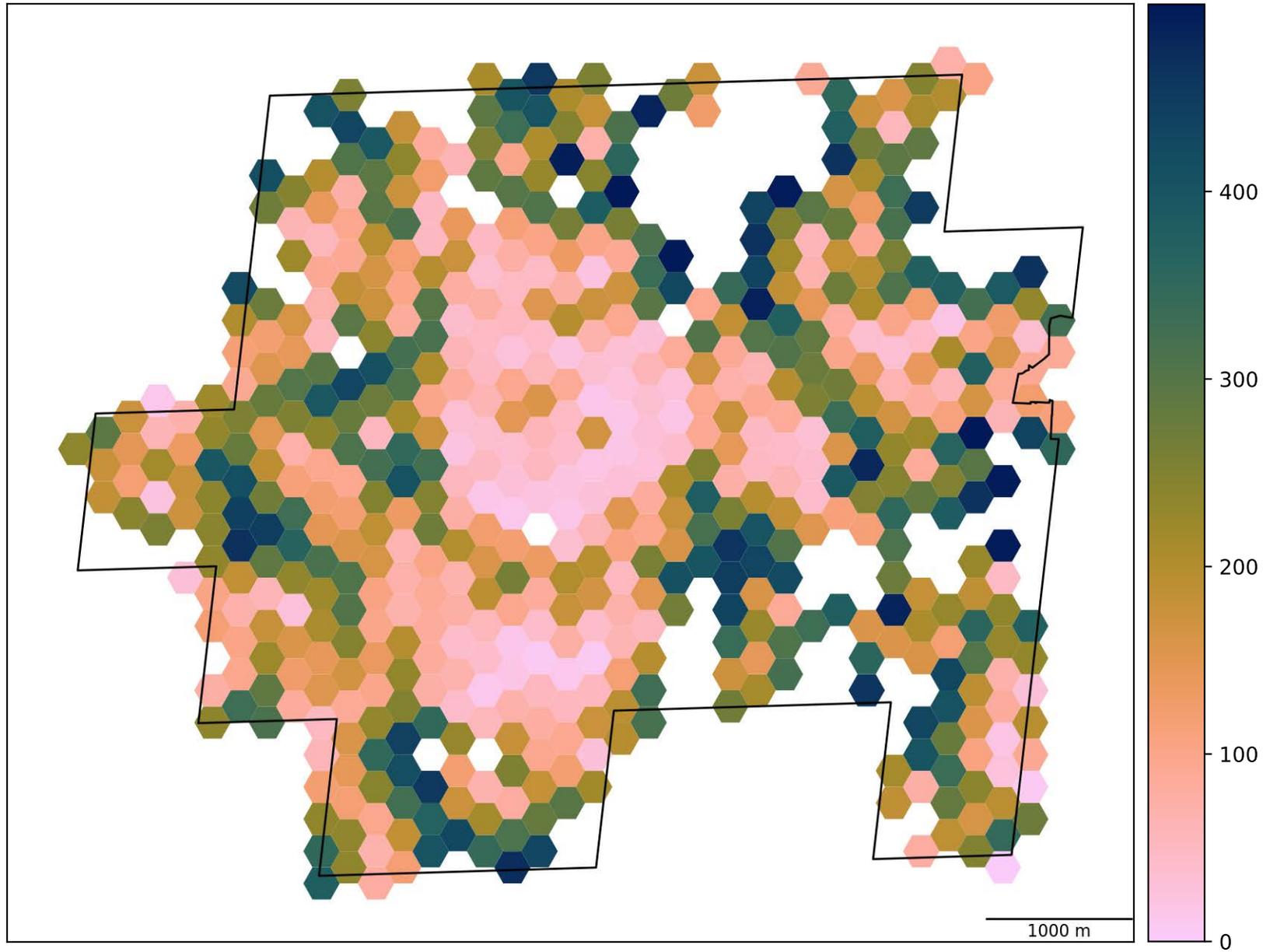

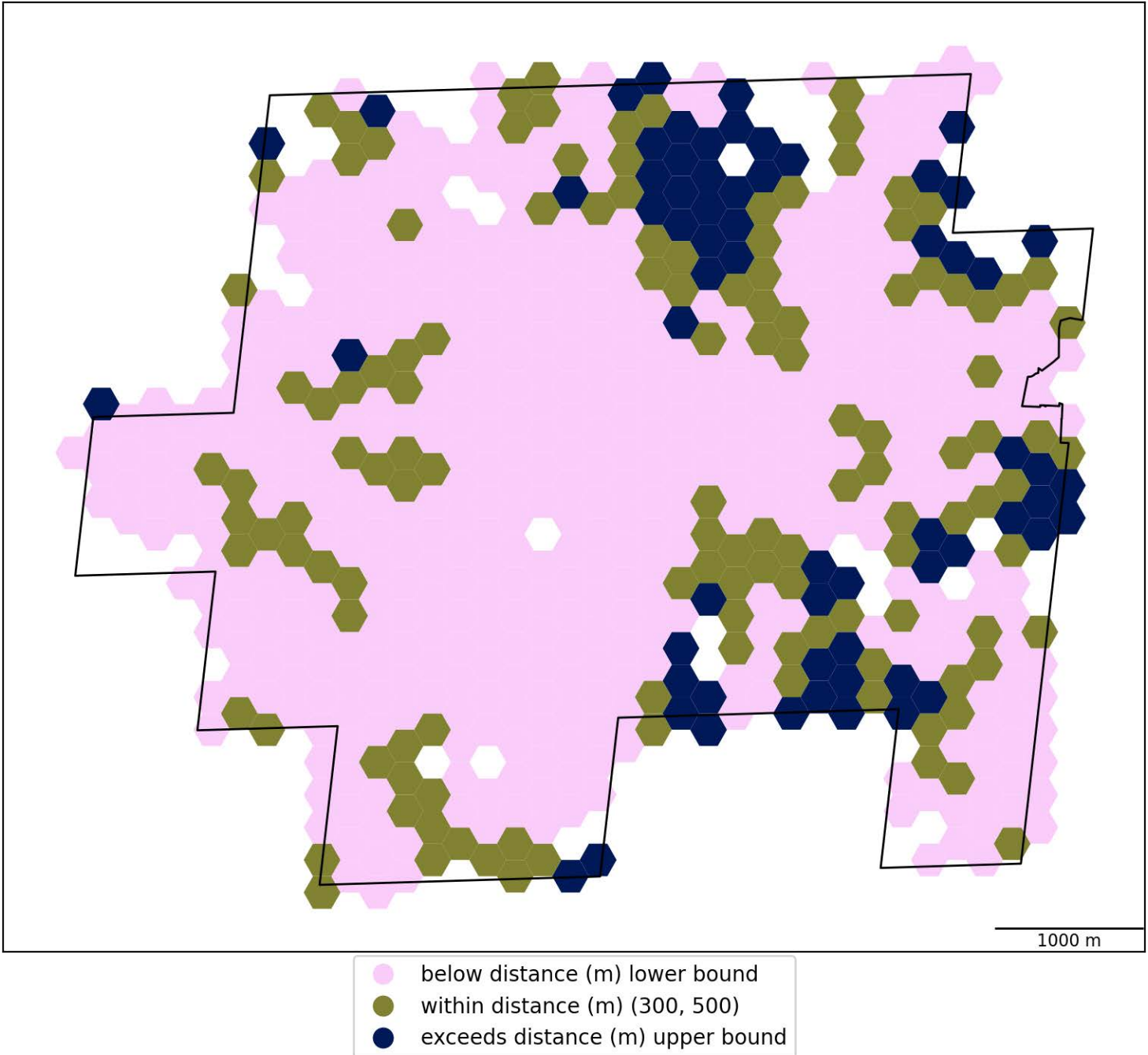

distances: Estimated Distance to nearest park (m; up to 500m) requirement for distances to destinations, measured up to a maximum distance target threshold of 500 metres

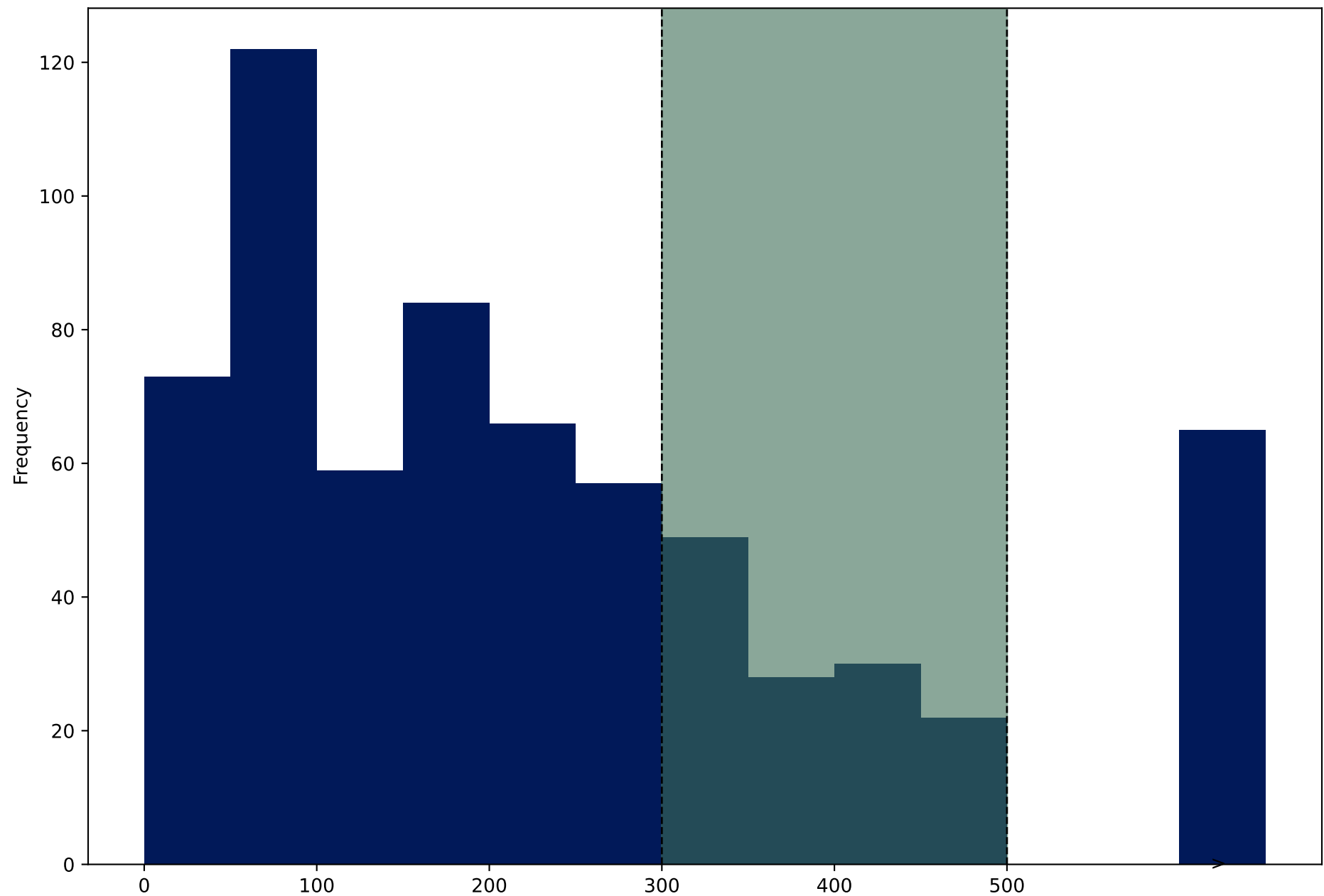

# Europe, Denmark, Odense

| Satellite imagery of urban study region (Bing) | Walkability, relative to city | Walkability, relative to 25 global cities |

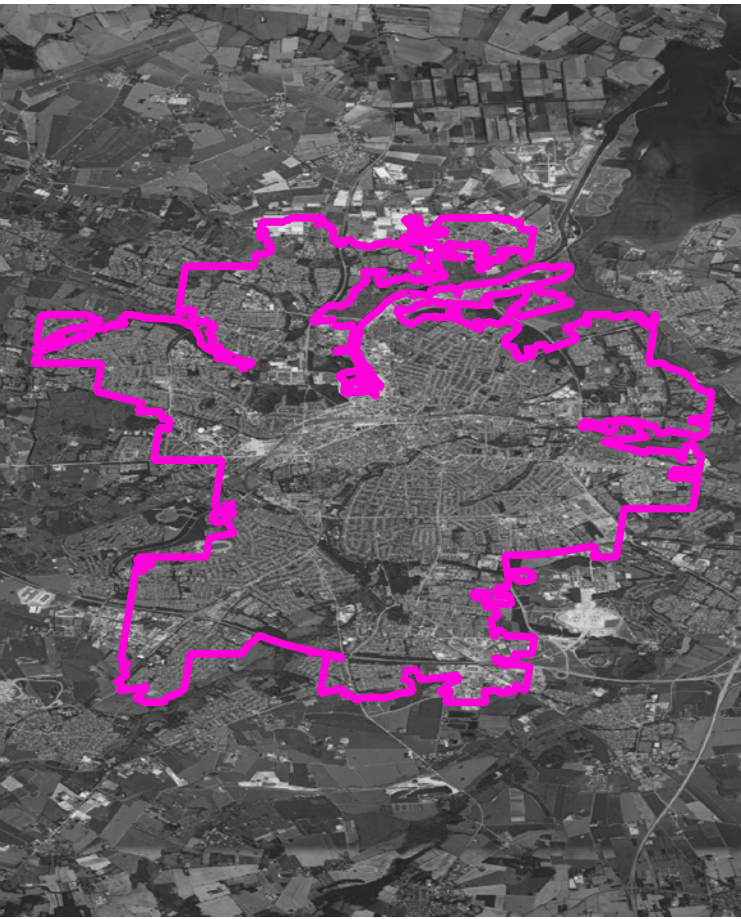

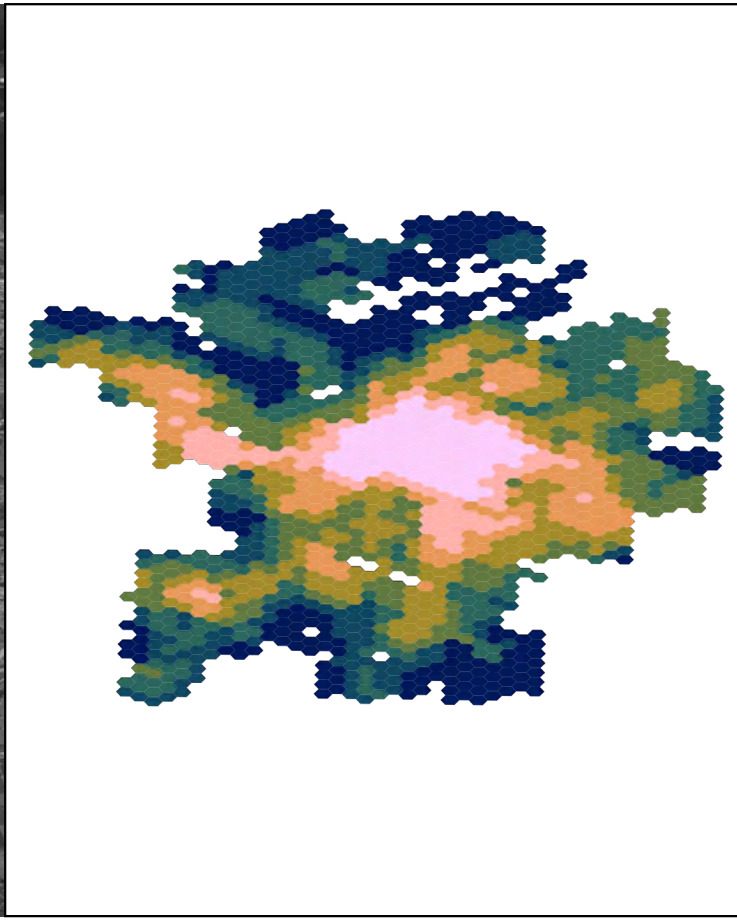
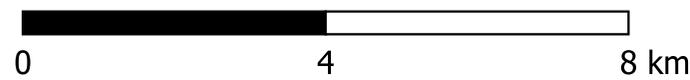
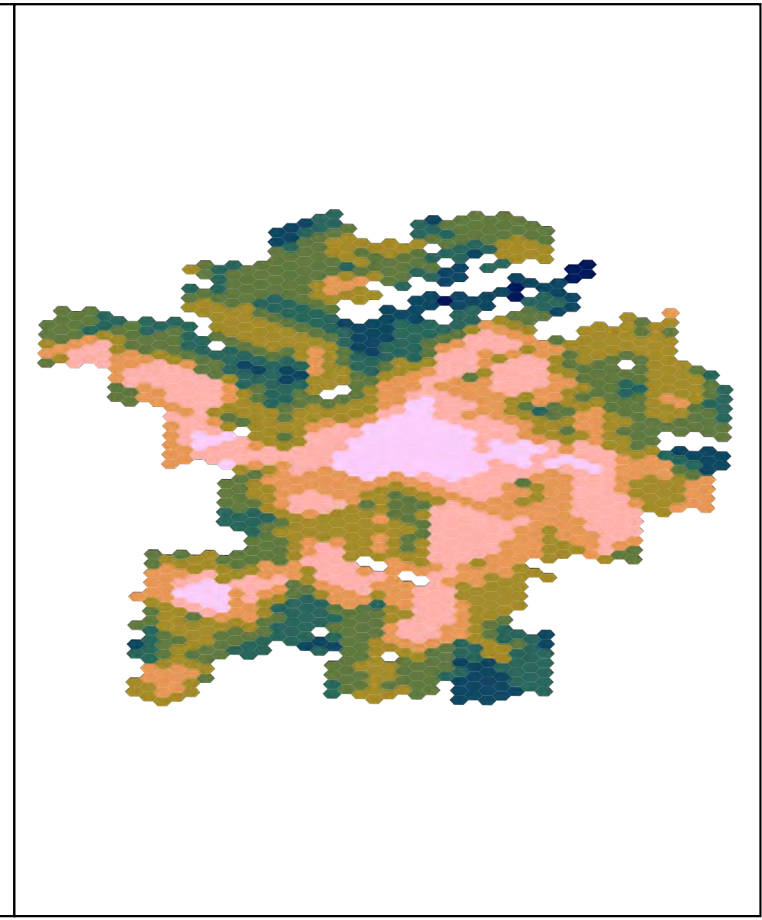
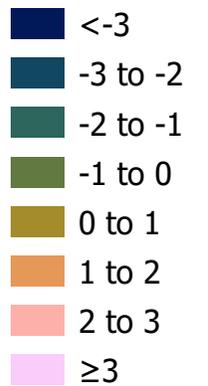

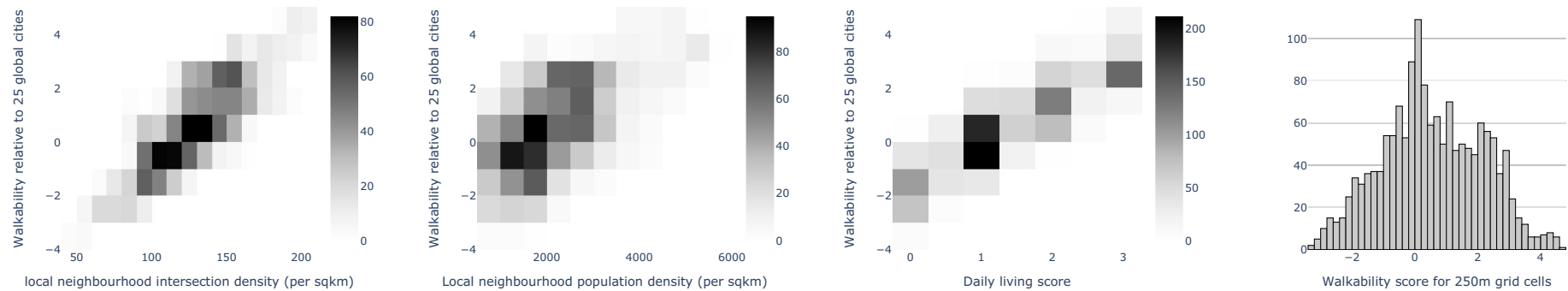

Walkability relative to all cities by component variables (2D histograms), and overall (histogram)

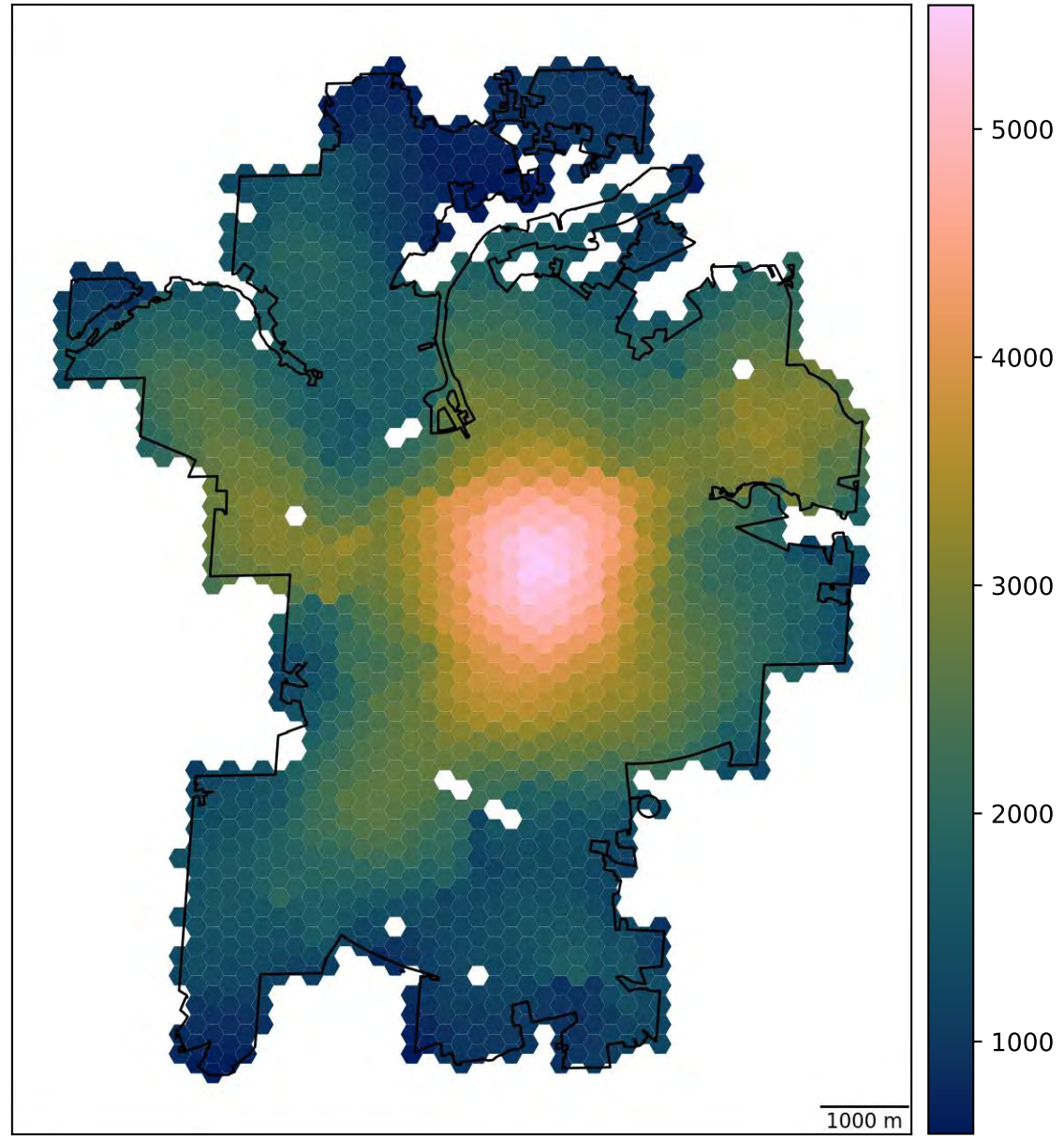

Mean 1000 m neighbourhood population per km²

A: Estimated Mean 1000 m neighbourhood population per km² requirement for ≥80% probability of engaging in walking for transport

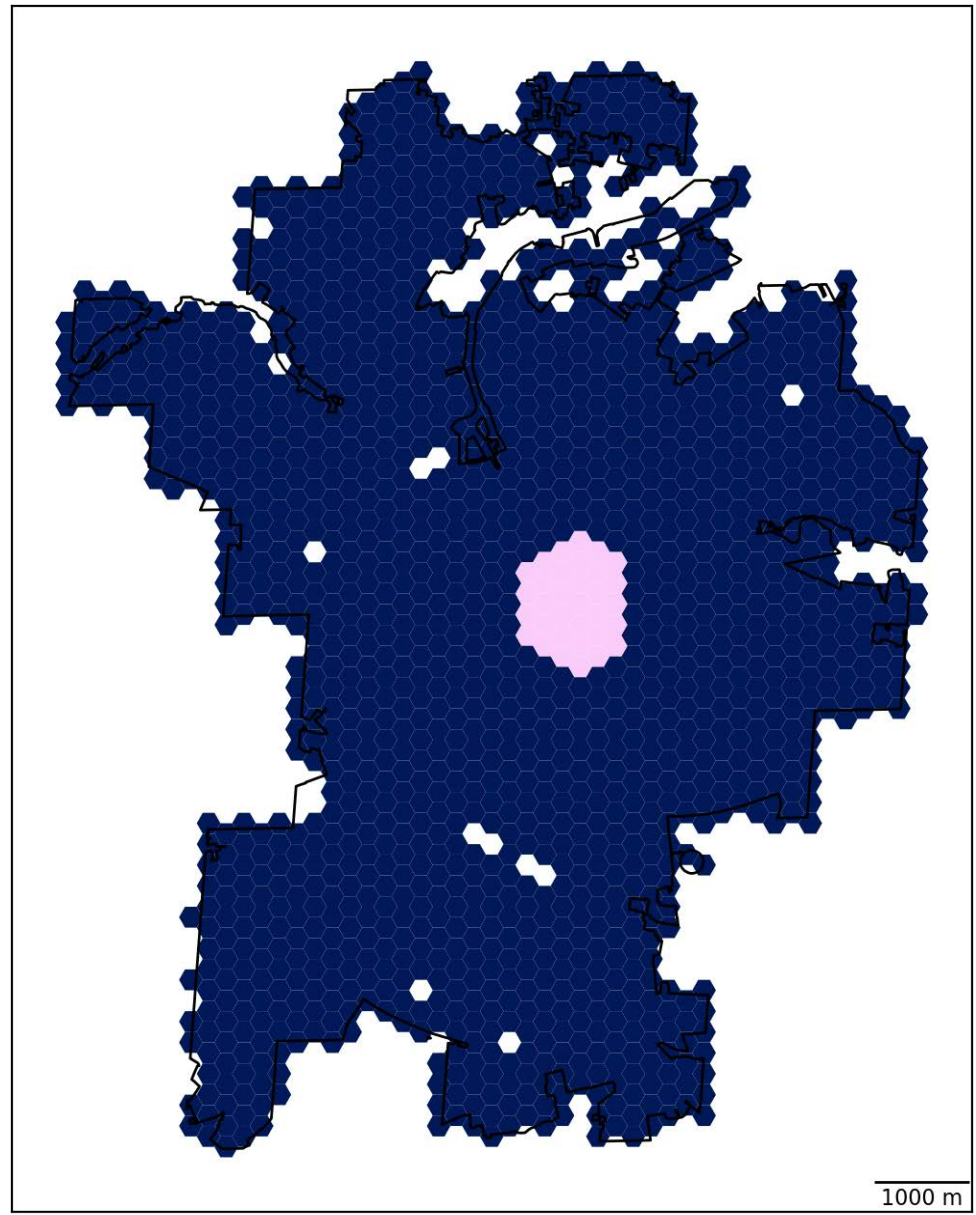

- below 95% CrI lower bound
- within 95% CrI (4790, 6750)

B: Estimated Mean 1000 m neighbourhood population per km² requirement for reaching the WHO's target of a ≥15% relative reduction in insufficient physical activity through walking

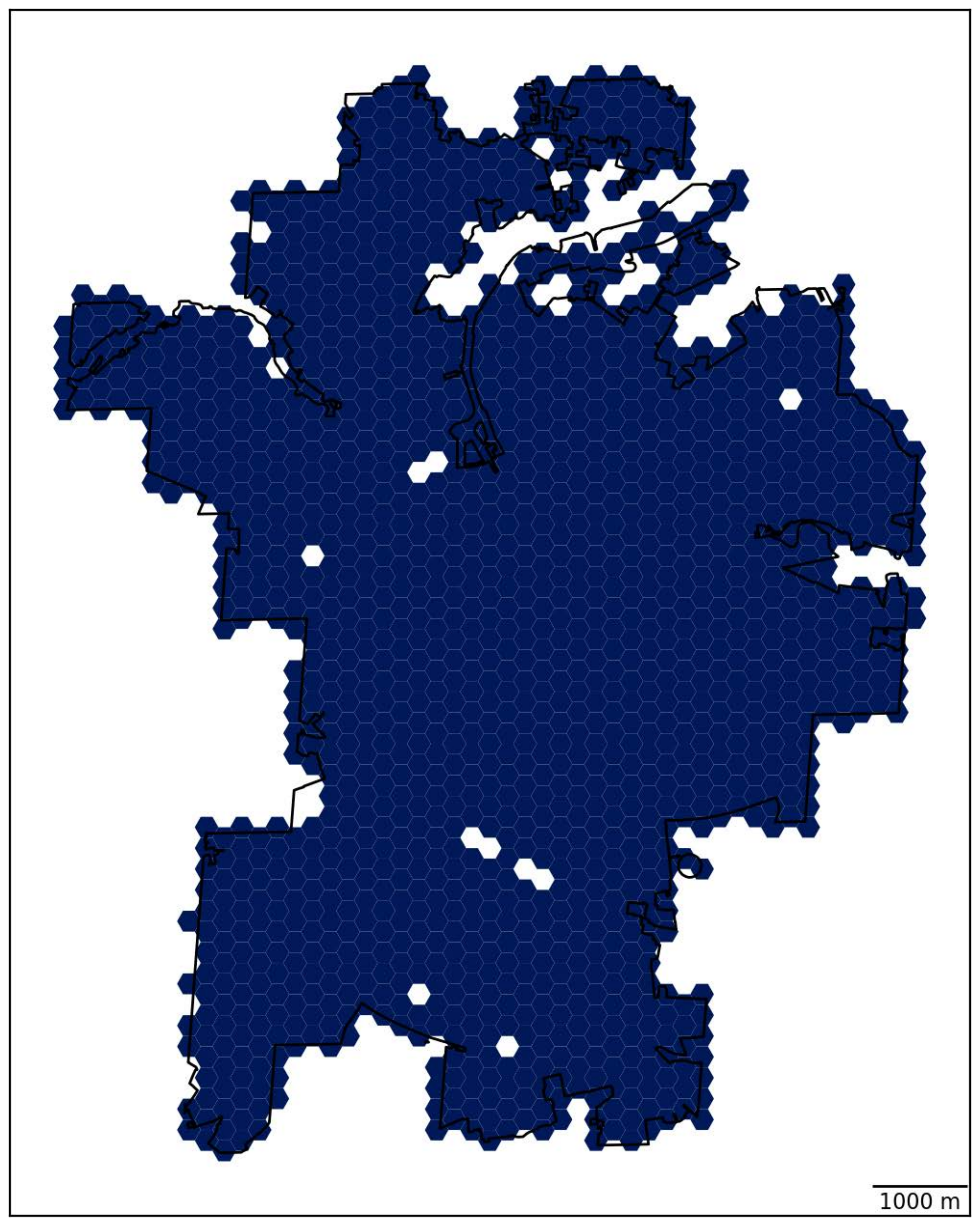

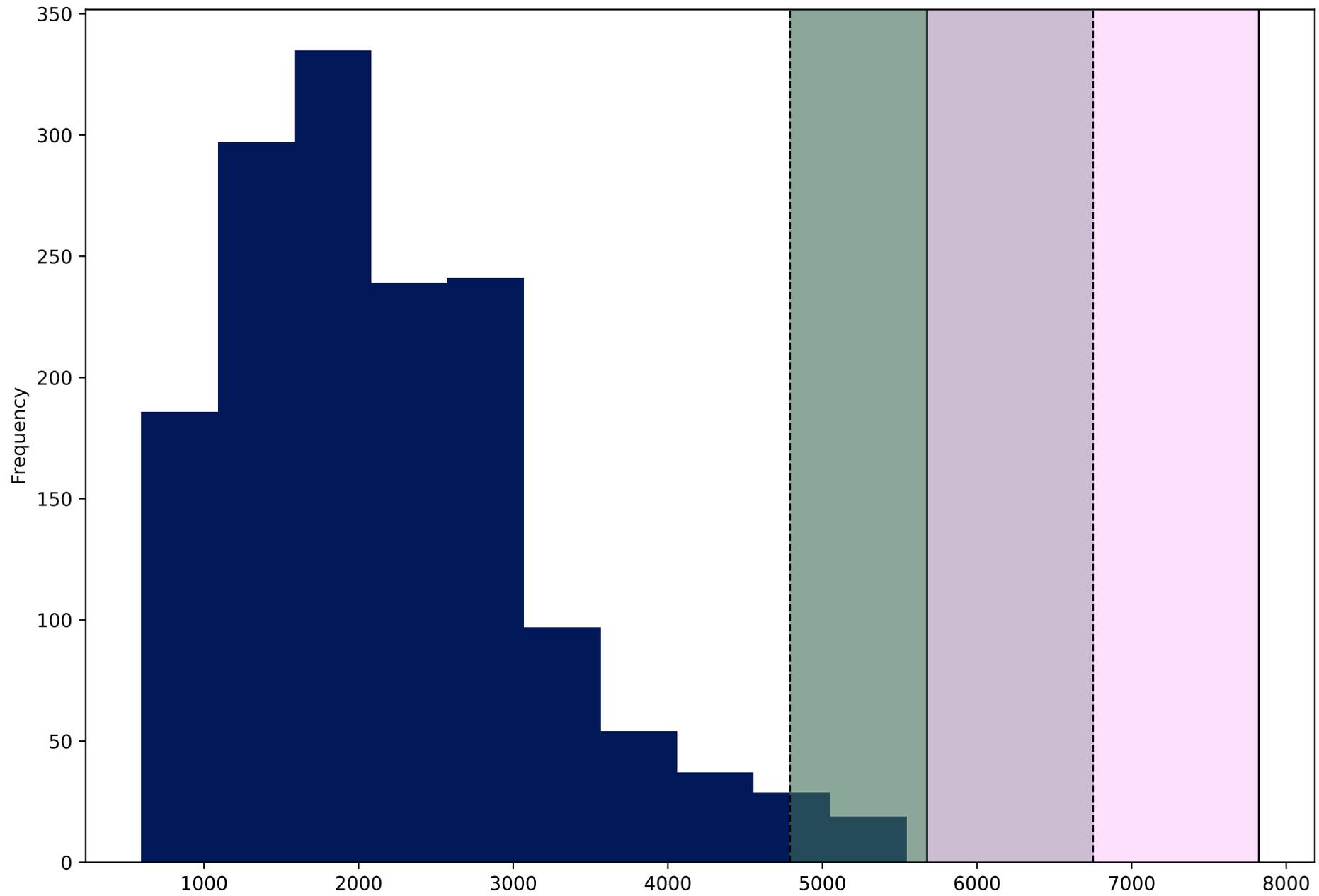

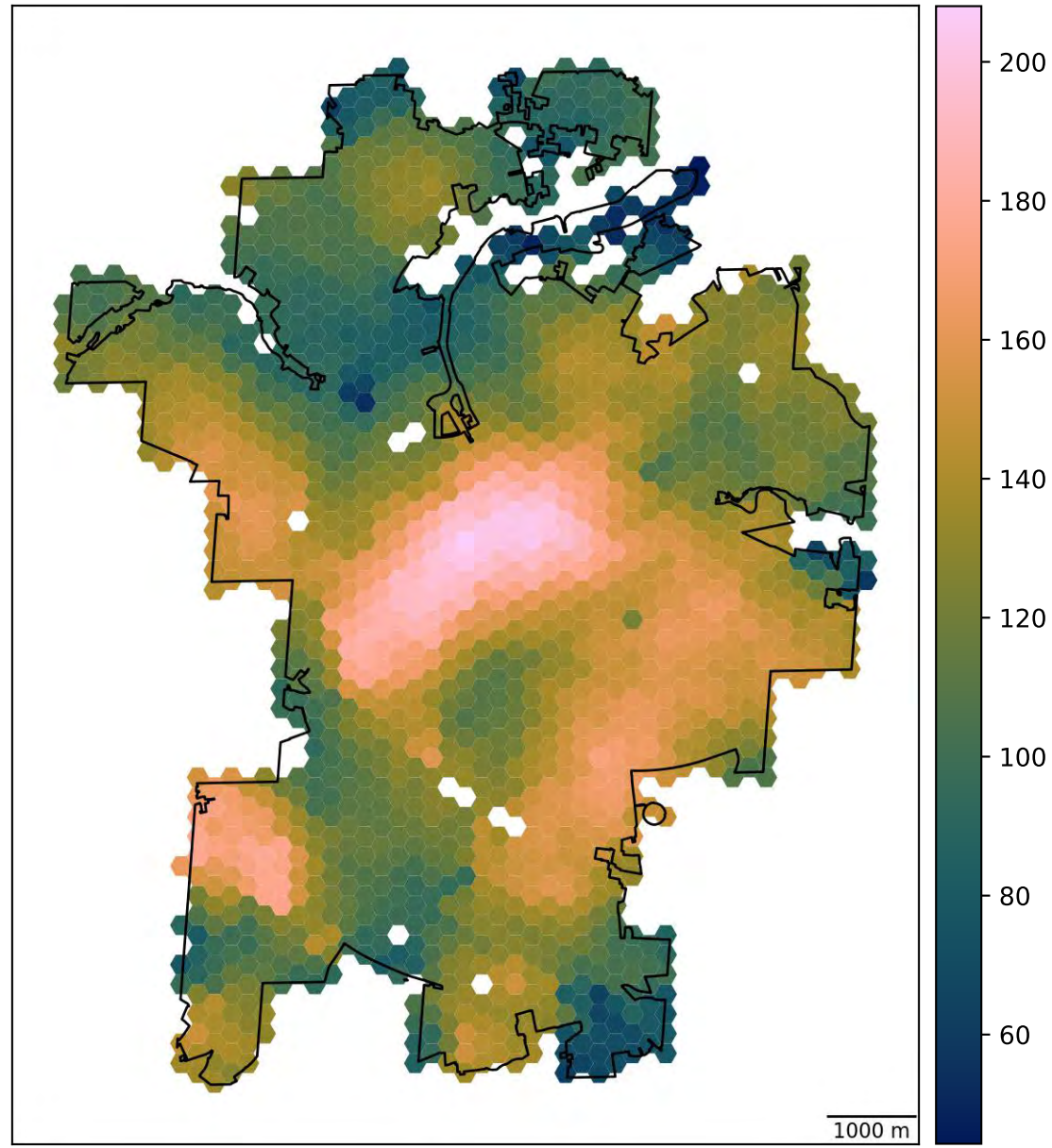

Mean 1000 m neighbourhood street intersections per km²

A: Estimated Mean 1000 m neighbourhood street intersections per km² requirement for ≥80% probability of engaging in walking for transport

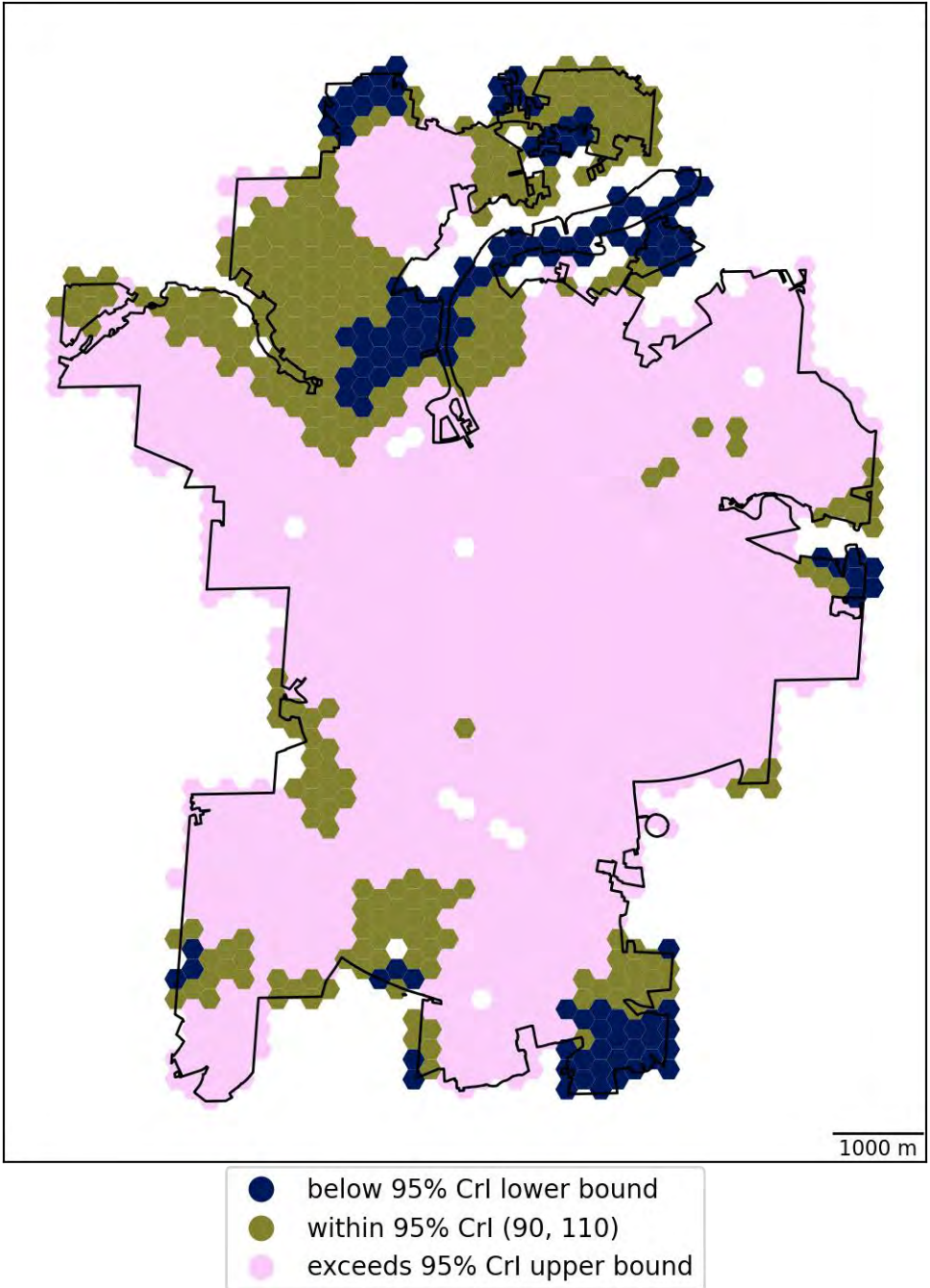

below 95% CrI lower bound
within 95% CrI (90, 110)
exceeds 95% CrI upper bound

B: Estimated Mean 1000 m neighbourhood street intersections per km² requirement for reaching the WHO's target of a ≥15% relative reduction in insufficient physical activity through walking

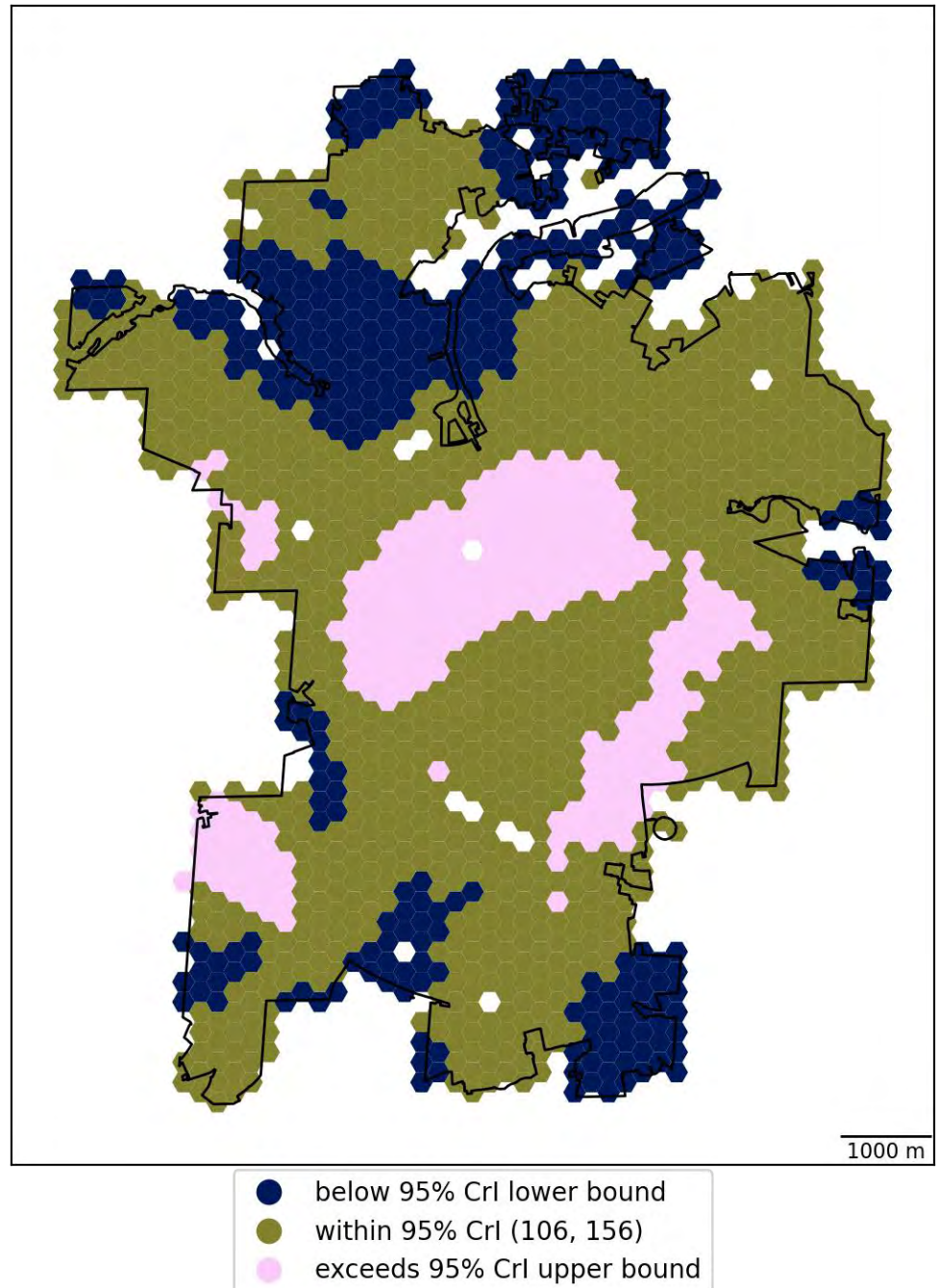

- below 95% CrI lower bound
- within 95% CrI (106, 156)
- exceeds 95% CrI upper bound

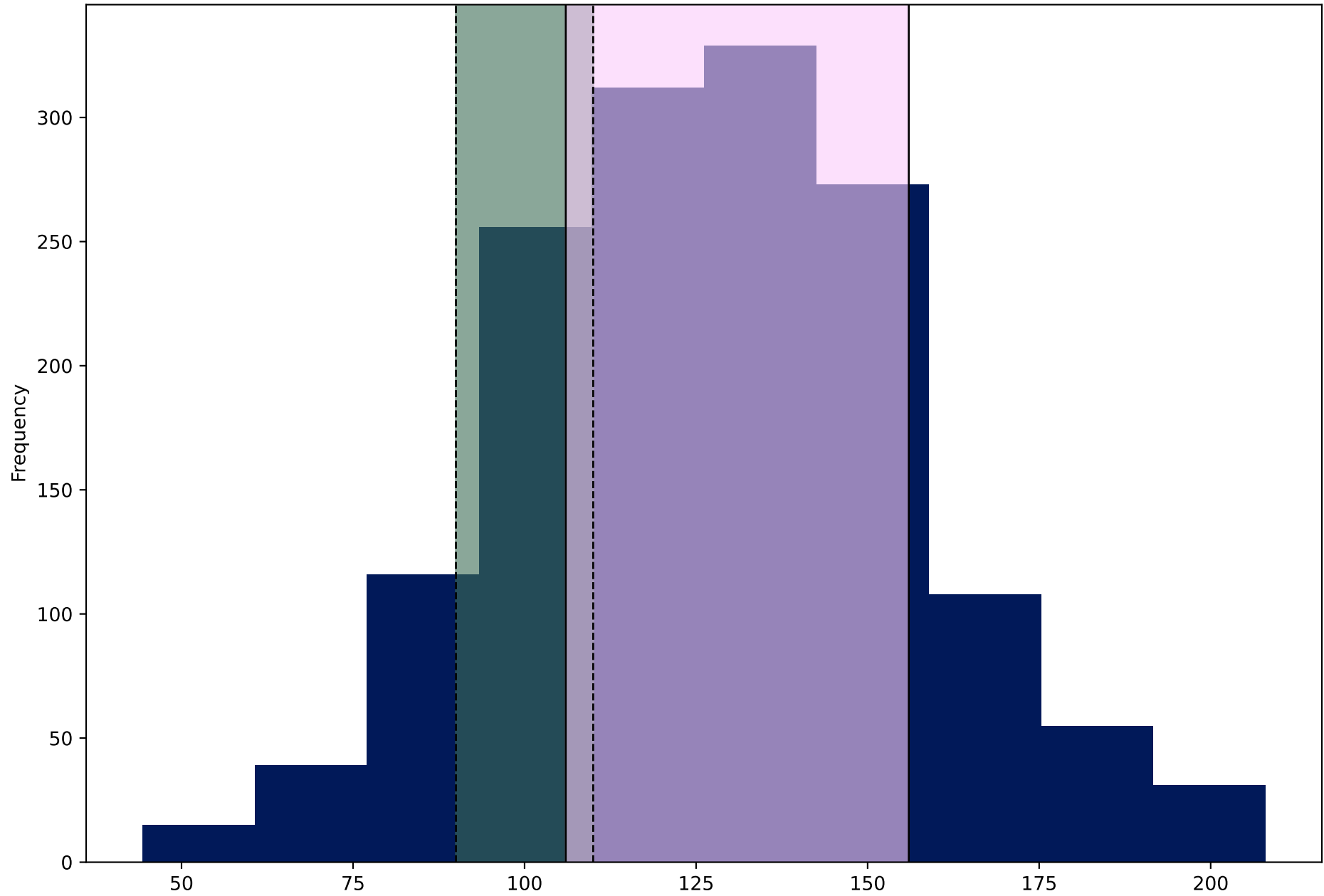

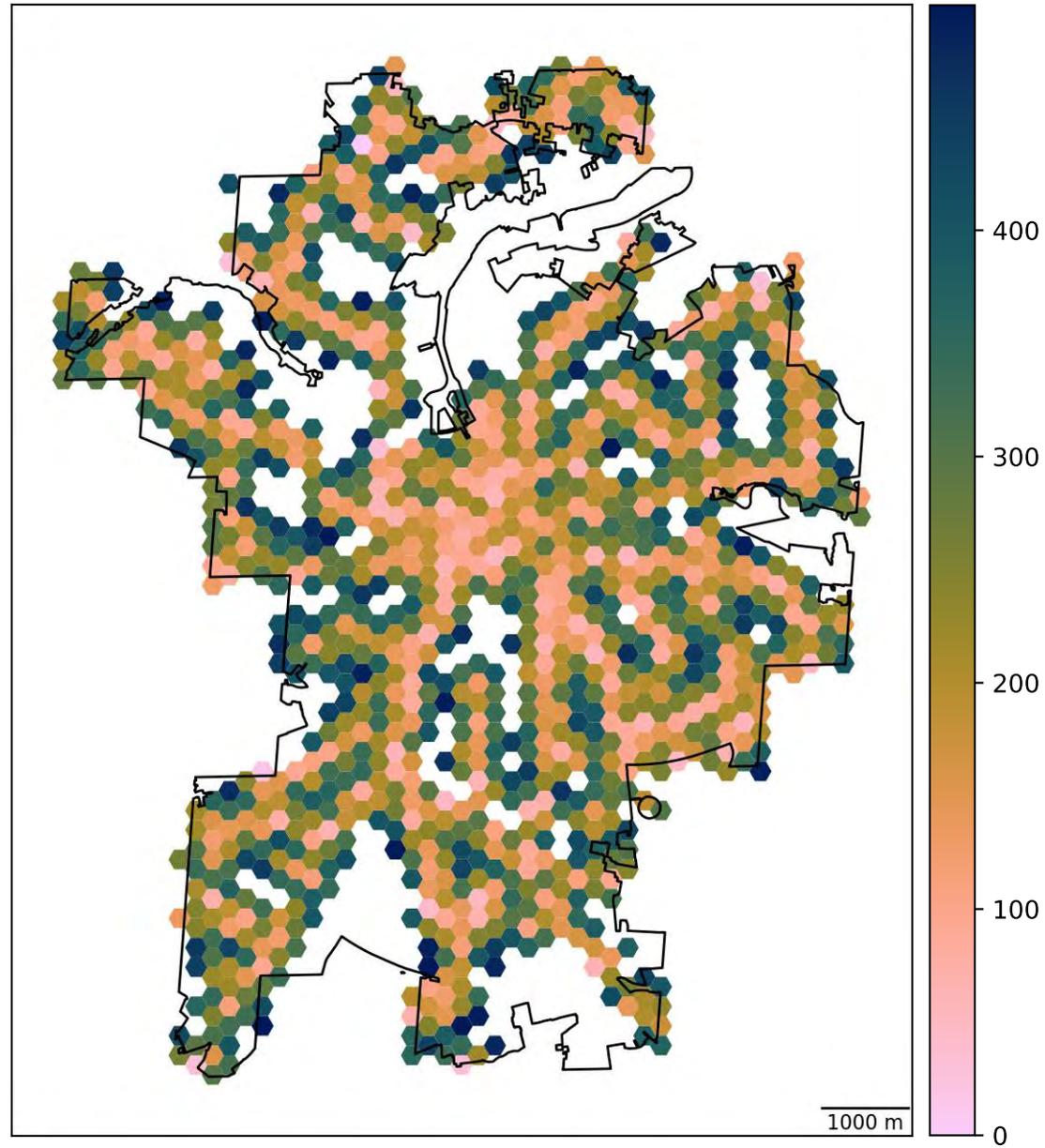

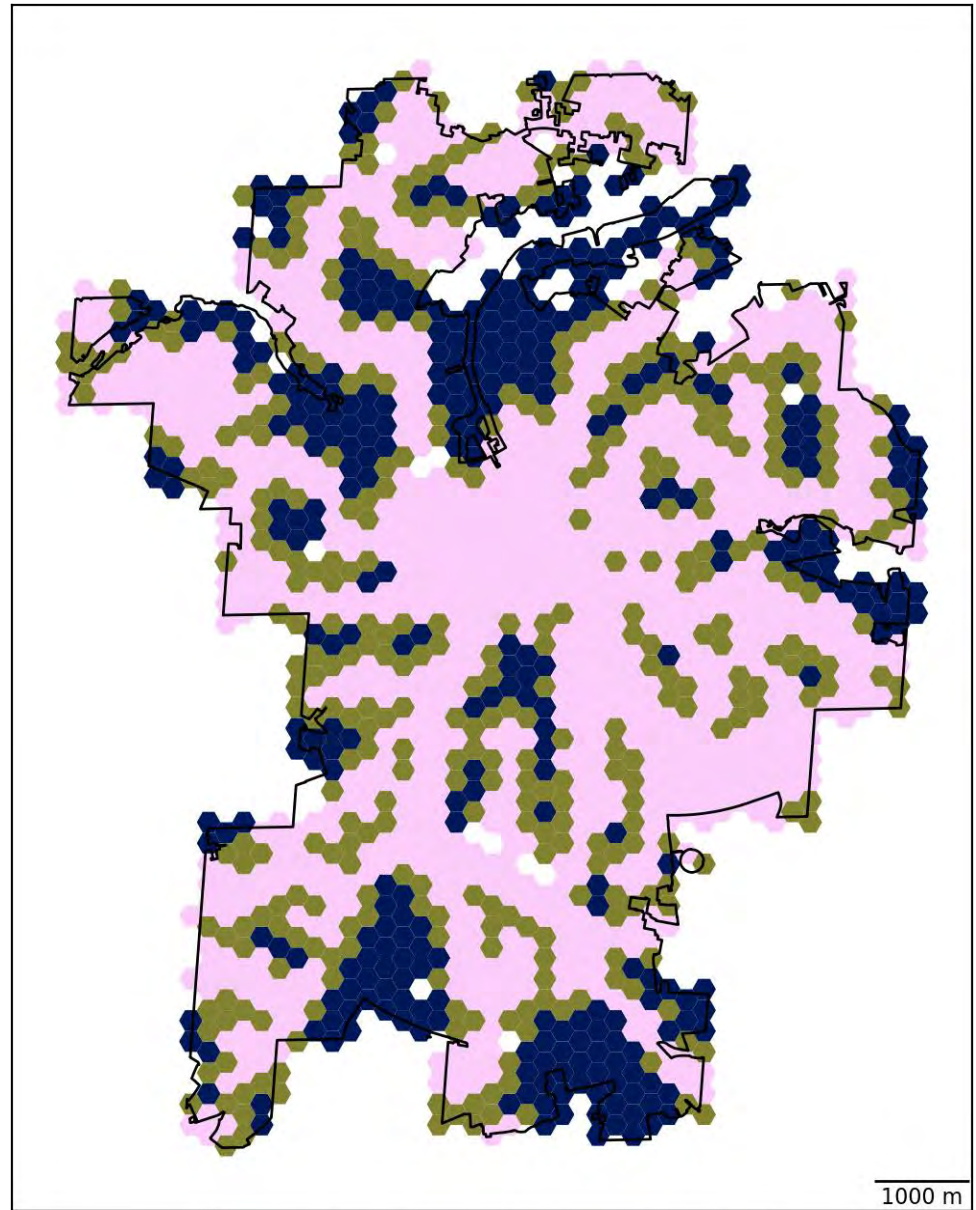

distances: Estimated Distance to nearest public transport stops (m; up to 500m) requirement for distances to destinations, measured up to a maximum distance target threshold of 500 metres

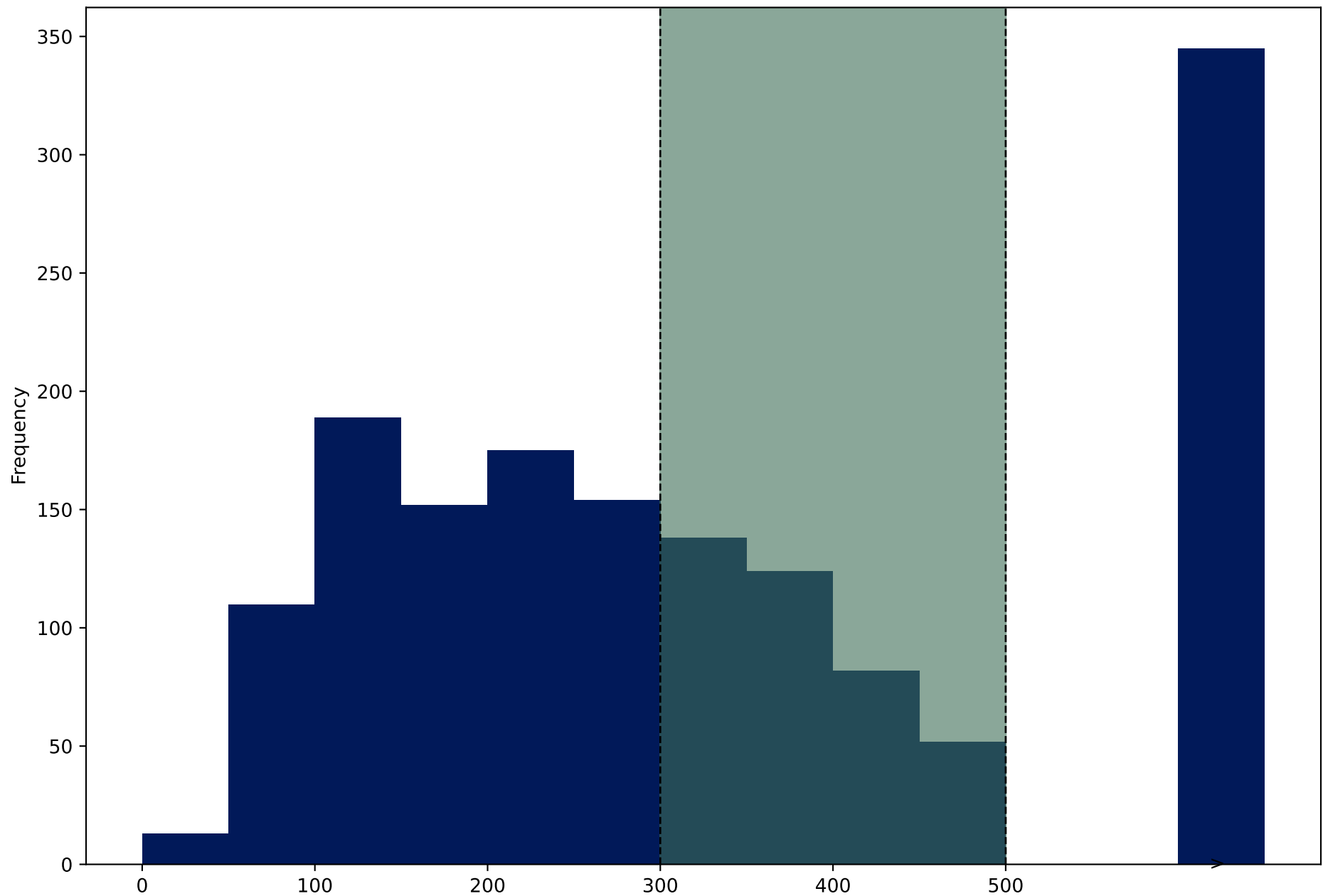

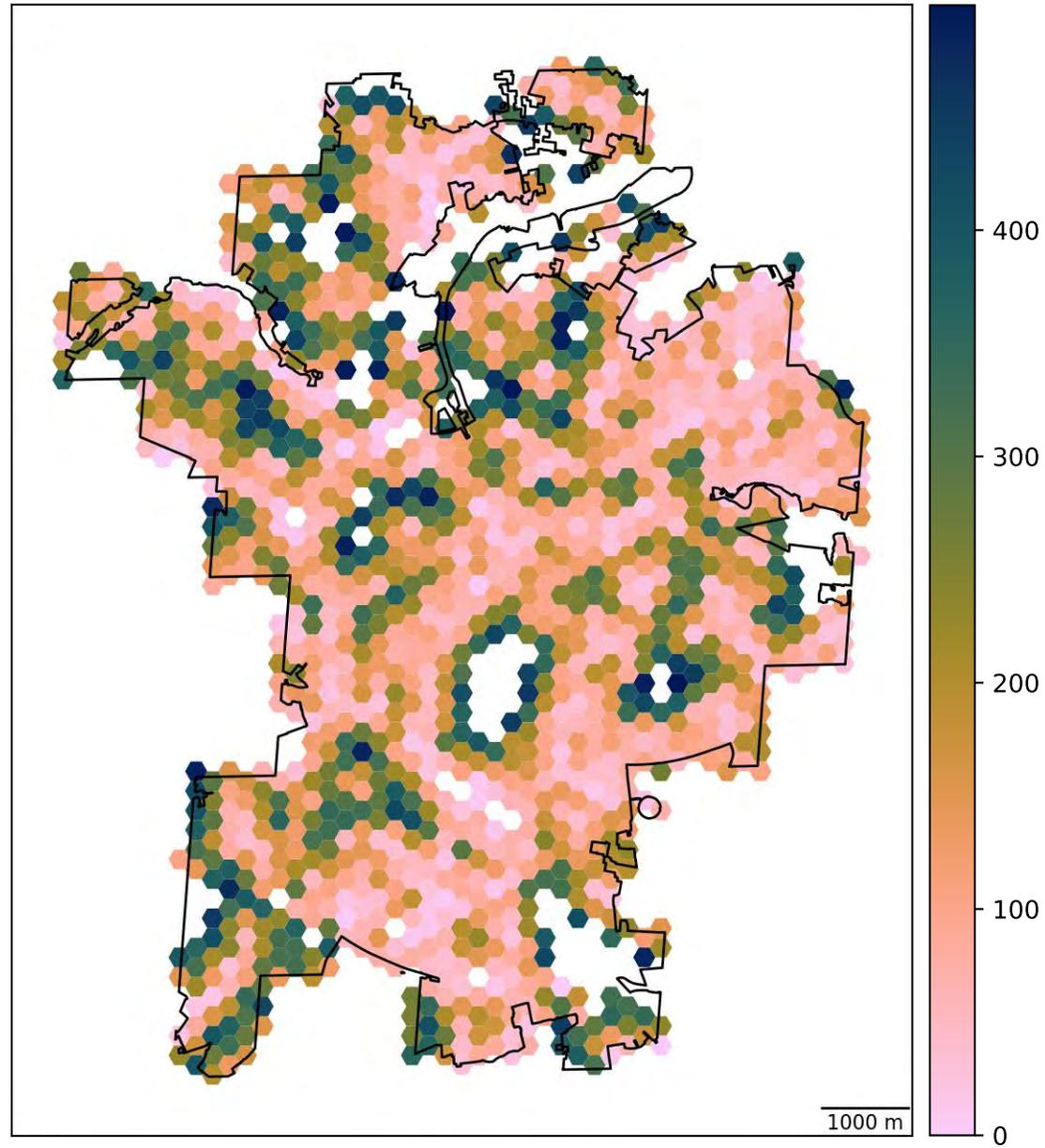

distances: Estimated Distance to nearest park (m; up to 500m) requirement for distances to destinations, measured up to a maximum distance target threshold of 500 metres

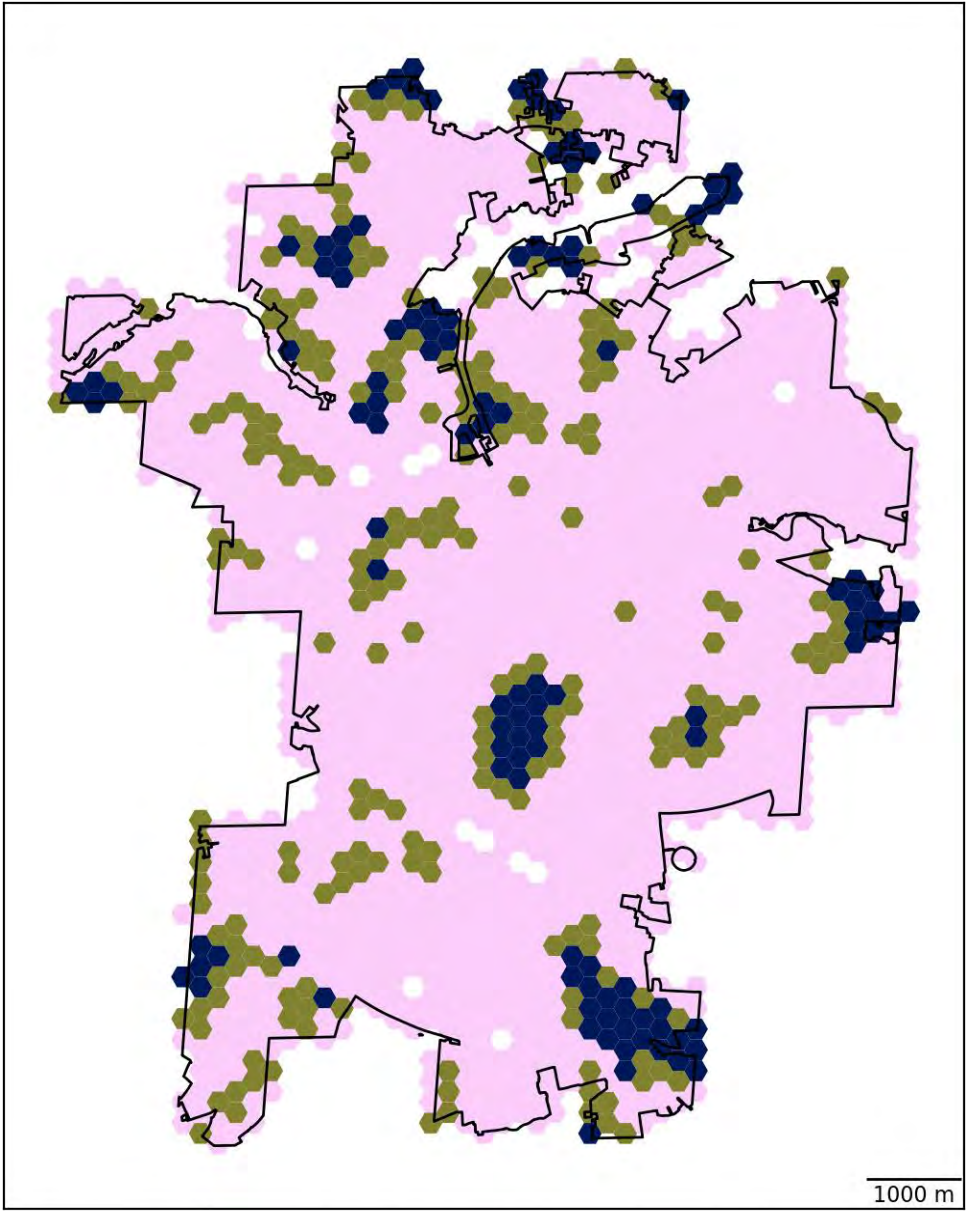

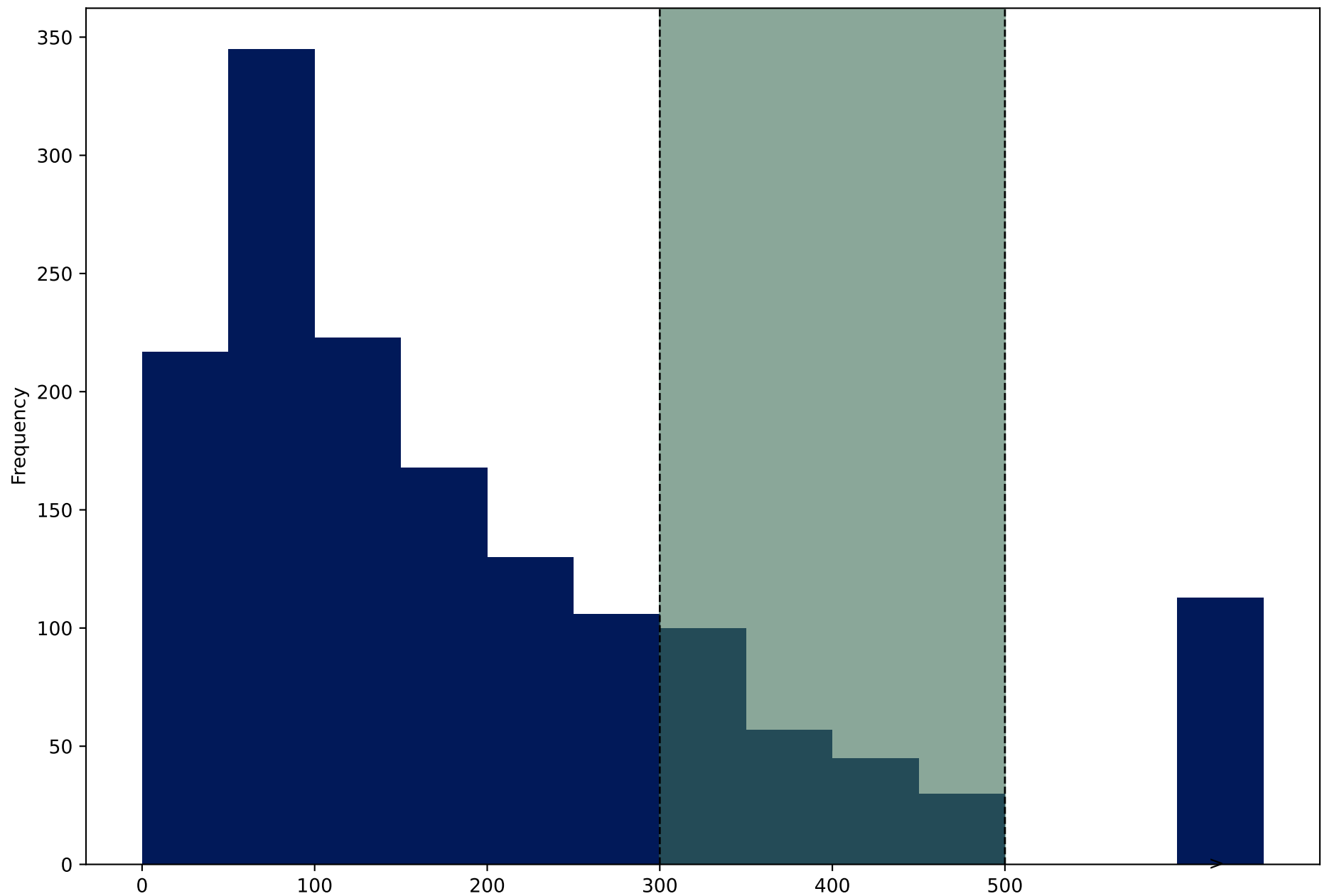

# Europe, Germany, Cologne

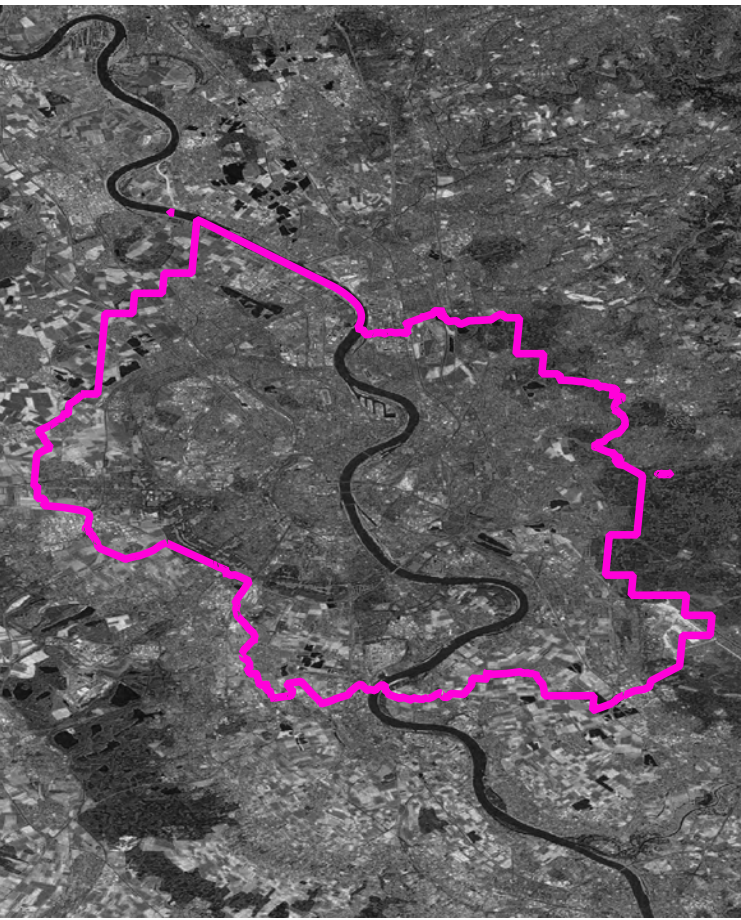
Satellite imagery of urban study region (Bing)

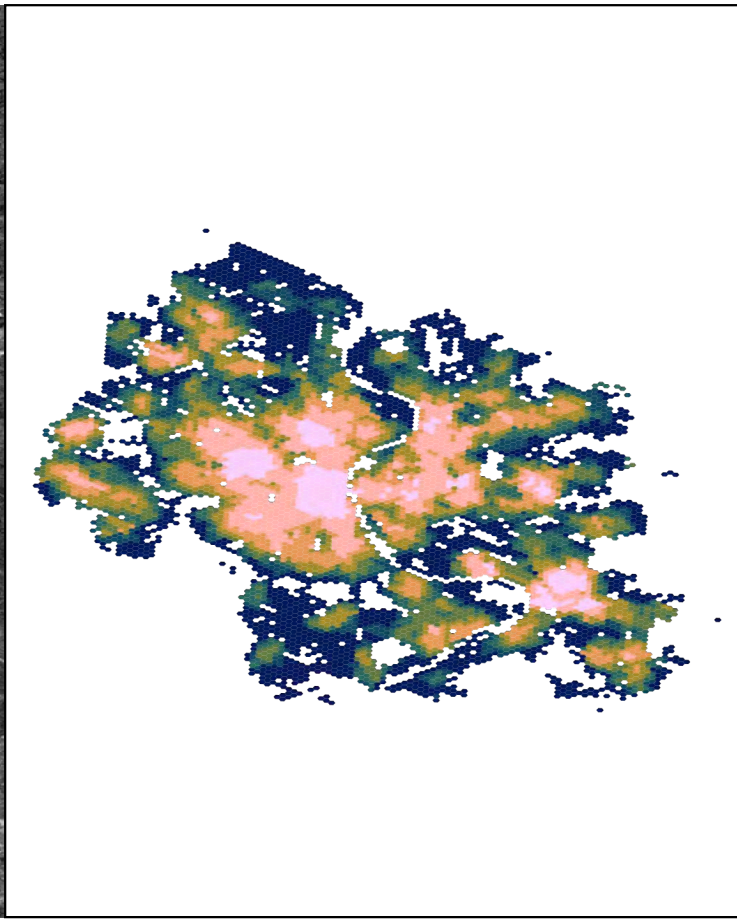
Walkability, relative to city

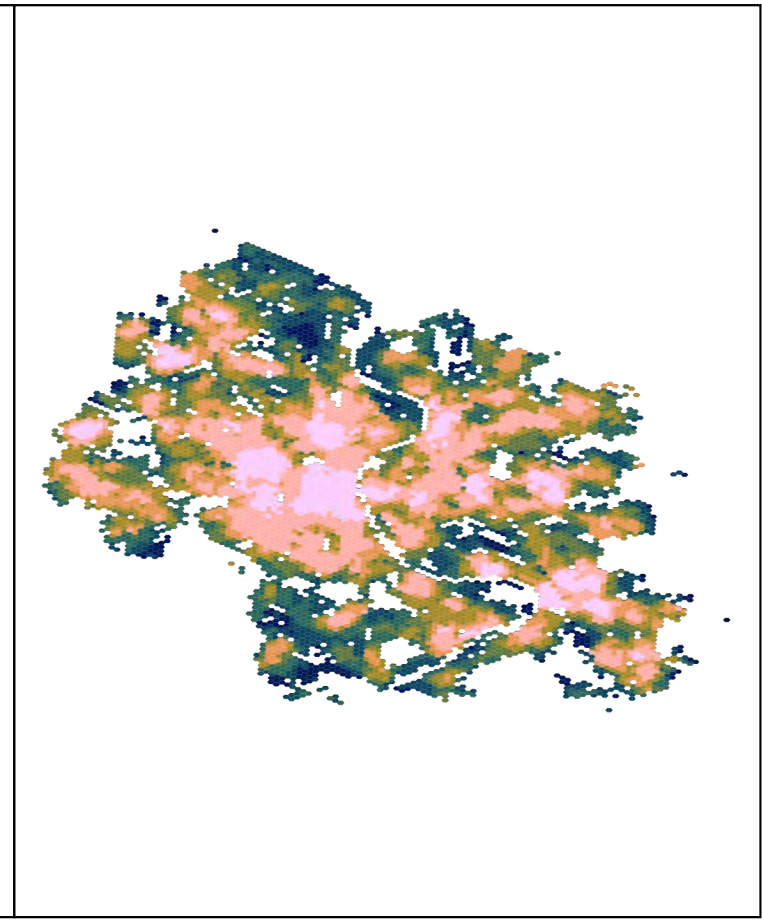
Walkability, relative to 25 global cities

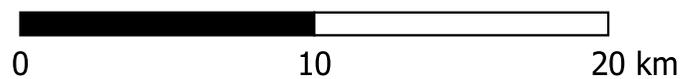

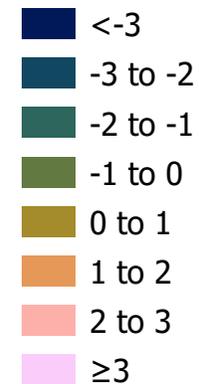

Walkability relative to all cities by component variables (2D histograms), and overall (histogram)

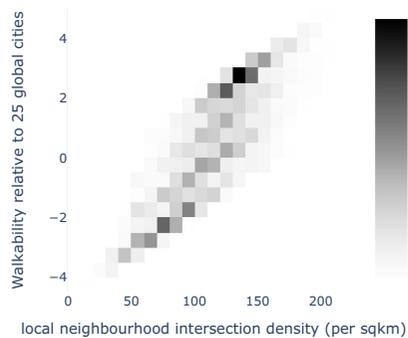
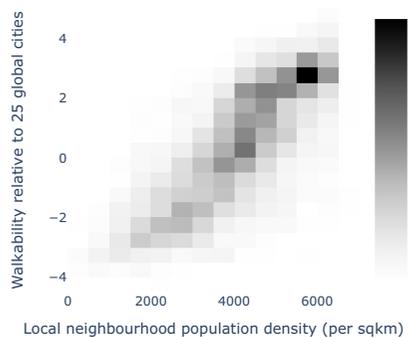
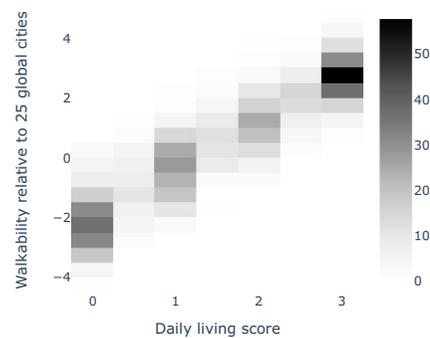
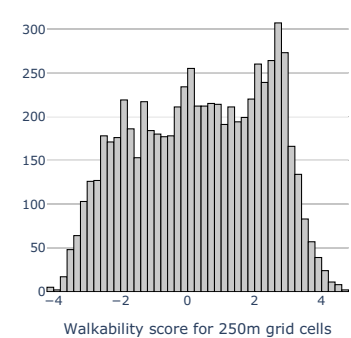

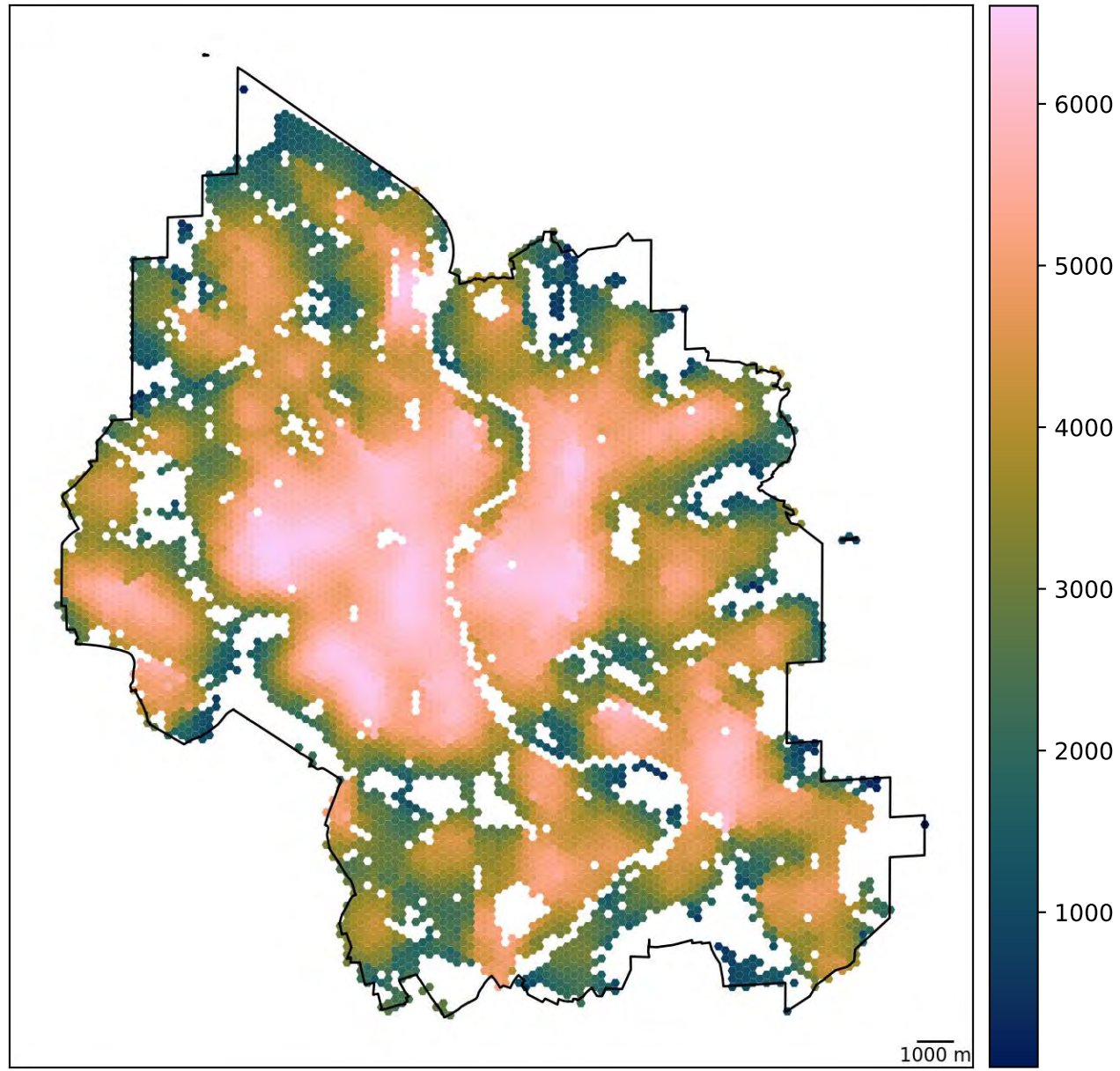

Mean 1000 m neighbourhood population per km²

A: Estimated Mean 1000 m neighbourhood population per km² requirement for ≥80% probability of engaging in walking for transport

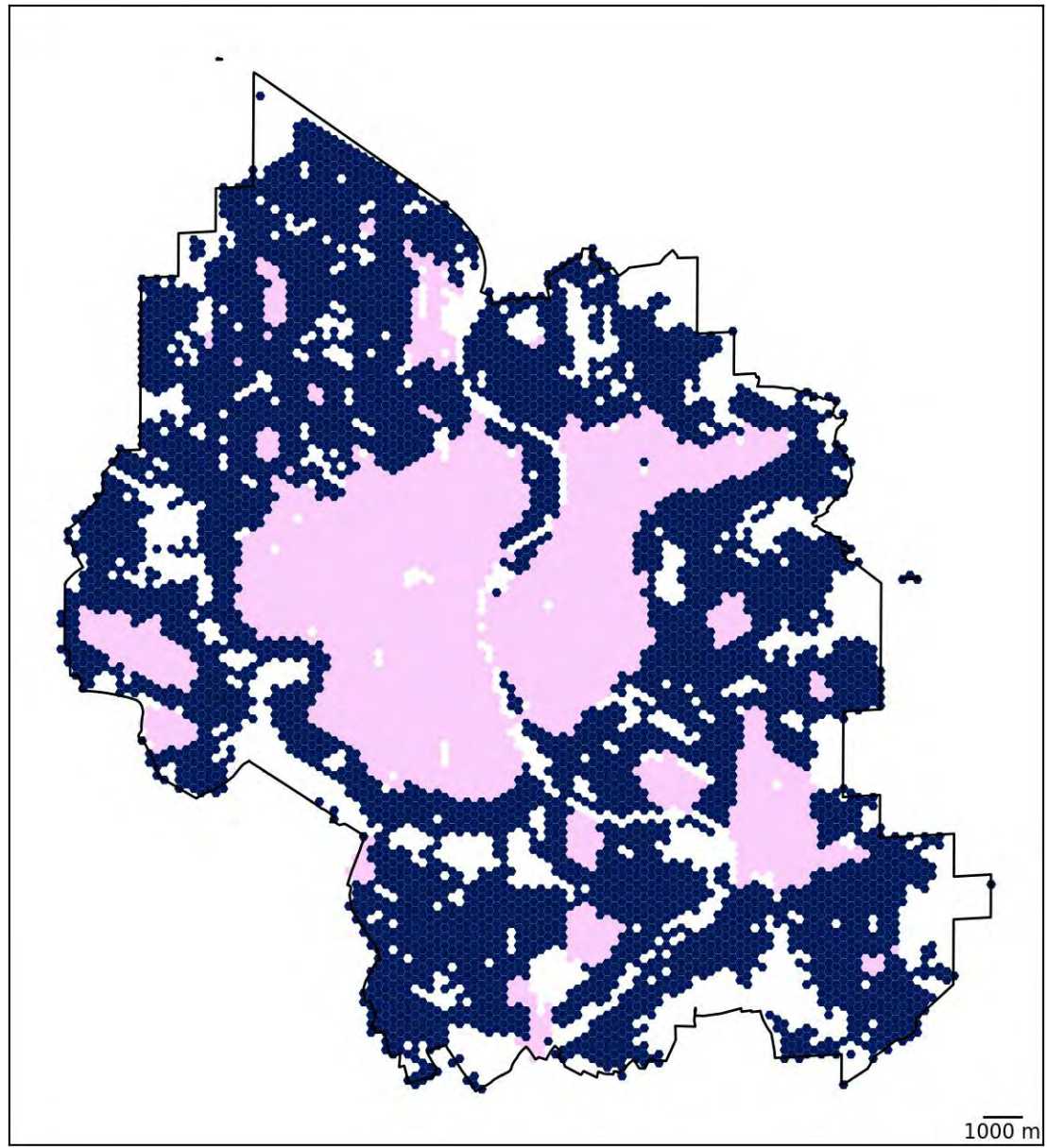

B: Estimated Mean 1000 m neighbourhood population per km² requirement for reaching the WHO's target of a ≥15% relative reduction in insufficient physical activity through walking

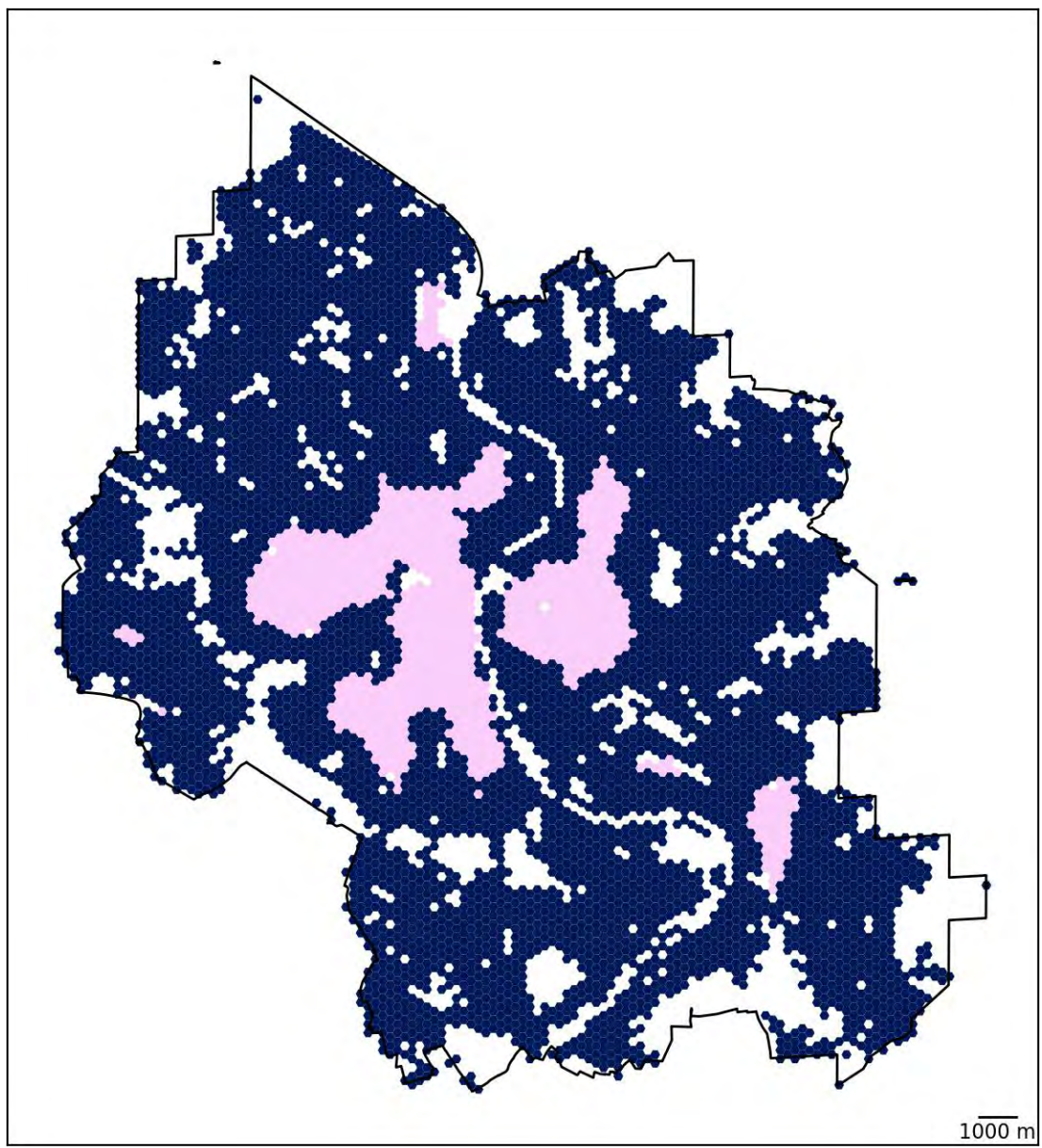

- below 95% CrI lower bound
- within 95% CrI (5677, 7823)

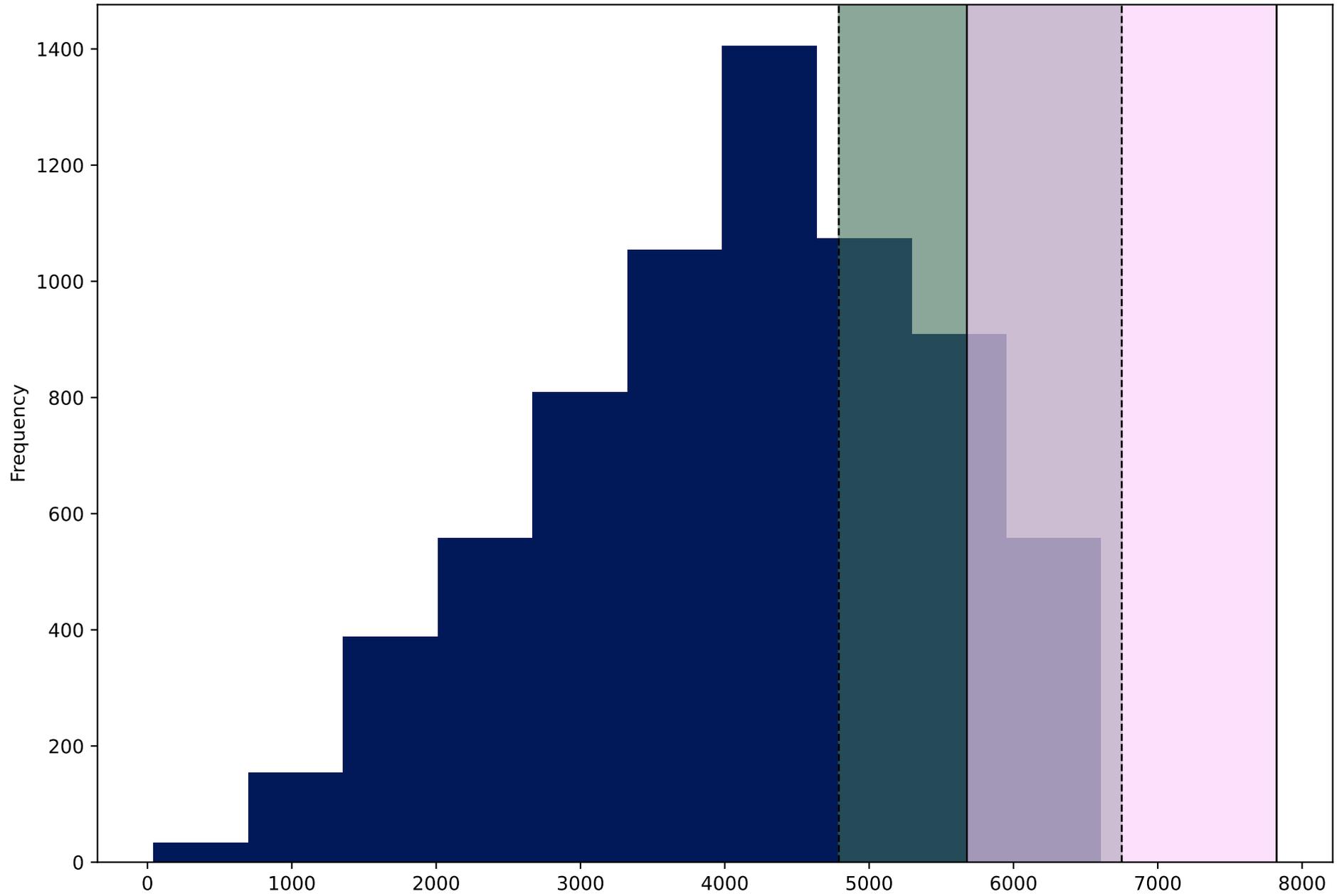

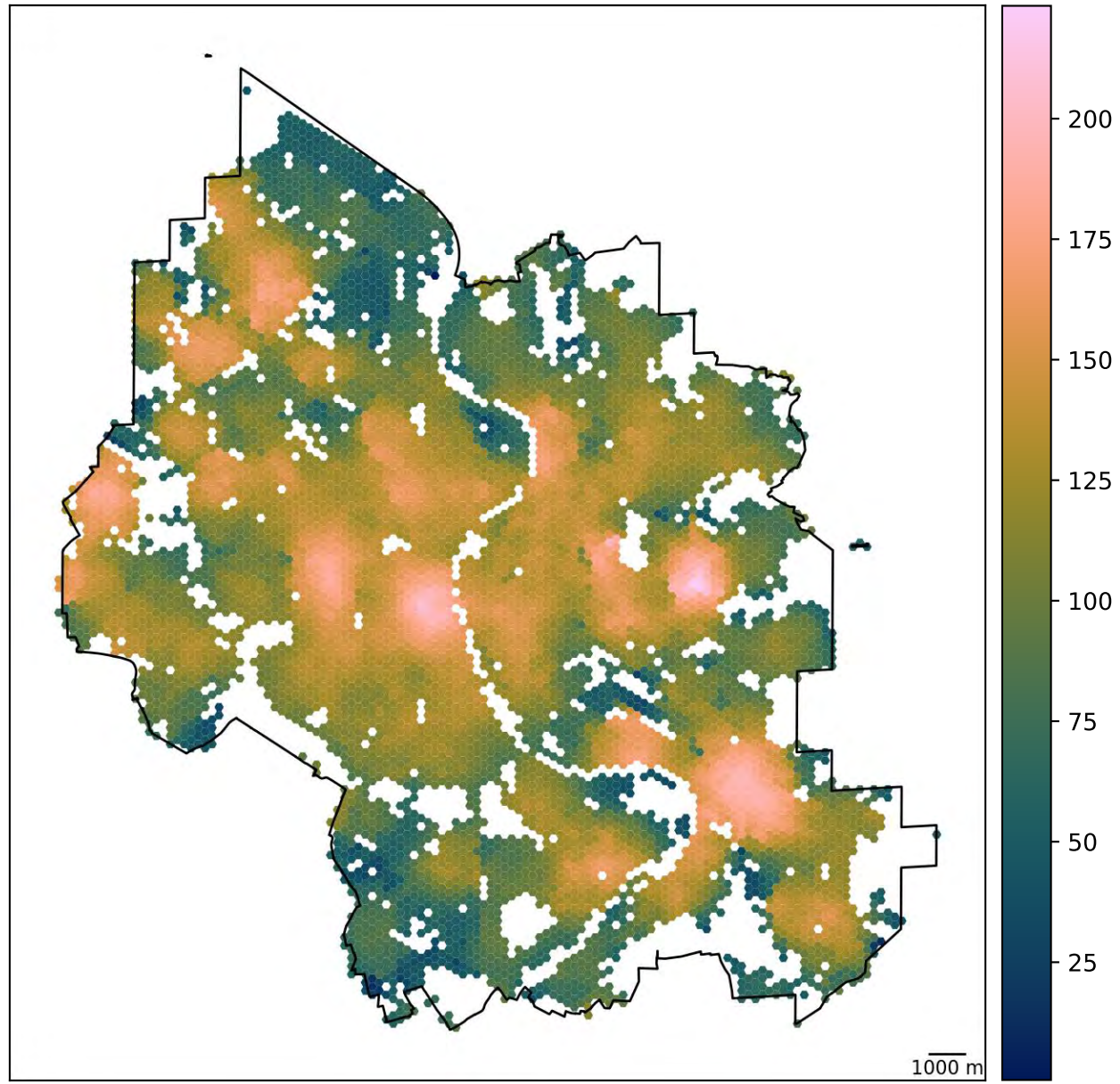

Mean 1000 m neighbourhood street intersections per km²

# A: Estimated Mean 1000 m neighbourhood street intersections per km² requirement for ≥80% probability of engaging in walking for transport

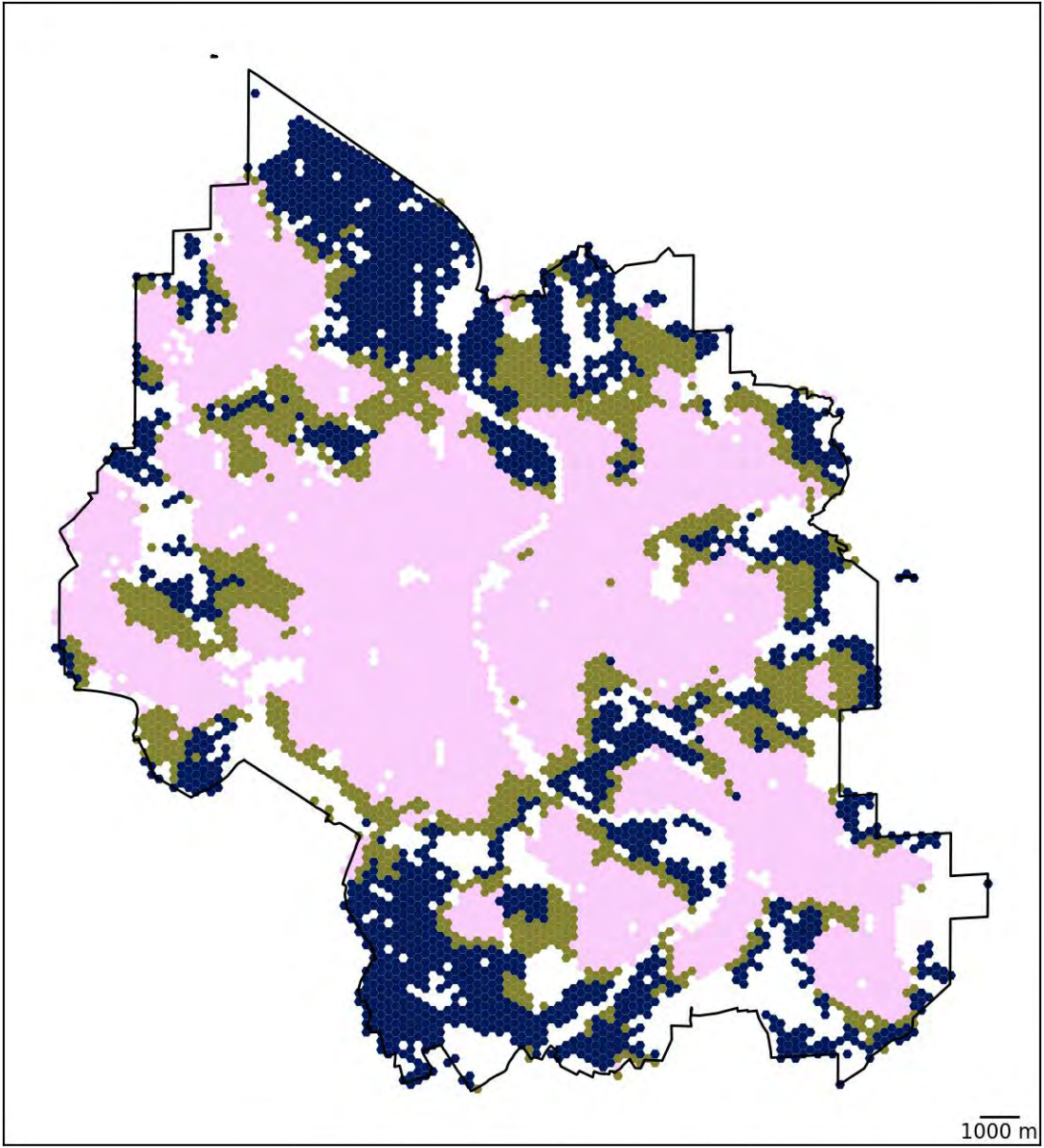

B: Estimated Mean 1000 m neighbourhood street intersections per km² requirement for reaching the WHO's target of a ≥15% relative reduction in insufficient physical activity through walking

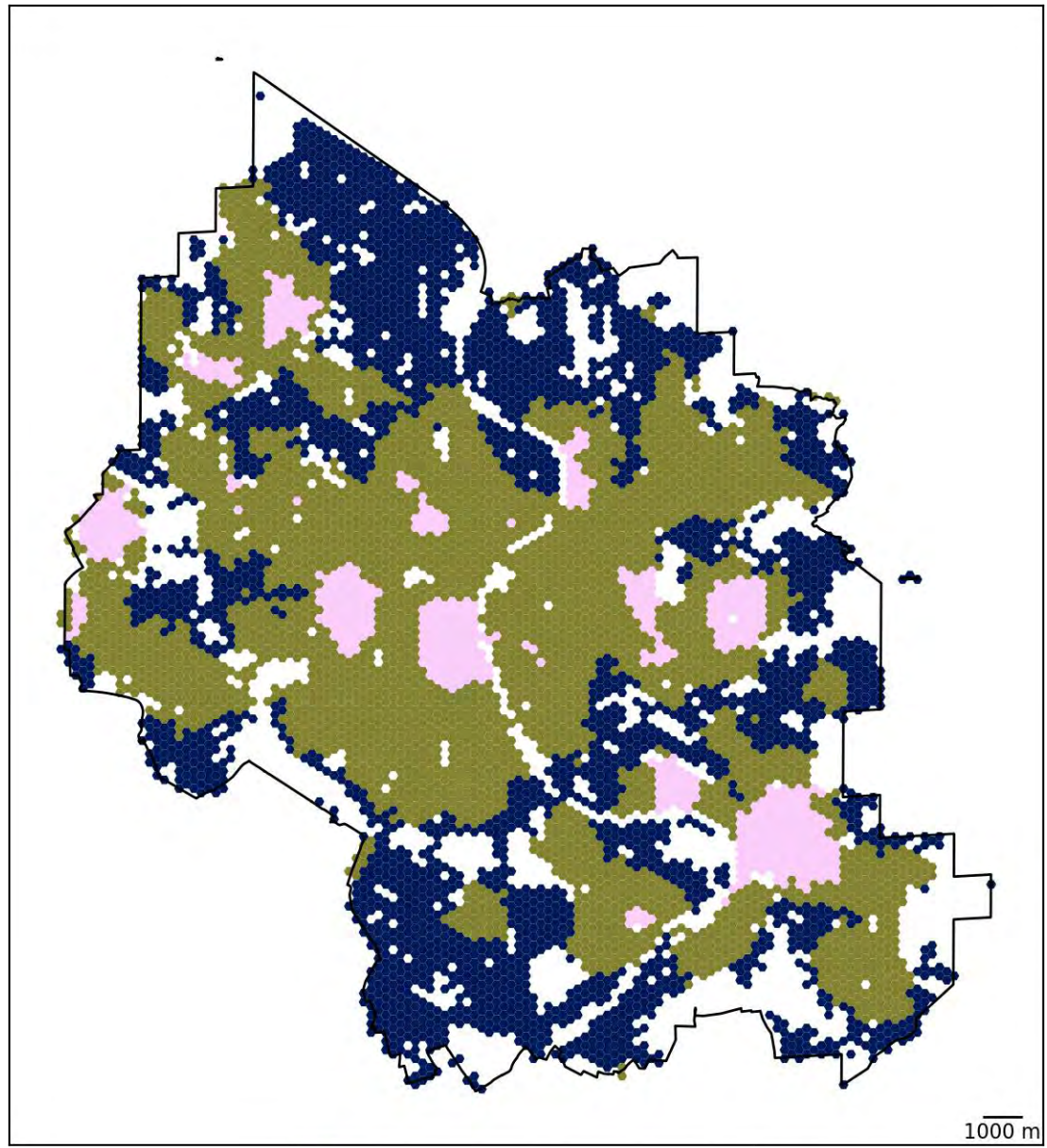

- below 95% CrI lower bound
- within 95% CrI (106, 156)
- exceeds 95% CrI upper bound

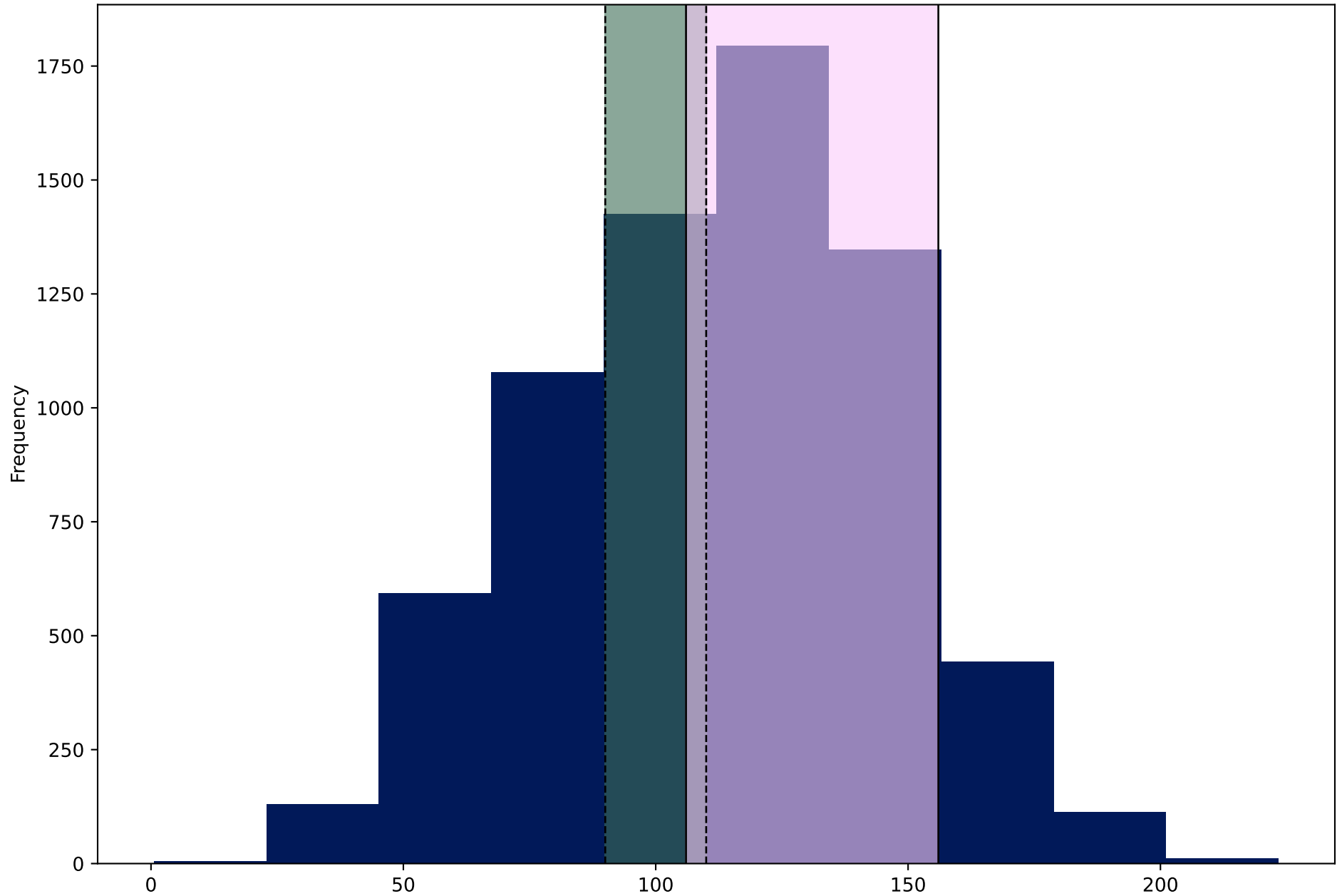

Distance to nearest public transport stops (m; up to 500m)

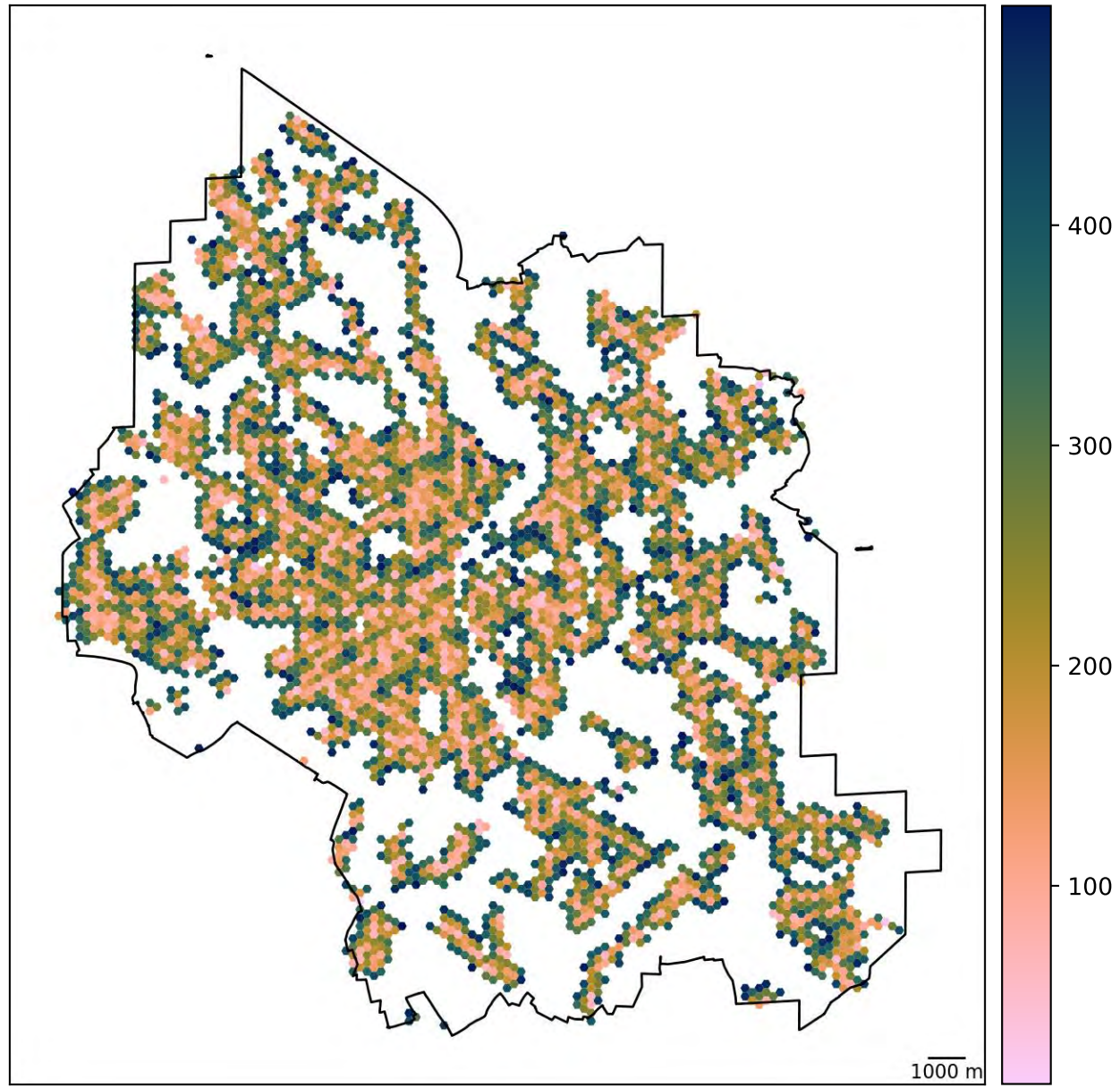

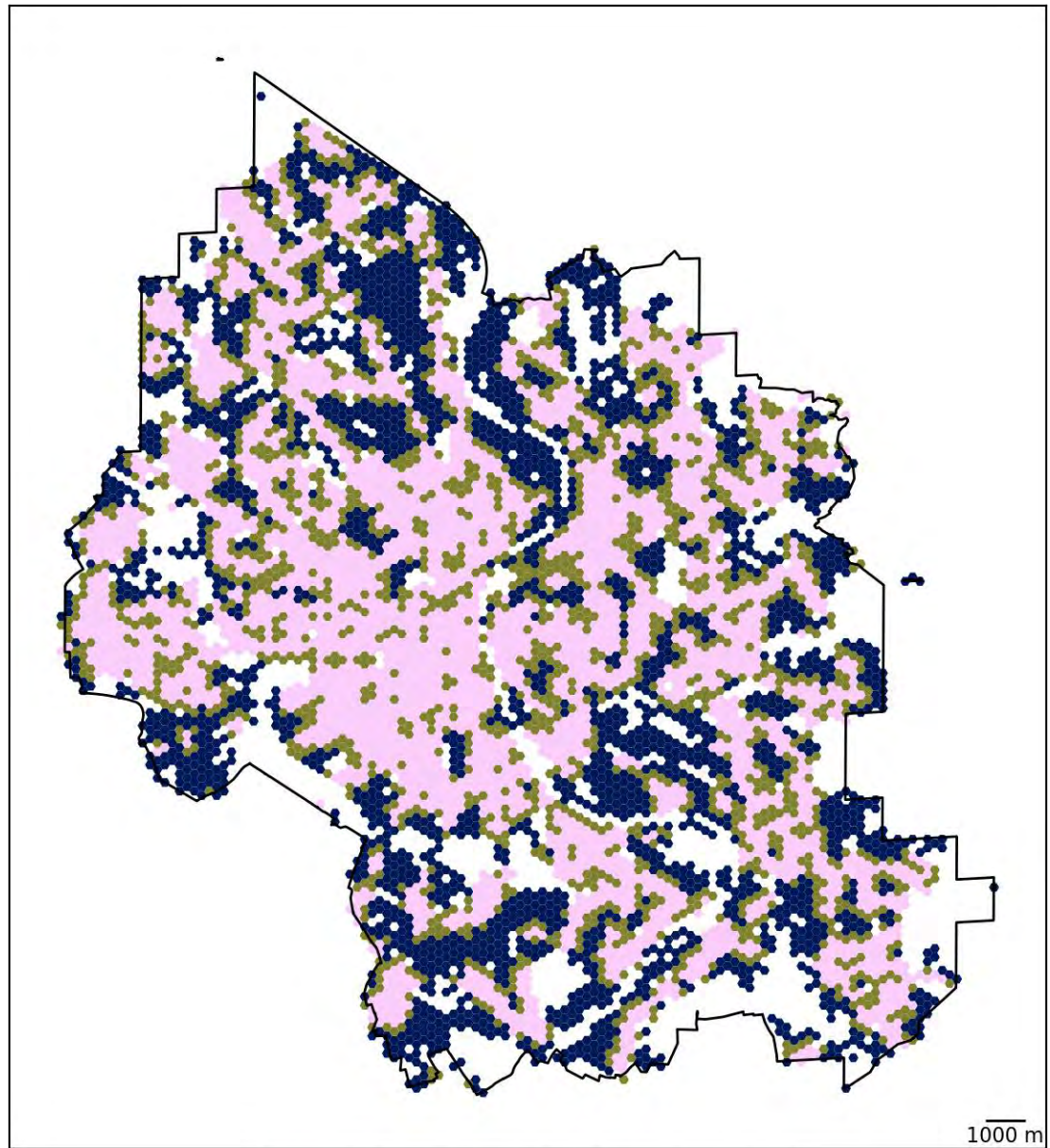

distances: Estimated Distance to nearest public transport stops (m; up to 500m) requirement for distances to destinations, measured up to a maximum distance target threshold of 500 metres

- below distance (m) lower bound
- within distance (m) (300, 500)
- exceeds distance (m) upper bound

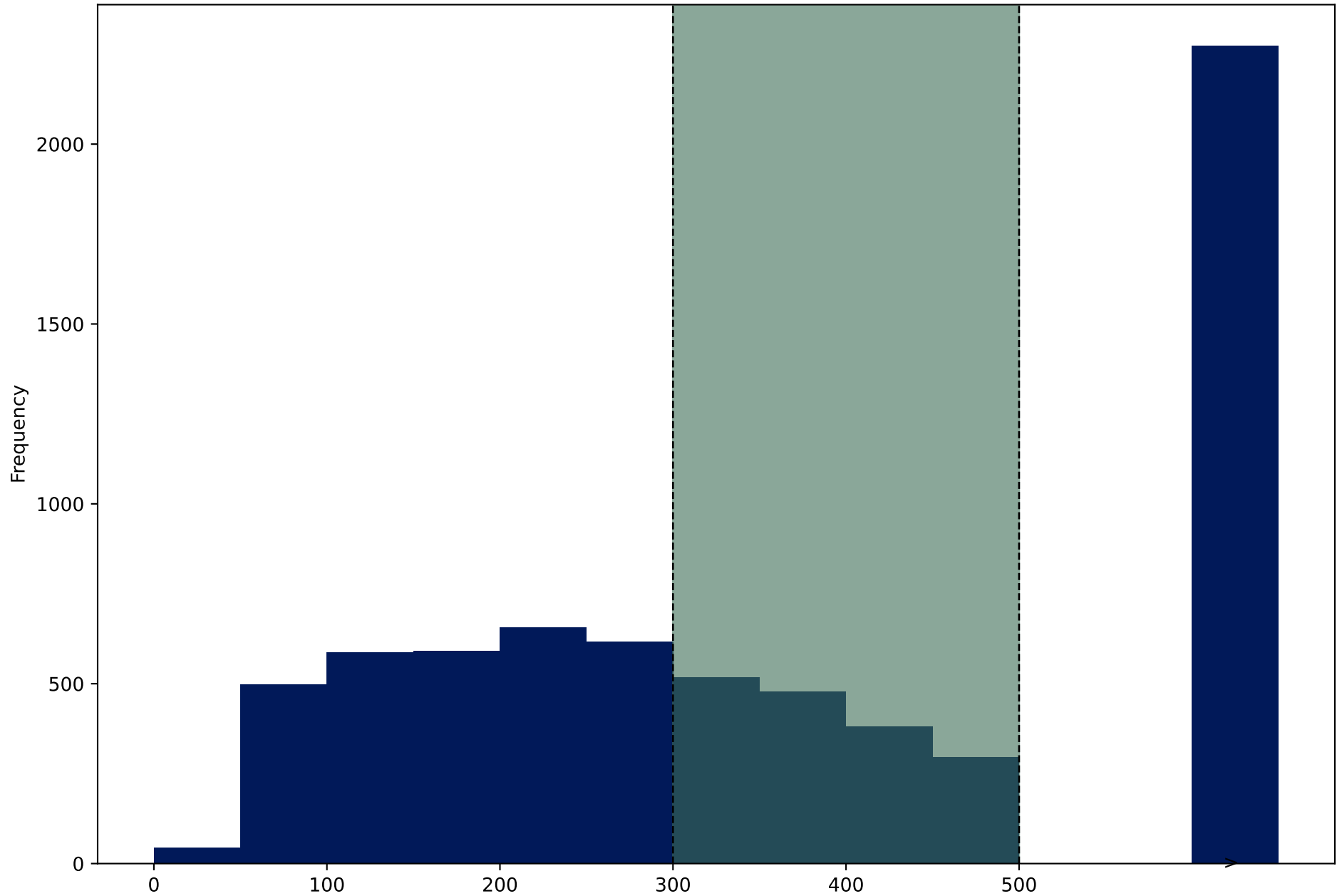

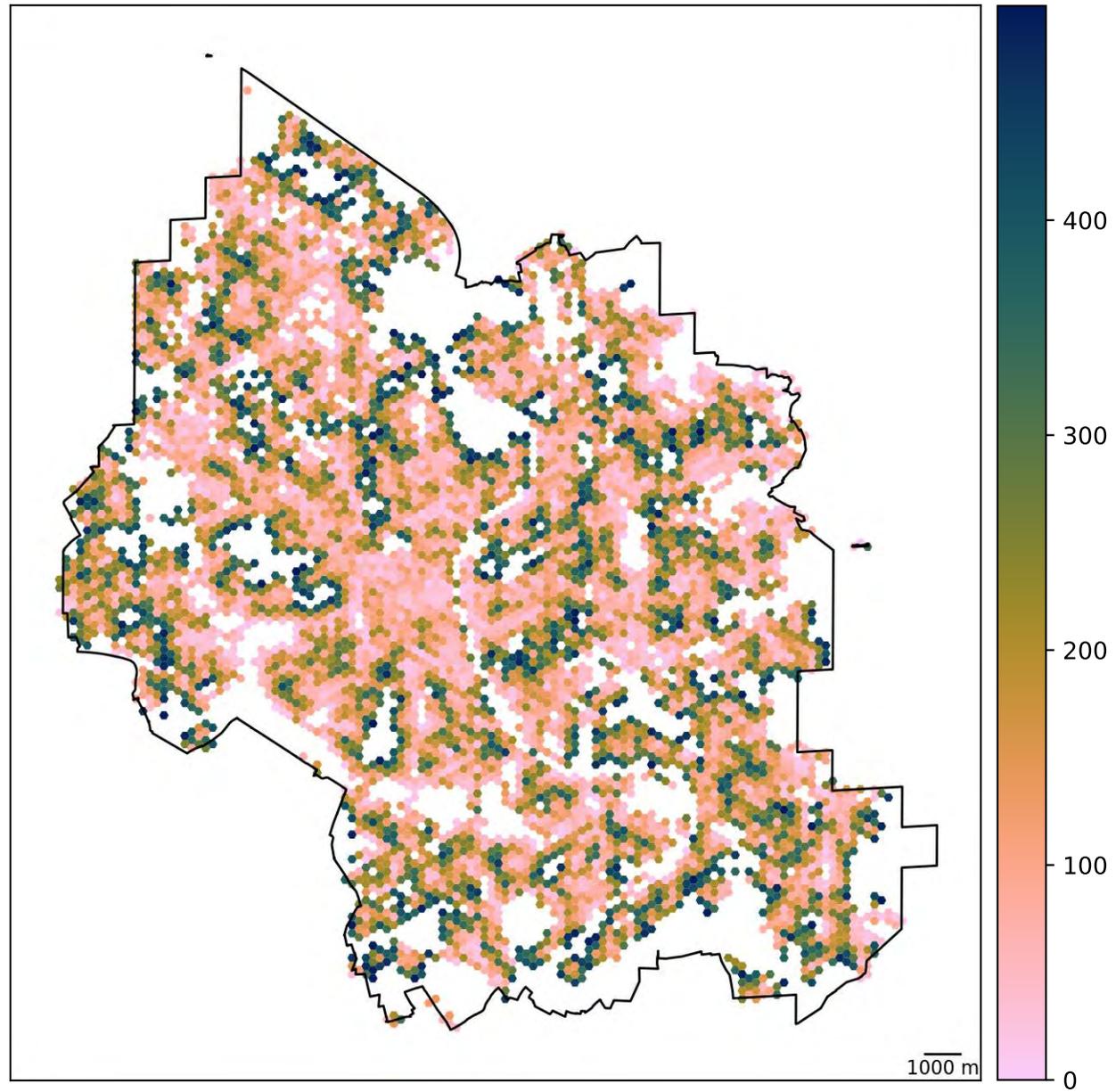

distances: Estimated Distance to nearest park (m; up to 500m) requirement for distances to destinations, measured up to a maximum distance target threshold of 500 metres

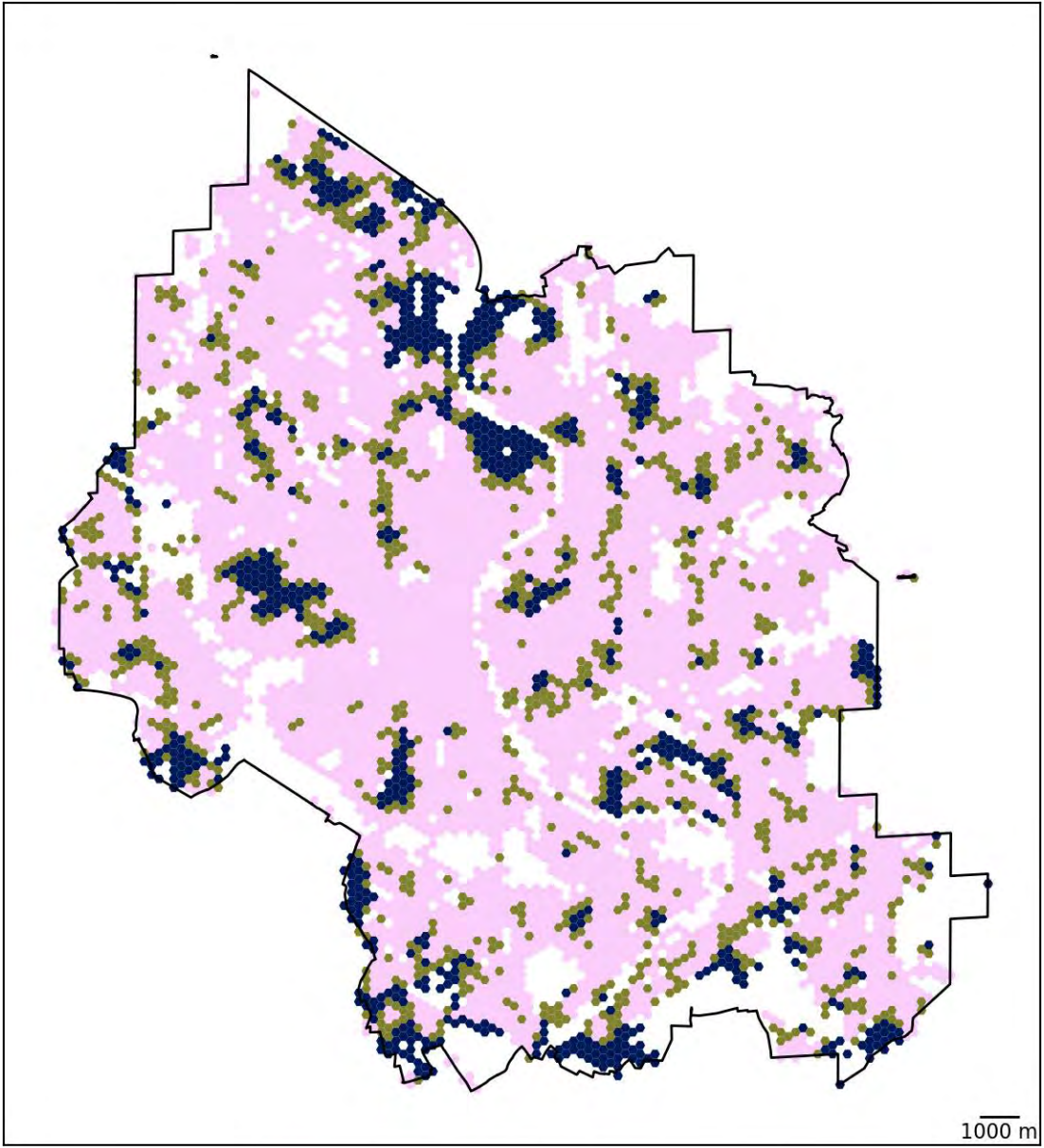

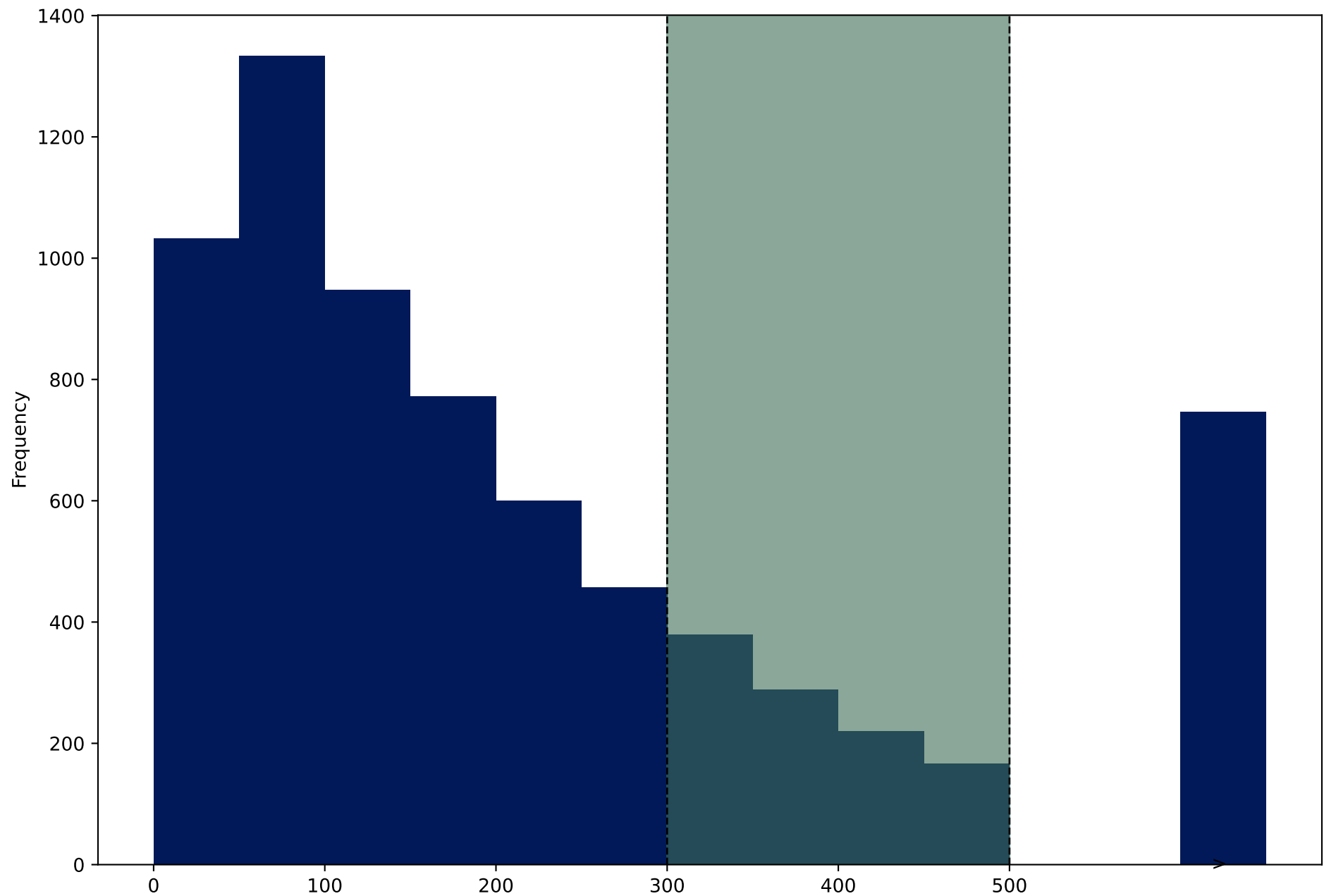

# Europe, Portugal, Lisbon

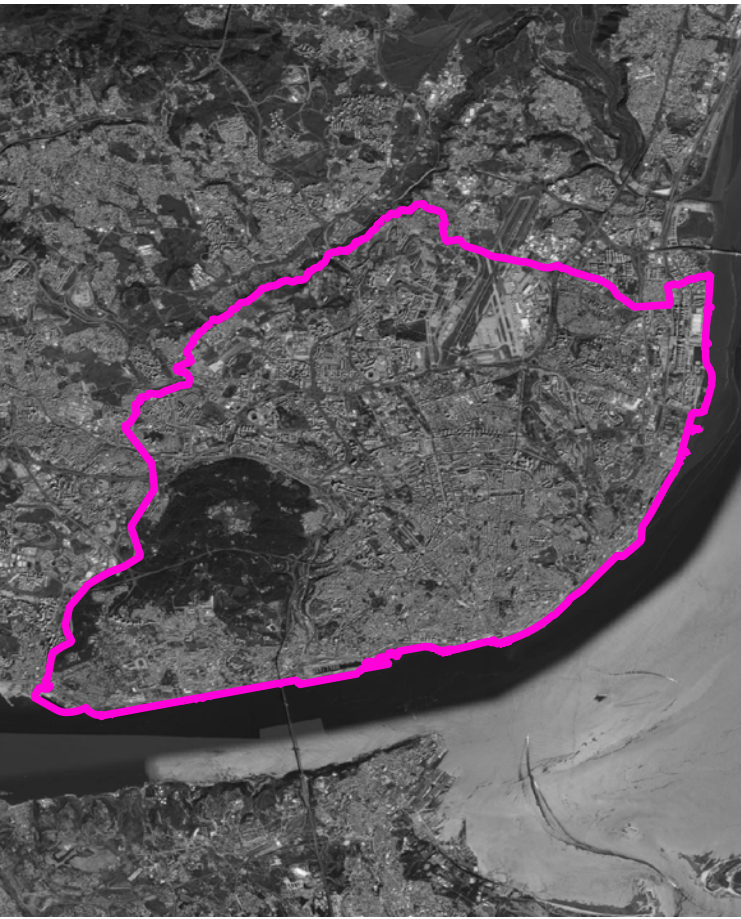
Satellite imagery of urban study region (Bing)

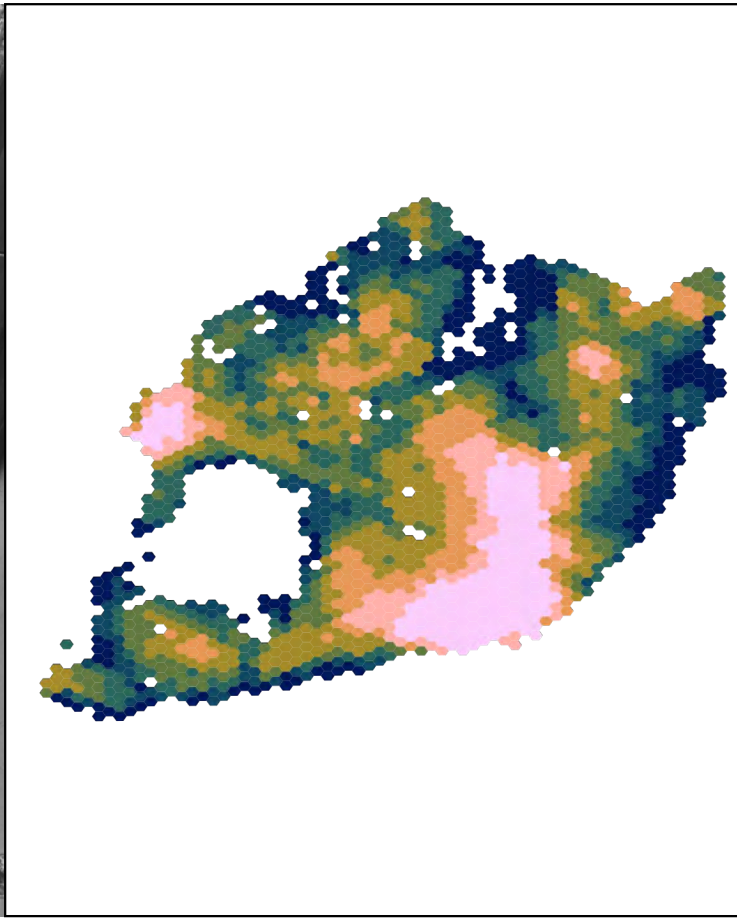
Walkability, relative to city

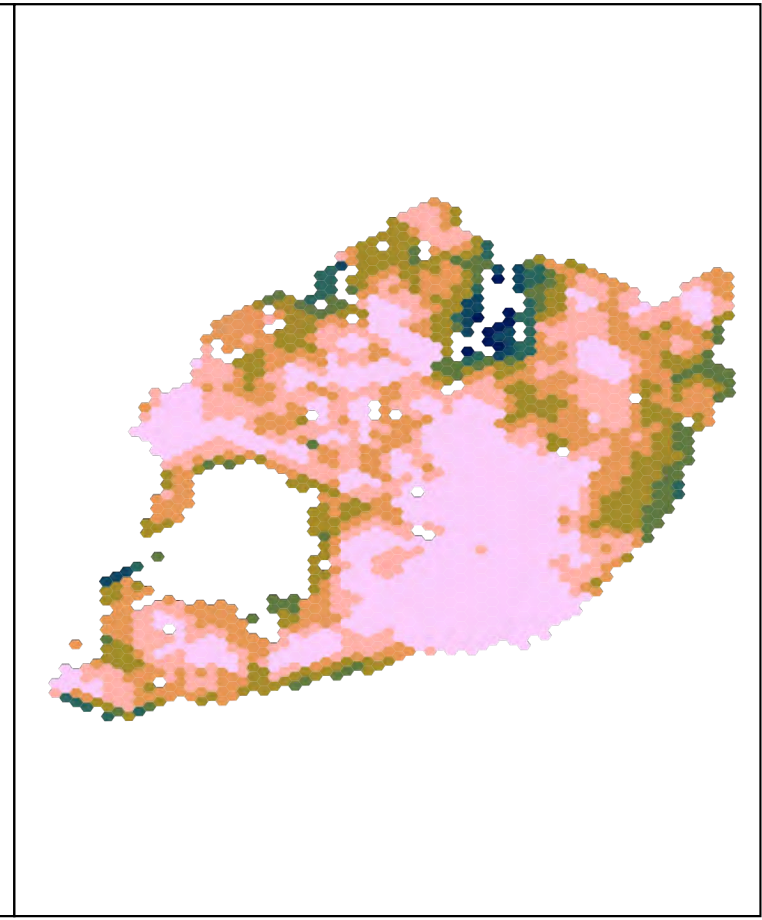
Walkability, relative to 25 global cities

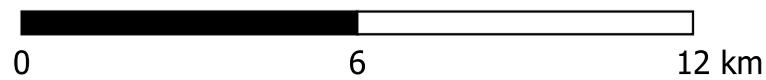

Walkability relative to all cities by component variables (2D histograms), and overall (histogram)

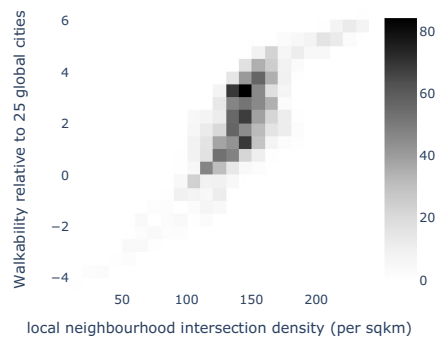
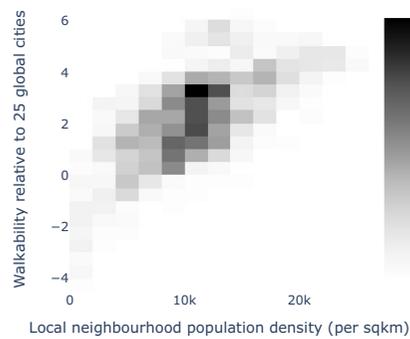
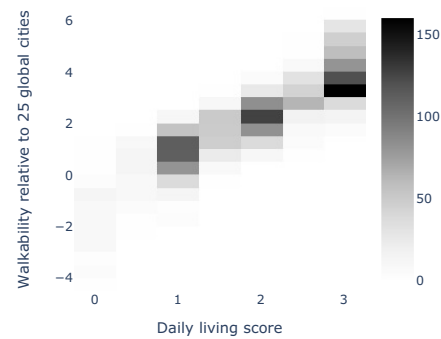
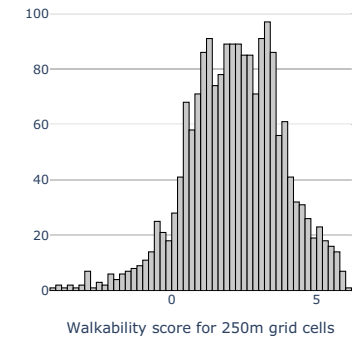

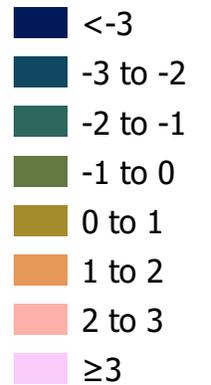

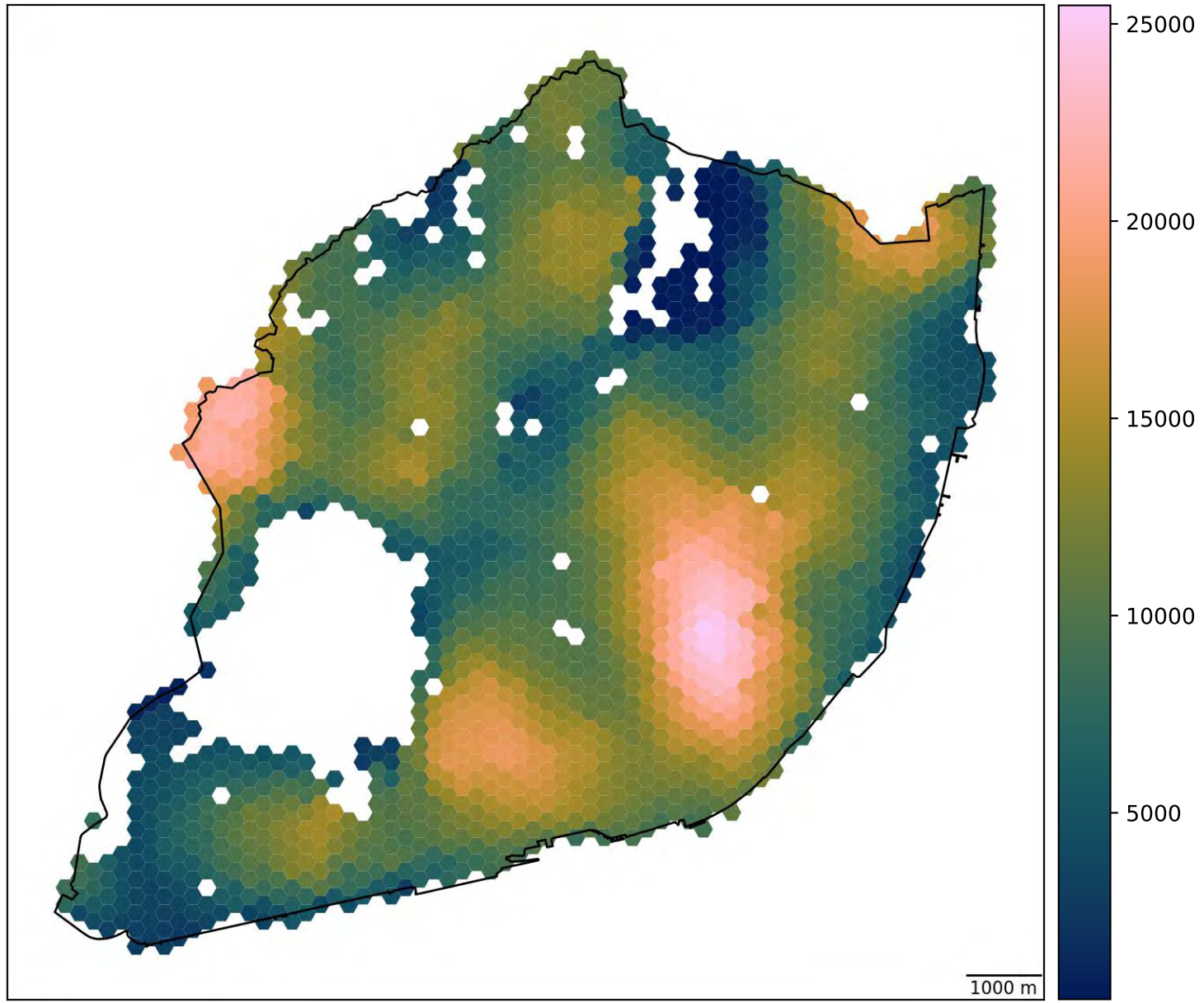

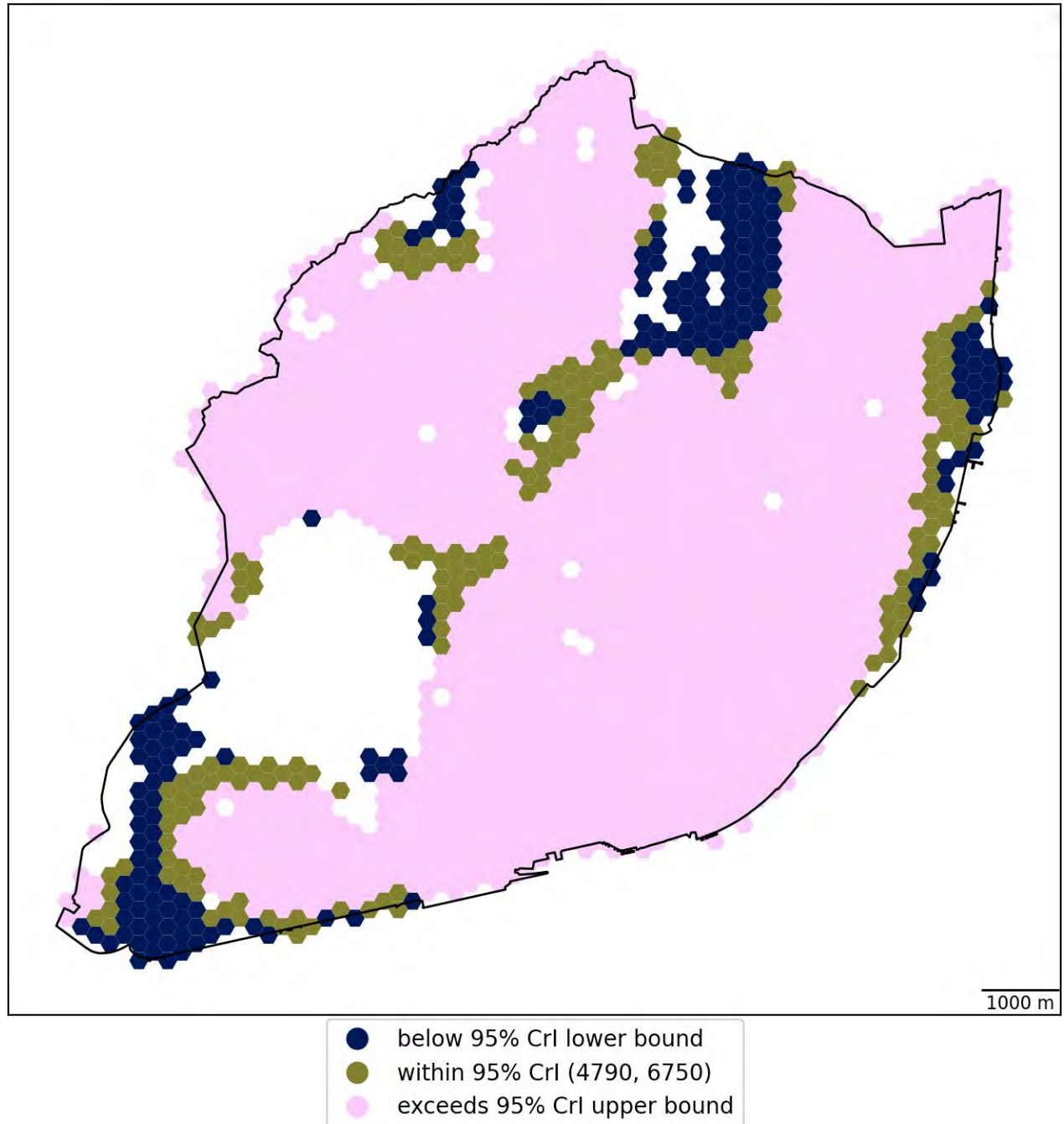

A: Estimated Mean 1000 m neighbourhood population per km² requirement for ≥80% probability of engaging in walking for transport

- below 95% CrI lower bound
- within 95% CrI (4790, 6750)
- exceeds 95% CrI upper bound

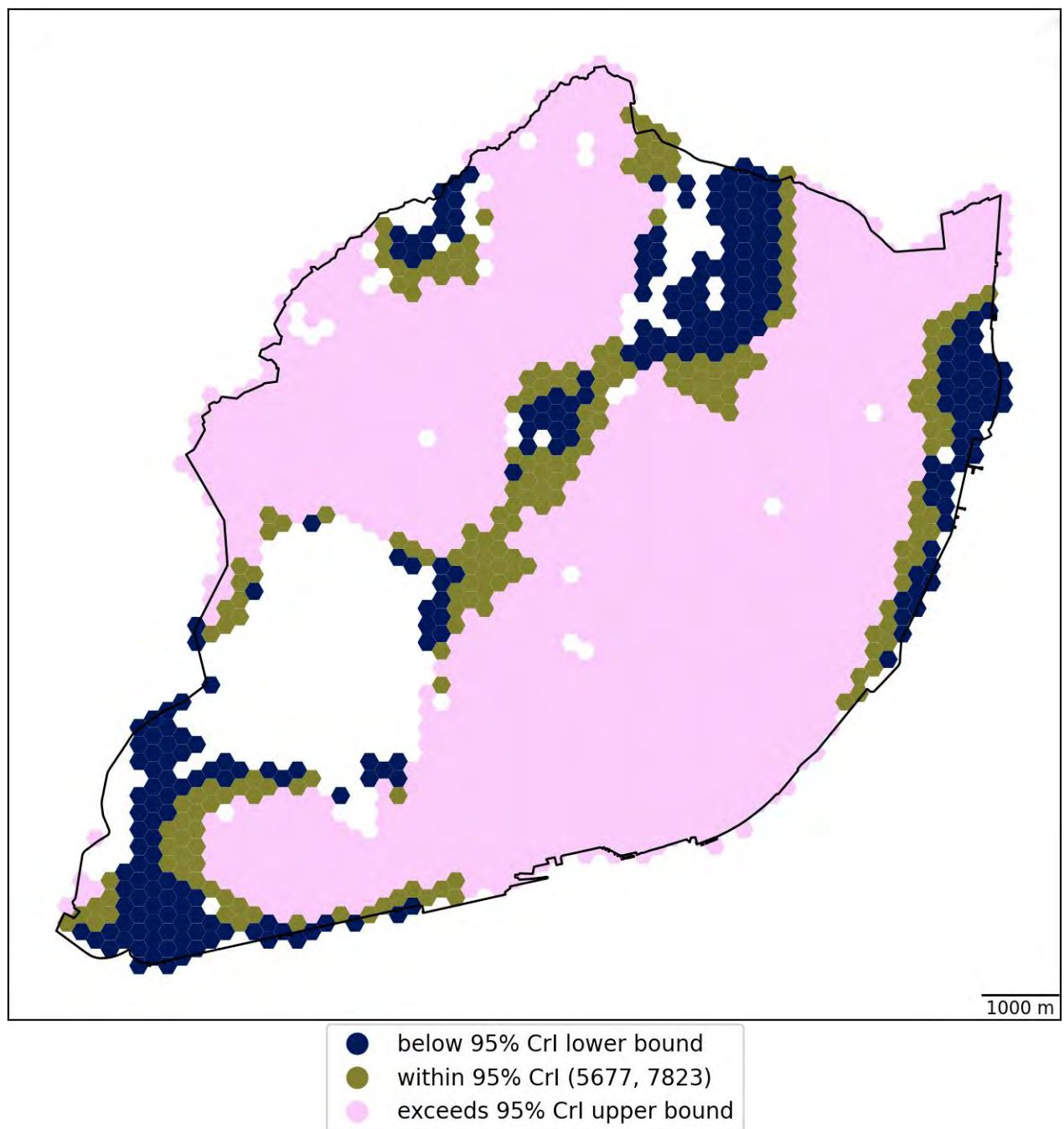

B: Estimated Mean 1000 m neighbourhood population per km² requirement for reaching the WHO's target of a ≥15% relative reduction in insufficient physical activity through walking

- below 95% CrI lower bound
- within 95% CrI (5677, 7823)
- exceeds 95% CrI upper bound

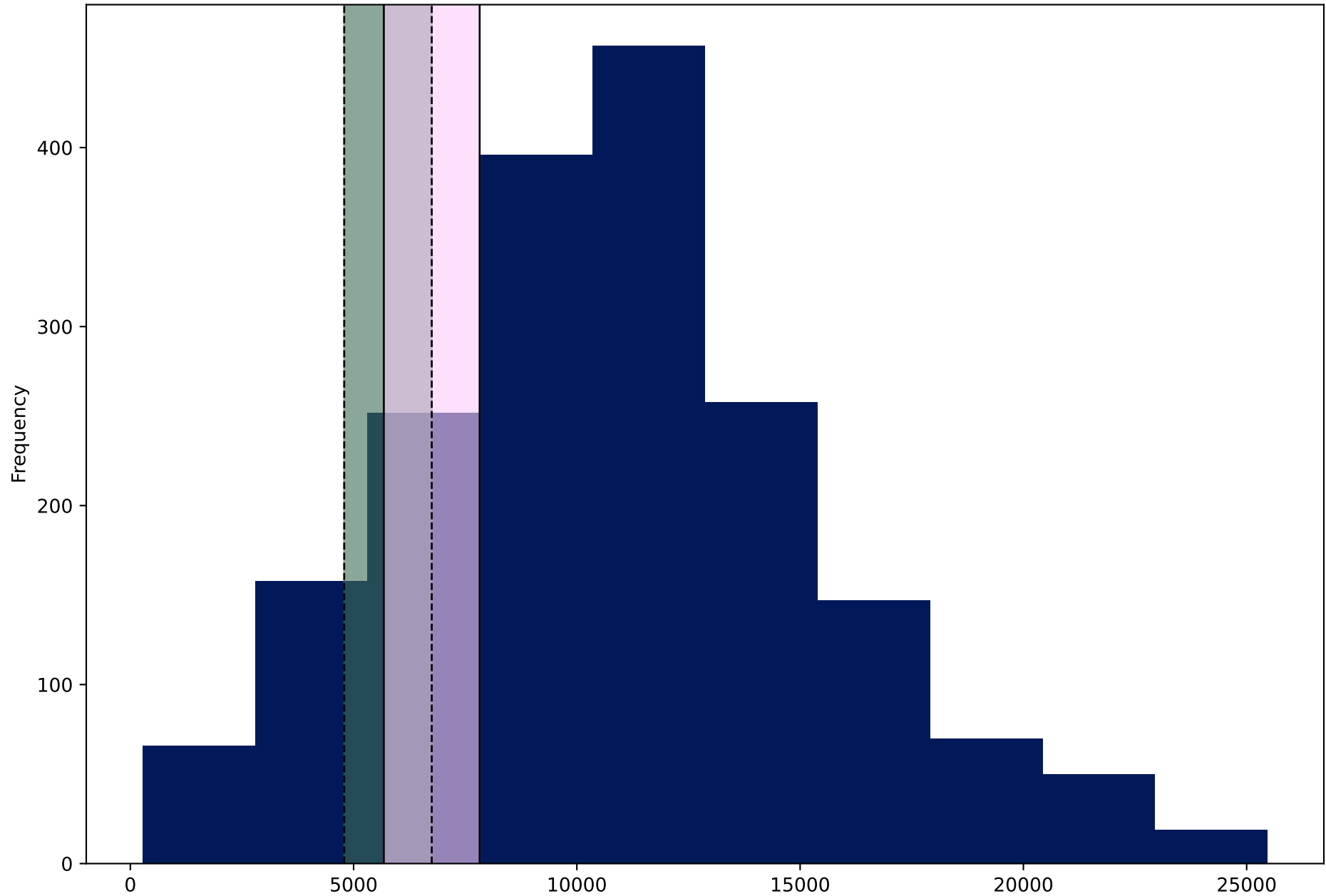

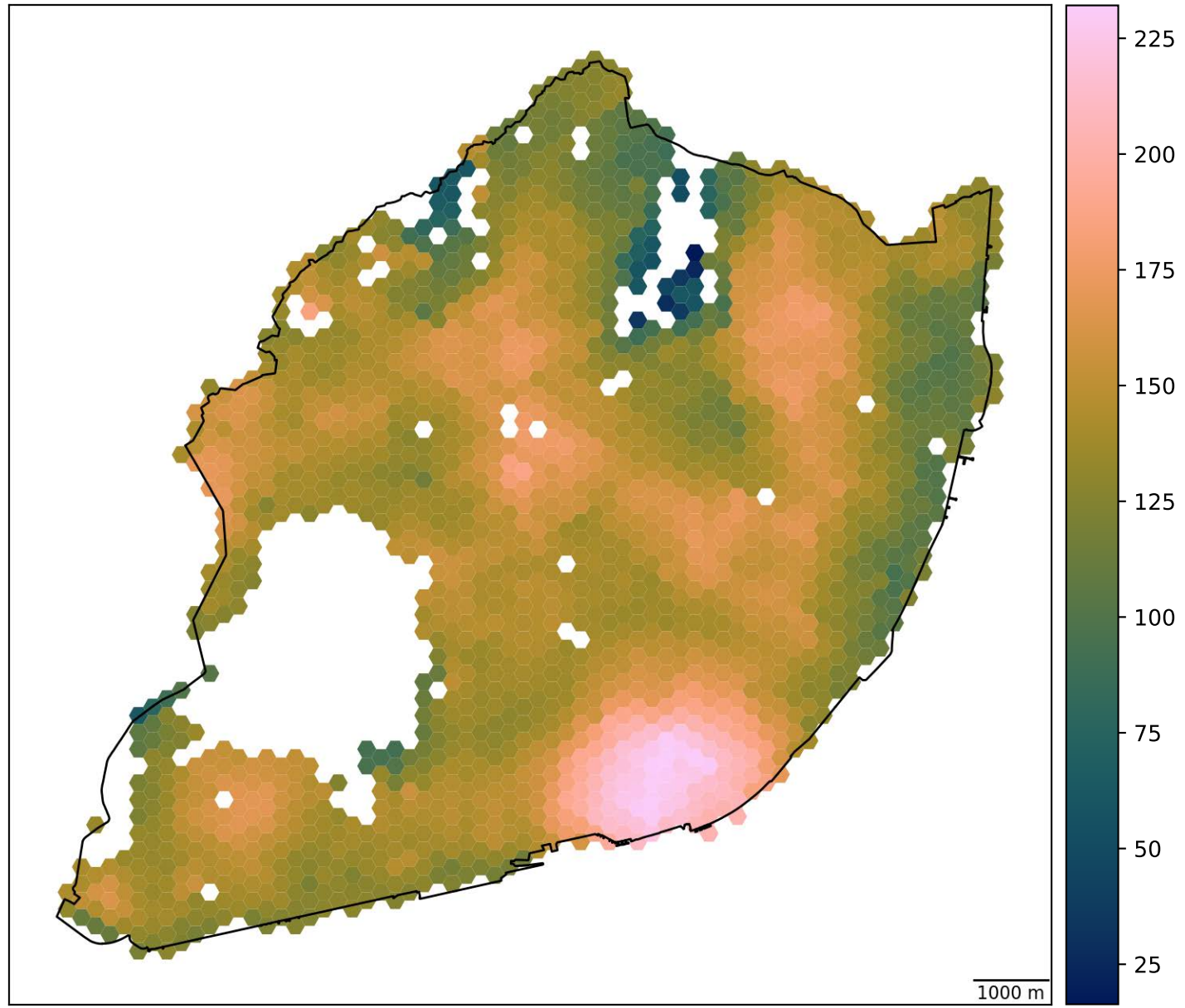

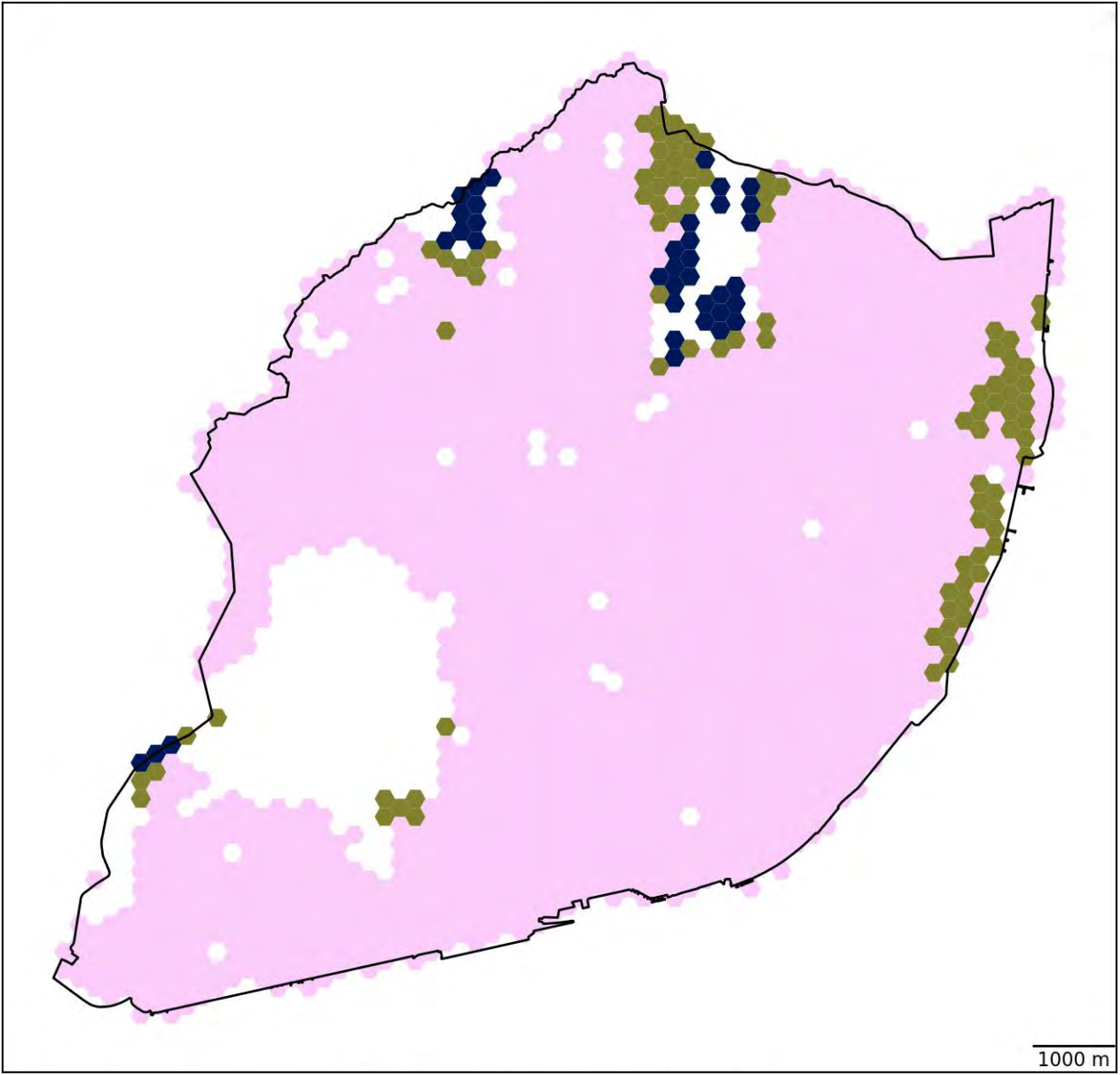

A: Estimated Mean 1000 m neighbourhood street intersections per km² requirement for ≥80% probability of engaging in walking for transport

below 95% CrI lower bound
within 95% CrI (90, 110)
exceeds 95% CrI upper bound

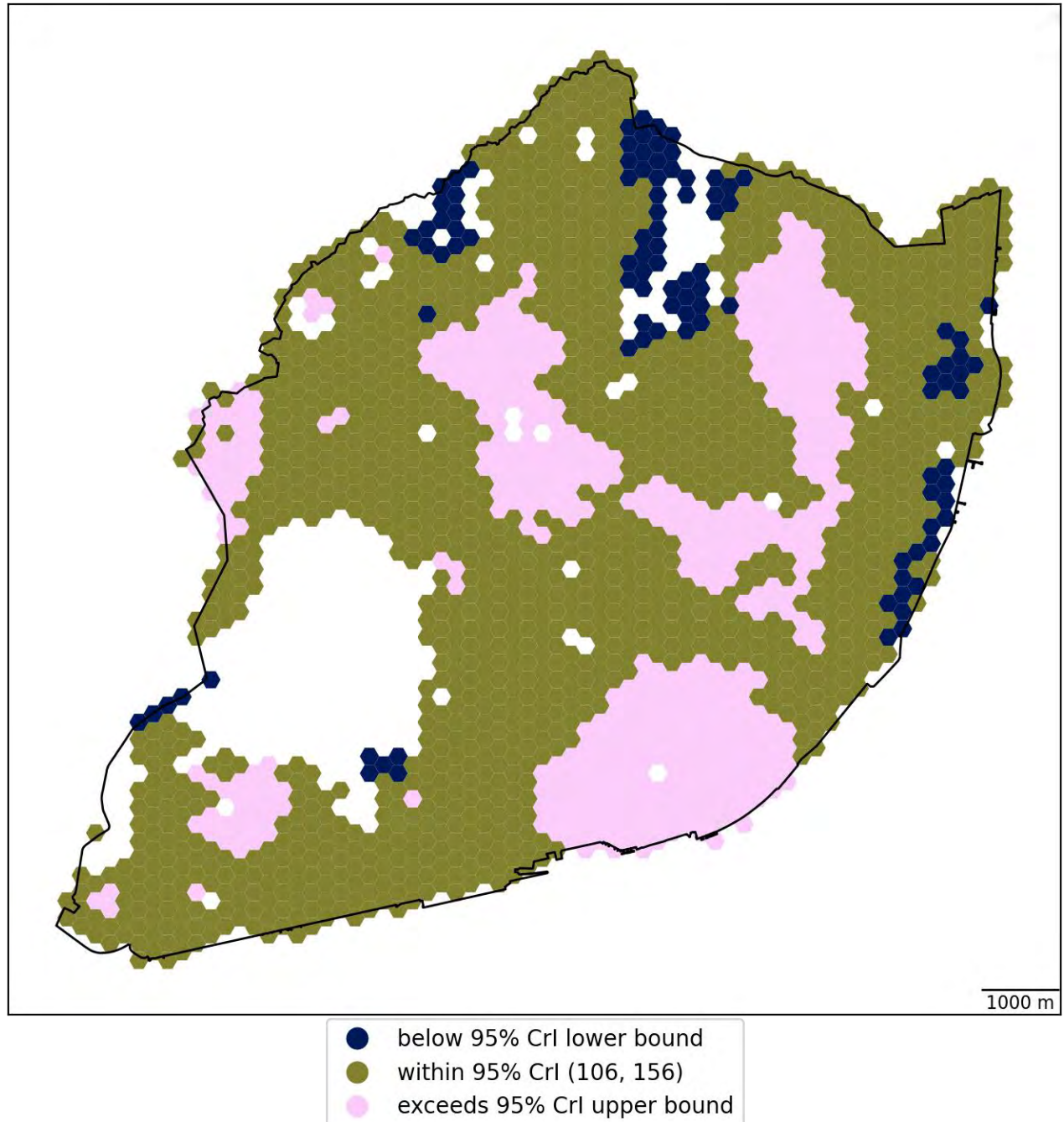

B: Estimated Mean 1000 m neighbourhood street intersections per km² requirement for reaching the WHO's target of a ≥15% relative reduction in insufficient physical activity through walking

below 95% CrI lower bound
within 95% CrI (106, 156)
exceeds 95% CrI upper bound

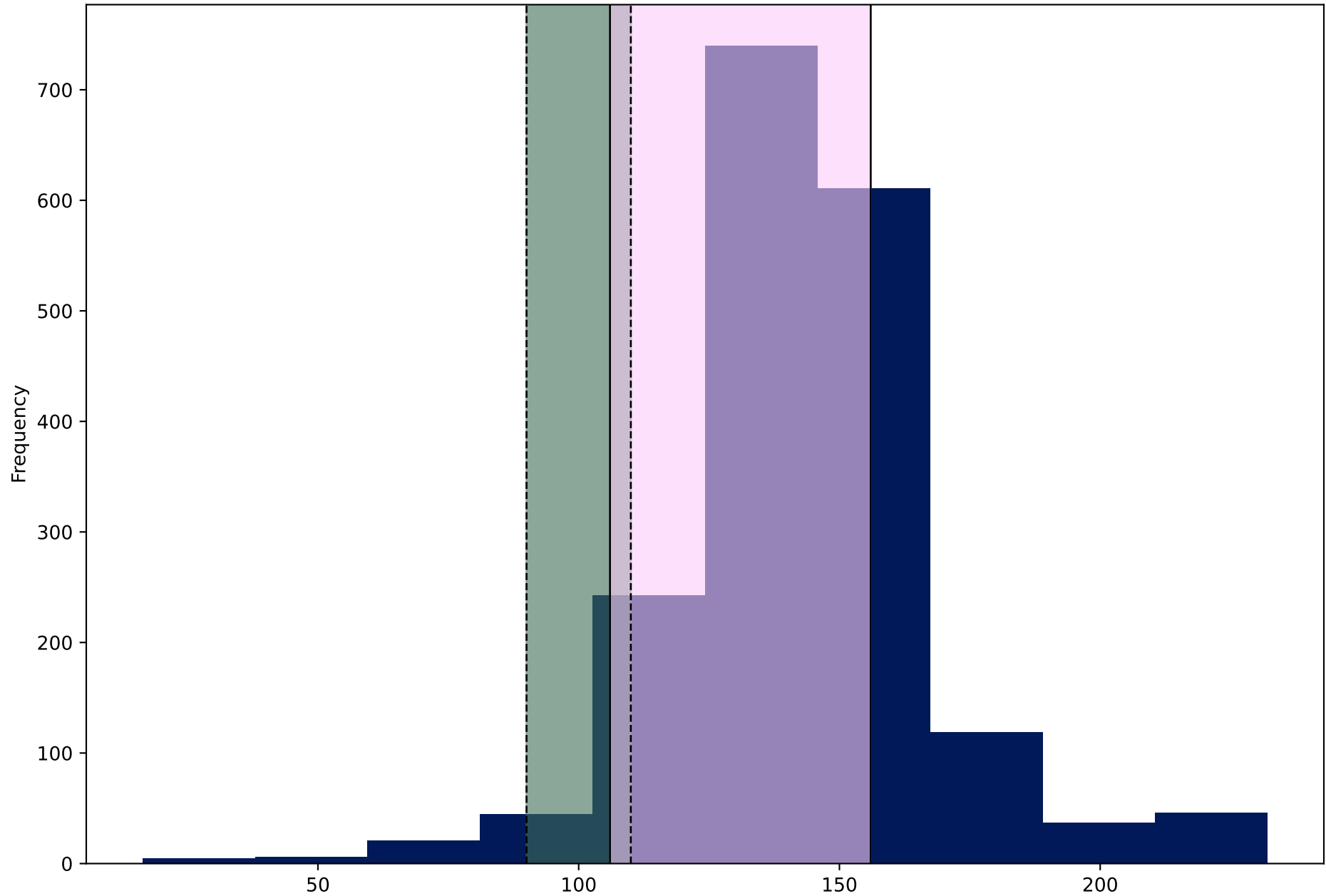

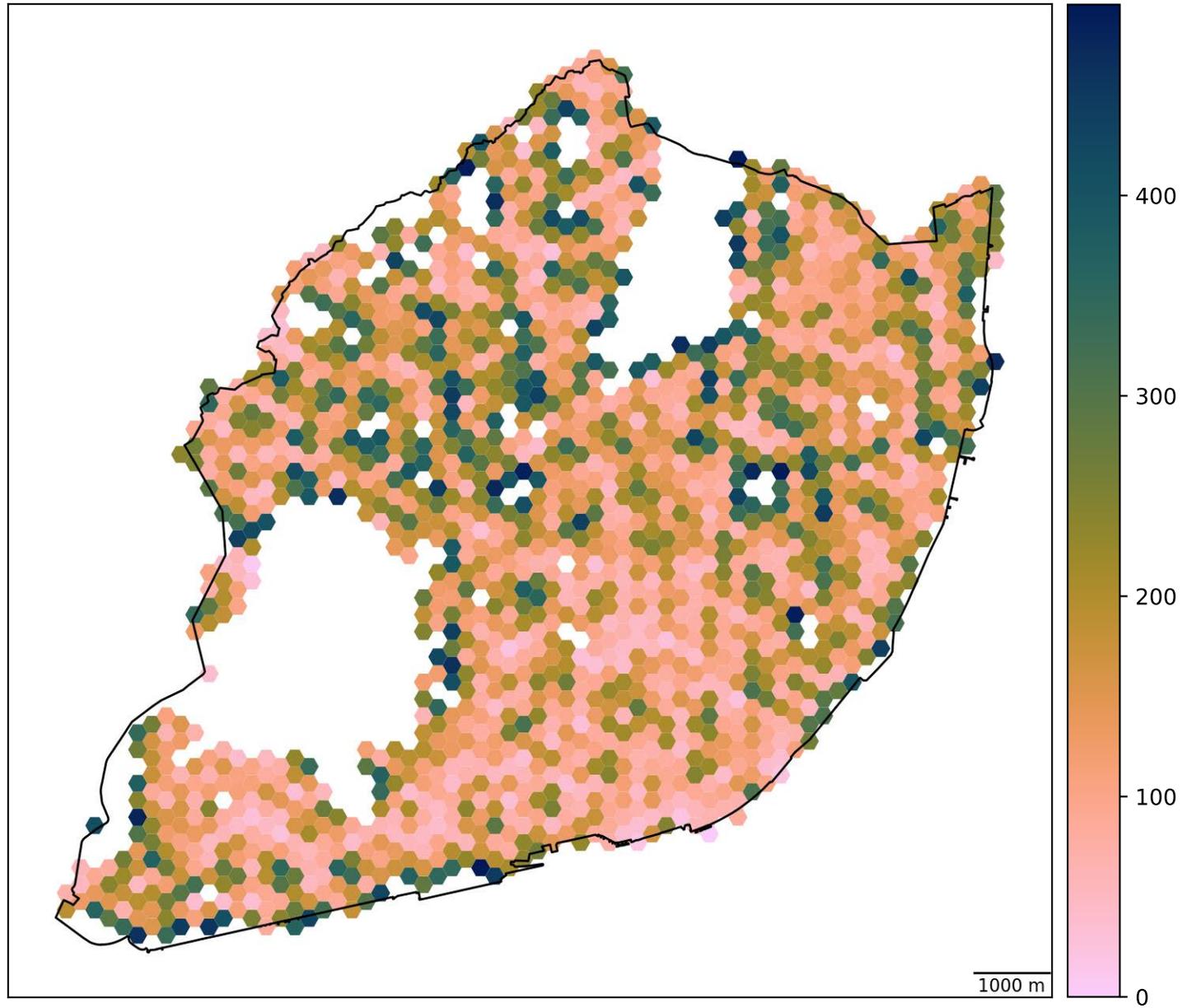
327

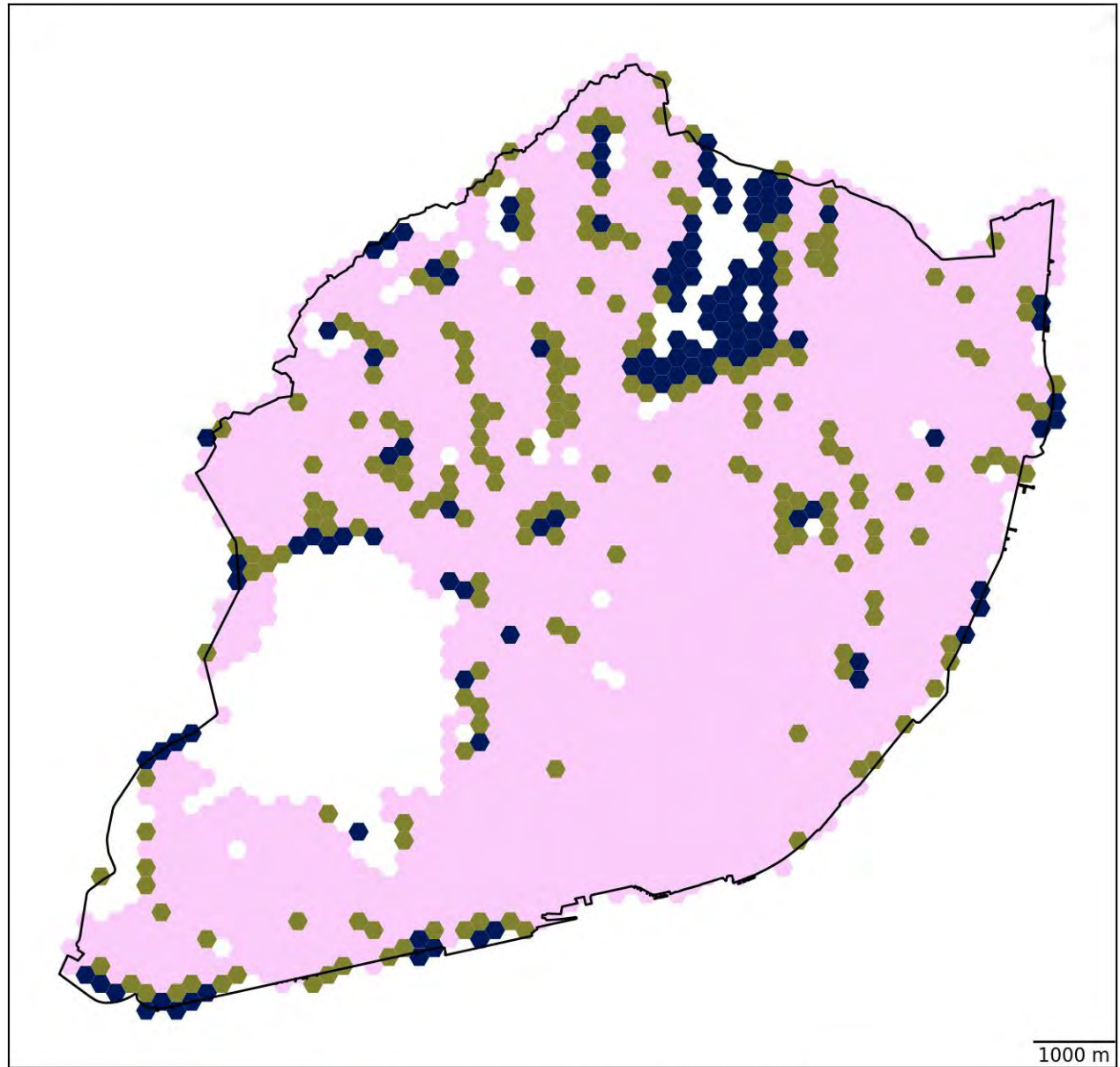

distances: Estimated Distance to nearest public transport stops (m; up to 500m) requirement for distances to destinations, measured up to a maximum distance target threshold of 500 metres

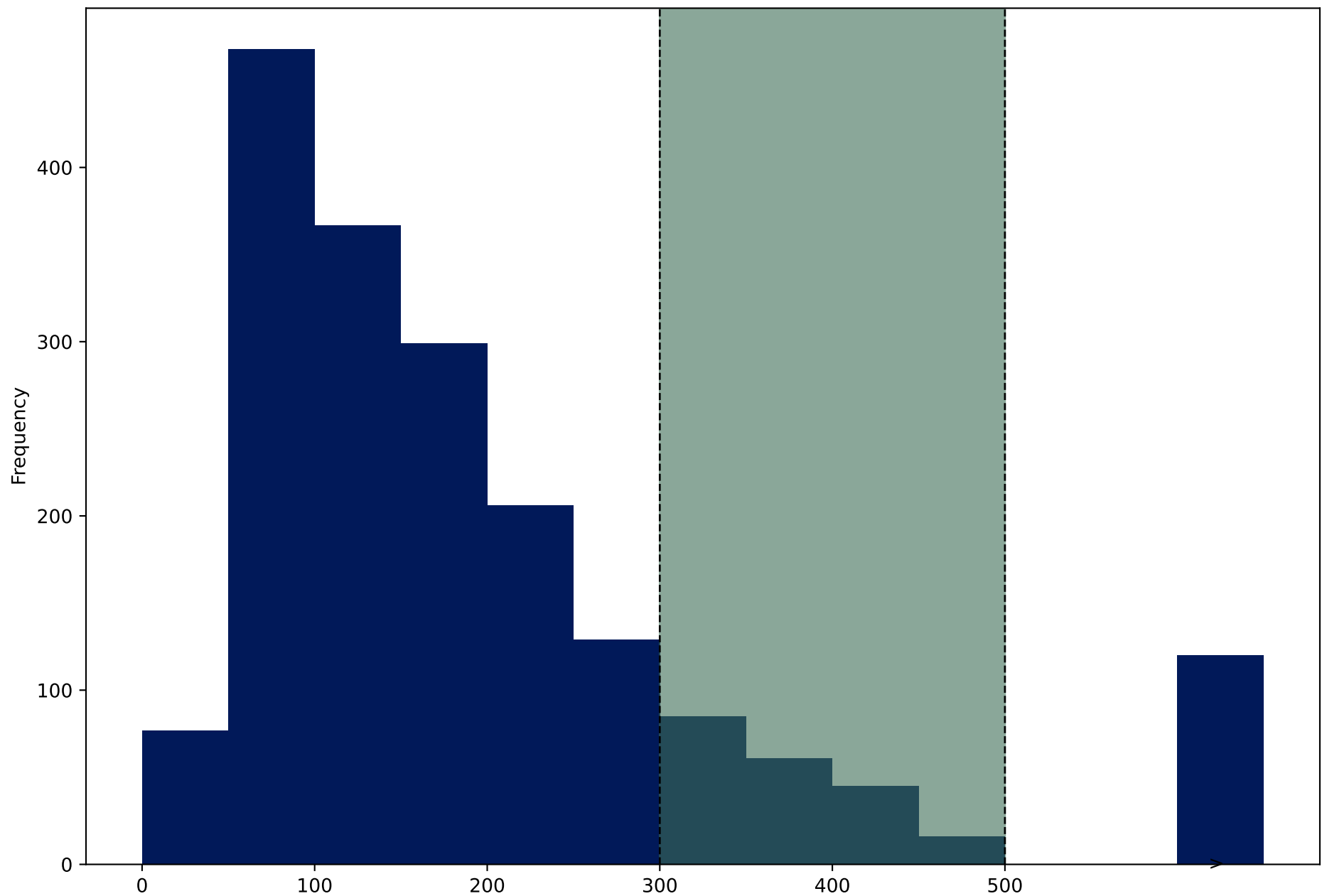

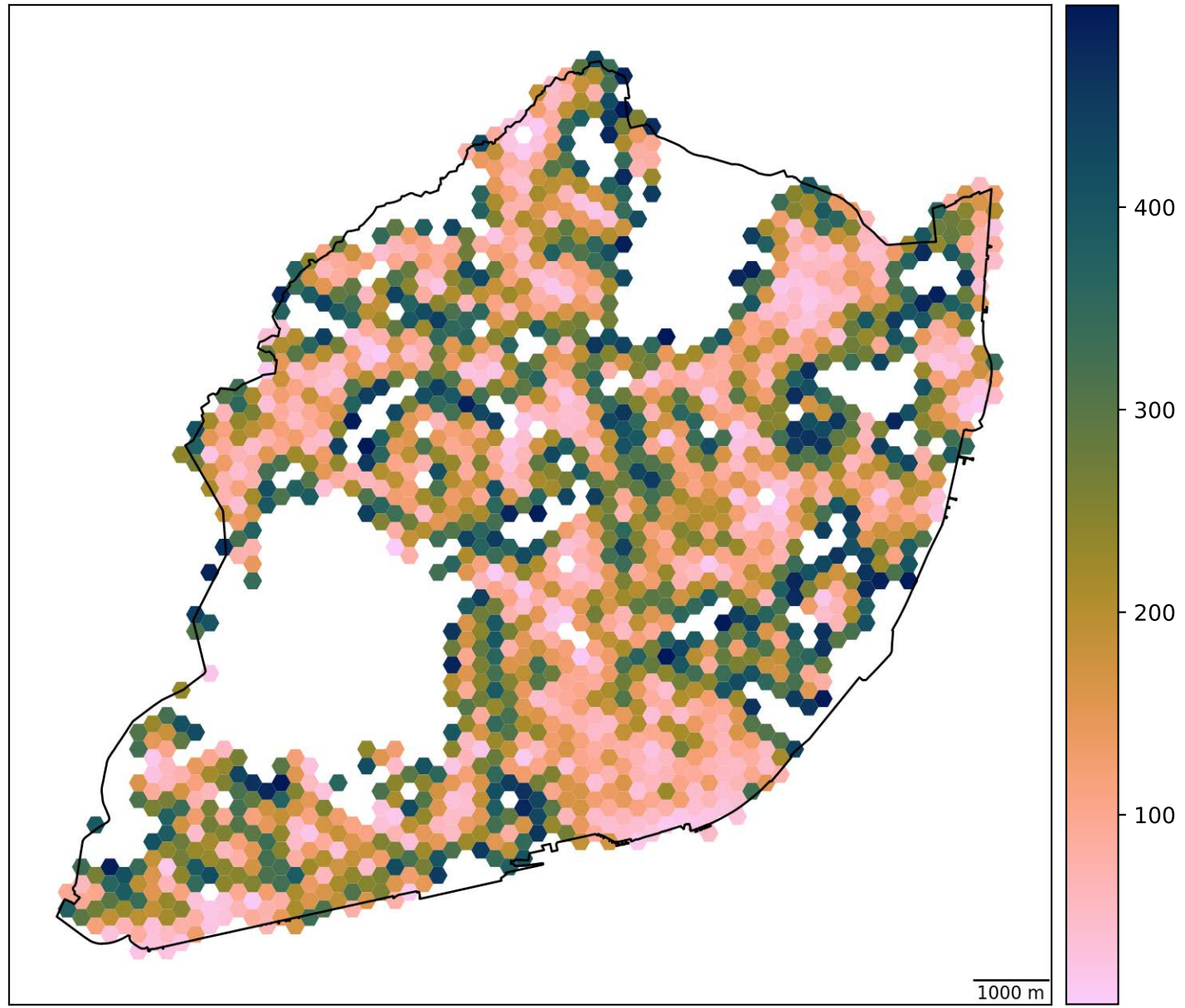

distances: Estimated Distance to nearest park (m; up to 500m) requirement for distances to destinations, measured up to a maximum distance target threshold of 500 metres

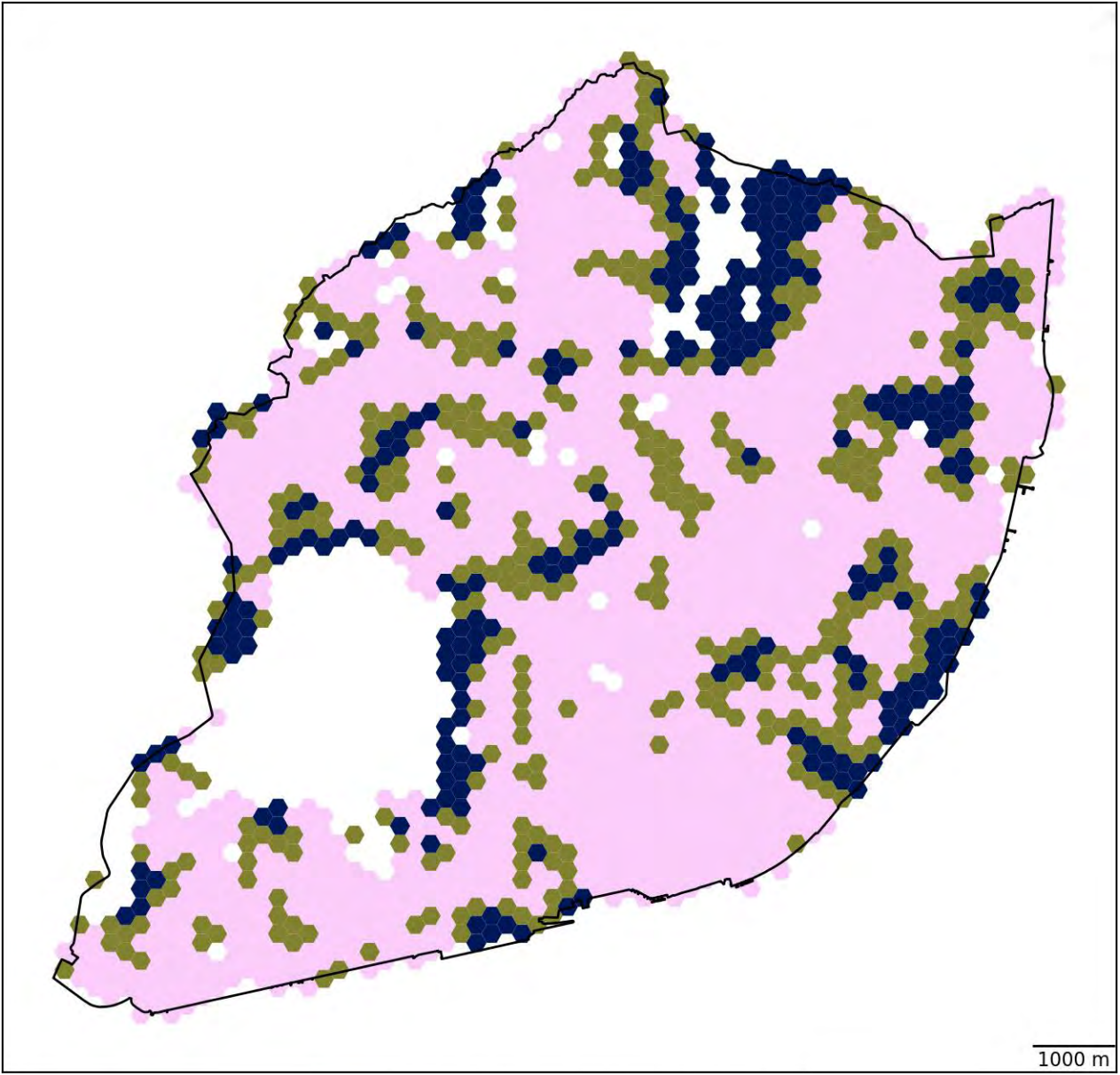

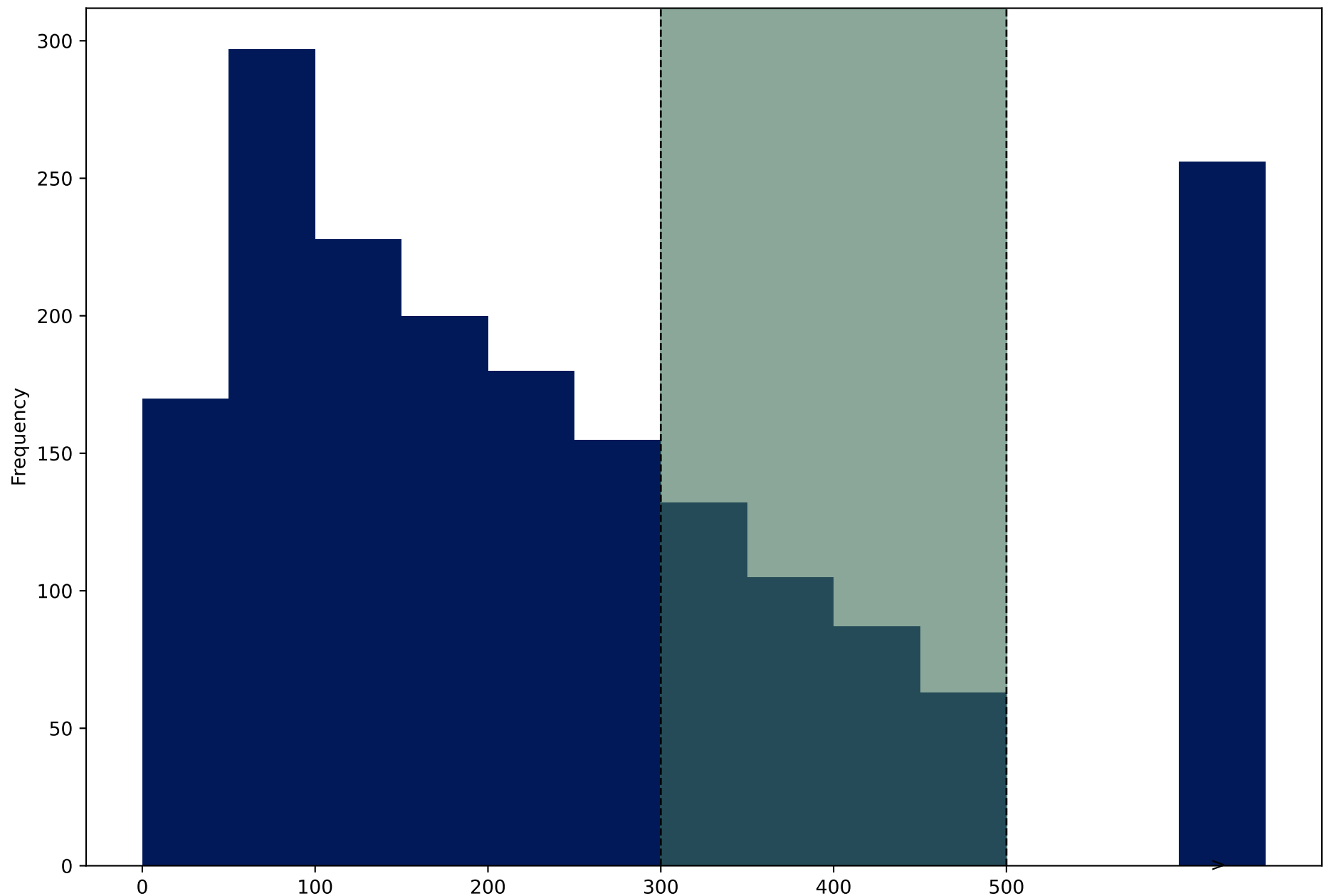

# Europe, Spain, Barcelona

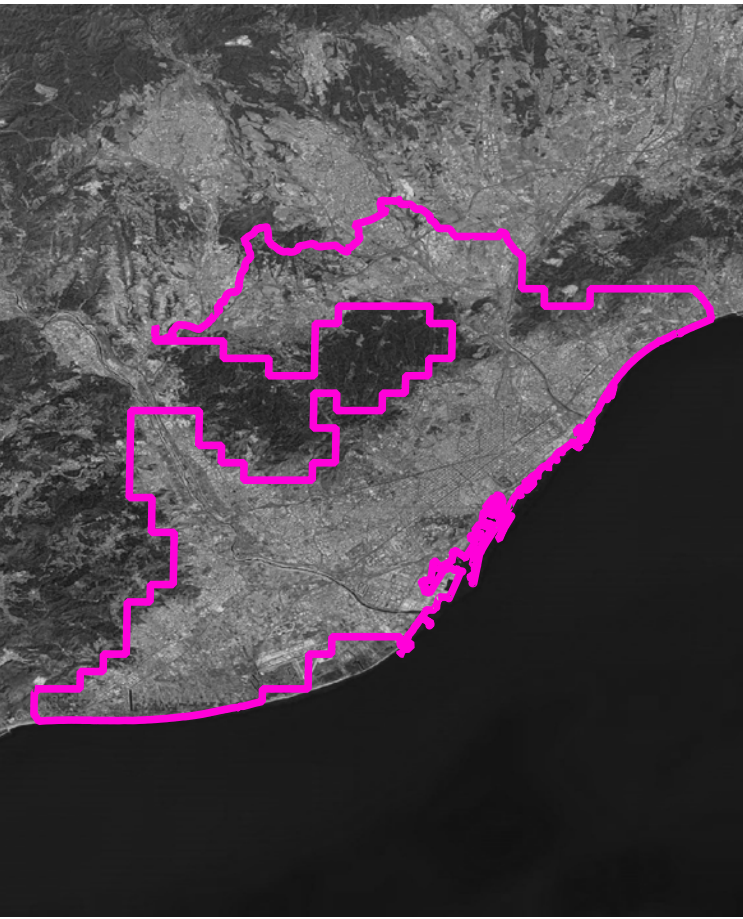
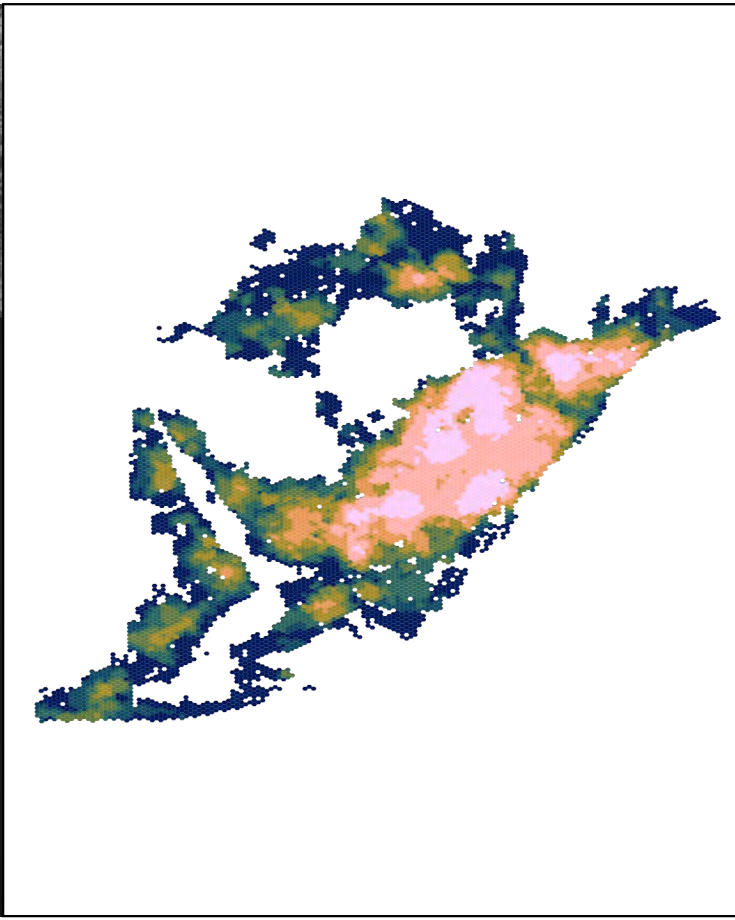
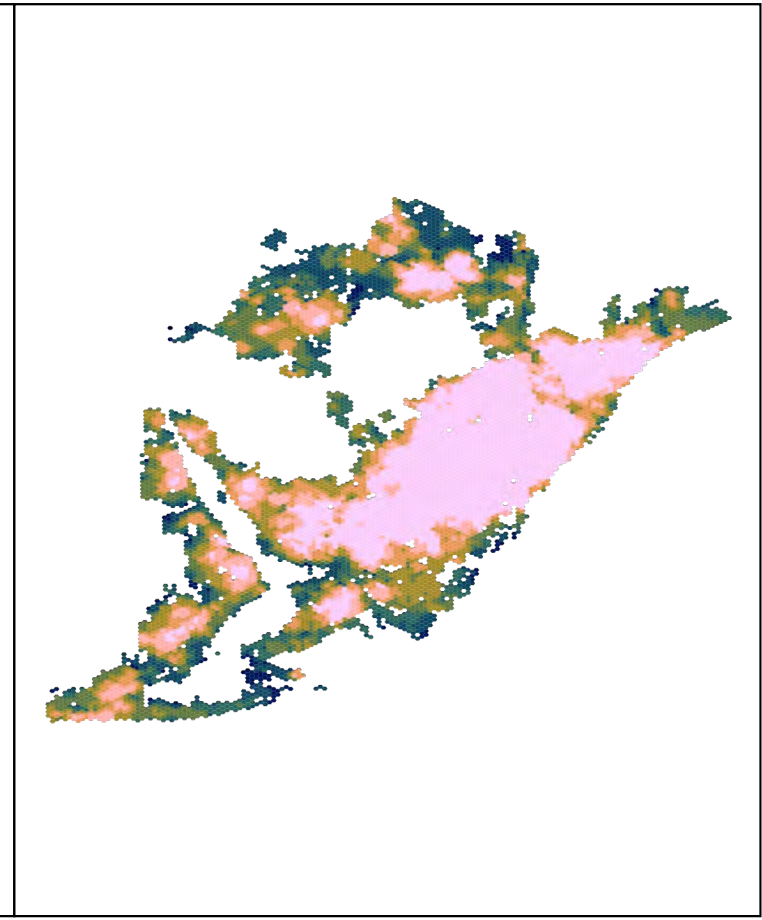
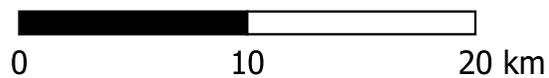
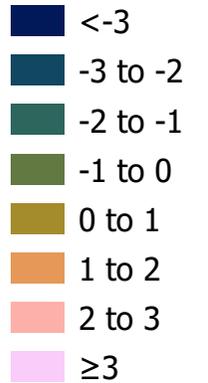
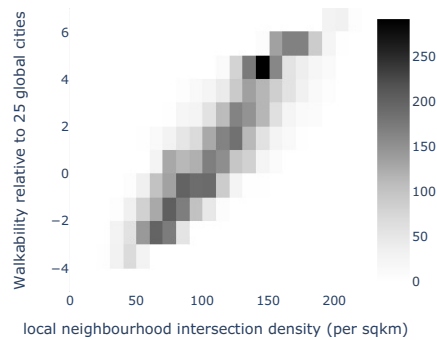
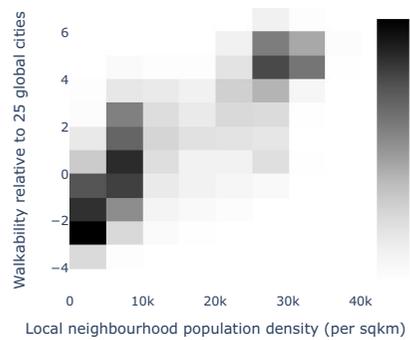
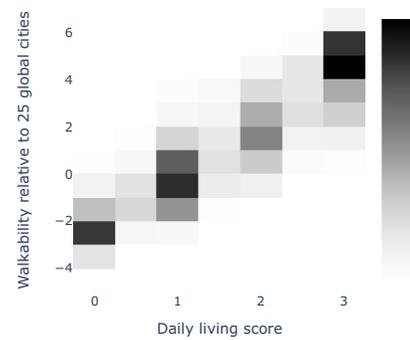
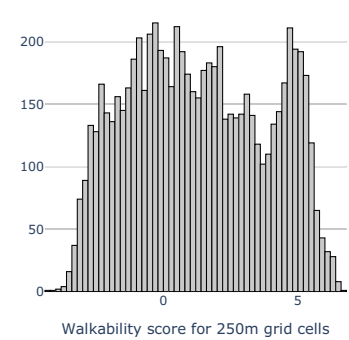

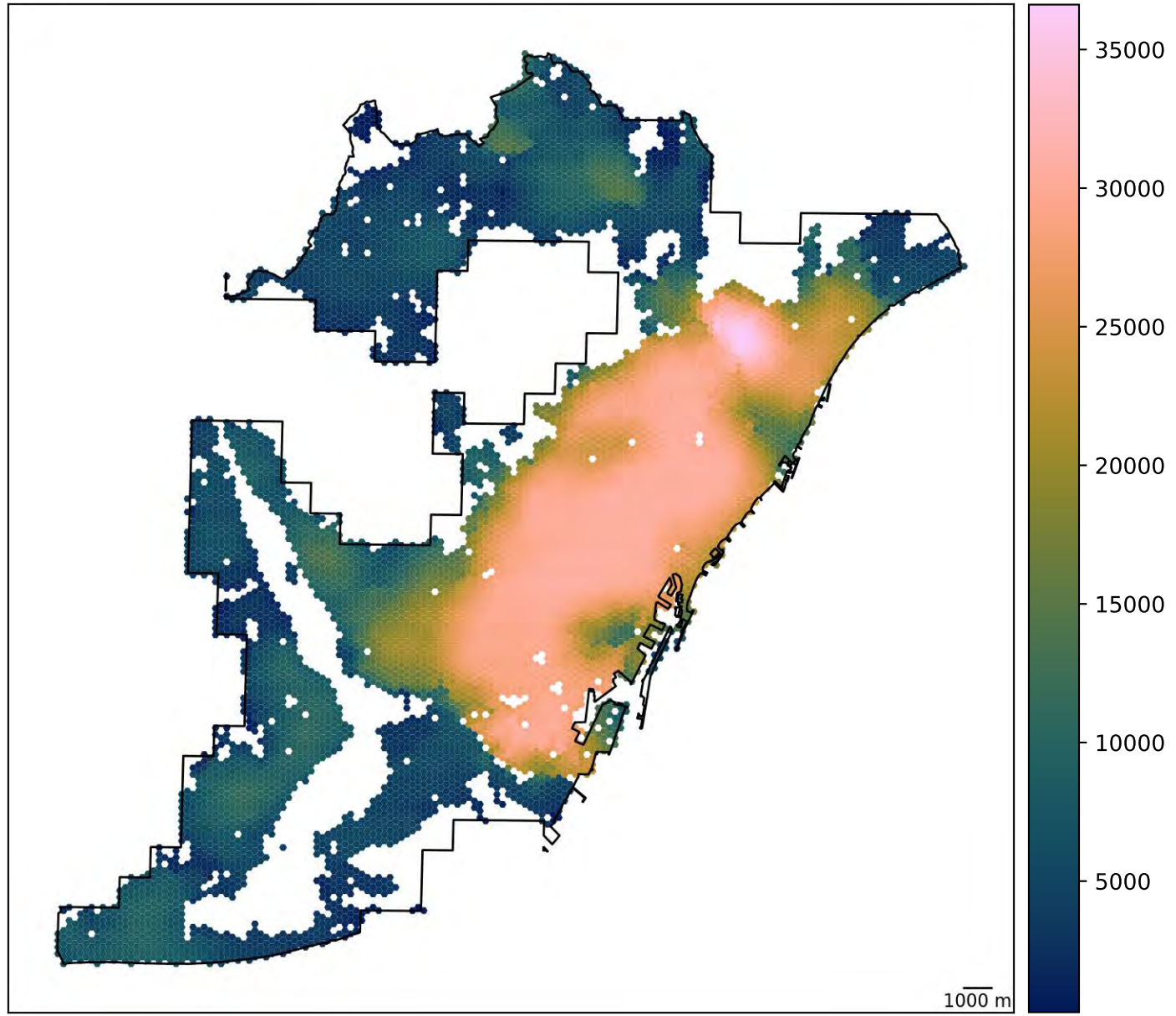

A: Estimated Mean 1000 m neighbourhood population per km² requirement for ≥80% probability of engaging in walking for transport

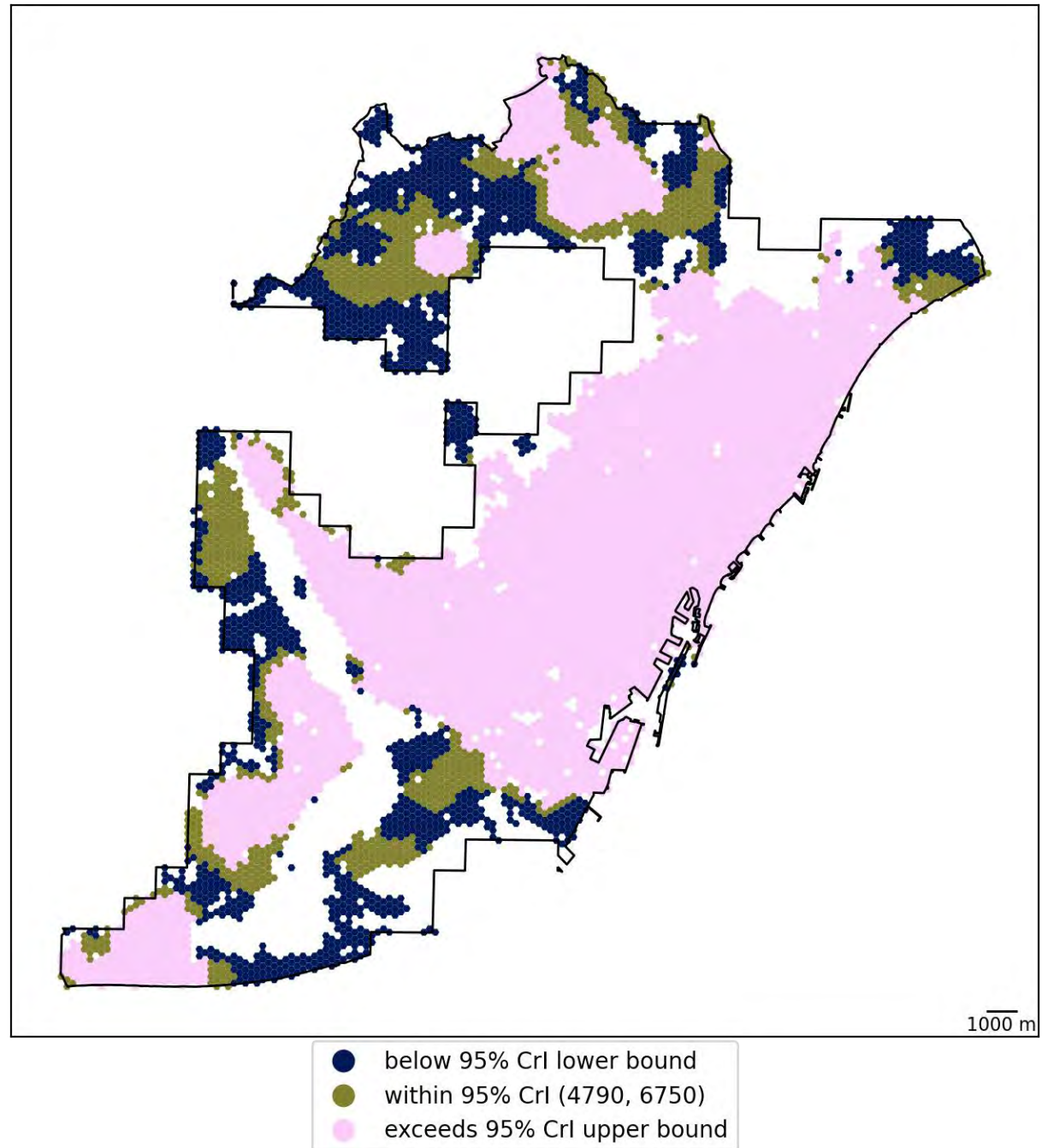

B: Estimated Mean 1000 m neighbourhood population per km² requirement for reaching the WHO's target of a ≥15% relative reduction in insufficient physical activity through walking

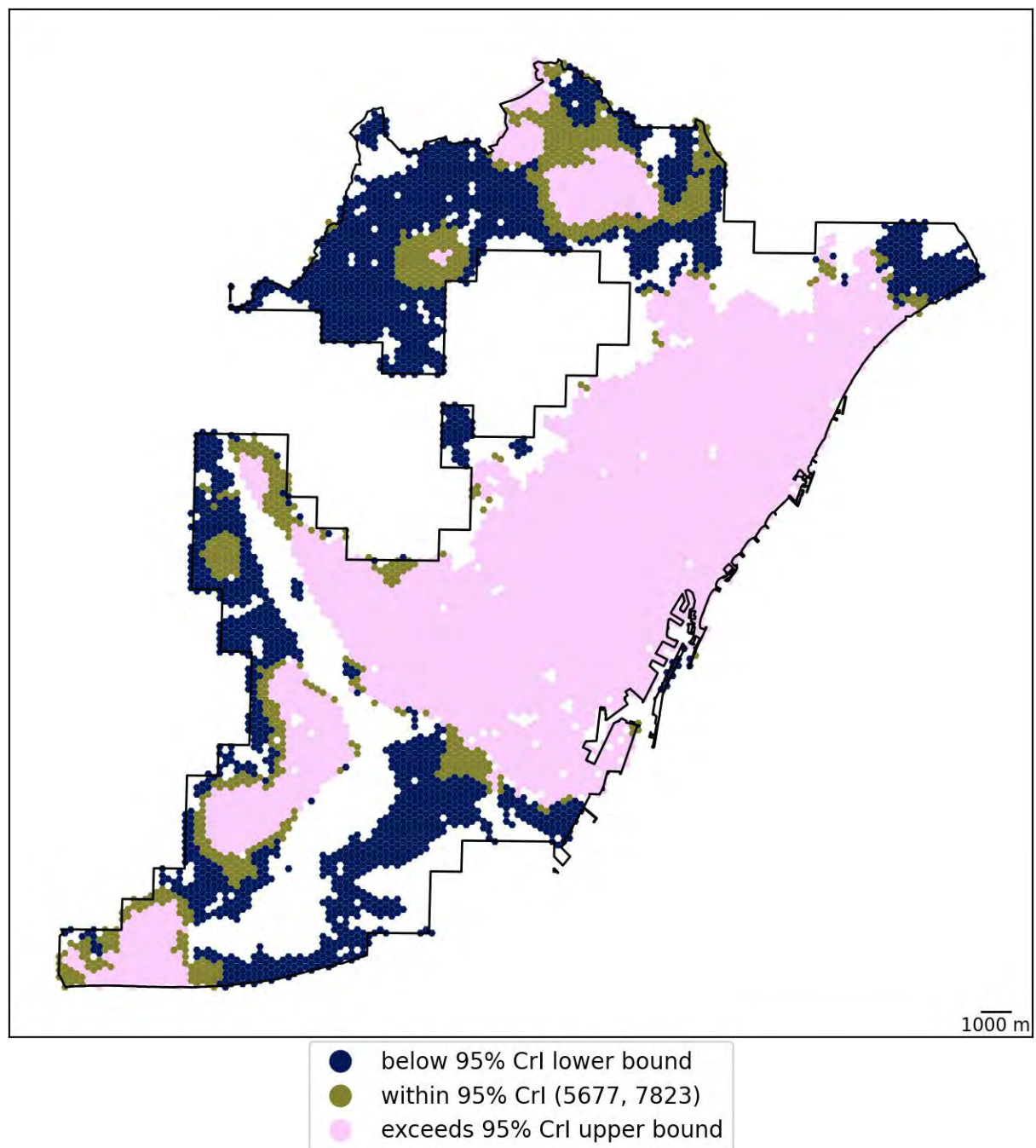

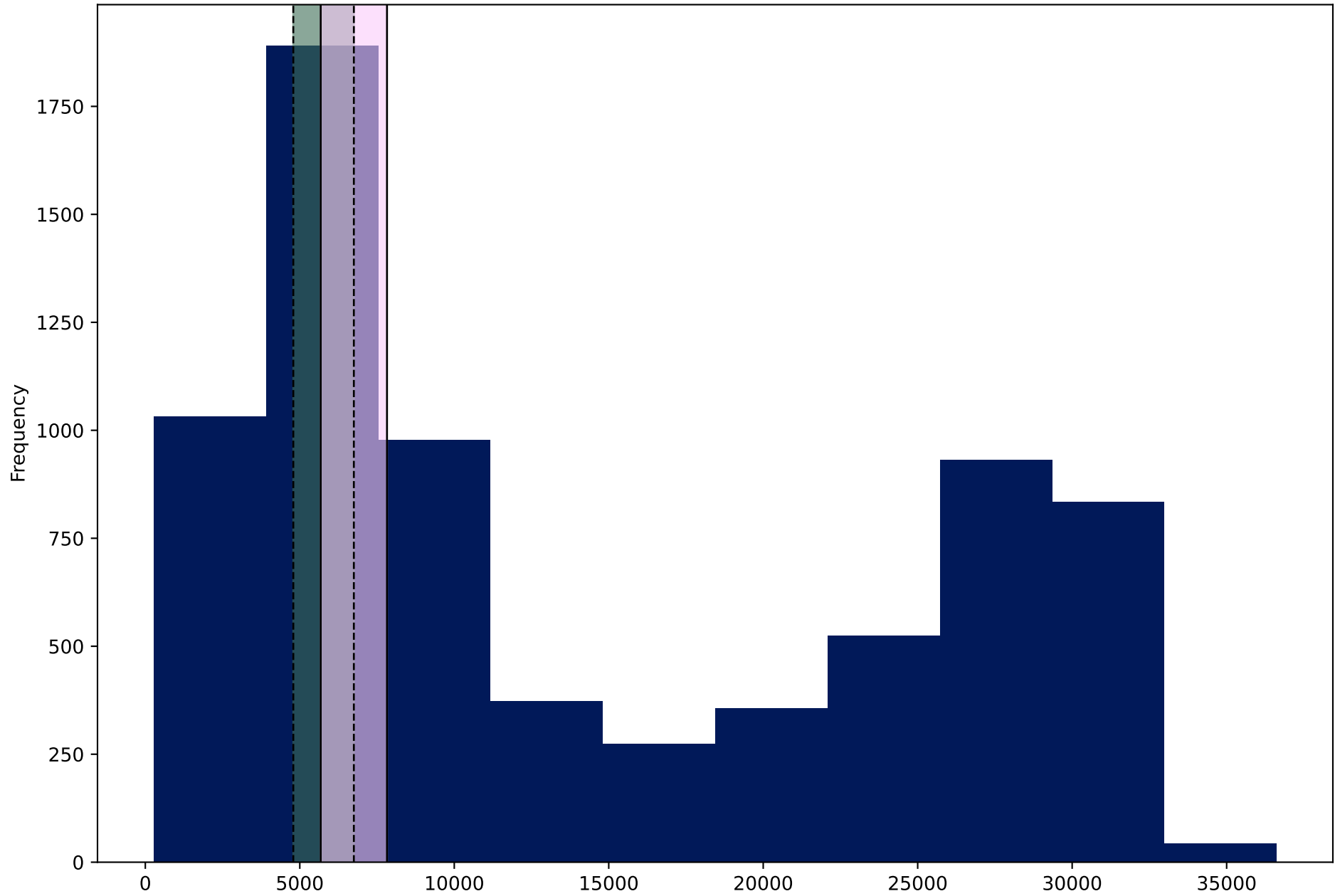

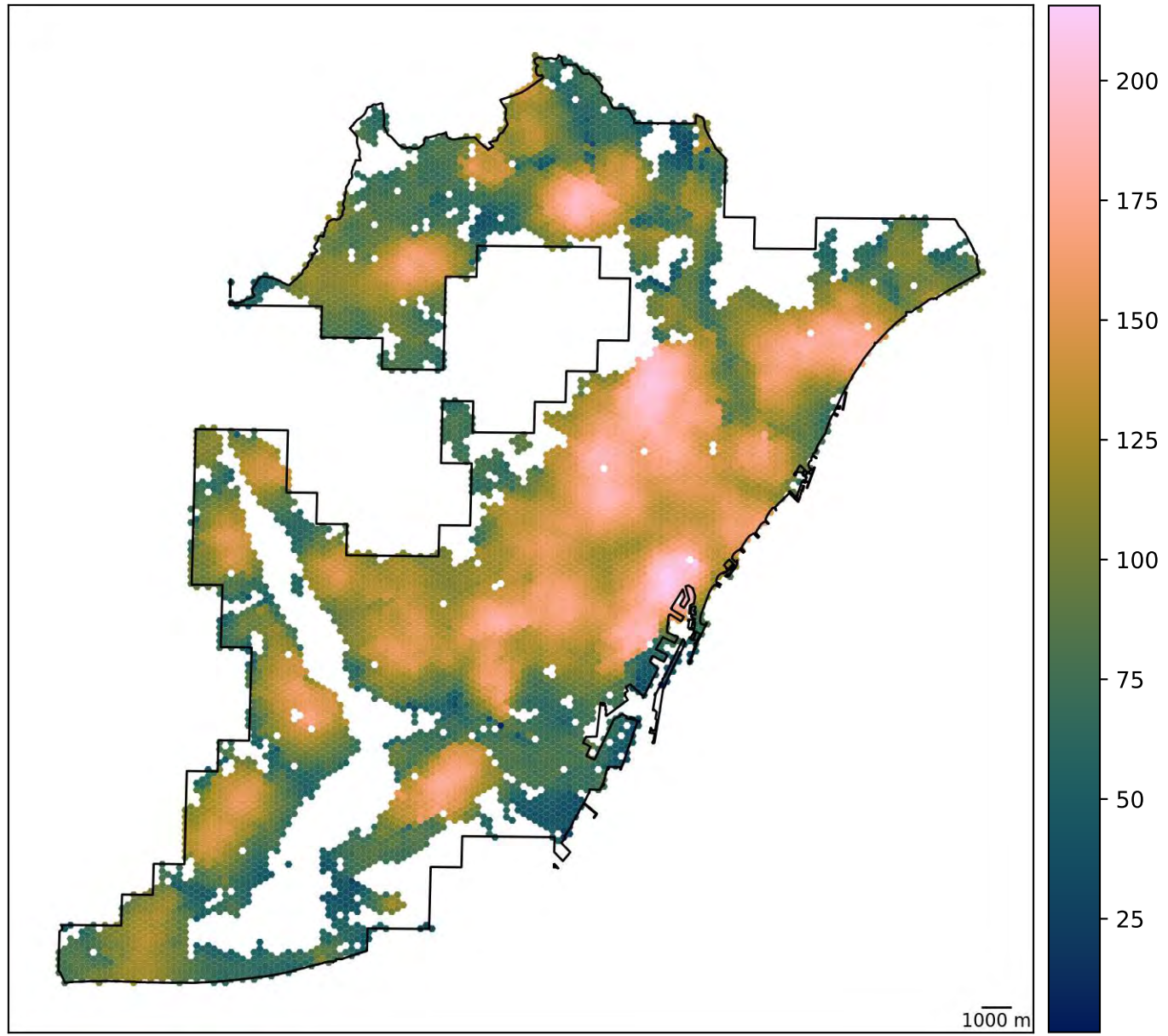

Mean 1000 m neighbourhood street intersections per km²

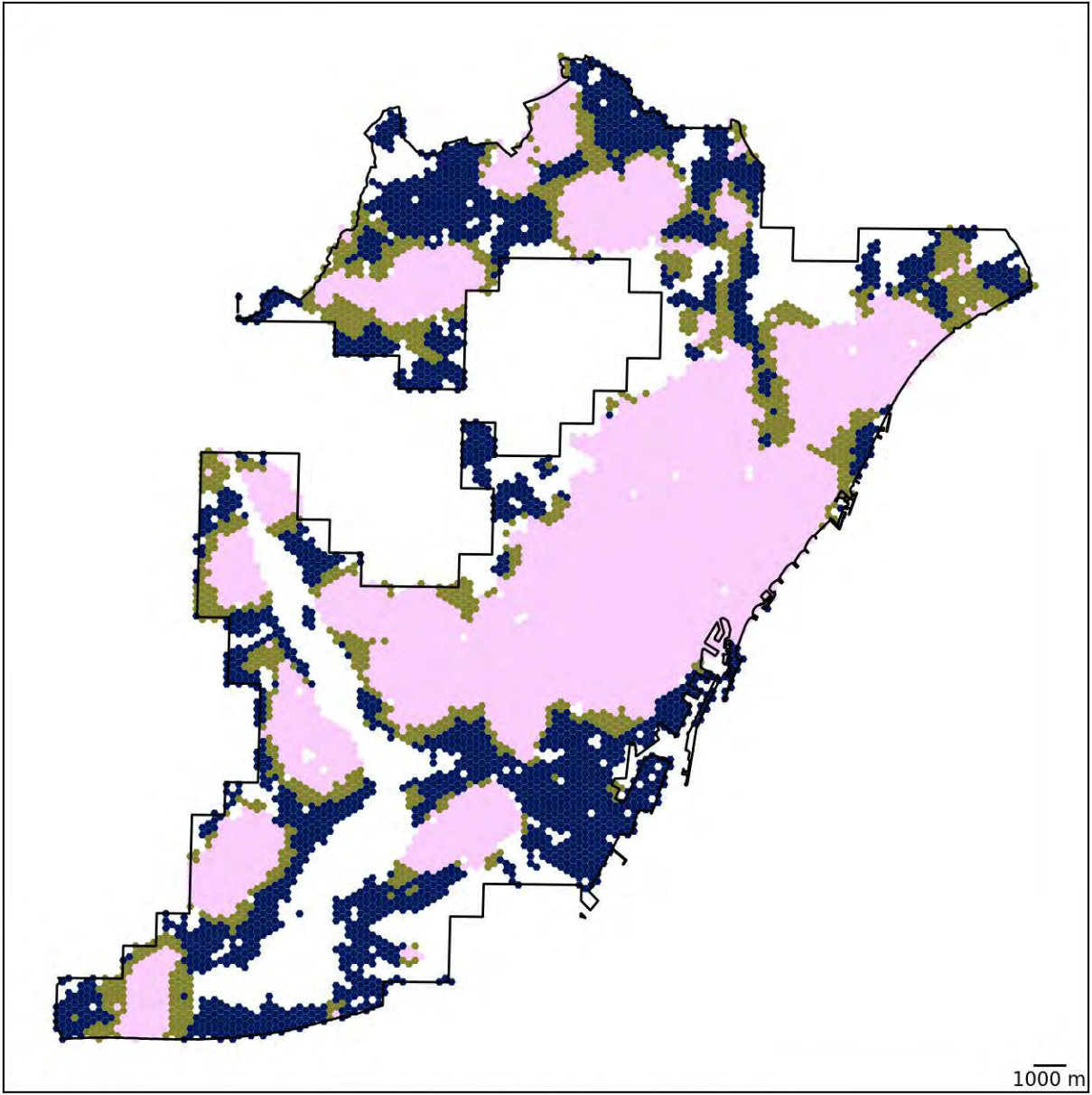

A: Estimated Mean 1000 m neighbourhood street intersections per km² requirement for ≥80% probability of engaging in walking for transport

B: Estimated Mean 1000 m neighbourhood street intersections per km² requirement for reaching the WHO's target of a ≥15% relative reduction in insufficient physical activity through walking

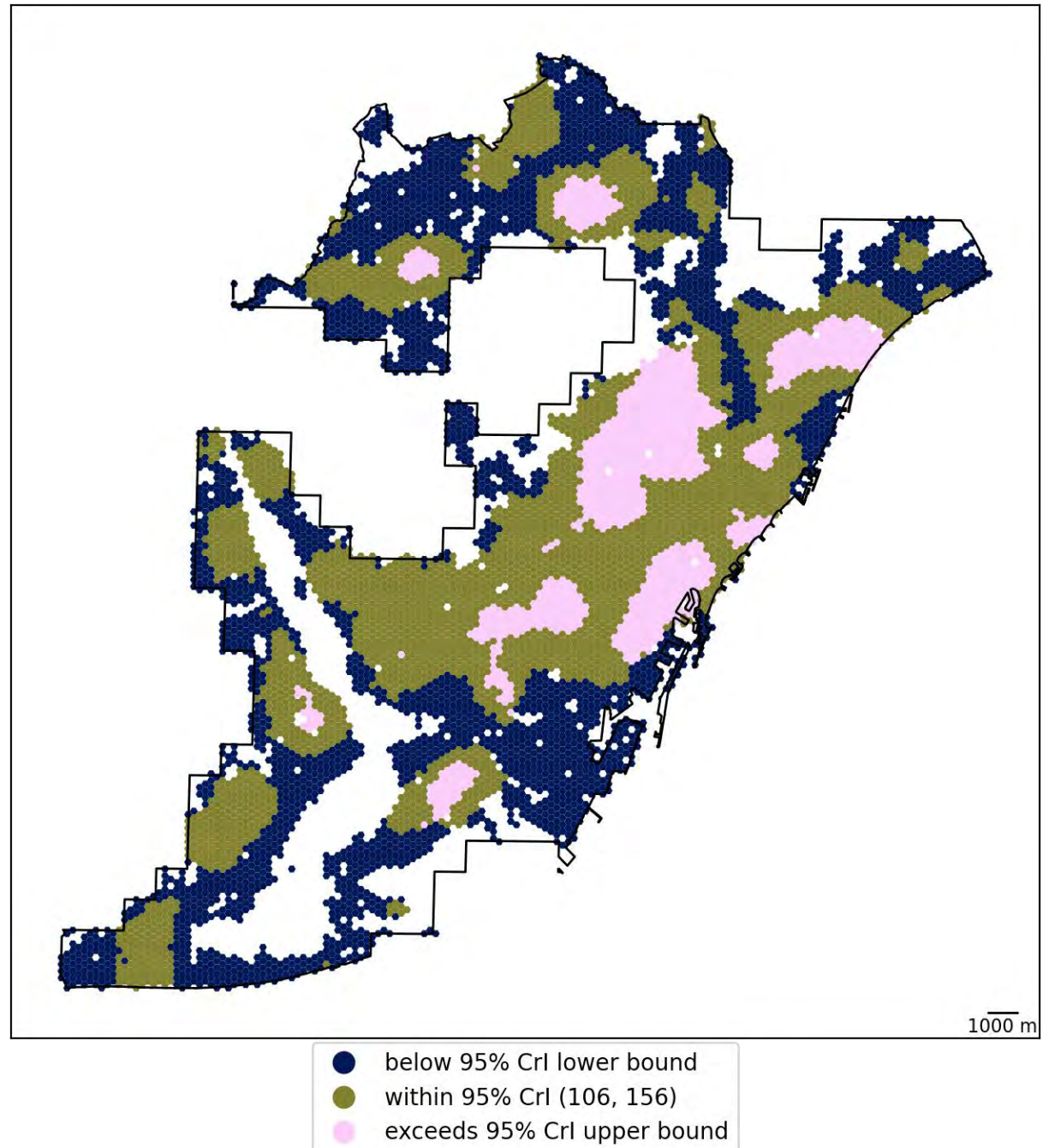

- below 95% CrI lower bound
- within 95% CrI (106, 156)
- exceeds 95% CrI upper bound

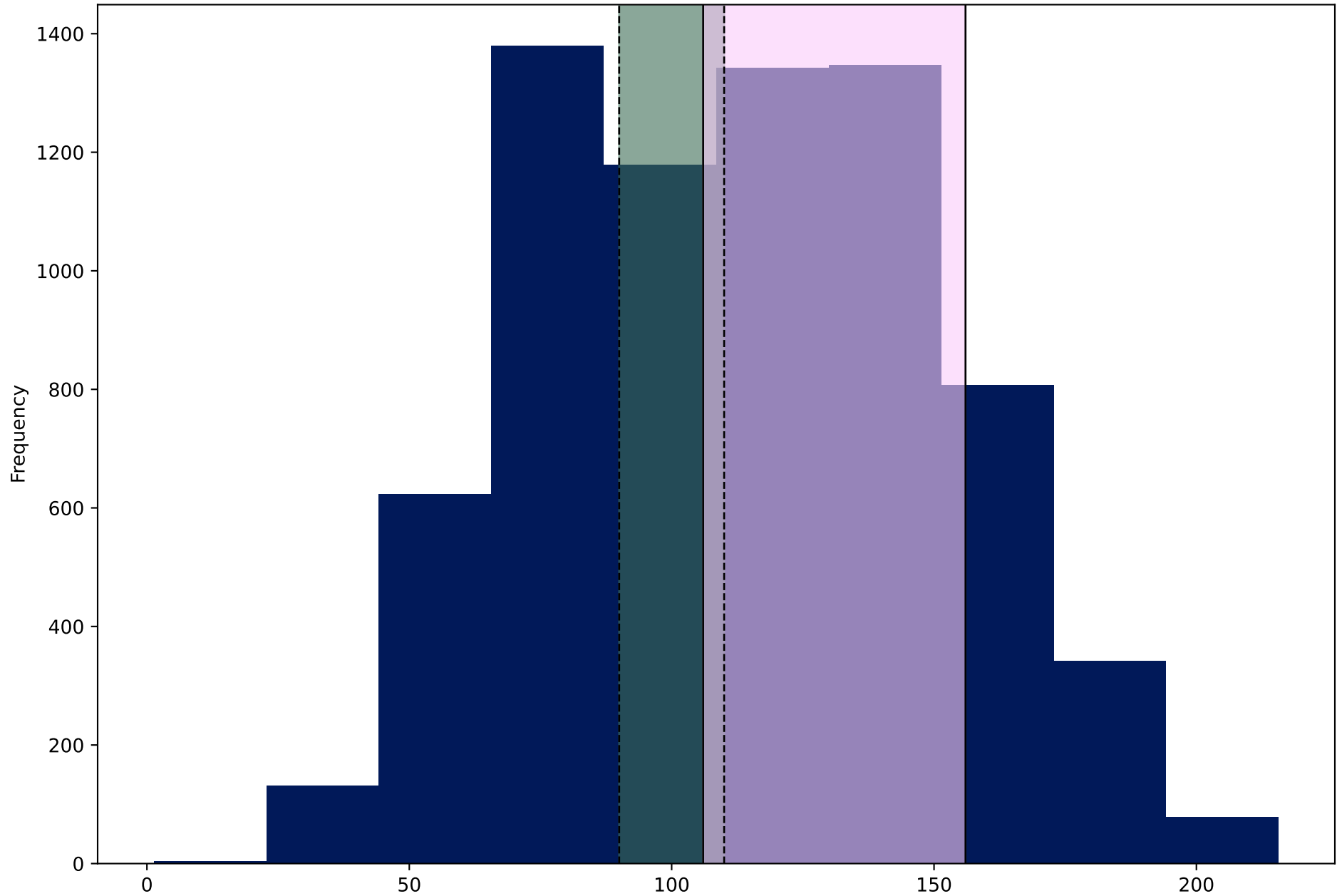

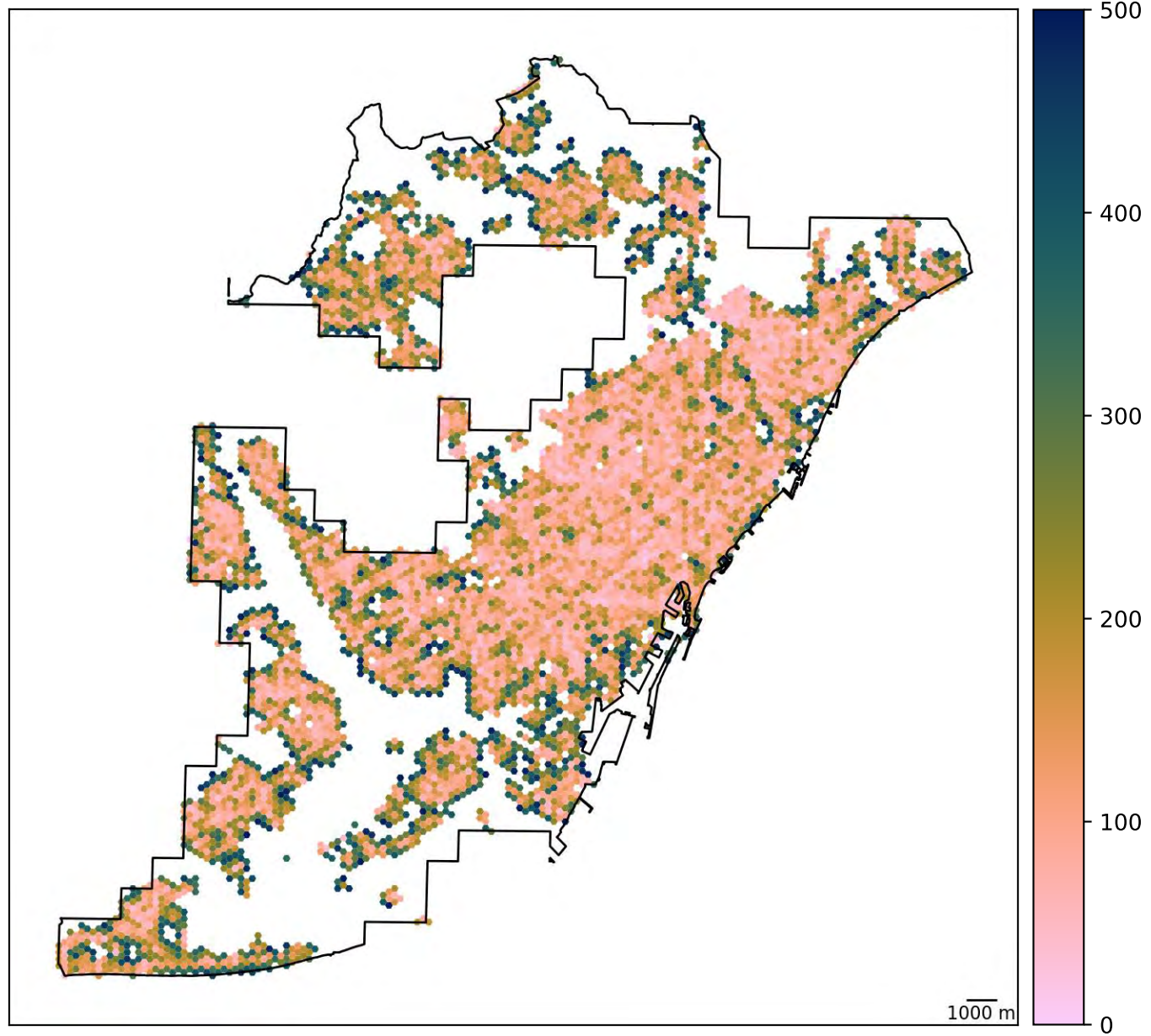

distances: Estimated Distance to nearest public transport stops (m; up to 500m) requirement for distances to destinations, measured up to a maximum distance target threshold of 500 metres

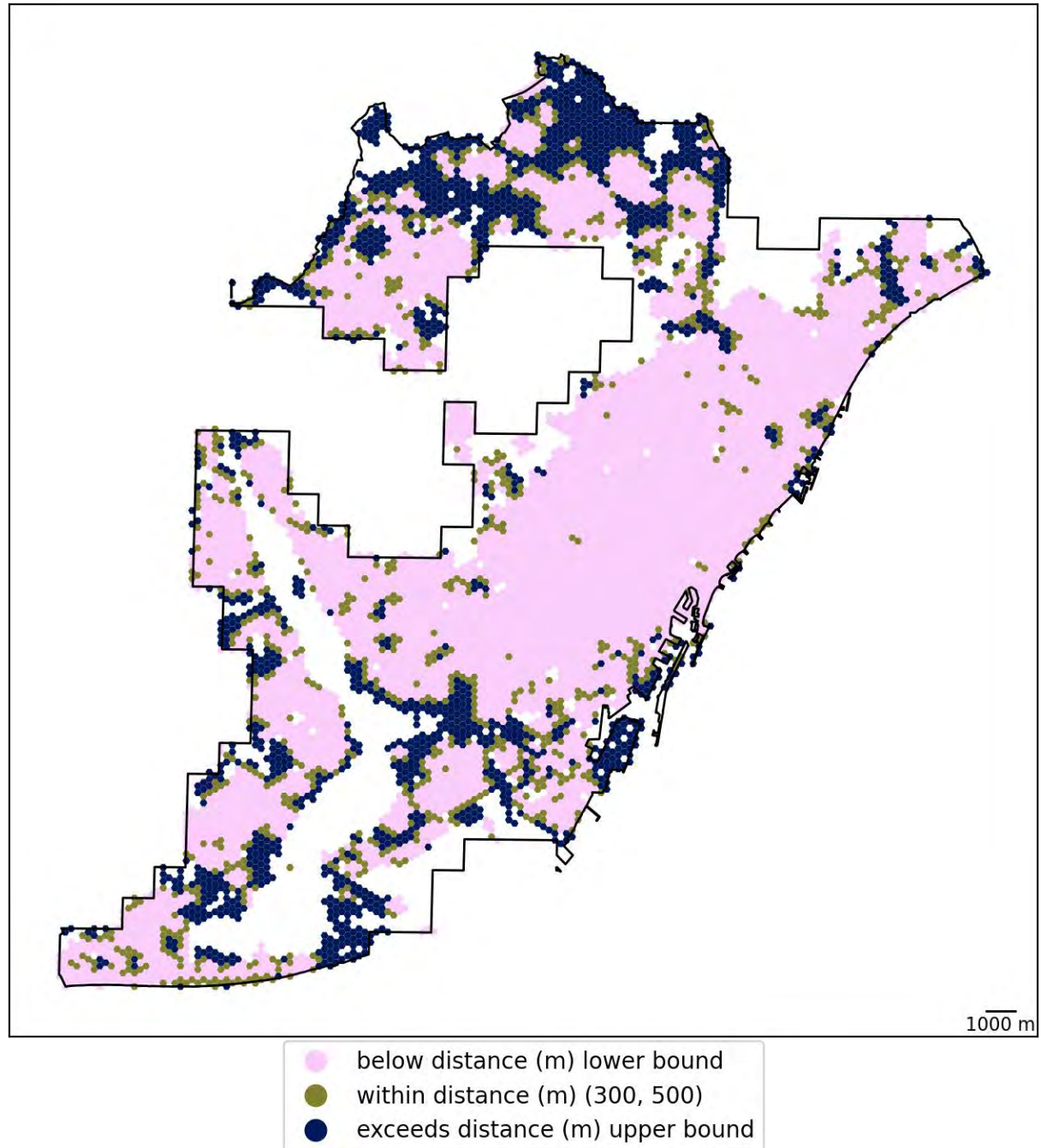

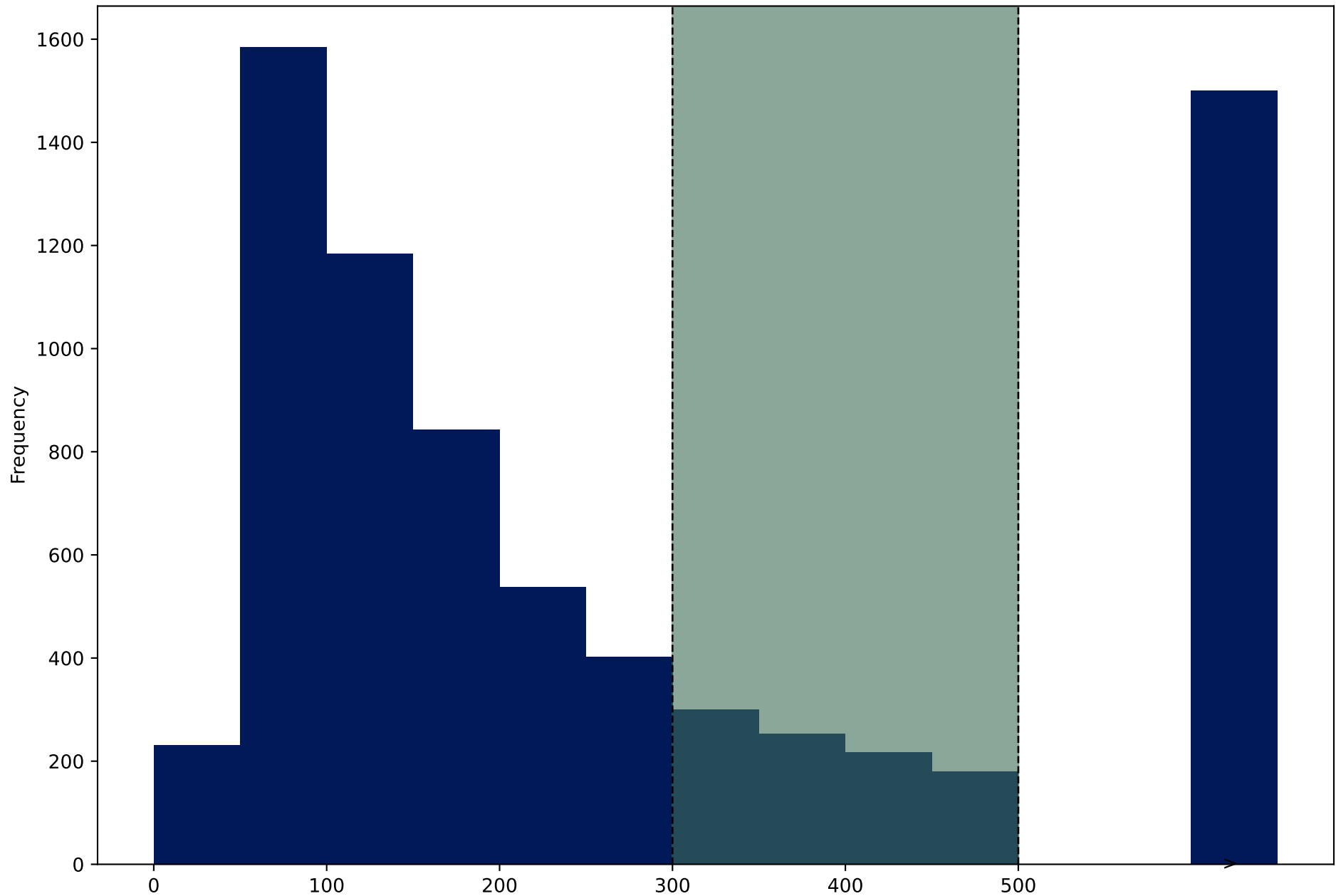

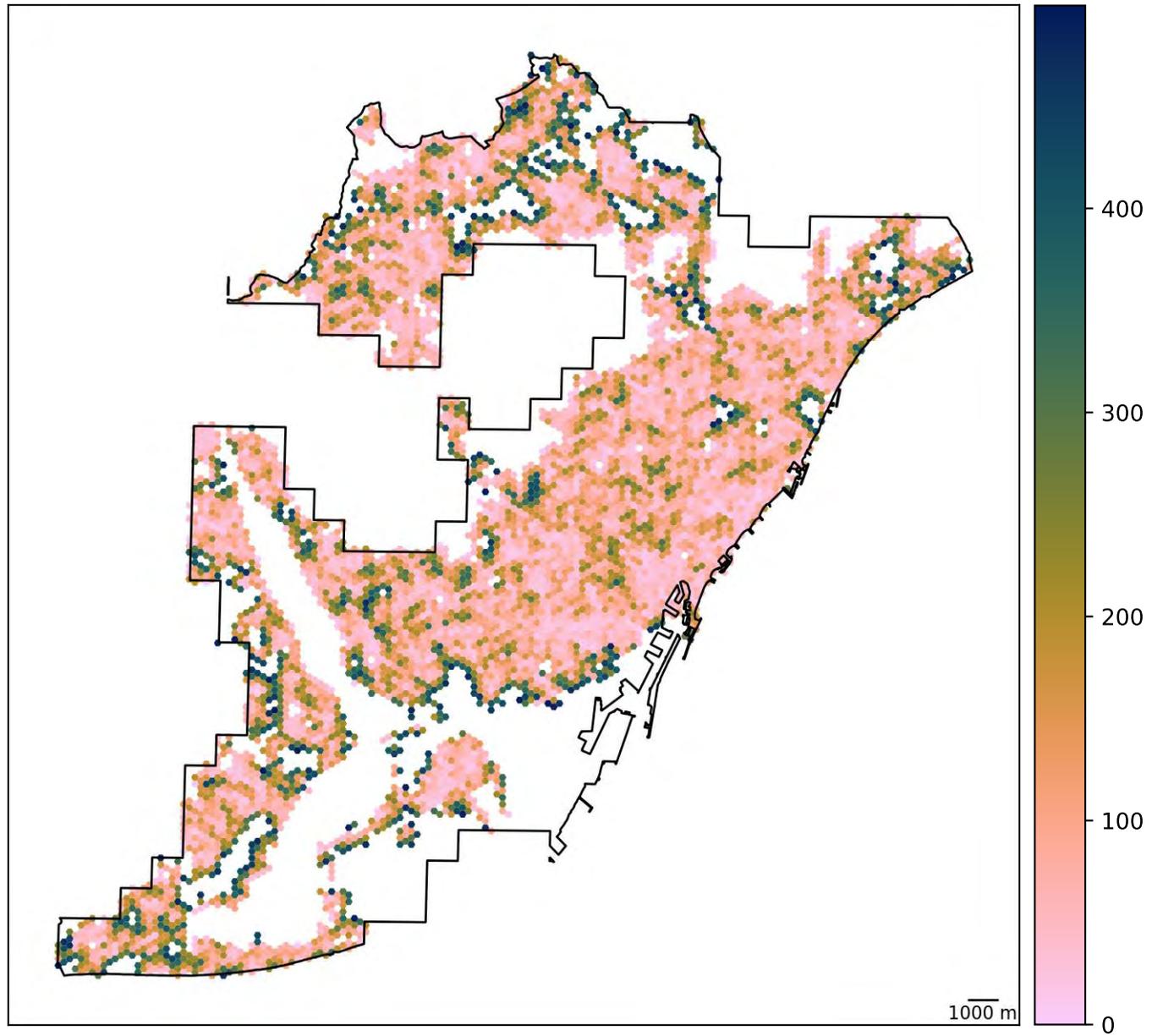

distances: Estimated Distance to nearest park (m; up to 500m) requirement for distances to destinations, measured up to a maximum distance target threshold of 500 metres

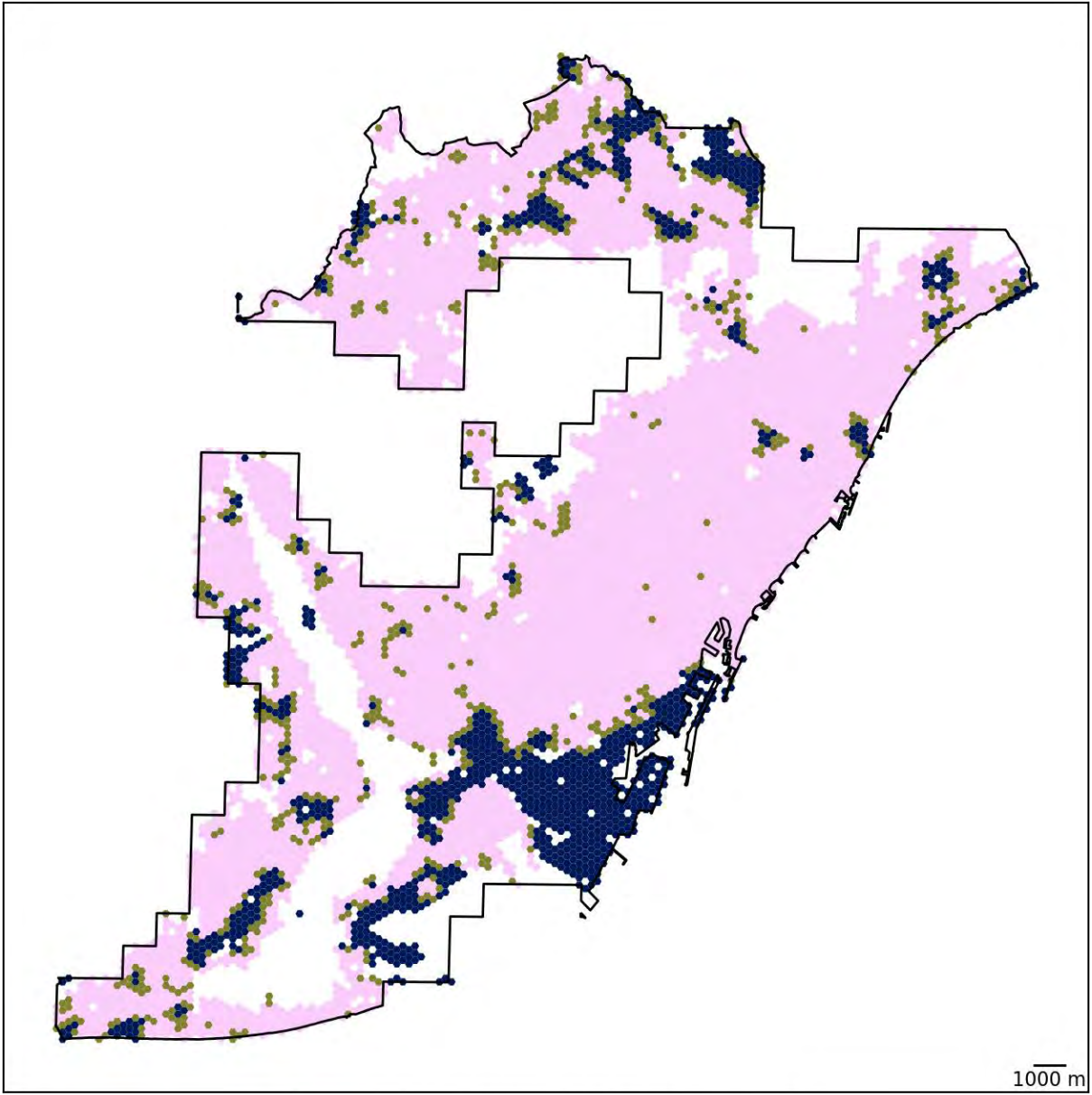

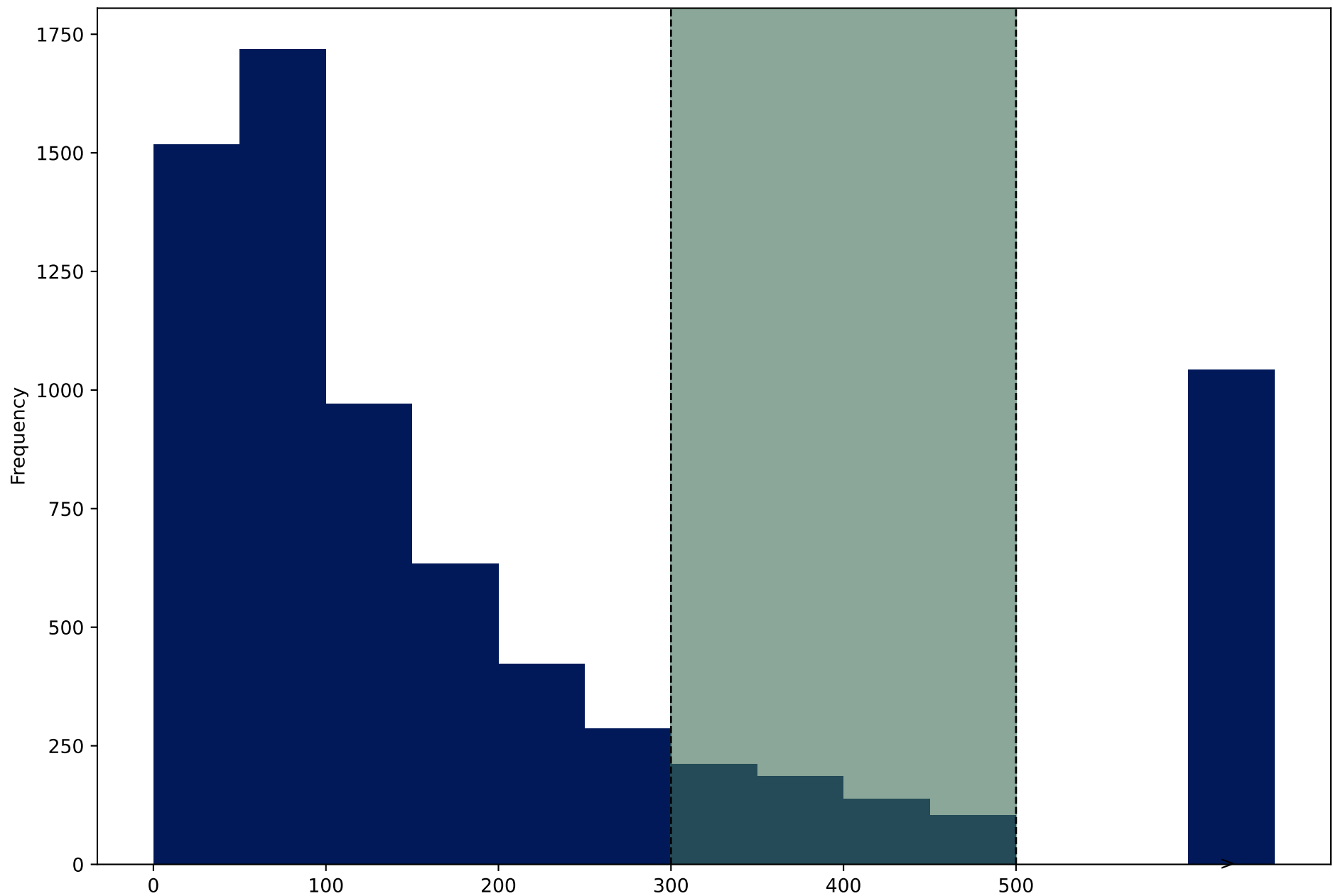

# Europe, Spain, Valencia

| Satellite imagery of urban study region (Bing) | Walkability, relative to city | Walkability, relative to 25 global cities |
|---|---|---|

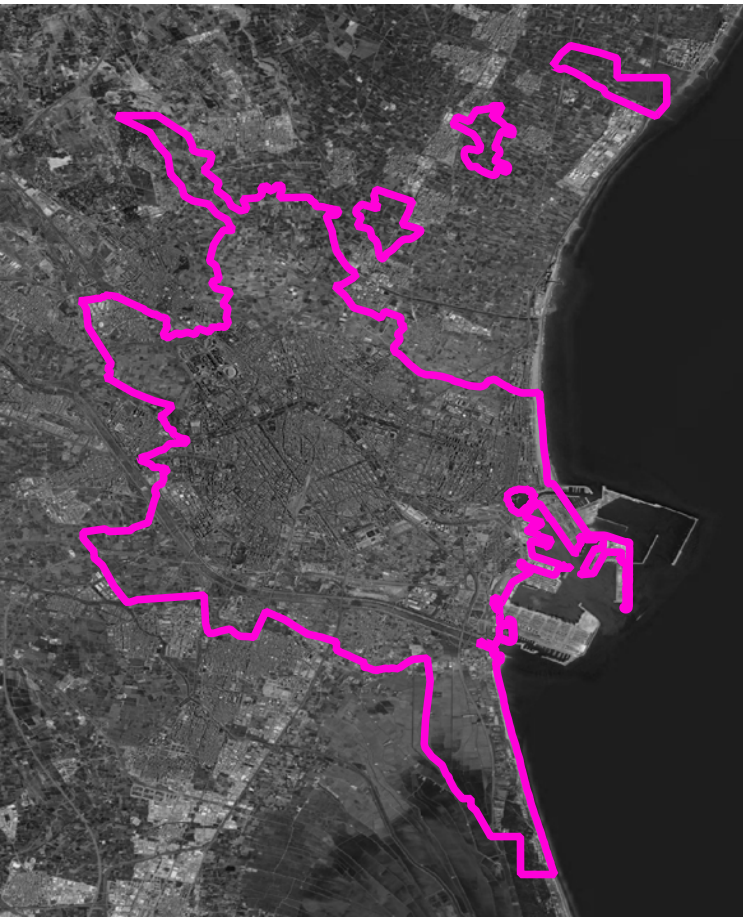
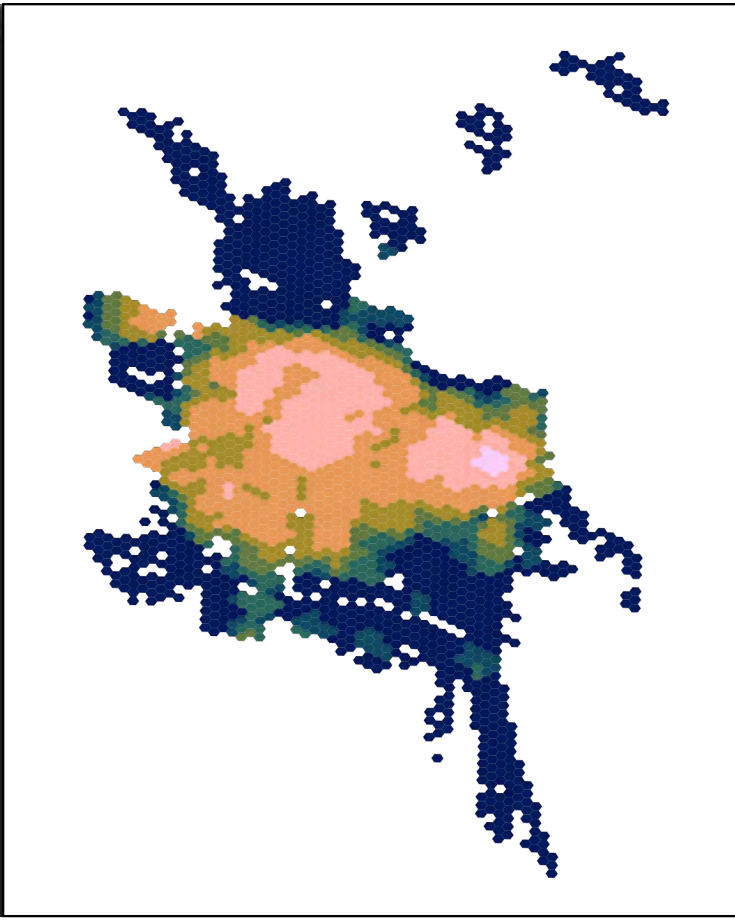
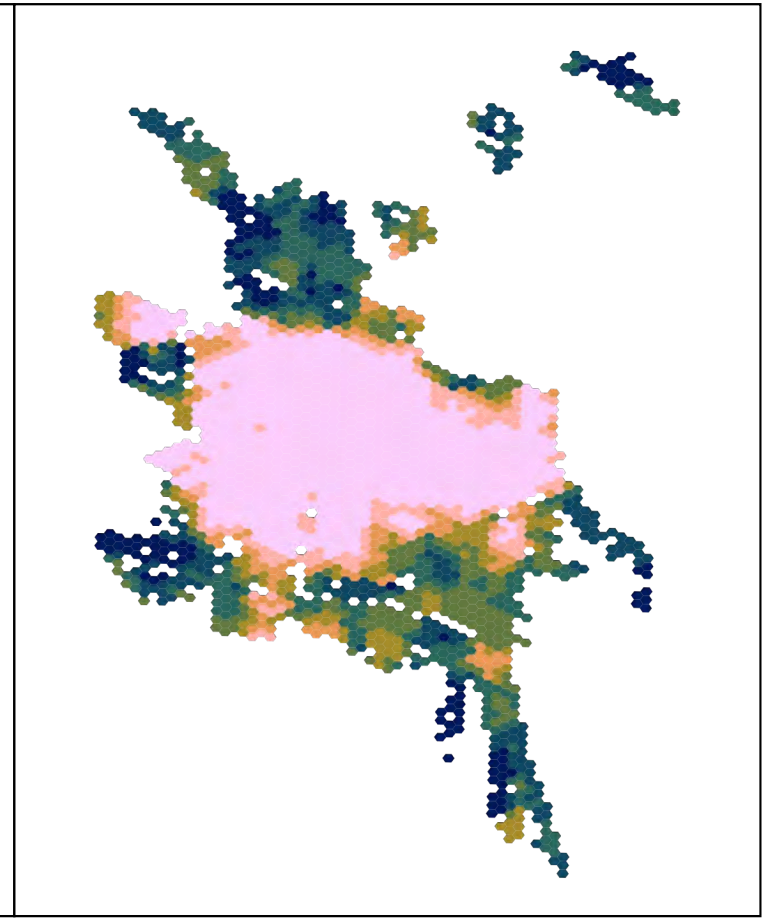

Urban boundary

0    6    12 km

**Walkability score**
- <-3
- -3 to -2
- -2 to -1
- -1 to 0
- 0 to 1
- 1 to 2
- 2 to 3
- ≥3

Walkability relative to all cities by component variables (2D histograms), and overall (histogram)

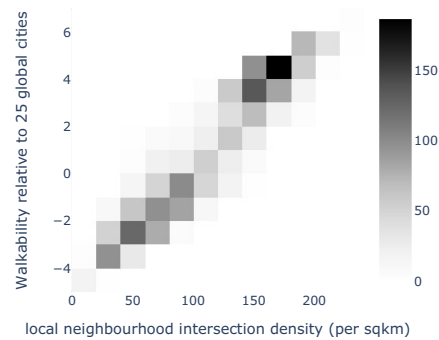
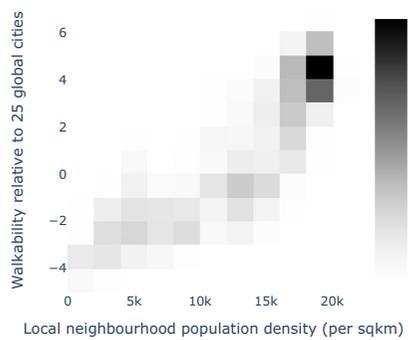
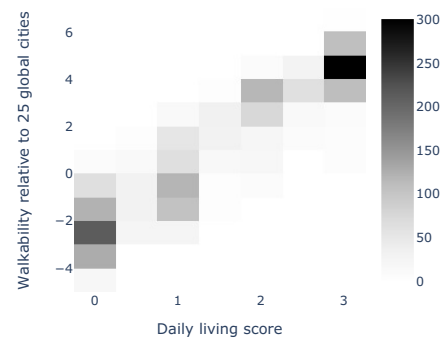
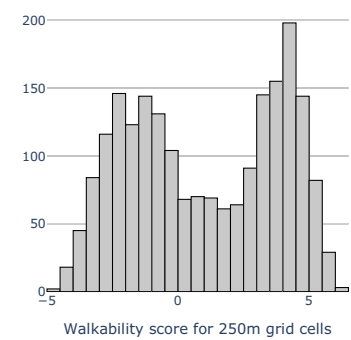

Mean 1000 m neighbourhood population per km²

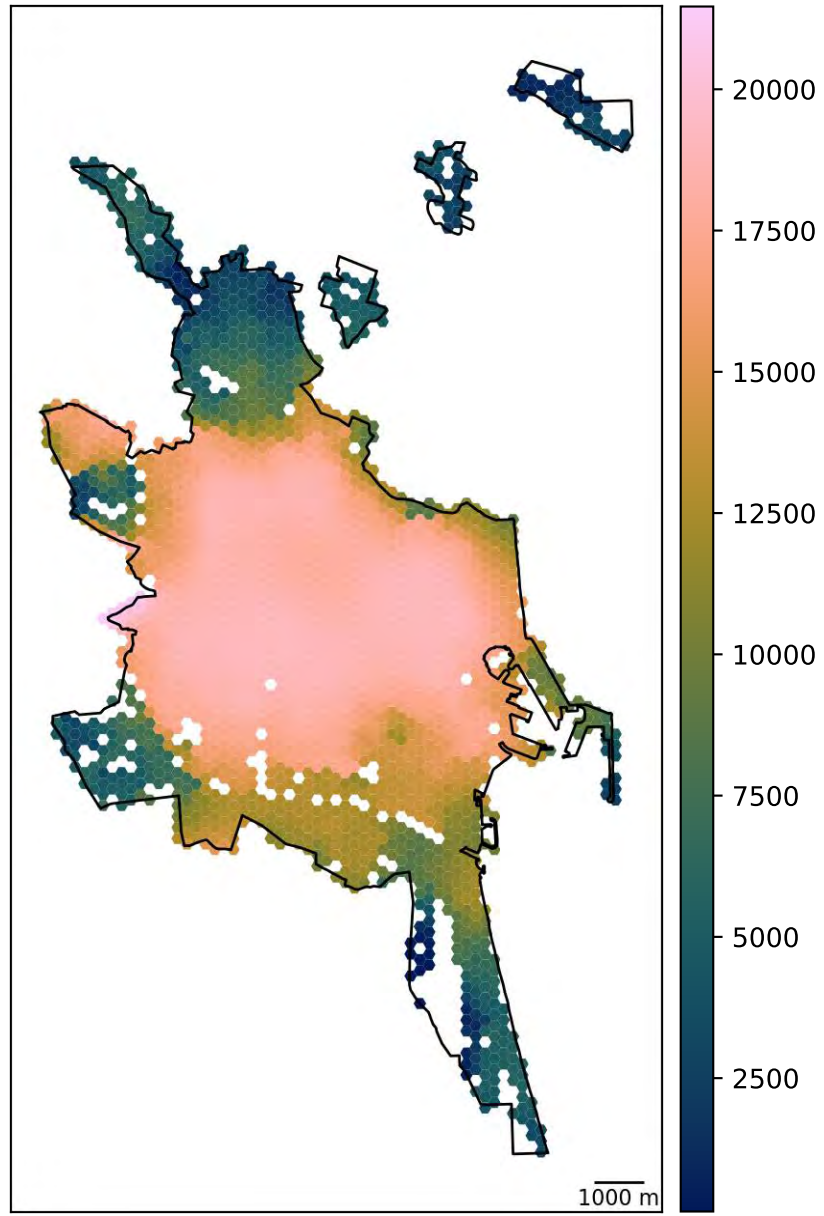

A: Estimated Mean 1000 m neighbourhood population per km² requirement for ≥80% probability of engaging in walking for transport

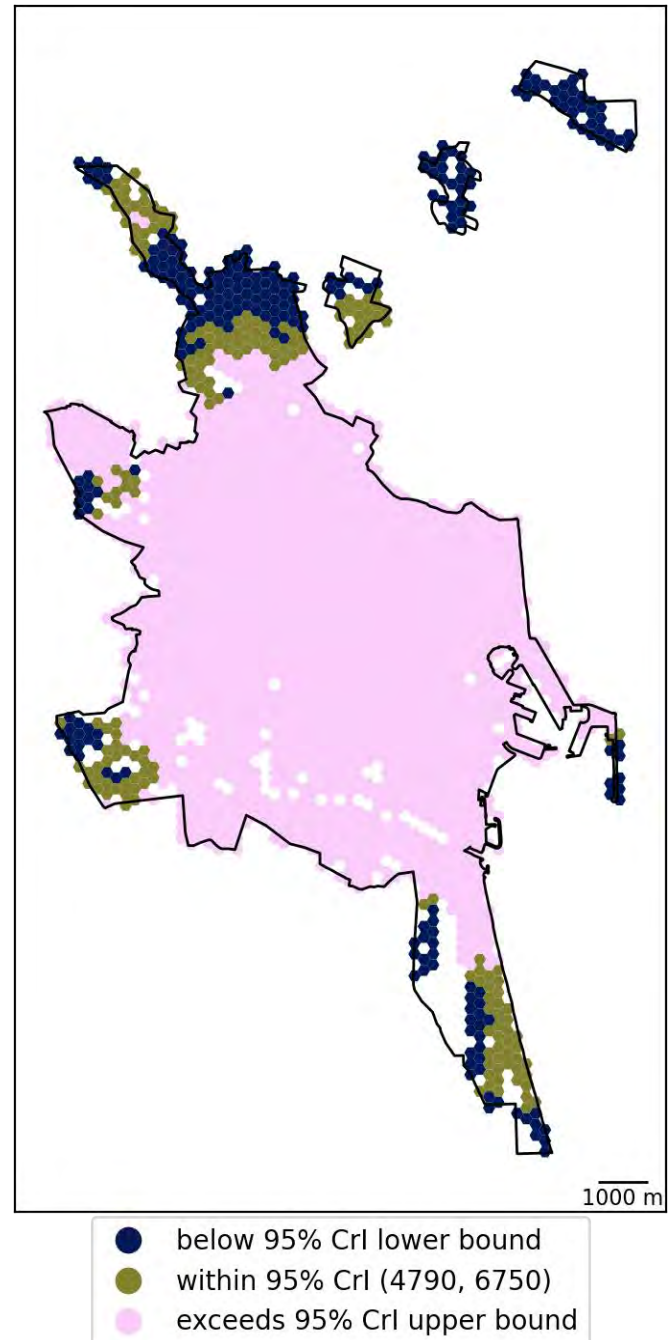

B: Estimated Mean 1000 m neighbourhood population per km² requirement for reaching the WHO's target of a ≥15% relative reduction in insufficient physical activity through walking

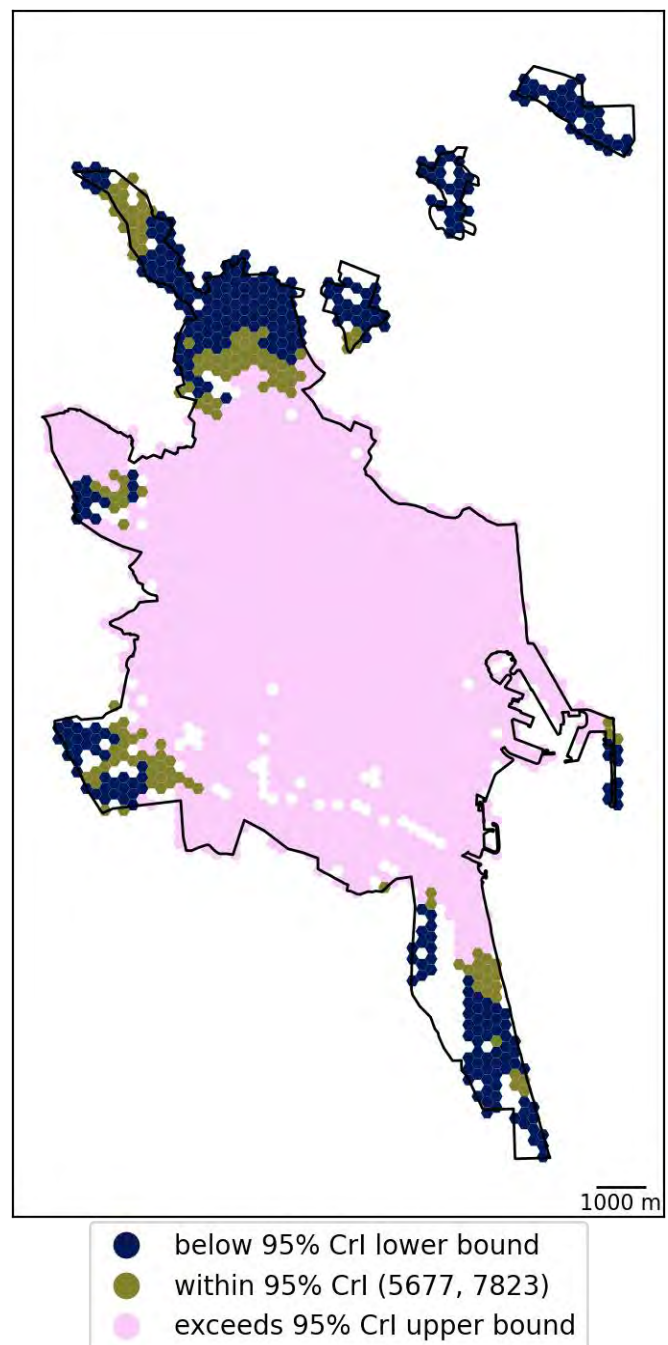

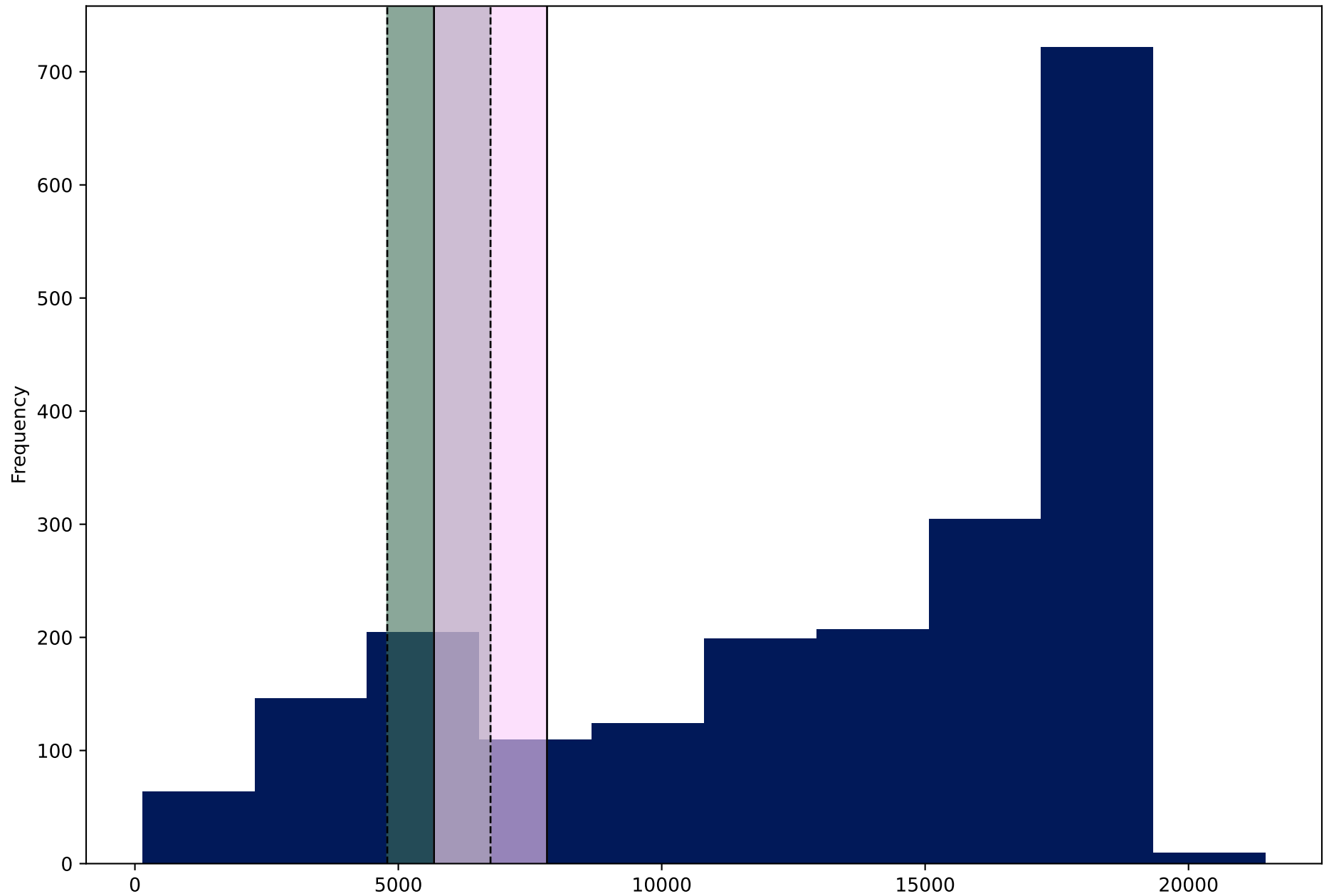

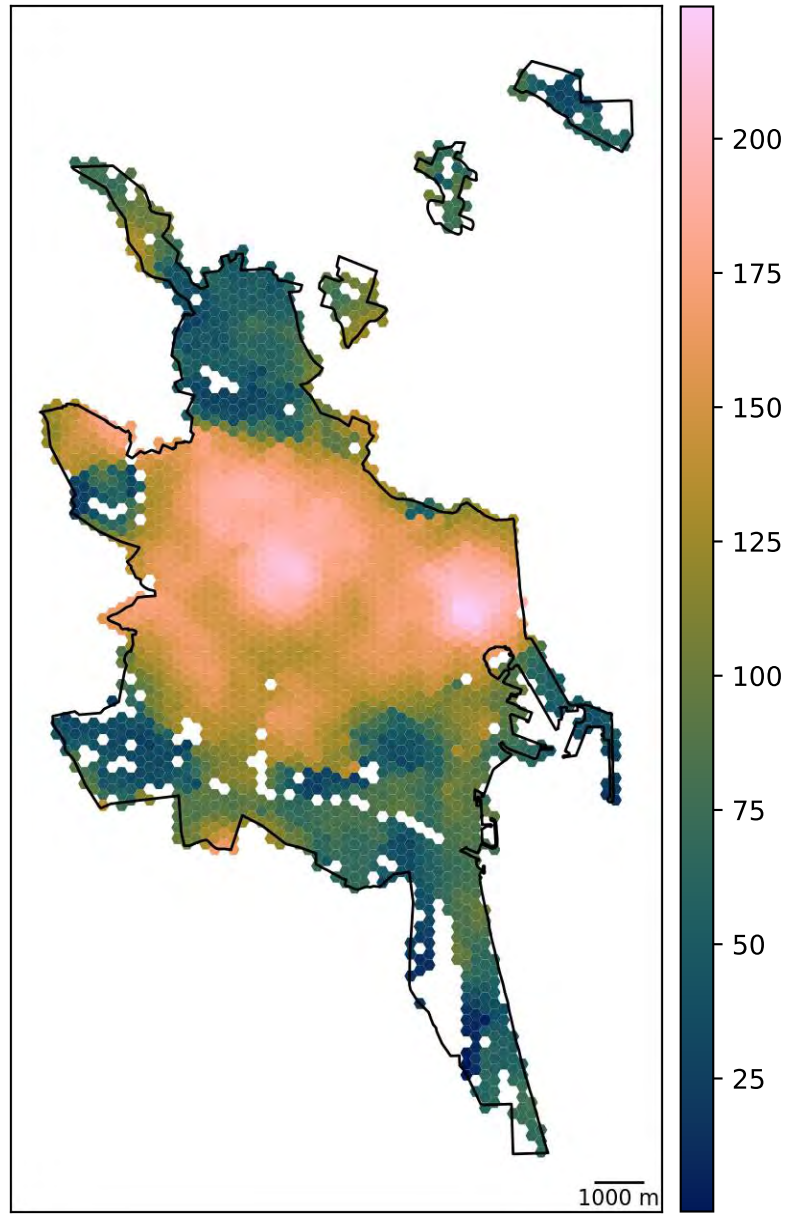

Mean 1000 m neighbourhood street intersections per km²

A: Estimated Mean 1000 m neighbourhood street intersections per km² requirement for ≥80% probability of engaging in walking for transport

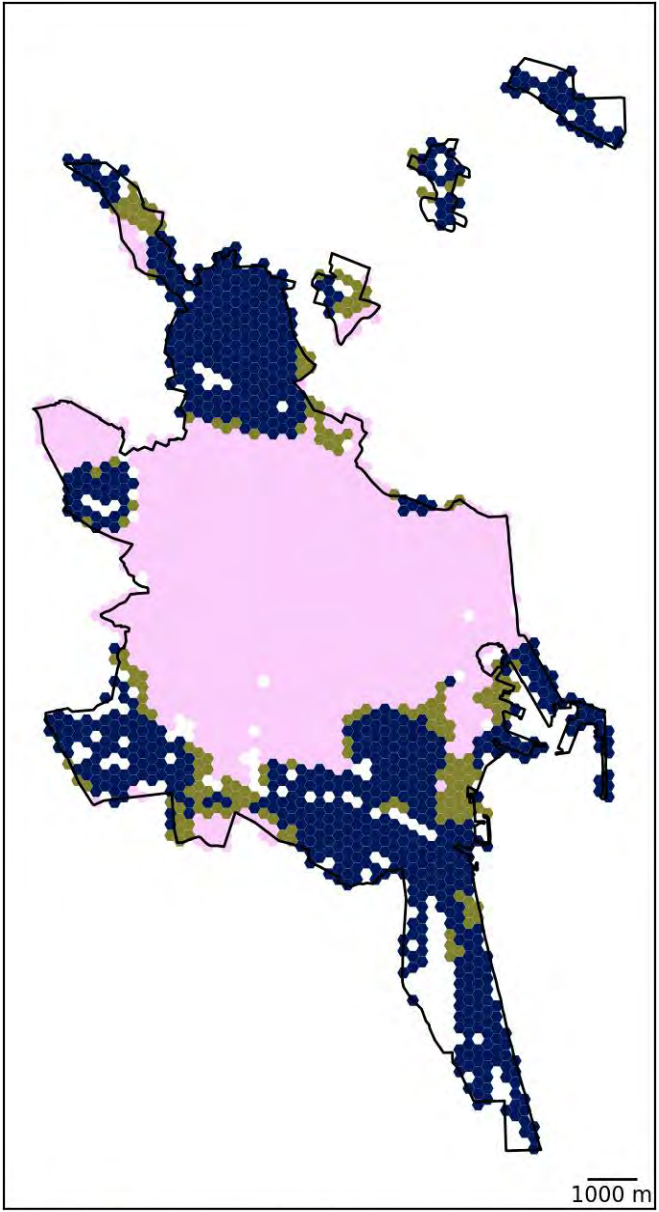

- below 95% CrI lower bound
- within 95% CrI (90, 110)
- exceeds 95% CrI upper bound

B: Estimated Mean 1000 m neighbourhood street intersections per km² requirement for reaching the WHO's target of a ≥15% relative reduction in insufficient physical activity through walking

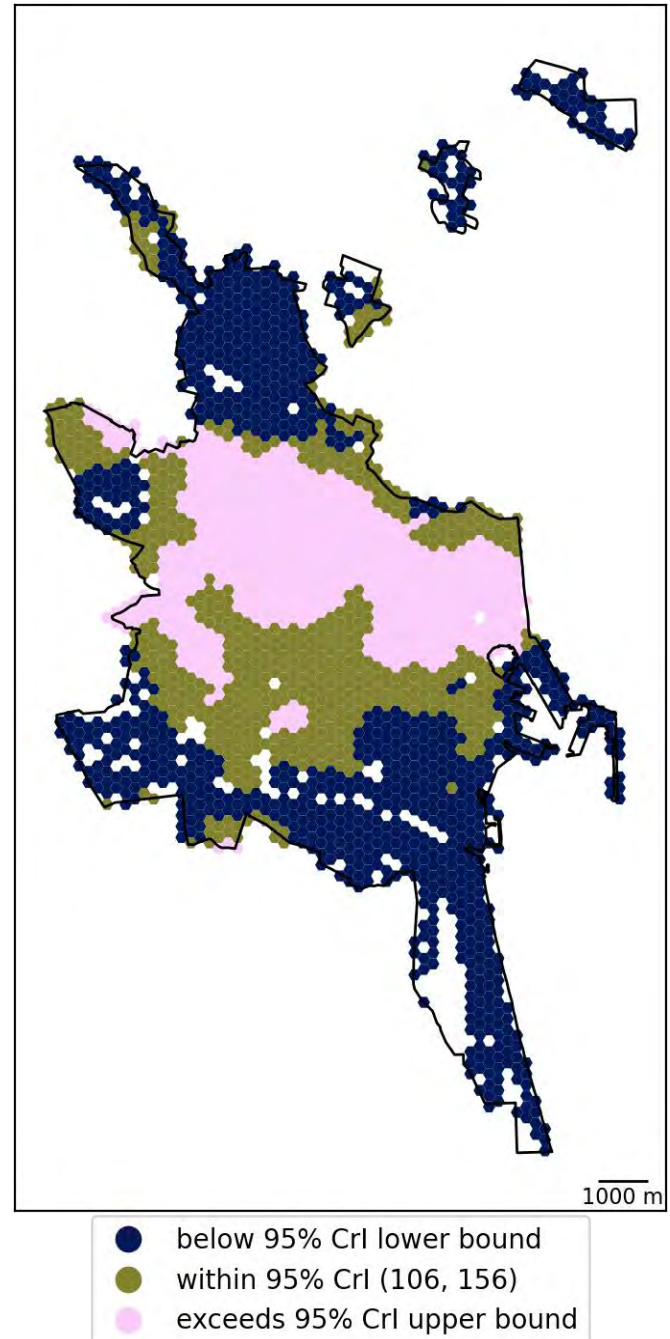

- below 95% CrI lower bound
- within 95% CrI (106, 156)
- exceeds 95% CrI upper bound

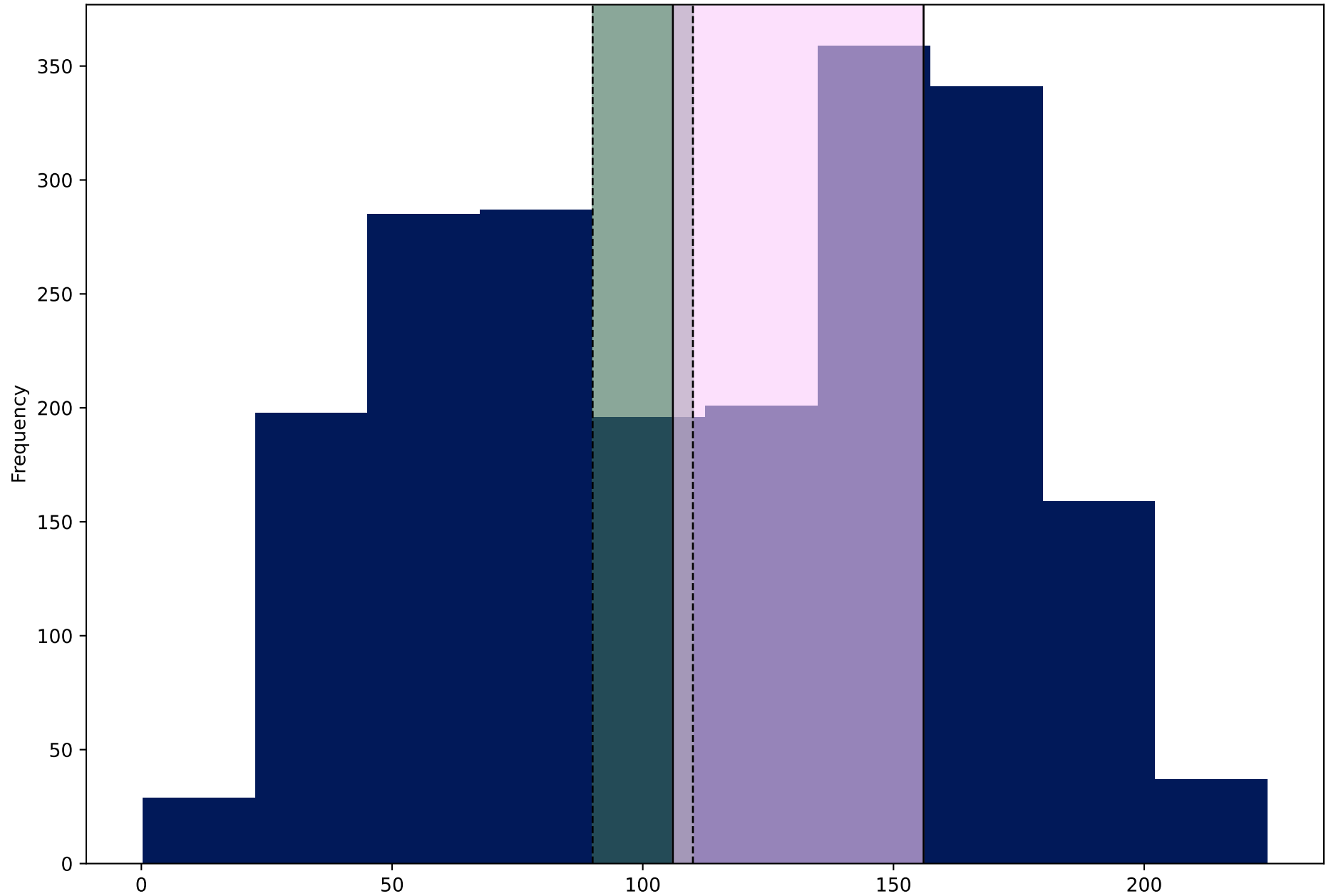

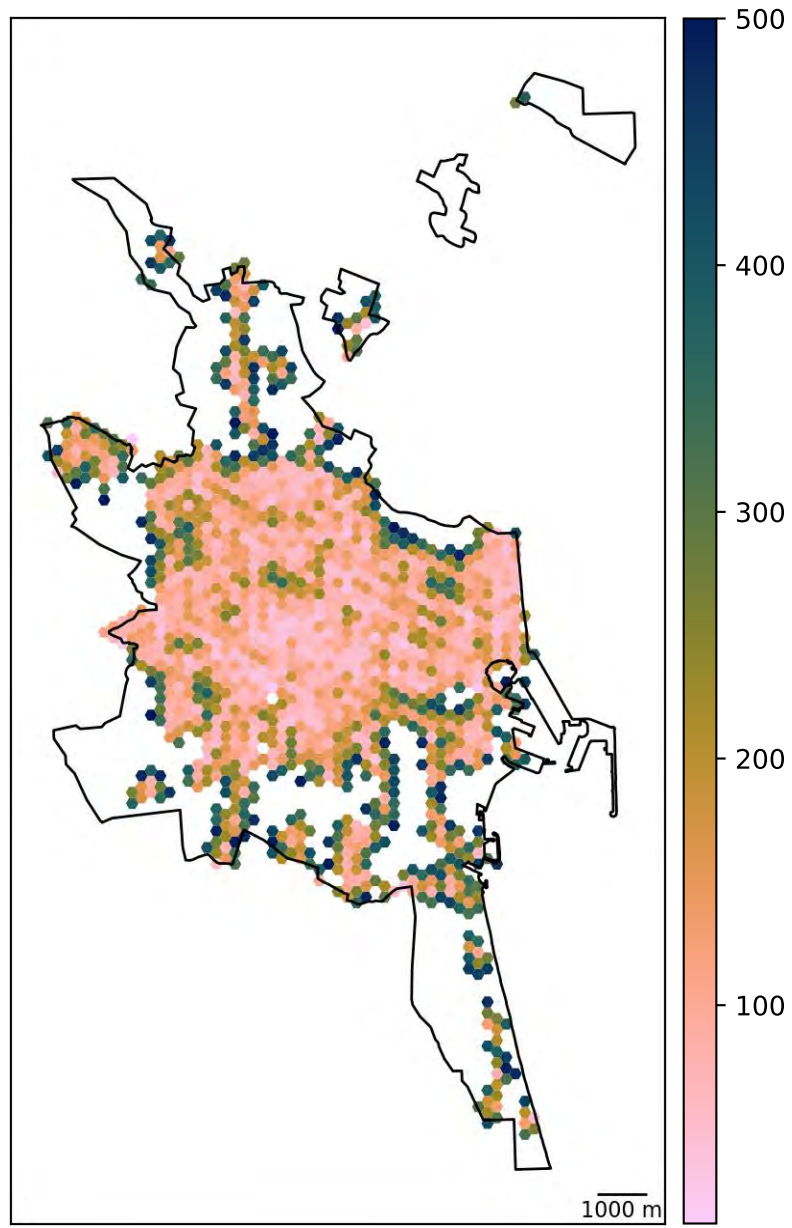

Distance to nearest public transport stops (m; up to 500m)

distances: Estimated Distance to nearest public transport stops (m; up to 500m) requirement for distances to destinations, measured up to a maximum distance target threshold of 500 metres

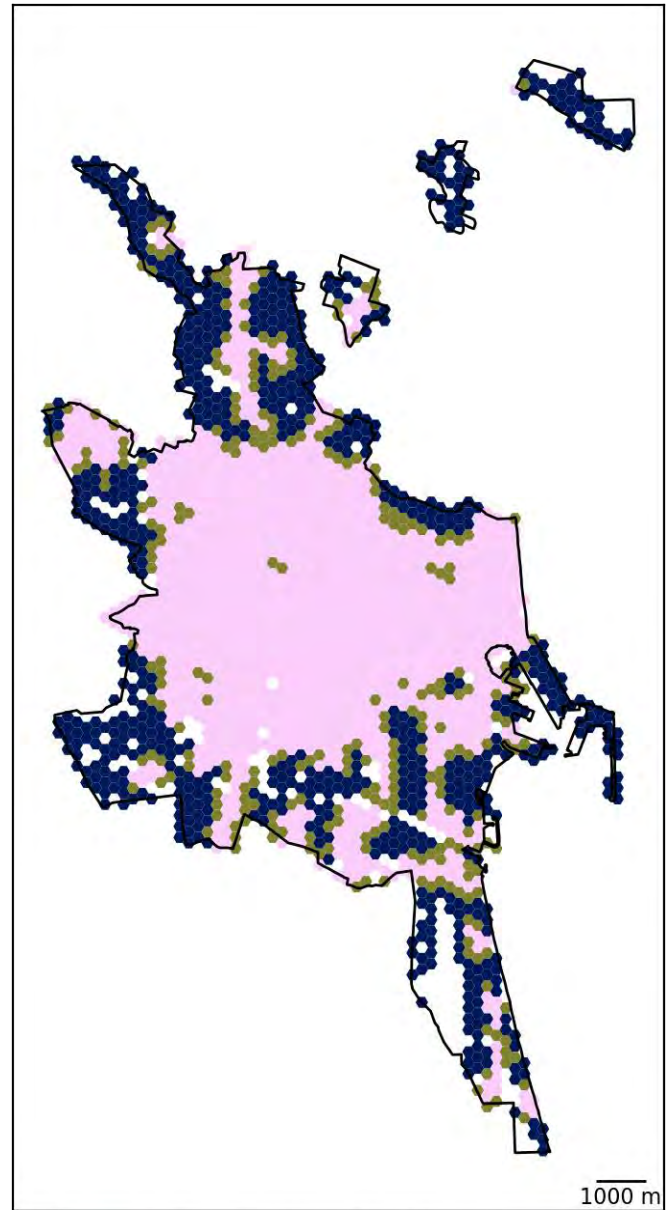

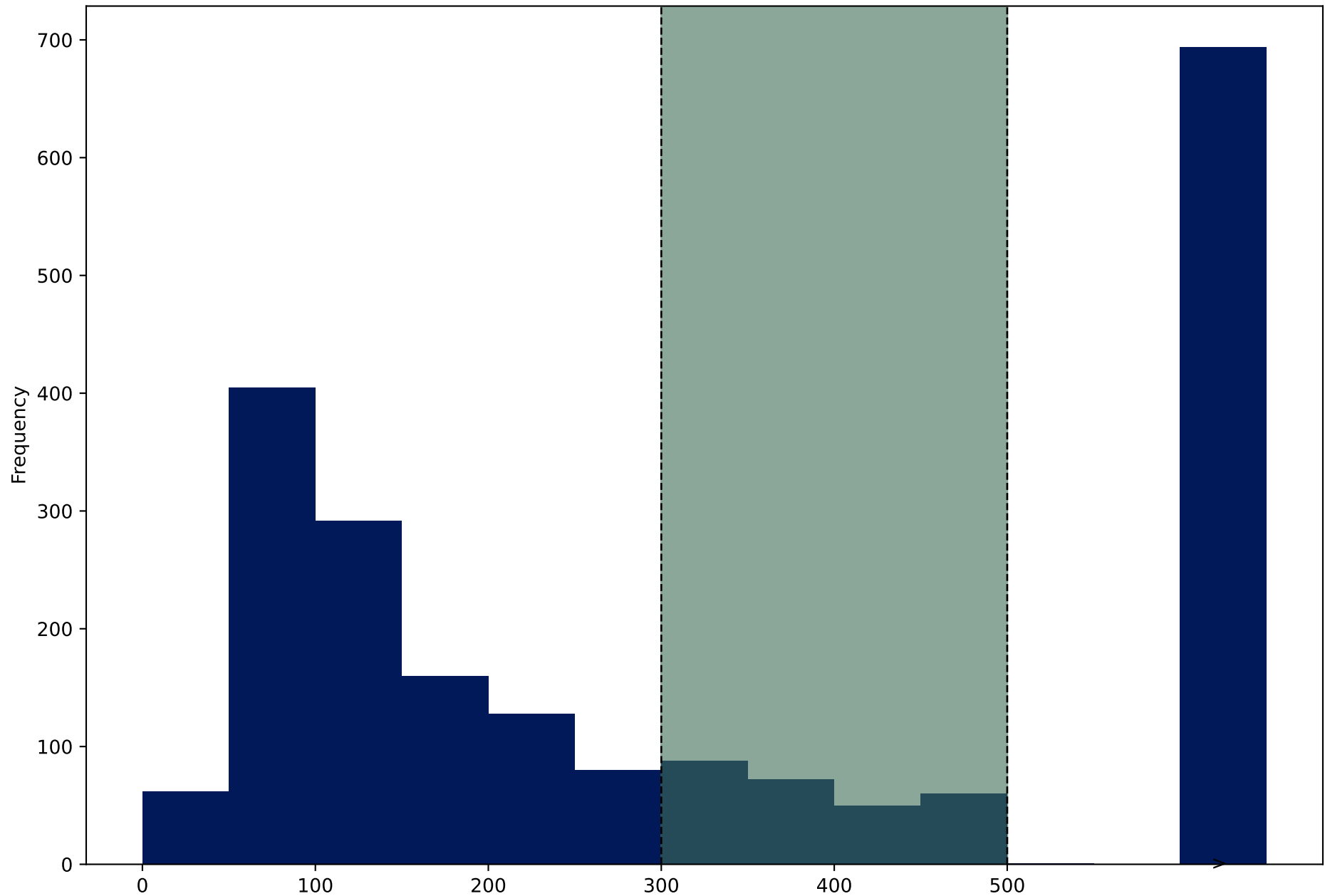

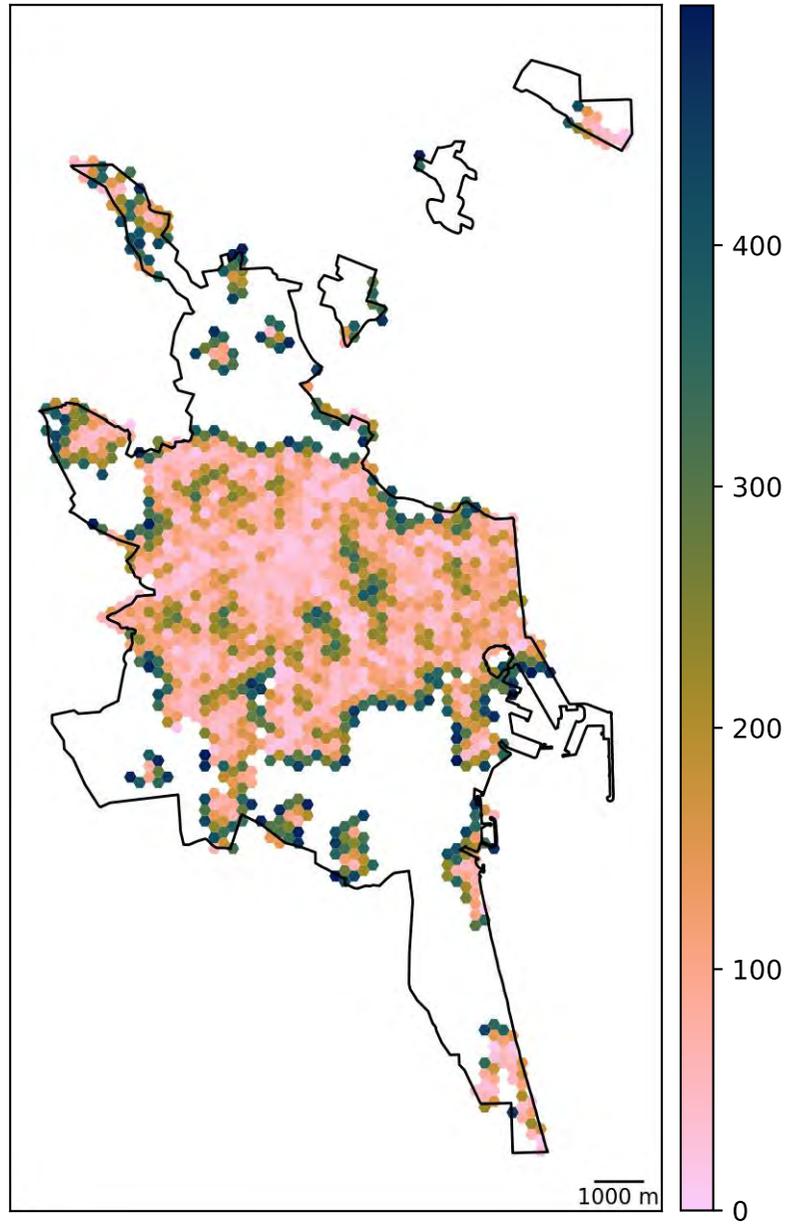

distances: Estimated Distance to nearest park (m; up to 500m) requirement for distances to destinations, measured up to a maximum distance target threshold of 500 metres

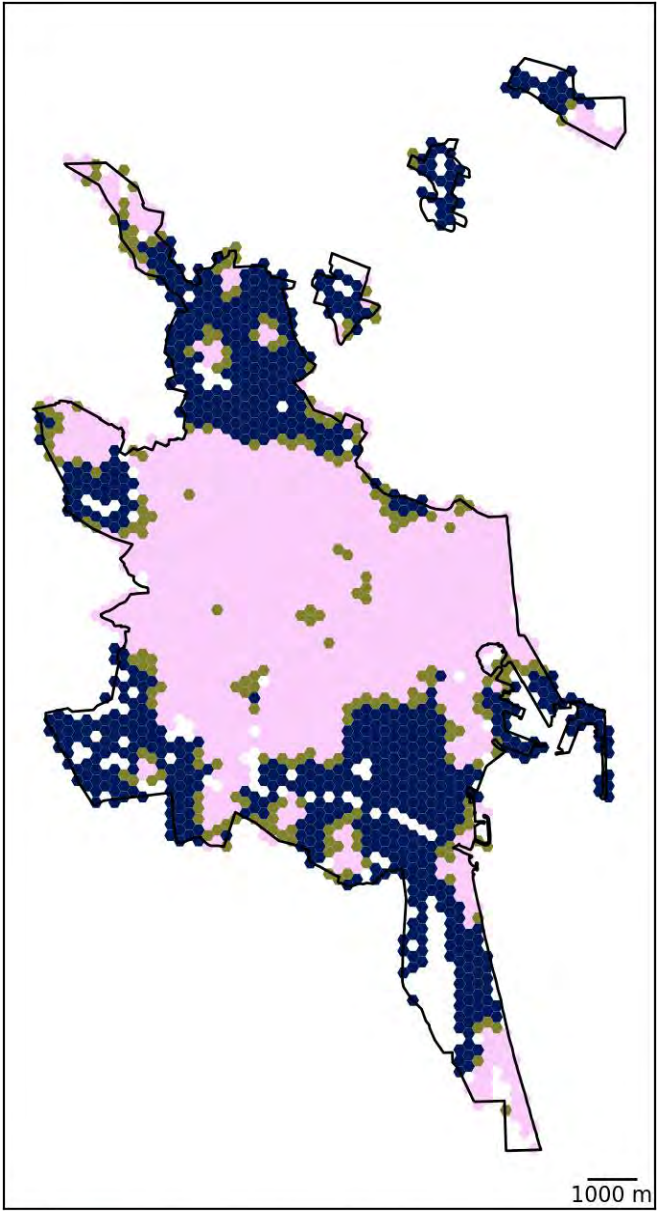

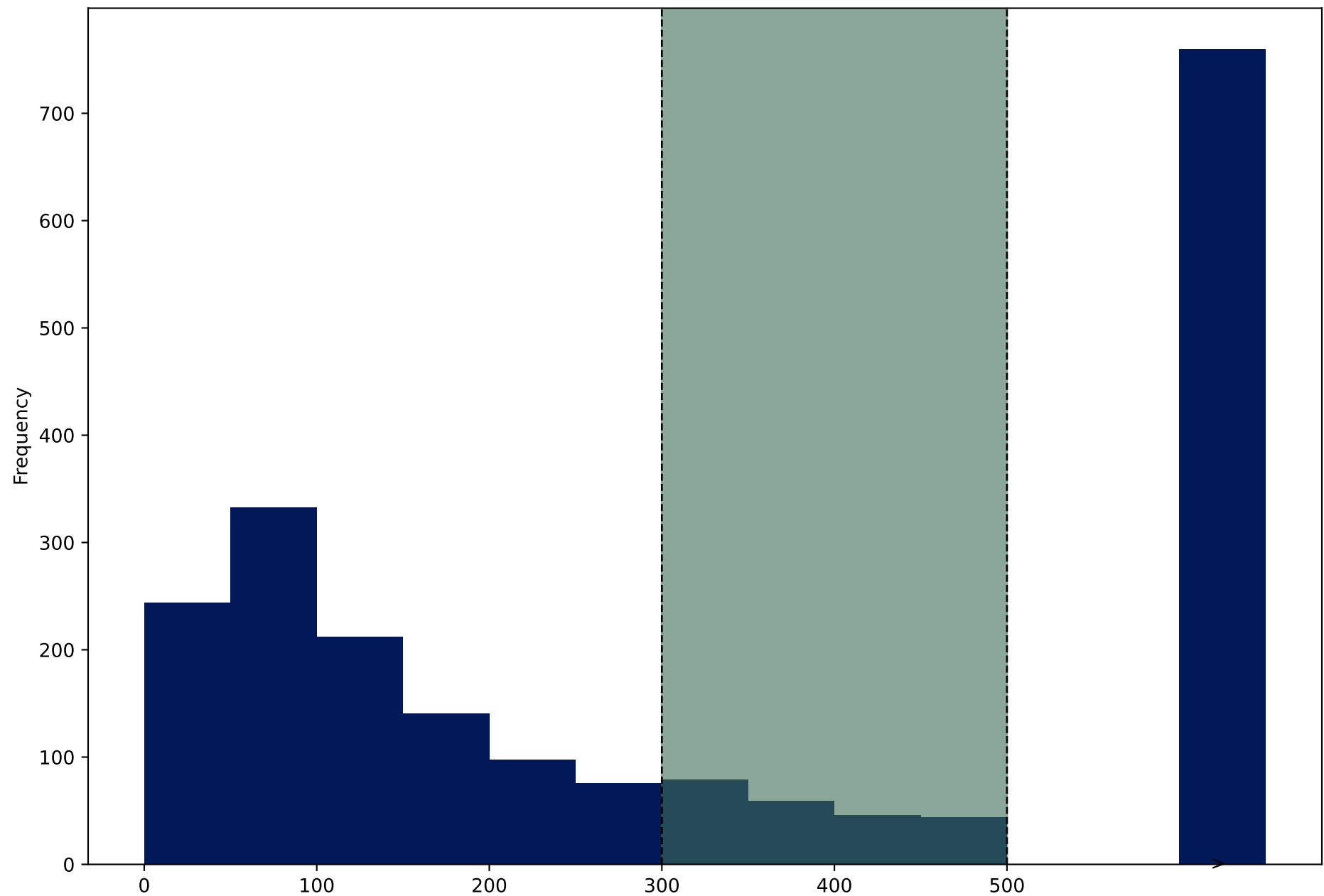

# Europe, Spain, Vic

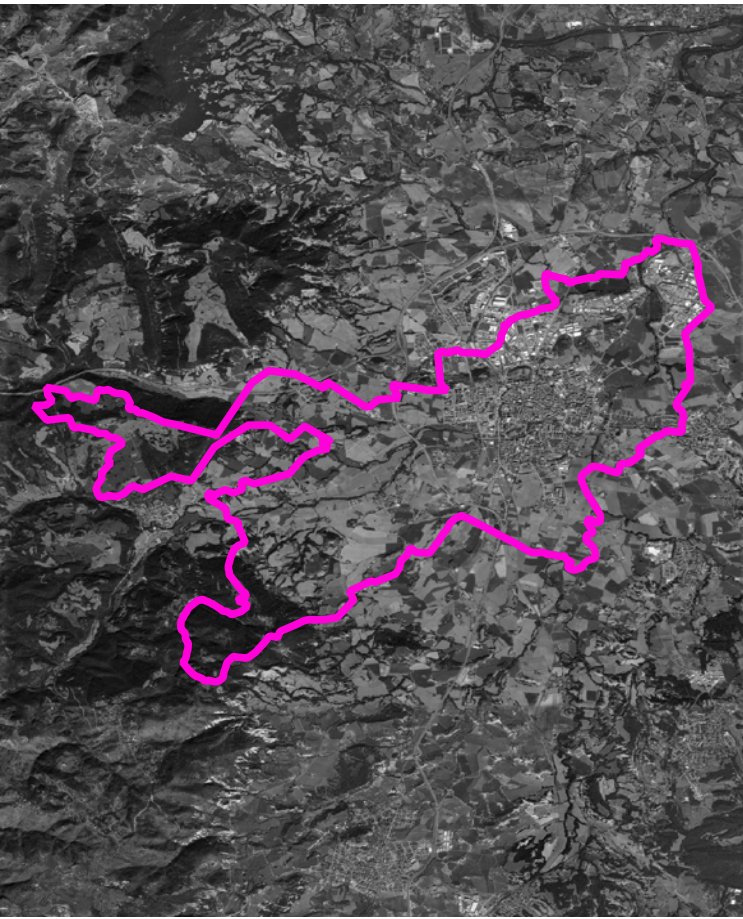
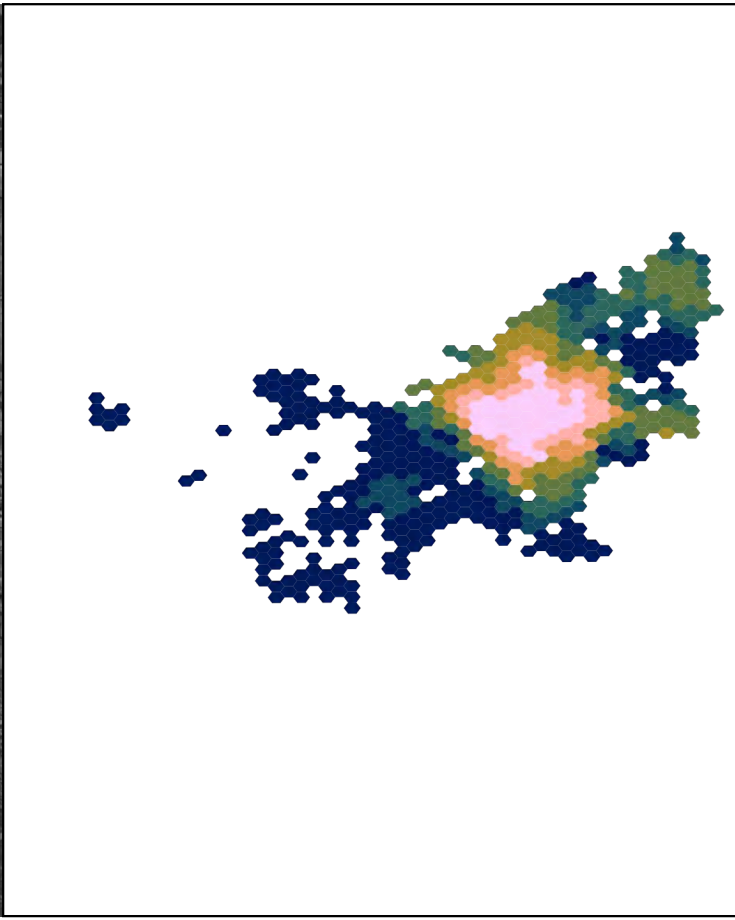
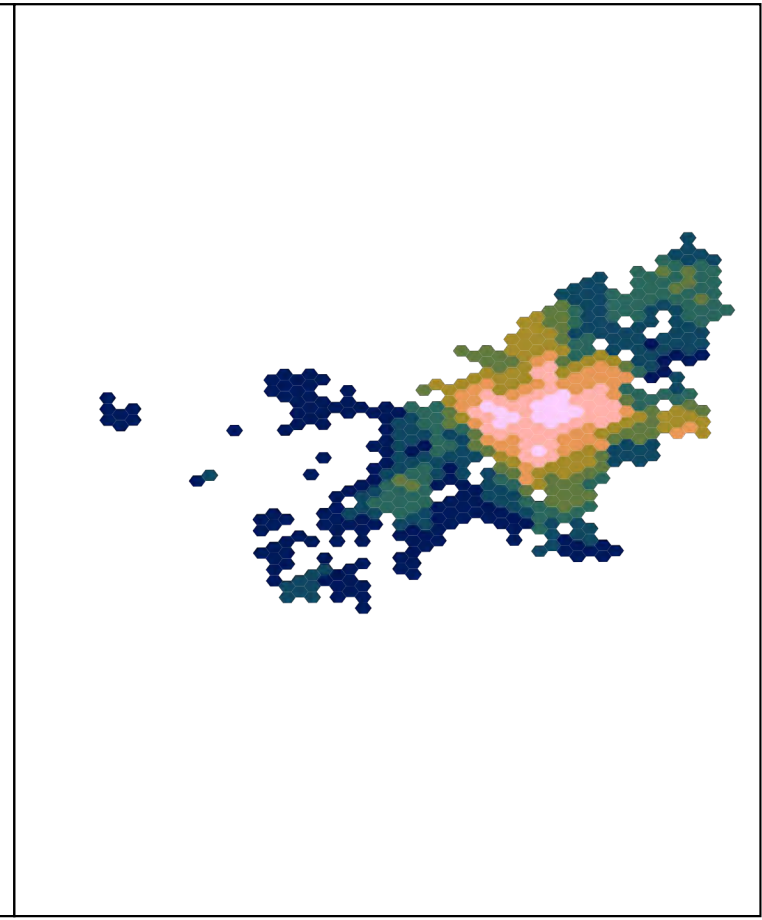
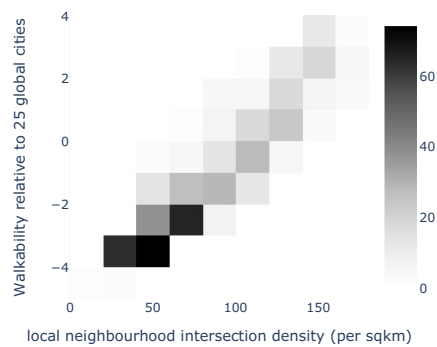
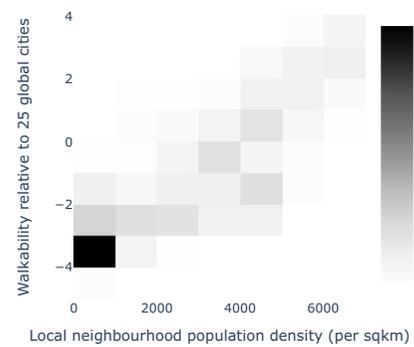
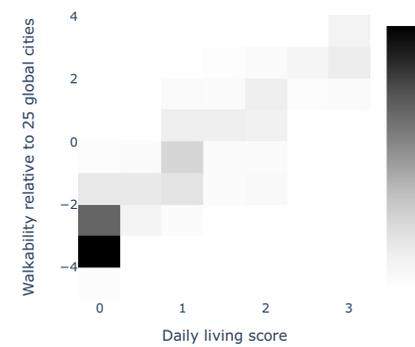
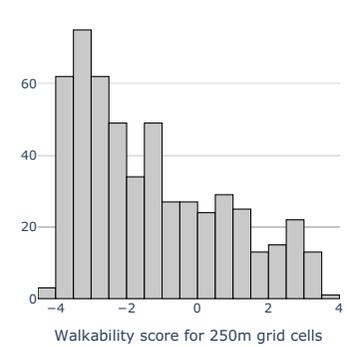

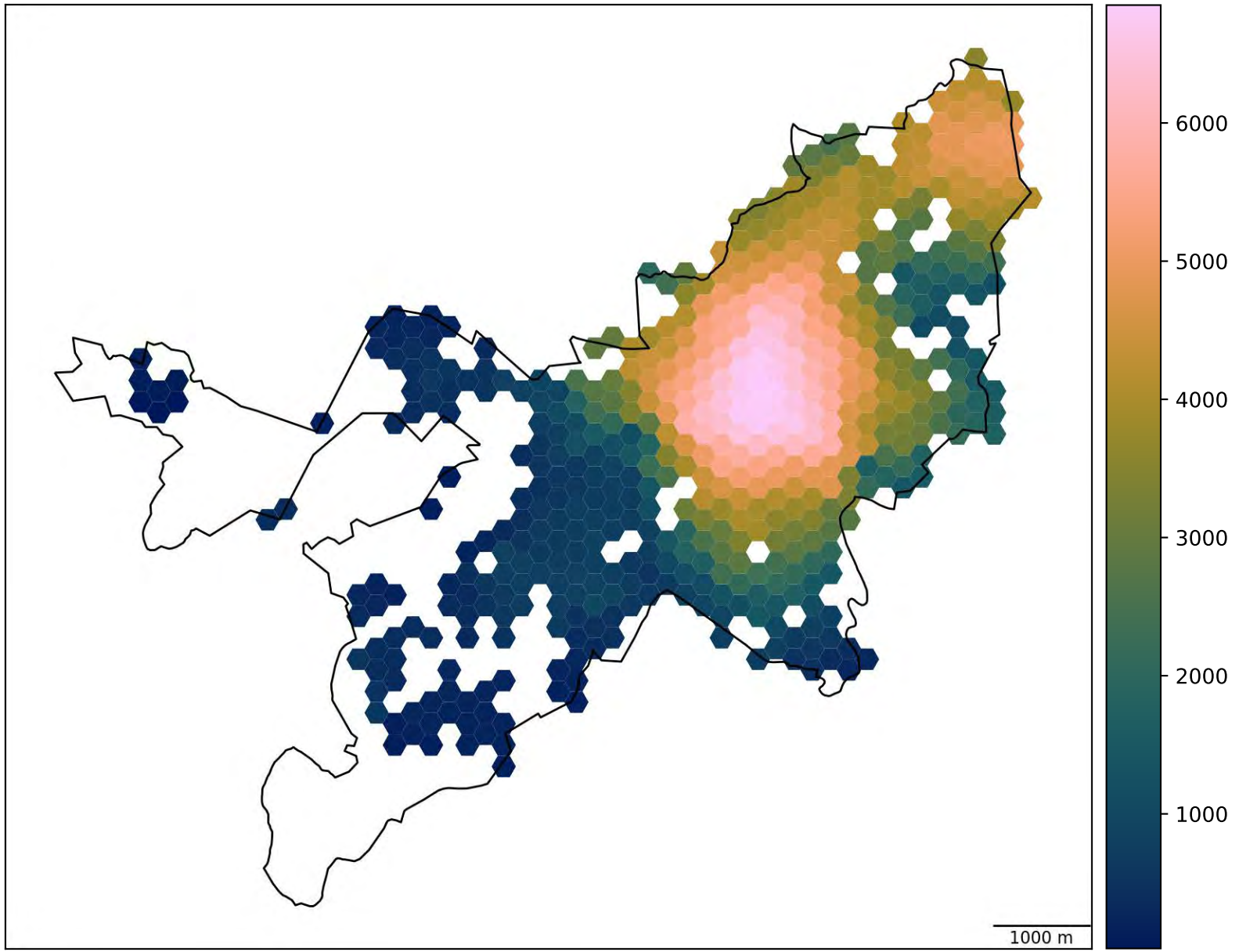

Mean 1000 m neighbourhood population per km²

A: Estimated Mean 1000 m neighbourhood population per km² requirement for ≥80% probability of engaging in walking for transport

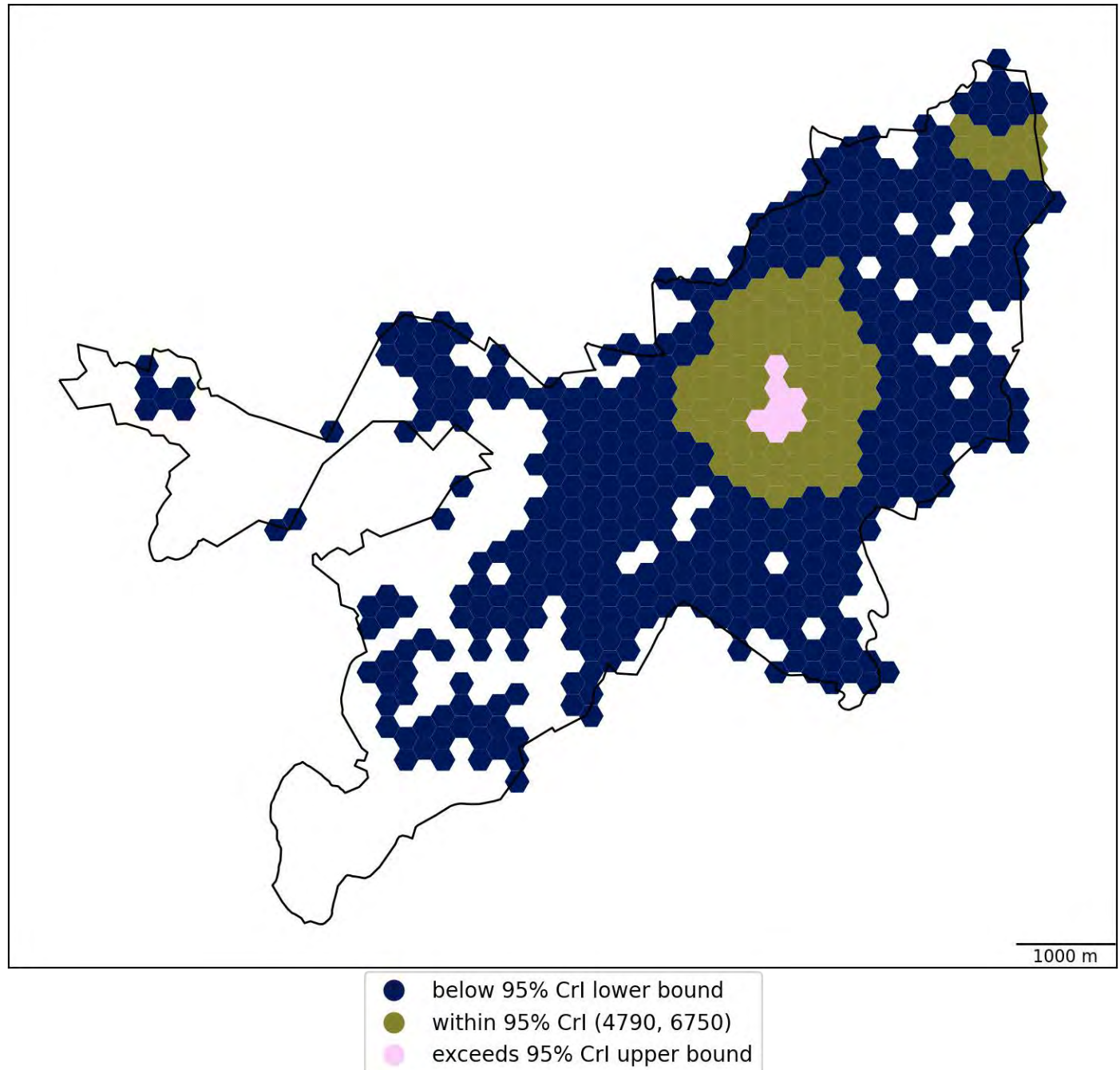

B: Estimated Mean 1000 m neighbourhood population per km² requirement for reaching the WHO's target of a ≥15% relative reduction in insufficient physical activity through walking

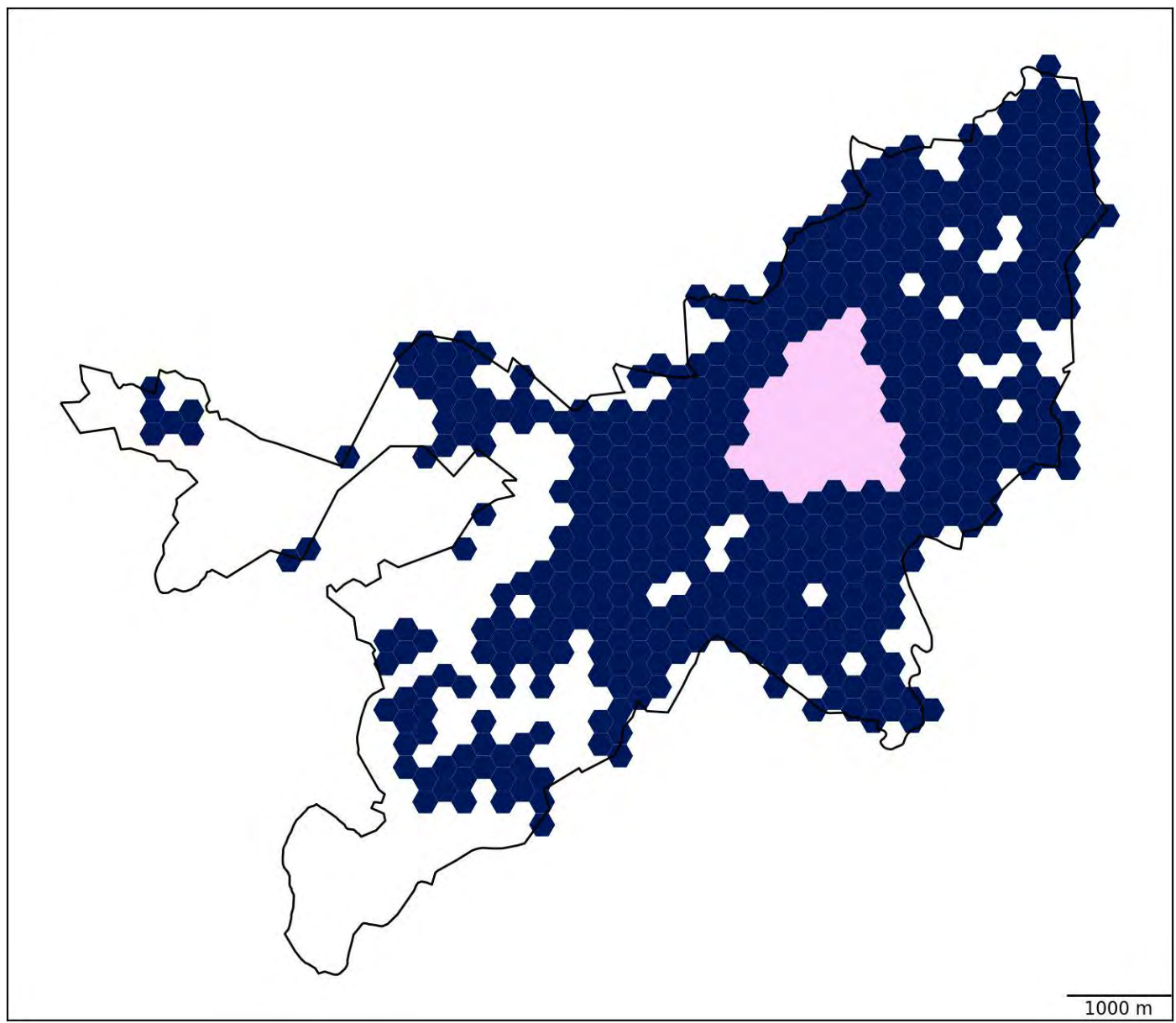

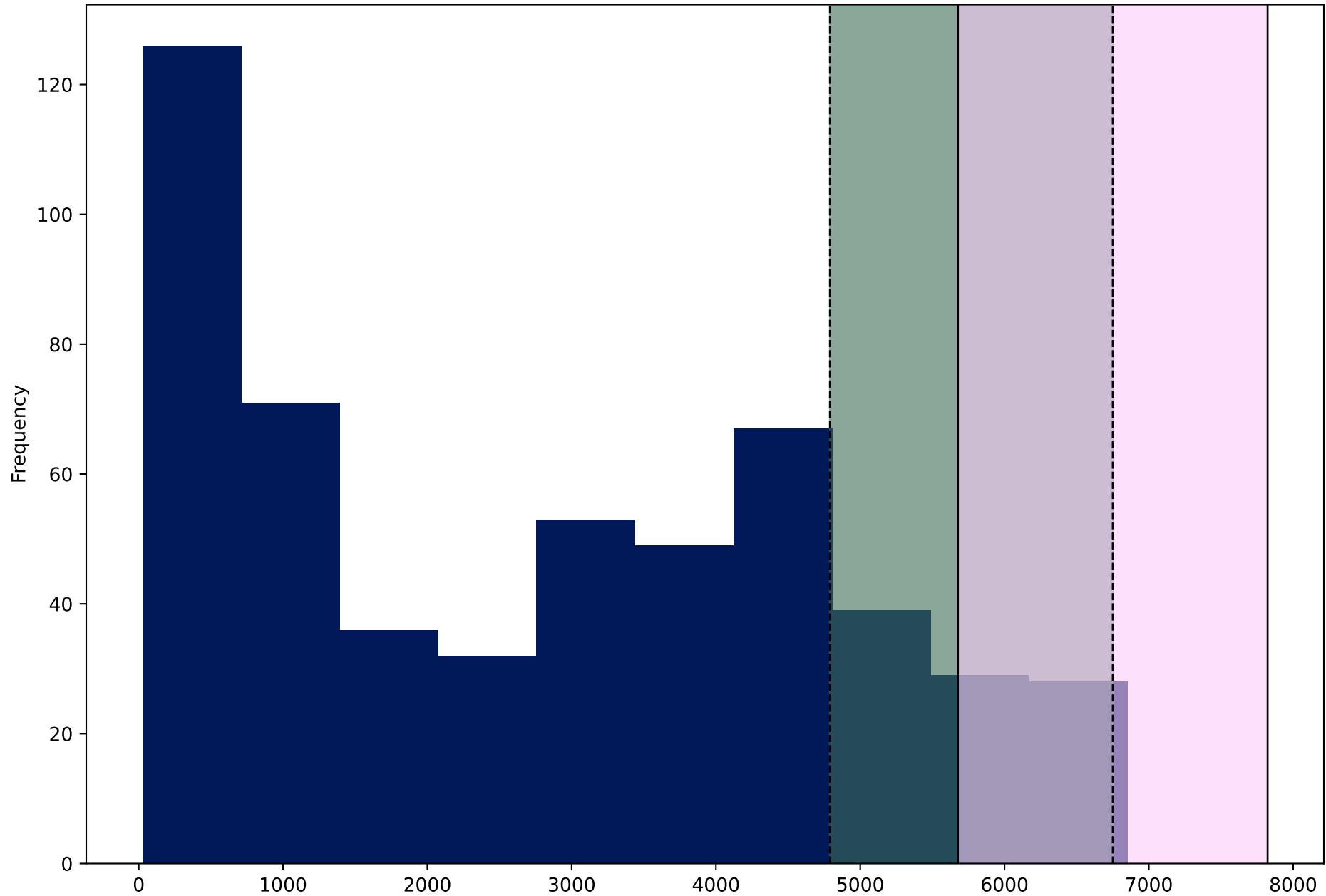

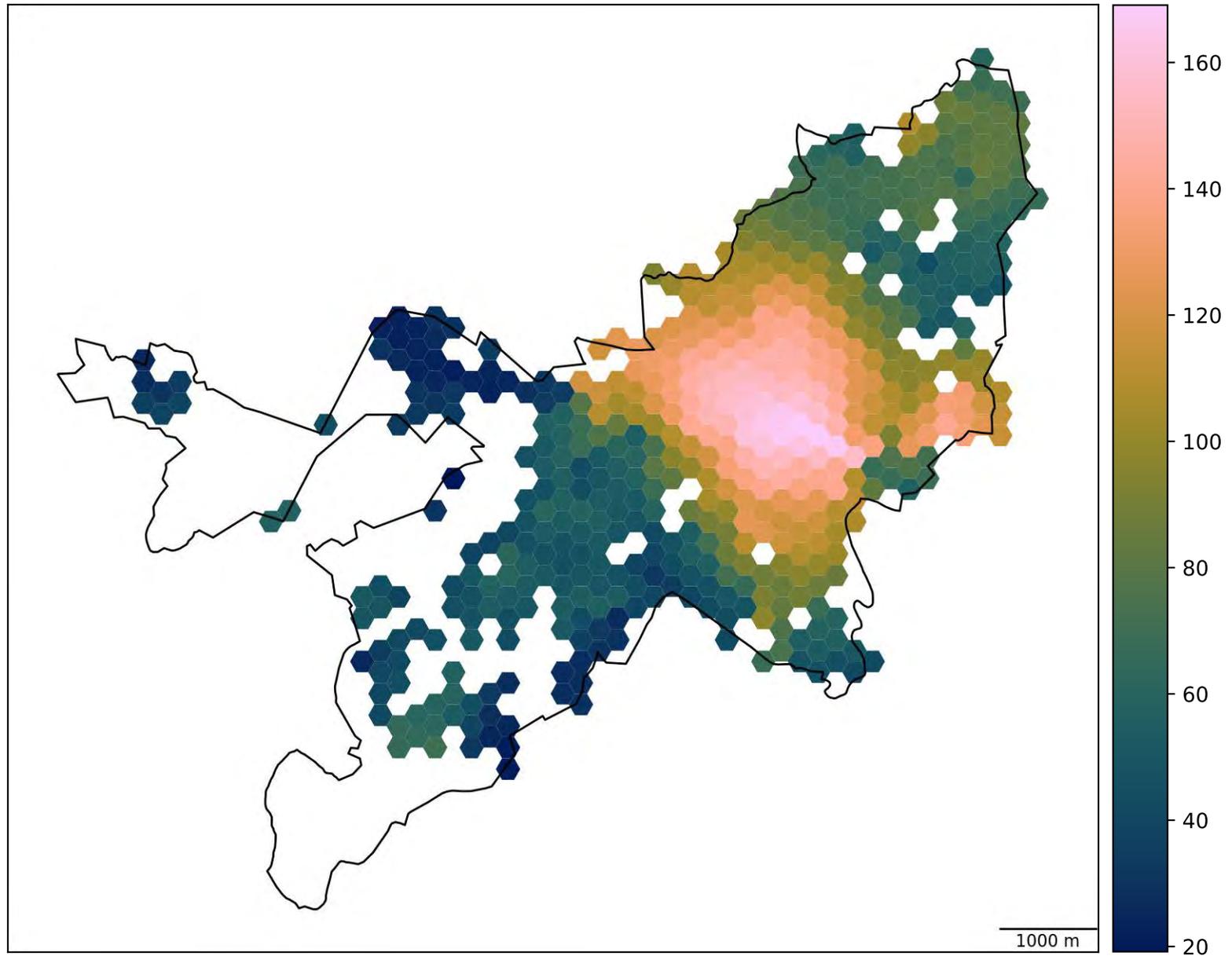

Mean 1000 m neighbourhood street intersections per km²

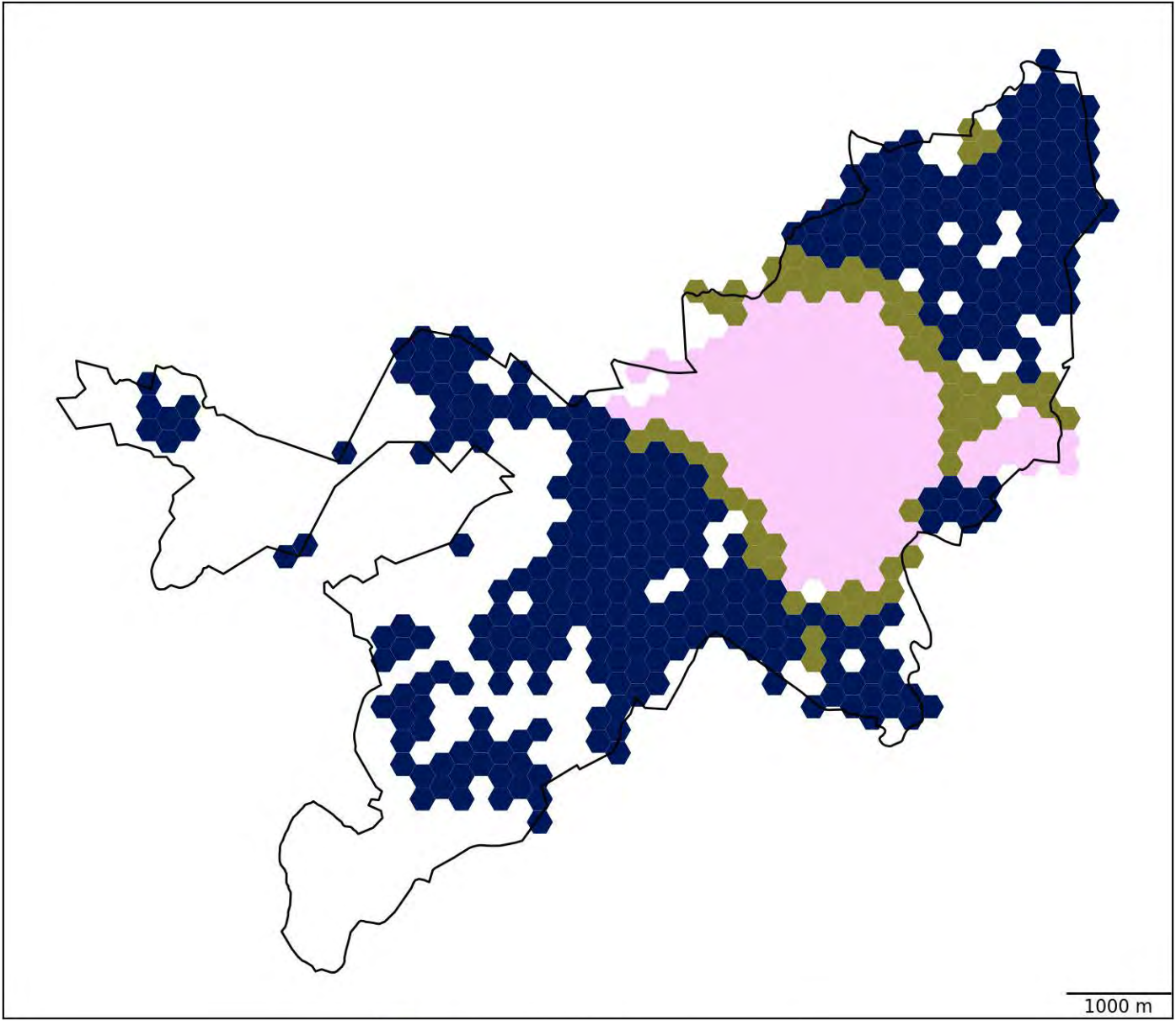

A: Estimated Mean 1000 m neighbourhood street intersections per km² requirement for ≥80% probability of engaging in walking for transport

- below 95% CrI lower bound
- within 95% CrI (90, 110)
- exceeds 95% CrI upper bound

B: Estimated Mean 1000 m neighbourhood street intersections per km² requirement for reaching the WHO's target of a ≥15% relative reduction in insufficient physical activity through walking

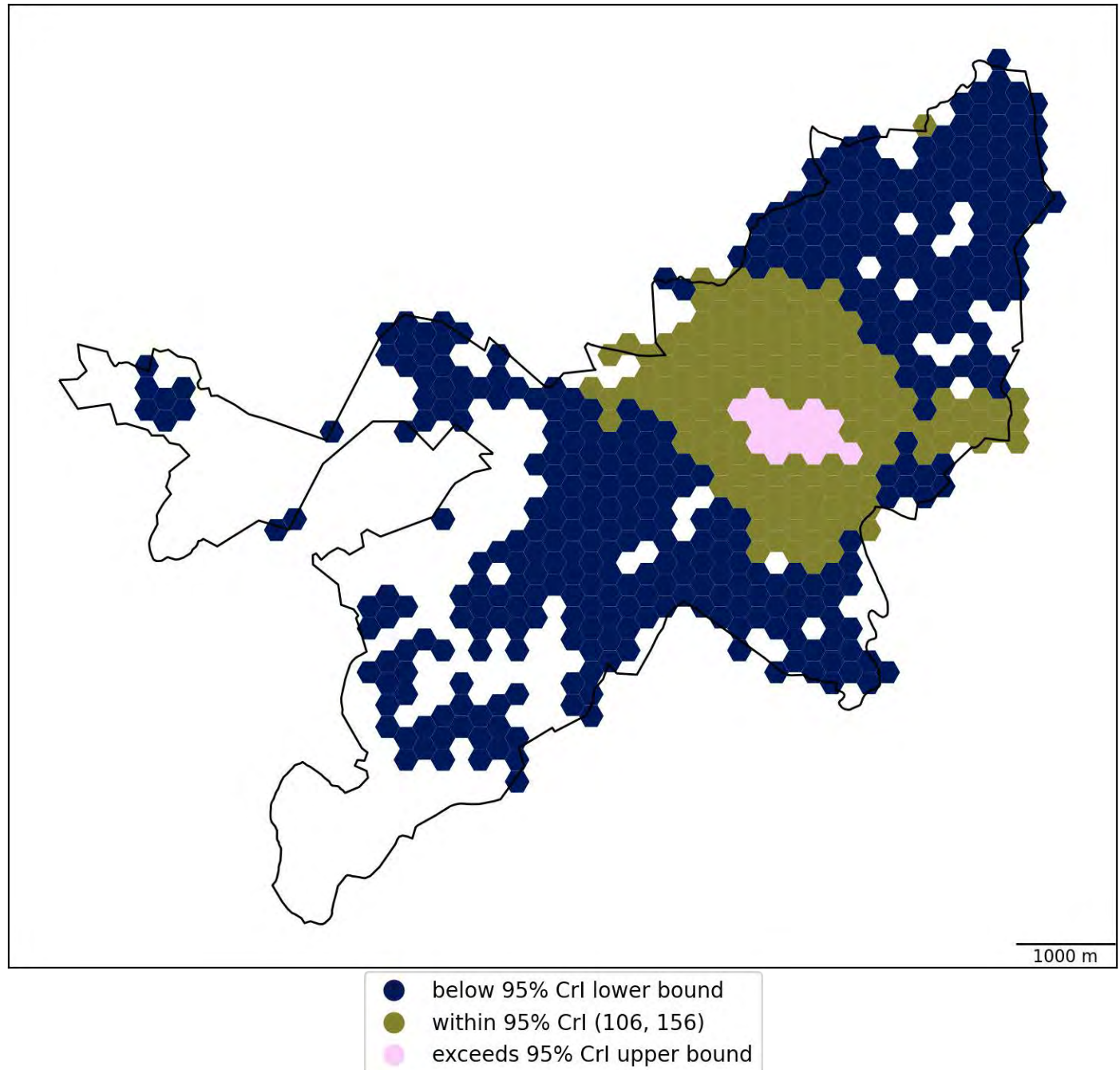

- below 95% CrI lower bound
- within 95% CrI (106, 156)
- exceeds 95% CrI upper bound

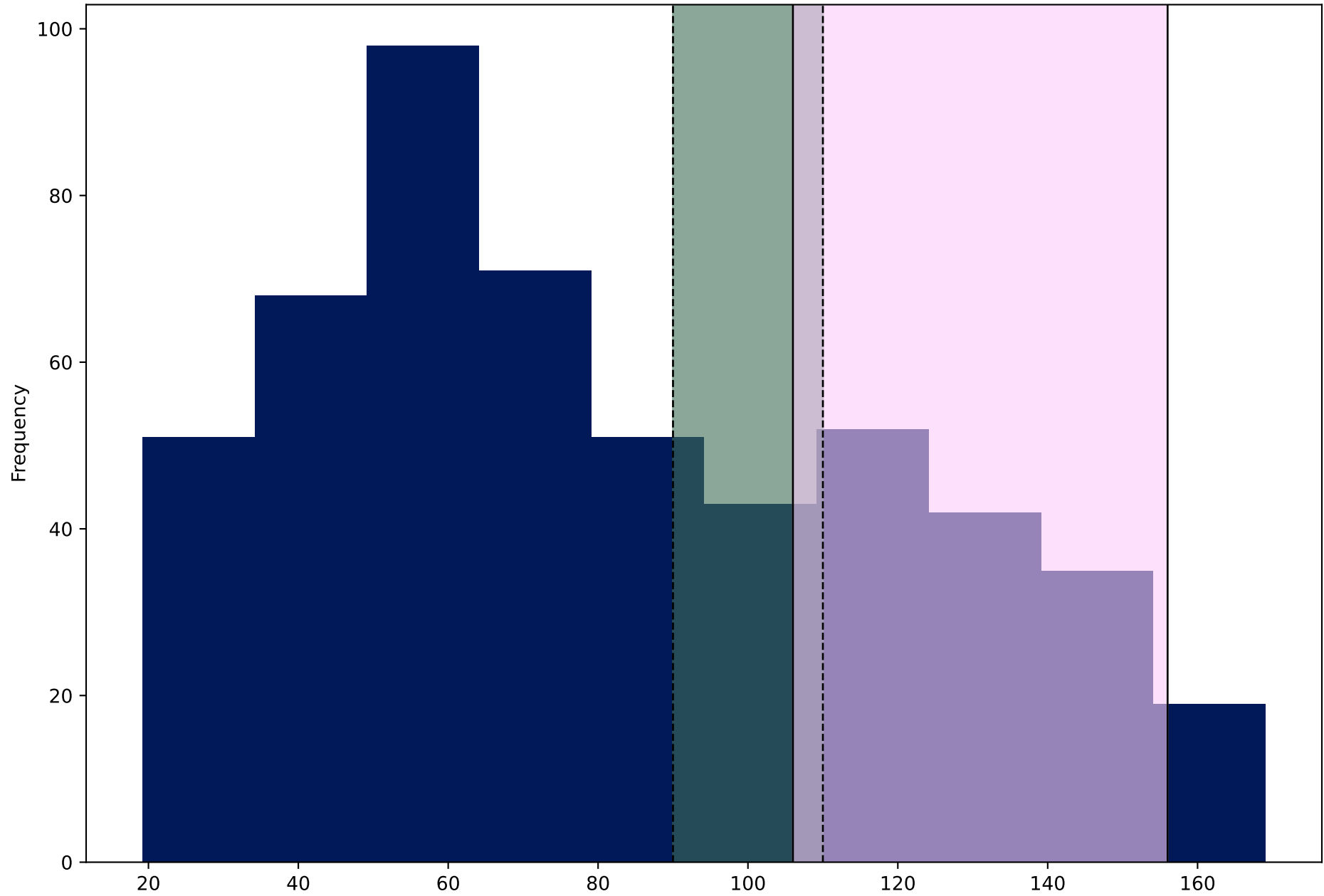

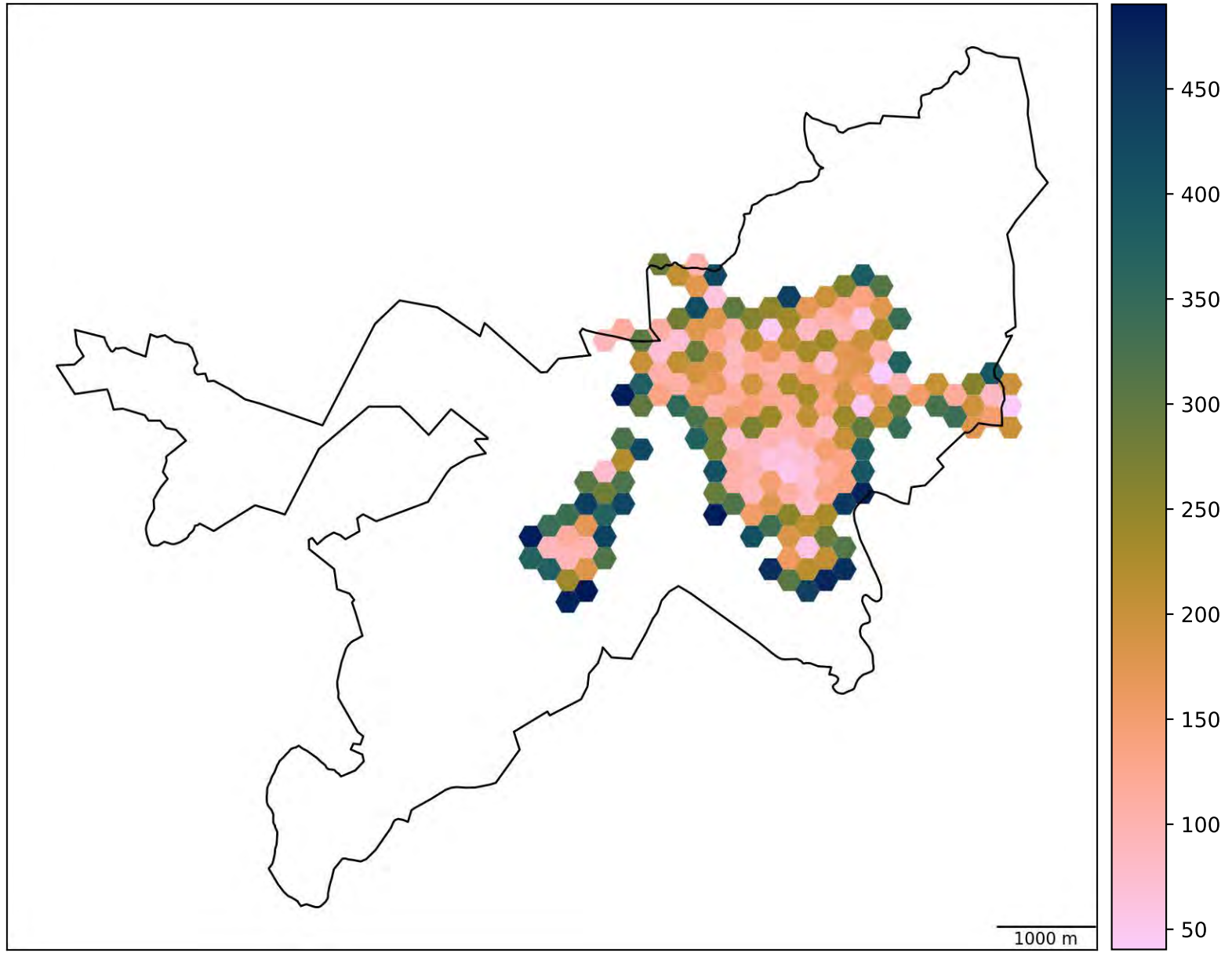

Distance to nearest public transport stops (m; up to 500m)

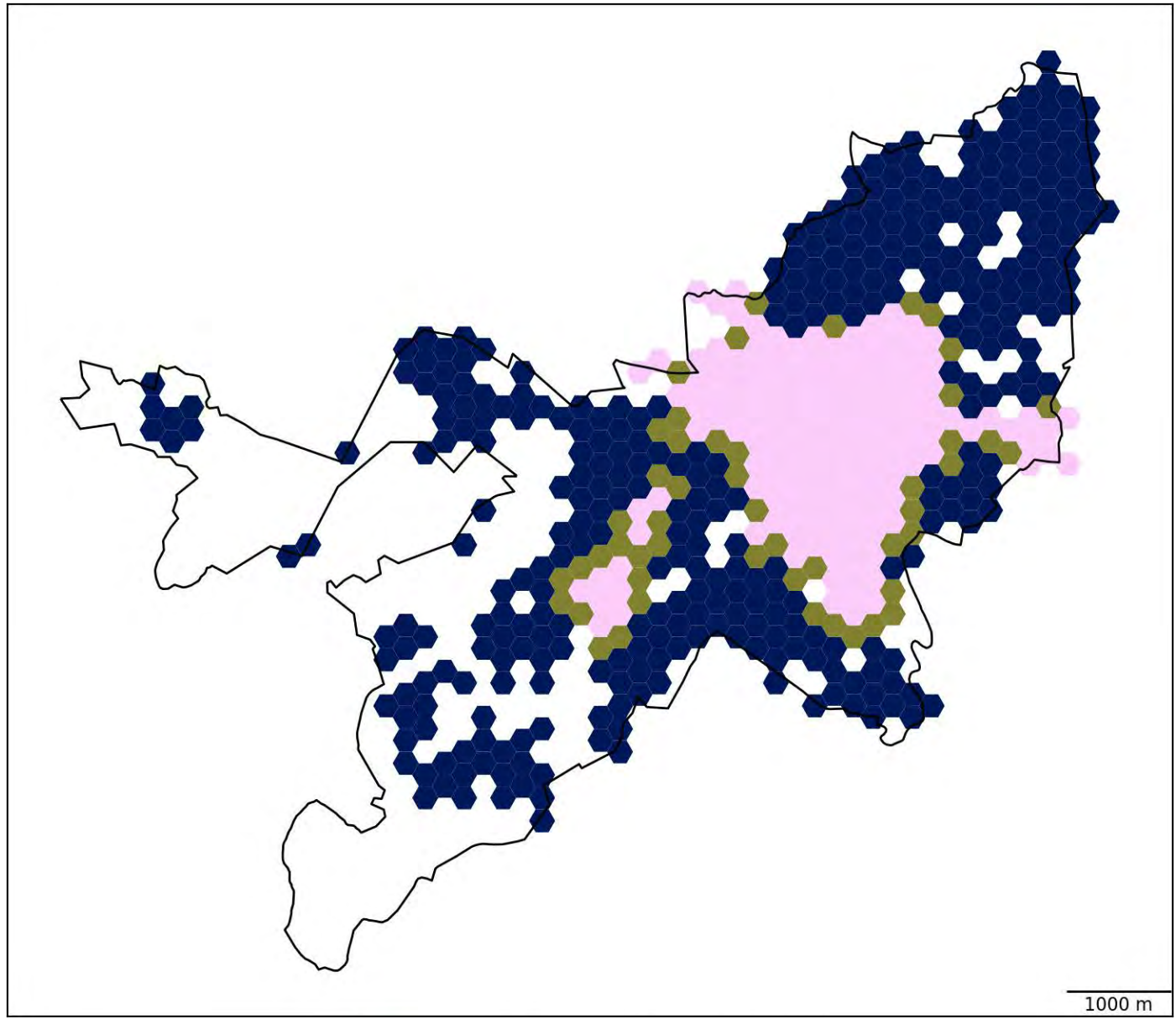

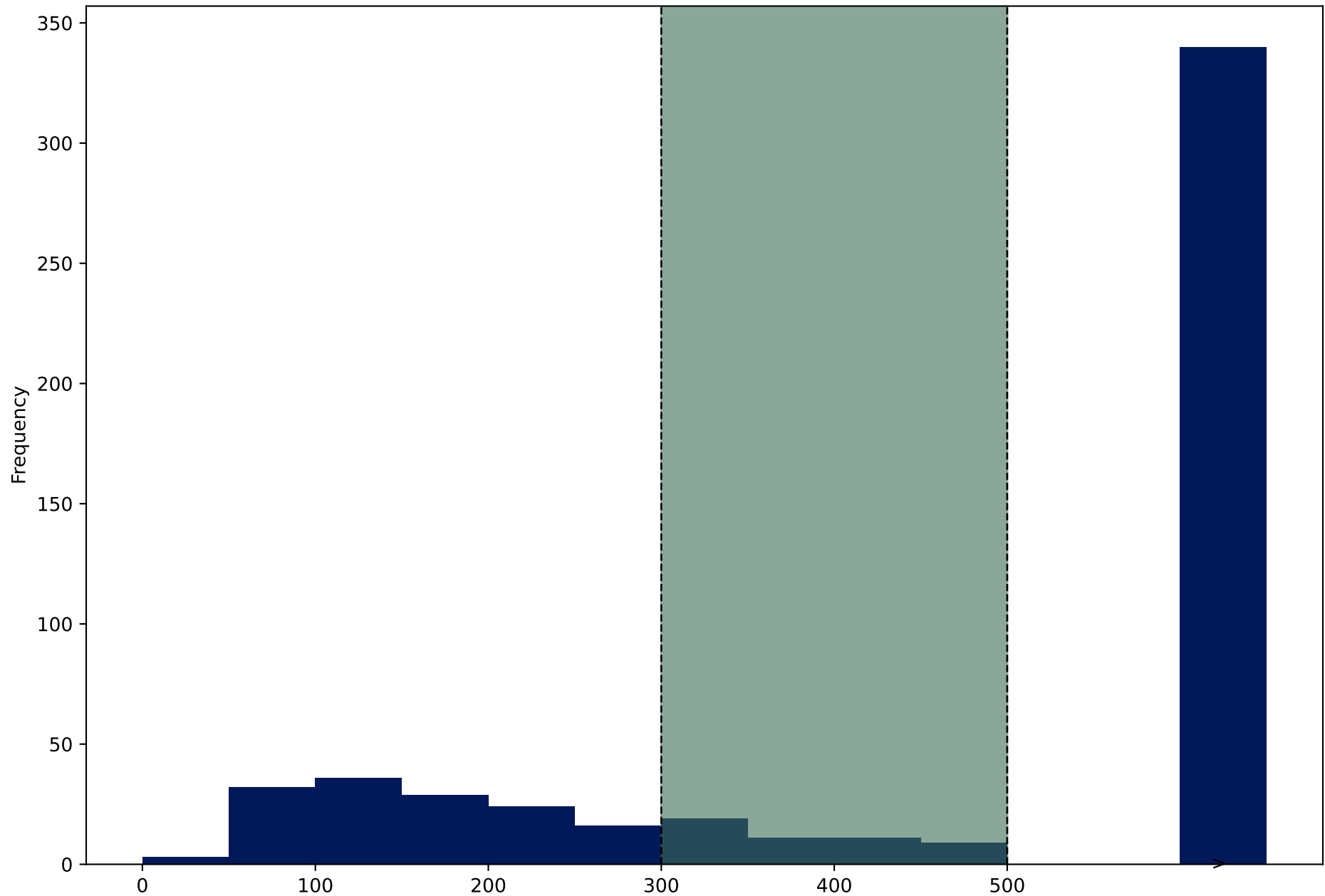

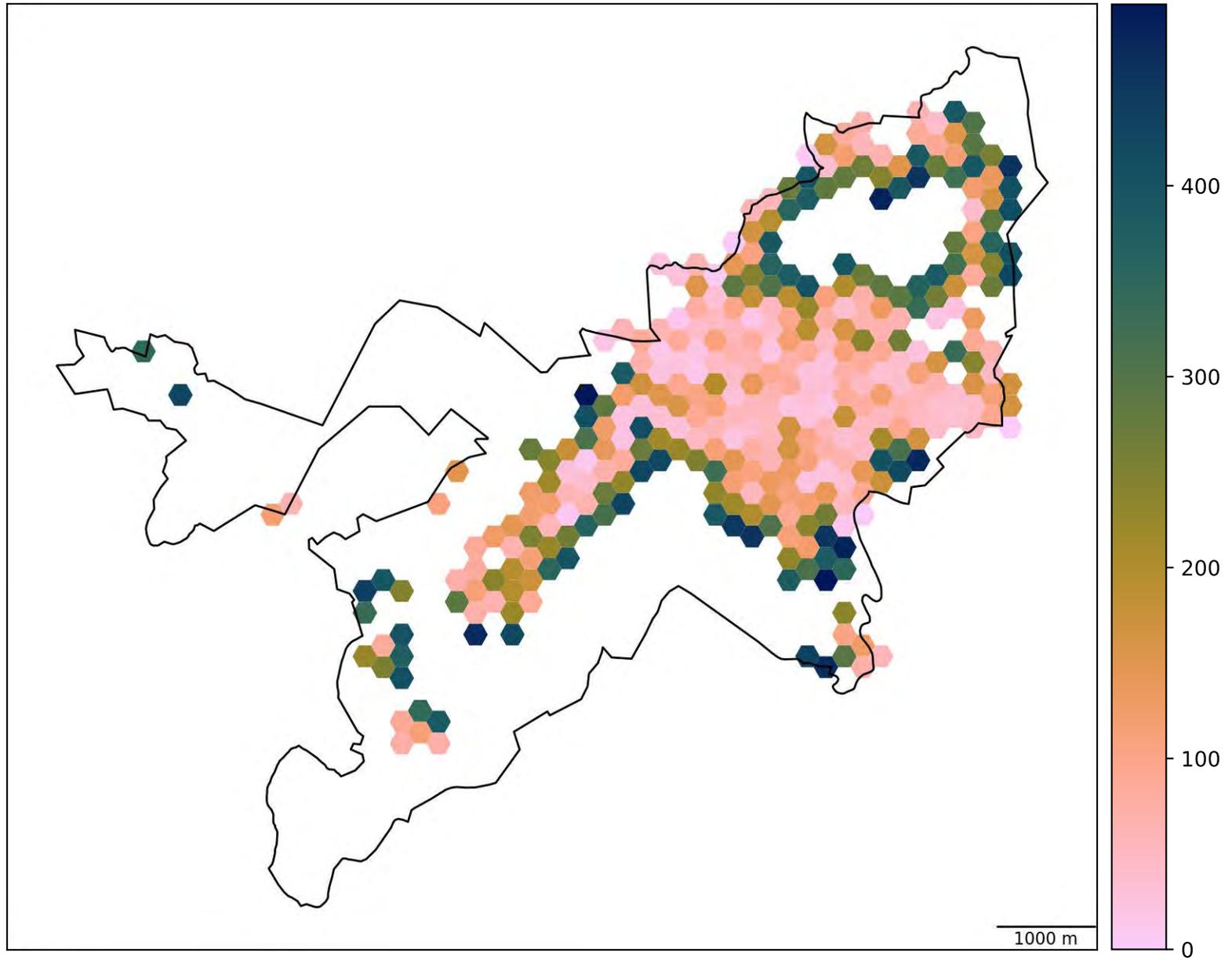

distances: Estimated Distance to nearest park (m; up to 500m) requirement for distances to destinations, measured up to a maximum distance target threshold of 500 metres

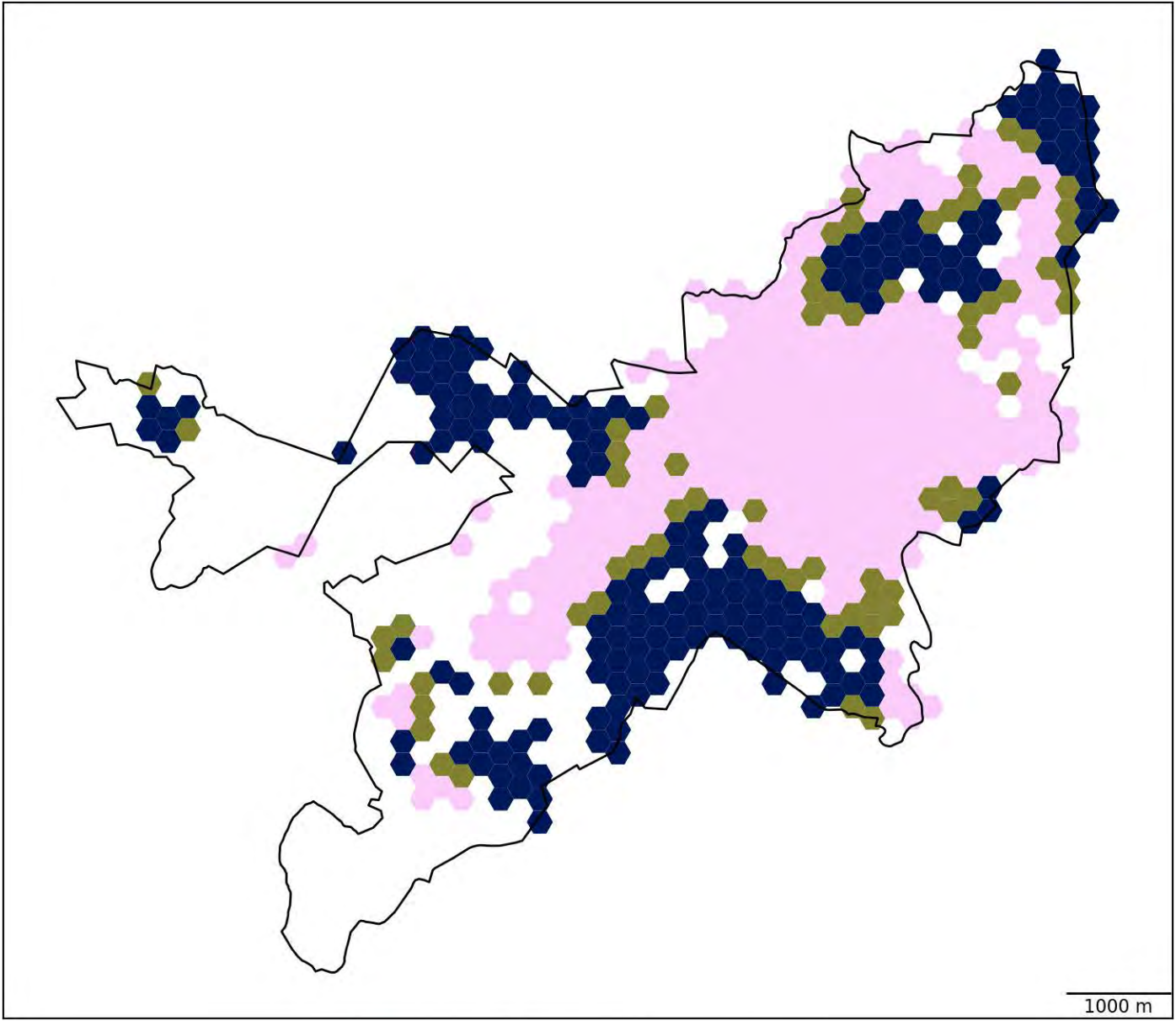

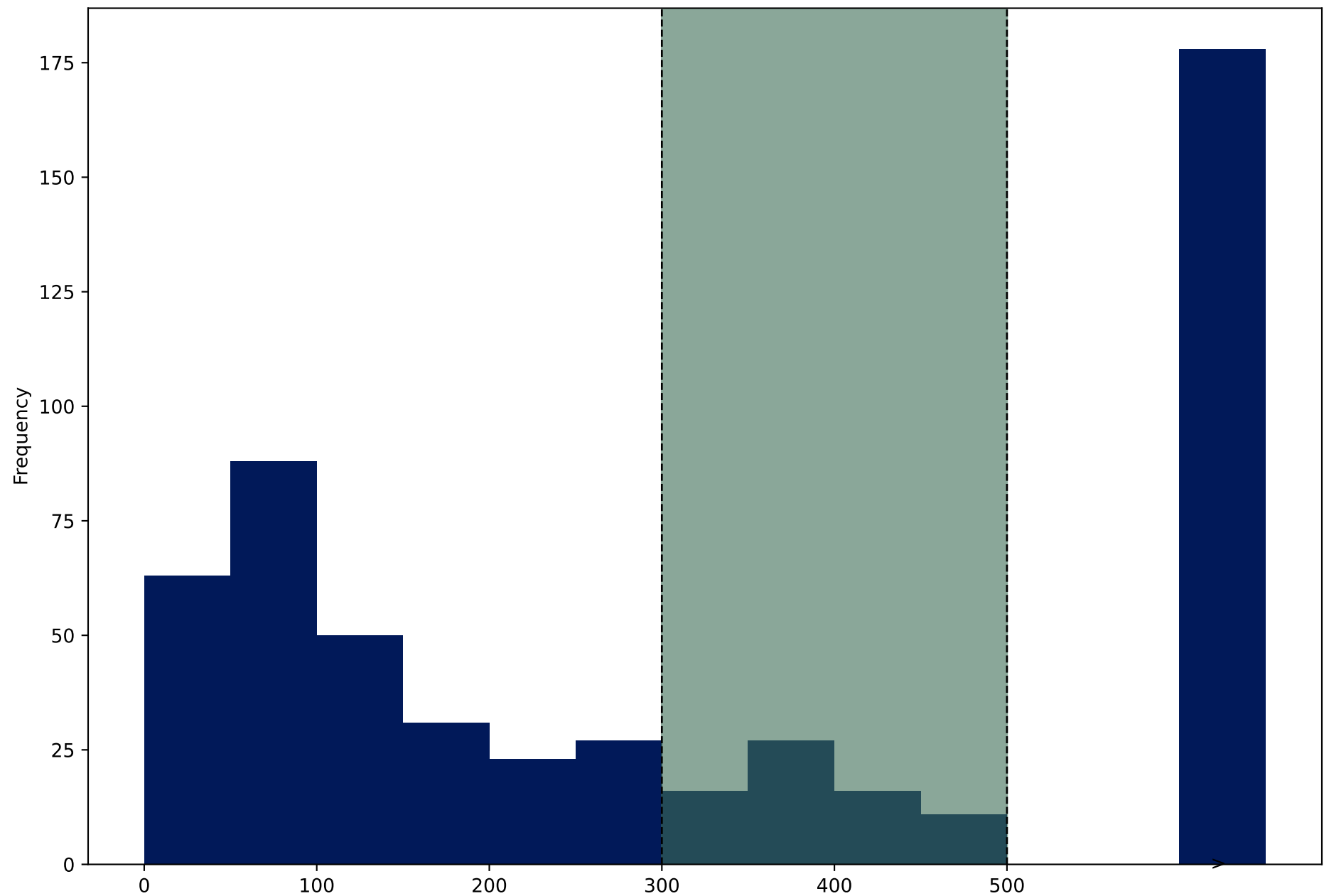

# Europe, Switzerland, Bern

| Satellite imagery of urban study region (Bing) | Walkability, relative to city | Walkability, relative to 25 global cities |
|---|---|---|

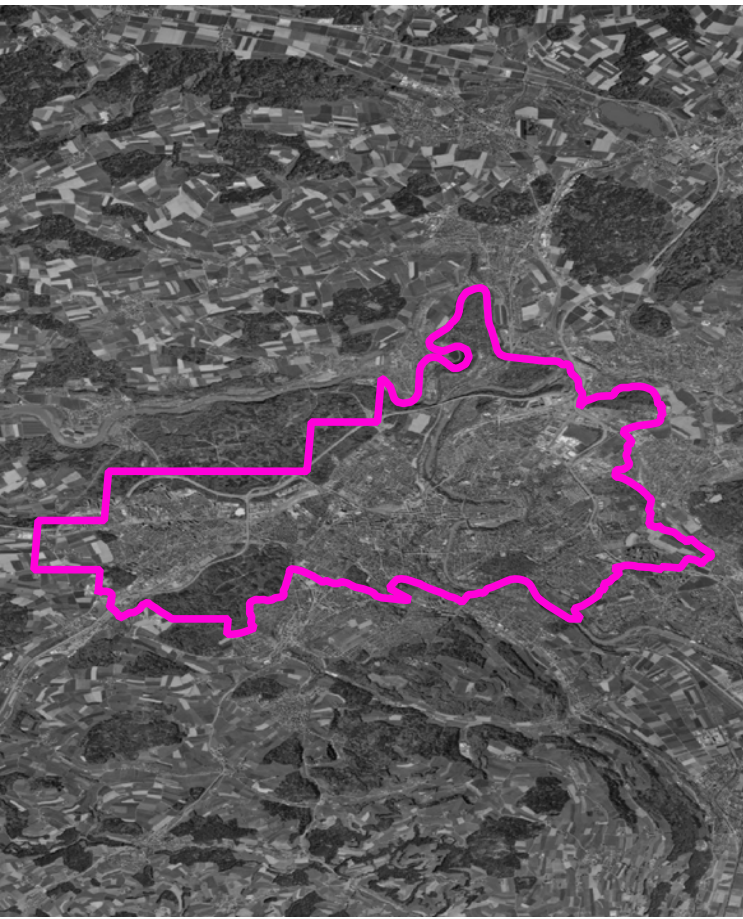
Urban boundary

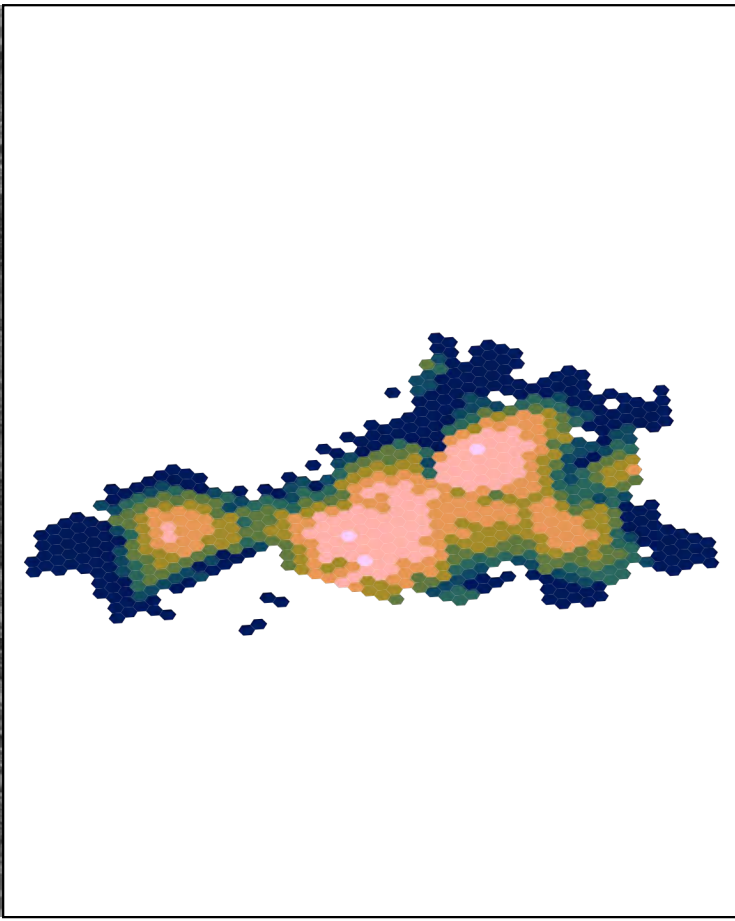

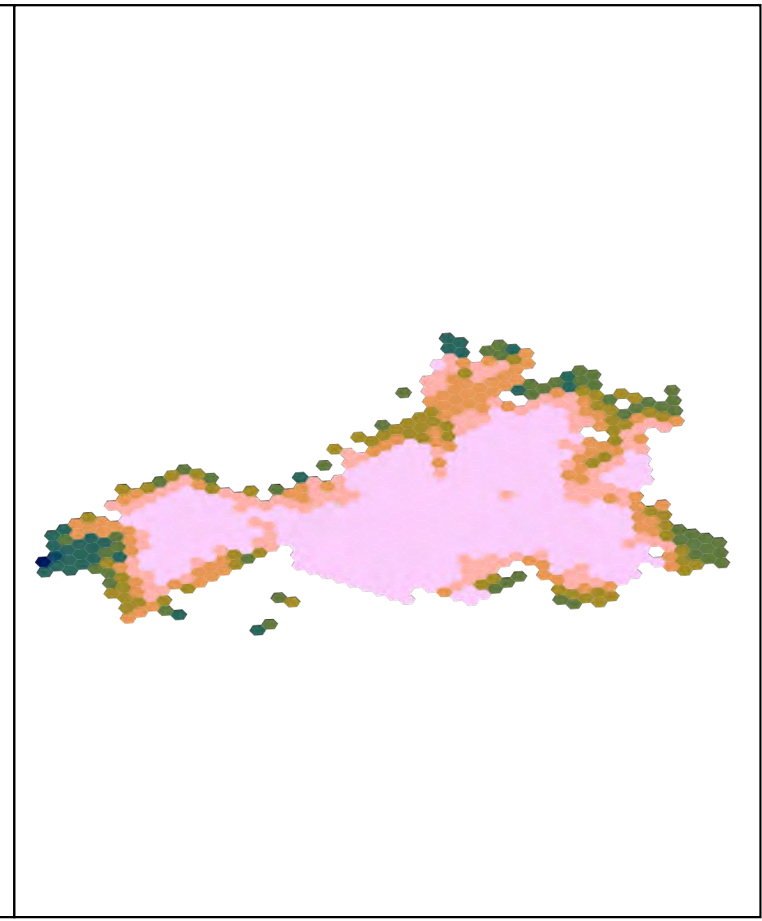

Walkability score
- <-3
- -3 to -2
- -2 to -1
- -1 to 0
- 0 to 1
- 1 to 2
- 2 to 3
- ≥3

Walkability relative to all cities by component variables (2D histograms), and overall (histogram)

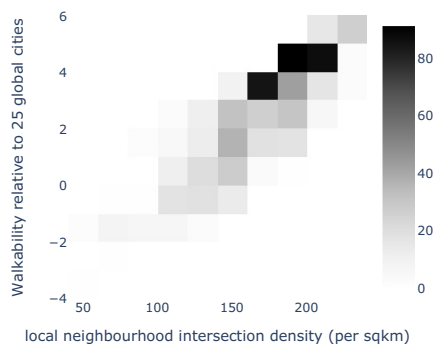
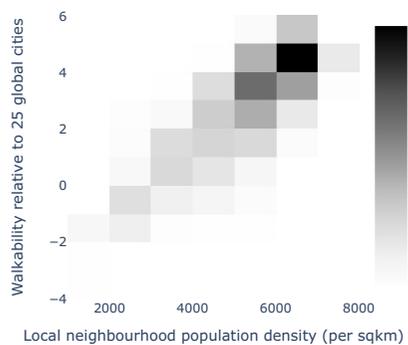
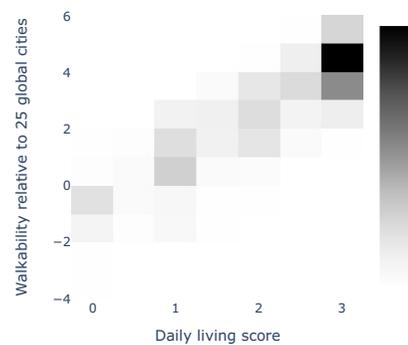
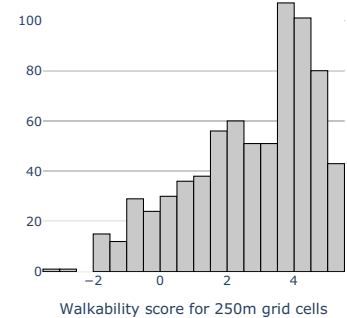

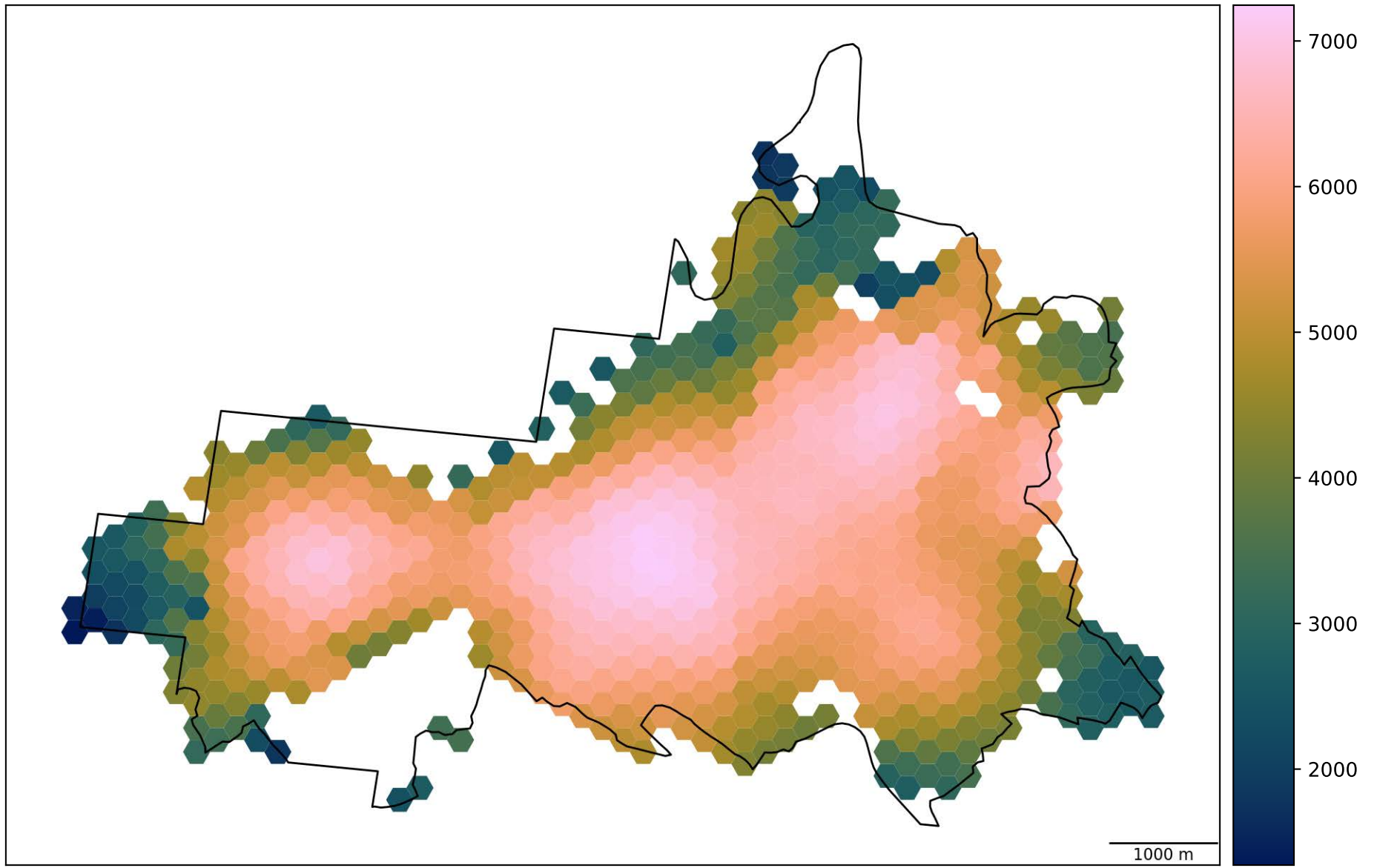

Mean 1000 m neighbourhood population per km²

A: Estimated Mean 1000 m neighbourhood population per km² requirement for ≥80% probability of engaging in walking for transport

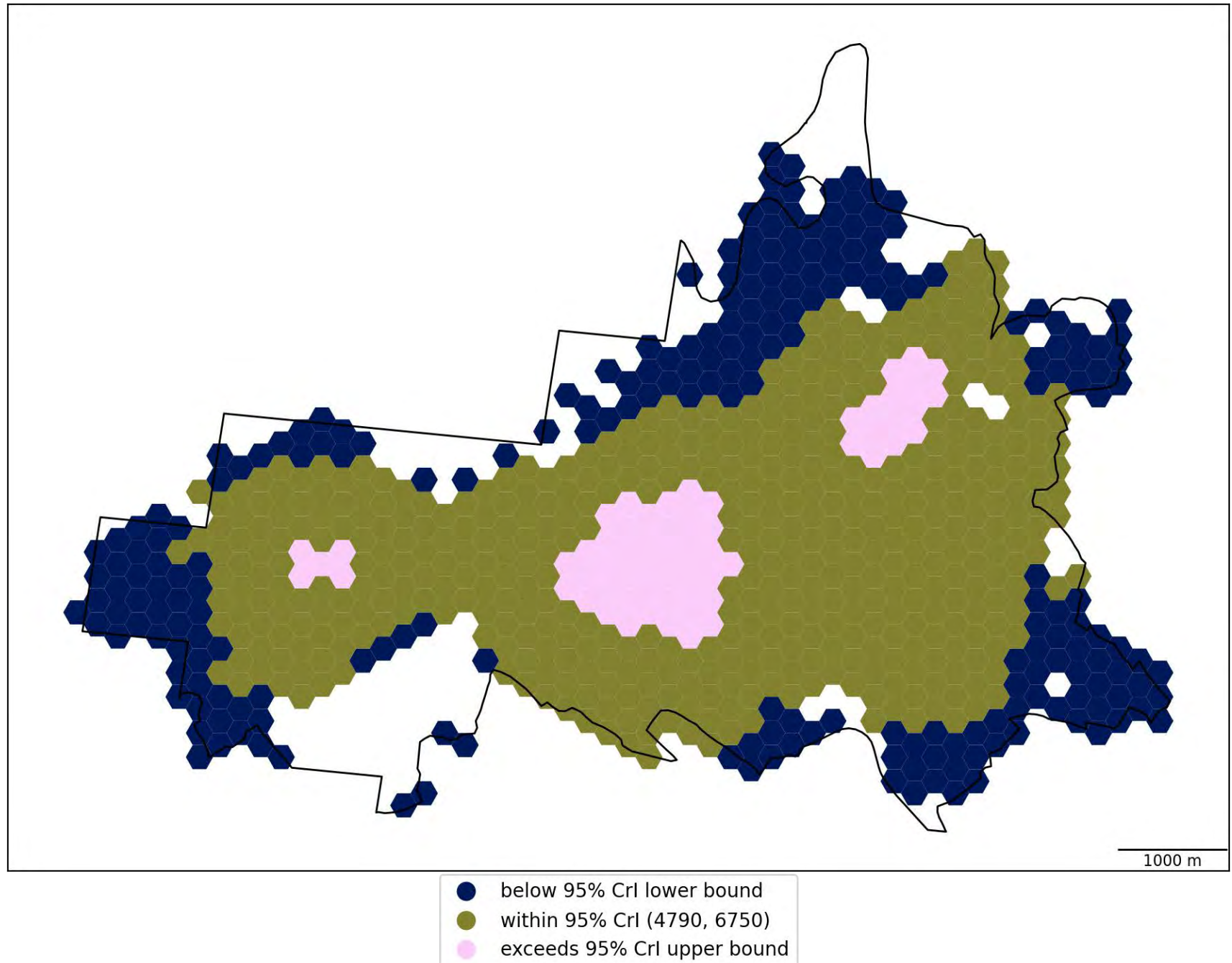

- below 95% CrI lower bound
- within 95% CrI (4790, 6750)
- exceeds 95% CrI upper bound

B: Estimated Mean 1000 m neighbourhood population per km² requirement for reaching the WHO's target of a ≥15% relative reduction in insufficient physical activity through walking

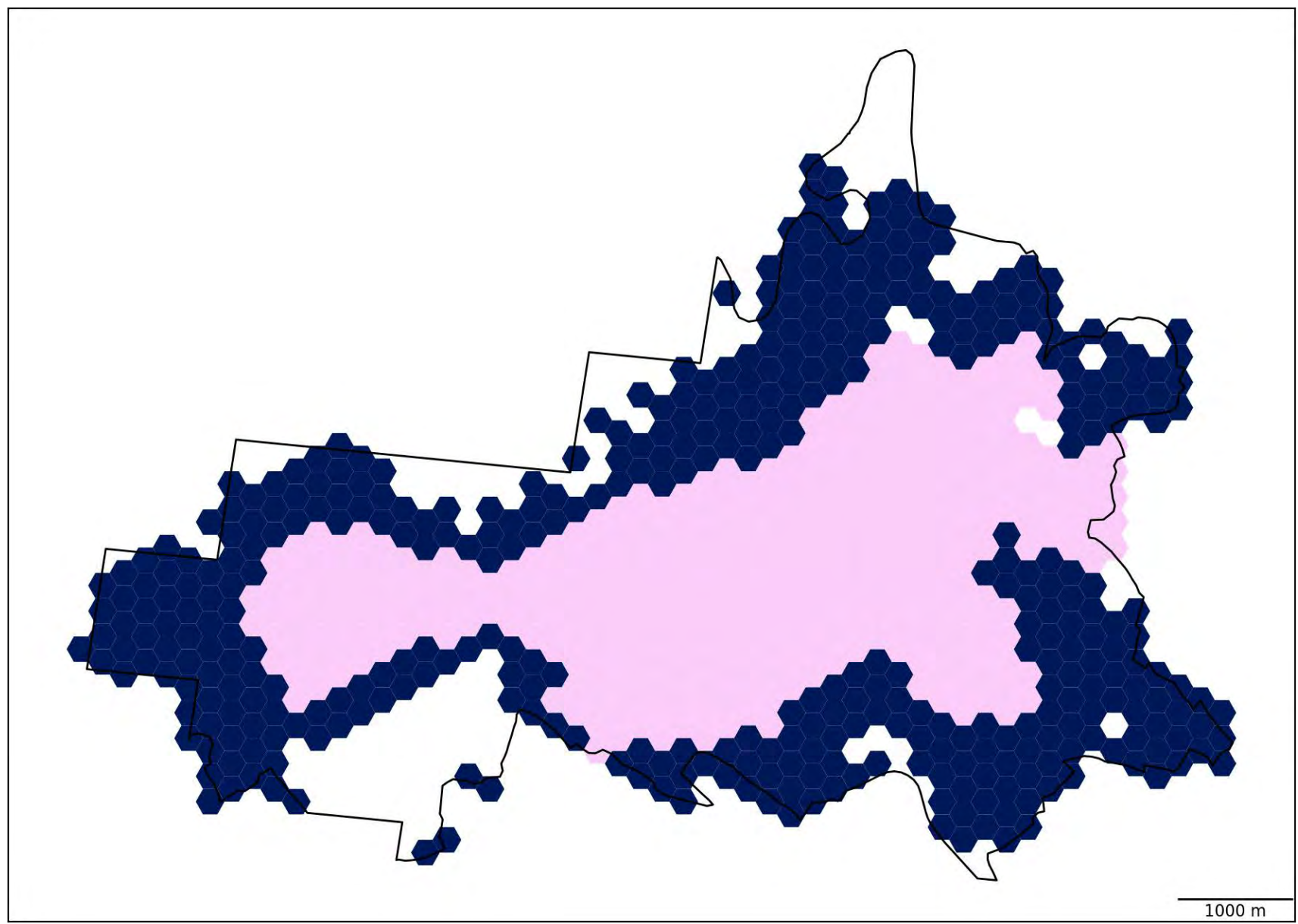

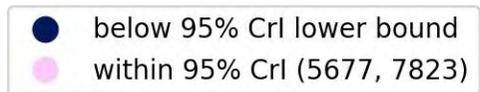

below 95% CrI lower bound
within 95% CrI (5677, 7823)

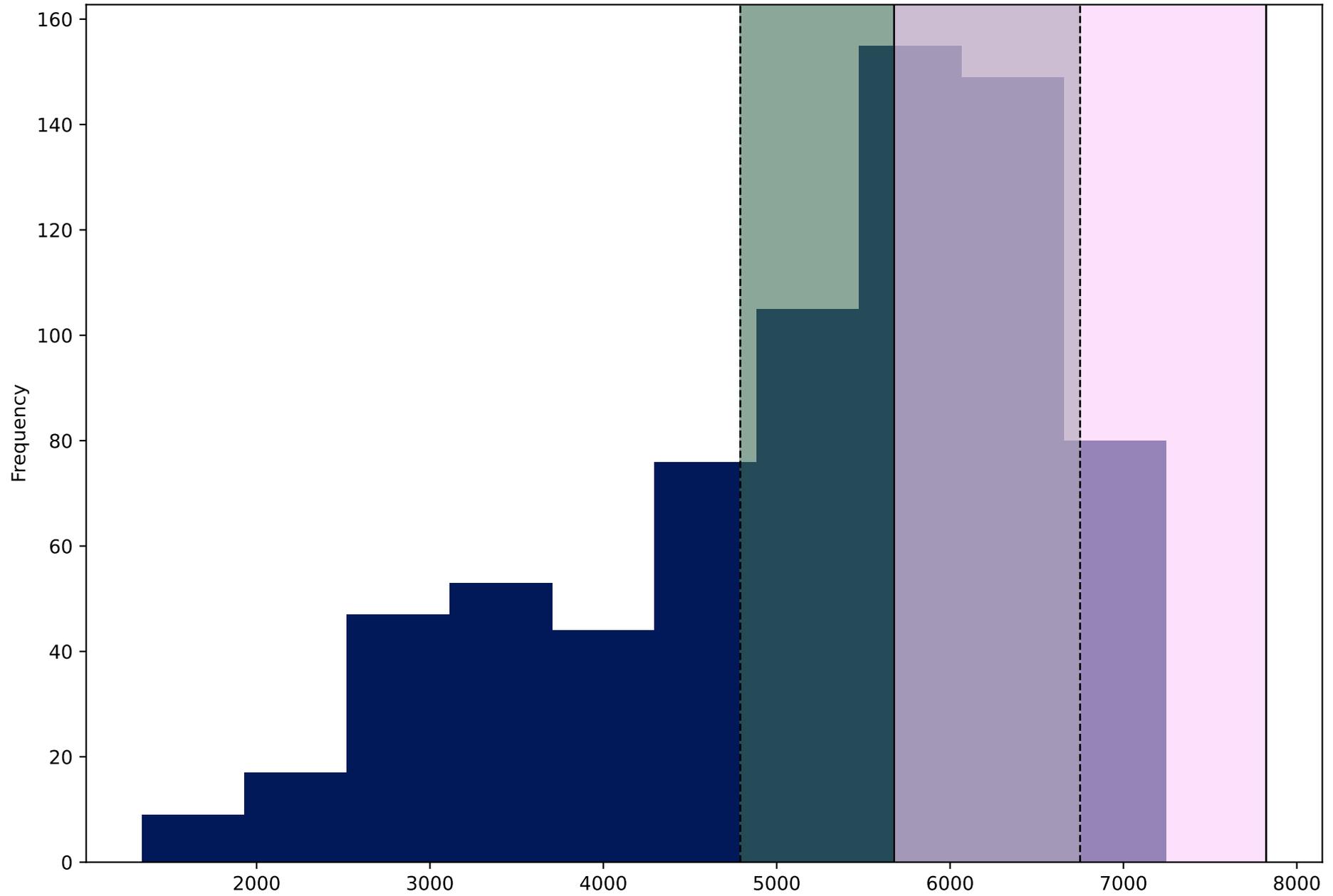

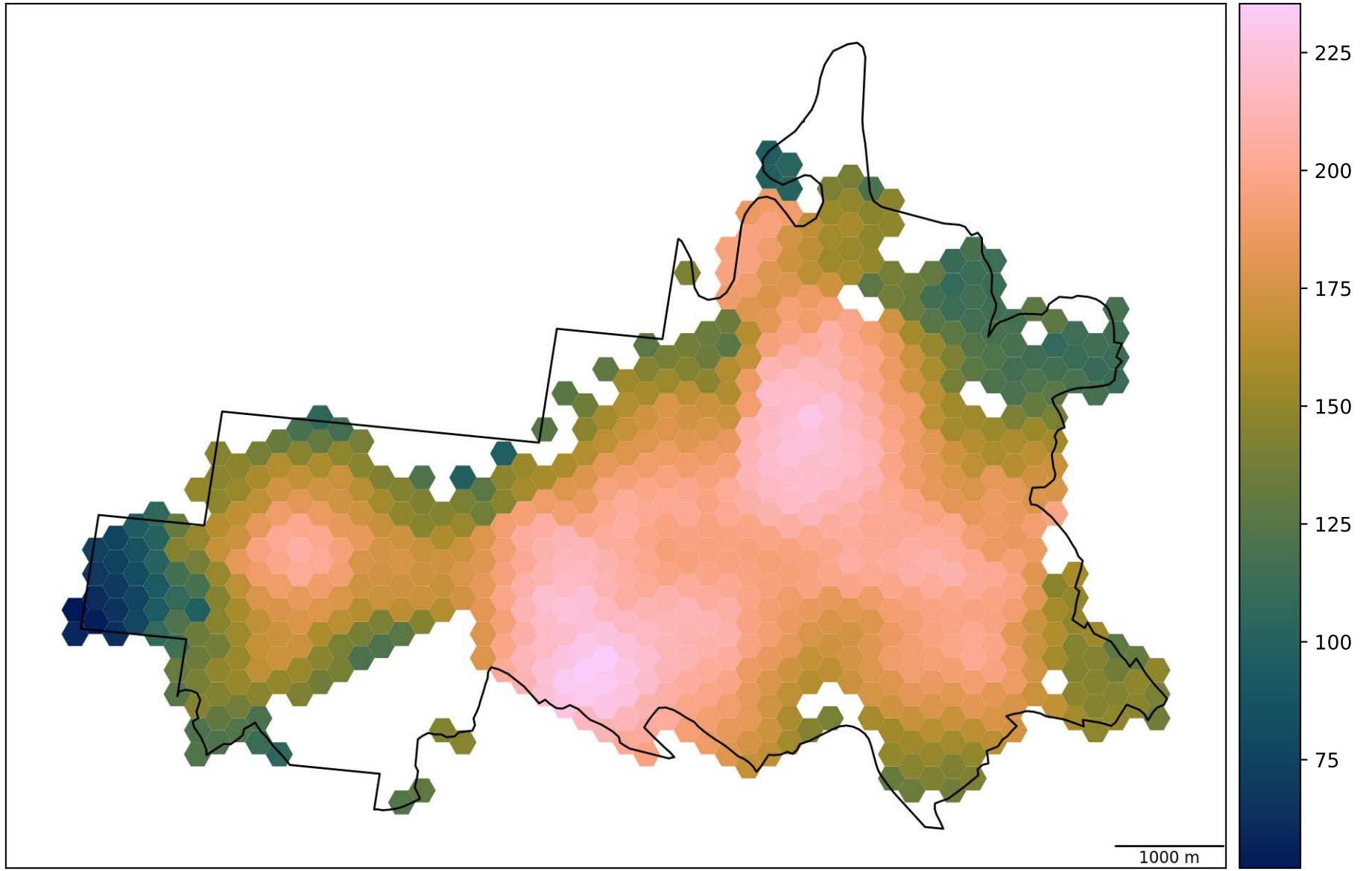

Mean 1000 m neighbourhood street intersections per km²

A: Estimated Mean 1000 m neighbourhood street intersections per km² requirement for ≥80% probability of engaging in walking for transport

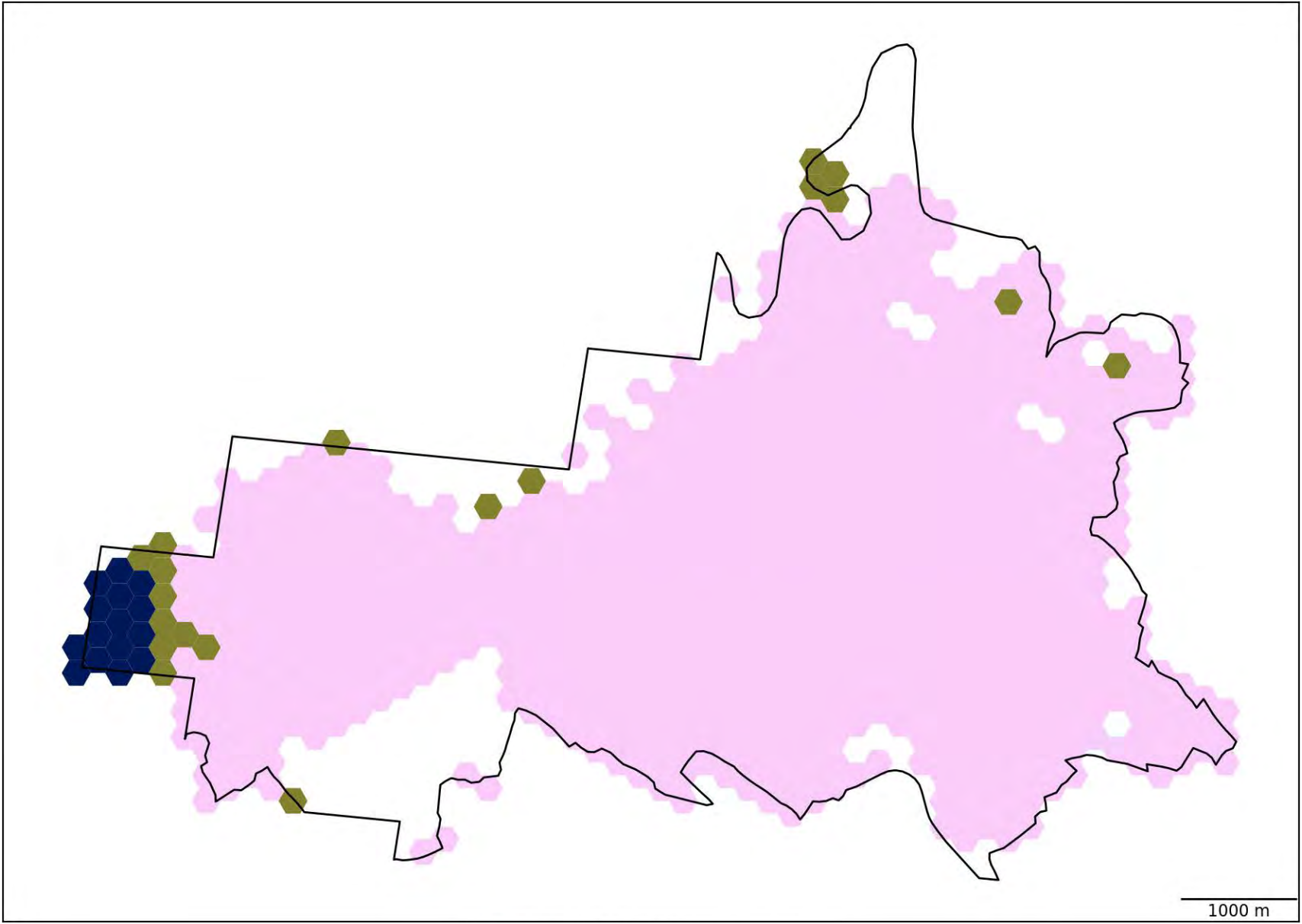

- below 95% CrI lower bound
- within 95% CrI (90, 110)
- exceeds 95% CrI upper bound

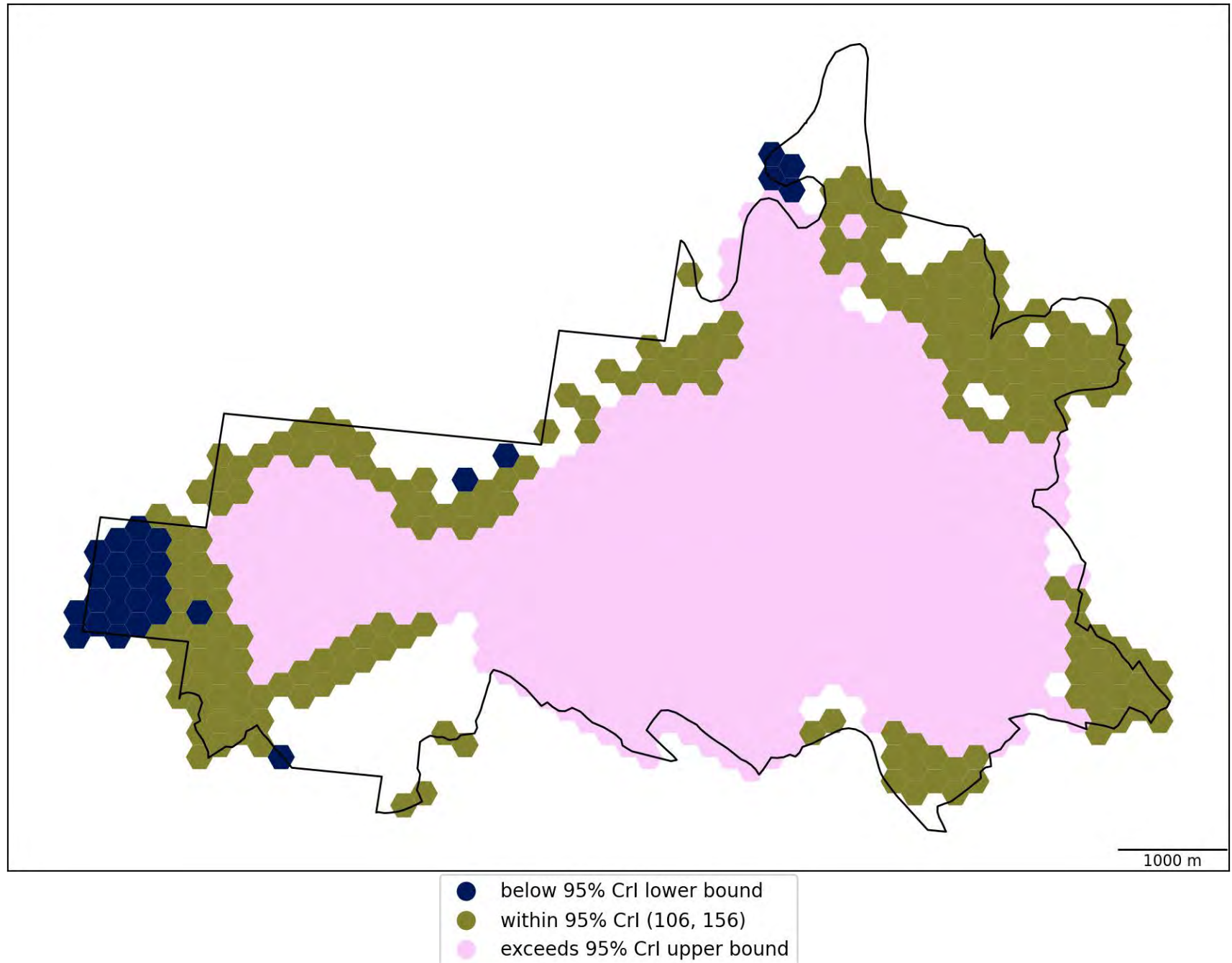

B: Estimated Mean 1000 m neighbourhood street intersections per km² requirement for reaching the WHO's target of a ≥15% relative reduction in insufficient physical activity through walking

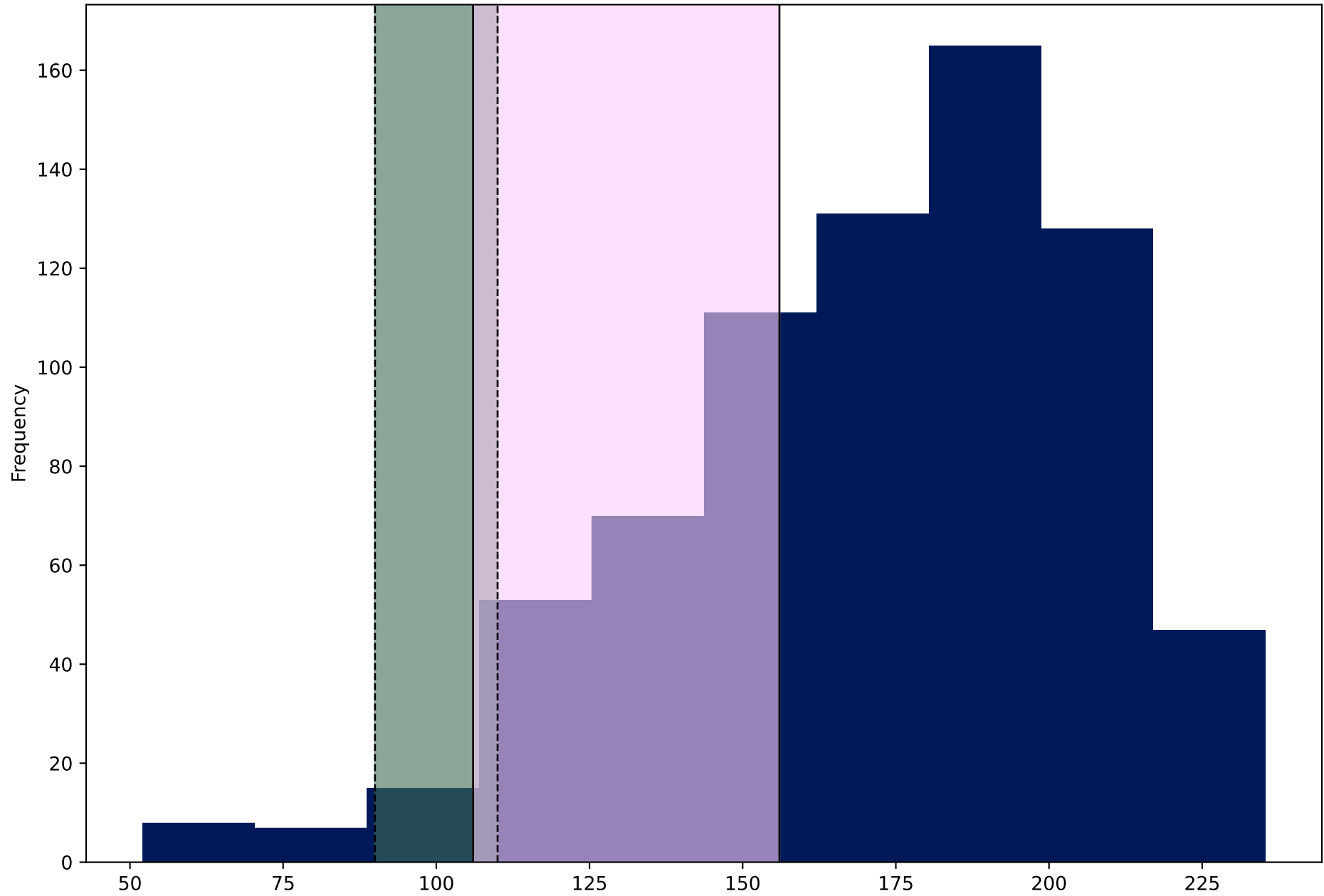

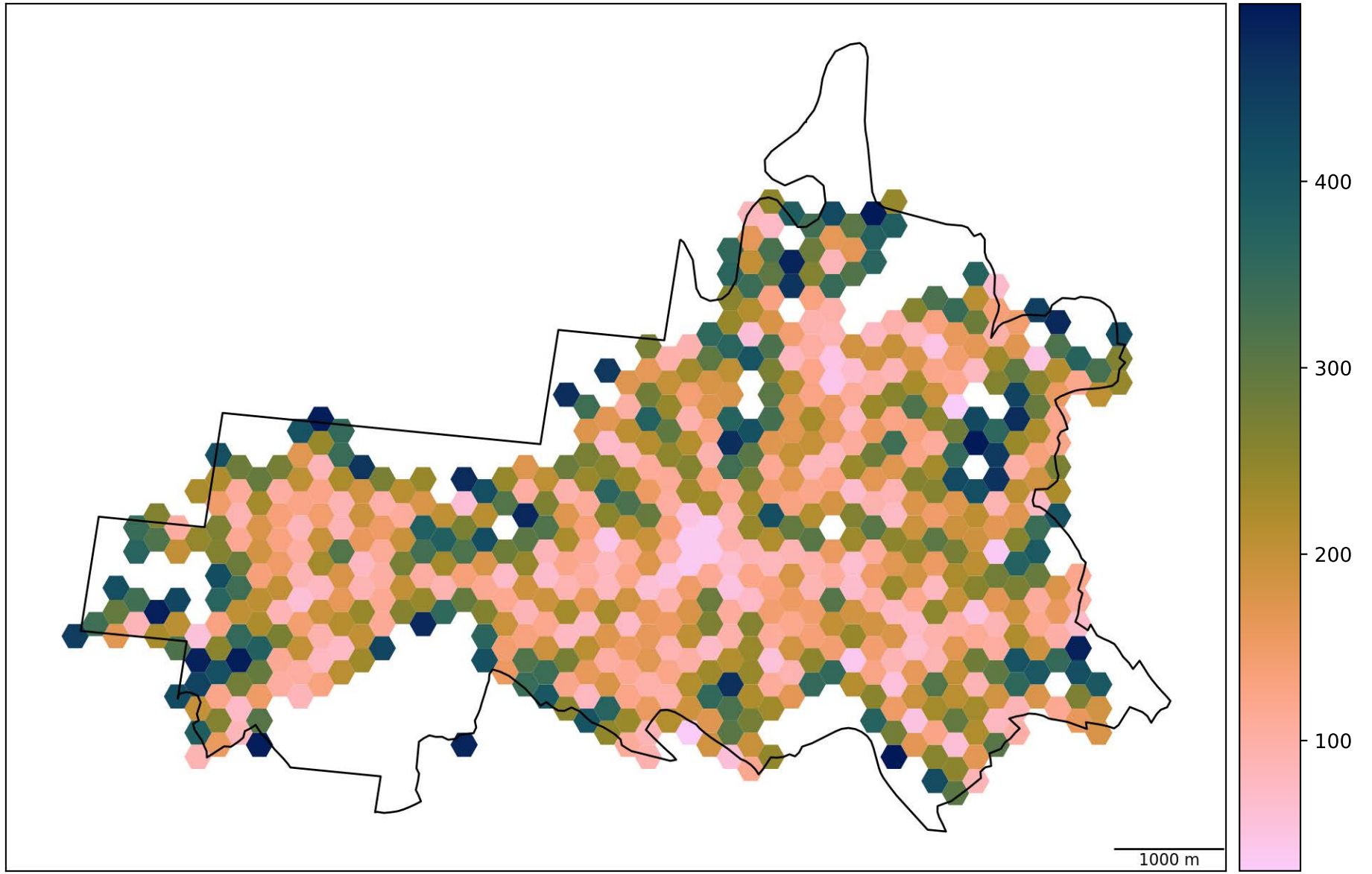

distances: Estimated Distance to nearest public transport stops (m; up to 500m) requirement for distances to destinations, measured up to a maximum distance target threshold of 500 metres

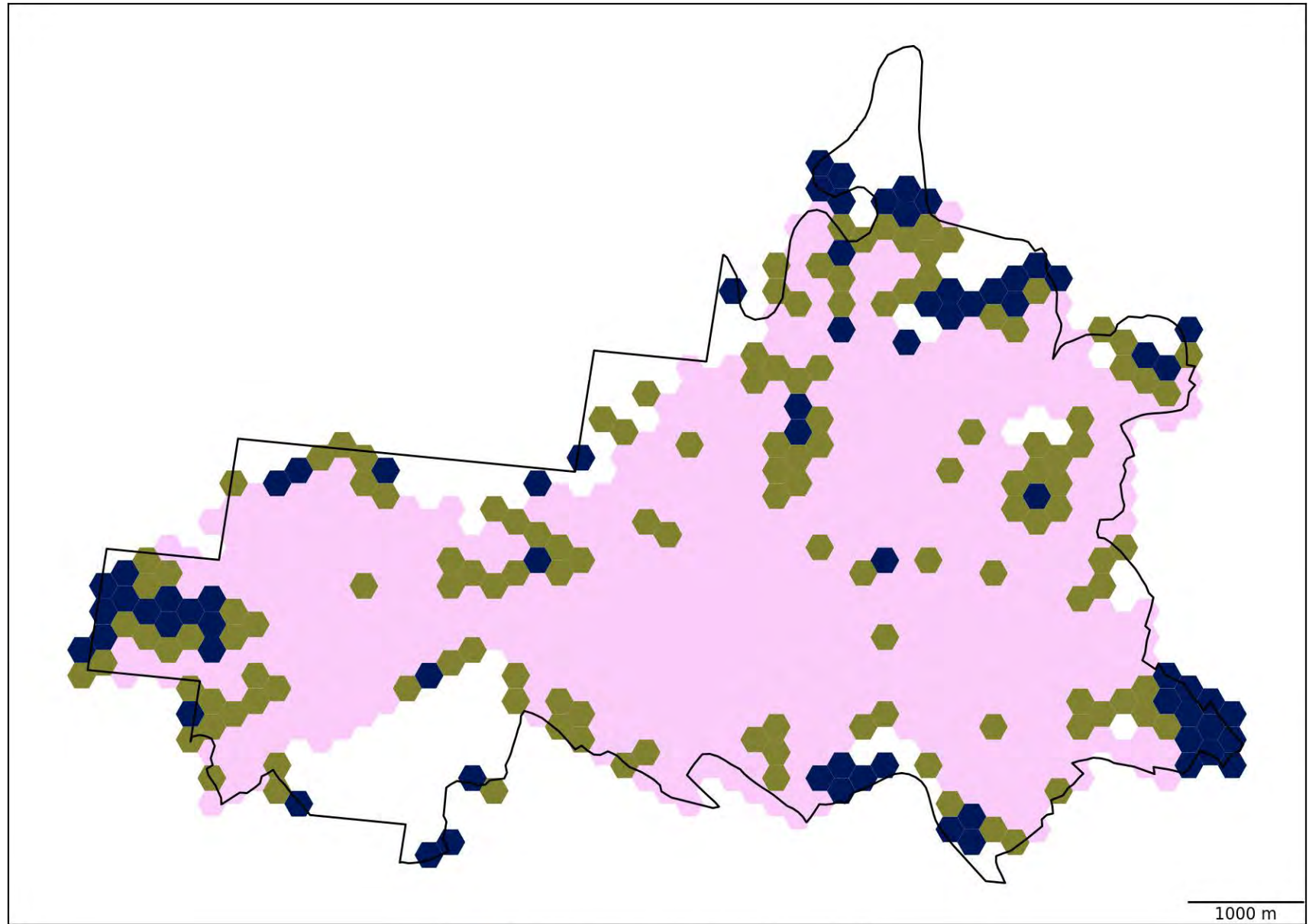

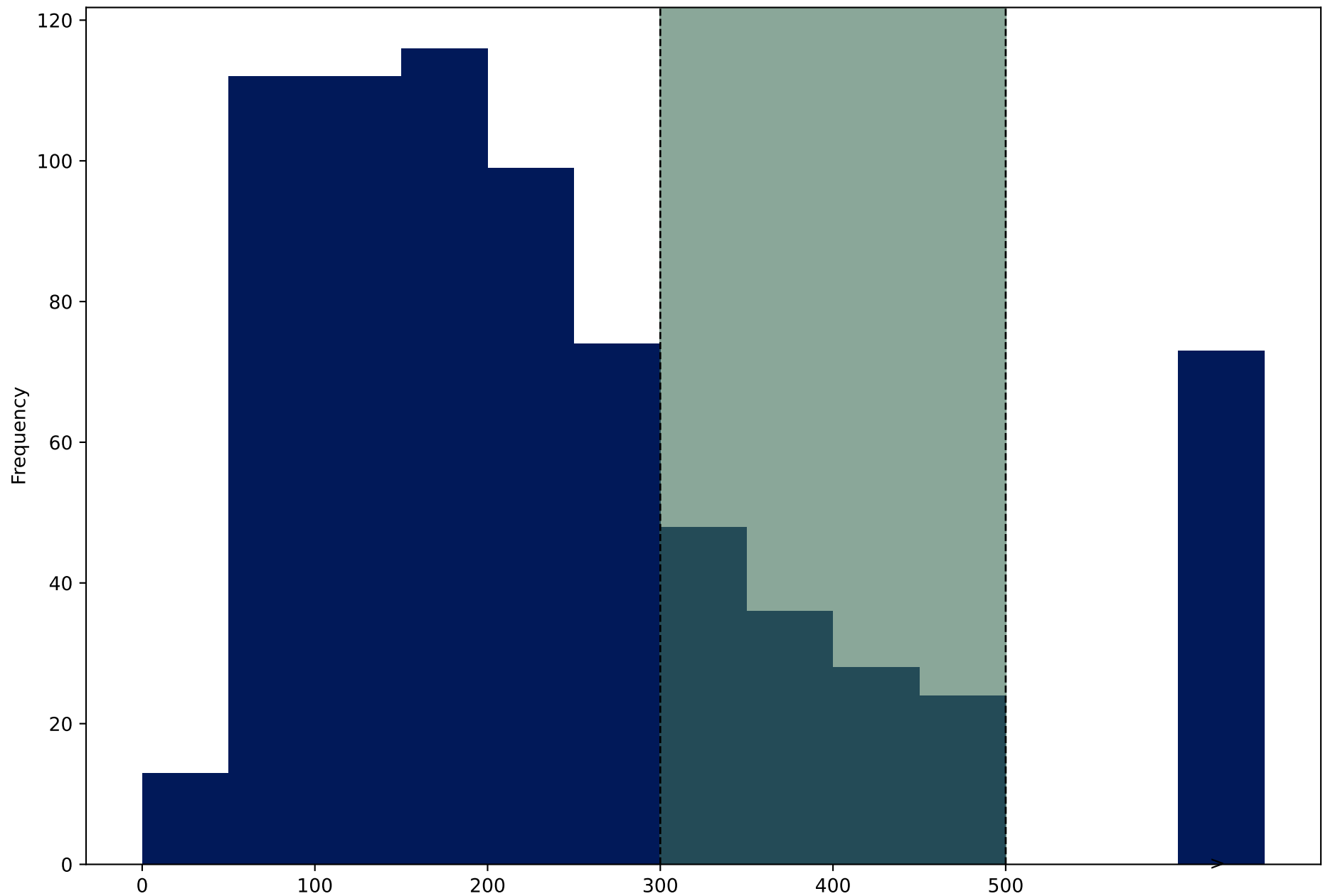

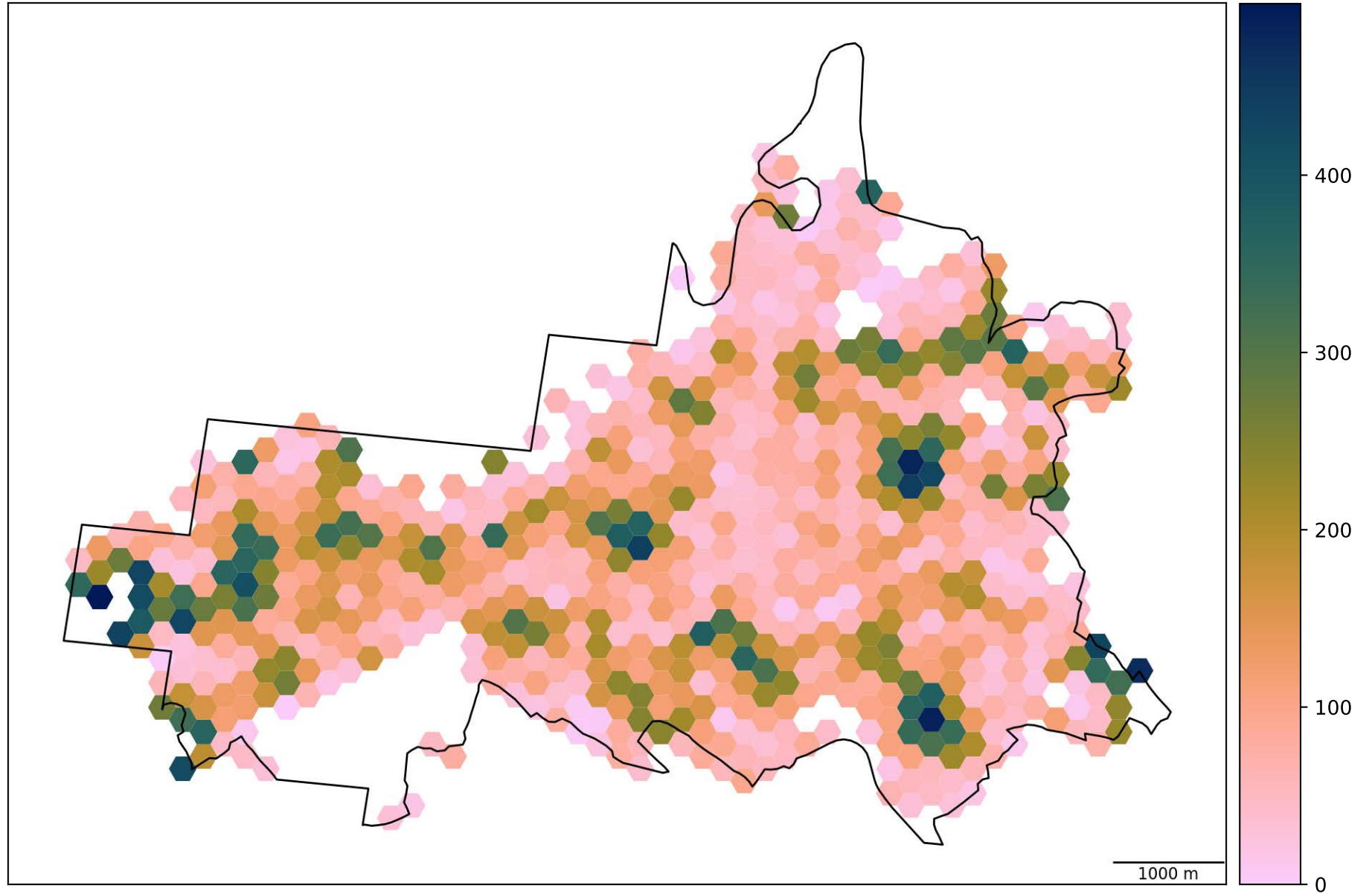

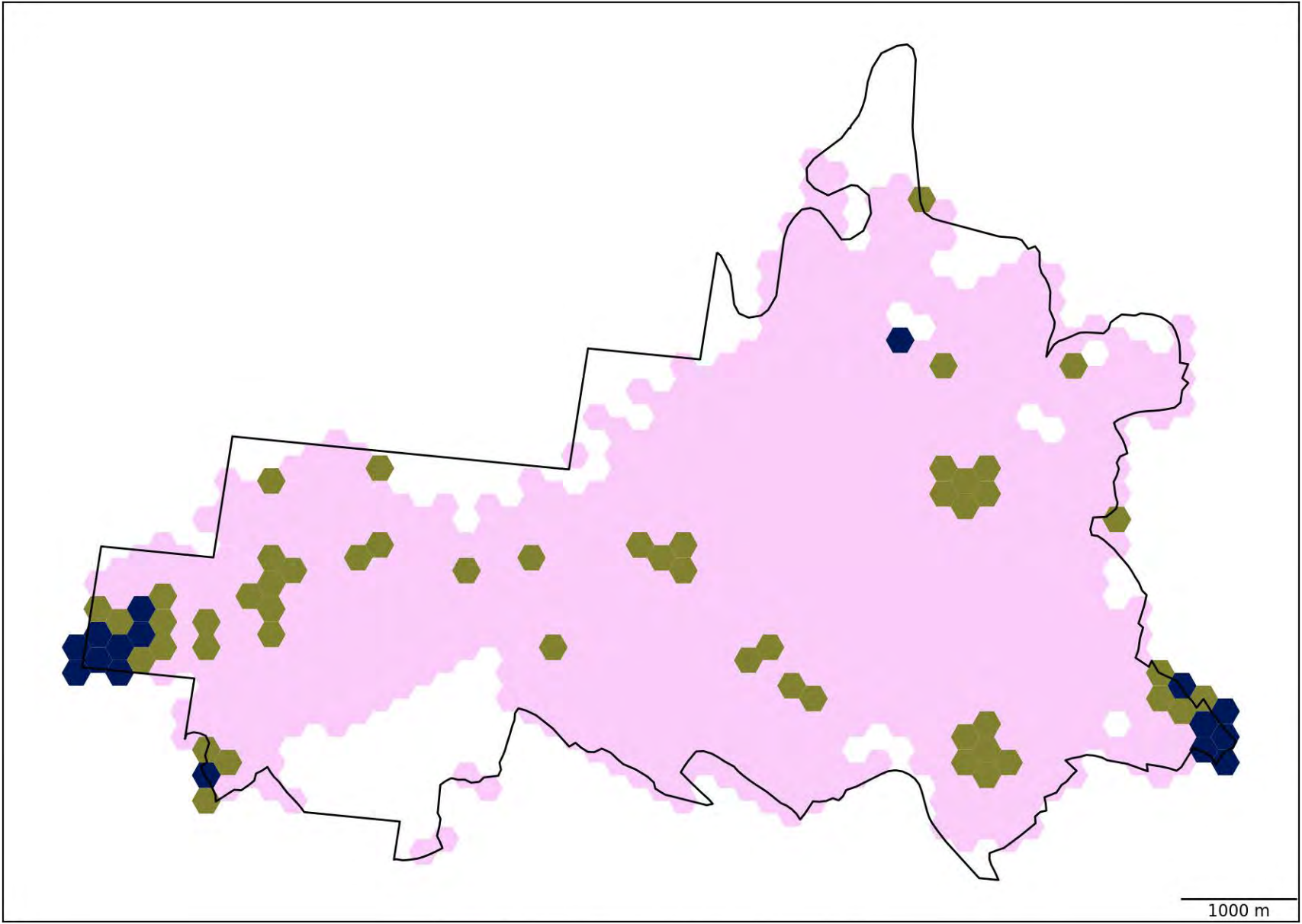

distances: Estimated Distance to nearest park (m; up to 500m) requirement for distances to destinations, measured up to a maximum distance target threshold of 500 metres

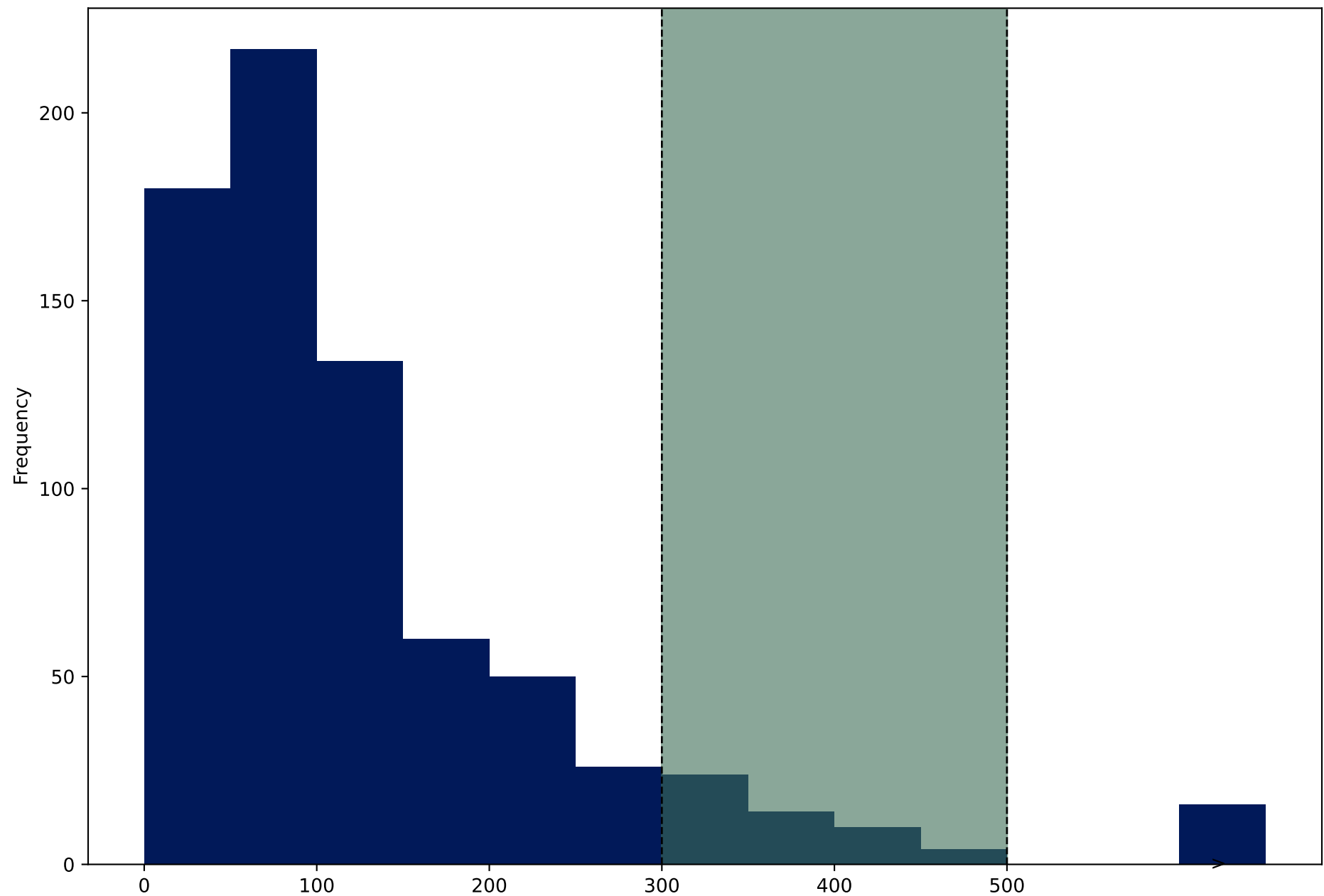

# Europe, United Kingdom, Belfast

Satellite imagery of urban study region (Bing) | Walkability, relative to city | Walkability, relative to 25 global cities

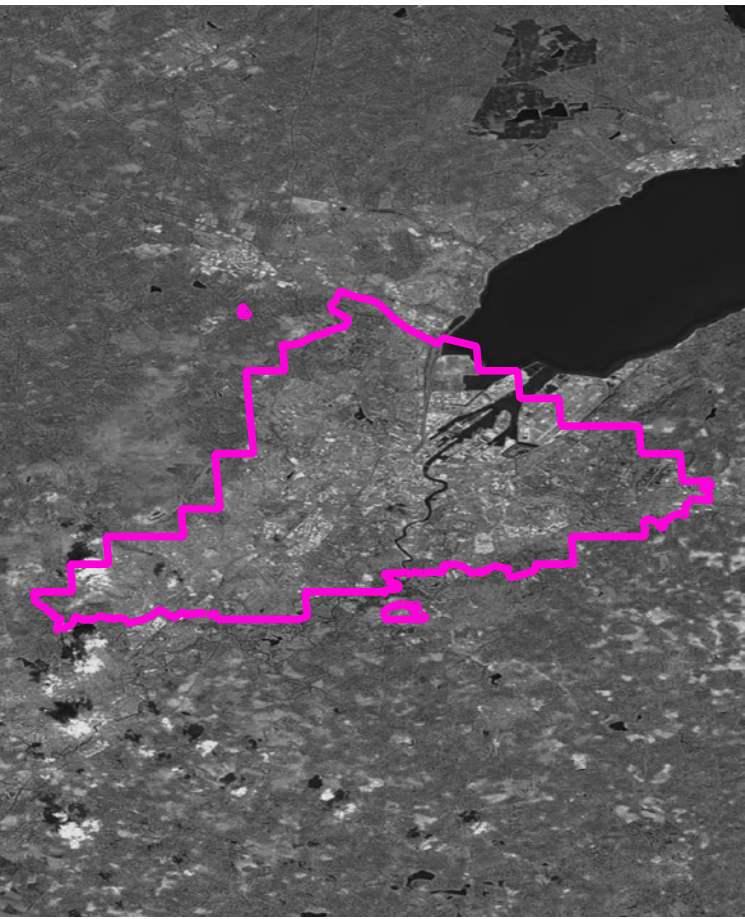
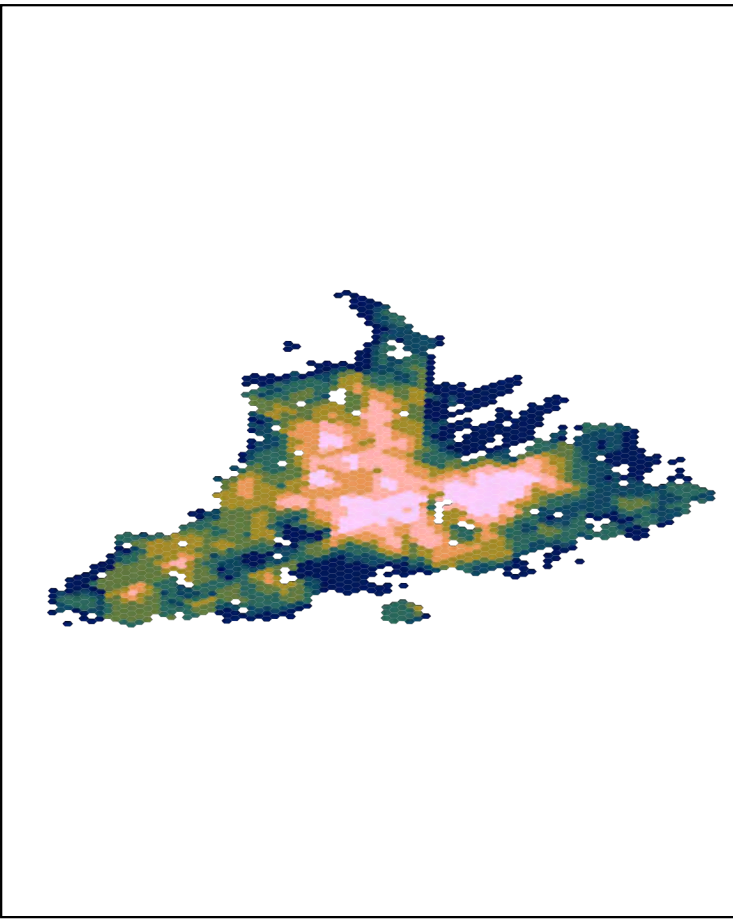
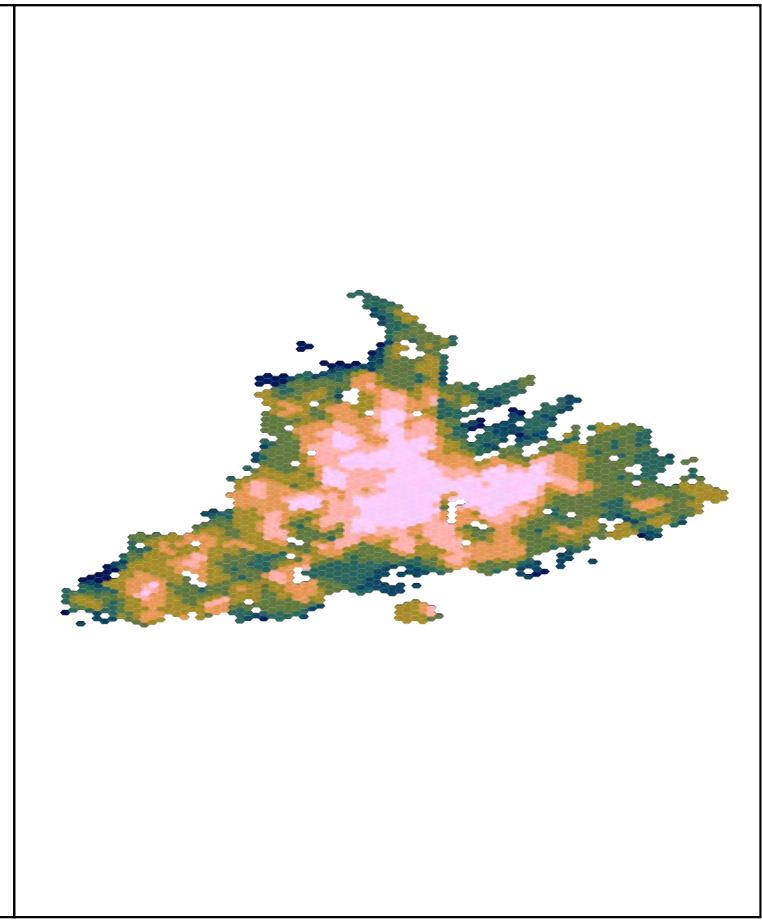

Urban boundary

0 — 7 — 14 km

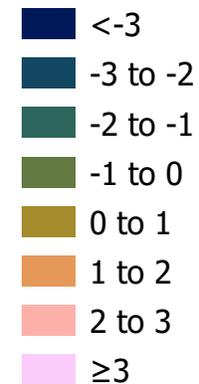

Walkability score
- <-3
- -3 to -2
- -2 to -1
- -1 to 0
- 0 to 1
- 1 to 2
- 2 to 3
- ≥3

Walkability relative to all cities by component variables (2D histograms), and overall (histogram)

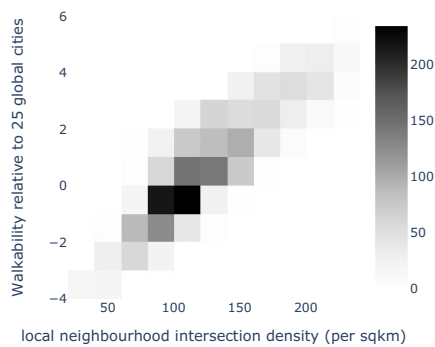
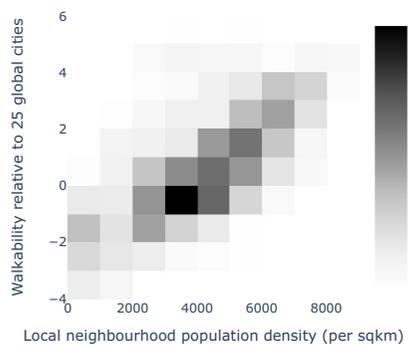
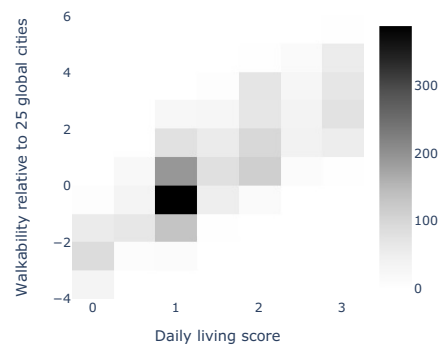
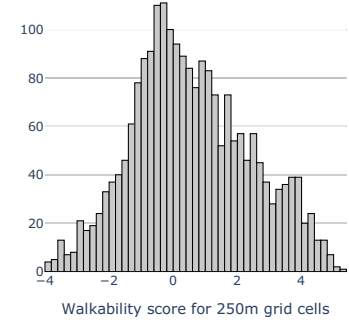

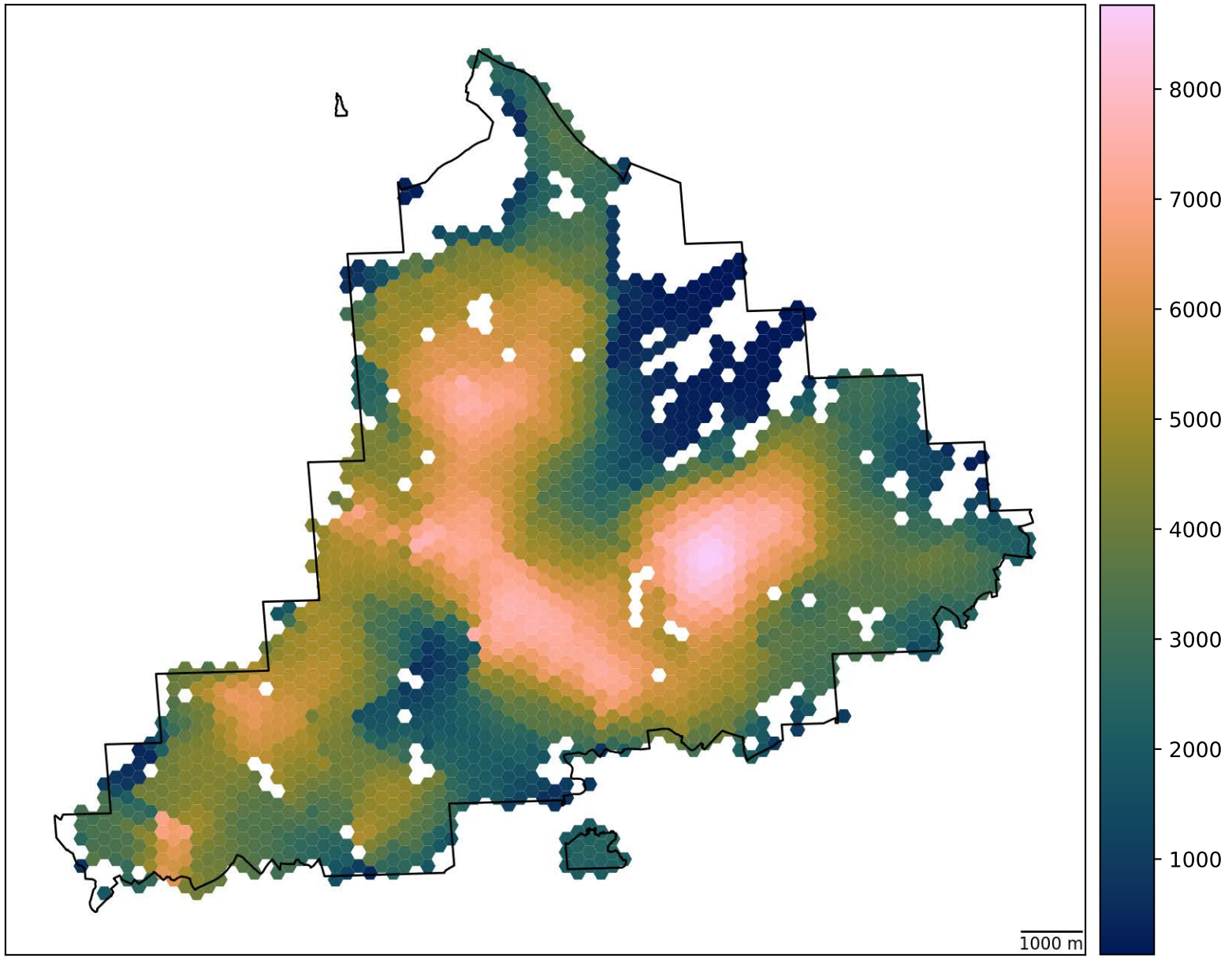

Mean 1000 m neighbourhood population per km²

A: Estimated Mean 1000 m neighbourhood population per km² requirement for ≥80% probability of engaging in walking for transport

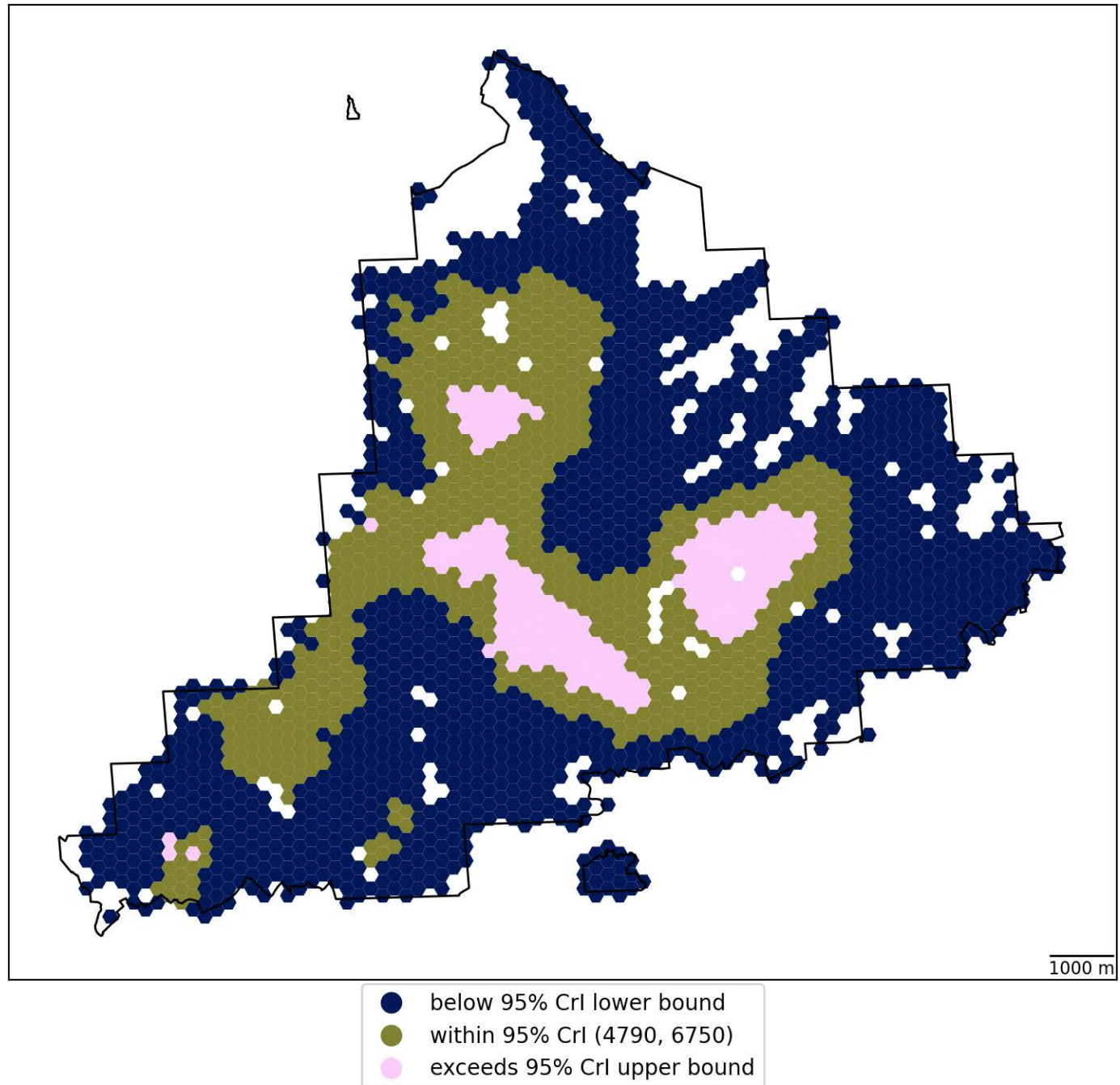

- below 95% CrI lower bound
- within 95% CrI (4790, 6750)
- exceeds 95% CrI upper bound

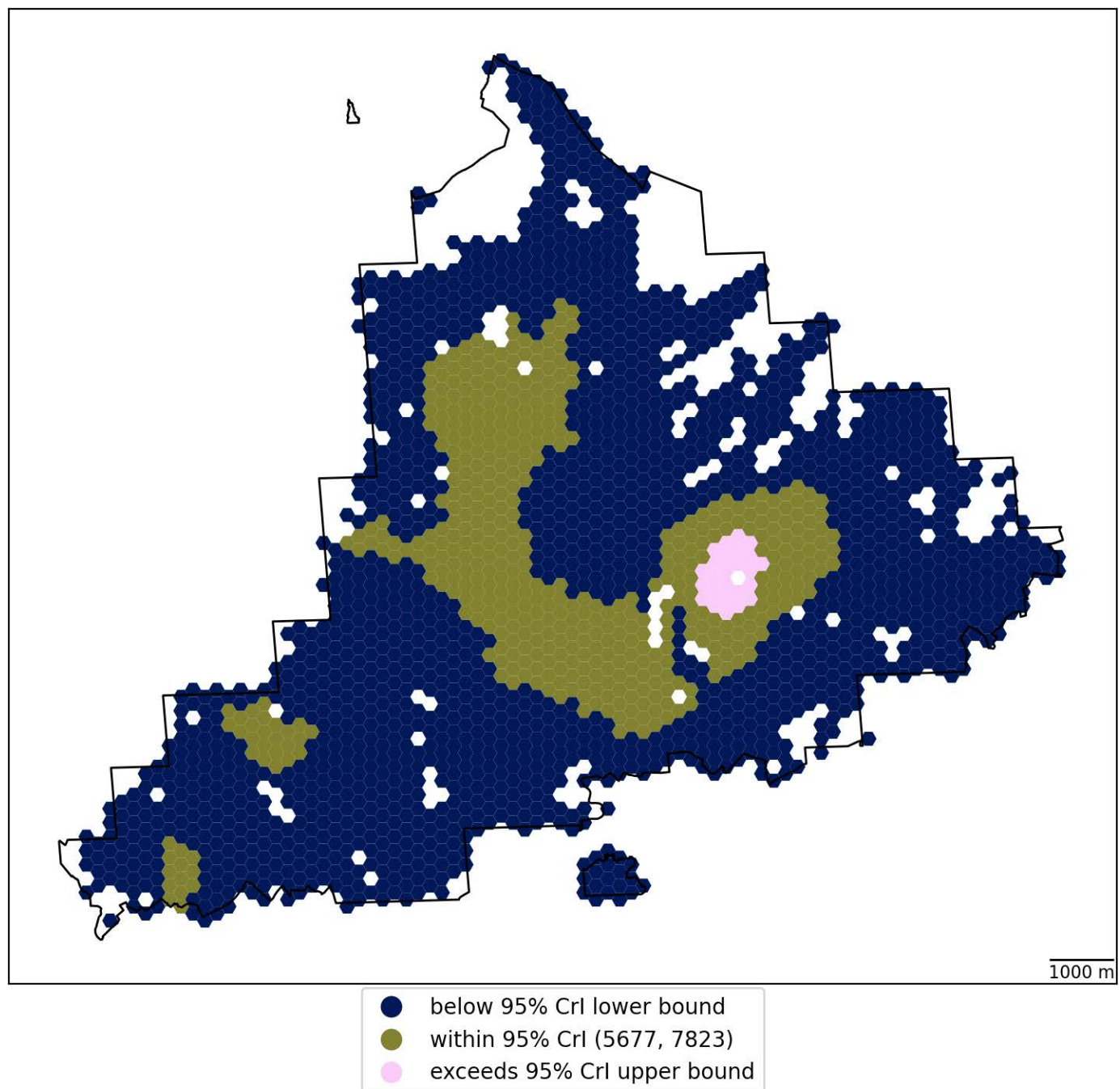

B: Estimated Mean 1000 m neighbourhood population per km² requirement for reaching the WHO's target of a ≥15% relative reduction in insufficient physical activity through walking

- below 95% CrI lower bound
- within 95% CrI (5677, 7823)
- exceeds 95% CrI upper bound

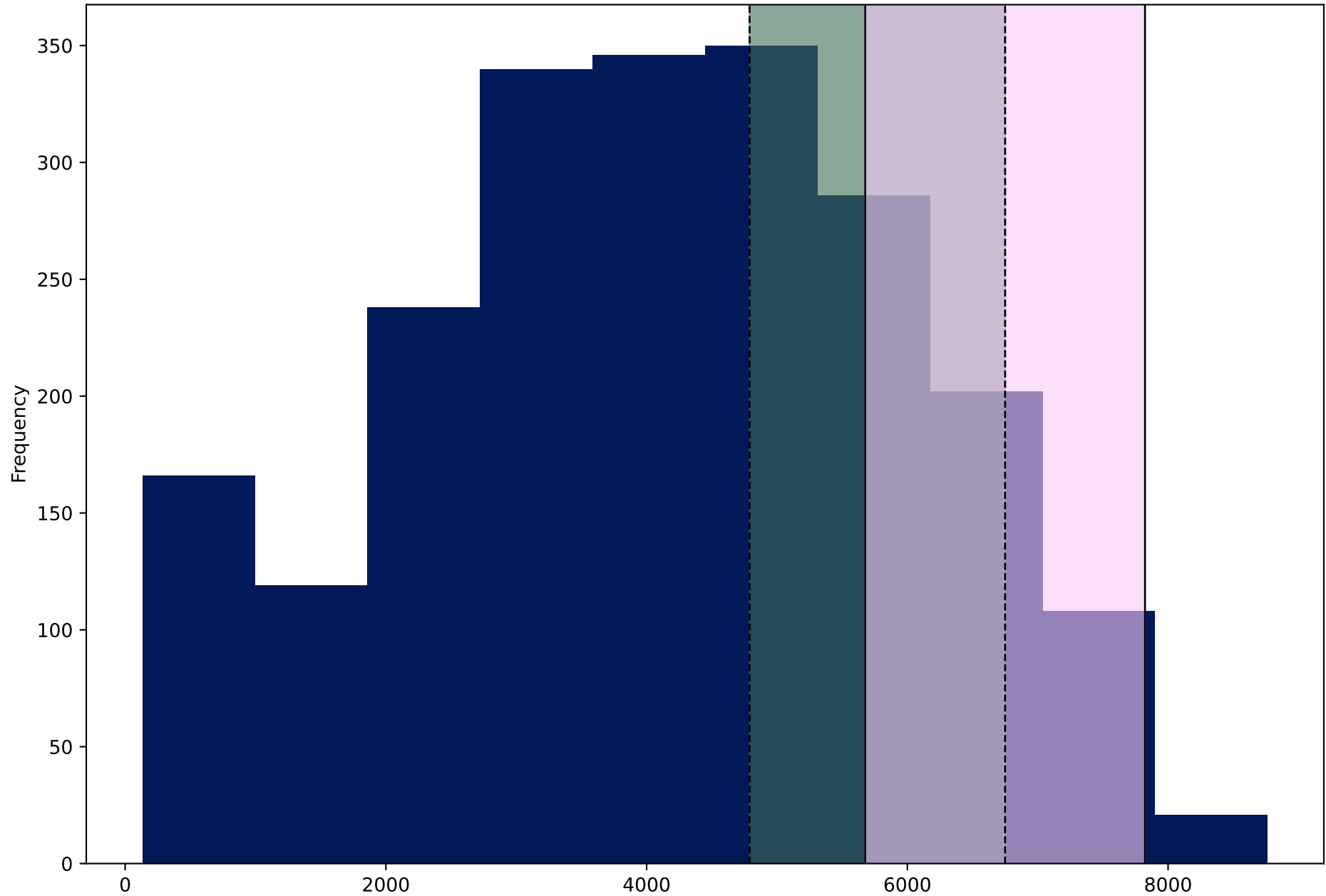

Mean 1000 m neighbourhood street intersections per km²

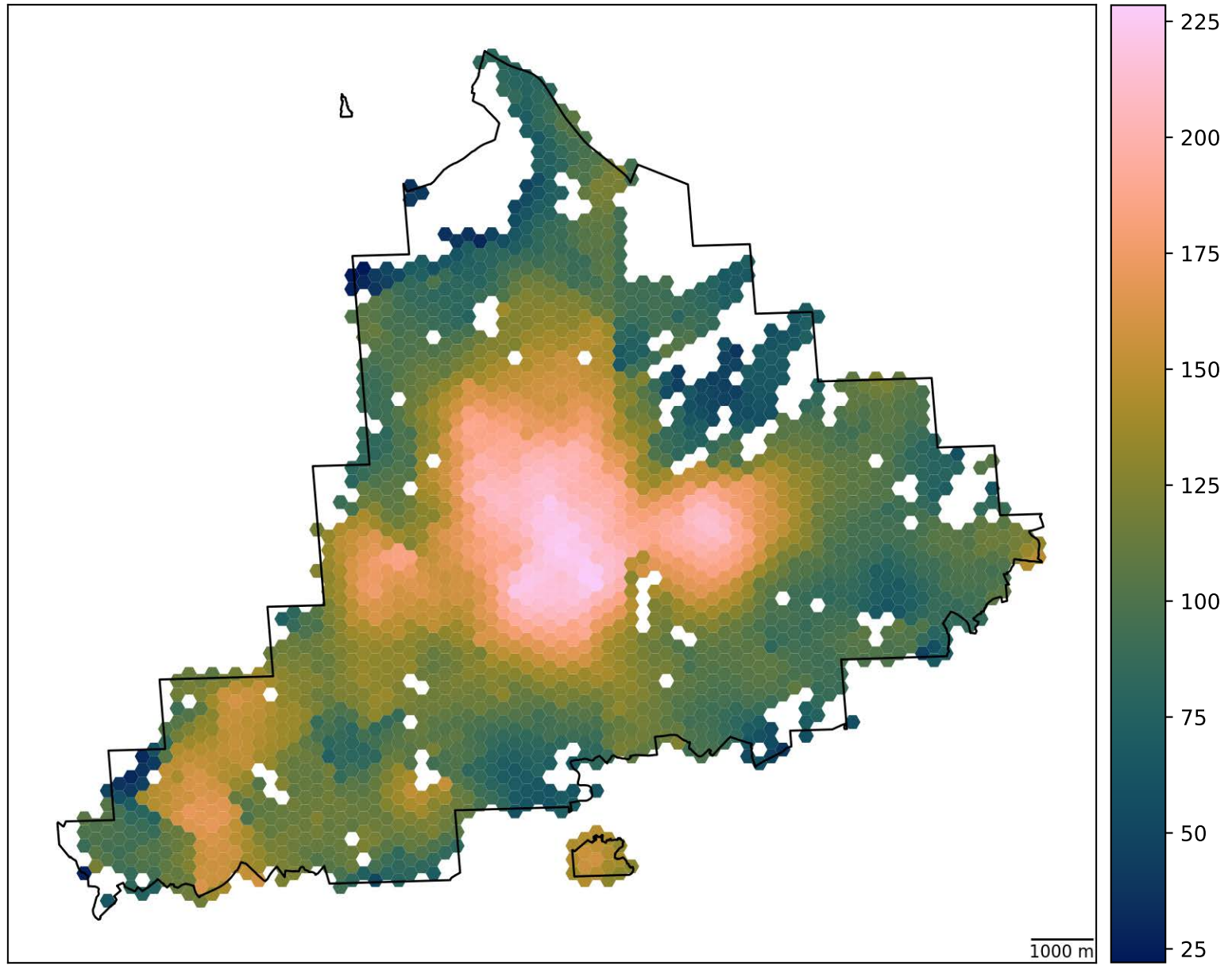

A: Estimated Mean 1000 m neighbourhood street intersections per km² requirement for ≥80% probability of engaging in walking for transport

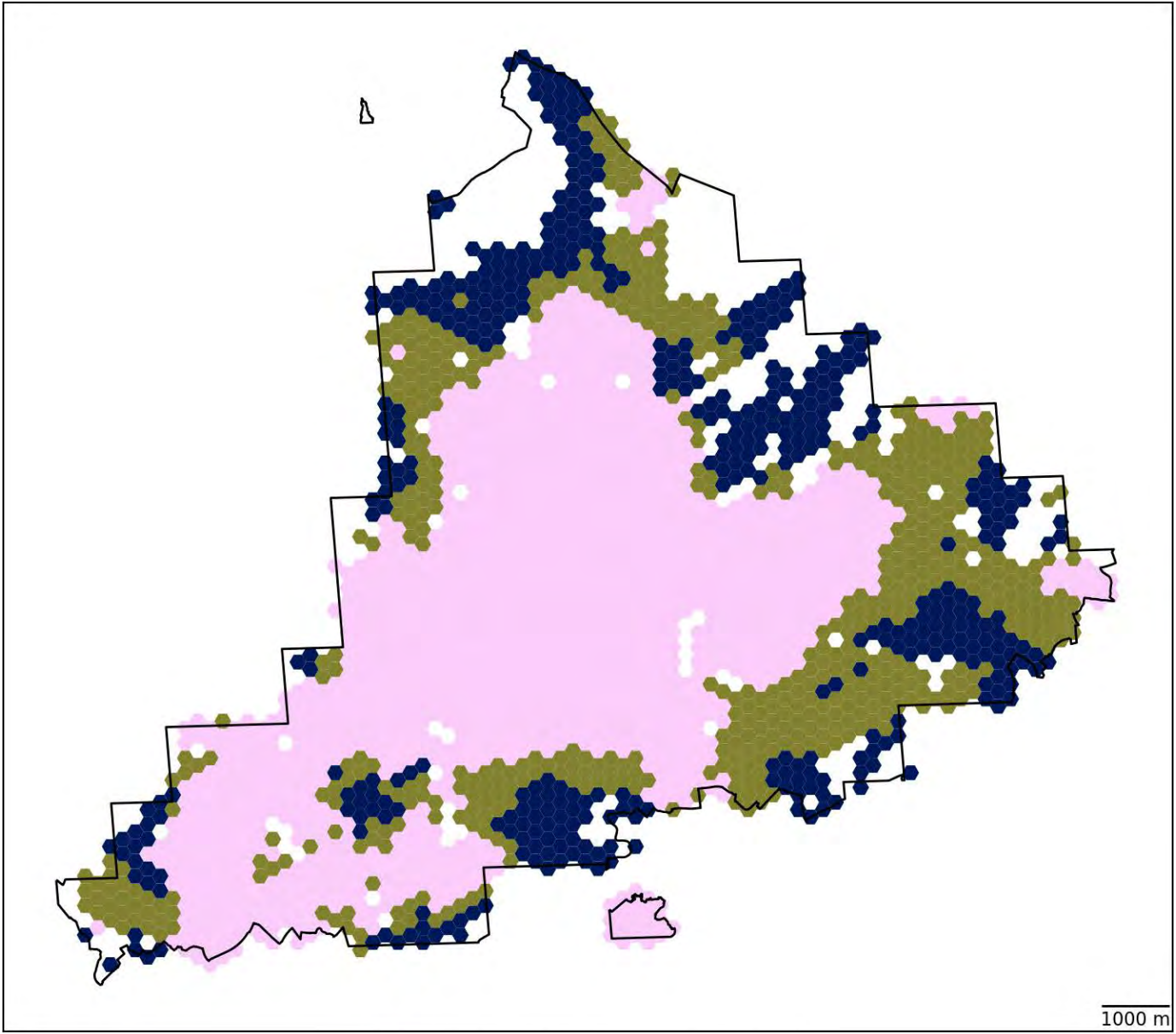

- below 95% CrI lower bound
- within 95% CrI (90, 110)
- exceeds 95% CrI upper bound

B: Estimated Mean 1000 m neighbourhood street intersections per km² requirement for reaching the WHO's target of a ≥15% relative reduction in insufficient physical activity through walking

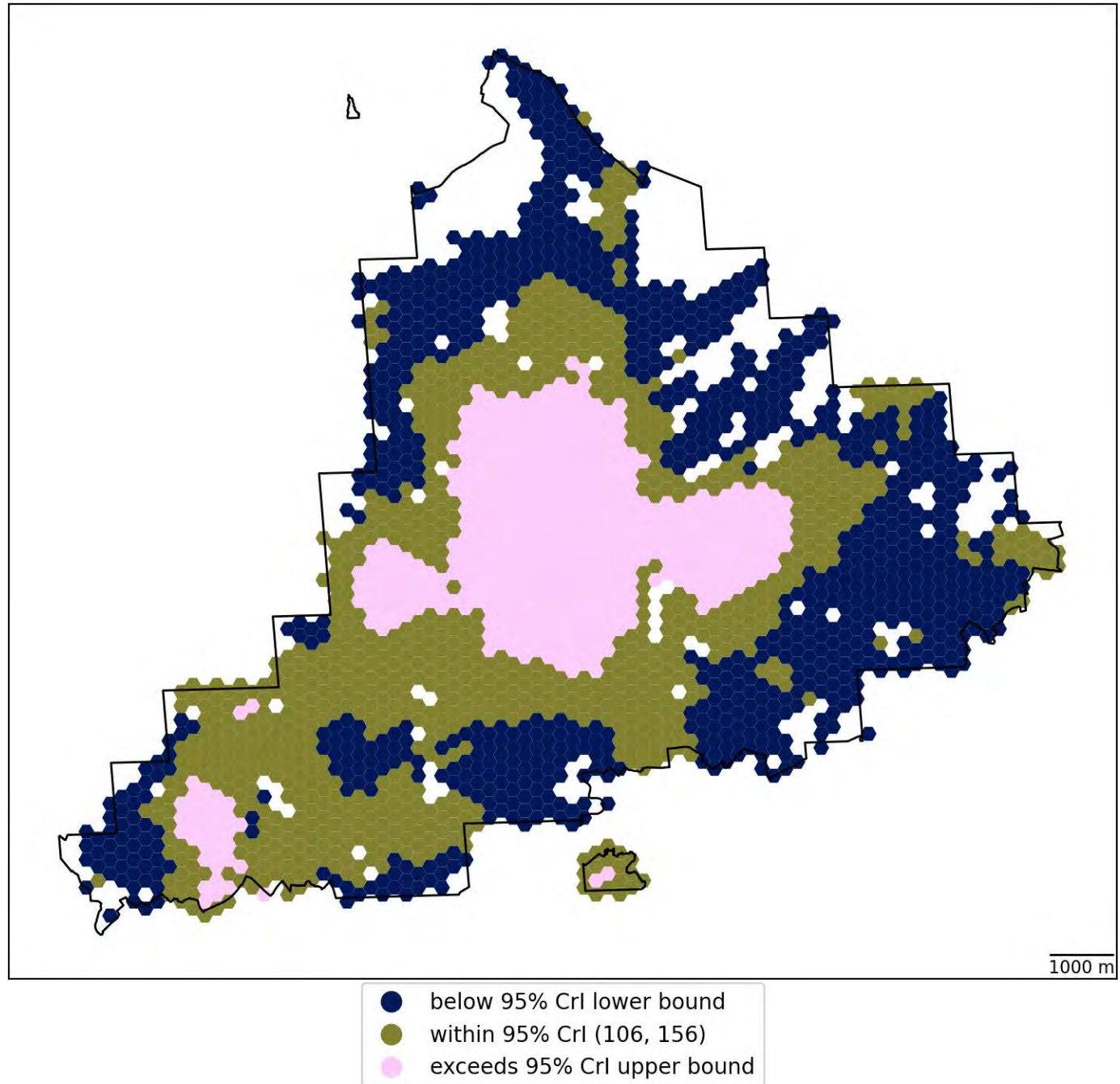

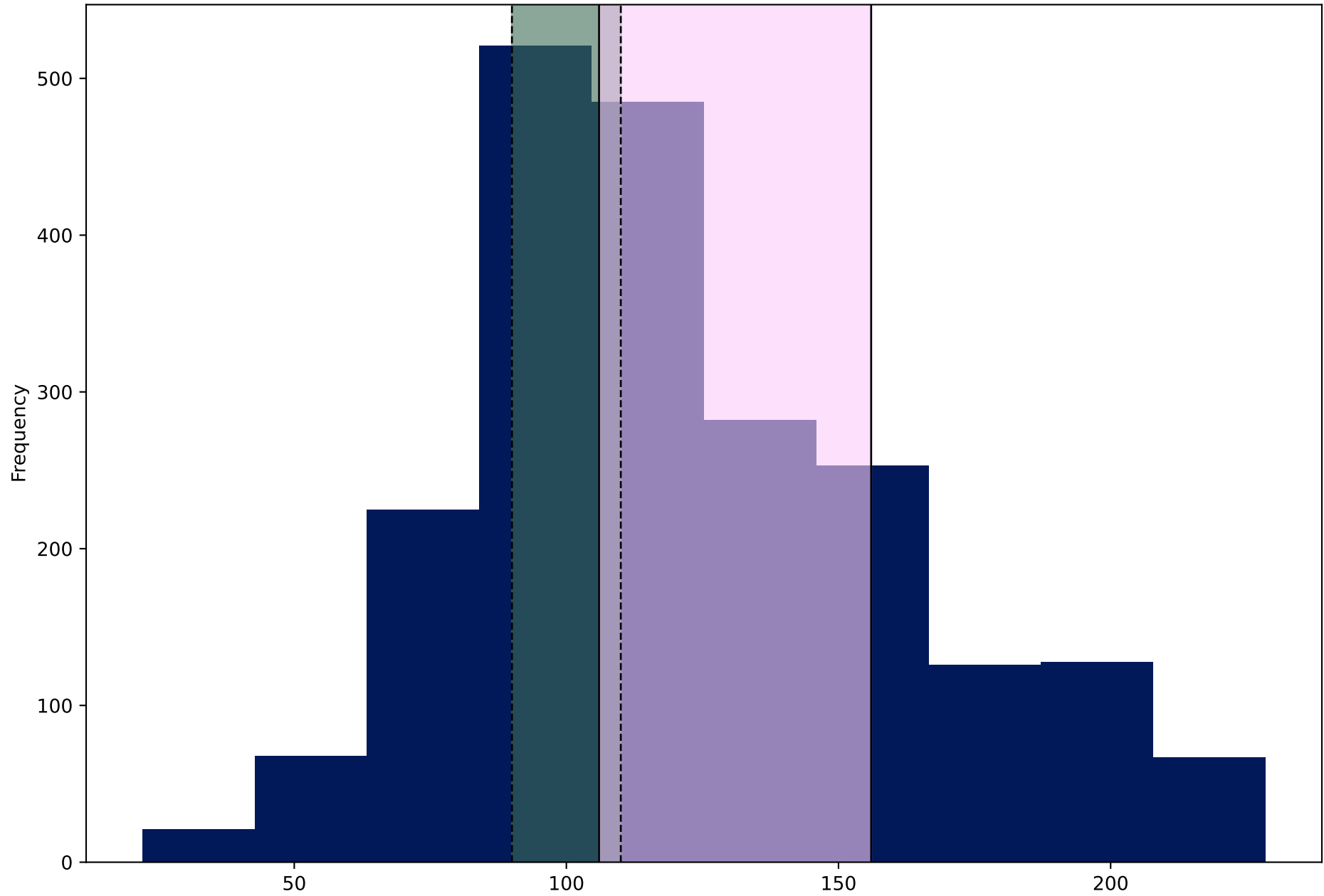

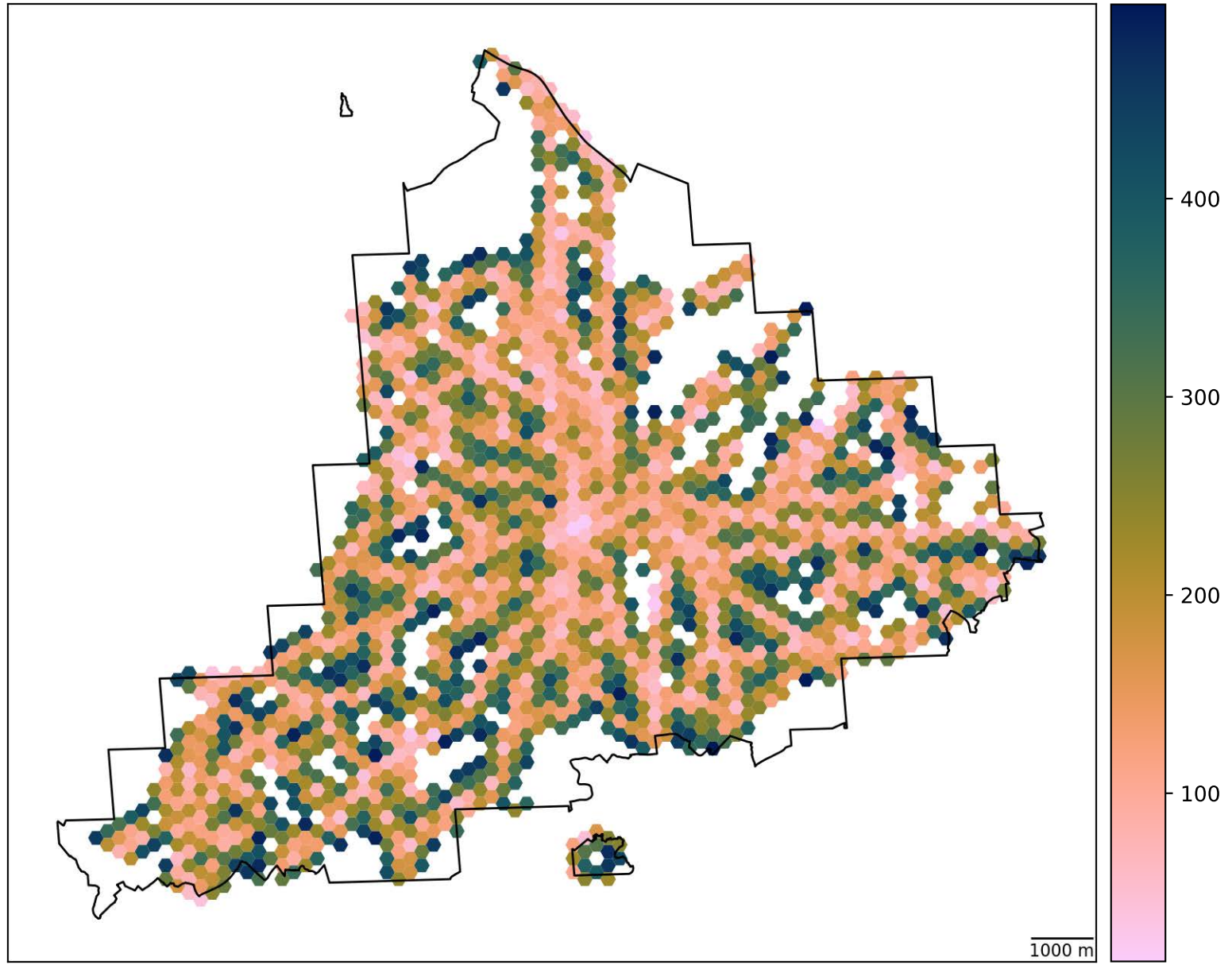

distances: Estimated Distance to nearest public transport stops (m; up to 500m) requirement for distances to destinations, measured up to a maximum distance target threshold of 500 metres

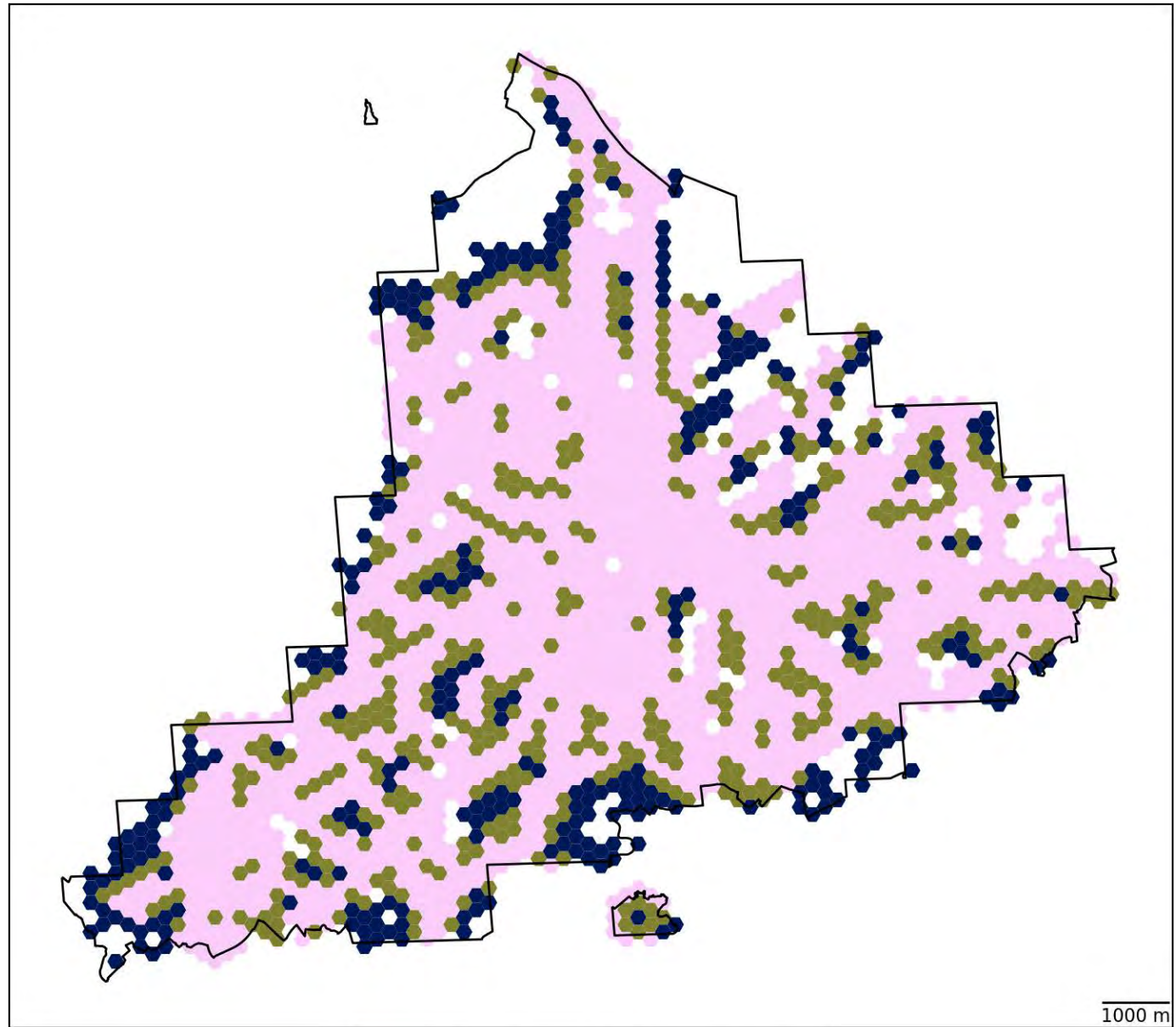

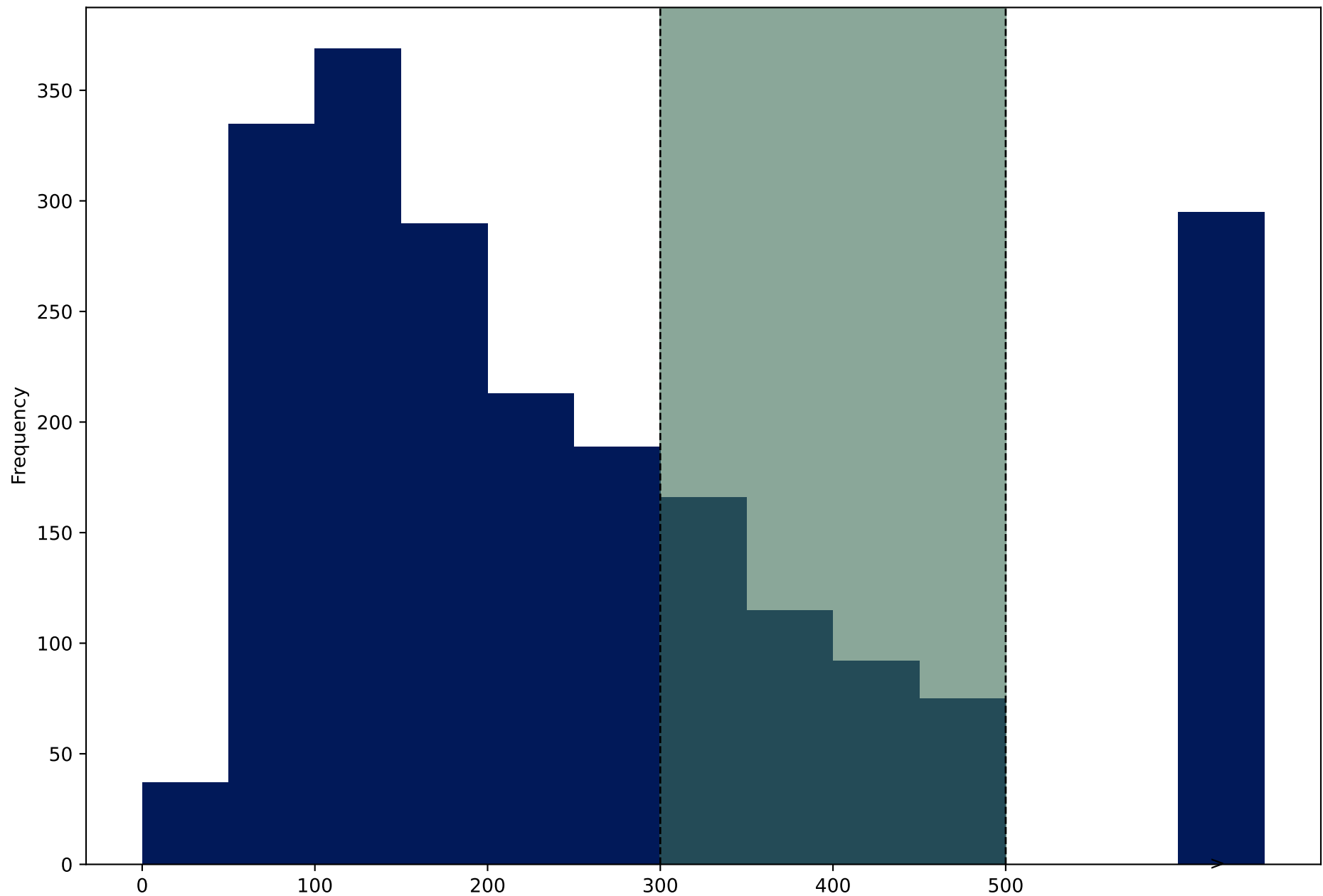

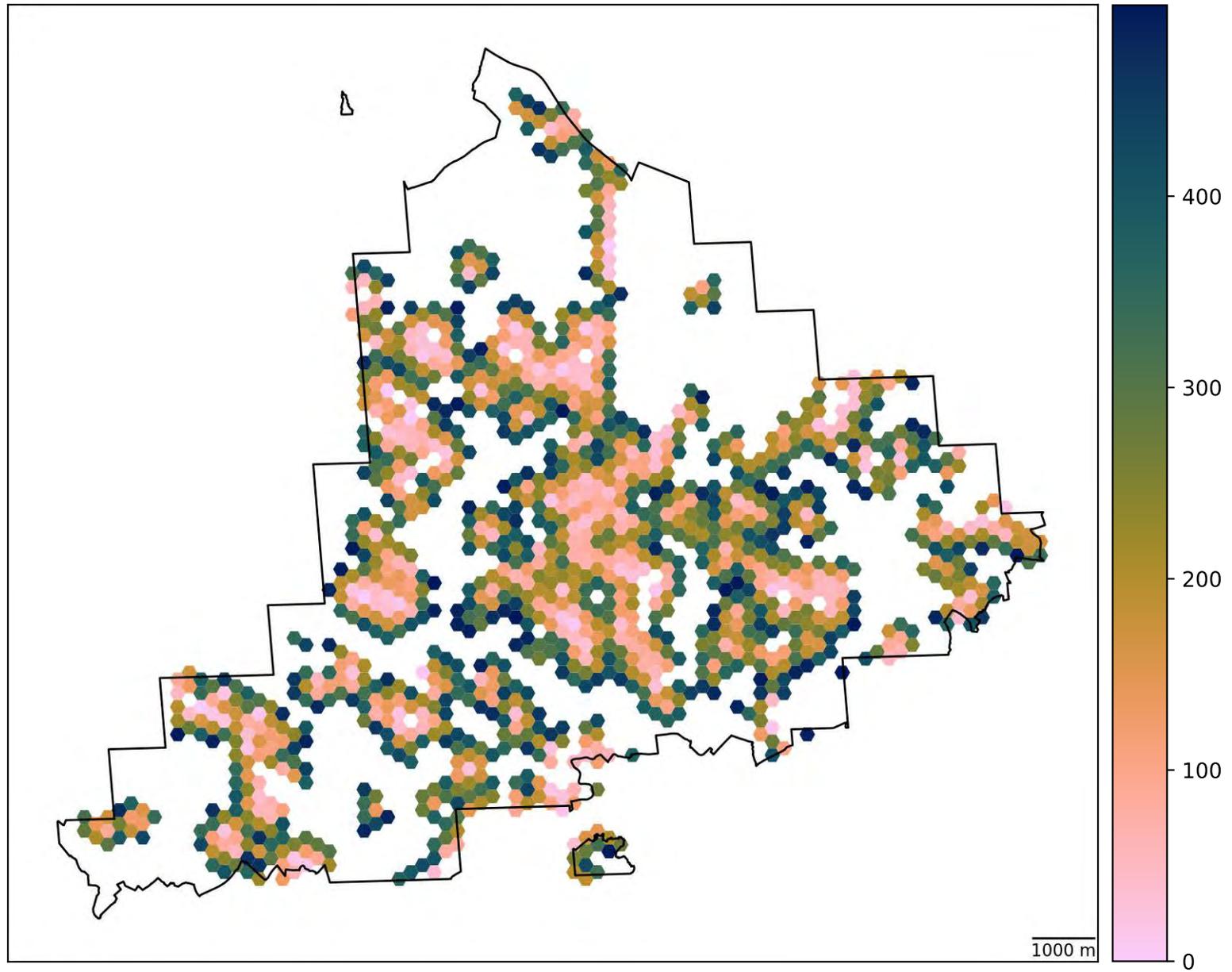

distances: Estimated Distance to nearest park (m; up to 500m) requirement for distances to destinations, measured up to a maximum distance target threshold of 500 metres

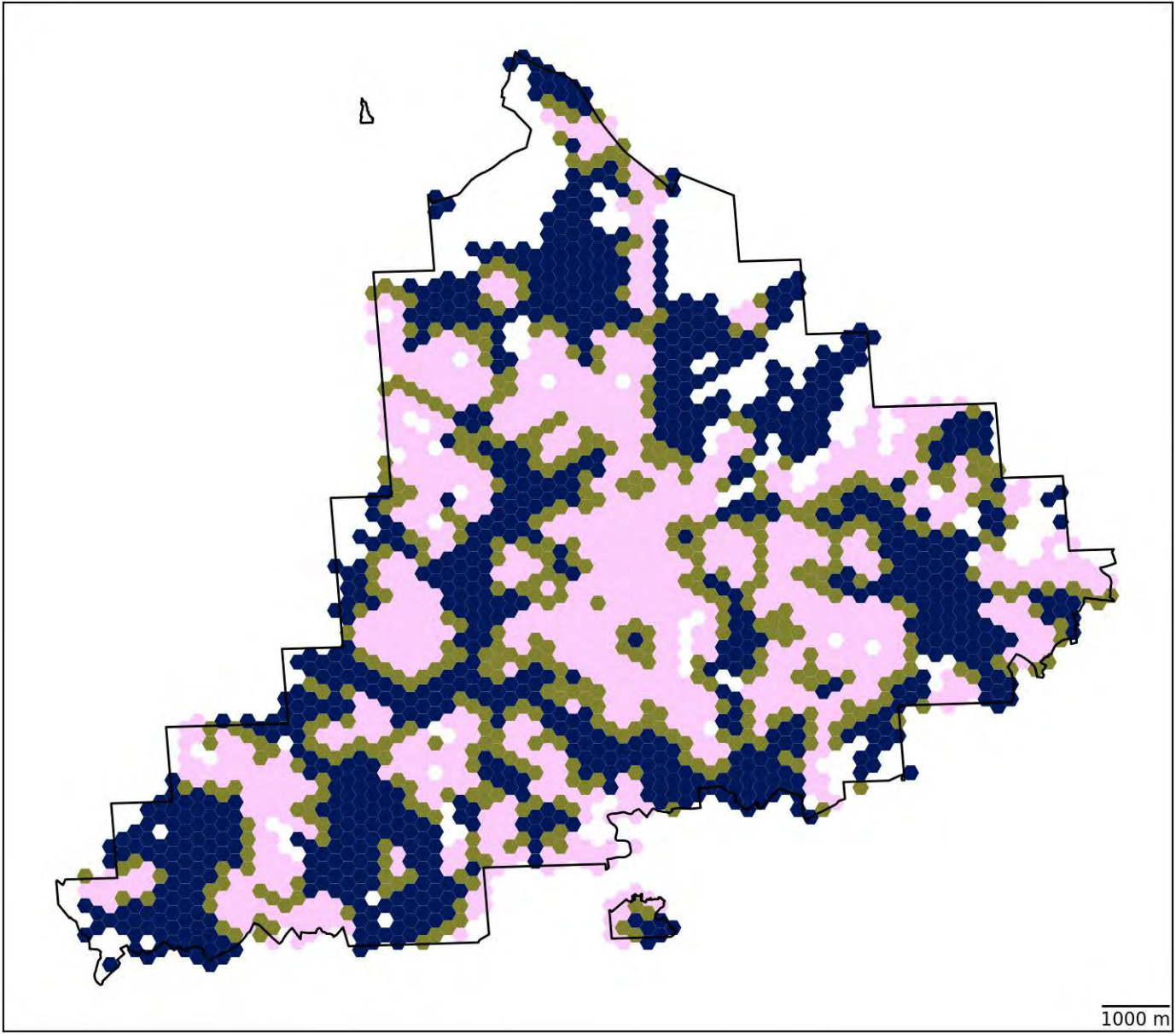

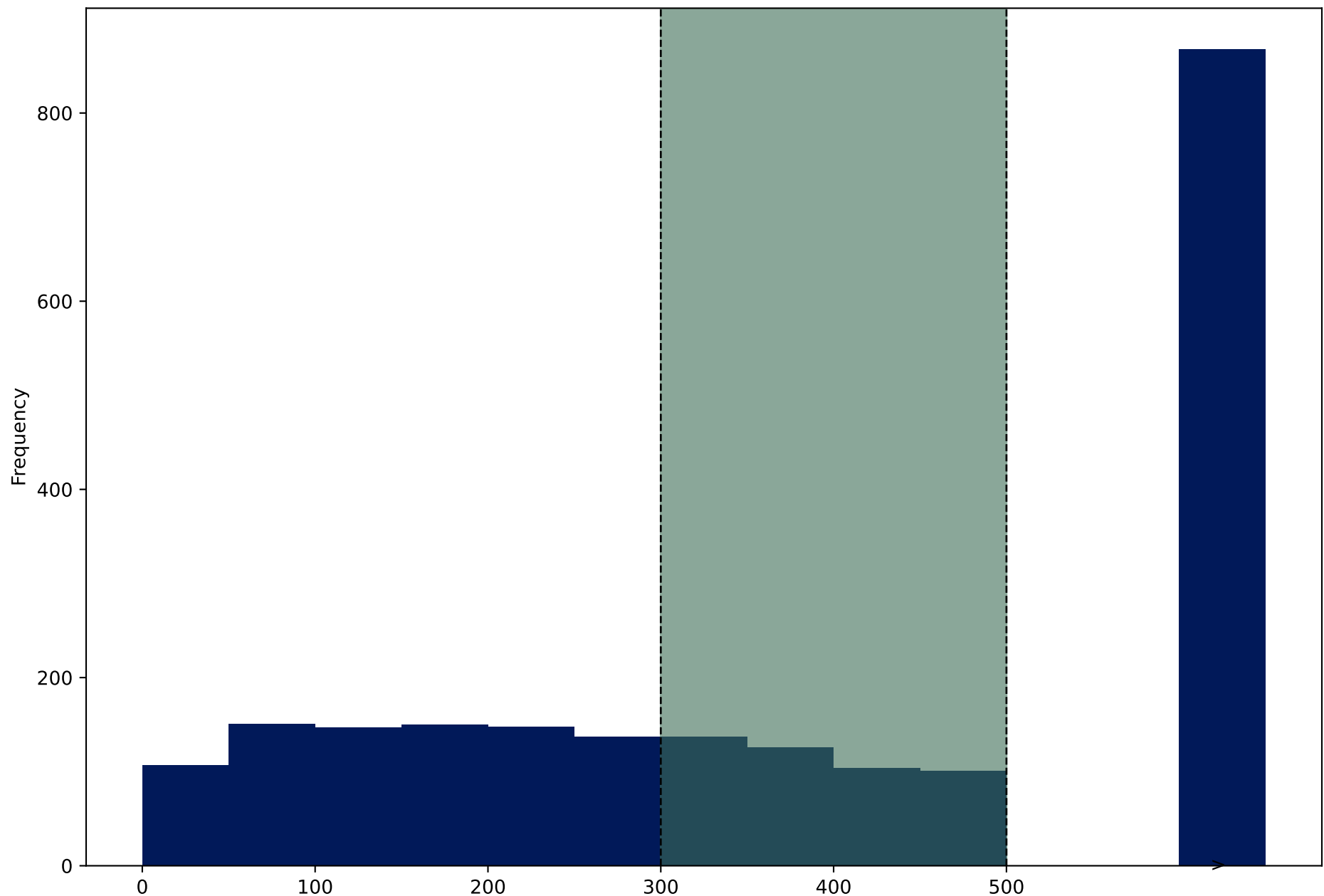